\documentclass[prx,aps,amssymb,twocolumn,superscriptaddress]{revtex4-1}
\usepackage{amsmath}
\usepackage{amssymb}
\usepackage{amsthm}
\usepackage{amsfonts}
\usepackage{listings}
\usepackage{longtable}
\usepackage{enumerate}
\usepackage{latexsym}
\usepackage{xcolor}
\usepackage{setspace} 
\usepackage{blindtext}
\usepackage{afterpage}

\extrafloats{52}

\usepackage{array,etoolbox}
\usepackage[titles]{tocloft}
\preto\tabular{\setcounter{magicrownumbers}{0}}
\newcounter{magicrownumbers}

\usepackage[french,english]{babel}

\usepackage{bm}
\usepackage{hyperref}
\hypersetup{
 pdfnewwindow=true, colorlinks=true,
 linkcolor=blue, anchorcolor=blue,
 citecolor=blue, filecolor=blue,
 menucolor=blue, urlcolor=blue}

\usepackage{psfrag}

\usepackage{bm}
\usepackage{graphicx}

\RequirePackage[normalem]{ulem} 
\RequirePackage{color}\definecolor{RED}{rgb}{1,0,0}\definecolor{BLUE}{rgb}{0,0,1} 

\newcommand{\braket}[2]{\left\langle #1 | #2 \right\rangle}
\newcommand{\bra}[1]{\left\langle#1\right|}
\newcommand{\ket}[1]{\left|#1\right\rangle}

\input{epsf}

\begin{document}


\newcommand{\icsdwebshort}[1]{\href{https://www.topologicalquantumchemistry.com/\#/detail/#1}{#1}}
\newcommand{\icsdweb}[1]{\href{https://www.topologicalquantumchemistry.com/\#/detail/#1}{ICSD #1}}
\newcommand{\webNoICSD}{\url{https://www.topologicalquantumchemistry.com/}}
\newcommand{\webTQC}{\href{https://www.topologicalquantumchemistry.com/}{Topological Materials Database}}

\newcommand{\webBCSfull}{\href{https://www.cryst.ehu.es/}{Bilbao Crystallographic Server}}

\newcommand{\webBCSshort}{\href{https://www.cryst.ehu.es/}{BCS}}

\newcommand{\webchecktopmat}{\href{https://www.cryst.ehu.es/cryst/checktopologicalmat}{Check Topological Mat}}
\newcommand{\identify}{\href{www.cryst.ehu.es/cryst/identify_group}{IDENTIFY GROUP}}

\newcommand{\vasptotrace}{\href{https://github.com/zjwang11/irvsp}{VASP2Trace}}
\newcommand{\checktopmat}{\href{https://www.cryst.ehu.es/cryst/checktopologicalmat}{Check Topological Mat}}

\newcommand{\TQCDTotICSDs}{193,426}
\newcommand{\TQCDTotICSDsExp}{181,218}
\newcommand{\TQCDTotICSDsTheo}{12,208}
\newcommand{\ThresholdNbrAtoms}{60}
\newcommand{\TQCDstoichiometric}{96,196}
\newcommand{\TQCDstoichiometricPercent}{49.73\%}
\newcommand{\TQCDstoichiometricExp}{85,701}
\newcommand{\TQCDstoichiometricExpPercent}{89.09\%}
\newcommand{\TQCDstoichiometricTheo}{10,495}
\newcommand{\TQCDstoichiometricTheoPercent}{10.91\%}
\newcommand{\TQCDNostoichiometric}{97,230}
\newcommand{\TQCDNostoichiometricPercent}{50.27\%}
\newcommand{\TQCDNostoichiometricExp}{95,517}
\newcommand{\TQCDNostoichiometricExpPercent}{98.24\%}
\newcommand{\TQCDNostoichiometricTheo}{1,713}
\newcommand{\TQCDNostoichiometricTheoPercent}{1.76\%}
\newcommand{\TQCDstoichiometricLEQatoms}{85,361}
\newcommand{\TQCDstoichiometricLEQatomsPercent}{88.74\%}
\newcommand{\TQCDstoichiometricExpLEQatoms}{75,375}
\newcommand{\TQCDstoichiometricExpLEQatomsPercent}{87.95\%}
\newcommand{\TQCDstoichiometricTheoLEQatoms}{9,986}
\newcommand{\TQCDstoichiometricTheoLEQatomsPercent}{95.15\%}
\newcommand{\TQCDstoichiometricGTatoms}{10,835}
\newcommand{\TQCDstoichiometricGTatomsPercent}{11.26\%}
\newcommand{\TQCDstoichiometricExpGTatoms}{10,326}
\newcommand{\TQCDstoichiometricExpGTatomsPercent}{12.05\%}
\newcommand{\TQCDstoichiometricTheoGTatoms}{509}
\newcommand{\TQCDstoichiometricTheoGTatomsPercent}{4.85\%}

\newcommand{\TQCDBHighConnectivityNbrBandsThreshold}{130}
\newcommand{\TQCDBNbrHighConnectivityMaterials}{244}
\newcommand{\TQCDBNbrNoSOCHighConnectivityMaterials}{972}
\newcommand{\TQCDBNbrMaterialsBndLowerEf}{508}
\newcommand{\TQCDBNbrMaterialsBndLowerEfICSDs}{1,047}
\newcommand{\TQCDBNbrNoSOCMaterialsBndLowerEf}{10,313}
\newcommand{\TQCDBNbrNoSOCMaterialsBndLowerEfICSDs}{20,899}

\newcommand{\vaspcputimelabel}[1]{\begin{tabular}{c}VASP#1\\(CPU hours)\end{tabular}}

\newcommand{\supappref}[1]{\ref{#1}}

\newcommand{\webmaterialsproject}{\href{https://materialsproject.org}{Materials Project}}
\newcommand{\TQCDBNbrICSDs}{73,234}
\newcommand{\TQCDBNbrNoSOCICSDs}{69,730}
\newcommand{\TQCDBPercentNoSOCICSDs}{95.22\%}
\newcommand{\TQCDBNbrFailedNoSOCICSDs}{3,504}
\newcommand{\TQCDBPercentFailedNoSOCICSDs}{4.78\%}
\newcommand{\TQCDBNbrUniqueMaterials}{38,298}
\newcommand{\TQCDBNbrNoSOCUniqueMaterials}{36,163}
\newcommand{\TQCDBPercentageNoSOCUniqueMaterials}{94.43\%}
\newcommand{\TQCDBNbrFailedNoSOCUniqueMaterials}{2,135}
\newcommand{\TQCDBPercentageFailedNoSOCUniqueMaterials}{5.57\%}

\newcommand{\TQCDBNbrMaterialsFElectrons}{10,987}
\newcommand{\TQCDBNbrMaterialsFElectronsPercent}{28.69\%}
\newcommand{\TQCDBNbrNoSOCICSDsNoFElectrons}{52,517}
\newcommand{\TQCDBNbrNoSOCICSDsNoFElectronsPercent}{75.31\%}
\newcommand{\TQCDBNbrMaterialsMagneticMP}{7,124}
\newcommand{\TQCDBNbrMaterialsMagneticMPPercent}{18.60\%}
\newcommand{\TQCDBNbrMaterialsMagneticMPVASP}{13,718}
\newcommand{\TQCDBNbrMaterialsMagneticMPVASPPercent}{35.82\%}
\newcommand{\TQCDBNbrMaterialsMagneticMPVASPFElectrons}{19,987}
\newcommand{\TQCDBNbrMaterialsMagneticMPVASPFElectronsPercent}{52.19\%}

\newcommand{\TQCDBNbrMaterialsTI}{6,128}
\newcommand{\TQCDBNbrMaterialsTIPercent}{16.00\%}
\newcommand{\TQCDBNbrMaterialsSM}{14,037}
\newcommand{\TQCDBNbrMaterialsSMPercent}{36.65\%}
\newcommand{\TQCDBNbrMaterialstrivial}{18,133}
\newcommand{\TQCDBNbrMaterialstrivialPercent}{47.35\%}

\newcommand{\TQCDBNbrMaterialsNLC}{3,000}
\newcommand{\TQCDBNbrMaterialsNLCPercent}{7.83\%}
\newcommand{\TQCDBNbrMaterialsSEBR}{3,128}
\newcommand{\TQCDBNbrMaterialsSEBRPercent}{8.17\%}
\newcommand{\TQCDBNbrMaterialsES}{4,102}
\newcommand{\TQCDBNbrMaterialsESPercent}{10.71\%}
\newcommand{\TQCDBNbrMaterialsESFD}{9,935}
\newcommand{\TQCDBNbrMaterialsESFDPercent}{25.94\%}
\newcommand{\TQCDBNbrMaterialsLCEBR}{18,133}
\newcommand{\TQCDBNbrMaterialsLCEBRPercent}{47.35\%}
\newcommand{\TQCDBNbrTopologicalMaterials}{20,165}
\newcommand{\TQCDBNbrTopologicalMaterialsPercent}{52.65\%}

\newcommand{\TQCDBNbrNoSOCMaterialsSM}{20,298}
\newcommand{\TQCDBNbrNoSOCMaterialsSMPercent}{56.13\%}
\newcommand{\TQCDBNbrNoSOCMaterialstrivial}{15,865}
\newcommand{\TQCDBNbrNoSOCMaterialstrivialPercent}{43.87\%}
\newcommand{\TQCDBNbrNoSOCMaterialsES}{6,006}
\newcommand{\TQCDBNbrNoSOCMaterialsESPercent}{16.61\%}
\newcommand{\TQCDBNbrNoSOCMaterialsESFD}{13,997}
\newcommand{\TQCDBNbrNoSOCMaterialsESFDPercent}{38.71\%}
\newcommand{\TQCDBNbrNoSOCMaterialsLCEBR}{15,865}
\newcommand{\TQCDBNbrNoSOCMaterialsLCEBRPercent}{43.87\%}
\newcommand{\TQCDBNbrNoSOCMaterialsNLCSM}{251}
\newcommand{\TQCDBNbrNoSOCMaterialsNLCSMPercent}{0.69\%}
\newcommand{\TQCDBNbrNoSOCMaterialsSEBRSM}{44}
\newcommand{\TQCDBNbrNoSOCMaterialsSEBRSMPercent}{0.12\%}
\newcommand{\TQCDBNbrNoSOCMaterialsNLCSMES}{6,257}
\newcommand{\TQCDBNbrNoSOCMaterialsSEBRSMESFD}{14,041}
\newcommand{\TQCDBNbrNoSOCMaterialsNLCSMSEBRSM}{295}

\newcommand{\TQCDBNbrMaterialsWithSOCNoSOCTI}{5,382}
\newcommand{\TQCDBNbrMaterialsWithSOCNoSOCTIPercent}{14.88\%}
\newcommand{\TQCDBNbrMaterialsWithSOCNoSOCSM}{13,270}
\newcommand{\TQCDBNbrMaterialsWithSOCNoSOCSMPercent}{36.69\%}
\newcommand{\TQCDBNbrMaterialsWithSOCNoSOCtrivial}{17,511}
\newcommand{\TQCDBNbrMaterialsWithSOCNoSOCtrivialPercent}{48.42\%}

\newcommand{\TQCDBNbrMaterialsWithSOCNoSOCNLC}{2,568}
\newcommand{\TQCDBNbrMaterialsWithSOCNoSOCNLCPercent}{7.10\%}
\newcommand{\TQCDBNbrMaterialsWithSOCNoSOCSEBR}{2,814}
\newcommand{\TQCDBNbrMaterialsWithSOCNoSOCSEBRPercent}{7.78\%}
\newcommand{\TQCDBNbrMaterialsWithSOCNoSOCES}{3,785}
\newcommand{\TQCDBNbrMaterialsWithSOCNoSOCESPercent}{10.47\%}
\newcommand{\TQCDBNbrMaterialsWithSOCNoSOCESFD}{9,485}
\newcommand{\TQCDBNbrMaterialsWithSOCNoSOCESFDPercent}{26.23\%}
\newcommand{\TQCDBNbrMaterialsWithSOCNoSOCLCEBR}{17,511}
\newcommand{\TQCDBNbrMaterialsWithSOCNoSOCLCEBRPercent}{48.42\%}
\newcommand{\TQCDBNbrMaterialsWithSOCNoSOCNLCSEBR}{5,382}
\newcommand{\TQCDBNbrMaterialsWithSOCNoSOCSEBRESFD}{12,299}
\newcommand{\TQCDBNbrMaterialsWithSOCNoSOCNLCES}{6,353}

\newcommand{\TQCDBNbrSTopo}{769}
\newcommand{\TQCDBNbrSTopoPercentage}{2.01\%}
\newcommand{\TQCDBNbrSTopoNLC}{37}
\newcommand{\TQCDBNbrSTopoNLCPercent}{0.10\%}
\newcommand{\TQCDBNbrSTopoSEBR}{109}
\newcommand{\TQCDBNbrSTopoSEBRPercent}{0.28\%}
\newcommand{\TQCDBNbrSTopoES}{178}
\newcommand{\TQCDBNbrSTopoESPercent}{0.46\%}
\newcommand{\TQCDBNbrSTopoESFD}{339}
\newcommand{\TQCDBNbrSTopoESFDPercent}{0.89\%}
\newcommand{\TQCDBNbrSTopoLCEBR}{106}
\newcommand{\TQCDBNbrSTopoLCEBRPercent}{0.28\%}

\newcommand{\TQCDBNbrTopoBand}{33,698}
\newcommand{\TQCDBNbrTopoBandPercent}{87.99\%}

\newcommand{\TQCDBNbrSMetal}{17}
\newcommand{\TQCDBNbrSMetalPercent}{0.04\%}
\newcommand{\TQCDBNbrSMetalICSDs}{65}
\newcommand{\TQCDBNbrSMetalPercentICSDs}{0.09\%}

\newcommand{\TQCDBBandSetsUniqueMaterials}{1,996,728}
\newcommand{\TQCDBNbrLCEBRBandSetsUniqueMaterials}{750,504}
\newcommand{\TQCDBPercentLCEBRBandSetsUniqueMaterials}{37.59\%}
\newcommand{\TQCDBNbrNLCBandSetsUniqueMaterials}{859,606}
\newcommand{\TQCDBPercentNLCBandSetsUniqueMaterials}{43.05\%}
\newcommand{\TQCDBNbrSEBRBandSetsUniqueMaterials}{379,321}
\newcommand{\TQCDBPercentSEBRBandSetsUniqueMaterials}{19.00\%}
\newcommand{\TQCDBNbrStrongBandSetsUniqueMaterials}{1,238,927}
\newcommand{\TQCDBPercentStrongBandSetsUniqueMaterials}{62.05\%}
\newcommand{\TQCDBNbrFragileBandSetsUniqueMaterials}{7,297}
\newcommand{\TQCDBPercentFragileBandSetsUniqueMaterials}{0.37\%}

\newcommand{\TQCDBNbrNoSOCSTopo}{28}
\newcommand{\TQCDBNbrNoSOCSTopoPercentage}{0.08\%}
\newcommand{\TQCDBNbrNoSOCSTopoNLC}{0}
\newcommand{\TQCDBNbrNoSOCSTopoNLCPercent}{0.00\%}
\newcommand{\TQCDBNbrNoSOCSTopoSEBR}{0}
\newcommand{\TQCDBNbrNoSOCSTopoSEBRPercent}{0.00\%}
\newcommand{\TQCDBNbrNoSOCSTopoES}{5}
\newcommand{\TQCDBNbrNoSOCSTopoESPercent}{0.01\%}
\newcommand{\TQCDBNbrNoSOCSTopoESFD}{14}
\newcommand{\TQCDBNbrNoSOCSTopoESFDPercent}{0.04\%}
\newcommand{\TQCDBNbrNoSOCSTopoLCEBR}{8}
\newcommand{\TQCDBNbrNoSOCSTopoLCEBRPercent}{0.02\%}

\newcommand{\TQCDBNbrNoSOCTopoBand}{10,001}
\newcommand{\TQCDBNbrNoSOCTopoBandPercent}{27.66\%}

\newcommand{\TQCDBNbrNoSOCSMetal}{1,138}
\newcommand{\TQCDBNbrNoSOCSMetalPercent}{3.15\%}
\newcommand{\TQCDBNbrNoSOCSMetalICSDs}{3,495}
\newcommand{\TQCDBNbrNoSOCSMetalPercentICSDs}{5.01\%}

\newcommand{\TQCDBVASPICPUhSOC}{9,500,801.70}
\newcommand{\TQCDBVASPIICPUhSOC}{170,633.80}
\newcommand{\TQCDBVASPIIICPUhSOC}{2,560,882.60}
\newcommand{\TQCDBVASPIVCPUhSOC}{5,804,948.80}
\newcommand{\TQCDBVASPTotalCPUhSOC}{18,037,266.90}

\newcommand{\TQCDBVASPICPUhNoSOC}{2,298,583.30}
\newcommand{\TQCDBVASPIICPUhNoSOC}{56,520.70}
\newcommand{\TQCDBVASPIIICPUhNoSOC}{773,019.50}
\newcommand{\TQCDBVASPIVCPUhNoSOC}{1,397,550.30}
\newcommand{\TQCDBVASPTotalCPUhNoSOC}{4,525,673.80}

\newcommand{\TQCDBVASPTotalCPUTimeAllHours}{22,562,940.70}
\newcommand{\TQCDBVASPTotalCPUTimeAllMillionHours}{22.60}

\newcommand{\TQCDBTotalStorage}{2,037.80Gb}

\newcommand{\TQCDBNoSOCTotalStorage}{343.30Gb}
\newcommand{\TQCDBPercentNoSOCTotalStorage}{16.85\%}
\newcommand{\TQCDBNoSOCPROCARStorage}{173.60Gb}
\newcommand{\TQCDBNoSOCCHGCARStorage}{119.00Gb}

\newcommand{\TQCDBTotalSOCStorage}{1,694.50Gb}
\newcommand{\TQCDBPercentSOCTotalStorage}{83.15\%}
\newcommand{\TQCDBPROCARSOCStorage}{1,013.80Gb}
\newcommand{\TQCDBPercentPROCARSOCStorage}{49.75\%}
\newcommand{\TQCDBCHGCARSOCStorage}{559.30Gb}
\newcommand{\TQCDBPercentCHGCARSOCStorage}{27.45\%}

\newcommand{\TQCDBNbrSkippedICSDs}{22,996}
\newcommand{\TQCDBNbrFailedComputedICSDs}{11,315}
\newcommand{\TQCDBNbrFailedVASPToTraceICSDs}{5,014}
\newcommand{\TQCDBNbrFailedVASPToTraceBadTRICSDs}{1,703}
\newcommand{\TQCDBNbrFailedVASPToTraceBadTracesICSDs}{655}
\newcommand{\TQCDBNbrFailedVASPToTraceAccidentalFermiICSDs}{2,647}

\tolerance 10000

\newcommand{\vk}{{\bf k}}

\newcommand{\AppAppendixOverviewAppendix}{\supappref{App:AppendixOverview_appendix}}
\newcommand{\AppTQCReviewAppendix}{\supappref{App:TQCReview_appendix}}
\newcommand{\AppCheckTopoAppendix}{\supappref{App:Check_Topo_appendix}}
\newcommand{\AppVASPAppendix}{\supappref{App:VASP_appendix}}
\newcommand{\AppCPUTime}{\supappref{App:CPUtime}}
\newcommand{\AppTopologicalMaterialsNoSOC}{\supappref{App:TopologicalMaterialsNoSOC}}
\newcommand{\AppWebsite}{\supappref{App:Website}}
\newcommand{\AppTopoBands}{\supappref{App:TopoBands}}
\newcommand{\AppSupertopological}{\supappref{App:Supertopological}}
\newcommand{\AppPhaseTransitionsNoSOCSOC}{\supappref{App:PhaseTransitionsNoSOCSOC}}
\newcommand{\AppMaterialsSelection}{\supappref{App:MaterialsSelection}}

\newcommand{\AppVASPToTrace}{\supappref{App:VASP2Trace}}
\newcommand{\AppPhysicalMeaningnoSOCNLCSEBR}{\supappref{App:physicalMeaningnoSOCNLCSEBR}}
\newcommand{\AppDefRTopo}{\supappref{App:DefRTopo}}
\newcommand{\AppDefSTopo}{\supappref{App:DefSTopo}}
\newcommand{\AppStatForStopo}{\supappref{App:StatForStopo}}
\newcommand{\AppZFourTrivial}{\supappref{App:z4trivial}}
\newcommand{\AppRTopoMaterials}{\supappref{App:RTopoMaterials}}
\newcommand{\AppES}{\supappref{App:ES}}
\newcommand{\AppTransitions}{\supappref{App:transitions}}
\newcommand{\AppFragileBands}{\supappref{App:fragileBands}}

\newcommand{\TabBandCharacterizationBiTwoMgThree}{Table~\ref{tab:BandCharacterizationBiTwoMgThree}}

\title{All Topological Bands of All Nonmagnetic Stoichiometric Materials}

\author{Maia G. Vergniory$^\dag$} \thanks{These authors contributed equally to this work.}
\affiliation{Donostia International Physics Center, P. Manuel de Lardizabal 4, 20018 Donostia-San Sebastian, Spain}
\affiliation{IKERBASQUE, Basque Foundation for Science, Bilbao, Spain}
\affiliation{Max Planck Institute for Chemical Physics of Solids, 01187 Dresden, Germany}
\author{Benjamin J. Wieder$^\dag$} \thanks{These authors contributed equally to this work.}
\affiliation{Department of Physics, Massachusetts Institute of Technology, Cambridge, MA 02139, USA}
\affiliation{Department of Physics, Northeastern University, Boston, MA 02115, USA}
\affiliation{Department of Physics, Princeton University, Princeton, New Jersey 08544, USA}
\author{Luis Elcoro}
\affiliation{Department of Condensed Matter Physics, University of the Basque Country UPV/EHU, Apartado 644, 48080 Bilbao, Spain}
\author{Stuart S.~P. Parkin}
\affiliation{Max Planck Institute of Microstructure Physics, 06120 Halle, Germany}
\author{Claudia Felser}
\affiliation{Max Planck Institute for Chemical Physics of Solids, 01187 Dresden, Germany}
\author{B. Andrei Bernevig}
\affiliation{Department of Physics, Princeton University, Princeton, New Jersey 08544, USA}
\author{Nicolas Regnault}\thanks{To whom correspondence should be addressed: \href{mailto:regnault@princeton.edu}{regnault@princeton.edu} (N. R.), \href{mailto:maiagv@gmail.com}{maiagv@gmail.com} (M. G. V.), \href{mailto:bwieder@mit.edu}{bwieder@mit.edu} (B. J. W.)}
\affiliation{Department of Physics, Princeton University, Princeton, New Jersey 08544, USA}
\affiliation{Laboratoire de Physique de l’\'Ecole Normale Sup\'erieure, PSL University, CNRS, Sorbonne Universit\'e, Universit\'e Paris Diderot, Sorbonne Paris Cit\'e, Paris, France}

\date{\today}

\begin{abstract}
  Topological Quantum Chemistry and Symmetry-Based Indicators have facilitated large-scale searches for materials with topological properties at the Fermi energy ($E_{F}$). We report the completion of a publicly accessible catalog of stable and fragile topology in all of the bands both at and away from $E_{F}$ in the~\TQCDstoichiometric~processable entries in the Inorganic Crystal Structure Database.  Our calculations represent the completion of the symmetry-indicated band topology of known nonmagnetic materials, and enable the discovery of repeat-topological and supertopological materials, which include rhombohedral bismuth and Bi$_2$Mg$_3$. We find that~\TQCDBNbrTopologicalMaterialsPercent~of all materials are topological at $E_{F}$, roughly $2/3$ of bands across all materials exhibit symmetry-indicated stable topology, and that~\emph{\TQCDBNbrTopoBandPercent}~of all materials contain at least one stable or fragile topological band.
\end{abstract}

\maketitle

\addtocontents{toc}{\protect\setcounter{tocdepth}{0}}
\addtocontents{lot}{\protect\setcounter{lotdepth}{-1}}

The field of solid-state physics has over the past 15 years advanced in large part by the discovery of nontrivial electronic band topology in nonmagnetic crystalline solids~\cite{WiederReview,CharlieZahidReview}.  Since the conceptual proposal and theoretical prediction of the first 2D~\cite{KaneMeleZ2,AndreiTI} and 3D~\cite{FuKaneMele} topological insulators (TIs), researchers have predicted solid-state realizations of topological phases of matter at a rapid pace.  Notable examples include symmetry-protected topological semimetals (TSMs)~\cite{ZJDirac,DDP,NewFermions,CoSiArc,WeylReview} and topological crystalline insulators (TCIs)~\cite{LiangTCIOriginal,HsiehTCI,HourglassInsulator,DiracInsulator,WladPump,ChenTCI,AshvinTCI}. Owing to negligible electron-electron interactions, many of the theoretical predictions of TI, TCI, and TSM phases in solid-state materials were shortly afterwards confirmed in spectroscopic and transport experiments~\cite{HsiehBismuthSelenide,CavaDirac1,ZahidWeyl,AndreiWeyl,HOTIBismuth,CoSiObserveHasan,CoSiObserveChina,PdGaObserve,CDWWeyl,AxionCDWExperiment}.

Initially, solid-state TIs and TSMs were believed to be rare and esoteric phases of matter.  However over the past four years, the theories of Topological Quantum Chemistry (TQC)~\cite{QuantumChemistry,MTQC} and symmetry-based indicators (SIs)~\cite{FuKaneInversion,YoungkukLineNode,SlagerSymmetry,AshvinIndicators,ChenTCI,AshvinTCI,TMDHOTI,AshvinMagnet,ZhidaSemimetals} have facilitated high-throughput searches for magnetic~\cite{MTQCmaterials} and nonmagnetic~\cite{AndreiMaterials,ChenMaterials,AshvinMaterials} topological materials, revealing solid-state TIs and TSMs to be ubiquitous in nature. For example, Ref.~\cite{AndreiMaterials} reported high-throughput analyses of the symmetry-indicated stable band topology at the Fermi level ($E_{F}$) in $\sim$ 26,000 stoichiometric, inorganic, ``high-quality'' crystal structures obtained from the Inorganic Crystal Structure Database (ICSD)~\cite{ICSD}.  Of the $\sim$ 26,000 ICSD entries, roughly $\sim$ 7,000 were found to be topological at $E_{F}$~\cite{AndreiMaterials}.  

\begin{figure}
\includegraphics[width=0.95\columnwidth,angle=0]{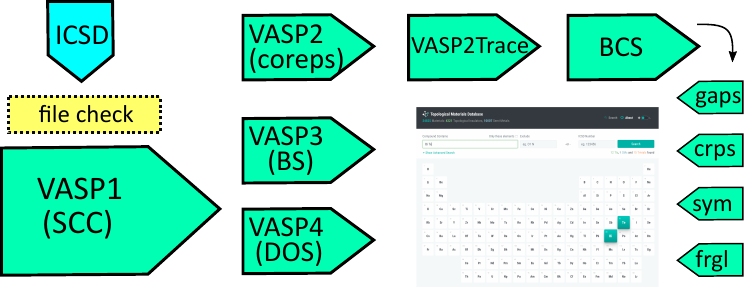}
\caption{Workflow schematic for generating the~\webTQC~(\webNoICSD).  For each entry in the Inorganic Crystal Structure Database (ICSD)~\cite{ICSD}, we first determine if the entry is stoichiometric, contains coordinates for all of the atoms listed in the chemical formula, and lists atomic coordinates compatible with the crystallographic space group (SG) of the material.  We then perform self-consistent density functional theory (DFT) calculations, both with and without incorporating the effects of spin-orbit coupling (SOC). Irrespective of whether SOC is taken into account, each DFT calculation consists of four steps (labeled VASP1 -- VASP4, see \AppVASPAppendix for further details) in which we perform a self-consistent calculation (SCC) of the charge density, obtain the symmetry data [small corepresentation (corep) multiplicities] at all integer electronic fillings above the core shell, and compute the electronic band structure (BS) at $E_{F}$ and the density of states (DOS). Lastly, using the~\vasptotrace~program \cite{V2TREF} previously implemented for Ref.~\cite{AndreiMaterials} and an updated version of the~\checktopmat~program on the~\webBCSfull~\cite{CHECKTOPOREF} implemented for this work, we compute for each DFT calculation the stable and fragile topological classifications of all isolated groupings of bands above the core shell as determined by the compatibility relations in Topological Quantum Chemistry (TQC)~\cite{QuantumChemistry} (see~\AppTQCReviewAppendix~and~\AppCheckTopoAppendix for further details).}
\label{Fig3}
\end{figure}

This result raises two questions.  First, the complete ICSD contains \TQCDstoichiometric~stoichiometric ICSD entries with processable (non-corrupt) structural data as defined in Section~\AppVASPAppendix~of the Supplementary Material (SM), raising the question of whether a high percentage of \emph{all} of the known inorganic materials are topological at $E_{F}$.  Second, energetically isolated groupings of bands may exhibit nontrivial topology, independent of the electronic filling.  Hence, the distribution of topological bands \emph{away from $E_{F}$} in real materials is also a major outstanding question in the study of quantum matter.  In this work, we answer both questions by performing a complete study of symmetry-indicated band topology in all known, stoichiometric, inorganic crystalline solids without magnetic order. In contrast to the previous high-throughput topological materials searches in Refs.~\cite{AndreiMaterials,ChenMaterials,AshvinMaterials}, our calculations include a complete diagnosis of symmetry-indicated band topology away from $E_{F}$.  Using TQC and nonmagnetic SIs, we compute the symmetry-indicated, nonmagnetic band topology of all bands in all \TQCDstoichiometric~stoichiometric ICSD entries with valid structure files from electronic fillings ranging from the core shell up to at least $2N_{e}$ above, where $N_{e}$ is the number of valence electrons.  Unlike in Refs.~\cite{AndreiMaterials,ChenMaterials,AshvinMaterials}, the SIs used in this work include both the SIs of stable band topology -- which indicate familiar strong and weak TI and TCI phases -- as well as the recently introduced SIs of \emph{fragile} band topology~\cite{JenFragile1,AshvinFragile,ZhidaFragile,KoreanFragileInversion} -- which represent more exotic TCI phases with topological corner modes~\cite{TMDHOTI} and twisted-boundary edge states~\cite{ZhidaFragile2}.  Fragile topological phases have emerged as an area of intense interest after recent theoretical studies indicated that the superconducting and correlated-insulating states in magic-angle twisted bilayer graphene may originate from nearly-flat fragile topological bands~\cite{ZhidaBLG,AshvinBLG2}.

Through our calculations, we discover the existence of previously unrecognized classes of topological materials, including enforced TSMs with energetically isolated fragile bands at $E_{F}$, \emph{repeat-topological} (RTopo) materials with stable topological insulating gaps at and just below $E_{F}$, and \emph{supertopological} (STopo) materials in which \emph{every} energetically isolated set of bands above the core shell is stable topological.  We have upgraded the~\webTQC~\cite{DBREF} -- a publicly available online catalog of topological materials -- for accessing and intuitively searching the results of this study.  In the SM, we present detailed statistics for our computations, lists of idealized materials in each topological class, and highlight the features of the~\webTQC~\cite{DBREF} implemented for this work, which include dynamical zoom options, density-of-states calculations, electronic-structure calculations in the absence of spin-orbit coupling (SOC), and advanced search options (a complete list of added features is provided in~\AppWebsite).  We find that~\TQCDBNbrTopologicalMaterialsPercent~of all materials are topological at $E_{F}$, and roughly $2/3$ of bands across all materials exhibit the symmetry-indicated topology of 3D strong TIs, weak TIs, TCIs, and higher-order TIs, which are in this work together classified as topologically \emph{stable}, as they are robust to the addition of trivial or fragile bands (though sensitive to the relaxation of time-reversal and crystal symmetries)~\cite{WiederReview,CharlieZahidReview,KaneMeleZ2,AndreiTI,FuKaneMele,FuKaneInversion,LiangTCIOriginal,HsiehTCI,HourglassInsulator,DiracInsulator,WladPump,ChenTCI,AshvinTCI,MTQC}.  Most surprisingly, we find that~\emph{\TQCDBNbrTopoBandPercent}~of all materials contain at least one stable or fragile topological band in their energy spectrum, even if away from $E_{F}$.  Our discovery of ubiquitous electronic band topology in solid-state materials motivates the formulation of a new periodic table of chemical compounds in which electronic bands in materials are sorted by a \emph{combination} of common topological features and chemical and structural properties.

Lastly, our characterization of electronic band topology \emph{away from $E_{F}$} is immediately useful in numerous experimental settings, including angle-resolved photoemission spectroscopy (ARPES) experiments -- which measure states at and below $E_{F}$ -- and pump-probe experiments -- in which electrons can be excited to observe bands above $E_{F}$.  As will be discussed below, our theoretical calculations provides robust explanations for previously puzzling ARPES data.  Beyond photoemission experiments, states away from $E_{F}$ may also be accessed via (electro)chemical doping, electrostatic gating, hydrostatic pressure, and nonequilibrium photoexcitation, and are relevant to Floquet engineering and nonlinear optical experiments~\cite{BasovNatMaterReview}.  Additionally, even in topologically trivial insulating materials, exotic interaction effects may be accessed by doping or gating the Fermi level into an isolated topological band.  For example, independent of the cumulative band topology at higher energies, if a superconducting state is induced from a partially occupied, isolated band, then the nontrivial stable or fragile topology of the isolated band has been shown to provide a lower bound on the superfluid weight~\cite{SuperfluidBoundChern,SuperfluidBoundFragileAndrei}.

\section*{Data Set Generation}

In this work, we apply TQC~\cite{QuantumChemistry,MTQC} and stable and fragile SIs~\cite{FuKaneInversion,YoungkukLineNode,SlagerSymmetry,AshvinIndicators,ChenTCI,AshvinTCI,TMDHOTI,AshvinMagnet,ZhidaSemimetals,MTQCmaterials,JenFragile1,AshvinFragile,ZhidaFragile,KoreanFragileInversion} to diagnose the symmetry-indicated topology of all isolated bands above the core shell in the stoichiometric materials in the ICSD.  To obtain the topological classification of each separated group of bands in the energy spectrum, we employ the methods previously used in the high-throughput material searches in Refs.~\cite{AndreiMaterials,MTQCmaterials}.  Uniquely in this work, after computing the electronic structure using the intrinsic electronic filling $\nu=N_{e}$, we then vary $\nu$ to compute the topological properties above and below $E_{F}$.  Below, we will summarize the numerical and topological details of our calculations (Fig.~\ref{Fig3}) -- further specific details are provided in~\AppTQCReviewAppendix,~\AppCheckTopoAppendix, and~\AppVASPAppendix.

We begin by selecting one of the~\TQCDstoichiometric~processable entries in the ICSD with a stoichiometric chemical formula.  We next convert the crystal structure listed in the ICSD into an input file for DFT calculations, excluding cases in which the ICSD structure file is missing atoms listed in the chemical formula or reports atomic positions incompatible with the symmetries of the material space group (SG, see~\AppVASPAppendix).  We then perform self-consistent, \emph{ab initio} calculations of the electronic band structure and density of states with $\nu$ set to charge neutrality ($\nu=N_{e}$), filtering out cases in which the calculations did not converge or converged to a magnetic (meta)stable state.

Next, as detailed in~\AppCheckTopoAppendix, we use the \vasptotrace~\cite{V2TREF} and \checktopmat~\cite{CHECKTOPOREF} programs to determine the unitary symmetry eigenvalues of each of the Bloch-state multiplets at each of the maximal (high-symmetry) ${\bf k}$ vectors.  For each degenerate multiplet of Bloch states, the combination of extracted symmetry eigenvalues establishes a correspondence to an irreducible small corepresentation (corep) of the little group at ${\bf k}$~\cite{BigBook}.  Then, choosing increasing integer values of the valence electronic filling $\nu$ from $0$ to at least $2N_{e}$ (or up to the first filling $\nu > N_{e}$ at which the electronic band structure is not symmetric and convergent, see~\AppVASPToTrace), we identify fillings at which all of the Bloch states in each maximal ${\bf k}$ vector are fully occupied, such that a gap is present in the energy spectrum at each maximal ${\bf k}$ vector (though not necessarily at the same energy, as states are taken to be separately filled up to $\nu$ at each ${\bf k}$ point).  Lastly, at $\nu=N_{e}$, as well as at each $\nu$ at which there is a gap at all maximal ${\bf k}$ points, we extract the \emph{symmetry data vector}~\cite{AndreiMaterials,MTQC,MTQCmaterials} -- defined as the multiplicities of the small coreps corresponding to the filled Bloch-eigenstate multiplets across each of the maximal ${\bf k}$ vectors.

Crucially, the symmetry data vector at each $\nu$ facilitates the topological classification of filled or energetically isolated groupings of bands.  Specifically, in each SG, the \emph{elementary band coreps} (EBRs) correspond to the independent, topologically trivial (\emph{i.e.} Wannierizable~\cite{AlexeyVDBWannier}) bands~\cite{QuantumChemistry,Bandrep1}.  Hence, if a set of bands does not transform in an integer-valued linear combination of EBRs, then the band exhibits nontrivial stable topology.  As established in Refs.~\cite{AndreiMaterials,MTQCmaterials} and detailed in~\AppTQCReviewAppendix, a symmetry data vector may be classified into one of five possible topological classes.  First, if the symmetry data vector corresponds to a set of Bloch states in which a degenerate multiplet at a maximal ${\bf k}$ point is partially filled, then the system is an \emph{enforced semimetal with Fermi degeneracy} (ESFD).  As an example, 3D HgTe [\icsdweb{31845}, SG 216 ($F\bar{4}3m$)] is an experimentally-established, ESFD-classified TSM that realizes a 2D TI phase in few-layer quantum-well geometries~\cite{AndreiTI}.  Next, if the symmetry data vector corresponds to a set of high-symmetry-point Bloch eigenstates that are implied by the compatibility relations to connect to other states outside of the symmetry data along a high-symmetry line or plane, then the system is an \emph{enforced semimetal} (ES).  A well-studied ES material is the archetypal higher-order-topological Dirac semimetal Cd$_3$As$_2$ [\icsdweb{107918}, SG 137 ($P4_{2}/nmc$)]~\cite{ZJDirac,CavaDirac1,HingeSM,HingeSMExp}.  Conversely, if the symmetry data vector satisfies the compatibility relations along all high-symmetry lines and planes, then the bands that transform in the symmetry data either correspond to a stable or fragile TI or TCI, a TSM phase, or are topologically trivial.  In the case in which the symmetry data vector corresponds to a stable topological set of bands, the bands described by the symmetry data may be classified into two categories.  First, if stable topological bands transform in an integer-valued linear combination of pieces of a \emph{disconnected} EBRs, then the bands are classified as a split EBR (SEBR)~\cite{QuantumChemistry,Bandrep1}. Alternatively, if stable topological bands do not transform in an integer-valued linear combination of disconnected EBR pieces, then the bands are classified as ``not equal to a linear combination’’ (NLC).  The archetypal symmetry-indicated stable topological material is the experimentally established 3D TI Bi$_2$Se$_3$ [\icsdweb{617079}, SG 166 ($R\bar{3}m$)]~\cite{HsiehBismuthSelenide}, which is classified as SEBR.  Lastly, if the symmetry data of an isolated grouping of bands does correspond to an integer-valued linear combination of EBRs (LCEBR), then the symmetry eigenvalues of the occupied bands are compatible with either fragile or trivial topology (though the bands may also exhibit non-symmetry-indicated stable topology~\cite{TMDHOTI}).

In the above cases in which the symmetry data vector satisfies the compatibility relations (SEBR, NLC, and LCEBR), a finer topological classification may be obtained using stable and fragile topological SIs.  First, in the presence of non-negligible SOC, nonmagnetic SEBR and NLC bands correspond to stable TIs and TCIs with anomalous surface Dirac cones and helical hinge states~\cite{FuKaneInversion,ChenTCI,AshvinTCI}.  Conversely, as shown in Refs.~\cite{YoungkukLineNode,ZhidaSemimetals,TMDHOTI,MTQC} and detailed in~\AppTopologicalMaterialsNoSOC, in the absence of SOC and magnetism, bands with nontrivial stable SIs correspond to TSM (specifically SEBR-SM and NLC-SM) phases with bulk nodal degeneracies in the Brillouin zone (BZ) interior and topological surface Fermi arcs or flat-band like surface and intrinsic hinge states.  Lastly, it was recently discovered that LCEBR bands with or without SOC may in some cases be further classified as fragile topological~\cite{JenFragile1,AshvinFragile} through the fragile SIs elucidated in Refs.~\cite{ZhidaFragile,KoreanFragileInversion}.  For all possible classes of symmetry-indicated stable and fragile TI, TCI, and TSM phases in the presence of SOC, we provide in~\AppMaterialsSelection a detailed enumeration of the most idealized material realizations across all stoichiometric materials in the ICSD.  Although we find that there do not exist materials in which the entire valence manifold is fragile topological, we do find numerous examples of experimentally accessible materials with well-isolated fragile bands close to $E_{F}$ (see~\AppFragileBands).

\section*{Results}

We will now summarize the symmetry-indicated topological properties of the stoichiometric materials in the ICSD.  First, we will discuss statistical trends uncovered across the materials analyzed in this work, including the distribution of topological features at $E_{F}$ and statistics for SOC-driven topological phase transitions, which are defined by comparing the stable topology at $E_{F}$ for each ICSD entry in calculations performed with and without incorporating the effects of SOC.  We will then highlight material candidates in which our investigations have uncovered previously unrecognized topological features at experimentally accessible energies.

\subsection*{Material Statistics}

\begin{figure*}
\includegraphics[width=1\textwidth,angle=0]{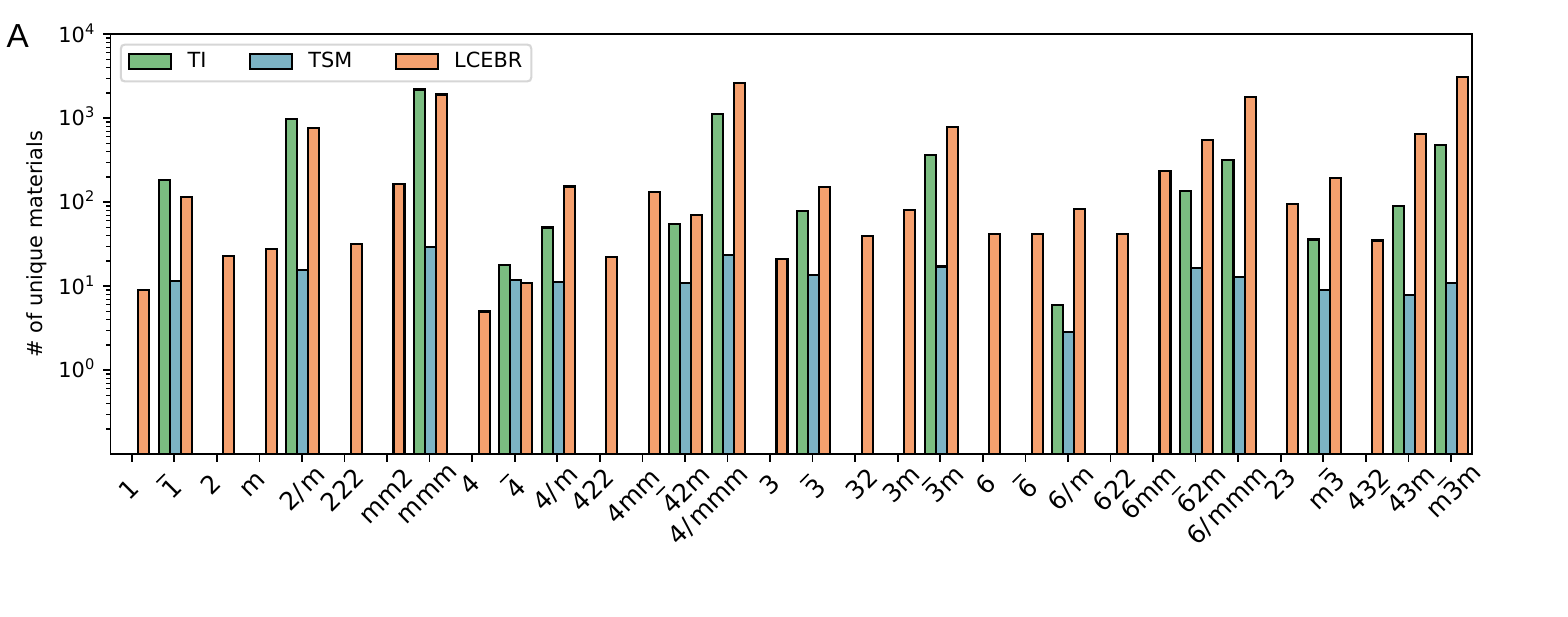}
\includegraphics[width=0.9\textwidth,angle=0]{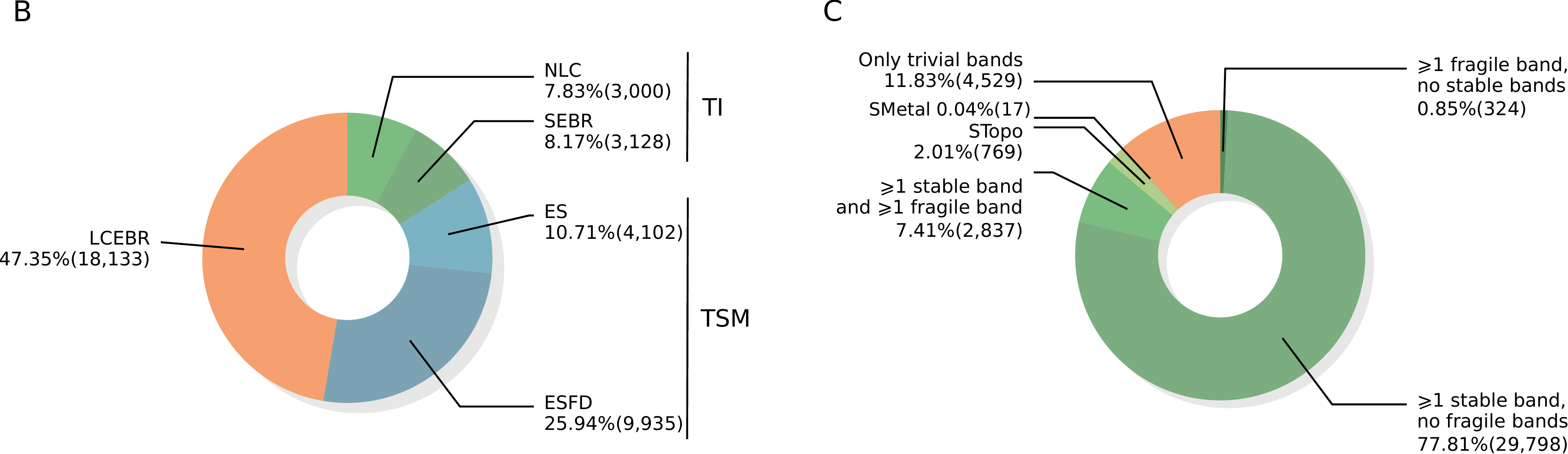}
\caption{Topology of the stoichiometric materials in the ICSD.  (A) Distribution of the topology at $E_{F}$ with SOC among the~\TQCDBNbrUniqueMaterials~stoichiometric unique materials in the ICSD as defined in~\AppVASPAppendix, subdivided by crystallographic point group (PG).  For each of the 32 PGs of the 230 SGs~\cite{BigBook}, we list the number of symmetry-indicated stable topological (crystalline) insulators (NLC- or SEBR-classified TIs and TCIs), TSMs (ES or ESFD), or unique materials with trivial symmetry-indicated topology (LCEBR).  (B) The distribution of the topology at $E_{F}$ across all PGs and topological classes (see Refs.~\cite{AndreiMaterials,MTQCmaterials} and~\AppTQCReviewAppendix). ~\TQCDBNbrMaterialsTIPercent~of the unique materials are TIs or TCIs at $E_{F}$ and~\TQCDBNbrMaterialsSMPercent~are TSMs, implying that a remarkable~\TQCDBNbrTopologicalMaterialsPercent~of the stoichiometric materials in nature are topological at intrinsic filling.  (C) Distribution of the symmetry-indicated band topology away from $E_{F}$ across the stoichiometric unique materials in the ICSD.  An overwhelming~\emph{\TQCDBNbrTopoBandPercent~of materials} contain at least one symmetry-indicated stable or fragile topological band across the energy range of our first-principles calculations (see \AppVASPAppendix for calculation details), which is even more notable when considering that many materials (\emph{e.g.} noncentrosymmetric crystals) have SGs without either stable or fragile SIs (see Refs.~\cite{WiederReview,MTQC,FuKaneInversion,YoungkukLineNode,SlagerSymmetry,AshvinIndicators,ChenTCI,AshvinTCI,TMDHOTI,AshvinMagnet,ZhidaSemimetals,ZhidaFragile,KoreanFragileInversion}).     Because previous works have demonstrated the existence of \emph{non-symmetry-indicated} TI and TCI phases in materials with trivial stable SIs~\cite{WiederReview,KaneMeleZ2,HourglassInsulator,DiracInsulator,TMDHOTI}, then the percentage of materials in nature with topological bands at and away from $E_{F}$ is necessarily \emph{even larger than}~\TQCDBNbrTopoBandPercent, suggesting an intriguing direction for future study.  In addition to the repeat-topological and supertopological materials (see Fig.~\ref{fig:STopoMain} and~\AppSupertopological), we have also discovered the existence of supermetallic (SMetal) materials, in which all of the bands above the core shell are connected up to at least a filling of $2N_{e}$, where $N_{e}$ is the number of valence electrons. (see~\AppTopologicalMaterialsNoSOC~and~\AppTopoBands).  Although SMetal materials are relatively rare in the presence of SOC [$\TQCDBNbrSMetal$~unique materials, see (C)], there are $\TQCDBNbrNoSOCSMetal$ unique SMetal materials without SOC, which is consistent with the general trend of increased band connectivity in materials when neglecting the effects of SOC (see Table~\ref{tb:statistics_nosoc_to_soc_uniquematerials_summary} and~\AppTopologicalMaterialsNoSOC,~\AppTopoBands, and~\AppPhaseTransitionsNoSOCSOC).  In \AppTopoBands, we provide further detailed statistics for the symmetry-indicated band topology at and away from $E_{F}$ across all of the stoichiometric materials in the ICSD.}
\label{fig:histogramPerPG}
\end{figure*}

At the time of our investigations, the ICSD~\cite{ICSD} contained~\TQCDTotICSDs~entries, of which~\TQCDstoichiometric~characterized stoichiometric chemical compounds with processable (non-corrupt) CIF structure files.  We performed first-principles calculations on all~\TQCDstoichiometric~processable stoichiometric entries, resulting in convergent electronic structures for~\TQCDBNbrICSDs~entries in the presence of SOC, which we grouped into~\TQCDBNbrUniqueMaterials~unique materials by common chemical formulas and stable topology at $E_{F}$ (see~\AppVASPToTrace).  In Fig.~\ref{fig:histogramPerPG}A, we show the distribution of the topology at $E_{F}$ for the unique materials within each of the 32 crystallographic point groups of the 230 nonmagnetic SGs, and in Fig.~\ref{fig:histogramPerPG}B, we show the total number of unique materials in each topological class.  We find that at intrinsic filling,~\TQCDBNbrMaterialsTI~unique materials are symmetry-indicated stable TIs or TCIs, and~\TQCDBNbrMaterialsSM~unique materials are TSMs.  Our findings represent a doubling of the number of known topological materials from those identified in previous high-throughput calculations~\cite{AndreiMaterials,ChenMaterials,AshvinMaterials}.  Furthermore, we find that~\TQCDBNbrTopologicalMaterialsPercent~of unique materials are symmetry-indicated stable TIs, TCIs, or TSMs at $E_{F}$, which also represents a doubling from previous estimates of the percentage of 3D topological materials in nature~\cite{AndreiMaterials,ChenMaterials}.  This percentage is considerably higher than the percentage of topological stoichiometric 2D materials, which has previously been computed to lie within the range of a few to 20\%~\cite{Marzari2DTIAbundantDB,Ashvin2DMaterials}.

In~\AppCPUTime~and~\AppTopoBands, we respectively provide detailed statistics for the computational resources expended for the calculations performed for this work, and for the topology \emph{at and away} from $E_{F}$ across the stoichiometric materials in the ICSD.  In the presence of SOC, we find that an overwhelming~\TQCDBNbrTopoBandPercent~of unique materials contain at least one symmetry-indicated stable or fragile topological band [Fig.~\ref{fig:histogramPerPG}C].  This percentage is all the more surprising, as we have included materials whose SGs have trivial stable SI groups (\emph{e.g.} noncentrosymmetric crystals without high-fold rotoinversion symmetries~\cite{AshvinTCI,ChenTCI,MTQC}), such that all isolated bands in the SG are classified by SIs as trivial or fragile.  Taking all~\TQCDBNbrUniqueMaterials~unique materials into consideration -- including the materials in SGs with trivial stable SI groups~\cite{AshvinTCI,ChenTCI,MTQC} -- we find that nearly $2/3$ of all energetically isolated bands in nature exhibit symmetry-indicated stable topology.  Hence, our investigations reveal that even away from $E_{F}$, the \emph{majority} of electronic features in solid-state materials can only be robustly modeled by incorporating topological band theory.  

\begin{table}
{\small
\begin{tabular}{|c||c|c|c|c|c|c|}
\hline
 & NLC & SEBR & ES & ESFD & LCEBR  \\
\hline
\hline
ES & \begin{tabular}{c}2,104\\(35\%)\end{tabular} & \begin{tabular}{c}1,585\\(26.4\%)\end{tabular} & \begin{tabular}{c}944\\(15.7\%)\end{tabular} & \begin{tabular}{c}38\\(0.6\%)\end{tabular} & \begin{tabular}{c}1,335\\(22.2\%)\end{tabular} \\ 
 \hline 
ESFD & \begin{tabular}{c}207\\(1.5\%)\end{tabular} & \begin{tabular}{c}1,089\\(7.8\%)\end{tabular} & \begin{tabular}{c}2,792\\(19.9\%)\end{tabular} & \begin{tabular}{c}9,423\\(67.3\%)\end{tabular} & \begin{tabular}{c}486\\(3.5\%)\end{tabular} \\ 
 \hline 
LCEBR & \begin{tabular}{c}49\\(0.3\%)\end{tabular} & \begin{tabular}{c}78\\(0.5\%)\end{tabular} & \begin{tabular}{c}42\\(0.3\%)\end{tabular} & \begin{tabular}{c}24\\(0.2\%)\end{tabular} & \begin{tabular}{c}15,672\\(98.8\%)\end{tabular} \\ 
 \hline 
NLC-SM & \begin{tabular}{c}208\\(82.9\%)\end{tabular} & \begin{tabular}{c}24\\(9.6\%)\end{tabular} & \begin{tabular}{c}2\\(0.8\%)\end{tabular} & ---  & \begin{tabular}{c}17\\(6.8\%)\end{tabular} \\ 
 \hline 
SEBR-SM & ---  & \begin{tabular}{c}38\\(86.4\%)\end{tabular} & \begin{tabular}{c}5\\(11.4\%)\end{tabular} & ---  & \begin{tabular}{c}1\\(2.3\%)\end{tabular}\\
\hline
\hline
\end{tabular}
\caption[SOC-driven topological phase transitions in unique materials]{SOC-driven topological phase transitions in the unique stoichiometric materials in the ICSD.  We show statistics for the band topology at $E_{F}$ for the unique stoichiometric materials in the ICSD with and without incorporating the effects of SOC in first-principles calculations.  The rows (columns) list the topological classification at $E_{F}$ in the absence (presence) of SOC.  We find that nearly $2/3$ of ES-classified unique materials without SOC become stable TIs and TCIs with SOC, whereas most ESFD TSMs without SOC remain semimetallic when SOC is incorporated.  Conversely, over 85\% of NLC-SM- and SEBR-SM-classified TSMs without SOC (see Refs.~\cite{YoungkukLineNode,ZhidaSemimetals,TMDHOTI} and~\AppTopologicalMaterialsNoSOC) become stable TIs and TCIs after incorporating SOC.  Complete statistics for the SOC-driven topological phase transitions at $E_{F}$ in the ICSD are provided in~\AppPhaseTransitionsNoSOCSOC, and representative materials for each class of TSM-insulator phase transition are provided in~\AppTransitions.}
\label{tb:statistics_nosoc_to_soc_uniquematerials_summary}
}
\end{table}

Lastly, we have also computed detailed statistics for the relative topology of each stoichiometric ICSD entry with and without incorporating the effects of SOC.  As discussed above and in~\AppPhaseTransitionsNoSOCSOC~and demonstrated in Refs.~\cite{YoungkukLineNode,ZhidaSemimetals,TMDHOTI,MTQC}, all symmetry-indicated, nonmagnetic stable topological phases without SOC are TSMs.  Hence in the absence of SOC, there are both ESFD- and ES-classified TSMs -- in which the nodal degeneracies respectively lie at high-symmetry points (ESFD) and along high-symmetry lines and planes (ES) -- as well as SEBR-SM- and NLC-SM-classified TSMs -- in which the bands along all high-symmetry lines and planes satisfy the insulating compatibility relations~\cite{QuantumChemistry,Bandrep1}, and the nodal degeneracies lie in the BZ interior.  In Table~\ref{tb:statistics_nosoc_to_soc_uniquematerials_summary} we show the number of unique materials in each topological class with and without SOC.  Table~\ref{tb:statistics_nosoc_to_soc_uniquematerials_summary} indicates that the majority of ES-, NLC-SM-, and SEBR-SM-classified TSMs without SOC become stable TIs and TCIs (NLC- and SEBR-classified) when SOC is introduced, whereas ESFD-classified TSMs without SOC overwhelmingly remain TSMs (ES- and ESFD-classified) when the effects of SOC are incorporated.  In~\AppPhaseTransitionsNoSOCSOC, we provide complete statistics for the SOC-driven topological phase transitions at $E_{F}$ in the ICSD, and in~\AppTransitions, we identify representative materials for each class of SOC-driven TSM-insulator transition in Table~\ref{tb:statistics_nosoc_to_soc_uniquematerials_summary}.

\subsection*{Material Candidates}

\begin{figure*}[h!]
\centering
\begin{tabular}{c c c}
A\hspace{0.2cm} \scriptsize{$\rm{Mo} \rm{Ge}_{2}$ - \icsdweb{76139} - SG 139 ($I4/mmm$) - ES} & \hspace{0.5cm} &B\hspace{0.2cm} \scriptsize{$\rm{Hg} \rm{Ba}_{2} \rm{Cu} \rm{O}_{4}$ - \icsdweb{75720} - SG 123 ($P4/mmm$) - ESFD}\\
 \hspace{0.5cm} & \hspace{0.5cm} & \\
\includegraphics[height=4cm]{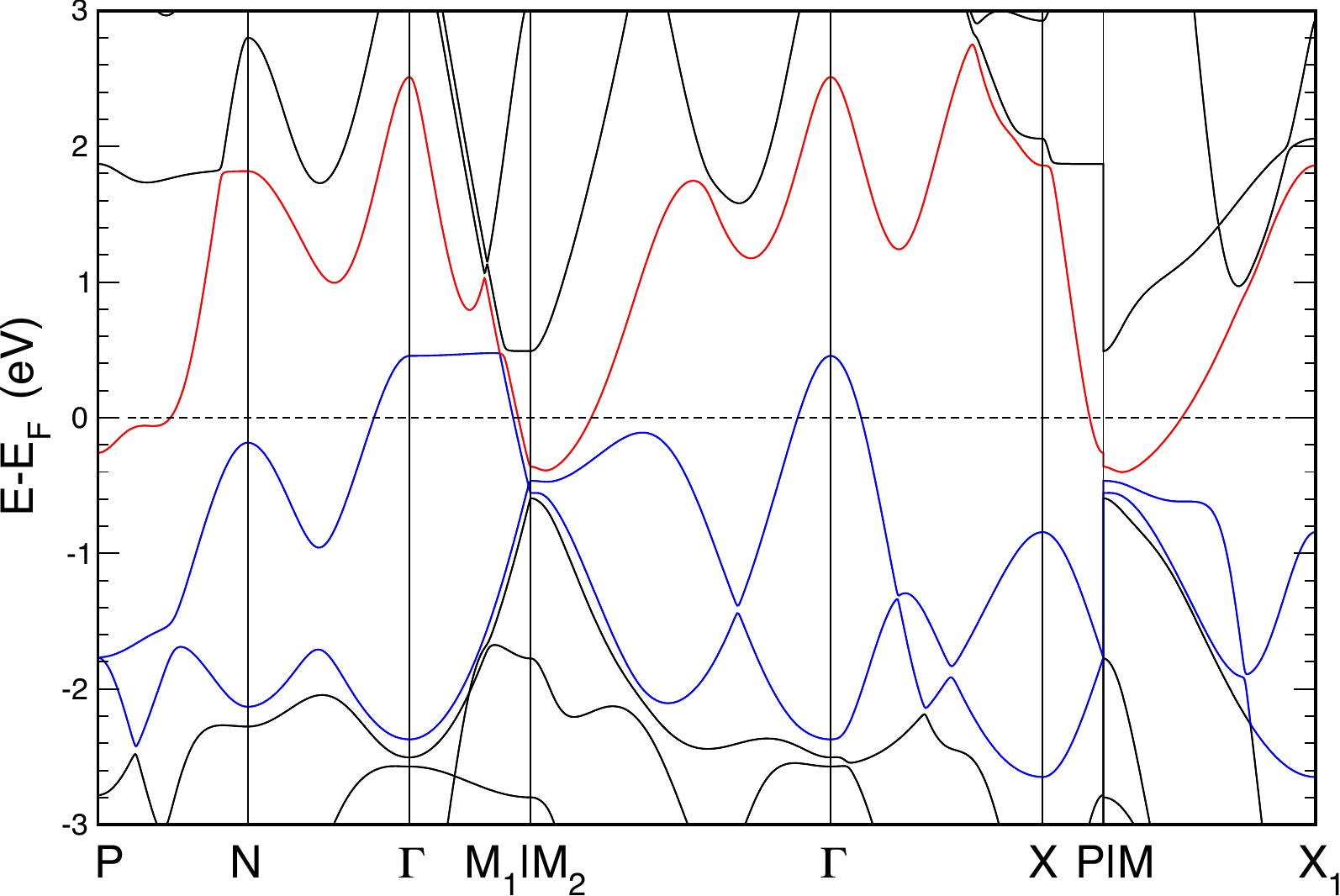} & \hspace{0.5cm} & \includegraphics[height=4cm]{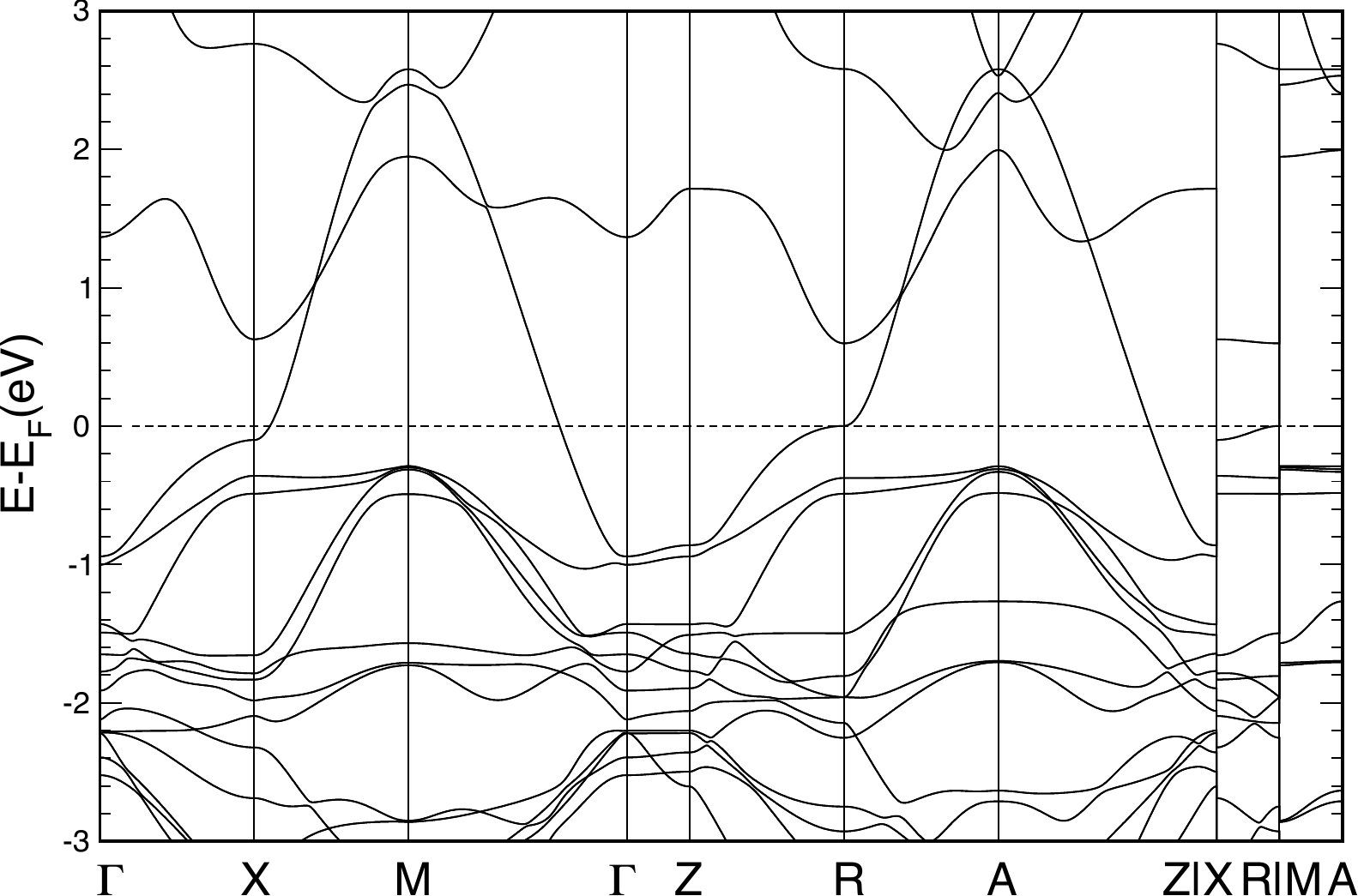}\\
\null & \hspace{0.5cm} & \null \\
C\hspace{0.2cm} \scriptsize{$\rm{Ta} \rm{Se}_{2}$ - \icsdweb{24313} - SG 164 ($P\bar{3}m1$) - ESFD} & \hspace{0.5cm} &D\hspace{0.2cm} \scriptsize{$\rm{Ti} \rm{S}_{2}$ - \icsdweb{72042} - SG 227 ($Fd\bar{3}m$) - SEBR}\\
\hspace{0.5cm} & \hspace{0.5cm} & \tiny{ $\;Z_{2,1}=1\;Z_{2,2}=1\;Z_{2,3}=1\;Z_4=1\;Z_2=1$ }\\
\includegraphics[height=4cm]{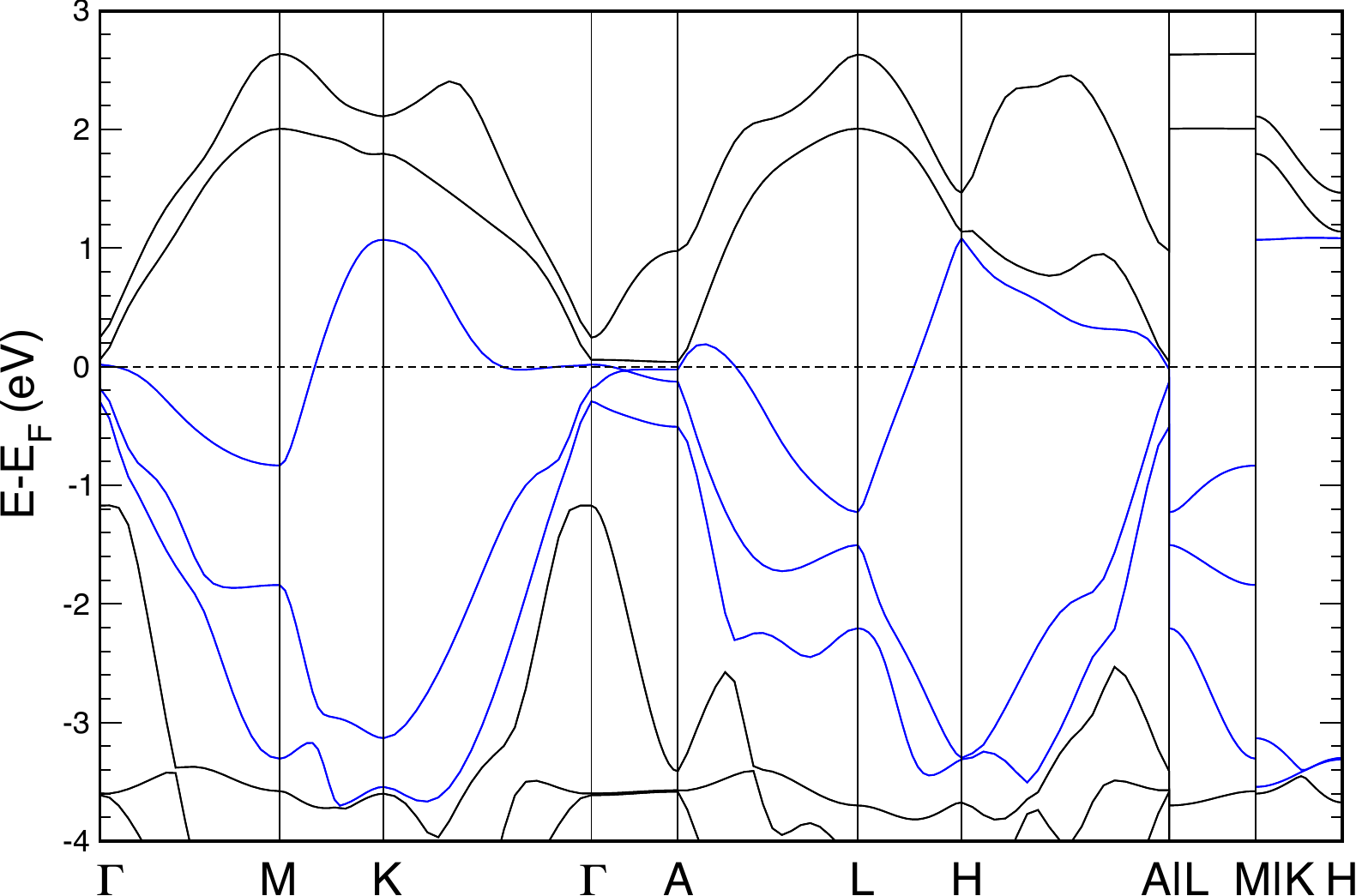} & \hspace{0.5cm} & \includegraphics[height=4cm]{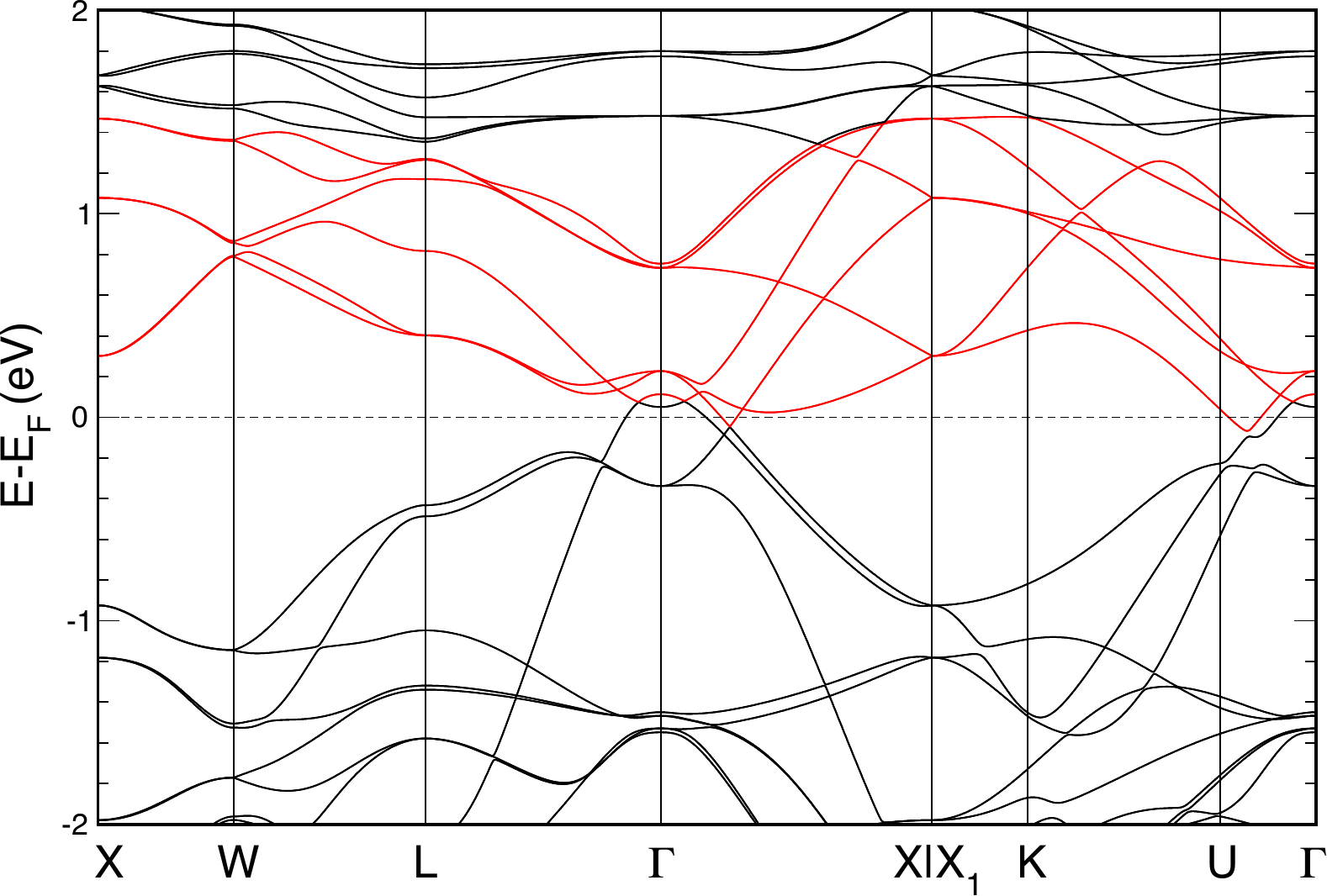}\\
\null & \hspace{0.5cm} & \null \\
E\hspace{0.2cm} \scriptsize{$\rm{Ta}_{2} \rm{Ni} \rm{Se}_{5}$ - \icsdweb{61148} - SG 15 ($C2/c$) - NLC}  & \hspace{0.5cm} & F\hspace{0.2cm} \scriptsize{$\rm{Ta}_2 \rm{Ni} \rm{Se}_7$ - \icsdweb{61352} - SG 12 ($C2/m$) - NLC}\\
\tiny{ $\;Z_{2,1}=0\;Z_{2,2}=0\;Z_{2,3}=0\;Z_4=3$ } & \hspace{0.5cm}  & \tiny{ $\;Z_{2,1}=0\;Z_{2,2}=0\;Z_{2,3}=0\;Z_4=3$ }\\
\includegraphics[height=4cm]{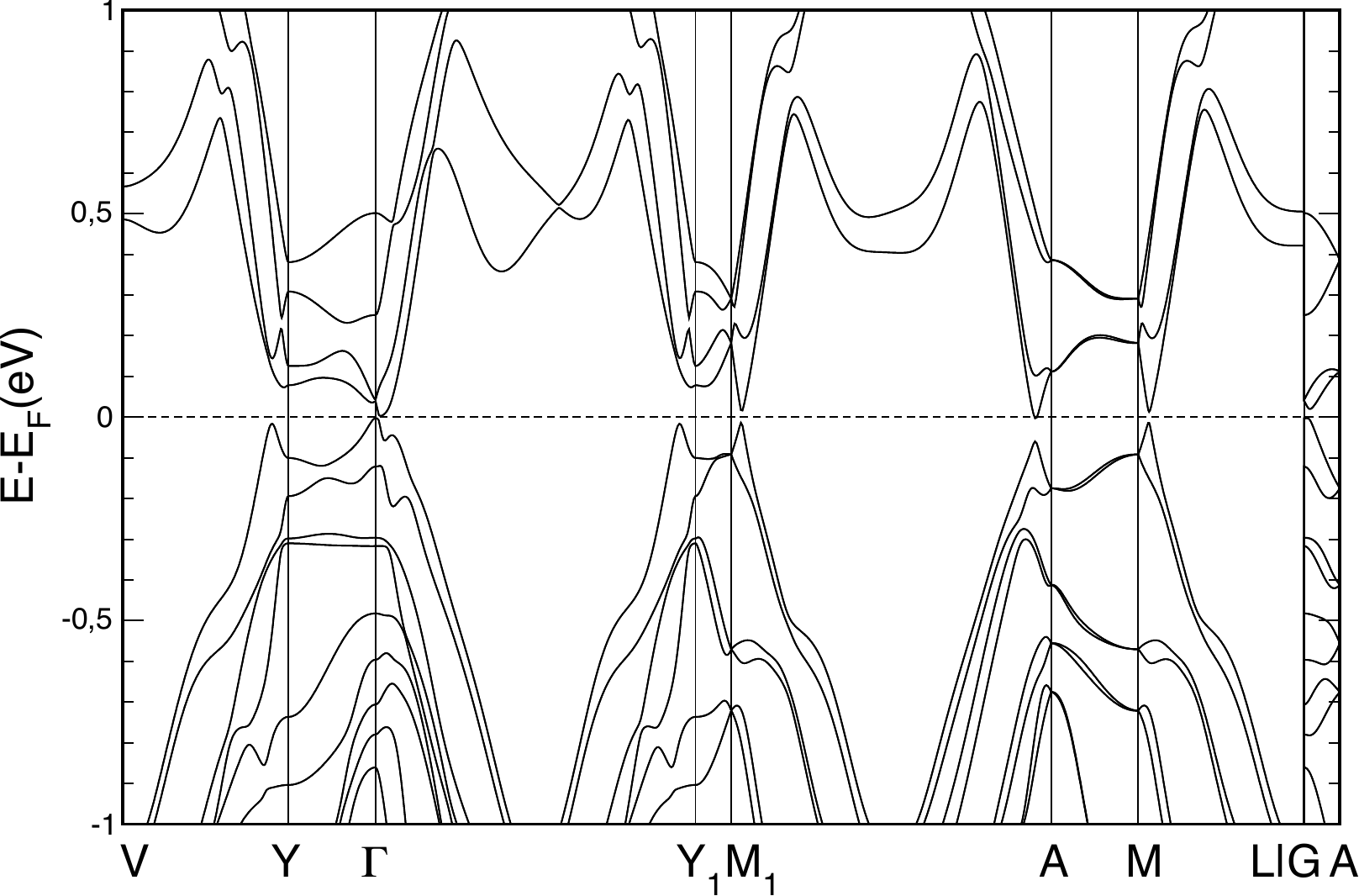} & \hspace{0.5cm} & \includegraphics[height=4cm]{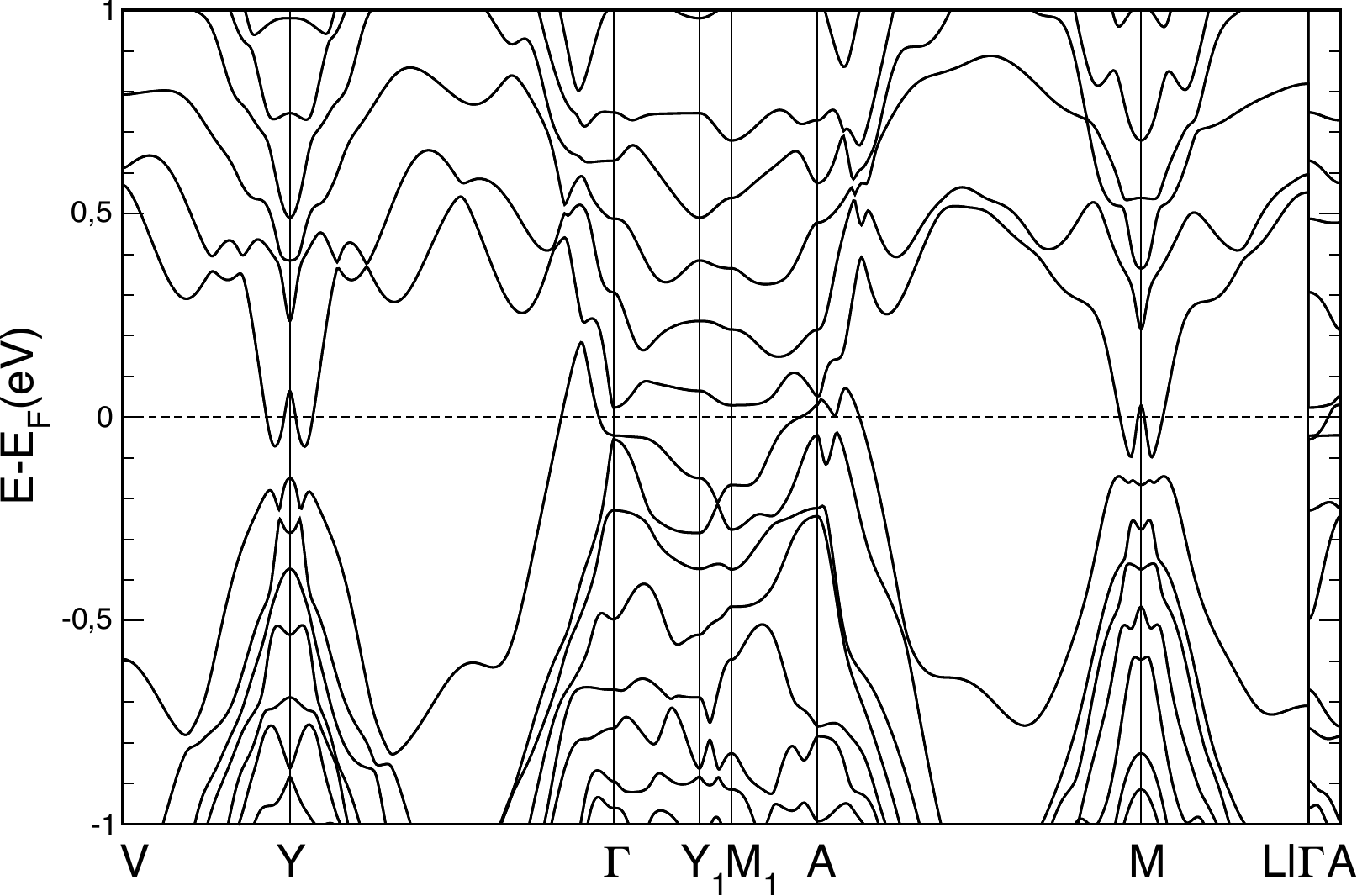}\\
\end{tabular}
\caption{Material candidates with topological features at $E_{F}$.  We highlight materials with previously unidentified or recontextualized topological features at $E_{F}$.  In all panels, bands with symmetry-indicated stable and trivial topology are plotted in black, and symmetry-indicated fragile valence (conduction) bands are plotted in blue (red).  (A) MoGe$_2$ [\icsdweb{76139}, SG 139 ($I4/mmm$)] is a higher-order-topological, centrosymmetric Dirac TSM~\cite{HingeSM} in which the doubly-degenerate valence and conduction bands that meet along $\Gamma M_{1}$ in fourfold Dirac points at $E_{F}$, along with the next-highest doubly-degenerate valence bands, are as a set fragile topological.  (B) HgBa$_2$CuO$_4$ [\icsdweb{75720}, SG 123 ($P4/mmm$)] -- an established high-$T_{C}$ superconductor~\cite{HgBaCuONature} -- is found to be an ideal ESFD-classified metal with a doubly-degenerate, half-filled band at $E_{F}$.  (C,D) In the transition-metal chalcogenides (TMCs) TaSe$_2$ [\icsdweb{24313}, SG 164 ($P\bar{3}m1$)] and TiS$_2$ [\icsdweb{72042}, SG 227 ($Fd\bar{3}m$)] -- previously determined to host 2D charge-density-wave phases~\cite{TaSe2SynthesizeNature,TaSe2CDW,TiS2CDW} -- the highest valence and lowest conduction bands are respectively fragile topological.  (E) Ta$_2$NiSe$_5$ [\icsdweb{61148}, SG 15 ($C2/c$)] is a layered TMC that has been demonstrated to host an exciton-insulator phase~\cite{TaNiSeExciton1,TaNiSeExciton2}; we find that the narrow-gap semiconducting state of Ta$_2$NiSe$_5$ is in fact a 3D TI.  (F)  Conversely, the closely-related quasi-1D TMC Ta$_2$NiSe$_7$ [\icsdweb{61352}, SG 12 ($C2/m$)] -- which we find to also be a 3D TI -- exhibits a charge-density-wave instability~\cite{Ta2NiSe7OriginalCDW}.  In Fig.~\ref{fig:TaNiSeparity}, we provide further analysis of the band ordering in Ta$_2$NiSe$_5$ and Ta$_2$NiSe$_7$.}
\label{fig:good-bands}
\end{figure*}

\begin{figure*}[h!]
\centering

\begin{tabular}{c c c}
A\hspace{0.2cm} \scriptsize{$\rm{Ta}_{2} \rm{Ni} \rm{Se}_{5}$ w/o SOC - \icsdweb{61148} - SG 15 ($C2/c$) - LCEBR} & \hspace{0.5cm} &B\hspace{0.2cm}  \scriptsize{$\rm{Ta}_{2} \rm{Ni} \rm{Se}_{5}$ with SOC - \icsdweb{61148} - SG 15 ($C2/c$) - NLC}\\
 & \hspace{0.5cm} & \tiny{ $\;Z_{2,1}=0\;Z_{2,2}=0\;Z_{2,3}=0\;Z_4=3$ } \\
  \includegraphics[height=4.5cm]{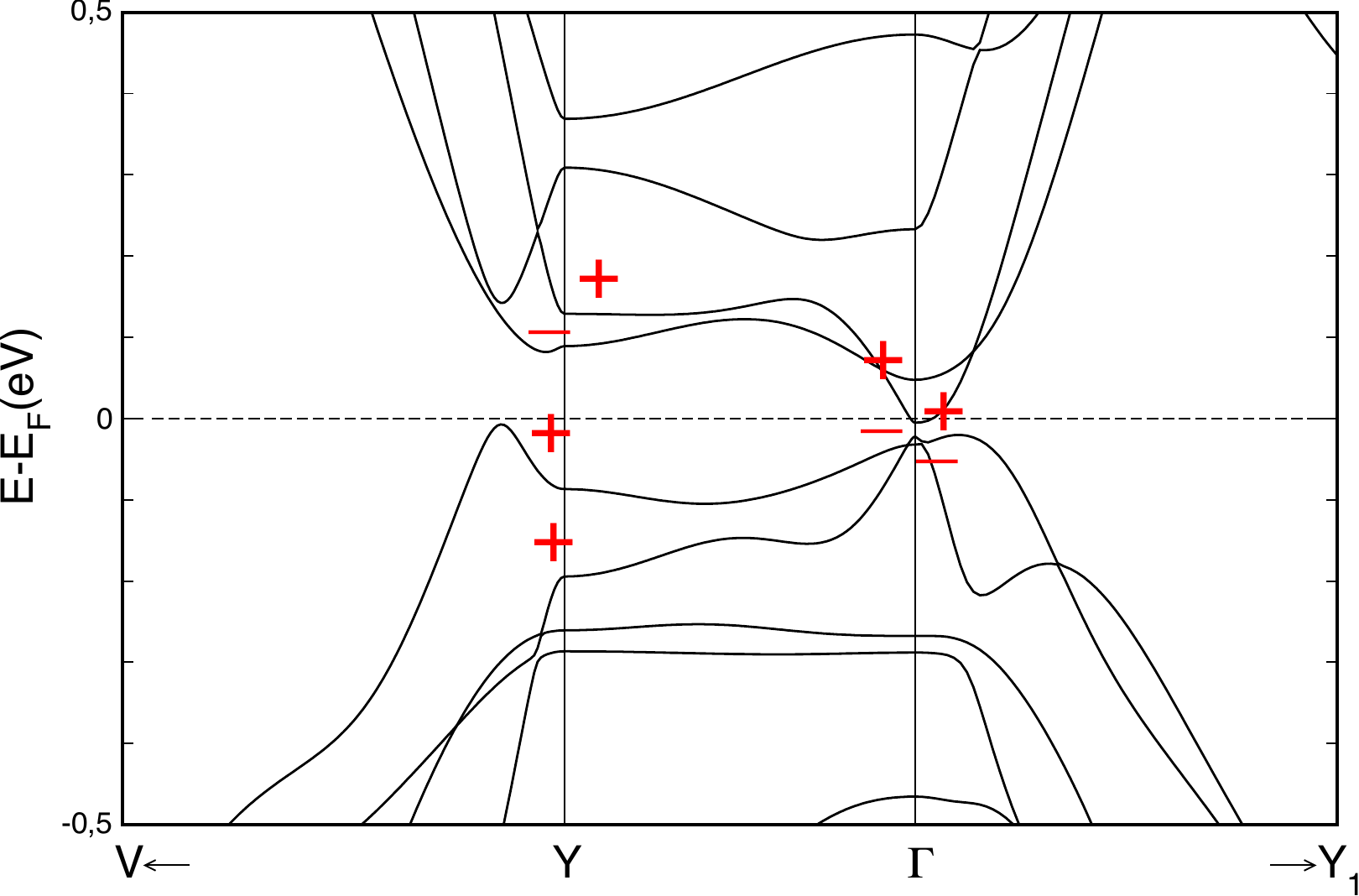} & \hspace{0.5cm} & \includegraphics[height=4.5cm]{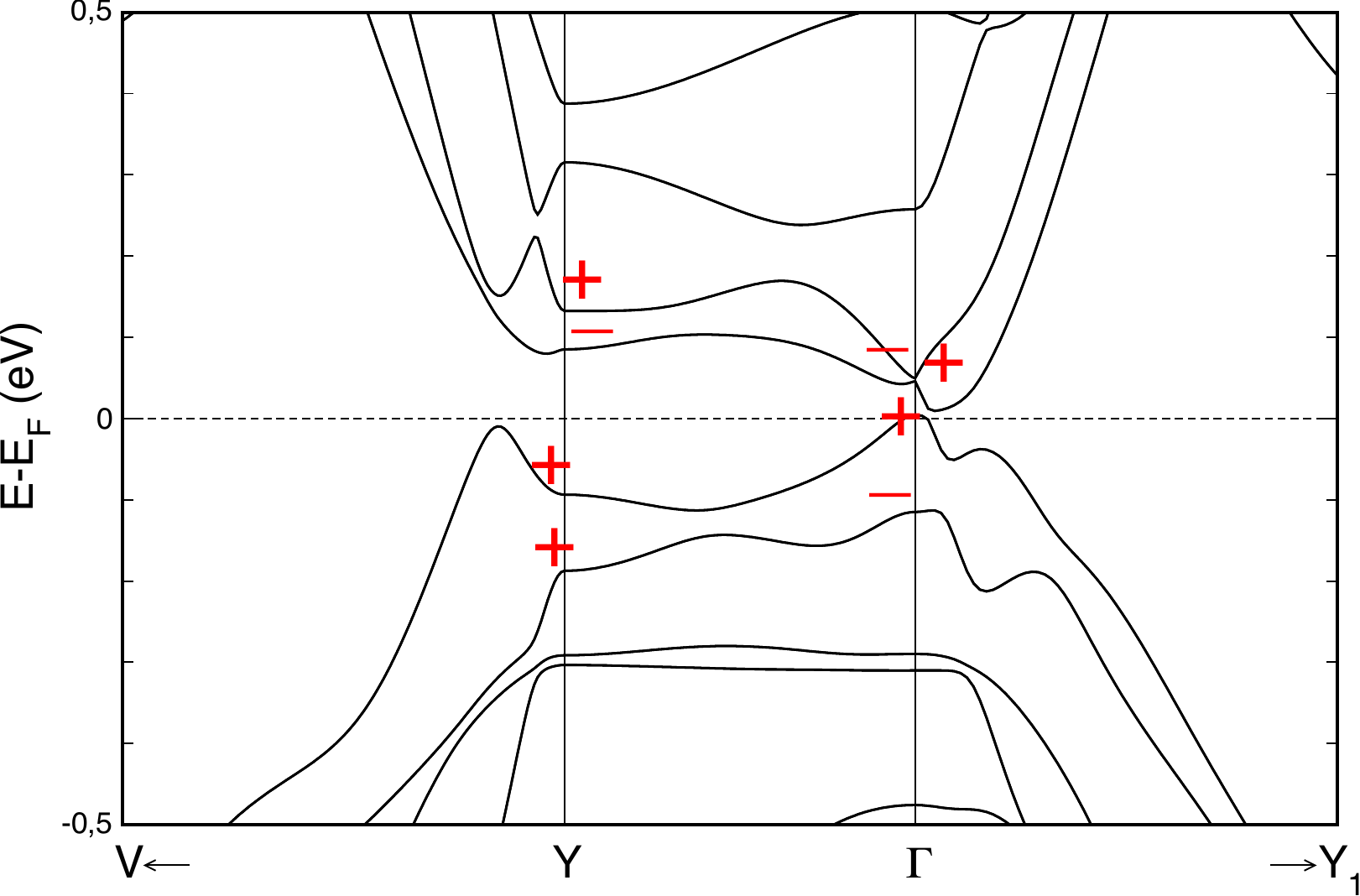}\\
\null & \hspace{0.5cm} & \null \\
C\hspace{0.2cm} \scriptsize{$\rm{Ta}_2 \rm{Ni} \rm{Se}_7$ w/o SOC - \icsdweb{61352} - SG 12 ($C2/m$) - ES}  & \hspace{0.5cm} & D\hspace{0.2cm} \scriptsize{$\rm{Ta}_2 \rm{Ni} \rm{Se}_7$ with SOC - \icsdweb{61352} - SG 12 ($C2/m$) - NLC}\\
  & \hspace{0.5cm} & \tiny{ $\;Z_{2,1}=0\;Z_{2,2}=0\;Z_{2,3}=0\;Z_4=3$ }\\
\includegraphics[height=4.5cm]{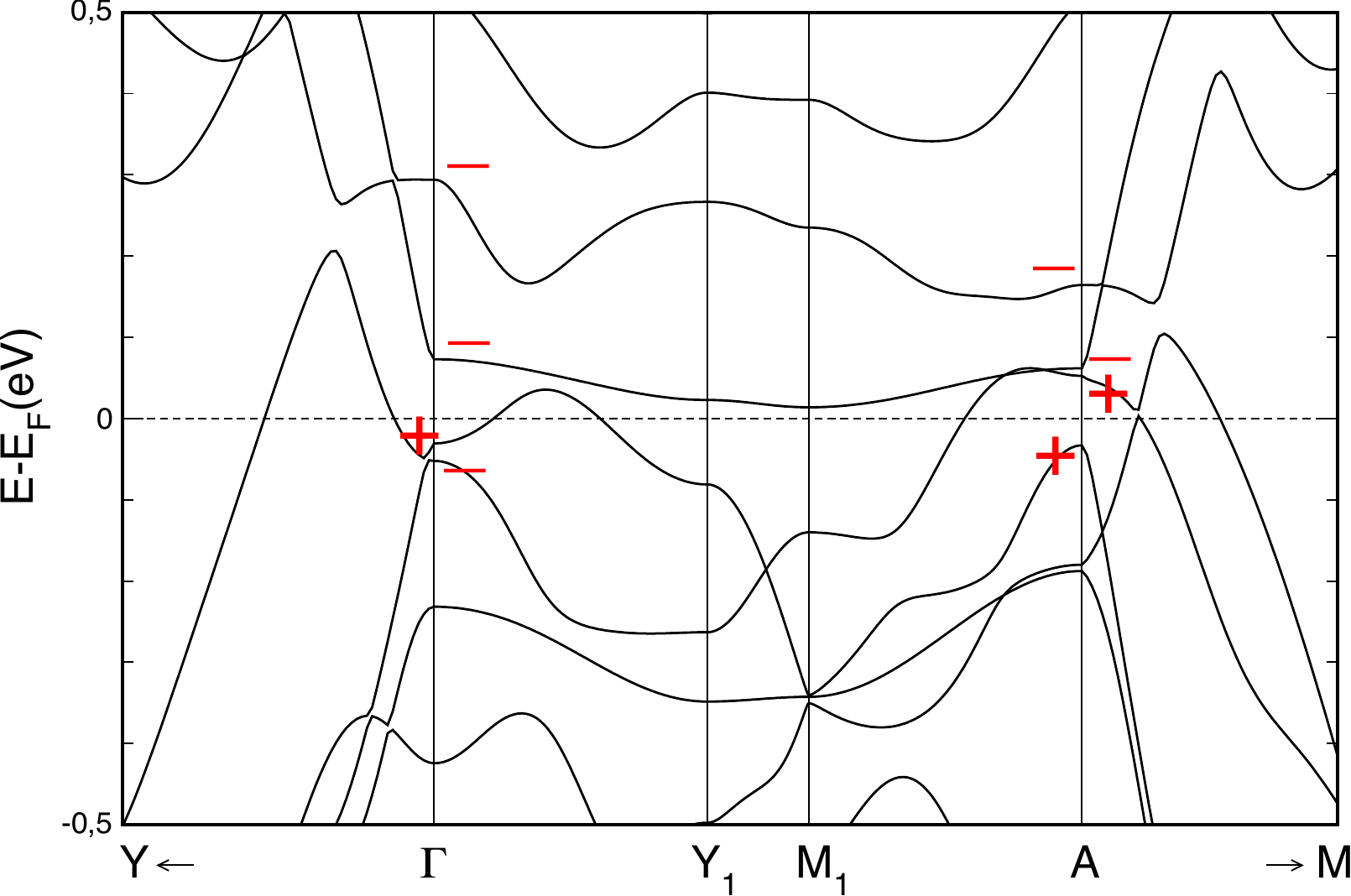} & \hspace{0.5cm} & \includegraphics[height=4.5cm]{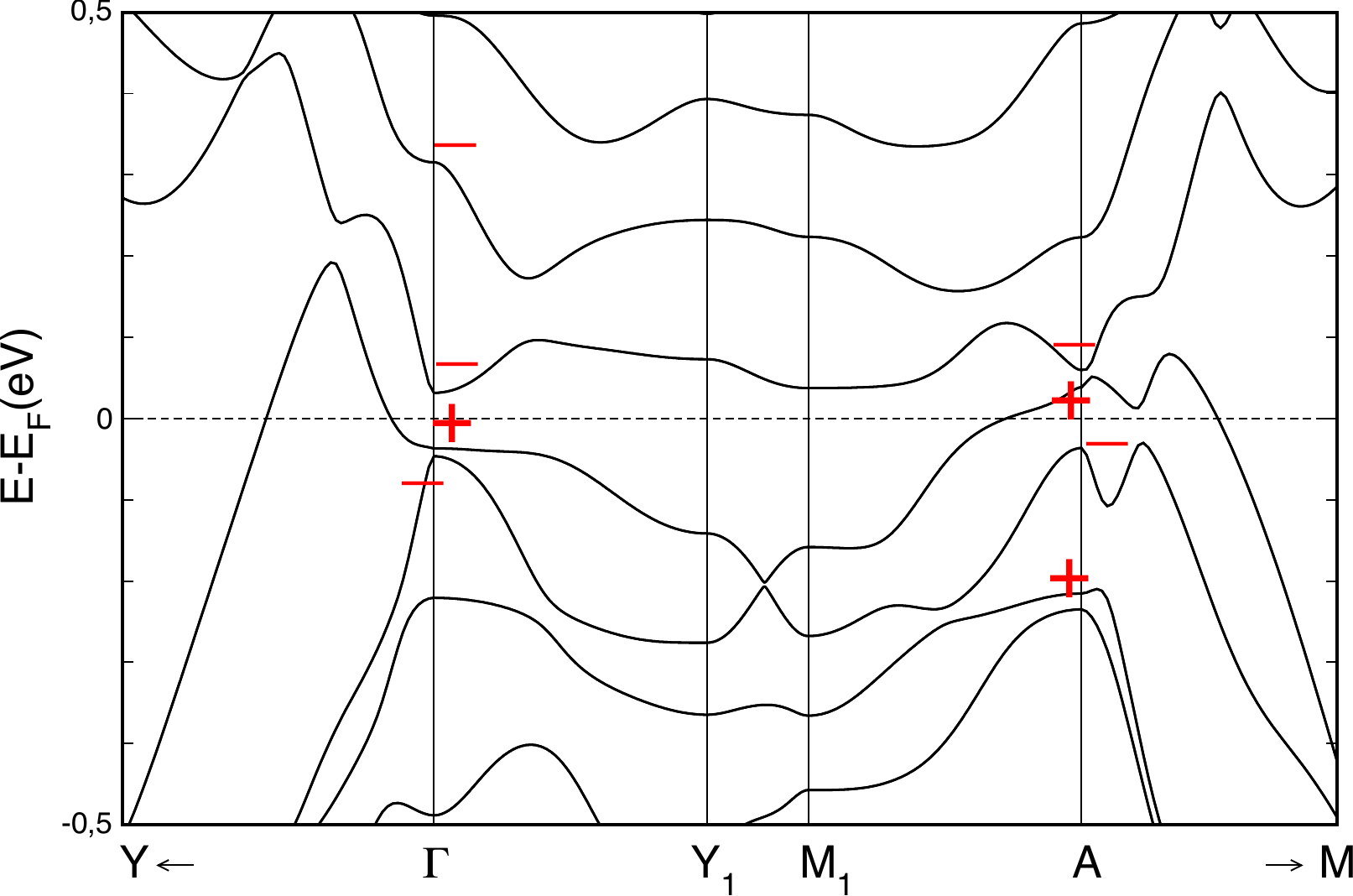}\\
\end{tabular}
\caption{Topological band-inversion transitions in Ta$_2$NiSe$_5$ and Ta$_2$NiSe$_7$.  We analyze the band ordering of Ta$_2$NiSe$_5$ [\icsdweb{61148}, SG 15 ($C2/c$)] and Ta$_2$NiSe$_7$ [\icsdweb{61352}, SG 12 ($C2/m$)] with and without SOC.  In all panels in this figure, the red $\pm$ symbols respectively indicate Kramers pairs of states with positive and negative parity (inversion) eigenvalues.  (A,B) The electronic band structure of Ta$_2$NiSe$_5$ respectively neglecting and incorporating the effects of SOC.  Without SOC, (A) Ta$_2$NiSe$_5$ exhibits trivial symmetry-indicated stable topology at $E_{F}$ (\emph{i.e.} Ta$_2$NiSe$_5$ is LCEBR-classified without SOC).  When SOC is reintroduced in (B), it drives a band inversion at $\Gamma$, transitioning Ta$_2$NiSe$_5$ into a $Z_{4}=3$ 3D TI phase.  (C,D) The electronic band structure of Ta$_2$NiSe$_7$ respectively neglecting and incorporating the effects of SOC.  Unlike in Ta$_2$NiSe$_5$ (A), the bands at $\Gamma$ and $A$ [$k_{x,y}=0$, $k_{z}=0,\pi$, respectively] in Ta$_2$NiSe$_7$ (C) are already inverted by orbital coupling in the $xy$- ($ab$-) plane, taking the Ta$_2$NiSe$_7$ chains to be oriented along the $z$- ($c$-axis) direction (see~\icsdweb{61352}).  Hence without SOC, Ta$_2$NiSe$_7$ realizes an ES-classified nodal-line-semimetal phase with the same parity eigenvalues as a $z$-directed weak TI [$(z_{2,1},z_{2,2},z_{2,3},z_{4})=(0012)$ when subduced onto SG 2 ($P\bar{1}$), see~\AppPhysicalMeaningnoSOCNLCSEBR and Refs.~\cite{FuKaneMele,FuKaneInversion,YoungkukLineNode,ZhidaSemimetals}].  When SOC is reintroduced, the nodal lines at $E_{F}$ in Ta$_2$NiSe$_7$ become gapped, the band inversion at $A$ is removed by SOC, and the bands at $\Gamma$ remain inverted, overall driving Ta$_2$NiSe$_7$ into a $Z_{4}=3$ 3D TI phase.}
\label{fig:TaNiSeparity}
\end{figure*}

We next highlight the materials in which our investigations have revealed or provided a different context for topological features within an experimentally accessible range from $E_{F}$ -- an extensive tabulation of the most idealized realizations in nature of all previously-established classes of symmetry-indicated TI, TCI, and TSM phases in the presence of SOC is further provided in~\AppMaterialsSelection.  To begin, in Fig.~\ref{fig:good-bands}, we present experimentally favorable material candidates with stable and fragile topology at $E_{F}$.  First, in Fig.~\ref{fig:good-bands}A, we show the electronic band structure of MoGe$_2$ [\icsdweb{76139}, SG 139 ($I4/mmm$)] -- a higher-order-topological, centrosymmetric TSM~\cite{HingeSM} in which the doubly-degenerate valence and conduction bands that meet along $\Gamma M_{1}$ in fourfold Dirac points at $E_{F}$, along with the next-highest doubly-degenerate valence bands, are as a set fragile topological [\emph{i.e.} the two conduction bands and four valence bands closest to $E_{F}$ in Fig.~\ref{fig:good-bands}A as a set exhibit nontrivial fragile SIs].  Next, in Fig.~\ref{fig:good-bands}B, we highlight HgBa$_2$CuO$_4$ [\icsdweb{75720}, SG 123 ($P4/mmm$)] -- a well-studied high temperature cuprate superconductor~\cite{HgBaCuONature}, which we find to be an ideal ESFD-classified metal with a doubly-degenerate, half-filled band at $E_{F}$.  Then, in Fig.~\ref{fig:good-bands}C and D, we respectively show the band structures of the transition-metal chalcogenides (TMCs) TaSe$_2$ [\icsdweb{24313}, SG 164 ($P\bar{3}m1$)] and TiS$_2$ [\icsdweb{72042}, SG 227 ($Fd\bar{3}m$)].  Both TaSe$_2$ and TiS$_2$ host well-established 2D charge-density-wave phases~\cite{TaSe2SynthesizeNature,TaSe2CDW,TiS2CDW}; in this work, we discover that the valence bands extending to $\sim 3$~eV below (conduction bands extending to $\sim 1.5$~eV above) $E_{F}$ in 3D TaSe$_2$ (TiS$_2$) are fragile topological.

In Figs.~\ref{fig:good-bands}E and F, we next respectively show the electronic band structures of the closely-related TMCs Ta$_2$NiSe$_5$ [\icsdweb{61148}, SG 15 ($C2/c$)] and Ta$_2$NiSe$_7$ [\icsdweb{61352}, SG 12 ($C2/m$)], which show particular promise for experimental investigations owing to their relatively simple normal-state Fermi surfaces.  Ta$_2$NiSe$_5$ and Ta$_2$NiSe$_7$ belong to a larger family of materials previously highlighted for hosting metal-insulator transitions and strongly-correlated phases of matter. Previous investigations have demonstrated that the layered TMC Ta$_2$NiSe$_5$ hosts an exotic exciton-insulator phase~\cite{TaNiSeExciton1,TaNiSeExciton2}.  By comparing the parity (inversion) eigenvalues of Ta$_2$NiSe$_5$ with and without incorporating the effects of SOC [Fig.~\ref{fig:TaNiSeparity}A and B], we find that the narrow-gap semiconducting state of Ta$_2$NiSe$_5$ in fact realizes a 3D TI phase originating from weak, SOC-driven band inversion at the $\Gamma$ point.  Unlike Ta$_2$NiSe$_5$, Ta$_2$NiSe$_7$ [Fig.~\ref{fig:good-bands}F] is instead a quasi-1D TMC, and has been previously highlighted for exhibiting a charge-density-wave instability~\cite{Ta2NiSe7OriginalCDW}.  Ta$_2$NiSe$_7$ is closely related to the ES-classified, structurally chiral, quasi-1D TMC Weyl semimetal Ta$_2$ISe$_8$ [\icsdweb{35190}, SG 97 ($I422$), see~\AppES], which was shown in recent experimental works to exhibit a topological (axionic) charge-density wave that competes with a superconducting phase under applied pressure~\cite{CDWWeyl,AxionCDWExperiment,TaSeIpressureSC}.  In this work, we find that the room-temperature phase of Ta$_2$NiSe$_7$ is a 3D TI originating from a combination of anisotropic orbital coupling and SOC-driven band inversion.  Specifically, we find that without SOC, Ta$_2$NiSe$_7$ is an ES-classified nodal-line semimetal with band inversions at $k_{x,y}=0$, $k_{z}=0,\pi$ [respectively the $\Gamma$ and $A$ points in Fig.~\ref{fig:TaNiSeparity}C] driven by orbital coupling in the $xy$- ($ab$-) plane [taking the Ta$_2$NiSe$_7$ chains to be oriented along the $z$- ($c$-axis) direction, see~\icsdweb{61352}].  This can be seen by recognizing that Ta$_2$NiSe$_7$ without SOC [Fig.~\ref{fig:TaNiSeparity}C] exhibits the same parity eigenvalues as a $z$-directed weak TI [$(z_{2,1},z_{2,2},z_{2,3},z_{4})=(0012)$ when subduced onto SG 2 ($P\bar{1}$), see~\AppPhysicalMeaningnoSOCNLCSEBR and Refs.~\cite{FuKaneMele,FuKaneInversion,YoungkukLineNode,ZhidaSemimetals}].  When SOC is reintroduced, the nodal lines at $E_{F}$ in Ta$_2$NiSe$_7$ become gapped, and the band inversion at $A$ is removed, while the bands at $\Gamma$ remain inverted [see Fig.~\ref{fig:TaNiSeparity}D].

\begin{figure*}
\centering
\includegraphics[width=0.95\textwidth,angle=0]{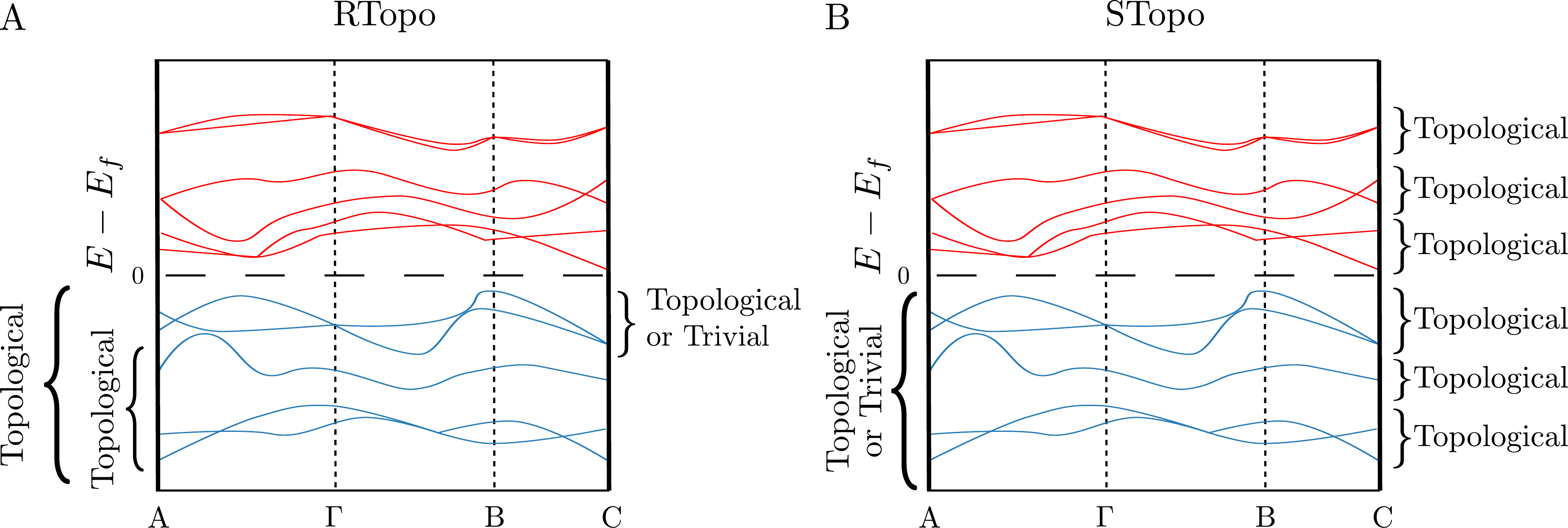}
\vspace{0.5cm}\\

\begin{tabular}{ccc}
C\hspace{0.2cm} \footnotesize{$\rm{Bi}_2 \rm{Mg}_{3}$ - \icsdweb{659569} - SG 164 ($P\bar{3}m1$) - SEBR} & \hspace{0.5cm} & D\hspace{0.2cm} \footnotesize{$(0001)$-Surface Spectrum of Bi$_2$Mg$_3$}\\
\scriptsize{ $\;Z_{2,1}=0\;Z_{2,2}=0\;Z_{2,3}=0\;Z_4=3$} & \hspace{0.5cm} &\\
\includegraphics[height=5.2cm,angle=0]{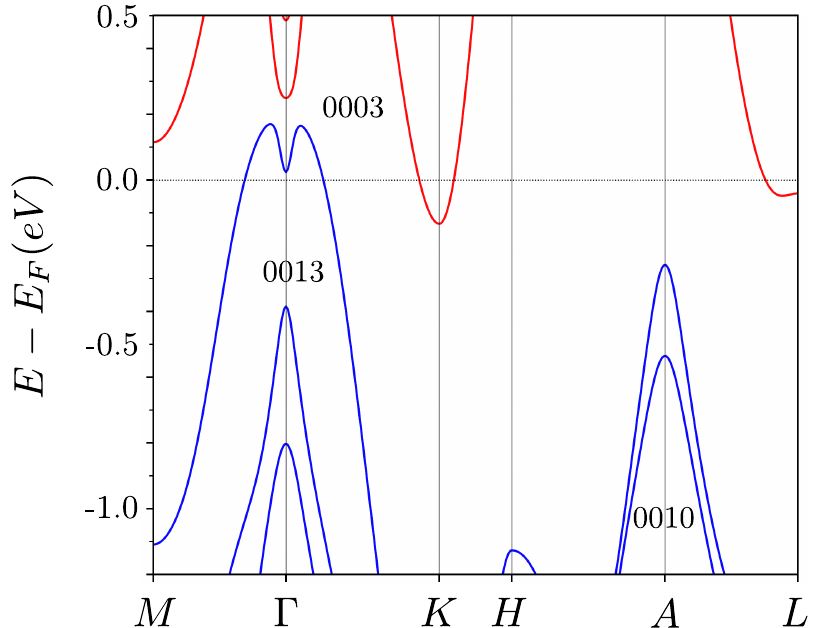} & \hspace{0.5cm} & \includegraphics[height=5.2cm,angle=0]{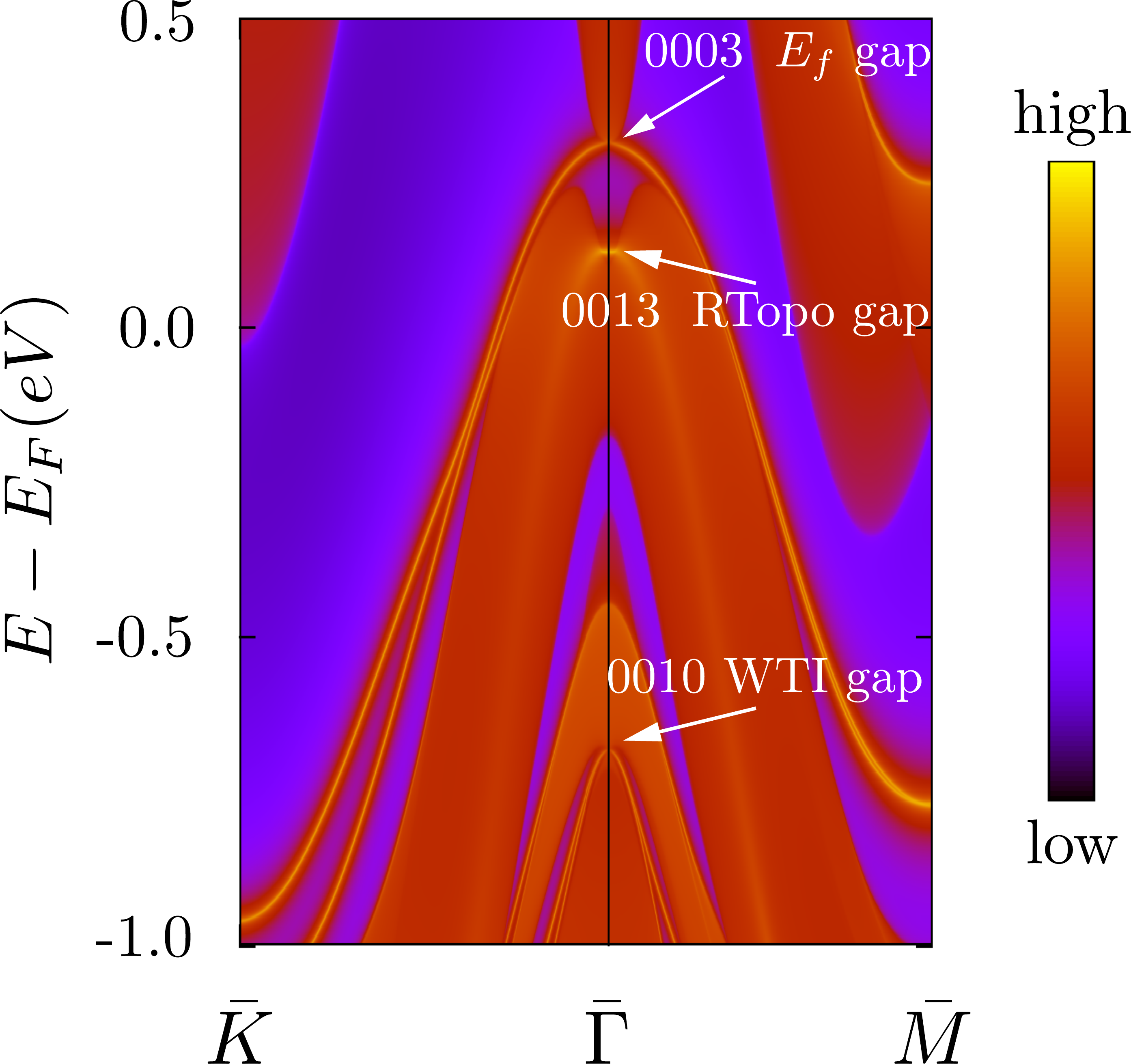}
\end{tabular}
\caption{Repeat-topology and supertopology in Bi$_2$Mg$_3$.  (A) A schematic depiction of a repeat-topological (RTopo) band structure.  In RTopo materials, the gap at the Fermi level, as well as the next gap below $E_{F}$ as measured by band connectivity through TQC (see~\AppTQCReviewAppendix), exhibit cumulative symmetry-indicated stable topology (see \AppDefRTopo).  By this definition -- which is motivated by the experimental accessibility of topological gaps and boundary states below $E_{F}$ -- we note that it is possible for the isolated bands between $E_{F}$ and the next-highest gap below $E_{F}$ to be topologically trivial.  (B) A schematic depiction of a supertopological (STopo) band structure.  In STopo materials, \emph{every} energetically isolated set of bands in the spectrum exhibits symmetry-indicated stable topology (though the system at $E_{F}$ is still free to be a TSM or exhibit cumulative trivial topology, see \AppDefSTopo).  (C) Bulk band structure of Bi$_2$Mg$_3$ [\icsdweb{659569}, SG 164 ($P\bar{3}m1$)], an experimentally-established 3D TI~\cite{Mg3Bi2RTopo1} revealed by our investigations to additionally be RTopo and STopo.  The numbers between the stable topological bands in (C) indicate the cumulative stable SIs of each band gap in the order $Z_{2w,1}$,$Z_{2w,2}$,$Z_{2w,3}$,$Z_{4}$ [see Refs.~\cite{FuKaneInversion,AshvinIndicators,ChenTCI,AshvinTCI,TMDHOTI,MTQC} and \AppMaterialsSelection for the physical meaning of the stable SIs in SG 164 ($P\bar{3}m1$)].  (D) $(0001)$-surface states of Bi$_2$Mg$_3$ obtained from surface Green's functions (see \AppDefRTopo for calculation details).  We have labeled the surface states using the cumulative stable SIs of
  the projected bulk gaps (as determined by band connectivity, see \TabBandCharacterizationBiTwoMgThree in~\AppCheckTopoAppendix for further details).  Previous ARPES investigations of Bi$_2$Mg$_3$ have revealed the existence of ``surface resonance bands'' below $E_{F}$ at the $(0001)$-surface $\Gamma$ point~\cite{Mg3Bi2RTopo1}.  We find that the surface resonance bands in Bi$_2$Mg$_3$ are in fact RTopo Dirac-cone surface states in the first gap below $E_{F}$ [the ``$0013$ RTopo gap''].  The surface Green's function calculations exhibit additional twofold-Dirac cone surface states within the projected bulk gap $\sim 0.8$~eV below $E_{F}$ [the ``0010 WTI gap''].  Although the earlier ARPES experiments also observed surface states in the projected 0010 gap in Bi$_2$Mg$_3$~\cite{Mg3Bi2RTopo1}, we emphasize that the surface states in the 0010 gap in (D) are either trivial or originate from non-symmetry-indicated stable topology, because the cumulative stable SIs $(0010)$ characterize an (obstructed) weak TI phase that does not generically exhibit anomalous twofold Dirac cones on $z$-normal [$(0001)$-] surfaces (see~\AppZFourTrivial).}
\label{fig:STopoMain}
\end{figure*}

Finally, in this work, we have additionally performed a high-throughput search for topological bands \emph{away from $E_{F}$}.  Our calculations have revealed the existence of two additional classes of topological materials: RTopo materials with stable topological insulating gaps at and just below $E_{F}$ [see Fig.~\ref{fig:STopoMain}A and~\AppDefRTopo], and STopo materials in which \emph{every} energetically isolated set of bands above the core shell exhibits symmetry-indicated stable topology [see Fig.~\ref{fig:STopoMain}B and~\AppDefSTopo].  From a physical perspective, STopo phases result from a combination of EBR splitting, band backfolding from unit-cell enlargement, and favorable hopping parameters -- hence, the identification of an STopo phase in a particular material can be sensitive to sample and calculation details.  In~\AppDefSTopo, we highlight the example of the experimentally-established higher-order TI rhombohedral bismuth [\icsdweb{64703}, SG 166 ($R\bar{3}m$)]~\cite{HOTIBismuth}, for which some ICSD entries (\emph{e.g.} \icsdweb{64703}) are STopo, whereas others (\emph{e.g.} \icsdweb{53797}) are not.  Crucially, Bi (\icsdweb{64703}) and Bi (\icsdweb{53797}) differ only by weak band inversion away from $E_{F}$, and both exhibit topologically nontrivial gaps at $E_{F}$ and just below $E_{F}$.  This provides a physical motivation for introducing a second topological class -- RTopo -- for materials like Bi (\icsdweb{64703}).  In idealized RTopo materials, there exist two sets of surface or hinge states within consecutive bulk gaps that both lie at energies accessible to ARPES probes without doping ($\sim 1.5$~eV below $E_{F}$), a property that is crucially \emph{insensitive} to band inversions at experimentally inaccessible energies far from $E_{F}$.  The repeated topological surface and hinge states of RTopo materials are analogous to the Fermi-arc ``quantum ladder'' recently observed in the unconventional chiral semimetal alloy Rh$_x$Ni$_y$Si~\cite{FermiArcQuantumLadder} (see~\AppDefRTopo for further details).  Lastly, we emphasize that in principle, the STopo and RTopo classification of materials can be extended from stable topological states to fragile topology -- however, we did not in this study encounter any ICSD entries with super- or repeat-fragile topology.

In Fig.~\ref{fig:STopoMain}C and D, we show the bulk and surface states of the RTopo and STopo material Bi$_2$Mg$_3$ [\icsdweb{659569}, SG 164 ($P\bar{3}m1$)].  Previous theoretical investigations and spectroscopic experiments have determined Bi$_2$Mg$_3$ to be a 3D TI with ``surface resonance bands'' below $E_{F}$ at the $(0001)$-surface $\Gamma$ point~\cite{Mg3Bi2RTopo1}. We find that the surface resonance bands detected in Ref. ~\cite{Mg3Bi2RTopo1} are in fact the RTopo Dirac-cone surface states of Bi$_2$Mg$_3$ in the first gap below $E_{F}$ [the ``$0013$ RTopo gap'' in Fig.~\ref{fig:STopoMain}D].  In~\AppRTopoMaterials~and \AppStatForStopo respectively, we provide detailed tabulations of RTopo and STopo materials in the ICSD.  Taken together, the large number of experimentally accessible materials in the ICSD with stable topological bands away from $E_{F}$ (see~\AppTopoBands), as well as the relative preponderance of RTopo and STopo materials (see~\AppRTopoMaterials~and \AppStatForStopo, respectively), suggest that many of the surface resonances detected in previous ARPES investigations may in fact be topological surface states protected by spectral flow below $E_{F}$.   

\section*{Discussion}

We have performed a complete calculation of symmetry-indicated stable and fragile band topology at and away from $E_{F}$ across all known stoichiometric, solid-state, nonmagnetic materials.  We have discovered that over half of the materials in nature are stable TIs and TSMs at $E_{F}$, and nearly $2/3$ of bands across all materials exhibit symmetry-indicated stable topology.  Our investigations raise several questions.  First, there exist well-established classes of noncentrosymmetric TI and TCI phases that cannot be diagnosed through SIs.  For example, depending on experimental or first-principles calculation details, WTe$_2$ [\icsdweb{14348}, SG 31 ($Pmn2_{1}$)] can realize an LCEBR-classified TSM phase with tilted Weyl cones~\cite{AlexeyType2}, or a non-symmetry-indicated twofold-rotation-anomaly TCI phase~\cite{TMDHOTI}.  Additionally, like stable topological phases with trivial SIs, there also exist non-symmetry-indicated fragile phases that evade the methods employed in this work~\cite{HingeSM}.  It remains an important outstanding question whether computationally-efficient methods beyond TQC and SIs can be formulated for the high-throughput discovery of non-symmetry-indicated topological materials.  Several promising studies have recently introduced alternative, scalable methods for high-throughput topological materials identification~\cite{S4HighThroughput,SOCSpillTavazzaMag}, suggesting a future in which an even larger percentage of materials in nature are recognized as topologically nontrivial.  Furthermore, while we have focused our efforts on nonmagnetic topological materials, recent studies have expanded the methods of TQC and SIs to magnetic systems~\cite{MTQC,AshvinMagnet}, facilitating the high-throughput identification of over 100 magnetic materials with nontrivial topology at $E_{F}$~\cite{MTQCmaterials}.  Following the results of this work, future investigations may be directed towards identifying nontrivial electronic band topology away from $E_{F}$ in magnetic materials, in particular targeting bands in which the topology and magnetism are coupled and experimentally tunable.

Lastly, we note that during the final stages of preparing this work, the 3D TMCs Ta$_2$M$_3$Te$_5$ (M$=$Pd,Ni) -- which are closely related to the TMCs Ta$_2$NiSe$_5$ and Ta$_2$NiSe$_7$ discovered to be 3D TIs in this work -- were theoretically predicted in Ref.~\cite{ZhijunTMCTheory} to be non-symmetry-indicated mirror TCIs.  Experimental evidence of a 2D TI phase in Ta$_2$Pd$_3$Te$_5$ monolayers was also reported in Ref.~\cite{ZhijunTMCExp}.

\section*{Materials and Methods}
	
The first-principles calculations in this work were performed using the Vienna Ab-initio Simulation Package (VASP) \cite{vasp1,vasp2,PhysRevB.48.13115}. We treated the interaction between the ion cores and the valence electrons using the projector augmented-wave method~\cite{paw1}, and we used the generalized gradient approximation (GGA) with the Perdew-Burke-Ernzerhof parameterization for the exchange-correlation potential~\cite{PhysRevLett.77.3865}. For each material calculation, we used as input the structural parameters reported on the ICSD\cite{ICSD}. For calculations incorporating the effects of SOC, we accounted for the effects of SOC using the second variation method~\cite{PhysRevB.62.11556}. For the self-consistent calculations (SCC) of the charge density and the density of states, we used a grid of $11\times 11 \times 11 \;k$ points centered at the $\Gamma$  point ($k=0$).  For the band structure calculations along high-symmetry lines and planes, we then used the input from the SCC.  Each segment of the $k$ path contained 20 $k$ points. At each $k$ point along each path segment, we specifically calculated the energies of at least $2N_e$ Bloch states, where $N_e$, is the number of valence electrons in the primitive cell; we did not include bands originating from core-shell atomic orbitals. To identify the symmetry-indicated stable and fragile topological bands in each analyzed material, we then used the VASP2Trace \cite{V2TREF}\cite{Irvsp2020} program and an implementation of the Check Topological Mat \cite{CHECKTOPOREF} program updated for this work (see \AppCheckTopoAppendix for further details).  Lastly, to compute the $(0001)$-surface spectrum of Bi$_2$Mg$_3$ in Fig.~\ref{fig:STopoMain}D, we used WANNIER90~\cite{Pizzi_2020} to construct a Wannier-based tight-binding model from the $s$ and $p$ orbitals of Mg and the $p$ orbitals of Bi, which we found to accurately reproduce the bulk electronic band structure of Bi$_2$Mg$_3$ (see \AppDefRTopo for further calculation details).  

\section*{Acknowledgments}

We are grateful to Zhijun Wang for collaboration on previous related works and for writing the~\vasptotrace~program~\cite{Irvsp2020} extensively used in this study (see~\AppCheckTopoAppendix).  We further acknowledge helpful discussions with Mois I. Aroyo, Barry Bradlyn, Jennifer Cano, Leslie Schoop, Zhida Song, and Yuanfeng Xu. M.G.V. and N.R. additionally thank B.Y. Scarpin for enlightening discussions on VASP and CIF files. We acknowledge the computational resources Cobra/Draco in the Max Planck Computing and Data Facility (MPCDF) and Atlas in the Donostia International Physics Center (DIPC).  This research was conducted using the resources of the National Energy Research Scientific Computing Center (NERSC), a U.S. Department of Energy Office of Science User Facility operated under Contract No. DE-AC02-05CH11231.  B.J.W., B.A.B., and N.R. were primarily supported by the U.S. Department of Energy (Grant No. DE-SC0016239), and were partially supported by the National Science Foundation (EAGER Grant No. DMR 1643312), a Simons Investigator grant (No. 404513), the Office of Naval Research (ONR Grant No. N00014-20-1-2303), the NSF-MRSEC (Grant No. DMR-142051), the Packard Foundation, the Schmidt Fund for Innovative Research, the BSF Israel US foundation (Grant No. 2018226), the Gordon and Betty Moore Foundation through Grant No. GBMF8685 towards the Princeton theory program, and a Guggenheim Fellowship from the John Simon Guggenheim Memorial Foundation.  L.E. was supported by the Government of the Basque Country (Project IT1301-19) and the Spanish Ministry of Science and Innovation (PID2019-106644GB-I00).  M.G.V. acknowledges the Spanish Ministerio de Ciencia e Innovacion (Grant No. PID2019-109905GB-C21).  C.F. was supported by the European Research Council (ERC) Advanced Grant No.  742068 ``TOP-MAT'', Deutsche Forschungsgemeinschaft (DFG) through SFB 1143, and the Würzburg-Dresden Cluster of Excellence on Complexity and Topology in Quantum Matter-ct.qmat (EXC 2147, Project No. 390858490).

\section*{Author contributions}

All authors contributed equally to the intellectual content of this work.  N.R.and M.G.V.performed the high-throughput first-principles calculations with help from L.E., B.A.B., and B.J.W. The topological analysis of the material calculations was performed by B.J.W.and M.G.V.with help from L.E. and B.A.B. B.J.W., M.G.V., and N.R. proposed the existence of RTopo and STopo phases and identified material candidates. The representative topological materials listed in SM 11 were manually selected by B.J.W. and M.G.V. with help from N.R., and the most experimentally relevant candidates were identified by C.F., S.S.P., M.G.V., and B.J.W. Upgrades to the \webTQC~\cite{DBREF} were implemented by N.R. The main text was written by M.G.V., B.J.W., B.A.B., and N.R. with help from all of the authors. The Supplementary Material was written by B.J.W., M.G.V., and N.R. This study was conceived by B.A.B., M.G.V. and N.R. N.R. was responsible for the overall research direction. 

\section*{Competing interests:}

The authors declare that they have no competing interests.

\section*{Data and materials availability} 

The data supporting the findings of this study are available within the paper and on the \webTQC~\cite{DBREF}. We provide the data of every figure and table within the paper in an open format on Zenodo\cite{Zenodo}. Additional information regarding the data generated for this study is available from the corresponding authors upon reasonable request. the Vienna Ab-initio Simulation Package (VASP)\cite{vasp1,vasp2,PhysRevB.48.13115} is a commercial software VASP availalable at \href{https://www.vasp.at}{https://www.vasp.at}.

\clearpage
\onecolumngrid

\bibliographystyle{Science}
\bibliography{140KMaterials_20210519_NR.bib,internal.bib}

\begin{thebibliography}{100}

\bibitem{WiederReview}
B.~J. Wieder, {\it et~al.\/}, {\it Nature Reviews Materials\/}  (2021).

\bibitem{CharlieZahidReview}
M.~Z. Hasan, C.~L. Kane, {\it Rev. Mod. Phys.\/} {\bf 82}, 3045 (2010).

\bibitem{KaneMeleZ2}
C.~L. Kane, E.~J. Mele, {\it Phys. Rev. Lett.\/} {\bf 95}, 146802 (2005).

\bibitem{AndreiTI}
B.~A. Bernevig, T.~L. Hughes, S.-C. Zhang, {\it Science\/} {\bf 314}, 1757
  (2006).

\bibitem{FuKaneMele}
L.~Fu, C.~L. Kane, E.~J. Mele, {\it Phys. Rev. Lett.\/} {\bf 98}, 106803
  (2007).

\bibitem{ZJDirac}
Z.~Wang, H.~Weng, Q.~Wu, X.~Dai, Z.~Fang, {\it Phys. Rev. B\/} {\bf 88}, 125427
  (2013).

\bibitem{DDP}
B.~J. Wieder, Y.~Kim, A.~M. Rappe, C.~L. Kane, {\it Phys. Rev. Lett.\/} {\bf
  116}, 186402 (2016).

\bibitem{NewFermions}
B.~Bradlyn, {\it et~al.\/}, {\it Science\/} {\bf 353} (2016).

\bibitem{CoSiArc}
P.~Tang, Q.~Zhou, S.-C. Zhang, {\it Phys. Rev. Lett.\/} {\bf 119}, 206402
  (2017).

\bibitem{WeylReview}
N.~P. Armitage, E.~J. Mele, A.~Vishwanath, {\it Rev. Mod. Phys.\/} {\bf 90},
  015001 (2018).

\bibitem{LiangTCIOriginal}
L.~Fu, {\it Phys. Rev. Lett.\/} {\bf 106}, 106802 (2011).

\bibitem{HsiehTCI}
T.~H. Hsieh, {\it et~al.\/}, {\it Nature Communications\/} {\bf 3}, 982 (2012).

\bibitem{HourglassInsulator}
Z.~Wang, A.~Alexandradinata, R.~J. Cava, B.~A. Bernevig, {\it Nature\/} {\bf
  532}, 189 (2016).

\bibitem{DiracInsulator}
B.~J. Wieder, {\it et~al.\/}, {\it Science\/} {\bf 361}, 246 (2018).

\bibitem{WladPump}
W.~A. Benalcazar, B.~A. Bernevig, T.~L. Hughes, {\it Phys. Rev. B\/} {\bf 96},
  245115 (2017).

\bibitem{ChenTCI}
Z.~Song, T.~Zhang, Z.~Fang, C.~Fang, {\it Nature Communications\/} {\bf 9},
  3530 (2018).

\bibitem{AshvinTCI}
E.~Khalaf, H.~C. Po, A.~Vishwanath, H.~Watanabe, {\it Phys. Rev. X\/} {\bf 8},
  031070 (2018).

\bibitem{HsiehBismuthSelenide}
D.~Hsieh, {\it et~al.\/}, {\it Nature\/} {\bf 460}, 1101 (2009).

\bibitem{CavaDirac1}
S.~Borisenko, {\it et~al.\/}, {\it Phys. Rev. Lett.\/} {\bf 113}, 027603
  (2014).

\bibitem{ZahidWeyl}
S.-Y. Xu, {\it et~al.\/}, {\it Science\/} {\bf 349}, 613 (2015).

\bibitem{AndreiWeyl}
H.~Weng, C.~Fang, Z.~Fang, B.~A. Bernevig, X.~Dai, {\it Phys. Rev. X\/} {\bf
  5}, 011029 (2015).

\bibitem{HOTIBismuth}
F.~Schindler, {\it et~al.\/}, {\it Nature Physics\/} {\bf 14}, 918 (2018).

\bibitem{CoSiObserveHasan}
D.~S. Sanchez, {\it et~al.\/}, {\it Nature\/} {\bf 567}, 500 (2019).

\bibitem{CoSiObserveChina}
Z.~Rao, {\it et~al.\/}, {\it Nature\/} {\bf 567}, 496 (2019).

\bibitem{PdGaObserve}
N.~B.~M. Schr{\"o}ter, {\it et~al.\/}, {\it Science\/} {\bf 369}, 179 (2020).

\bibitem{CDWWeyl}
W.~Shi, {\it et~al.\/}, {\it Nature Physics\/} {\bf 17}, 381 (2021).

\bibitem{AxionCDWExperiment}
J.~Gooth, {\it et~al.\/}, {\it Nature\/} {\bf 575}, 315 (2019).

\bibitem{QuantumChemistry}
B.~Bradlyn, {\it et~al.\/}, {\it Nature\/} {\bf 547}, 298 (2017).

\bibitem{MTQC}
L.~Elcoro, {\it et~al.\/}, {\it Nature Communications\/} {\bf 12}, 5965 (2021).

\bibitem{FuKaneInversion}
L.~Fu, C.~L. Kane, {\it Phys. Rev. B\/} {\bf 76}, 045302 (2007).

\bibitem{YoungkukLineNode}
Y.~Kim, B.~J. Wieder, C.~L. Kane, A.~M. Rappe, {\it Phys. Rev. Lett.\/} {\bf
  115}, 036806 (2015).

\bibitem{SlagerSymmetry}
J.~Kruthoff, J.~de~Boer, J.~van Wezel, C.~L. Kane, R.-J. Slager, {\it Phys.
  Rev. X\/} {\bf 7}, 041069 (2017).

\bibitem{AshvinIndicators}
H.~C. Po, A.~Vishwanath, H.~Watanabe, {\it Nature Communications\/} {\bf 8}, 50
  (2017).

\bibitem{TMDHOTI}
Z.~Wang, B.~J. Wieder, J.~Li, B.~Yan, B.~A. Bernevig, {\it Phys. Rev. Lett.\/}
  {\bf 123}, 186401 (2019).

\bibitem{AshvinMagnet}
H.~Watanabe, H.~C. Po, A.~Vishwanath, {\it Science Advances\/} {\bf 4} (2018).

\bibitem{ZhidaSemimetals}
Z.~Song, T.~Zhang, C.~Fang, {\it Phys. Rev. X\/} {\bf 8}, 031069 (2018).

\bibitem{MTQCmaterials}
Y.~Xu, {\it et~al.\/}, {\it Nature\/} {\bf 586}, 702 (2020).

\bibitem{AndreiMaterials}
M.~G. Vergniory, {\it et~al.\/}, {\it Nature\/} {\bf 566}, 480 (2019).

\bibitem{ChenMaterials}
T.~Zhang, {\it et~al.\/}, {\it Nature\/} {\bf 566}, 475 (2019).

\bibitem{AshvinMaterials}
F.~Tang, H.~C. Po, A.~Vishwanath, X.~Wan, {\it Nature\/} {\bf 566}, 486 (2019).

\bibitem{ICSD}
G.~Bergerhoff, R.~Hundt, R.~Sievers, I.~D. Brown, {\it Journal of Chemical
  Information and Computer Sciences\/} {\bf 23}, 66 (1983).

\bibitem{JenFragile1}
J.~Cano, {\it et~al.\/}, {\it Phys. Rev. Lett.\/} {\bf 120}, 266401 (2018).

\bibitem{AshvinFragile}
H.~C. Po, H.~Watanabe, A.~Vishwanath, {\it Phys. Rev. Lett.\/} {\bf 121},
  126402 (2018).

\bibitem{ZhidaFragile}
Z.-D. Song, L.~Elcoro, Y.-F. Xu, N.~Regnault, B.~A. Bernevig, {\it Phys. Rev.
  X\/} {\bf 10}, 031001 (2020).

\bibitem{KoreanFragileInversion}
Y.~Hwang, J.~Ahn, B.-J. Yang, {\it Phys. Rev. B\/} {\bf 100}, 205126 (2019).

\bibitem{ZhidaFragile2}
Z.-D. Song, L.~Elcoro, B.~A. Bernevig, {\it Science\/} {\bf 367}, 794 (2020).

\bibitem{ZhidaBLG}
Z.~Song, {\it et~al.\/}, {\it Phys. Rev. Lett.\/} {\bf 123}, 036401 (2019).

\bibitem{AshvinBLG2}
H.~C. Po, L.~Zou, T.~Senthil, A.~Vishwanath, {\it Phys. Rev. B\/} {\bf 99},
  195455 (2019).

\bibitem{DBREF}
\webTQC~\webNoICSD.

\bibitem{BasovNatMaterReview}
D.~N. Basov, R.~D. Averitt, D.~Hsieh, {\it Nature Materials\/} {\bf 16}, 1077
  (2017).

\bibitem{SuperfluidBoundChern}
S.~Peotta, P.~T{\"o}rm{\"a}, {\it Nature Communications\/} {\bf 6}, 8944
  (2015).

\bibitem{SuperfluidBoundFragileAndrei}
F.~Xie, Z.~Song, B.~Lian, B.~A. Bernevig, {\it Phys. Rev. Lett.\/} {\bf 124},
  167002 (2020).

\bibitem{V2TREF}
VASP2Trace~\href{https://github.com/zjwang11/irvsp}{https://github.com/zjwang11/irvsp}.

\bibitem{CHECKTOPOREF}
Check Topological
  Mat~\href{https://www.cryst.ehu.es/cryst/checktopologicalmat}{https://www.cryst.ehu.es/cryst/checktopologicalmat}.

\bibitem{BigBook}
C.~Bradley, A.~Cracknell, {\it The Mathematical Theory of Symmetry in Solids:
  Representation Theory for Point Groups and Space Groups\/} (Clarendon Press,
  1972).

\bibitem{AlexeyVDBWannier}
A.~A. Soluyanov, D.~Vanderbilt, {\it Phys. Rev. B\/} {\bf 83}, 035108 (2011).

\bibitem{Bandrep1}
L.~Elcoro, {\it et~al.\/}, {\it Journal of Applied Crystallography\/} {\bf 50},
  1457 (2017).

\bibitem{HingeSM}
B.~J. Wieder, {\it et~al.\/}, {\it Nature Communications\/} {\bf 11}, 627
  (2020).

\bibitem{HingeSMExp}
C.-Z. Li, {\it et~al.\/}, {\it Phys. Rev. Lett.\/} {\bf 124}, 156601 (2020).

\bibitem{Marzari2DTIAbundantDB}
A.~Marrazzo, M.~Gibertini, D.~Campi, N.~Mounet, N.~Marzari, {\it Nano
  Letters\/} {\bf 19}, 8431 (2019).

\bibitem{Ashvin2DMaterials}
D.~Wang, {\it et~al.\/}, {\it Phys. Rev. B\/} {\bf 100}, 195108 (2019).

\bibitem{HgBaCuONature}
Y.~Li, {\it et~al.\/}, {\it Nature\/} {\bf 455}, 372 (2008).

\bibitem{TaSe2SynthesizeNature}
J.~Zhou, {\it et~al.\/}, {\it Nature\/} {\bf 556}, 355 (2018).

\bibitem{TaSe2CDW}
Y.~Chen, {\it et~al.\/}, {\it Nature Physics\/} {\bf 16}, 218 (2020).

\bibitem{TiS2CDW}
K.~Dolui, S.~Sanvito, {\it {EPL} (Europhysics Letters)\/} {\bf 115}, 47001
  (2016).

\bibitem{TaNiSeExciton1}
Y.~F. Lu, {\it et~al.\/}, {\it Nature Communications\/} {\bf 8}, 14408 (2017).

\bibitem{TaNiSeExciton2}
G.~Mazza, {\it et~al.\/}, {\it Phys. Rev. Lett.\/} {\bf 124}, 197601 (2020).

\bibitem{Ta2NiSe7OriginalCDW}
R.~M. Fleming, S.~A. Sunshine, C.~H. Chen, L.~F. Schneemeyer, J.~V. Waszczak,
  {\it Phys. Rev. B\/} {\bf 42}, 4954 (1990).

\bibitem{TaSeIpressureSC}
Q.-G. Mu, {\it et~al.\/}, {\it Phys. Rev. Materials\/} {\bf 5}, 084201 (2021).

\bibitem{FermiArcQuantumLadder}
T.~A. {Cochran}, {\it et~al.\/}, {\it arXiv e-prints\/} p. arXiv:2004.11365
  (2020).

\bibitem{Mg3Bi2RTopo1}
T.-R. Chang, {\it et~al.\/}, {\it Advanced Science\/} {\bf 6}, 1800897 (2019).

\bibitem{AlexeyType2}
A.~A. Soluyanov, {\it et~al.\/}, {\it Nature\/} {\bf 527}, 495 (2015).

\bibitem{S4HighThroughput}
J.~Gao, {\it et~al.\/}, {\it Science Bulletin\/} {\bf 66}, 667 (2021).

\bibitem{SOCSpillTavazzaMag}
K.~Choudhary, K.~F. Garrity, N.~J. Ghimire, N.~Anand, F.~Tavazza, {\it Phys.
  Rev. B\/} {\bf 103}, 155131 (2021).

\bibitem{ZhijunTMCTheory}
Z.~Guo, {\it et~al.\/}, {\it Phys. Rev. B\/} {\bf 103}, 115145 (2021).

\bibitem{ZhijunTMCExp}
X.~{Wang}, {\it et~al.\/}, {\it arXiv e-prints\/} p. arXiv:2012.07293 (2020).

\bibitem{vasp1}
G.~Kresse, J.~Furthmüller, {\it Computational Materials Science\/} {\bf 6}, 15
   (1996).

\bibitem{vasp2}
G.~Kresse, J.~Furthm\"uller, {\it Phys. Rev. B\/} {\bf 54}, 11169 (1996).

\bibitem{PhysRevB.48.13115}
G.~Kresse, J.~Hafner, {\it Phys. Rev. B\/} {\bf 48}, 13115 (1993).

\bibitem{paw1}
G.~Kresse, D.~Joubert, {\it Phys. Rev. B\/} {\bf 59}, 1758 (1999).

\bibitem{PhysRevLett.77.3865}
J.~P. Perdew, K.~Burke, M.~Ernzerhof, {\it Phys. Rev. Lett.\/} {\bf 77}, 3865
  (1996).

\bibitem{PhysRevB.62.11556}
D.~Hobbs, G.~Kresse, J.~Hafner, {\it Phys. Rev. B\/} {\bf 62}, 11556 (2000).

\bibitem{Irvsp2020}
J.~Gao, Q.~Wu, C.~Persson, Z.~Wang, {\it Computer Physics Communications\/}
  {\bf 261}, 107760 (2021).

\bibitem{Pizzi_2020}
G.~Pizzi, {\it et~al.\/}, {\it Journal of Physics: Condensed Matter\/} {\bf
  32}, 165902 (2020).

\bibitem{Zenodo}
Data for "all topological bands of all nonmagnetic stoichiometric materials",
  10.5281/zenodo.6123281.

\bibitem{HOTIChen}
Z.~Song, Z.~Fang, C.~Fang, {\it Phys. Rev. Lett.\/} {\bf 119}, 246402 (2017).

\bibitem{Michel1999}
L.~Michel, J.~Zak, {\it Phys. Rev. B\/} {\bf 59}, 5998 (1999).

\bibitem{Michel2001}
L.~Michel, J.~Zak, {\it Physics Reports\/} {\bf 341}, 377  (2001).

\bibitem{JenOAL}
J.~{Cano}, L.~{Elcoro}, M.~I. {Aroyo}, B.~A. {Bernevig}, B.~{Bradlyn}, {\it
  arXiv e-prints\/} p. arXiv:2107.00647 (2021).

\bibitem{PhysRevLett.45.1025}
J.~Zak, {\it Phys. Rev. Lett.\/} {\bf 45}, 1025 (1980).

\bibitem{Bandrep2}
M.~G. Vergniory, {\it et~al.\/}, {\it Phys. Rev. E\/} {\bf 96}, 023310 (2017).

\bibitem{Bandrep3}
J.~Cano, {\it et~al.\/}, {\it Phys. Rev. B\/} {\bf 97}, 035139 (2018).

\bibitem{BCS1}
M.~I. Aroyo, {\it et~al.\/}, {\it Zeitschrift f{\"u}r Kristallographie -
  Crystalline Materials\/} {\bf 221}, 15  (2006).

\bibitem{BCS2}
M.~I. Aroyo, A.~Kirov, C.~Capillas, J.~M. Perez-Mato, H.~Wondratschek, {\it
  Acta Crystallographica Section A\/} {\bf 62}, 115 (2006).

\bibitem{AshvinFragile2}
S.~Liu, A.~Vishwanath, E.~Khalaf, {\it Phys. Rev. X\/} {\bf 9}, 031003 (2019).

\bibitem{BarryFragile}
B.~Bradlyn, Z.~Wang, J.~Cano, B.~A. Bernevig, {\it Phys. Rev. B\/} {\bf 99},
  045140 (2019).

\bibitem{AdrianFragile}
A.~Bouhon, A.~M. Black-Schaffer, R.-J. Slager, {\it Phys. Rev. B\/} {\bf 100},
  195135 (2019).

\bibitem{KoreanFragile}
J.~Ahn, S.~Park, B.-J. Yang, {\it Phys. Rev. X\/} {\bf 9}, 021013 (2019).

\bibitem{FragileFlowMeta}
V.~Peri, {\it et~al.\/}, {\it Science\/} {\bf 367}, 797 (2020).

\bibitem{DelicateAris}
A.~Nelson, T.~Neupert, T.~c.~v. Bzdu\ifmmode~\check{s}\else \v{s}\fi{}ek,
  A.~Alexandradinata, {\it Phys. Rev. Lett.\/} {\bf 126}, 216404 (2021).

\bibitem{WiederAxion}
B.~J. {Wieder}, B.~A. {Bernevig}, {\it ArXiv e-prints\/}  (2018).

\bibitem{KitaevClass}
A.~Kitaev, {\it AIP Conference Proceedings\/} {\bf 1134}, 22 (2009).

\bibitem{QHZ}
X.-L. Qi, T.~L. Hughes, S.-C. Zhang, {\it Phys. Rev. B\/} {\bf 78}, 195424
  (2008).

\bibitem{ChenRotation}
C.~Fang, L.~Fu, {\it Science Advances\/} {\bf 5} (2019).

\bibitem{HermeleSymmetry}
S.-J. Huang, H.~Song, Y.-P. Huang, M.~Hermele, {\it Phys. Rev. B\/} {\bf 96},
  205106 (2017).

\bibitem{WladCorners}
W.~A. Benalcazar, T.~Li, T.~L. Hughes, {\it Phys. Rev. B\/} {\bf 99}, 245151
  (2019).

\bibitem{Fidkowski2011}
L.~Fidkowski, T.~S. Jackson, I.~Klich, {\it Phys. Rev. Lett.\/} {\bf 107},
  036601 (2011).

\bibitem{AndreiXiZ2}
R.~Yu, X.~L. Qi, A.~Bernevig, Z.~Fang, X.~Dai, {\it Phys. Rev. B\/} {\bf 84},
  075119 (2011).

\bibitem{Cohomological}
A.~Alexandradinata, Z.~Wang, B.~A. Bernevig, {\it Phys. Rev. X\/} {\bf 6},
  021008 (2016).

\bibitem{multipole}
W.~A. Benalcazar, B.~A. Bernevig, T.~L. Hughes, {\it Science\/} {\bf 357}, 61
  (2017).

\bibitem{S4Weyl}
Y.~Qian, {\it et~al.\/}, {\it Phys. Rev. B\/} {\bf 101}, 155143 (2020).

\bibitem{checktopwebsite}
\vasptotrace~may be accessed to compute the topological classification of a
  material by selecting the function~\textsc{CheckTopological}
  at~\url{https://cryst.ehu.es/cgi-bin/cryst/programs/topological.pl}.

\bibitem{LuisSubduction}
L.~Elcoro, Z.~Song, B.~A. Bernevig, {\it Phys. Rev. B\/} {\bf 102}, 035110
  (2020).

\bibitem{CavaDirac2}
M.~N. Ali, {\it et~al.\/}, {\it Inorganic Chemistry\/} {\bf 53}, 4062 (2014).

\bibitem{YulinCadmiumExp}
Z.~K. Liu, {\it et~al.\/}, {\it Nature Materials\/} {\bf 13}, 677 (2014).

\bibitem{ITCA}
M.~I. Aroyo, ed., {\it International Tables for Crystallography, Volume A:
  Space-Group Symmetry\/}, vol.~A (International Union of Crystallography,
  2016).

\bibitem{identifygroupwebsite}
J.~Perez-Mato, {\it et~al.\/}, {\it Annual Review of Materials Research\/} {\bf
  45}, 217 (2015).

\bibitem{Hohenberg-PR64}
P.~Hohenberg, W.~Kohn, {\it Phys. Rev.\/} {\bf 136}, B864 (1964).

\bibitem{Kohn-PR65}
W.~Kohn, L.~J. Sham, {\it Phys. Rev.\/} {\bf 140}, A1133 (1965).

\bibitem{BarryPbTe}
I.~n. Robredo, M.~G. Vergniory, B.~Bradlyn, {\it Phys. Rev. Materials\/} {\bf
  3}, 041202 (2019).

\bibitem{ASE}
A.~H. Larsen, {\it et~al.\/}, {\it Journal of Physics: Condensed Matter\/} {\bf
  29}, 273002 (2017).

\bibitem{PHONOPY}
A.~Togo, I.~Tanaka, {\it Scripta Materialia\/} {\bf 108}, 1  (2015).

\bibitem{BCTBZ}
W.~Setyawan, S.~Curtarolo, {\it Computational Materials Science\/} {\bf 49},
  299  (2010).

\bibitem{SEKPATH}
Y.~Hinuma, G.~Pizzi, Y.~Kumagai, F.~Oba, I.~Tanaka, {\it Computational
  Materials Science\/} {\bf 128}, 140  (2017).

\bibitem{MaterialsProject}
A.~Jain, {\it et~al.\/}, {\it APL Materials\/} {\bf 1}, 011002 (2013).

\bibitem{CharlieTI}
C.~L. Kane, E.~J. Mele, {\it Phys. Rev. Lett.\/} {\bf 95}, 226801 (2005).

\bibitem{HourglassExperiment}
J.~Ma, {\it et~al.\/}, {\it Science Advances\/} {\bf 3} (2017).

\bibitem{zeroHallExp}
S.~Liang, {\it et~al.\/}, {\it Nature Materials\/} {\bf 18}, 443 (2019).

\bibitem{WiederLayers}
B.~J. Wieder, C.~L. Kane, {\it Phys. Rev. B\/} {\bf 94}, 155108 (2016).

\bibitem{RhSiArc}
G.~Chang, {\it et~al.\/}, {\it Phys. Rev. Lett.\/} {\bf 119}, 206401 (2017).

\bibitem{KramersWeyl}
G.~Chang, {\it et~al.\/}, {\it Nature Materials\/} {\bf 17}, 978 (2018).

\bibitem{BarryMultifold}
F.~Flicker, {\it et~al.\/}, {\it Phys. Rev. B\/} {\bf 98}, 155145 (2018).

\bibitem{ChiralSGsRightandWrong}
M.~Nespolo, M.~I. Aroyo, B.~Souvignier, {\it Journal of Applied
  Crystallography\/} {\bf 51}, 1481 (2018).

\bibitem{CoSiObserveJapan}
D.~Takane, {\it et~al.\/}, {\it Phys. Rev. Lett.\/} {\bf 122}, 076402 (2019).

\bibitem{AlPtObserve}
N.~B.~M. Schr{\"o}ter, {\it et~al.\/}, {\it Nature Physics\/} {\bf 15}, 759
  (2019).

\bibitem{XiLineNode}
R.~Yu, H.~Weng, Z.~Fang, X.~Dai, X.~Hu, {\it Phys. Rev. Lett.\/} {\bf 115},
  036807 (2015).

\bibitem{FangWithWithout}
C.~Fang, Y.~Chen, H.-Y. Kee, L.~Fu, {\it Phys. Rev. B\/} {\bf 92}, 081201
  (2015).

\bibitem{YoungkukMonopole}
J.~Ahn, D.~Kim, Y.~Kim, B.-J. Yang, {\it Phys. Rev. Lett.\/} {\bf 121}, 106403
  (2018).

\bibitem{AdrianMonopole}
A.~{Bouhon}, A.~M. {Black-Schaffer}, {\it ArXiv e-prints\/}  (2017).

\bibitem{SigristMonopole}
T.~c.~v. Bzdu\ifmmode~\check{s}\else \v{s}\fi{}ek, M.~Sigrist, {\it Phys. Rev.
  B\/} {\bf 96}, 155105 (2017).

\bibitem{AndreiInversion}
T.~L. Hughes, E.~Prodan, B.~A. Bernevig, {\it Phys. Rev. B\/} {\bf 83}, 245132
  (2011).

\bibitem{AshvinWeyl1}
X.~Wan, A.~M. Turner, A.~Vishwanath, S.~Y. Savrasov, {\it Phys. Rev. B\/} {\bf
  83}, 205101 (2011).

\bibitem{ChenSubduction}
T.~Zhang, {\it et~al.\/}, {\it Phys. Rev. Research\/} {\bf 2}, 022066 (2020).

\bibitem{NodaLineMap1}
I.~Tateishi, {\it Phys. Rev. Research\/} {\bf 2}, 043112 (2020).

\bibitem{NodaLineMap2}
I.~Tateishi, {\it Phys. Rev. B\/} {\bf 102}, 155111 (2020).

\bibitem{AdyWeak}
Z.~Ringel, Y.~E. Kraus, A.~Stern, {\it Phys. Rev. B\/} {\bf 86}, 045102 (2012).

\bibitem{MooreBalentsWeak}
J.~E. Moore, L.~Balents, {\it Phys. Rev. B\/} {\bf 75}, 121306 (2007).

\bibitem{WiederBarryCDW}
B.~J. Wieder, K.-S. Lin, B.~Bradlyn, {\it Phys. Rev. Research\/} {\bf 2},
  042010 (2020).

\bibitem{JiabinCDW}
J.~Yu, B.~J. Wieder, C.-X. Liu, {\it Phys. Rev. B\/} {\bf 104}, 174406 (2021).

\bibitem{WPVZ}
H.~Watanabe, H.~C. Po, A.~Vishwanath, M.~Zaletel, {\it Proceedings of the
  National Academy of Sciences\/} {\bf 112}, 14551 (2015).

\bibitem{SteveMagnet}
S.~M. Young, B.~J. Wieder, {\it Phys. Rev. Lett.\/} {\bf 118}, 186401 (2017).

\bibitem{ARPESReviewYulin}
H.~Yang, {\it et~al.\/}, {\it Nature Reviews Materials\/} {\bf 3}, 341 (2018).

\bibitem{ARPESReviewHongDing}
B.~Lv, T.~Qian, H.~Ding, {\it Nature Reviews Physics\/} {\bf 1}, 609 (2019).

\bibitem{IlyaPumpProbe1}
I.~Belopolski, {\it et~al.\/}, {\it Nature Communications\/} {\bf 8}, 942
  (2017).

\bibitem{IlyaPumpProbe2}
I.~Belopolski, {\it et~al.\/}, {\it Nature Communications\/} {\bf 7}, 13643
  (2016).

\bibitem{WU2018405}
Q.~Wu, S.~Zhang, H.-F. Song, M.~Troyer, A.~A. Soluyanov, {\it Computer Physics
  Communications\/} {\bf 224}, 405  (2018).

\bibitem{Mg3Bi2RTopo2}
X.~Zhang, L.~Jin, X.~Dai, G.~Liu, {\it The Journal of Physical Chemistry
  Letters\/} {\bf 8}, 4814 (2017).

\bibitem{Mg3Bi2RTopo3}
T.~Zhou, {\it et~al.\/}, {\it Chinese Physics Letters\/} {\bf 36}, 117303
  (2019).

\bibitem{GrapheneReview}
A.~H. Castro~Neto, F.~Guinea, N.~M.~R. Peres, K.~S. Novoselov, A.~K. Geim, {\it
  Rev. Mod. Phys.\/} {\bf 81}, 109 (2009).

\bibitem{TeoFuKaneTCI}
J.~C.~Y. Teo, L.~Fu, C.~L. Kane, {\it Phys. Rev. B\/} {\bf 78}, 045426 (2008).

\bibitem{TanakaSnTeExp}
Y.~Tanaka, {\it et~al.\/}, {\it Nature Physics\/} {\bf 8}, 800 (2012).

\bibitem{HOTIBernevig}
F.~Schindler, {\it et~al.\/}, {\it Science Advances\/} {\bf 4} (2018).

\bibitem{FradkinPbTe}
E.~Fradkin, E.~Dagotto, D.~Boyanovsky, {\it Phys. Rev. Lett.\/} {\bf 57}, 2967
  (1986).

\bibitem{SpinlessWeylGuoqing}
G.~Chang, {\it et~al.\/}, {\it Science Advances\/} {\bf 2} (2016).

\bibitem{MobiusInsulator}
K.~Shiozaki, M.~Sato, K.~Gomi, {\it Phys. Rev. B\/} {\bf 91}, 155120 (2015).

\bibitem{HigherOrderTIPiet}
J.~Langbehn, Y.~Peng, L.~Trifunovic, F.~von Oppen, P.~W. Brouwer, {\it Phys.
  Rev. Lett.\/} {\bf 119}, 246401 (2017).

\bibitem{ZJDirac2}
Z.~Wang, {\it et~al.\/}, {\it Phys. Rev. B\/} {\bf 85}, 195320 (2012).

\bibitem{SteveDirac}
S.~M. Young, {\it et~al.\/}, {\it Phys. Rev. Lett.\/} {\bf 108}, 140405 (2012).

\bibitem{JuliaDirac}
J.~A. Steinberg, {\it et~al.\/}, {\it Phys. Rev. Lett.\/} {\bf 112}, 036403
  (2014).

\bibitem{NagaosaDirac}
B.-J. Yang, N.~Nagaosa, {\it Nature Communications\/} {\bf 5}, 4898 (2014).

\bibitem{NaDirac}
Z.~K. Liu, {\it et~al.\/}, {\it Science\/} {\bf 343}, 864 (2014).

\bibitem{SYDiracSurface}
S.-Y. {Xu}, {\it et~al.\/}, {\it Science\/} {\bf 347}, 294 (2015).

\bibitem{MBP_fragile}
M.~B. de~Paz, M.~G. Vergniory, D.~Bercioux, A.~Garc\'{\i}a-Etxarri, B.~Bradlyn,
  {\it Phys. Rev. Research\/} {\bf 1}, 032005 (2019).

\bibitem{ChenBernevigTCI}
C.~Fang, M.~J. Gilbert, B.~A. Bernevig, {\it Phys. Rev. B\/} {\bf 86}, 115112
  (2012).

\bibitem{ArisInversion}
A.~Alexandradinata, X.~Dai, B.~A. Bernevig, {\it Phys. Rev. B\/} {\bf 89},
  155114 (2014).

\bibitem{HsiehDiracInsulator}
D.~Hsieh, {\it et~al.\/}, {\it Nature\/} {\bf 452}, 970 (2008).

\bibitem{BismuthFacet}
C.-H. Hsu, {\it et~al.\/}, {\it Proceedings of the National Academy of
  Sciences\/} {\bf 116}, 13255 (2019).

\bibitem{BiBrFan}
C.-C. Liu, J.-J. Zhou, Y.~Yao, F.~Zhang, {\it Phys. Rev. Lett.\/} {\bf 116},
  066801 (2016).

\bibitem{AshvinFirstMaterials}
F.~Tang, H.~C. Po, A.~Vishwanath, X.~Wan, {\it Nature Physics\/} {\bf 15}, 470
  (2019).

\bibitem{BiBrSuyang}
C.-H. Hsu, {\it et~al.\/}, {\it 2D Materials\/} {\bf 6}, 031004 (2019).

\bibitem{BiBrFanHOTI}
C.~{Yoon}, C.-C. {Liu}, H.~{Min}, F.~{Zhang}, {\it arXiv e-prints\/} p.
  arXiv:2005.14710 (2020).

\bibitem{BiBrNatMater}
R.~Noguchi, {\it et~al.\/}, {\it Nature Materials\/} {\bf 20}, 473 (2021).

\bibitem{BismuthHaimDefect}
A.~K. Nayak, {\it et~al.\/}, {\it Science Advances\/} {\bf 5} (2019).

\bibitem{RaquelPartial}
R.~Queiroz, I.~C. Fulga, N.~Avraham, H.~Beidenkopf, J.~Cano, {\it Phys. Rev.
  Lett.\/} {\bf 123}, 266802 (2019).

\bibitem{DavidMoTe2Exp}
F.-T. Huang, {\it et~al.\/}, {\it Nature Communications\/} {\bf 10}, 4211
  (2019).

\bibitem{MazWTe2Exp}
Y.-B. Choi, {\it et~al.\/}, {\it Nature Materials\/} {\bf 19}, 974 (2020).

\bibitem{WTe2HingeStep}
A.~Kononov, {\it et~al.\/}, {\it Nano Letters\/} {\bf 20}, 4228 (2020).

\bibitem{PhuanOngMoTe2Hinge}
W.~Wang, {\it et~al.\/}, {\it Science\/} {\bf 368}, 534 (2020).

\bibitem{BiIWTIExp}
R.~Noguhi, {\it et~al.\/}, {\it Nature\/} {\bf 566}, 518 (2019).

\bibitem{SSH}
W.~P. Su, J.~R. Schrieffer, A.~J. Heeger, {\it Phys. Rev. Lett.\/} {\bf 42},
  1698 (1979).

\bibitem{SSHExp}
C.~K. Chiang, {\it et~al.\/}, {\it Phys. Rev. Lett.\/} {\bf 39}, 1098 (1977).

\bibitem{JeanNoelSSH}
J.-N. Fuchs, F.~Pi\'echon, {\it Phys. Rev. B\/} {\bf 104}, 235428 (2021).

\bibitem{AshvinScrewTI}
Y.~Ran, Y.~Zhang, A.~Vishwanath, {\it Nature Physics\/} {\bf 5}, 298 (2009).

\bibitem{KramersNodalLine}
Y.-M. Xie, {\it et~al.\/}, {\it Nature Communications\/} {\bf 12}, 3064 (2021).

\bibitem{Steve2D}
S.~M. Young, C.~L. Kane, {\it Phys. Rev. Lett.\/} {\bf 115}, 126803 (2015).

\bibitem{PdSb2}
N.~Kumar, {\it et~al.\/}, {\it Advanced Materials\/} {\bf 32}, 1906046 (2020).

\bibitem{ChiralTellurium}
G.~Gatti, {\it et~al.\/}, {\it Phys. Rev. Lett.\/} {\bf 125}, 216402 (2020).

\bibitem{FerNatCom}
F.~de~Juan, A.~G. Grushin, T.~Morimoto, J.~E. Moore, {\it Nature
  Communications\/} {\bf 8}, 15995 (2017).

\bibitem{ReesMoorePhotoExp}
D.~Rees, {\it et~al.\/}, {\it Science Advances\/} {\bf 6} (2020).

\bibitem{LiangPennOpticalCoSi}
B.~Xu, {\it et~al.\/}, {\it Proceedings of the National Academy of Sciences\/}
  {\bf 117}, 27104 (2020).

\bibitem{LiangPennOpticalCoSi2}
Z.~Ni, {\it et~al.\/}, {\it Nature Communications\/} {\bf 12}, 154 (2021).

\bibitem{AuBeChiralSC}
D.~J. Rebar, {\it et~al.\/}, {\it Phys. Rev. B\/} {\bf 99}, 094517 (2019).

\bibitem{KTLawMultifoldSC}
J.~Z.~S. {Gao}, {\it et~al.\/}, {\it arXiv e-prints\/} p. arXiv:2012.11287
  (2020).

\bibitem{TayRongKramersWeylSC}
E.~Emmanouilidou, {\it et~al.\/}, {\it Phys. Rev. B\/} {\bf 102}, 235144
  (2020).

\bibitem{DDPMott1}
G.~Sharma, {\it et~al.\/}, {\it J. Mater. Chem. A\/} {\bf 4}, 2936 (2016).

\bibitem{DDPMott2}
D.~Di~Sante, {\it et~al.\/}, {\it Physical Review B\/} {\bf 96}, 121106 (2017).

\bibitem{Khoury_JACS}
J.~F. Khoury, {\it et~al.\/}, {\it Journal of the American Chemical Society\/}
  {\bf 141}, 19130 (2019).

\bibitem{SchoopAFM}
L.~M. Schoop, {\it et~al.\/}, {\it Science Advances\/} {\bf 4} (2018).

\bibitem{ChenMultiWeyl}
C.~Fang, M.~J. Gilbert, X.~Dai, B.~A. Bernevig, {\it Phys. Rev. Lett.\/} {\bf
  108}, 266802 (2012).

\bibitem{ZhijunMultiWeyl}
G.~Xu, H.~Weng, Z.~Wang, X.~Dai, Z.~Fang, {\it Phys. Rev. Lett.\/} {\bf 107},
  186806 (2011).

\bibitem{ZahidMultiWeyl}
S.-M. Huang, {\it et~al.\/}, {\it Proceedings of the National Academy of
  Sciences\/} {\bf 113}, 1180 (2016).

\bibitem{StepanMultiWeyl}
S.~S. Tsirkin, I.~Souza, D.~Vanderbilt, {\it Phys. Rev. B\/} {\bf 96}, 045102
  (2017).

\bibitem{ScrewChiralWeyl}
W.~Wu, Z.-M. Yu, X.~Zhou, Y.~X. Zhao, S.~A. Yang, {\it Phys. Rev. B\/} {\bf
  101}, 205134 (2020).

\bibitem{TripleChen}
H.~Weng, C.~Fang, Z.~Fang, X.~Dai, {\it Phys. Rev. B\/} {\bf 93}, 241202
  (2016).

\bibitem{AlexeyTriple}
Z.~Zhu, G.~W. Winkler, Q.~Wu, J.~Li, A.~A. Soluyanov, {\it Phys. Rev. X\/} {\bf
  6}, 031003 (2016).

\bibitem{ZahidTriple}
G.~Chang, {\it et~al.\/}, {\it Scientific Reports\/} {\bf 7}, 1688 (2017).

\bibitem{WiederTripleSummary}
B.~J. Wieder, {\it Nature Physics\/} {\bf 14}, 329 (2018).

\bibitem{MoPTriple}
B.~Q. Lv, {\it et~al.\/}, {\it Nature\/} {\bf 546}, 627 (2017).

\bibitem{WCarc}
J.-Z. Ma, {\it et~al.\/}, {\it Nature Physics\/} {\bf 14}, 349 (2018).

\bibitem{DiracWeylYoungkuk}
H.~Gao, {\it et~al.\/}, {\it Phys. Rev. Lett.\/} {\bf 121}, 106404 (2018).

\bibitem{TaSeIPRL}
C.~Tournier-Colletta, {\it et~al.\/}, {\it Phys. Rev. Lett.\/} {\bf 110},
  236401 (2013).

\bibitem{TaSeIDFTDagotto}
Y.~Zhang, L.-F. Lin, A.~Moreo, S.~Dong, E.~Dagotto, {\it Phys. Rev. B\/} {\bf
  101}, 174106 (2020).

\bibitem{TaSeIZhijunDirac}
H.~Yi, {\it et~al.\/}, {\it Phys. Rev. Research\/} {\bf 3}, 013271 (2021).

\bibitem{TaylorToy}
M.~Lin, T.~L. Hughes, {\it Phys. Rev. B\/} {\bf 98}, 241103 (2018).

\bibitem{VladHOFA}
D.~C\ifmmode \u{a}\else \u{a}\fi{}lug\ifmmode~\u{a}\else \u{a}\fi{}ru,
  V.~Juri\ifmmode \check{c}\else \v{c}\fi{}i\ifmmode~\acute{c}\else \'{c}\fi{},
  B.~Roy, {\it Phys. Rev. B\/} {\bf 99}, 041301 (2019).

\bibitem{BitanHOFA}
A.~L. Szab\'o, R.~Moessner, B.~Roy, {\it Phys. Rev. B\/} {\bf 101}, 121301
  (2020).

\bibitem{BitanHOFA2}
A.~L. Szab\'o, B.~Roy, {\it Phys. Rev. Research\/} {\bf 2}, 043197 (2020).

\bibitem{TaylorWeylHOFA}
S.~A.~A. Ghorashi, T.~Li, T.~L. Hughes, {\it Phys. Rev. Lett.\/} {\bf 125},
  266804 (2020).

\bibitem{OtherWeylHOFA}
H.-X. Wang, Z.-K. Lin, B.~Jiang, G.-Y. Guo, J.-H. Jiang, {\it Phys. Rev.
  Lett.\/} {\bf 125}, 146401 (2020).

\bibitem{ZhijunAndreiElectride}
S.~Nie, B.~A. Bernevig, Z.~Wang, {\it Phys. Rev. Research\/} {\bf 3}, L012028
  (2021).

\end{thebibliography}


\clearpage

\onecolumngrid


\renewcommand{\thefigure}{S\arabic{figure}}
\renewcommand{\theequation}{S\arabic{equation}}
\renewcommand{\thetable}{S\arabic{table}}
\renewcommand{\thesection}{SM~\arabic{section}}
\setcounter{equation}{0}
\setcounter{figure}{0}
\setcounter{table}{0}

\clearpage
\begin{center}
{\bf Supplementary Material for ``All Topological Bands of All Non-Magnetic Stoichiometric Materials''}
\end{center}

\tableofcontents

\clearpage

\listoftables

\clearpage

\addtocontents{toc}{\protect\setcounter{tocdepth}{3}}
\addtocontents{lot}{\protect\setcounter{lotdepth}{3}}

\section{Introduction to the Supplementary Material}\label{App:AppendixOverview_appendix}

In this work, we have introduced a complete catalogue of the symmetry-indicated stable and fragile topology of all of the bands in all of the stoichiometric materials in the Inorganic Crystal Structure Database (ICSD)~\cite{ICSD}.  Below, we provide Supplementary Material (SM) containing a description of our methodology, as well as detailed statistics for the materials studied in this work.  First, in \supappref{App:TQCReview_appendix}, we will review the methods employed in this work, which derive from the recently introduced theory of \emph{Topological Quantum Chemistry} (TQC)~\cite{QuantumChemistry}.  Next, in \supappref{App:Check_Topo_appendix} and~\ref{App:VASP_appendix}, we will document the computational machinery that we developed to apply TQC to the ICSD database.  Specifically, in \supappref{App:Check_Topo_appendix} and~\ref{App:VASP_appendix}, we will respectively review the \vasptotrace~program previously implemented for Ref.~\onlinecite{AndreiMaterials}, and will detail updates to the~\checktopmat~program implemented for this work -- for both programs, we will additionally detail their interface with the density functional theory (DFT) Vienna Ab-initio Simulation Package (VASP)~\cite{vasp1,vasp2}.  In \supappref{App:CPUtime}, we will next discuss how we prepared the data used to generate this work, and will detail the CPU time invested per number of atoms in the unit cell of each material, and per each crystallographic space group (SG)~\cite{BigBook}.  In the following section -- \supappref{App:TopologicalMaterialsNoSOC}, we will then review the theory of TQC in the absence of spin-orbit coupling (SOC), and will specifically discuss the physical interpretation of symmetry-based indicators of band topology~\cite{AshvinIndicators,ChenTCI,AshvinTCI,HOTIChen,ZhidaSemimetals,MTQC,MTQCmaterials} in material calculations performed in the absence of SOC (w/o SOC).  In the following section -- \supappref{App:Website} -- we will then provide detailed descriptions of the significant~\webTQC~(\webNoICSD) updates implemented for this work.

In the remaining sections of the Supplementary Material, we will present detailed statistics, tables, and lists supporting the results shown in the main text.  First, in \supappref{App:TopoBands}, we will provide extensive statistics for all of the topological bands and band connectivity in the materials studied in this work.  Next, in \supappref{App:Supertopological}, we will rigorously define the \emph{repeat-topological} (RTopo) and \emph{supertopological} (STopo) material classes introduced in this work, and will provide a list of all of the symmetry-indicated STopo materials in the ICSD.  Then, in \supappref{App:PhaseTransitionsNoSOCSOC}, we will analyze the role of SOC in driving phase transitions between topological (semi)metals w/o SOC and topological (crystalline) insulators with SOC.  In \supappref{App:PhaseTransitionsNoSOCSOC}, we will additionally provide detailed statistics for all of the topological phase transitions observed in this work between materials with and w/o SOC.  Finally, in \supappref{App:MaterialsSelection}, we will present an extensive list of all of the highest-quality topological (crystalline) insulators and semimetals among all of the stoichiometric materials in the ICSD.  Though some of the materials in \supappref{App:MaterialsSelection} are well known -- such as the archetypal 3D topological insulator (TI) Bi$_2$Se$_3$~\cite{HsiehBismuthSelenide} [\icsdweb{617079}, SG 166 ($R\bar{3}m$)] -- others are relatively unstudied, and should inspire new experimental investigations as well as provide new context for previously unexplained experimental data.

\newpage

\section{Topological Quantum Chemistry Review}\label{App:TQCReview_appendix}

\subsection{Elementary Band Representations }
\label{App:EBRs}

Topological Quantum Chemistry (TQC)~\cite{QuantumChemistry} is a position-space theory of band topology in crystalline solids.  The building blocks of TQC are trivial atomic limits, which transform in elementary band representations (EBRs)~\cite{Michel1999,Michel2001,JenOAL}.  Most generally, a band representation (band rep), elementary or otherwise, is a collection of momentum-space states (bands) induced from (exponentially) localized orbitals placed throughout each unit cell of a crystal~\cite{PhysRevLett.45.1025,Michel2001}.  If the band rep is induced from a \emph{maximal} Wyckoff position $\omega$ and a set of atomic orbitals that transform in an irreducible corepresentation (corep) of the site-symmetry group $G_{\omega}$ [defined in detail in Refs.~\onlinecite{QuantumChemistry,Bandrep1,Bandrep2,Bandrep3}], then the band rep can be further classified as \emph{elementary} (\emph{i.e.}, as an EBR), provided that the band rep is not an \emph{exceptional case} (see Refs.~\onlinecite{Bandrep3,QuantumChemistry,MTQC}).  Conversely, if the band rep is not elementary, then it can be expressed as a sum of EBRs~\cite{Bandrep3}.  In this work and in TQC, we restrict analysis to nonmagnetic (paramagnetic) crystals [\emph{i.e.} Type-II Shubnikov space groups (SGs)~\cite{BigBook,MTQC}], which respect time-reversal ($\mathcal{T}$) symmetry.  The EBRs of the nonmagnetic SGs can be accessed through the~\href{https://www.cryst.ehu.es/cgi-bin/cryst/programs/bandrep.pl}{BANDREP} tool~\cite{QuantumChemistry,Bandrep1} on the~\webBCSfull~(\webBCSshort)~\cite{BCS1,BCS2}.  During the preparation of this work, the EBRs of the magnetic SGs were also computed in Ref.~\onlinecite{MTQC} (see~\href{https://www.cryst.ehu.es/cgi-bin/cryst/programs/mbandrep.pl}{MBANDREP} on the~\webBCSshort), and were used to perform the first high-throughput search for magnetic topological materials~\cite{MTQCmaterials}.

For the purposes of this work, we are primarily concerned with the small coreps at high-symmetry (maximal) ${\bf k}$ points.  Specifically, given a set of bands in a first-principles or tight-binding calculation, one can extract the coreps of the occupied bands at the maximal ${\bf k}$ points to construct a \emph{symmetry data vector}~\cite{AndreiMaterials}.  Given a symmetry data vector, the first step is to determine if the symmetry data vector is compatible with an insulating gap along all high-symmetry lines and planes -- this can be accomplished using the compatibility relations, which are explained in detail in \supappref{App:CR}.  If the symmetry data is incompatible with an insulating gap, then the bulk bands are necessarily (semi)metallic.  Semimetals that fail to satisfy the compatibility relations can then be further classified by whether a multiplet of Bloch states (transforming in an irreducible small corep of the little group) at a maximal ${\bf k}$ point is partially occupied [enforced semimetal with Fermi degeneracy (ESFD)], or whether the compatibility relations fail along a high-symmetry line or plane [enforced semimetal (ES)].  If the symmetry data vector \emph{is} compatible with an insulating gap (along high-symmetry lines and planes), we then determine if the symmetry data can be re-expressed as a linear combination of EBRs (LCEBR) with integer coefficients.  As we will shortly see, in order for a set of bands to be topologically trivial, it is a necessary (but not sufficient) condition that the bands are classified as LCEBR.

The EBRs also provide a means of characterizing the topology of a set of bands obtained from a DFT or tight-binding calculation.  Specifically, given an energetically isolated set of bands, the small coreps (symmetry eigenvalues) can imply one of two broad topological classes.  First, there are stable topological bands, which characterize strong and weak topological (crystalline) insulators (TIs and TCIs)~\cite{FuKaneMele,FuKaneInversion,AshvinIndicators,AshvinTCI,ChenTCI} that cannot be rendered topologically trivial through the addition of bands that are not themselves stable topological.  A subset of topological bands, known as \emph{symmetry-indicated} topological bands~\cite{FuKaneMele,FuKaneInversion,AshvinIndicators,HOTIChen,ChenTCI,AshvinTCI}, can be recognized as topologically stable because their symmetry data cannot be expressed as a linear combination of EBRs with integer coefficients.  In this work, we follow the nomenclature established in Ref.~\cite{AndreiMaterials} in which a symmetry-indicated topological band is classified as originating from a split EBR (SEBR) if its symmetry data can be expressed as a linear combination of connected pieces of EBRs that are themselves disconnected (while continuing to not match any linear combination of full EBRs).  Continuing to employ the nomenclature of Ref.~\cite{AndreiMaterials}, we classify symmetry-indicated topological bands as NLC if their symmetry data cannot be expressed as an integer-valued linear combination of pieces of EBRs.  Additionally, it was recently discovered that there also exist \emph{fragile} topological bands~\cite{AshvinFragile,JenFragile1,AshvinFragile2,BarryFragile,AdrianFragile,KoreanFragile,ZhidaFragile,FragileFlowMeta,ZhidaBLG,ZhidaFragile2,KoreanFragileInversion,DelicateAris}, which, unlike stable topological bands, can characterize a trivial insulator (or an ``obstructed'' atomic limit~\cite{TMDHOTI,WiederAxion,HingeSM}) if they are combined with trivial or other fragile bands.  TIs and TCIs characterized by stable topological bands can be classified using methods including K-theory~\cite{KitaevClass}, surface-state anomalies~\cite{QHZ,ChenTCI,AshvinTCI,DiracInsulator,ChenRotation,MTQC}, and layer constructions~\cite{ChenTCI,HermeleSymmetry,MTQC}, whereas fragile topological bands can be characterized by corner filling anomalies~\cite{TMDHOTI,WiederAxion,HingeSM,AshvinFragile2,KoreanFragile,KoreanFragileInversion} and quantized twisted-boundary~\cite{ZhidaFragile2,FragileFlowMeta} and defect~\cite{WladCorners} responses.  For both stable and fragile TCIs, a complete diagnosis of nontrivial band topology can be obtained by performing (nested) Wilson loop calculations~\cite{Fidkowski2011,AndreiXiZ2,HourglassInsulator,Cohomological,DiracInsulator,multipole,TMDHOTI,WiederAxion,S4Weyl}.

For each SG, symmetry-indicated topological bands labeled NLC or SEBR can be further characterized using symmetry-based indicators (SIs).  Specifically, following a prescription detailed in Ref.~\onlinecite{SlagerSymmetry}, the authors of Refs.~\onlinecite{AshvinIndicators,AshvinTCI,ChenTCI} compared the EBRs, small coreps, and compatibility relations in each SG to determine the linearly independent sets of NLC and SEBR symmetry data [we provide specific details of this calculation in the text below].  In each SG, the calculation prescribed in Ref.~\onlinecite{SlagerSymmetry} returned a set of positive-definite integers -- known as the \emph{SI group} (\emph{e.g.} $\{4,2,2,2\}$ in SG 2) -- that aligned with the symmetry-indicated classification of strong and weak stable topological (crystalline) insulators in that SG (\emph{e.g.} $\mathbb{Z}_{4}\otimes\mathbb{Z}_{2}^{3}$).  In Refs.~\onlinecite{AshvinTCI,ChenTCI}, the authors then matched the SI groups (\emph{e.g.} $\mathbb{Z}_{4}\otimes\mathbb{Z}_{2}^{3}$) to anomalous boundary states and occupied bands with specific combinations of crystal symmetry eigenvalues  -- known as \emph{SI formulas} (\emph{i.e.} generalized Fu-Kane formulas~\cite{FuKaneInversion}) -- to construct a complete enumeration of all possible symmetry-indicated stable TIs and TCIs in the nonmagnetic SGs.  This method was then expanded in Ref.~\onlinecite{ZhidaSemimetals} to the nonmagnetic SGs without SOC to compute the stable SIs for special classes of topological semimetals, which in this work we designate as SEBR-SM and NLC-SM (see \supappref{App:TopologicalMaterialsNoSOC} for further details).  Finally, building upon the first demonstration of symmetry-indicated fragile topology in TQC~\cite{JenFragile1}, the authors of Refs.~\onlinecite{KoreanFragileInversion,ZhidaFragile} discovered the existence of fragile SIs, specifically showing that if a set of bands is classified as LCEBR, then it can be further classified as fragile if some of the integer coefficients in the linear combination of EBRs are negative.

In summary, TQC provides a unique, deterministic method for diagnosing the topology of any set of energetically isolated bands.  Using TQC, we can analyze the full energy spectrum of any material up to any number of bands using the following procedure.  First, using \vasptotrace~\cite{AndreiMaterials,Irvsp2020,checktopwebsite} -- a program that was implemented by Zhijun Wang to be used in conjunction with VASP~\cite{vasp1,vasp2,paw1} -- we calculate the characters $\chi(g)$ of all of the unitary symmetries $g$ of the occupied Bloch wavefunctions at all high-symmetry (maximal) ${\bf k}$ points, where $\chi(g)$ is equal to the trace of the matrix representative of the symmetry element $g$ in the reducible small corep corresponding to the set of all of the occupied Bloch states at ${\bf k}$.  Next, using $\chi(g)$, we compute the multiplicity $m_{\alpha}$ of the $\alpha$-th \emph{irreducible} small corep of each little group $G_{\bf k}$ by systematically applying the so-called {\it magic formula}:
\begin{equation}
m_{\alpha}=\frac{1}{\|G_{\bf k}\|}\sum_{g\in G_{\bf k}}\chi_{\alpha}^{*}(g)\chi(g),
\label{eq:magicFormula}
\end{equation}
where $\chi_{\alpha}(g)$ is the character of $g$ in the $\alpha$-th irreducible small corep of $G_{\bf k}$, and where $\|G_{\bf k}\|$ is the order of the little group (\emph{i.e.}, the number of symmetry operations in the little group not including operations related by integer lattice translations).  The characters of the irreducible small coreps of the little groups of all ${\bf k}$ points in all 230 nonmagnetic SGs can be accessed using the~\href{http://www.cryst.ehu.es/cgi-bin/cryst/programs/representations.pl?tipogrupo=dbg}{REPRESENTATIONS DSG} tool on the~\webBCSshort~implemented for TQC~\cite{QuantumChemistry,Bandrep1,Bandrep2,Bandrep3,JenFragile1,BarryFragile}.  Next, using an upgraded implementation of the~\checktopmat~\cite{checktopwebsite} tool on the~\webBCSshort~(see \supappref{App:Check_Topo_appendix} for a detailed description of new features), we analyze each set of isolated bands.  Specifically, we first obtain a symmetry data vector $B$, defined as:
\begin{equation}
B=(m(\rho_{G_{K_{1}}}^{1}),m(\rho_{G_{K_{1}}}^{1}),...,m(\rho_{G_{K_{1}}}^{2}),m(\rho_{G_{K_{1}}}^{2})...)^{T},
\end{equation}
where $m(\rho_{G_{K_{i}}}^{j})$ is the multiplicity of the $j$-th small corep of the little group ${G_{K_{i}}}$ at the maximal momentum $K_{i}$.  Next, we denote the symmetry data vector of the $i$-th EBR as $EBR_{i}$. Then, for any isolated subset of bands characterized by a symmetry vector that satisfies the compatibly relations, we solve a linear system of equations given by:
\begin{equation}
B = EBR \cdot X,
\label{B_vect_eq}
\end{equation}
in which the solution $X=(X_1,X_2...X_N)$ indicates the coefficients of the linear combination of EBRs corresponding to $B$.  The presence of non-integer coefficients in $\{ X_1,X_2,...,X_N\}$ in Eq.~(\ref{B_vect_eq}) implies that some or all of the stable SIs are nontrivial for the bands characterized by $B$.  For each SG, the possible existence of non-integer solutions to Eq.~(\ref{B_vect_eq}), and hence the existence of nontrivial stable SIs, may be predicted through the Smith decomposition of the EBR matrix~\cite{SlagerSymmetry}.  Specifically, because an EBR is a non-square matrix with integer coefficients, then its Smith decomposition can be expressed as a product of three matrices with integer coefficients:
\begin{equation}
EBR=L\cdot D\cdot R,
\end{equation}
where $L$ and $R$ are unimodular matrices and $D$ is a diagonal matrix.  The number of nonzero diagonal elements in $D$ is the rank of the stable SI group, and the matrices $L$ and $R$ can be chosen such that the (diagonal) entries in $D$ are positive integers~\cite{MTQC,LuisSubduction}.  If some of diagonal entries $\{n_1,n_2...\}$ are greater than one, then the stable SI group is nontrivial, and is given by $\mathbb{Z}_{n_1}\otimes\mathbb{Z}_{n_2}\otimes ... $ [\emph{e.g.}, the nonzero elements in $D$ in SG 2 ($P\bar{1}$) are $\{4,2,2,2\}$, corresponding to a stable SI group $\mathbb{Z}_{4}\otimes\mathbb{Z}_{2}^{3}$].  By then systematically applying \vasptotrace~and \checktopmat~to each isolated set of bands in the electronic structure of each material in the ICSD~\cite{ICSD}, we obtain the symmetry-indicated stable and fragile band topology of every known stoichiometric, nonmagnetic material.

\subsection{The Compatibility Relations}
\label{App:CR}

To determine the connectivity of electronic bands characterized by a given symmetry data vector [Eq.~(\ref{B_vect_eq})], one must use the compatibility relations between ${\bf k}$ points throughout the BZ.  Specifically, in a given SG, the compatibility relations are defined as the dependencies between the multiplicity of the small coreps of the little groups of a ${\bf k}$ point in the BZ and the multiplicities of the small coreps of the little groups along a line or plane in which the point sits~\cite{Bandrep2}.  The compatibility relations exist because the little group of a line (or a plane) is necessarily a subgroup of the little group of a high-symmetry point along the line (or plane)~\cite{BigBook}.

A set of bands can be \emph{partially} identified by its decomposition into integer-valued linear combinations of the irreducible small coreps of the little group of each ${\bf k}$ point in the BZ.  However, in general, it is not necessary to consider the multiplicities of the coreps in the whole BZ to analyze the connectivity of a given symmetry data vector, due to the compatibility relations between the coreps of two connected subsets of ${\bf k}$-vectors.  This renders the problem of decomposing bands into the coreps of the little group of the continuous variable ${\bf k}$ feasible, because the corep decomposition in the whole BZ is fully fixed by the corep decomposition at only a finite (and in practice very small) number of ${\bf k}$ points known as the \emph{maximal} ${\bf k}$ vectors~\cite{QuantumChemistry,Bandrep1,Bandrep2,Bandrep3}.  For example, consider a ${\bf k}$ point that belongs to a line ($l$) ${\bf k}_l$ for which the little group $\mathcal{G}^{\bf{k}}$ of ${\bf k}$ is a supergroup of the little group of ${\bf k}_l$, $\mathcal{G}^{\bf{k}_l}\leq \mathcal{G}^{\bf{k}}$.  The matrix representatives of a small corep $\rho^i_{\mathcal{G}^{\bf{k}}}$ of the little group $\mathcal{G}^{\bf{k}}$ associated with the symmetry operations that belong to $\mathcal{G}^{\bf{k}_l}$ transform in a small representation, not necessarily irreducible, of $\mathcal{G}^{\bf{k}_l}$.  This representation, in general, is equivalent to a direct sum of the irreducible small coreps $\rho_{\mathcal{G}^{\bf{k}_l}}^j$ of $\mathcal{G}^{\bf{k}_l}$:
 \begin{equation}
 \rho_{\mathcal{G}^{\bf{k}}}^i\downarrow \mathcal{G}^{\bf{k}_l}=\bigoplus_{i=1}^sm_{ij}^{\bf{k},\bf{k}_l}\rho_{\mathcal{G}^{\bf{k}_l}}^j \label{representationreduction}
 \end{equation}
 where $s$ is the number of irreducible small coreps of $\mathcal{G}^{\bf{k}_l}$ and $m_{ij}^{\bf{k},\bf{k}_l}$ is the (integer) multiplicity of $\rho_{\mathcal{G}^{\bf{k}_l}}^j$ in the decomposition of $\rho^i_{\mathcal{G}^{\bf{k}}}$.  The same arguments can also be employed to calculate the compatibility relations between the multiplicities of the coreps in the symmetry data at a high-symmetry point and the corep multiplicities in a plane containing the high-symmetry point, or between the corep multiplicities along a line and the corep multiplicities in a plane containing the line.  For TQC, we previously implemented the program~\href{https://www.cryst.ehu.es/cgi-bin/cryst/programs/dcomprel.pl}{DCOMPREL}~\cite{QuantumChemistry,Bandrep1,Bandrep2,Bandrep3,JenFragile1,BarryFragile} for obtaining the compatibility relations between each pair of connected ${\bf k}$-vectors in the BZ of each nonmagnetic double (spinful) SG.

The compatibility relations provide a means of analyzing a given symmetry data vector obtained from first-principles or tight-binding calculations.  Specifically, as shown in Refs.~\onlinecite{AndreiMaterials,MTQCmaterials}, there are three cases for the connectivity of a set of Bloch eigenstates characterized by a symmetry data vector, given a specified number of filled Bloch eigenstates (electrons):
\begin{enumerate}
\item{We observe that a set of small coreps at a maximal point ${\bf k}$ corresponds to a set of Bloch states in which a band degeneracy is partially occupied.  Numerically using~\checktopmat, this occurs when the number of bands (small coreps corresponding to the occupied Bloch states) is not the same at every maximal ${\bf k}$-vector (\emph{i.e.}, the sum of the dimensions of the coreps is different from the number of occupied bands in at least at one maximal ${\bf k}$-vector, which can occur because \checktopmat~does not output fractional numbers of coreps).  In this case, the bands below the Fermi level characterized by the symmetry data vector are necessarily connected to states above the Fermi energy, and therefore fail to satisfy the compatibility relations at a maximal ${\bf k}$ point.  Following the nomenclature established in Ref.~\onlinecite{AndreiMaterials}, we label a material with a partially filled multiplet of Bloch eigenstates at a maximal ${\bf k}$ point as an ``enforced semimetal with Fermi degeneracy'' (ESFD).  A well-known example of an ESFD-classified material is HgTe in SG 216 ($F\bar{4}3m$)~\cite{AndreiTI} (\icsdweb{31845}, see \supappref{App:ESFD}).}

\item{The symmetry data vector corresponds to fully occupied Bloch-state multiplets at all maximal ${\bf k}$ points, but the small corep multiplicities in the symmetry data vector do not satisfy the compatibility relations along a line or a plane connecting at least one pair of maximal ${\bf k}$ points.  Specifically, using~\checktopmat, this numerically occurs when the number of bands (small coreps corresponding to the occupied Bloch states) at a maximal vector ${\bf k}$ is \emph{different than} the corep multiplicities at ${\bf k}$ implied through the compatibility relations by the Bloch states that transform in the symmetry data at a different maximal vector ${\bf k}'$.  This implies that the bands described by the symmetry data vector necessarily connect along a high-symmetry line or plane to bands above the Fermi energy not included in the symmetry data.  In this case, because the symmetry data necessarily characterizes a topological semimetal, then we label the partially occupied connected set of electronic bands characterized by the symmetry data as an ``enforced semimetal'' (ES).  A well-studied example of an ES material is the archetypal Dirac semimetal Cd$_3$As$_2$ in SG 137 ($P4_{2}/nmc$)~\cite{ZJDirac,CavaDirac1,CavaDirac2,YulinCadmiumExp} (\icsdweb{107918}, see \supappref{App:ES}), which was recently confirmed in theory~\cite{HingeSM} and experiment~\cite{HingeSMExp} to be a \emph{higher-order} topological semimetal.}

\item{The symmetry data vector corresponds to fully occupied Bloch-state multiplets at all maximal ${\bf k}$ points, and the small corep multiplicities in the symmetry data vector \emph{satisfy} the compatibility relations along all lines and planes connecting the maximal ${\bf k}$ points.  Specifically, using~\checktopmat, this numerically occurs when the number of bands (small coreps corresponding to the occupied Bloch states) at all maximal ${\bf k}$ vectors is \emph{the same} as the corep multiplicities at ${\bf k}$ implied through the compatibility relations by the Bloch states that transform in the symmetry data at all other maximal vectors ${\bf k}'$.

The bands described by the symmetry data are therefore compatible with an insulating gap along all high-symmetry lines and planes, and the symmetry data vector is labeled either LCEBR, NLC, or SEBR (see \supappref{App:EBRs} for further details).  It is important to emphasize that LCEBR, NLC, and SEBR bands may still be connected to bands at other energies not described by the symmetry data by topological Weyl points or nodal lines in the BZ interior (\emph{i.e.} away from all high-symmetry lines and planes)~\cite{YoungkukLineNode,ZhidaSemimetals,S4Weyl,MTQC}; NLC- and SEBR-classified topological semimetal phases are discussed in further detail in \supappref{App:TopologicalMaterialsNoSOC} for further details).  In particular, as established in Refs.~\onlinecite{YoungkukLineNode,ZhidaSemimetals} and discussed in \supappref{App:TopologicalMaterialsNoSOC}, \emph{all} NLC and SEBR bands in nonmagnetic crystals w/o SOC characterize topological semimetal phases with nodal degeneracies in the BZ interior (respectively termed NLC-SM and SEBR-SM phases in this work).

}

\end{enumerate}

Finally, we note that in \checktopmat, it is occasionally possible for a set of characters in the symmetry data to appear to violate time-reversal ($\mathcal{T}$) symmetry.  We attribute this situation to numerical error, and in all cases, we have confirmed that the characters match the small irreps of the little groups of the Type-I magnetic (unitary) SG~\cite{BigBook,MTQC} that is a subgroup of the nonmagnetic SG for the symmetry data.  In the calculations performed for this work on nonmagnetic materials, we have merged all Bloch states that exhibit $\mathcal{T}$-breaking splitting on an energy scale of less than $1~meV$ to ensure that the symmetry data and band structures displayed on~\webNoICSD~are $\mathcal{T}$-symmetric.

\section{Overview of the Previous Functionality of the~\vasptotrace~Program and New Features of the~\checktopmat~Program}\label{App:Check_Topo_appendix} 

In a recent previous work~\cite{AndreiMaterials}, the authors introduced the~\vasptotrace~\cite{Irvsp2020} and \webchecktopmat~programs on the~\webBCSshort~to diagnose the topology of materials whose electronic structure has been calculated using VASP~\cite{vasp1,vasp2}.  The input files for \webchecktopmat~are the output files of \vasptotrace, which calculates the traces (eigenvalues) of the unitary symmetries of the electronic Bloch wavefunctions obtained from VASP~\cite{vasp1,vasp2}.  For this work, we have for the first time used~\vasptotrace~to compute band topology away from $E_{F}$, and have introduced new features for~\webchecktopmat.  In this section, we will first provide in~\supappref{App:VASP2Trace} an overview of the previous functionality of the~\vasptotrace~and \webchecktopmat~programs implemented for Ref.~\onlinecite{AndreiMaterials}.  We will then in~\supappref{App:CheckTopoMat} subsequently detail the new features in~\webchecktopmat~implemented for this work.

\subsection{Overview of the Previous Functionality of the~\vasptotrace~and~\webchecktopmat~Programs}
\label{App:VASP2Trace}

In this section, we will briefly summarize the algorithm and output of \vasptotrace~and the previous functionality of~\webchecktopmat.  To begin, in the SG of a crystalline solid, each unitary symmetry operation is defined as ${\cal O}=\{R|{\bf t}\}$, where $R$ is a point group symmetry operation (rotation or rotoinversion) and ${\bf t}$ is a translation vector (which may have a fractional value in the units of the lattice constants)~\cite{BigBook}.  For each maximal ${\bf k}$ point (defined in Refs.~\onlinecite{QuantumChemistry,Bandrep1}), we first obtain the single-particle Bloch wavefunctions $\psi_{n\bf k}(\vec r)\equiv \braket{r}{\psi_{n\bf k}}$ using the DFT package VASP~\cite{vasp1,vasp2} and identify a set of orthonormal wavefunctions $\{\psi_{n\bf k}^1(\vec r),\psi_{n\bf k}^2(\vec r),\ldots,\psi_{n\bf k}^m(\vec r)\}$ with a degeneracy $m$ that each have the same energy eigenvalue $E_n({\bf k})$.  If the symmetry operation ${\cal O}$ belongs to the little group $G_{\bf k}$ of {\bf k}, then acting on any of the wavefunctions $\psi_{n\bf k}^i(\vec r)$ with ${\cal O}$ results in a linear combination of all of the $m$ degenerate Bloch states at $E_{n}$ and ${\bf k}$:
\begin{equation}
\mathcal{O}\ket{\psi_{n\bf k}^i}=\sum_{j=1}^m O_{ij}^{\bf k}\ket{\psi_{n\bf k}^j},
\end{equation}
where $O^{\bf k}$ is the matrix representative of symmetry operation $\mathcal{O}$.  The $ij$-th element of $O^{\bf k}$ is in turn given by:
\begin{equation}
O_{ij}^{\bf k}=\bra{\psi_{n\bf k}^j}\mathcal{O}\ket{\psi_{n\bf k}^i}.
\label{eq:matrixrep}
\end{equation}
The matrix representatives $O^{\bf k}$ form a (generically reducible) small corep of $G_{\bf k}$.

It is important to emphasize that in the output of a DFT calculation, numerical precision and symmetry-breaking effects (\emph{e.g.} weak magnetism) may lead to several nondegenerate states appearing as degenerate, or may lead to degenerate states appearing weakly split.  Hence, it is necessary to implement a numerical tolerance factor for the energies of states at the same ${\bf k}$ point, such that states within the energy range of the tolerance factor are taken to be degenerate.  To avoid accidental degeneracies, it is desirable to have as small a tolerance factor as possible up to the energy scale at which numerical precision issues lead to symmetry-breaking splitting.  For this work, we have chosen a tolerance factor of $0.002$~eV.

As discussed in~\supappref{App:TQCReview_appendix}, the multiplicities of the irreducible small coreps in the (generically reducible) representation corresponding to the degenerate states $\{\psi_{n\bf k}^1(\vec r),\psi_{n\bf k}^2(\vec r),\ldots,\psi_{n\bf k}^m(\vec r)\}$ can be obtained by computing the symmetry characters (traces) of the matrix representatives $O^{\bf k}$ in Eq.~(\ref{eq:matrixrep}).  Hence, it is sufficient to calculate:
\begin{equation}
\chi_{n{\bf k}}^{O}=\sum_{i=1}^m\bra{\psi_{n\bf k}^i}\mathcal{O}\ket{\psi_{n\bf k}^i},
\label{eq:trace}
\end{equation}
for the matrix representative $\mathcal{O}$ of each unitary symmetry operation $g\in G_{\bf k}$.  The traces (characters) $\chi_{n{\bf k}}^{O}$ in Eq.~(\ref{eq:trace}) may straightforwardly be obtained from the~\textsc{WAVECAR} output file of VASP, in which the Bloch wavefunctions at each ${\bf k}$ point are each expressed as a linear combination of plane waves:
\begin{equation}
\psi_{n\bf k}^i({\bf r})=\sum_{j}C_{j,{n\bf k}}^ie^{i({\bf k}+{\bf G}_j)\cdot{\bf r}}.
\label{eq:wavefunction}
\end{equation}
In Eq.~(\ref{eq:wavefunction}), each ${\bf G}_j$ is a reciprocal lattice vector, and states in different BZs related by reciprocal lattice vectors are orthonormal such that $\braket{{\bf k}+{\bf G}_i}{{\bf k}+{\bf G}_j}=\delta_{ij}$.  Although the reciprocal lattice contains an infinite number of BZs related by the vectors ${\bf G}_j$, in practice, VASP only considers the truncated finite set of ${\bf k}$ points for which $\frac{\hbar^2}{2m_e}({\bf k}+{\bf G}_j)^2<E_{\text{cutoff}}$, where the energy cutoff $E_{\text{cutoff}}$ is an input parameter.

Given $\psi_{n\bf k}^i({\bf r})$ in the plane-wave basis in Eq.~(\ref{eq:wavefunction}), the action of the symmetry operation $O$ is given by:
\begin{eqnarray}
\mathcal{O}\psi_{n\bf k}^i({\bf r})&=&\{R|{\bf t}\}\psi_{n\bf k}^i({\bf r})=\psi_{n\bf k}^i(\{R|{\bf t}\}^{-1}{\bf r})=\sum_{j}C_{j,n{\bf k}}^ie^{i({\bf k}+{\bf G}_j)\cdot R^{-1}({\bf r}-{\bf t})}=\nonumber\\
&=&\sum_{j}C_{j,n{\bf k}}^ie^{iR({\bf k}+{\bf G}_j)\cdot({\bf r}-{\bf t})}=e^{-i{\bf k}\cdot{\bf t}}\sum_{j}C_{j,n{\bf k}}^ie^{i({\bf k}+{\bf G}_{j^{'}})\cdot{\bf r}}e^{-i{\bf G}_{j^{'}}\cdot{\bf t}},
\end{eqnarray}
where ${\bf G}_{j^{'}}$ is the reciprocal lattice vector for which:
\begin{equation}
{\bf k}+{\bf G}_{j^{'}}=R{\bf k}+{\bf G}_{j}.
\label{eq:relGvecs}
\end{equation}
Hence, the $i$-th element of the diagonal of the matrix representative $O^{\bf k}$ is given by:
\begin{equation}
\bra{\psi_{n\bf k}^i}\mathcal{O}\ket{\psi_{n\bf k}^i}=e^{-i{\bf k}\cdot{\bf t}}\sum_{j}(C_{j^{'},n{\bf k}}^i)^{*}C_{j,n{\bf k}}^ie^{-i{\bf G}_{j^{'}}\cdot{\bf t}},
\end{equation}
where in each term in the sum over $j$, the index $j^{'}$ is individually determined through Eq.~(\ref{eq:relGvecs}).  The trace of the matrix representative $O^{\bf k}$ defined in Eq.~(\ref{eq:trace}) is thus given by:
\begin{equation}
\chi_{n,{\bf k}}^{O}=e^{-i{\bf k}\cdot{\bf t}}\sum_{i=i}^m \sum_{j}(C_{j^{'},n{\bf k}}^i)^{*}C_{j,n{\bf k}}^ie^{-i{\bf G}_{j^{'}}\cdot{\bf t}},
\label{eq:finalChars}
\end{equation}
such that the trace of the identity operation $\mathcal{O}=\{E|0\}$ is equal to the band degeneracy $m$ at $E_{n}(\bf k)$.

In summary, \vasptotrace~obtains the plane-wave coefficients $C_{j,n{\bf k}}^{i}$ from the~\textsc{WAVECAR} output file generated by VASP for the $m$ degenerate states at an energy $E_{n}$ and a point ${\bf k}$. ~\vasptotrace~then produces the following output data:
\begin{enumerate}
	\item{The symmetry operations of the SG (one symmetry operation for each element of the point group of the SG), given in a primitive basis that in general does not correspond to the standard setting of the SG listed in the International Tables for Crystallography~\cite{ITCA} or the~\webBCSshort.}
	\item{The coordinates of the maximal {\bf k} vectors of the SG expressed in a reciprocal basis consistent with the (generically non-standard) setting of the outputted symmetry operations.}
	\item{For each maximal {\bf k} vector and band $n$,~\vasptotrace~outputs the traces of the symmetry operations of the little group $G_{\bf k}$ (one trace for each element of the point group of $G_{\bf k}$).}
\end{enumerate}

In Ref.~\onlinecite{AndreiMaterials}, together with the~\vasptotrace~program written by Zhijun Wang~\cite{Irvsp2020}, the authors also introduced the \webchecktopmat~program to diagnose the topology of a material using the output of~\vasptotrace. To diagnose the bulk topology,~\webchecktopmat~first compares the symmetry operation traces computed {\it ab-initio} to the little group character tables on the~\webBCSshort, allowing the determination of the little group small coreps in the symmetry data at a set of (typically, but not necessarily) maximal ${\bf k}$ points.

To identify the multiplicities of the irreducible small coreps in the symmetry data at each maximal ${\bf k}$ point, the previous implementation of~\webchecktopmat~first converts the data given by~\vasptotrace~into the standard setting of the SG used by the programs on the~\webBCSshort.  The conversion between the SG setting in the output of~\vasptotrace~and the standard setting of the~\webBCSshort~is performed using the~\identify~program on the~\webBCSshort~\cite{identifygroupwebsite} to calculate a transformation matrix between the output of~\vasptotrace~and the standard setting.  The transformation matrix additionally allows the identification of the maximal ${\bf k}$ vector labels used by VASP given the output of~\vasptotrace.  Next, given the characters of the SG symmetry elements in the standard setting [Eq.~(\ref{eq:finalChars}) after using the~\identify~program] and the ${\bf k}$ vector labels, the small corep multiplicities are obtained through the Schur orthogonality relation (magic formula), as detailed in the text surrounding Eq.~(\ref{eq:magicFormula}) in~\supappref{App:TQCReview_appendix}.  Lastly, the irreducible small corep multiplicities corresponding to the symmetry data are then analyzed using the machinery of TQC to obtain the topological classification of the occupied bands, as detailed in Ref.~\onlinecite{AndreiMaterials} and~\supappref{App:TQCReview_appendix}.

However, it is important to emphasize that due to numerical precision issues with bands far above $E_{F}$, or symmetry-breaking effects such as magnetism, the Bloch wavefunctions at ${\bf k}$ in Eq.~(\ref{eq:wavefunction}) may exhibit unitary symmetry eigenvalues (characters) that are not possible given the symmetry elements in the little group $G_{\bf k}$.  Specifically, for $m$ degenerate Bloch states at a point ${\bf k}$ and an energy $E_{n}$ in the output of a VASP calculation, Eq.~(\ref{eq:finalChars}) may return impossible combinations of characters for the set of symmetry-operation matrix representatives $\{O^{\bf k}\}$ if symmetries have become broken, or if the DFT calculations did not converge at $E_{n}$ and ${\bf k}$.  We refer to this situation, in which a group of degenerate states exhibits a set of characters that are together incompatible with the little group $G_{\bf k}$, as a \emph{bad trace}.  At each of the maximal ${\bf k}$ vectors,~\webchecktopmat~searches in increasing integer electronic fillings for a filling $\nu$ at which there is a bad trace at ${\bf k}$.  If~\webchecktopmat~at any ${\bf k}$ vector encounters a bad trace with $\nu \leq N_{e}$, where $N_{e}$ is the number of valence electrons, then the topological classification at $E_{F}$ cannot be determined.  In this work, we have discarded the~\TQCDBNbrFailedVASPToTraceBadTracesICSDs~ICSD entries for which VASP calculations converged to a (meta)stable state with bad traces at or below $E_{F}$.

\subsection{New Features of the~\webchecktopmat~Program Introduced in this Work}
\label{App:CheckTopoMat}

In addition to the previous functionality of the~\vasptotrace~and~\webchecktopmat~programs, we have introduced several new features in this work:
\begin{enumerate}
	\item{The most recent version of~\vasptotrace~\cite{Irvsp2020} now allows users to input a number of bands that differs from the number of electrons (\emph{i.e.} a number of bands that differs from the intrinsic electronic filling).  In this work, we have used this new feature to perform the first high-throughput analysis of band topology \emph{away from $E_{F}$} in nonmagnetic materials, specifically varying the number of electrons used in the input for~\vasptotrace~to analyze the topology all of the energetically isolated bands in the electronic structure of each ICSD entry.}
	\item{\webchecktopmat~can now identify small corep multiplicities both with and w/o SOC -- the previous iteration of~\webchecktopmat~implemented for Ref.~\onlinecite{AndreiMaterials} could only identify small corep multiplicities in the presence of SOC.}
	\item{\webchecktopmat~now generates a file containing topological data for all energetically isolated sets of bands.  For each set of energetically isolated bands (\emph{i.e.} bands that are separated from all other bands at all high-symmetry ${\bf k}$ points and along all high-symmetry BZ lines and planes), \webchecktopmat~now indicates if the set of bands is trivial, fragile topological, or stable topological (further differentiating between SEBR or NLC).  In the case in which the bands are stable topological, \webchecktopmat~additionally provides the values of the stable SIs using the convention established in Refs.~\onlinecite{ChenTCI,ZhidaSemimetals}.  As an example, we have reproduced in Table~\ref{tab:BandCharacterizationBiTwoMgThree} the output of \webchecktopmat~for the repeat-topological (RTopo) and supertopological (STopo) compound Bi$_2$Mg$_3$ [\icsdweb{659569}, SG 164 ($P\bar{3}m1$)] discussed in the main text (see \supappref{App:Supertopological} for precise definitions of RTopo and STopo compounds). }
	\item{When \webchecktopmat~diagnoses a partially filled set of bands in a material to be an ES or ESFD semimetal, \webchecktopmat~now provides the compatibility relations along the high-symmetry BZ lines and planes, thus identifying the location(s) of the protected crossing point(s).}
\end{enumerate}
For the present work, we have also made the updated version of~\webchecktopmat~detailed in this section publicly available on the~\webBCSshort~at~\href{http://www.cryst.ehu.es/cryst/checktopologicalmat}{www.cryst.ehu.es/cryst/checktopologicalmat}.

\begin{table}[h!]
	
{\small
	\begin{tabular}{ccccccccccccc}
		\hline 
		A & $\Gamma$ & H &K &L &M & dim & top. type  & ind/band & filling $\nu$ & top. type/all & ind/all  \\
		\hline
		\hline
		-A8(2)  &       -GM8(2)   &    -H4-H5(2)   &    -K4-K5(2)   &    -L3-L4(2)   &    -M5-M6(2) &  2&  SEBR&0 0 0 3 &     2 &SEBR&0 0 0 3\\
		-A9(2)  &       -GM9(2)   &       -H6(2)   &       -K6(2)   &    -L5-L6(2)   &    -M3-M4(2) &  2 &  SEBR& 0 0 0 1&      4 &LCEBR& 0 0 0 0 \\
		\hline
		-A9(2)  &       -GM8(2)   &    -H4-H5(2)   &       -K6(2)   &    -L3-L4(2)   &    -M3-M4(2) &  4&  SEBR&0 0 0 3 &     8 & SEBR&0 0 0 3\\
		-A8(2)  &       -GM8(2)   &       -H6(2)   &    -K4-K5(2)   &    -L5-L6(2)   &    -M5-M6(2) &   &        &        &       &       & \\
		\hline
		-A9(2)  &       -GM9(2)   &       -H6(2)   &       -K6(2)   &    -L3-L4(2)   &    -M5-M6(2) &  2&  SEBR&0 0 1 1 &    10 & SEBR&0 0 1 0\\
		\hline
		-A8(2)  &   -GM4-GM5(2)   &       -H6(2)   &       -K6(2)   &    -L5-L6(2)   &    -M3-M4(2) &  4&  SEBR&0 0 0 3 &    14 & SEBR&0 0 1 3\\
		-A6-A7(2)  &       -GM8(2)   &    -H4-H5(2)   &    -K4-K5(2)   &    -L3-L4(2)   &    -M5-M6(2) &   &        &        &       &       & \\
		\hline
		-A4-A5(2)  &   -GM6-GM7(2)   &       -H6(2)   &       -K6(2)   &    -L5-L6(2)   &    -M3-M4(2) &  2&  SEBR&0 0 1 0 &    16 & SEBR&0 0 0 3 \\
		\hline
		-A8(2)  &       -GM9(2)   &    -H4-H5(2)   &       -K6(2)   &    -L3-L4(2)   &    -M3-M4(2) &  6&  SEBR&0 0 0 2 &    22 & SEBR&0 0 0 1\\
		-A9(2)  &       -GM9(2)   &       -H6(2)   &       -K6(2)   &    -L5-L6(2)   &    -M5-M6(2) &   &        &        &       &       &        \\
		-A6-A7(2)  &   -GM6-GM7(2)   &       -H6(2)   &    -K4-K5(2)   &    -L5-L6(2)   &    -M3-M4(2) &   &        &        &       &       &        \\
		\hline
		-A9(2)  &       -GM9(2)   &       -H6(2)   &       -K6(2)   &    -L3-L4(2)   &    -M5-M6(2)&  2 & SEBR& 0 0 1 1  &   24  &SEBR& 0 0 1 2\\
		\hline
		-A8(2)  &       -GM8(2)   &       -H6(2)   &    -K4-K5(2)   &    -L3-L4(2)   &    -M3-M4(2)&  8 & SEBR& 0 0 1 1  &   32  &SEBR& 0 0 0 3\\
		-A4-A5(2)  &       -GM8(2)   &    -H4-H5(2)   &       -K6(2)   &    -L5-L6(2)   &    -M5-M6(2) &   &       &          &       &      &       \\
		-A8(2)  &   -GM4-GM5(2)   &       -H6(2)   &    -K4-K5(2)   &    -L5-L6(2)   &    -M5-M6(2) &   &     &          &       &      &       \\
		-A9(2)  &       -GM8(2)   &    -H4-H5(2)   &       -K6(2)   &    -L3-L4(2)   &    -M3-M4(2) &   &     &          &       &      &       \\
		\hline
		
		\hline

	\end{tabular}
}
	\caption[Topology of all of the bands in the RTopo and STopo compound Bi$_2$Mg$_3$]{A typical table generated by~\webchecktopmat.  Here, we have used as an example the repeat-topological (RTopo) and supertopological (STopo) compound Bi$_2$Mg$_3$ [\icsdweb{659569}, SG 164 ($P\bar{3}m1$)] discussed in the main text (see \supappref{App:Supertopological} for precise definitions of RTopo and STopo compounds).  In this table, the horizontal lines indicate the electronic fillings $\nu$ at which the occupied bands satisfy the compatibility relations.  Columns $A$, $\Gamma$, $H$, $K$, $L$, and $M$ contain the coreps at each maximal ${\bf k}$ point for each energetically isolated set of bands that satisfy the compatibility relations.  Column ``dim'' shows the dimension of each set of bands, Column ``top. type'' indicates the topology of the isolated bands [which can either be LCEBR, FRAGILE, NLC, or SEBR], and Column ``ind/band'' contains the stable symmetry-based indicators (SIs) ($Z_{2,1}$ $Z_{2,2}$ $Z_{2,3}$ $Z_{4}$) that result from subducing onto SG 2 ($P\bar{1}$).  The remaining three columns contain the cumulative information of several combined sets of energetically isolated bands that are filled up to a total number of valence electrons specified by the column ``filling $\nu$.''  Column ``top. type/all'' provides the cumulative topology (LCEBR, FRAGILE, NLC, or SEBR) at the gap specified by the electronic filling in the ``filling $\nu$'' column, and Column ``ind/all'' provides the cumulative stable SIs of all of the filled bands up to the same gap.  Hence, the values of ``ind/all'' at each filling $\nu$ listed in this table can be obtained by summing the values in the ``ind/band'' column up to the same filling $\nu$.  In Bi$_2$Mg$_3$, the Fermi level lies at a valence filling $\nu=16$, indicating that the gap at $E_{F}$ exhibits symmetry-indicated stable (SEBR) topology characterized by the subduced stable SIs ($0003$).}
\label{tab:BandCharacterizationBiTwoMgThree}
\end{table}

\afterpage{\clearpage}

\section {VASP Calculation Details and Data Set Preparation}\label{App:VASP_appendix}

In this work, we performed ab-initio calculations using {{Density Functional Theory (DFT)~\cite{Hohenberg-PR64,Kohn-PR65} as implemented in the Vienna Ab-initio Simulation Package (VASP)~\cite{vasp1,PhysRevB.48.13115}.  For each material calculation, we used as input the structural parameters reported on the ICSD~\cite{ICSD}.  We treated the interaction between the ion cores and the valence electrons using the projector augmented-wave method~\cite{paw1}.  For the exchange-correlation potential, we used the generalized gradient approximation (GGA) with the Perdew-Burke-Ernzerhof parameterization for solids~\cite{PhysRevLett.77.3865}.  For calculations incorporating the effects of spin-orbit coupling (SOC), we accounted for the effects of SOC using the second variation method~\cite{PhysRevB.62.11556}.  For the plane-wave expansion, we employed a $\Gamma$-centered ${\bf k}$-point grid of size (11$\times$11$\times$11) for reciprocal space integration and a 550 eV energy cutoff.  Because small changes in the structural parameters of a material with a similarly small gap size can drive a topological phase transition, then for this work, we have analyzed \emph{all} of the entries in the ICSD, including duplicate entries for the same compound that feature only minor variation in the reported parameters.  For example in PbTe, there are 42 total entries in the ICSD, of which only 5 have a negative (inverted) gap corresponding to a higher-order topological crystalline insulating (TCI) phase (see Ref.~\onlinecite{BarryPbTe} and \supappref{App:rotationAnomaly}).  Therefore, rather than simply report PbTe as a TCI, we provide on~\webNoICSD~a complete analysis of all 42 ICSD entries for PbTe, allowing users to choose between entries with and without nontrivial symmetry-indicated topology.

For each material, we prepared input files for VASP using the CIF structure files provided on the ICSD.  Specifically, we first used the Atomic Simulation Environment (ASE)~\cite{ASE} to transform the CIF file from the ICSD into a VASP POSCAR structure input file.  This step must be performed with caution, because some CIF files on the ICSD are missing atoms that are listed in the chemical formula (typically, but not always, hydrogen atoms), and some unit-cell coordinates are reported in a non-standard basis.  To account for the latter issue of differing unit-cell bases, we have employed the PHONOPY~\cite{PHONOPY} package to convert the coordinates of the atoms in the CIF files into positions in a primitive cell chosen in a consistent (standard) basis for each SG.  In Table~\ref{tab:NbrAnalyzedCompounds}, we provide statistics regarding the number of ICSD entries considered in our analysis.  In this work, we began by considering the complete list of $\TQCDTotICSDs$ entries in the ICSD.  Of the $\TQCDTotICSDs$ ICSD entries, $\TQCDTotICSDsExp$ contain experimentally-obtained atomic positions, and the remaining $\TQCDTotICSDsTheo$ entries contain theoretically-obtained atomic positions.  Separately, of the $\TQCDTotICSDs$ ICSD entries, $\TQCDstoichiometric$ $(\TQCDstoichiometricPercent)$ have stoichiometric chemical formulas and processable (non-corrupt) CIF structure files, whereas the remaining $\TQCDNostoichiometric$ $(\TQCDNostoichiometricPercent)$ entries are either not stoichiometric or exhibit corrupted CIF structure files.  In general, non-stoichiometric materials cannot be analyzed using TQC without constructing exceedingly large unit cells representing the average structure, and corrupted CIF files can only processed if repaired using a case-by-case methodology.

Next, having established the primitive cell, we construct the pseudopotential file for each compound and execute VASP.  This process is performed in an automated, high-throughput fashion through a script that uses each ICSD CIF file as an input, and then outputs the topological analysis, electronic band structure, and density of states.  As shown in Fig.~\ref{Fig3} of the main text, our topological analysis is performed by computing the characters of the unitary symmetry operations of the little groups at the maximal ${\bf k}$ vectors (see \supappref{App:Check_Topo_appendix}).  For each SG, the list of maximal ${\bf k}$ vectors may be obtained by choosing an EBR of the SG using the~\href{www.cryst.ehu.es/cryst/bandrep}{BANDREP} tool on the~\webBCSshort~(\url{www.cryst.ehu.es/cryst/bandrep}).  The first column in the output of~\href{www.cryst.ehu.es/cryst/bandrep}{BANDREP} contains the maximal ${\bf k}$ vectors of the SG.

Beyond the Bloch states at the maximal ${\bf k}$ vectors, we also compute for each material the electronic structure (bands) along an SG-dependent ${\bf k}$ path to generate band structure plots, and the density of states.  For each band structure calculation, we specifically calculate the energies of the Bloch eigenstates at 20 ${\bf k}$ points along each ${\bf k}$ path segment.  At each ${\bf k}$ point along each path segment, we calculate the energies of at least $2N_{e}$ Bloch states, where $N_{e}$ is the number of valence electrons in the primitive cell; we do not include bands originating from core-shell atomic orbitals.  We note that because VASP can be parallelized over bands, the number of bands included in some material calculations is larger than $2N_{e}$.  For completeness and consistency with other previous works~\cite{BCTBZ,ChenMaterials}, we have included in our electronic structure calculations both the minimal paths connecting all ${\bf k}$ vectors, as well as additional paths that are not required by band connectivity~\cite{QuantumChemistry,Bandrep2}, but are nevertheless commonly employed in earlier works.  For each of the 14 Bravias classes and crystallographic point groups, we have constructed ${\bf k}$ paths in reduced coordinates (independent of the lattice parameters) using a combination of symmetry considerations and band-structure compatibility relations~\cite{LuisSubduction}.  We emphasize that this approach differs from other programs, such as the~\emph{Seek-path} package~\cite{SEKPATH}, which use different criteria to generate ${\bf k}$ paths in non-reduced units of the lattice vectors, causing the ${\bf k}$ paths to become structure-dependent and vary for materials within the same SG.  For example, the high-throughput first-principles calculations performed in Ref.~\onlinecite{ChenMaterials} were performed without rescaled ${\bf k}$ vectors.  Hence, in the~\href{http://materiae.iphy.ac.cn/}{Catalogue of Topological Materials} introduced in Ref.~\onlinecite{ChenMaterials}, there exist materials with the same SG and different displayed ${\bf k}$ paths, whereas the electronic structures of all of the ICSD entries on~\webNoICSD~with the same SG are plotted along the same (reduced-coordinate) ${\bf k}$ paths.  We additionally emphasize that in order to locate all of the band crossings in topological semimetal phases, we found that it was necessary to search for crossing points not just along the minimal set of high-symmetry lines between pairs of maximal ${\bf k}$ vectors, but also along lines connecting the first and second BZs.  In some SGs, we also include additional ${\bf k}$ paths beyond the first BZ in order to avoid plotting discontinuous band structures.  For each ICSD entry on~\webNoICSD, we have made the ${\bf k}$ paths and POSCAR files available for users to download directly from the page for each ICSD entry on~\webNoICSD.

As stated in the main text, we define a unique material as the set of ICSD entries that share the same stoichiometric formula, topological subclassification with SOC, SG, and, in the case of a topological semimetal, type of crossing at the Fermi level in the presence of SOC.  We emphasize that our definition of a unique material does not account for the topological subclassification of a material w/o SOC, or account for variation in the topology of energetically isolated bands away from $E_{F}$.  Specifically, for simplicity, two ICSD entries with the same stoichiometric formula, SG, and topological classification (and crossing points if semimetallic) at $E_{F}$ in the presence of SOC are grouped together as a single unique material, whether or not the topology of the two ICSD entries is different at $E_{F}$ w/o SOC, or away from $E_{F}$ with or w/o SOC.  Hence as discussed in~\supappref{App:Supertopological}, there may exist variation in the topology away from $E_{F}$ across ICSD entries associated to the same unique material.  Additionally, we note that for a fraction of the analyzed ICSD entries, calculations only converged with SOC, and for the same material did not converge w/o SOC.  Specifically, calculations performed w/o SOC failed to converge for~\TQCDBNbrFailedNoSOCICSDs~ICSD entries, representing~\TQCDBPercentFailedNoSOCICSDs~of the~\TQCDBNbrICSDs~total materials for which calculations converged with SOC.  We have further classified our data by tagging materials that are either listed as magnetic on the~\webmaterialsproject~(\url{https://materialsproject.org/})~\cite{MaterialsProject}, or display nonzero magnetic moments in the output of our VASP calculations.  Finally, we have also identified the compounds that contain valence $f$ electrons as determined by VASP (\emph{i.e.} the compounds containing atoms with partially filled $f$ orbitals, see~\url{https://www.smcm.iqfr.csic.es/docs/vasp/node248.html} for further details); these materials may exhibit non-negligible electron-electron correlation effects.  Throughout this work, we follow the VASP pseudopotential documentation on~\url{https://www.smcm.iqfr.csic.es/docs/vasp/node248.html} in determining whether a material does not carry valence $f$ electrons, which we denote in this work using the shorthand expression ``without $f$ electrons.''  In Table~\ref{tab:NbrCompoundsWithTM}, we list the percentage of unique materials in which any of the associated ICSD entries are magnetic (either as listed on the~\webmaterialsproject~or in our first-principles calculations), and the percentage of unique materials with $f$ electrons at $E_{F}$ as determined by VASP.

\begin{table}[h!]
\caption{Number of analyzed compounds.}
\begin{center}
\begin{tabular}{|c|c|c|c|}
  \hline
  & Total & Experimentally-Obtained & Theoretically-Obtained \\
  & & Atomic Positions & Atomic Positions \\
  \hline
  Entries &  \TQCDTotICSDs &  \TQCDTotICSDsExp & \TQCDTotICSDsTheo \\
  Stoichiometric and Valid CIF File & \TQCDstoichiometric & \TQCDstoichiometricExp & \TQCDstoichiometricTheo \\ 
  Non-Stoichiometric or Corrupted CIF File & \TQCDNostoichiometric & \TQCDNostoichiometricExp & \TQCDNostoichiometricTheo \\
  \hline  
\end{tabular}
\label{tab:NbrAnalyzedCompounds}
\end{center}
\end{table}
\begin{table}[h!]
\caption{Percentage of unique materials with at least one magnetic ICSD entry and with valence $f$ electrons as determined by VASP (see~\url{https://www.smcm.iqfr.csic.es/docs/vasp/node248.html} for further details).}
\begin{center}
\begin{tabular}{c|c}

  Unique Materials with Magnetism &  \TQCDBNbrMaterialsMagneticMPVASPPercent \\
  \hline
  Unique Materials with Valence $f$ Electrons & \TQCDBNbrMaterialsFElectronsPercent \\ 
\end{tabular}
\label{tab:NbrCompoundsWithTM}
\end{center}
\end{table}

Lastly, it is important to emphasize additional issues that arise regarding magnetic ground states in VASP.  In this work, all of the calculations were performed with a zero initial magnetic moment on each atom.  However for a given compound, there may exist a magnetic ground state that is lower in energy than a metastable non- (para-) magnetic state.  Hence in this case, the nonmagnetic solution is a local minimum in the total energy landscape of the system, but may still correspond to the final state in the VASP calculation.  Depending on the details of the calculation, then VASP will either remain close to the initialized paramagnetic state, or may find a magnetic state that is lower in energy, but not necessarily the correct magnetic ground state (\emph{i.e.}, there may exist a different magnetic state, or a structurally distorted nonmagnetic state, that is lower in energy).  In VASP, the magnetism of a (meta)stable final state is most dependent on the following parameters:
\begin{itemize}
\item The total number of bands (NBANDS) in the calculation.  Because VASP is parallelized, then the number of bands must be divisible by the number of cores (NPAR) for an accurate and convergent calculation. Hence, VASP adjusts NBANDS to be a multiple of the number of cores.  Because NPAR depends on the details of the VASP configuration and hardware used to perform the calculation, then NPAR may vary for the same VASP calculation performed across two different (super)computers.  
\item The block size in the blocked Davidson algorithm (NSIM).
\end{itemize}
In most VASP calculations, NBANDs only affects the rate of convergence, and not the quantitative features (\emph{e.g.} magnetic configuration) of the final state.  However, in the presence of multiple nearby local minima in the total energy landscape, NBANDS, NPAR, and NSIM may (uncontrollably) bias the final local minimum of the VASP calculation.  In general, when using VASP to perform topological material calculations and to reproduce the calculations performed for this work, we recommend performing several tests with varying NBANDS, NPAR, NSIM, and other parameters, in particular confirming that all calculations converge to the same (ideally global) minimum.

In the high-throughput calculations performed for this work, we found it to be prohibitively difficult to systematically evaluate each ICSD entry for the appearance of metastable (potentially magnetic) final states.  However, by manually comparing intermediate self-consistent and band-structure output files in our VASP calculations, we were able to detect only $\sim$ 100 compounds with magnetic and nonmagnetic local minima that were close in energy.  Specifically, we compared the magnetization in the self-consistent and band structure output files of each VASP calculation, finding only $\sim$ 100 ICSD entries with large magnetization discrepancies (though we detected slight variations when our tests were performed on different supercomputers, due to varying NPAR, as discussed earlier in this section).  This makes us confident that the large majority of our material calculations exhibit reliable (\emph{i.e.} physically realistic) nonmagnetic final states.  

\afterpage{\clearpage}

\section{Computation Time and Resources}\label{App:CPUtime}

In this section, we detail the computational resources used in this work.  First, in \supappref{sec:SC}, we list the supercomputers used to perform the ab-initio calculations in this work, and the number CPU hours expended.  Then, in \supappref{sub:SG}, we provide a more detailed breakdown of our computational resource usage, focusing specifically on the number of CPU hours expended per SG and per number of atoms in the primitive cell.  Lastly, in \supappref{sub:Storage}, we detail the disk storage usage requested for the calculations performed in this work, focusing specifically on the disk storage requested per SG and per number of atoms in the primitive cell.

\subsection{Supercomputers and General Computational Resource Usage Statistics}
\label{sec:SC}

Our calculations were primarily executed on the Draco and Cobra supercomputers of the Max Planck Society (MPG) and on the Cori and Edison supercomputers of the United States Department of Energy (DOE).  Below, we provide information about the node architecture, communication network, and peak performance of each supercomputer:
\begin{itemize}
\item{Draco: HPC system with Intel 'Haswell' Xeon E5-2698 processors (880 nodes with 32 cores @ 2.3 GHz each). 106 of the nodes are equipped with accelerator cards (each with two PNY GTX980 GPUs).  Draco also contains 64 Intel 'Broadwell' processors that each have 40 cores and a main memory of 256 GB.  In total Draco has 30,688 cores and a total main memory of 128 TB, and provides a peak performance of 1.12 petaflop/s.  Beyond the compute nodes used for calculations, Draco has 4 login nodes and 8 I/O nodes, each with 1.5 petabytes of disk storage.}

\item{Cobra: HPC system with 3424 compute nodes, 136,960 CPU cores, 128 Tesla V100-32 GPUs, 240 Quadro RTX 5000 GPUs, 529 TB RAM DDR4, 7.9 TB HBM2, 11.4 petaflop/s peak DP, and 2.64 petaflop/s peak SP.}

\item{Cori: Cray XC40 system with 622,336 Intel processor cores and a theoretical peak performance of 30 petaflop/s (30 quadrillion operations per second).}

\item{Edison: Cray XC30 with 133,824 compute cores for running scientific applications, 357 TB of memory, and 7.56 petabytes of online disk storage with a peak I/O bandwidth of 168 GB/s.}
\end{itemize}

We performed first-principles calculations on a total of $\TQCDBNbrICSDs$ ICSD entries.  As shown in Fig.~\ref{Fig3} of the main text and discussed in \supappref{App:VASP_appendix}, the calculations that we performed can be divided into four steps, which we respectively label as VASP1, VASP2, VASP3, and VASP4: 
\begin{itemize}
\item{VASP1: Self-consistent DFT calculations. The calculations performed in VASP1 generate charge density (CHGCAR) files, which served as input for the remaining three calculation steps.}

\item{VASP2: Generating the Bloch wavefunctions at the maximal ${\bf k}$ vectors, using the input from VASP1.  VASP2 generates the output files used to calculate the characters of the unitary symmetry operations of the little groups at the maximal ${\bf k}$ points, which we then used to diagnose the material topology, as detailed in \supappref{App:Check_Topo_appendix}.}

\item{VASP3: Calculating the band structure along high-symmetry lines and planes, using the input from VASP1.  Each segment of the ${\bf k}$ path contains 20 ${\bf k}$ points.  At each ${\bf k}$ point along each path segment, we calculated the energies of at least $2N_{e}$ Bloch states, where $N_{e}$ is the number of valence electrons in the primitive cell; we did not include bands originating from core-shell atomic orbitals.  As discussed in~\supappref{App:VASP_appendix}, because VASP is parallelized, then the number of bands must be divisible by the number of cores (NPAR) for an accurate and convergent calculation.  Hence, VASP adjusts the number of bands to be a multiple of the number of cores.  Because NPAR depends on the details of the VASP configuration and hardware used to perform the calculation, then NPAR may vary for the same VASP calculation performed across two different (super)computers.}

\item{VASP4: Self-consistent calculations of the density of states, using the input from VASP1.  To calculate the density of states in VASP4, we employed an (11$\times$11$\times$11) ${\bf k}$-point grid.} 
\end{itemize}

In Table~\ref{Table:CPU}, we show the number of CPU hours expended in the calculations performed in this work divided into each of the four calculation steps.

\begin{table}[h!]
\begin{tabular}{c|c|c}
Job & CPU h. with SOC & CPU h. w/o SOC \\
\hline
    VASP1 &\TQCDBVASPICPUhSOC &\TQCDBVASPICPUhNoSOC \\
    VASP2 &\TQCDBVASPIICPUhSOC & \TQCDBVASPIICPUhNoSOC\\
    VASP3 &\TQCDBVASPIIICPUhSOC & \TQCDBVASPIICPUhNoSOC\\
    VASP4 &\TQCDBVASPIVCPUhSOC & \TQCDBVASPIVCPUhNoSOC \\
\hline
    Total Time & \TQCDBVASPTotalCPUhSOC & \TQCDBVASPTotalCPUhNoSOC\\
\end{tabular}
\caption{Total CPU hours expended in each of the four VASP calculation steps, subdivided by calculations performed with and without incorporating the effects of SOC (with SOC and w/o SOC, respectively).}
\label{Table:CPU}
\end{table}

\subsection{Computational Resources Expended per SG and per Number of Atoms in the Primitive Cell}
\label{sub:SG}

In this section, we provide tables containing a detailed statistical breakdown of the computational resources expended for this work.  First, in Table~\ref{tb:statistics_soc_compounds_cputime} (Table~\ref{tb:statistics_nosoc_compounds_cputime}), we list the CPU hours expended to perform ab-initio calculations with SOC (w/o SOC), subdivided by SG.  Next, in Table~\ref{tb:statistics_cputime_nbratoms_primitivecell} (Table~\ref{tb:statistics_cputime_nbratoms_primitivecell_nosoc}), we list the CPU hours expended to perform ab-initio calculations with SOC (w/o SOC), subdivided by the number of atoms in the primitive cell of each material.  Most interestingly, we observe that there are vanishingly few stoichiometric materials in the ICSD with greater than $\ThresholdNbrAtoms$ atoms in the primitive cell.  Specifically, we find that there are $\TQCDstoichiometricGTatoms$ ICSD entries with greater than $\ThresholdNbrAtoms$ in the primitive cell, representing $\TQCDstoichiometricGTatomsPercent$ of the $\TQCDstoichiometric$ total stoichiometric entries in the ICSD. 

{\tiny
\begin{longtable}{|c|c|c|c|c|c|c|}
\caption[Total CPU hours expended for ab-initio calculations with SOC per SG]{Total CPU time expended for all of the ab-initio calculations performed in this work incorporating the effects of SOC (with SOC), subdivided by space group (SG).  In order, the columns in this table list the SG, the number of ICSDs successfully analyzed in the SG, the CPU hours expended per each of the four calculation steps (VASP1 through VASP4) detailed in \supappref{sec:SC}, and the total CPU hours expended for materials in the SG.
\label{tb:statistics_soc_compounds_cputime}}\\
\hline
SG & \# ICSDs & \vaspcputimelabel{1} & \vaspcputimelabel{2} & \vaspcputimelabel{3} & \vaspcputimelabel{4} & Total (CPU hours) \\
\hline
1 & 225 & 67534.5 & 718.9 & 4560.6 & 38488.5 & 111302.5 \\ 
2 & 1885 & 572989.9 & 7439.1 & 81830.9 & 322951.5 & 985211.3 \\ 
3 & 19 & 6489.8 & 26.9 & 400.9 & 2675.3 & 9592.9 \\ 
4 & 305 & 97974.7 & 576 & 9153.7 & 61414.5 & 169118.9 \\ 
5 & 195 & 73484.8 & 305.5 & 4328.2 & 56460.5 & 134579 \\ 
6 & 39 & 7746 & 32.5 & 429.6 & 3391.4 & 11599.5 \\ 
7 & 143 & 52621.6 & 267.8 & 4131.8 & 29000.3 & 86021.5 \\ 
8 & 173 & 58561.7 & 188.4 & 2737.3 & 36156.2 & 97643.7 \\ 
9 & 278 & 194291.8 & 662.5 & 9442.9 & 123863.5 & 328260.7 \\ 
10 & 44 & 7407.7 & 59.6 & 938.5 & 3597 & 12002.9 \\ 
11 & 790 & 172295.1 & 1916.7 & 30243.2 & 126423.2 & 330878.2 \\ 
12 & 1844 & 473531.1 & 3539 & 46170.5 & 360788.5 & 884029.2 \\ 
13 & 314 & 76186.3 & 801.5 & 12216.9 & 39632.1 & 128836.8 \\ 
14 & 3787 & 1799616 & 16665.5 & 259458.6 & 1067200.4 & 3142940.5 \\ 
15 & 2526 & 1372458.9 & 8181.1 & 124624.6 & 830990.6 & 2336255.2 \\ 
16 & 2 & 73 & 24.3 & 49.2 & 70.7 & 217.1 \\ 
17 & 8 & 909.6 & 27.1 & 510.5 & 618.8 & 2066 \\ 
18 & 42 & 10938.8 & 300.9 & 5175.5 & 7024 & 23439.2 \\ 
19 & 364 & 84374.7 & 2351.3 & 41287.4 & 55194.6 & 183208 \\ 
20 & 85 & 15247.2 & 134.8 & 2707 & 8951.7 & 27040.7 \\ 
21 & 21 & 1548.9 & 13.9 & 244.3 & 863.6 & 2670.6 \\ 
22 & 17 & 1404.3 & 9.9 & 181.6 & 775.1 & 2370.9 \\ 
23 & 16 & 2789.7 & 17.4 & 259.2 & 1087.5 & 4153.9 \\ 
24 & 3 & 168.2 & 2.4 & 28.7 & 101.3 & 300.6 \\ 
25 & 59 & 2666.1 & 58 & 957.1 & 1603.2 & 5284.4 \\ 
26 & 85 & 21846.5 & 391.8 & 6848.2 & 11396.8 & 40483.4 \\ 
27 & 1 & 89.2 & 2.2 & 40.4 & 67.5 & 199.3 \\ 
28 & 18 & 1498.6 & 37.6 & 635.1 & 1114.7 & 3286 \\ 
29 & 89 & 19338.2 & 522.8 & 9292.9 & 14563.9 & 43717.7 \\ 
30 & 6 & 2866.8 & 58.1 & 925.4 & 1629.4 & 5479.7 \\ 
31 & 247 & 37185.5 & 889.3 & 15277.5 & 23724.4 & 77076.7 \\ 
32 & 12 & 3779.4 & 108.2 & 1962.8 & 2783.8 & 8634.2 \\ 
33 & 406 & 102540.7 & 2171.8 & 38634.4 & 60282.2 & 203629.2 \\ 
34 & 30 & 8458.9 & 208.5 & 3488.3 & 5763.3 & 17919.1 \\ 
35 & 8 & 1291 & 11 & 213.4 & 824.7 & 2340.2 \\ 
36 & 412 & 101801.2 & 649.6 & 12947.5 & 58890.4 & 174288.7 \\ 
37 & 7 & 1654.4 & 15.9 & 338 & 1094.4 & 3102.7 \\ 
38 & 173 & 15730.8 & 118 & 2542.3 & 11316.3 & 29707.4 \\ 
39 & 15 & 3302.6 & 68.5 & 887.3 & 2430 & 6688.3 \\ 
40 & 67 & 11630.1 & 122.5 & 3051.6 & 10744.8 & 25549 \\ 
41 & 44 & 14186.7 & 102 & 2590 & 7891 & 24769.7 \\ 
42 & 4 & 500.1 & 7.2 & 96.2 & 374.8 & 978.3 \\ 
43 & 126 & 27111.4 & 224.5 & 4999.9 & 21284 & 53619.8 \\ 
44 & 100 & 10278.7 & 85.1 & 1294.8 & 7763.2 & 19421.7 \\ 
45 & 11 & 7084.3 & 52.7 & 1032.5 & 3797.5 & 11967 \\ 
46 & 91 & 23841.1 & 177.4 & 3333.6 & 14095.6 & 41447.6 \\ 
47 & 95 & 4588.5 & 161 & 3059.2 & 2737.8 & 10546.5 \\ 
48 & 3 & 153.7 & 7.4 & 121.3 & 110.8 & 393.2 \\ 
49 & 1 & 98.5 & 4 & 65.5 & 59.1 & 227.2 \\ 
50 & 11 & 2279.1 & 79.4 & 1386.9 & 1409.5 & 5154.9 \\ 
51 & 147 & 10885.8 & 449.7 & 8318.7 & 7566.8 & 27220.9 \\ 
52 & 51 & 9752 & 378.4 & 6759.2 & 6091.3 & 22981 \\ 
53 & 27 & 1699.2 & 77.7 & 1410.5 & 1275.7 & 4463.1 \\ 
54 & 25 & 5666.9 & 261.5 & 4925.9 & 4427.9 & 15282.2 \\ 
55 & 548 & 118860.4 & 3849.4 & 67835.6 & 61494.5 & 252039.9 \\ 
56 & 31 & 5994.3 & 212.2 & 3642.1 & 3354.3 & 13202.9 \\ 
57 & 208 & 49081.8 & 1486.3 & 27007.4 & 25043.3 & 102618.7 \\ 
58 & 493 & 37096.3 & 1289.4 & 22742.3 & 20676.6 & 81804.6 \\ 
59 & 338 & 37551.5 & 1166.3 & 20022.8 & 18167.1 & 76907.7 \\ 
60 & 252 & 36258.7 & 1677.5 & 31542.6 & 28260.9 & 97739.8 \\ 
61 & 233 & 54377.6 & 2172.4 & 36147.1 & 32512.4 & 125209.5 \\ 
62 & 5664 & 905383.9 & 28981.3 & 523829.7 & 469273.2 & 1927468.1 \\ 
63 & 2188 & 279088.8 & 2394.2 & 43507.4 & 169019 & 494009.4 \\ 
64 & 435 & 131710.8 & 1377.2 & 26477.8 & 91983.7 & 251549.5 \\ 
65 & 372 & 31828 & 893.7 & 8655.2 & 31429.7 & 72806.6 \\ 
66 & 51 & 7580.5 & 124.5 & 2536.3 & 6356.4 & 16597.8 \\ 
67 & 63 & 1465.3 & 25.8 & 331.3 & 873.2 & 2695.6 \\ 
68 & 21 & 5123.5 & 70.1 & 1519.6 & 3280.9 & 9994 \\ 
69 & 75 & 6718.6 & 81.2 & 1652.5 & 4493.9 & 12946.1 \\ 
70 & 208 & 30036.8 & 431.6 & 10250.5 & 21787.5 & 62506.4 \\ 
71 & 666 & 68696.9 & 797.7 & 13262.9 & 44634.7 & 127392.2 \\ 
72 & 240 & 26125.3 & 312 & 5526.8 & 18592.2 & 50556.3 \\ 
73 & 29 & 18882.4 & 222.6 & 4164.3 & 8777.1 & 32046.4 \\ 
74 & 370 & 28410 & 362.5 & 6227.7 & 18657.3 & 53657.5 \\ 
75 & 5 & 732.2 & 12.3 & 234.5 & 451.8 & 1430.8 \\ 
76 & 12 & 3354.8 & 81.1 & 1256.1 & 2268.4 & 6960.4 \\ 
77 & 2 & 42.3 & 1.4 & 14.9 & 28.8 & 87.4 \\ 
78 & 2 & 942.2 & 19.3 & 259.8 & 523.5 & 1744.8 \\ 
79 & 14 & 2200.4 & 19.3 & 285 & 1418.3 & 3923 \\ 
80 & 3 & 1735.2 & 15 & 270.9 & 1017.7 & 3038.9 \\ 
81 & 19 & 1657.2 & 48 & 656.3 & 1147.8 & 3509.3 \\ 
82 & 275 & 23777.8 & 230.4 & 2910.7 & 17937 & 44855.9 \\ 
83 & 11 & 2863.5 & 59.8 & 1052.8 & 1052.8 & 5029 \\ 
84 & 35 & 2992.7 & 123 & 2087.3 & 2218.5 & 7421.5 \\ 
85 & 54 & 11358 & 381.4 & 6312.3 & 6706.7 & 24758.4 \\ 
86 & 71 & 8051.1 & 262.8 & 4838.5 & 5006.1 & 18158.5 \\ 
87 & 298 & 20912.8 & 278.2 & 4524.8 & 16391.6 & 42107.3 \\ 
88 & 332 & 30525.1 & 529.7 & 8039.4 & 24842.2 & 63936.4 \\ 
90 & 5 & 2335.3 & 65 & 1021.4 & 994 & 4415.7 \\ 
91 & 9 & 534.8 & 20.6 & 367.3 & 368.7 & 1291.5 \\ 
92 & 135 & 12743 & 495.2 & 7427.4 & 7669.5 & 28335.1 \\ 
94 & 1 & 22.3 & 2.3 & 12.6 & 14.1 & 51.4 \\ 
95 & 3 & 196.6 & 6.2 & 105.9 & 106.8 & 415.6 \\ 
96 & 37 & 1705 & 94.1 & 1286.5 & 1349.4 & 4435.1 \\ 
97 & 2 & 215.7 & 3.6 & 58.2 & 175.9 & 453.3 \\ 
98 & 7 & 685.8 & 10 & 158.4 & 547.3 & 1401.6 \\ 
99 & 195 & 1644.6 & 74.9 & 729.8 & 960.1 & 3409.3 \\ 
100 & 38 & 6227.6 & 180.1 & 2927.2 & 3610.2 & 12945.1 \\ 
102 & 15 & 3196.3 & 73.5 & 1364.5 & 1754.2 & 6388.5 \\ 
103 & 9 & 258.2 & 15.3 & 298.5 & 405.9 & 977.9 \\ 
104 & 4 & 458.6 & 17 & 397.8 & 521.2 & 1394.7 \\ 
105 & 4 & 393 & 16 & 228 & 310.3 & 947.3 \\ 
106 & 1 & 139.3 & 6.9 & 54 & 70.3 & 270.4 \\ 
107 & 159 & 4325.5 & 51.9 & 653.4 & 3190 & 8220.7 \\ 
108 & 15 & 2502.9 & 32.7 & 551.2 & 2098.6 & 5185.4 \\ 
109 & 54 & 2504.3 & 29.3 & 408 & 2427.1 & 5368.7 \\ 
110 & 36 & 11290.9 & 117.1 & 2034.2 & 6727.7 & 20169.9 \\ 
111 & 27 & 541.8 & 22.5 & 389.6 & 444.1 & 1398 \\ 
112 & 4 & 102.9 & 7 & 125.2 & 142.5 & 377.5 \\ 
113 & 167 & 13435.6 & 527.2 & 7351.6 & 8437.9 & 29752.3 \\ 
114 & 42 & 5869.9 & 160.9 & 2567.3 & 2881.2 & 11479.3 \\ 
115 & 24 & 1012.6 & 39.6 & 520.7 & 588.4 & 2161.3 \\ 
116 & 25 & 1923.3 & 73.6 & 1298 & 1485 & 4779.8 \\ 
117 & 12 & 2826.3 & 87.2 & 1414.2 & 1608 & 5935.8 \\ 
118 & 12 & 1583.2 & 59.4 & 990.9 & 1129 & 3762.4 \\ 
119 & 50 & 2606.9 & 40.3 & 595.8 & 2441.6 & 5684.5 \\ 
120 & 25 & 3680.8 & 58.2 & 953 & 2724.3 & 7416.4 \\ 
121 & 216 & 9964.4 & 182.4 & 2435.8 & 9008.1 & 21590.7 \\ 
122 & 515 & 18306.3 & 359.9 & 4071.3 & 14727.5 & 37464.9 \\ 
123 & 727 & 9138.4 & 527 & 7910.2 & 6013.3 & 23589 \\ 
124 & 34 & 3013.3 & 187.5 & 3373.3 & 2539.7 & 9113.9 \\ 
125 & 73 & 4167.4 & 193.3 & 4022.9 & 2916.9 & 11300.5 \\ 
126 & 12 & 910.5 & 58.8 & 1009 & 726.2 & 2704.5 \\ 
127 & 574 & 37869.2 & 1968.8 & 32145.3 & 22982.4 & 94965.6 \\ 
128 & 94 & 17718.4 & 723.2 & 12642.4 & 9418.7 & 40502.7 \\ 
129 & 1549 & 19001.8 & 1879.5 & 18025.9 & 13538.8 & 52446 \\ 
130 & 59 & 8942.4 & 418.6 & 7886.6 & 5681.1 & 22928.6 \\ 
131 & 58 & 651.5 & 274.5 & 676.8 & 536.3 & 2139 \\ 
132 & 10 & 341.5 & 74.1 & 423 & 312.5 & 1151.1 \\ 
133 & 5 & 238.3 & 14.1 & 269.8 & 195.2 & 717.4 \\ 
134 & 9 & 304.6 & 24.9 & 354.4 & 262.4 & 946.3 \\ 
135 & 91 & 6534.5 & 310.8 & 4810.7 & 3525.2 & 15181.1 \\ 
136 & 790 & 29152.5 & 3660.6 & 25817.1 & 18624.4 & 77254.6 \\ 
137 & 160 & 7048.9 & 632.4 & 6872.3 & 5063.7 & 19617.3 \\ 
138 & 26 & 5684.8 & 210.1 & 3361.3 & 2476 & 11732.2 \\ 
139 & 3533 & 107816.6 & 1729.5 & 24559.3 & 86811.5 & 220916.8 \\ 
140 & 959 & 56603.1 & 982.3 & 16694.7 & 41044.6 & 115324.6 \\ 
141 & 733 & 44661.2 & 895.1 & 14205.8 & 29961.4 & 89723.4 \\ 
142 & 104 & 38012.6 & 655.3 & 11937.5 & 21227.4 & 71832.8 \\ 
143 & 19 & 6765.7 & 103.9 & 1502.8 & 3837.2 & 12209.5 \\ 
144 & 32 & 9948.2 & 190.1 & 2751.8 & 6470.5 & 19360.6 \\ 
145 & 5 & 1075.4 & 20.9 & 308.2 & 760.5 & 2165.1 \\ 
146 & 113 & 25918.6 & 127.2 & 1764.4 & 16159.4 & 43969.5 \\ 
147 & 84 & 13832.8 & 382.6 & 5773.4 & 7442.1 & 27430.9 \\ 
148 & 822 & 91747.5 & 841 & 12476.9 & 67723.6 & 172789 \\ 
149 & 15 & 809.2 & 76.8 & 403.9 & 546.5 & 1836.4 \\ 
150 & 91 & 10043.7 & 534.5 & 3985.8 & 5418.4 & 19982.4 \\ 
151 & 5 & 881 & 29.6 & 494.5 & 691.1 & 2096.1 \\ 
152 & 273 & 6662.8 & 1740.9 & 3458.7 & 4898.2 & 16760.6 \\ 
153 & 2 & 276.7 & 12.4 & 125.3 & 170.4 & 584.8 \\ 
154 & 140 & 1804.1 & 140.6 & 1111.2 & 1481 & 4537 \\ 
155 & 70 & 8848.2 & 73.4 & 926.8 & 5156.8 & 15005.1 \\ 
156 & 255 & 128759 & 3557 & 39738.8 & 63758.3 & 235813.1 \\ 
157 & 29 & 1159.6 & 34.8 & 509 & 864.6 & 2568 \\ 
158 & 3 & 688.1 & 14.8 & 306.5 & 395 & 1404.3 \\ 
159 & 59 & 6786 & 438.6 & 2846 & 4625.1 & 14695.6 \\ 
160 & 306 & 44692.4 & 444.9 & 6463 & 35660.1 & 87260.5 \\ 
161 & 252 & 29282.3 & 237.9 & 3457.2 & 20565.9 & 53543.3 \\ 
162 & 73 & 1615.2 & 596.8 & 1122.7 & 994.2 & 4329 \\ 
163 & 88 & 17151.5 & 1898.2 & 10981.6 & 9491.5 & 39522.8 \\ 
164 & 1025 & 20013.8 & 5203.1 & 13803.1 & 11970.2 & 50990.2 \\ 
165 & 68 & 6490.7 & 1406.1 & 3987 & 3404.9 & 15288.7 \\ 
166 & 1795 & 159909.7 & 1392.2 & 18670.5 & 88486.6 & 268459 \\ 
167 & 959 & 63955.4 & 925.7 & 13857.5 & 43213.4 & 121952 \\ 
169 & 1 & 90.5 & 3.5 & 65.1 & 86.5 & 245.7 \\ 
173 & 196 & 37263 & 1038.7 & 15754.4 & 20274.7 & 74330.8 \\ 
174 & 72 & 12286.9 & 390.6 & 4588.3 & 6398.7 & 23664.5 \\ 
175 & 2 & 66.1 & 6.3 & 62.9 & 44.6 & 179.9 \\ 
176 & 369 & 28937.9 & 1519.5 & 27429.4 & 20322.5 & 78209.3 \\ 
177 & 3 & 242.9 & 13.7 & 250.1 & 171.2 & 677.9 \\ 
180 & 95 & 740.6 & 72.6 & 1000.4 & 709.4 & 2522.9 \\ 
181 & 15 & 241 & 20.9 & 275.1 & 186.8 & 723.9 \\ 
182 & 51 & 1957.6 & 146.9 & 2619.7 & 1891.9 & 6616.2 \\ 
183 & 3 & 3.7 & 0.3 & 1.6 & 1.6 & 7.1 \\ 
185 & 97 & 16176.6 & 429.3 & 7369.6 & 7161.5 & 31137 \\ 
186 & 877 & 45454.6 & 1691.8 & 32979.6 & 32104.5 & 112230.4 \\ 
187 & 251 & 2258.5 & 146.1 & 1794.5 & 1664.5 & 5863.6 \\ 
188 & 28 & 2412.5 & 84 & 1407.6 & 1314.6 & 5218.7 \\ 
189 & 871 & 25895.7 & 1193.9 & 21404.6 & 19445.2 & 67939.5 \\ 
190 & 73 & 7812.5 & 442.4 & 5770.9 & 5294.1 & 19319.9 \\ 
191 & 1401 & 21880.5 & 992.8 & 15193.9 & 8156.1 & 46223.3 \\ 
192 & 15 & 1522.4 & 134.8 & 2550.4 & 1385.5 & 5593.1 \\ 
193 & 498 & 36772.9 & 1806.1 & 39473.2 & 21158.7 & 99210.9 \\ 
194 & 3433 & 127489.3 & 5781.7 & 101337.4 & 57052.7 & 291661.1 \\ 
195 & 1 & 90.5 & 3.7 & 62 & 63.7 & 219.9 \\ 
196 & 1 & 78.3 & 0.9 & 12.1 & 21.7 & 112.9 \\ 
197 & 36 & 3709.5 & 75.2 & 1465.2 & 2454.8 & 7704.8 \\ 
198 & 327 & 17350.1 & 889.9 & 13603.1 & 13919 & 45762.1 \\ 
199 & 41 & 4559.4 & 86.5 & 1664.4 & 2976.9 & 9287.2 \\ 
200 & 27 & 1317.3 & 117.4 & 1819.9 & 1197.2 & 4451.8 \\ 
201 & 8 & 1462.7 & 82.1 & 1427.1 & 970.6 & 3942.5 \\ 
202 & 26 & 823.7 & 22.5 & 397.6 & 451 & 1694.9 \\ 
203 & 8 & 828.5 & 28 & 591.4 & 737.5 & 2185.4 \\ 
204 & 331 & 9737.2 & 206.2 & 4302.3 & 5233.5 & 19479.2 \\ 
205 & 368 & 8392.1 & 884.3 & 12607 & 8132 & 30015.4 \\ 
206 & 213 & 72674.7 & 1026.3 & 22110.1 & 30663 & 126474.2 \\ 
208 & 2 & 11.4 & 1.5 & 7 & 4.2 & 24 \\ 
210 & 1 & 32.1 & 1.2 & 19.1 & 26.7 & 79.1 \\ 
211 & 1 & 39.3 & 2.9 & 49.7 & 28.6 & 120.5 \\ 
212 & 29 & 1946.3 & 134.8 & 2104 & 1203.1 & 5388.2 \\ 
213 & 41 & 1574.5 & 167.6 & 1492 & 875.9 & 4110 \\ 
214 & 19 & 2204.1 & 73.9 & 1867.9 & 1419.5 & 5565.6 \\ 
215 & 98 & 4124.9 & 248.1 & 4278.8 & 3403 & 12054.8 \\ 
216 & 1826 & 18159.4 & 415.4 & 5503 & 9330 & 33407.9 \\ 
217 & 185 & 16382.1 & 354.9 & 6838.9 & 7300 & 30875.9 \\ 
218 & 98 & 8681.2 & 489.5 & 8172.1 & 6399.2 & 23742.1 \\ 
219 & 14 & 1595.8 & 34.1 & 716.4 & 1094 & 3440.3 \\ 
220 & 299 & 22985.3 & 501.7 & 9567.4 & 22312.4 & 55366.7 \\ 
221 & 3559 & 13622.7 & 2031 & 13444.2 & 7459 & 36556.8 \\ 
223 & 578 & 36140.1 & 2172.2 & 38604.6 & 18647.8 & 95564.8 \\ 
224 & 48 & 587.7 & 59.8 & 618.7 & 381 & 1647.1 \\ 
225 & 5740 & 51102.3 & 1755 & 26600 & 27342.9 & 106800.2 \\ 
226 & 110 & 3272.1 & 145.4 & 2975.3 & 2116.1 & 8508.9 \\ 
227 & 3108 & 53642.1 & 1892.3 & 32005.3 & 37696.5 & 125236.2 \\ 
229 & 589 & 7518.1 & 257.6 & 5123.9 & 4291.5 & 17191.2 \\ 
230 & 6 & 609.8 & 26.5 & 704.3 & 499.2 & 1839.8 \\ 
\hline
Total & 73234 & 9500790.9 & 170632.8 & 2560878.8 & 5804953.8 & 18037256.3
\\
\hline
\end{longtable}
}

{\tiny
\begin{longtable}{|c|c|c|c|c|c|c|}
\caption[Total CPU hours expended for ab-initio calculations w/o SOC per SG]{Total CPU time expended for all of the ab-initio calculations performed in this work without incorporating the effects of SOC (w/o SOC), subdivided by SG.  In order, the columns in this table list the SG, the number of ICSDs successfully analyzed in the SG, the CPU hours expended per each of the four calculation steps (VASP1 through VASP4) detailed in \supappref{sec:SC}, and the total CPU hours expended for materials in the SG.
\label{tb:statistics_nosoc_compounds_cputime}}\\
\hline
SG & \# ICSDs & \vaspcputimelabel{1} & \vaspcputimelabel{2} & \vaspcputimelabel{3} & \vaspcputimelabel{4} & Total (CPU hours) \\
\hline 
1 & 222 & 16640.5 & 263.4 & 2250.8 & 10564.9 & 29719.6 \\ 
2 & 1755 & 235667.3 & 3350.5 & 33564.7 & 152953.3 & 425535.9 \\ 
3 & 16 & 930.1 & 7 & 81.8 & 538.9 & 1557.8 \\ 
4 & 291 & 17447.8 & 156 & 2082.8 & 11934.1 & 31620.7 \\ 
5 & 186 & 6913.4 & 57.2 & 653.8 & 4689 & 12313.4 \\ 
6 & 39 & 1515.9 & 10.7 & 124 & 809.5 & 2460.1 \\ 
7 & 143 & 8217.1 & 130.3 & 1050.1 & 5396.9 & 14794.4 \\ 
8 & 168 & 5057.3 & 41.9 & 424.9 & 3167.5 & 8691.6 \\ 
9 & 276 & 14522.1 & 112 & 1485 & 10613.8 & 26732.8 \\ 
10 & 39 & 1873.5 & 14.9 & 185.8 & 1110.9 & 3185.1 \\ 
11 & 749 & 53377.6 & 554.4 & 6215.3 & 33315.3 & 93462.6 \\ 
12 & 1722 & 87000.4 & 614.7 & 7800 & 50440 & 145855.1 \\ 
13 & 274 & 19194 & 170.8 & 2384.8 & 12399.9 & 34149.5 \\ 
14 & 3602 & 391541.5 & 3358.7 & 48706.8 & 272488.3 & 716095.2 \\ 
15 & 2419 & 177227 & 1461.2 & 18799.9 & 123027.7 & 320515.8 \\ 
16 & 2 & 16.1 & 2 & 14.2 & 12.1 & 44.5 \\ 
17 & 8 & 258.1 & 14.1 & 213.5 & 187.3 & 673 \\ 
18 & 42 & 2809.3 & 127.5 & 2381.5 & 2167.5 & 7485.8 \\ 
19 & 352 & 21449.7 & 940.1 & 17090.4 & 15336.1 & 54816.2 \\ 
20 & 77 & 2538.6 & 36.2 & 556.2 & 1810.3 & 4941.3 \\ 
21 & 20 & 306.5 & 5 & 65.1 & 165.5 & 542 \\ 
22 & 14 & 312.4 & 2.9 & 39.7 & 134.5 & 489.6 \\ 
23 & 15 & 510.5 & 5 & 58.8 & 204 & 778.3 \\ 
24 & 3 & 41.8 & 0.8 & 8.6 & 26.7 & 77.9 \\ 
25 & 58 & 872.9 & 42.9 & 476.7 & 406 & 1798.5 \\ 
26 & 84 & 4907.6 & 169.8 & 3080.8 & 2775.5 & 10933.7 \\ 
27 & 1 & 20.4 & 1 & 16.4 & 14.7 & 52.5 \\ 
28 & 18 & 403.1 & 20.4 & 281.8 & 255.4 & 960.7 \\ 
29 & 87 & 4800.4 & 237.4 & 4132.5 & 3720.1 & 12890.4 \\ 
30 & 6 & 693.3 & 24.8 & 465.5 & 421.4 & 1605.1 \\ 
31 & 245 & 8475 & 393.3 & 6672.2 & 5951 & 21491.5 \\ 
32 & 12 & 1042.7 & 58.7 & 844 & 763.6 & 2709 \\ 
33 & 399 & 22961 & 912.6 & 16815.8 & 15095.1 & 55784.5 \\ 
34 & 30 & 2285.8 & 93.4 & 1827.2 & 1636.7 & 5843 \\ 
35 & 8 & 253 & 16.4 & 67.5 & 166.3 & 503.1 \\ 
36 & 392 & 14360.2 & 181.5 & 2621.9 & 8920 & 26083.6 \\ 
37 & 7 & 328.9 & 14.2 & 99.8 & 250.9 & 693.7 \\ 
38 & 162 & 2810.1 & 193.8 & 876.2 & 1508.4 & 5388.5 \\ 
39 & 14 & 497.3 & 7.2 & 134.7 & 386.3 & 1025.5 \\ 
40 & 64 & 2449.7 & 32.1 & 682.7 & 2050.8 & 5215.3 \\ 
41 & 44 & 2926.4 & 30.2 & 600.1 & 1822.1 & 5378.8 \\ 
42 & 4 & 183.7 & 2.4 & 31.6 & 161.4 & 379.1 \\ 
43 & 125 & 4866 & 57.3 & 1071 & 3696.1 & 9690.4 \\ 
44 & 93 & 1896.8 & 21.7 & 270.3 & 878.4 & 3067.2 \\ 
45 & 11 & 1428.1 & 12.9 & 230.8 & 880.2 & 2552 \\ 
46 & 83 & 4135.5 & 40.4 & 620.8 & 2283 & 7079.7 \\ 
47 & 86 & 2117.4 & 71.1 & 869.9 & 787.8 & 3846.2 \\ 
48 & 3 & 86.4 & 5.6 & 64.6 & 55.3 & 211.9 \\ 
50 & 11 & 1020 & 38.6 & 773.3 & 700.8 & 2532.8 \\ 
51 & 138 & 4387.5 & 181.6 & 3072.2 & 2766.1 & 10407.3 \\ 
52 & 48 & 3899.3 & 186.1 & 2798.5 & 2550.8 & 9434.7 \\ 
53 & 27 & 798.4 & 37.4 & 594.6 & 557.5 & 1988 \\ 
54 & 23 & 1833.7 & 84.6 & 1722.7 & 1536.3 & 5177.3 \\ 
55 & 466 & 37375 & 1328.4 & 24747.9 & 22131.1 & 85582.4 \\ 
56 & 31 & 1967 & 89.6 & 1661.6 & 1522.9 & 5241 \\ 
57 & 185 & 21454.5 & 582.5 & 11391.2 & 10321 & 43749.3 \\ 
58 & 481 & 12890.5 & 576.2 & 9232.8 & 8317.6 & 31017.1 \\ 
59 & 317 & 14493.3 & 466 & 7550.1 & 6710.8 & 29220.3 \\ 
60 & 247 & 13852.8 & 686.1 & 11583.1 & 10368.2 & 36490.2 \\ 
61 & 227 & 17496.1 & 856.6 & 15266 & 13530 & 47148.7 \\ 
62 & 5327 & 302331.6 & 10405.5 & 187244.1 & 167174.9 & 667156.1 \\ 
63 & 2021 & 51822.1 & 601.2 & 7947.2 & 23375.5 & 83745.9 \\ 
64 & 421 & 19419.8 & 240.5 & 4187.2 & 13338.9 & 37186.4 \\ 
65 & 331 & 11135.8 & 102.4 & 1487.7 & 3952.3 & 16678.2 \\ 
66 & 49 & 2818.5 & 33.8 & 573.4 & 1907.5 & 5333.3 \\ 
67 & 48 & 900.2 & 9.6 & 108.1 & 372.4 & 1390.3 \\ 
68 & 21 & 1599.4 & 20.6 & 333.9 & 1121.1 & 3075 \\ 
69 & 73 & 2870.6 & 26.3 & 413.9 & 1294.6 & 4605.5 \\ 
70 & 197 & 7537.8 & 99.4 & 1998.5 & 5604.2 & 15239.8 \\ 
71 & 574 & 17088 & 168.7 & 2306.6 & 7159 & 26722.3 \\ 
72 & 227 & 7369.5 & 83.8 & 1192.4 & 3970.2 & 12615.9 \\ 
73 & 21 & 3113.5 & 36.8 & 535.6 & 2081.5 & 5767.5 \\ 
74 & 342 & 7896 & 104.2 & 1446.9 & 4571.6 & 14018.8 \\ 
75 & 5 & 192.3 & 7.1 & 112.5 & 119.3 & 431.2 \\ 
76 & 12 & 661.9 & 26.4 & 487.1 & 508.3 & 1683.7 \\ 
77 & 2 & 17.2 & 1.1 & 9.9 & 11 & 39.2 \\ 
78 & 2 & 289.9 & 8 & 148.1 & 155.6 & 601.5 \\ 
79 & 14 & 376.5 & 5.5 & 69.7 & 239 & 690.7 \\ 
80 & 3 & 362.8 & 3.8 & 58.2 & 234 & 658.8 \\ 
81 & 18 & 373.5 & 20.3 & 259.4 & 289.8 & 942.9 \\ 
82 & 265 & 3631.4 & 86.6 & 708.4 & 2588 & 7014.5 \\ 
83 & 10 & 716.2 & 20.7 & 345.2 & 370.8 & 1452.9 \\ 
84 & 35 & 1121.1 & 48.9 & 776.1 & 802.4 & 2748.6 \\ 
85 & 51 & 3973.6 & 130.9 & 2511 & 2619.6 & 9235.2 \\ 
86 & 68 & 2795.4 & 104.2 & 1759.5 & 1788.1 & 6447.3 \\ 
87 & 291 & 8338 & 89.3 & 1091.6 & 4117.1 & 13636 \\ 
88 & 314 & 9571.3 & 128.6 & 1716.2 & 6876.7 & 18292.9 \\ 
90 & 5 & 614.1 & 23 & 491.8 & 363.5 & 1492.5 \\ 
91 & 9 & 136.1 & 8.9 & 130.2 & 95.1 & 370.3 \\ 
92 & 133 & 3393.2 & 182.5 & 2995.1 & 2221.3 & 8792.1 \\ 
94 & 1 & 4.3 & 0.3 & 3.7 & 3 & 11.2 \\ 
95 & 3 & 44.4 & 3.2 & 43.1 & 31.6 & 122.3 \\ 
96 & 37 & 350.3 & 28.3 & 364.8 & 291.4 & 1034.8 \\ 
97 & 2 & 16.6 & 1.1 & 12.5 & 21.8 & 51.9 \\ 
98 & 7 & 144.1 & 2.9 & 37.6 & 111.9 & 296.5 \\ 
99 & 195 & 625.5 & 100.6 & 527.8 & 425.9 & 1679.8 \\ 
100 & 38 & 1376 & 74 & 1256.1 & 897.9 & 3604 \\ 
102 & 15 & 717.3 & 34 & 662.3 & 473.2 & 1886.8 \\ 
103 & 9 & 39.2 & 5.3 & 43.6 & 32.8 & 120.9 \\ 
104 & 4 & 100.9 & 8.4 & 127.8 & 96 & 333.1 \\ 
105 & 4 & 73.7 & 4.6 & 75.3 & 52.6 & 206.2 \\ 
106 & 1 & 24 & 1.7 & 25.6 & 18.5 & 69.8 \\ 
107 & 145 & 789.3 & 22.7 & 179.5 & 450.5 & 1442 \\ 
108 & 14 & 488.2 & 8.9 & 125.7 & 369.3 & 992.2 \\ 
109 & 50 & 436.9 & 12.2 & 103.3 & 241.3 & 793.8 \\ 
110 & 36 & 2058.5 & 29.2 & 446.8 & 1440.3 & 3974.8 \\ 
111 & 26 & 203.8 & 18.3 & 179.9 & 132.7 & 534.8 \\ 
112 & 4 & 22.6 & 2.5 & 20.4 & 15.6 & 61 \\ 
113 & 150 & 2409.9 & 164.4 & 2336.3 & 1741.1 & 6651.8 \\ 
114 & 42 & 1054.6 & 67.5 & 1108.7 & 791.4 & 3022.2 \\ 
115 & 24 & 298 & 19.7 & 289.5 & 198.4 & 805.7 \\ 
116 & 25 & 616.6 & 40.7 & 650.1 & 473.7 & 1781 \\ 
117 & 12 & 917.1 & 35.6 & 684.9 & 552.7 & 2190.2 \\ 
118 & 12 & 449.8 & 28.7 & 480.6 & 402.5 & 1361.6 \\ 
119 & 50 & 631.4 & 15.4 & 154.9 & 413.9 & 1215.5 \\ 
120 & 25 & 884.6 & 16.5 & 220.3 & 561.1 & 1682.5 \\ 
121 & 205 & 2258.1 & 62.4 & 631.7 & 1603 & 4555.2 \\ 
122 & 510 & 3995.8 & 123.2 & 1103.7 & 2803.2 & 8026 \\ 
123 & 699 & 4342.5 & 352 & 2844.7 & 2142 & 9681.2 \\ 
124 & 33 & 1119.5 & 62.1 & 1194.5 & 922 & 3298.1 \\ 
125 & 67 & 1841.9 & 79.5 & 1149.6 & 841.2 & 3912.1 \\ 
126 & 11 & 372.2 & 23.8 & 451.5 & 321.3 & 1168.7 \\ 
127 & 523 & 15351.3 & 591.5 & 9138.4 & 6490.3 & 31571.5 \\ 
128 & 89 & 8301.5 & 285.2 & 5630.7 & 4071.6 & 18289 \\ 
129 & 1506 & 10738.6 & 814.4 & 5929.9 & 4520.7 & 22003.5 \\ 
130 & 58 & 4060.1 & 179.9 & 3491.2 & 2537.5 & 10268.7 \\ 
131 & 57 & 240.7 & 27.9 & 179.6 & 138.2 & 586.4 \\ 
132 & 10 & 94.2 & 7.4 & 87.6 & 65.6 & 254.8 \\ 
133 & 5 & 100.6 & 5.9 & 104.2 & 74.5 & 285.2 \\ 
134 & 9 & 67.8 & 5.7 & 62.8 & 46.3 & 182.7 \\ 
135 & 84 & 1695.1 & 107 & 1590.2 & 1141.8 & 4534.1 \\ 
136 & 766 & 9562.1 & 608.4 & 8214.9 & 5921.6 & 24307 \\ 
137 & 158 & 3107.1 & 200.6 & 2993.2 & 2197.4 & 8498.4 \\ 
138 & 20 & 1175.6 & 44.3 & 776.5 & 560.3 & 2556.7 \\ 
139 & 3345 & 30000.9 & 577 & 5821.5 & 14249.1 & 50648.5 \\ 
140 & 916 & 16784.1 & 307.6 & 3577.2 & 8459.7 & 29128.5 \\ 
141 & 658 & 12790 & 221 & 2852.3 & 6060.4 & 21923.7 \\ 
142 & 98 & 8273.9 & 136.8 & 2139.7 & 5064 & 15614.4 \\ 
143 & 18 & 1468.6 & 54.8 & 792.1 & 1086 & 3401.5 \\ 
144 & 31 & 3401.4 & 88.3 & 1256.7 & 1685.7 & 6432 \\ 
145 & 5 & 254.5 & 10.9 & 143.7 & 192.5 & 601.7 \\ 
146 & 108 & 2882.8 & 29 & 293.2 & 1969.9 & 5174.9 \\ 
147 & 80 & 3872.6 & 108.9 & 1913.7 & 2397.8 & 8293 \\ 
148 & 794 & 26005.2 & 236.8 & 2812.5 & 16800 & 45854.5 \\ 
149 & 15 & 142.5 & 10.1 & 100.6 & 89 & 342.3 \\ 
150 & 88 & 2728.6 & 122.8 & 1604.5 & 1334.5 & 5790.4 \\ 
151 & 4 & 199.4 & 10.1 & 186.2 & 154.2 & 549.9 \\ 
152 & 269 & 1454.2 & 172 & 1229.5 & 1091 & 3946.7 \\ 
153 & 1 & 35.4 & 1.9 & 33.2 & 28.1 & 98.5 \\ 
154 & 138 & 617.1 & 83.1 & 535.7 & 460.5 & 1696.4 \\ 
155 & 66 & 1133 & 23.4 & 200.9 & 729.4 & 2086.7 \\ 
156 & 224 & 24561.7 & 714.8 & 13269.3 & 11261.8 & 49807.6 \\ 
157 & 27 & 320.9 & 27.2 & 263.3 & 226.9 & 838.4 \\ 
158 & 3 & 158.5 & 7.2 & 128.9 & 105.8 & 400.5 \\ 
159 & 57 & 1159.9 & 82 & 985.1 & 841 & 3068.1 \\ 
160 & 298 & 5234.3 & 116.7 & 1256.6 & 3419 & 10026.6 \\ 
161 & 251 & 5052.8 & 71.6 & 814.7 & 3616.4 & 9555.5 \\ 
162 & 70 & 461.9 & 43.7 & 358.9 & 317.6 & 1182.1 \\ 
163 & 80 & 4697 & 193.1 & 3418.7 & 2948.8 & 11257.6 \\ 
164 & 996 & 6051.4 & 596.2 & 4239 & 3709 & 14595.6 \\ 
165 & 66 & 2066.5 & 114.8 & 1803 & 1511.3 & 5495.5 \\ 
166 & 1679 & 43048.6 & 429.4 & 3973.1 & 15148.7 & 62599.9 \\ 
167 & 913 & 34970.4 & 517.1 & 4992.7 & 25149.1 & 65629.3 \\ 
169 & 1 & 26 & 2.7 & 34.2 & 24.8 & 87.7 \\ 
173 & 176 & 6442.7 & 300.5 & 5379.9 & 3825.6 & 15948.7 \\ 
174 & 62 & 1834.8 & 110.3 & 1473.5 & 1136.8 & 4555.4 \\ 
175 & 2 & 18.4 & 1.6 & 19.2 & 13.7 & 52.9 \\ 
176 & 339 & 10426.4 & 599 & 10057.4 & 7233.8 & 28316.6 \\ 
177 & 3 & 83.6 & 6.6 & 98.5 & 54.2 & 242.9 \\ 
180 & 95 & 302.6 & 63.2 & 380.9 & 220.5 & 967.3 \\ 
181 & 15 & 65.6 & 7.7 & 85.6 & 48 & 206.9 \\ 
182 & 49 & 578.1 & 61.7 & 816.5 & 442.3 & 1898.7 \\ 
183 & 3 & 3.2 & 1.5 & 3.8 & 2.4 & 10.9 \\ 
185 & 84 & 2742.7 & 174.9 & 3049.8 & 1572.5 & 7539.9 \\ 
186 & 842 & 8956.8 & 849.5 & 10271.5 & 5688.3 & 25766.1 \\ 
187 & 246 & 812.3 & 170.3 & 771.2 & 489.3 & 2243.1 \\ 
188 & 27 & 367 & 34.1 & 458.9 & 250.3 & 1110.3 \\ 
189 & 768 & 7479.3 & 674.7 & 6256.1 & 3510.8 & 17920.8 \\ 
190 & 67 & 1486.7 & 148.8 & 1859.8 & 1013.6 & 4508.9 \\ 
191 & 1269 & 10049.9 & 680.1 & 4765 & 2736.8 & 18231.8 \\ 
192 & 14 & 627.6 & 49.3 & 890.2 & 482.5 & 2049.6 \\ 
193 & 439 & 10935.5 & 515.5 & 7762.6 & 4193.8 & 23407.3 \\ 
194 & 3269 & 42642.6 & 2690.8 & 30749.5 & 16916.5 & 92999.5 \\ 
195 & 1 & 26.5 & 1.6 & 26 & 16.7 & 70.7 \\ 
196 & 1 & 12.7 & 0.4 & 3.3 & 4.5 & 21 \\ 
197 & 36 & 521.6 & 24.7 & 355 & 351.2 & 1252.6 \\ 
198 & 323 & 4050 & 333.9 & 4925.2 & 3259.5 & 12568.7 \\ 
199 & 40 & 823.9 & 22.4 & 341.6 & 496.3 & 1684.2 \\ 
200 & 27 & 526.7 & 47.7 & 693.7 & 452.5 & 1720.5 \\ 
201 & 7 & 496.5 & 26.8 & 473.5 & 306.3 & 1303.2 \\ 
202 & 26 & 236.9 & 7.3 & 100.6 & 151 & 495.8 \\ 
203 & 8 & 285.6 & 7 & 133.6 & 243.8 & 670.1 \\ 
204 & 323 & 2843.6 & 73.5 & 1114.7 & 1249.5 & 5281.4 \\ 
205 & 364 & 3012.8 & 299.7 & 3650 & 2415.6 & 9378.1 \\ 
206 & 208 & 14306.2 & 184.2 & 3907.3 & 6257.4 & 24655 \\ 
208 & 2 & 2.9 & 0.9 & 2.8 & 1.7 & 8.4 \\ 
210 & 1 & 10 & 0.5 & 4.6 & 8.1 & 23.1 \\ 
211 & 1 & 15.9 & 1.4 & 22.2 & 10.5 & 50.1 \\ 
212 & 29 & 609.5 & 47.2 & 738.8 & 363.7 & 1759.3 \\ 
213 & 41 & 328.5 & 36.8 & 497 & 244.6 & 1106.9 \\ 
214 & 18 & 367.9 & 16.8 & 373 & 268.2 & 1025.9 \\ 
215 & 95 & 1020.6 & 118 & 1625.3 & 810.9 & 3574.8 \\ 
216 & 1798 & 4916.7 & 210.6 & 1411.7 & 3321.6 & 9860.6 \\ 
217 & 178 & 2790.9 & 88 & 1441.5 & 1304 & 5624.4 \\ 
218 & 98 & 1925.3 & 194.4 & 3060.3 & 1496 & 6676.1 \\ 
219 & 13 & 261.8 & 8.1 & 146.6 & 213.9 & 630.4 \\ 
220 & 295 & 5456.1 & 151 & 2266.3 & 2167.5 & 10040.9 \\ 
221 & 3493 & 10189.3 & 3401 & 9912.1 & 6474.7 & 29977.1 \\ 
223 & 568 & 17296.6 & 978.4 & 14360.2 & 7033.1 & 39668.4 \\ 
224 & 45 & 271.5 & 45.9 & 265.3 & 142.2 & 724.8 \\ 
225 & 5619 & 21321.7 & 842.1 & 6760.4 & 12191.7 & 41116 \\ 
226 & 102 & 984.3 & 41.3 & 635.8 & 757.3 & 2418.7 \\ 
227 & 3012 & 17366.8 & 594.4 & 7540.4 & 8656.2 & 34157.8 \\ 
229 & 583 & 3066.4 & 107 & 1221.7 & 1359.9 & 5755 \\ 
230 & 6 & 219.5 & 8 & 159.2 & 199.3 & 586 \\ 
\hline
Total & 69730 & 2298583.3 & 56520.7 & 773019.5 & 1397550.3 & 4525673.8
\\
\hline
\end{longtable}
}

{\tiny
\begin{longtable}{|c|c|c|c|c|c|c|}
\caption[Total CPU hours expended for ab-initio calculations with SOC per number of atoms in the primitive cell]{Total CPU time expended for all of the ab-initio calculations performed in this work incorporating the effects of SOC (with SOC), subdivided by the number of atoms in the primitive cell in each material.  In order, the columns in this table list the number of atoms, the number of ICSDs successfully analyzed with the same number of atoms, the CPU hours expended per each of the four calculation steps (VASP1 through VASP4) detailed in \supappref{sec:SC}, and the total CPU hours expended for materials with the listed number of atoms in the primitive cell.
\label{tb:statistics_cputime_nbratoms_primitivecell}}\\
\hline
\# atoms & \# ICSDs & \vaspcputimelabel{1} & \vaspcputimelabel{2} & \vaspcputimelabel{3} & \vaspcputimelabel{4} & Total (CPU hours) \\
\hline 
1 & 1178 & 1692.2 & 133 & 455.7 & 1315.1 & 3596 \\ 
2 & 5838 & 8707.1 & 1229 & 4630.6 & 5155.7 & 19722.4 \\ 
3 & 2429 & 8457.2 & 586.1 & 2543.2 & 5180.9 & 16767.4 \\ 
4 & 4851 & 24909 & 1465 & 8512.3 & 17989.5 & 52875.8 \\ 
5 & 3912 & 38653.7 & 1437.9 & 9362.6 & 26276.9 & 75731.1 \\ 
6 & 6280 & 62655.4 & 3812.2 & 25308 & 46531.7 & 138307.2 \\ 
7 & 1169 & 20558.6 & 378.2 & 3953.5 & 21107.7 & 45997.9 \\ 
8 & 4239 & 69498.4 & 1969.7 & 26406.4 & 66879.8 & 164754.4 \\ 
9 & 1745 & 38455.6 & 1040.8 & 16788.1 & 36746.4 & 93030.9 \\ 
10 & 3498 & 101882.1 & 3068.6 & 38573.3 & 108096.9 & 251620.9 \\ 
11 & 488 & 37159.7 & 846.8 & 8064.9 & 39050.5 & 85121.8 \\ 
12 & 5023 & 271212.6 & 8527.4 & 131051 & 280299.7 & 691090.8 \\ 
13 & 748 & 59115.8 & 1356.4 & 12500.1 & 43733.6 & 116705.8 \\ 
14 & 2462 & 178253.4 & 4446.4 & 48149 & 209175.9 & 440024.7 \\ 
15 & 429 & 44783.7 & 1372.9 & 9509.1 & 57785.1 & 113450.9 \\ 
16 & 2980 & 291702.3 & 6978.2 & 101610.2 & 187394.1 & 587684.7 \\ 
17 & 336 & 32322.2 & 607.5 & 8352.8 & 25039.6 & 66322.1 \\ 
18 & 1607 & 187947 & 5758 & 57464.4 & 127823.4 & 378992.8 \\ 
19 & 271 & 59358.6 & 475.7 & 5950.7 & 24315.9 & 90100.9 \\ 
20 & 3618 & 480741.1 & 10167.6 & 149467.9 & 299831.8 & 940208.3 \\ 
21 & 199 & 30273.9 & 1085.7 & 6403.3 & 15711.2 & 53474.2 \\ 
22 & 1613 & 254127.1 & 3484 & 47223.3 & 131763.9 & 436598.3 \\ 
23 & 148 & 20209.7 & 636 & 4955.5 & 12606.8 & 38408 \\ 
24 & 2635 & 452531.4 & 11618.9 & 135134.3 & 232471.7 & 831756.3 \\ 
25 & 65 & 31170.7 & 200.7 & 2461.6 & 13554.2 & 47387.2 \\ 
26 & 686 & 151359.6 & 1566 & 26140.2 & 79825.2 & 258891 \\ 
27 & 154 & 26557.3 & 325.8 & 5019.3 & 15649.1 & 47551.5 \\ 
28 & 2663 & 520778.7 & 9095 & 168200.4 & 279719.3 & 977793.3 \\ 
29 & 441 & 64277.5 & 969.1 & 18818.8 & 28707.4 & 112772.8 \\ 
30 & 855 & 222765 & 3328.8 & 51435.4 & 120702.2 & 398231.4 \\ 
31 & 37 & 11691.3 & 96.4 & 1306.8 & 7565.8 & 20660.3 \\ 
32 & 1446 & 529701.1 & 5729.5 & 104171.6 & 335148.4 & 974750.5 \\ 
33 & 76 & 24907.1 & 274.1 & 5145.1 & 12343.3 & 42669.6 \\ 
34 & 341 & 171411.2 & 1259 & 20640.5 & 102531.1 & 295841.7 \\ 
35 & 44 & 34184 & 188.7 & 3082.9 & 21339.5 & 58795.1 \\ 
36 & 1644 & 674688.5 & 12436.8 & 220457.8 & 371435.6 & 1279018.7 \\ 
37 & 34 & 23767.8 & 224 & 2484.9 & 8439.8 & 34916.5 \\ 
38 & 494 & 227118.6 & 3134.9 & 53355.6 & 105738.9 & 389348.1 \\ 
39 & 67 & 41008.1 & 654.9 & 11049.4 & 16198.3 & 68910.7 \\ 
40 & 1533 & 590881.7 & 11528.7 & 209582.4 & 338425.5 & 1150418.3 \\ 
41 & 46 & 32255.1 & 242.9 & 4046 & 17630.1 & 54174.1 \\ 
42 & 495 & 217781.5 & 2799.9 & 47968.9 & 131825.1 & 400375.5 \\ 
43 & 16 & 6440.2 & 87.5 & 1193.6 & 3602.1 & 11323.3 \\ 
44 & 791 & 535425.1 & 6427.7 & 111650.2 & 327926 & 981429 \\ 
45 & 45 & 23017.7 & 390.7 & 6707.1 & 14180.8 & 44296.5 \\ 
46 & 390 & 211540.2 & 2535 & 43916.8 & 112491.1 & 370483 \\ 
47 & 10 & 15904.4 & 84.5 & 1334.3 & 5832.3 & 23155.4 \\ 
48 & 846 & 523845.7 & 8847.3 & 151462.9 & 312179.8 & 996335.7 \\ 
49 & 33 & 29334.5 & 229.2 & 3864.4 & 10371.8 & 43799.9 \\ 
50 & 183 & 121677.3 & 1375.6 & 22310.1 & 68542.9 & 213905.9 \\ 
51 & 12 & 9202.6 & 147.6 & 2251.9 & 5709.7 & 17311.7 \\ 
52 & 641 & 574723.4 & 6738.7 & 111780.4 & 280305 & 973547.6 \\ 
53 & 10 & 10657 & 54.7 & 688.1 & 2446.9 & 13846.7 \\ 
54 & 257 & 146755.9 & 2628.7 & 44400.6 & 90529.7 & 284314.9 \\ 
55 & 13 & 10828.9 & 128.6 & 2068.7 & 4207 & 17233.2 \\ 
56 & 552 & 398960.4 & 6814.9 & 116851.8 & 239838.5 & 762465.7 \\ 
57 & 11 & 8036.9 & 158.9 & 2340.2 & 3750.3 & 14286.3 \\ 
58 & 161 & 120980.8 & 2010.7 & 29671 & 57301 & 209963.4 \\ 
59 & 4 & 6722.4 & 45.1 & 706.8 & 4153.5 & 11627.8 \\ 
60 & 417 & 351268.4 & 5045.2 & 84244.6 & 208210.1 & 648768.3 \\ 
62 & 2 & 894.3 & 21.5 & 355.4 & 1717.5 & 2988.7 \\ 
64 & 2 & 2286.1 & 35.9 & 573.9 & 1571.7 & 4467.6 \\ 
68 & 2 & 1585.7 & 22.4 & 382.6 & 969 & 2959.6 \\ 
70 & 3 & 2061.2 & 28.3 & 477.9 & 1953.5 & 4520.9 \\ 
72 & 7 & 10535.4 & 119.5 & 2050 & 10071.1 & 22776 \\ 
74 & 2 & 1022.6 & 21.8 & 354.5 & 1834.5 & 3233.4 \\ 
76 & 6 & 4563.8 & 81.7 & 1275.2 & 4981.3 & 10902 \\ 
78 & 1 & 1007.7 & 14.1 & 227.9 & 1214.2 & 2463.9 \\ 
\hline
Total & 73234 & 9500790.9 & 170632.8 & 2560878.8 & 5804953.8 & 18037256.3
\\
\hline
\end{longtable}
}

{\tiny
\begin{longtable}{|c|c|c|c|c|c|c|}
\caption[Total CPU hours expended for ab-initio calculations w/o SOC per number of atoms in the primitive cell]{Total CPU time expended for all of the ab-initio calculations performed in this work without incorporating the effects of SOC (w/o SOC), subdivided by the number of atoms in the primitive cell in each material.  In order, the columns in this table list the number of atoms, the number of ICSDs successfully analyzed with the same number of atoms, the CPU hours expended per each of the four calculation steps (VASP1 through VASP4) detailed in \supappref{sec:SC}, and the total CPU hours expended for materials with the listed number of atoms in the primitive cell.
\label{tb:statistics_cputime_nbratoms_primitivecell_nosoc}}\\
\hline
\# atoms & \# ICSDs & \vaspcputimelabel{1} & \vaspcputimelabel{2} & \vaspcputimelabel{3} & \vaspcputimelabel{4} & Total (CPU hours) \\
\hline 
1 & 1174 & 1428.4 & 150.8 & 237.9 & 1464.8 & 3281.8 \\ 
2 & 5757 & 8494.6 & 1997.5 & 4335.5 & 6970.2 & 21797.9 \\ 
3 & 2375 & 4654.8 & 538.7 & 1930.2 & 3119.6 & 10243.2 \\ 
4 & 4724 & 13122.2 & 1848.2 & 6104.9 & 7951.1 & 29026.4 \\ 
5 & 3768 & 17086.8 & 1570.1 & 5693.8 & 8010.7 & 32361.4 \\ 
6 & 6005 & 27415.5 & 1914.6 & 11614.4 & 13057.6 & 54002.1 \\ 
7 & 1121 & 5835.6 & 302.4 & 1686 & 3342.6 & 11166.6 \\ 
8 & 4087 & 26286.6 & 1597.5 & 11208 & 14378.5 & 53470.6 \\ 
9 & 1634 & 11539.6 & 781.9 & 5615 & 5671 & 23607.5 \\ 
10 & 3366 & 29405.9 & 1002.8 & 8577.2 & 16863.4 & 55849.3 \\ 
11 & 442 & 9348.1 & 148.4 & 1500.1 & 4025.6 & 15022.2 \\ 
12 & 4721 & 74161.1 & 2387 & 25867.7 & 37434.5 & 139850.3 \\ 
13 & 687 & 13310.1 & 255.6 & 2598 & 5464.7 & 21628.4 \\ 
14 & 2346 & 36977.7 & 811.9 & 8930.2 & 19695.8 & 66415.6 \\ 
15 & 404 & 10900.2 & 161.3 & 1705.1 & 5488.7 & 18255.3 \\ 
16 & 2800 & 66554.1 & 1869.7 & 23967.2 & 35630.9 & 128021.8 \\ 
17 & 325 & 8368.3 & 160.3 & 1947.3 & 4687.1 & 15163 \\ 
18 & 1527 & 43958.1 & 1050.3 & 12785.7 & 21962.3 & 79756.5 \\ 
19 & 235 & 13056.8 & 108.7 & 1274.3 & 4504.2 & 18943.9 \\ 
20 & 3436 & 105041.5 & 2622.5 & 37317.1 & 60273.3 & 205254.4 \\ 
21 & 180 & 6459.4 & 161.4 & 2000.5 & 3555 & 12176.3 \\ 
22 & 1507 & 54674.8 & 925 & 11862.2 & 31129.2 & 98591.2 \\ 
23 & 143 & 4653.5 & 105 & 1261.1 & 2887.3 & 8906.8 \\ 
24 & 2445 & 107421.5 & 2878.4 & 44166.6 & 65128.8 & 219595.2 \\ 
25 & 56 & 6339.8 & 50.3 & 595.5 & 2730.8 & 9716.4 \\ 
26 & 636 & 27235.2 & 577.1 & 8100.2 & 15875.6 & 51788.1 \\ 
27 & 144 & 8156.7 & 144 & 1874.2 & 4699.9 & 14874.7 \\ 
28 & 2484 & 130370.6 & 3425.9 & 56219.6 & 75007.8 & 265023.9 \\ 
29 & 421 & 18100.3 & 276.5 & 4305.2 & 7713 & 30395 \\ 
30 & 771 & 44759.9 & 1138.5 & 17363.1 & 28079.2 & 91340.7 \\ 
31 & 36 & 2787.1 & 41.6 & 430.6 & 1939.6 & 5198.9 \\ 
32 & 1336 & 101776.4 & 2053.3 & 34289.6 & 63445.4 & 201564.7 \\ 
33 & 71 & 7223.7 & 112.9 & 1889.5 & 3001.4 & 12227.6 \\ 
34 & 315 & 28517.9 & 359 & 5176.1 & 18254 & 52307 \\ 
35 & 42 & 6647.6 & 57.5 & 746.3 & 3465 & 10916.3 \\ 
36 & 1535 & 180424.4 & 4037.4 & 73477.6 & 108486.3 & 366425.8 \\ 
37 & 33 & 5944.5 & 66.1 & 823.5 & 3392.1 & 10226.2 \\ 
38 & 438 & 58361.8 & 976.5 & 16280.2 & 30233.3 & 105851.8 \\ 
39 & 63 & 10970.5 & 200.3 & 3261.2 & 5283.2 & 19715.2 \\ 
40 & 1484 & 164516.7 & 3821.1 & 67853.9 & 103775.8 & 339967.5 \\ 
41 & 39 & 4761 & 59.6 & 888.3 & 3289 & 8997.9 \\ 
42 & 441 & 50165.2 & 858.1 & 13486.9 & 33587.9 & 98098.2 \\ 
43 & 14 & 2472.3 & 34.2 & 393.1 & 1583 & 4482.6 \\ 
44 & 708 & 115514.9 & 1865.9 & 32835.9 & 80554.8 & 230771.5 \\ 
45 & 38 & 7729.9 & 160.7 & 2177.1 & 4031.6 & 14099.4 \\ 
46 & 382 & 48077.9 & 773.4 & 12418.4 & 30508.1 & 91777.8 \\ 
47 & 10 & 2162 & 26.9 & 385.5 & 1425.3 & 3999.6 \\ 
48 & 815 & 132027.4 & 2797.2 & 51535.4 & 92872.3 & 279232.3 \\ 
49 & 33 & 7950.4 & 65.4 & 919.6 & 4103.8 & 13039.2 \\ 
50 & 176 & 29245.7 & 426.7 & 5948.3 & 19606.6 & 55227.2 \\ 
51 & 10 & 1858.8 & 24.8 & 458 & 1336.6 & 3678.3 \\ 
52 & 596 & 112774.2 & 1755.4 & 31895.2 & 73103.5 & 219528.3 \\ 
53 & 10 & 2535.4 & 20 & 249.8 & 1016.3 & 3821.5 \\ 
54 & 252 & 42895.2 & 844.6 & 14972.5 & 28809.5 & 87521.8 \\ 
55 & 12 & 3719.9 & 39.5 & 617.8 & 1600.3 & 5977.6 \\ 
56 & 529 & 103002.2 & 2019.1 & 38861.8 & 73748.8 & 217631.9 \\ 
57 & 10 & 1918.4 & 48.2 & 717.9 & 1319.2 & 4003.8 \\ 
58 & 151 & 28945.3 & 493.5 & 9623.3 & 18075 & 57137 \\ 
59 & 4 & 1545.6 & 13.2 & 182.4 & 1055.2 & 2796.4 \\ 
60 & 401 & 86389.7 & 1441.7 & 26559.6 & 61958.9 & 176349.8 \\ 
62 & 2 & 562.1 & 3.8 & 60.9 & 360.9 & 987.8 \\ 
64 & 2 & 1124.3 & 7.9 & 128.6 & 701.5 & 1962.3 \\ 
68 & 2 & 433.1 & 3.5 & 49.8 & 287.9 & 774.2 \\ 
70 & 3 & 928.2 & 5.9 & 95.4 & 583.8 & 1613.3 \\ 
72 & 7 & 4042.6 & 23.3 & 370.4 & 2161.8 & 6598.1 \\ 
74 & 2 & 471.9 & 3.6 & 57.1 & 333.6 & 866.2 \\ 
76 & 6 & 2508 & 14.5 & 222.7 & 1237.4 & 3982.6 \\ 
78 & 1 & 533.6 & 2.2 & 35.9 & 218.1 & 789.7 \\ 
\hline
Total & 73234 & 2298583.3 & 56520.7 & 773019.5 & 1397550.3 & 4525673.8
\\
\hline
\end{longtable}
}

\subsection{Disk Storage Requirements and Statistics}
\label{sub:Storage}

In this section, we provide tables detailing the amount of disk storage requested for this work.  Overall, we have used~\TQCDBTotalStorage~of storage for this work, which can be subdivided into \TQCDBTotalSOCStorage~(\TQCDBPercentSOCTotalStorage) for the SOC DFT calculations and \TQCDBNoSOCTotalStorage~(\TQCDBPercentNoSOCTotalStorage) for the DFT calculations performed w/o SOC.  To conserve disk space, we have only stored the OUTCAR files outputted by VASP, the CHGCAR files generated during the self-consistent DFT calculations, and the PROCAR files generated during the density of states (DOS) evaluation. All of these files are stored in a compressed \textsc{bzip2} format. The majority of the storage is dedicated to the SOC computation output -- specifically, the CHGCAR files use \TQCDBCHGCARSOCStorage~(\TQCDBPercentCHGCARSOCStorage), and the PROCAR files use \TQCDBPROCARSOCStorage~(\TQCDBPercentPROCARSOCStorage). For a more detailed overview, we list in Table~\ref{tb:statistics_compounds_storage} the disk storage expended to perform ab-initio calculations subdivided by SG, and in Table~\ref{tb:statistics_compounds_storage_nbratoms_primitivecell} we list the disk storage expended to perform ab-initio calculations subdivided by the number of atoms in the primitive cell of each material.

{\tiny
\begin{longtable}{|c|c|c|c|c|c|}
\caption[Total disk storage space for ab-initio calculations per SG]{Total disk storage space (in Gb) expended for all of the ab-initio calculations performed in this work, subdivided by SG.  In order, the columns in this table list the SG, the number of ICSDs successfully analyzed in the SG, the total storage space (in Gb) for the calculations performed with and w/o SOC, the storage for the compressed CHGCAR files generated during the SOC self-consistent DFT calculations, the storage for the compressed PROCAR files generated during the SOC DFT DOS calculations, and the total storage space for the calculations performed w/o SOC.
\label{tb:statistics_compounds_storage}}\\
\hline
SG & \# ICSDs & \begin{tabular}{c}Total Storage\\SOC+w/o SOC (Gb)\end{tabular} & \begin{tabular}{c}CHGCAR SCC Storage\\(Gb)\end{tabular} & \begin{tabular}{c}PROCAR DOS Storage\\(Gb)\end{tabular} & \begin{tabular}{c}Total w/o SOC Storage\\(Gb)\end{tabular}  \\
\hline 
1 & 225 & 14 & 1.54 & 8.25 & 2.05 \\ 
2 & 1885 & 119.7 & 19.8 & 67.9 & 27.05 \\ 
3 & 19 & 1.32 & 0.15 & 0.95 & 0.15 \\ 
4 & 305 & 18.92 & 3.13 & 11.85 & 2.9 \\ 
5 & 195 & 9.83 & 1.83 & 6.2 & 1.3 \\ 
6 & 39 & 1.8 & 0.17 & 1.32 & 0.21 \\ 
7 & 143 & 8.53 & 1.46 & 5.29 & 1.36 \\ 
8 & 173 & 5.86 & 1.15 & 3.61 & 0.79 \\ 
9 & 278 & 17.1 & 3.37 & 10.42 & 2.62 \\ 
10 & 44 & 1.76 & 0.22 & 1.12 & 0.25 \\ 
11 & 790 & 39.49 & 6.18 & 23.58 & 6.17 \\ 
12 & 1844 & 70.2 & 17.18 & 37 & 11.06 \\ 
13 & 314 & 16.93 & 3.31 & 9.41 & 2.84 \\ 
14 & 3787 & 244.55 & 55.26 & 119.39 & 51.72 \\ 
15 & 2526 & 135.25 & 33.73 & 63.14 & 25.98 \\ 
16 & 2 & 0.03 & 0.01 & 0.02 & ---  \\ 
17 & 8 & 0.43 & 0.06 & 0.29 & 0.05 \\ 
18 & 42 & 2.15 & 0.43 & 1.21 & 0.39 \\ 
19 & 364 & 20.31 & 4.55 & 10.45 & 3.36 \\ 
20 & 85 & 3.75 & 0.96 & 1.96 & 0.53 \\ 
21 & 21 & 0.51 & 0.11 & 0.29 & 0.07 \\ 
22 & 17 & 0.6 & 0.14 & 0.35 & 0.06 \\ 
23 & 16 & 0.73 & 0.22 & 0.38 & 0.09 \\ 
24 & 3 & 0.11 & 0.04 & 0.05 & 0.02 \\ 
25 & 59 & 1.04 & 0.12 & 0.7 & 0.12 \\ 
26 & 85 & 3.5 & 0.71 & 1.99 & 0.57 \\ 
27 & 1 & 0.05 & 0.01 & 0.03 & 0.01 \\ 
28 & 18 & 0.65 & 0.07 & 0.49 & 0.06 \\ 
29 & 89 & 4.38 & 0.96 & 2.44 & 0.78 \\ 
30 & 6 & 0.36 & 0.08 & 0.17 & 0.09 \\ 
31 & 247 & 11.18 & 1.73 & 7.45 & 1.46 \\ 
32 & 12 & 0.44 & 0.12 & 0.14 & 0.14 \\ 
33 & 406 & 19.5 & 3.84 & 11.47 & 3.27 \\ 
34 & 30 & 1.22 & 0.27 & 0.56 & 0.3 \\ 
35 & 8 & 0.26 & 0.06 & 0.12 & 0.05 \\ 
36 & 412 & 18.25 & 4.43 & 10.31 & 2.32 \\ 
37 & 7 & 0.3 & 0.09 & 0.11 & 0.07 \\ 
38 & 173 & 3.47 & 0.8 & 1.86 & 0.45 \\ 
39 & 15 & 0.81 & 0.22 & 0.41 & 0.11 \\ 
40 & 67 & 2.7 & 0.67 & 1.42 & 0.41 \\ 
41 & 44 & 2.21 & 0.76 & 0.81 & 0.49 \\ 
42 & 4 & 0.2 & 0.07 & 0.09 & 0.03 \\ 
43 & 126 & 6.35 & 2.4 & 2.53 & 1.11 \\ 
44 & 100 & 3.18 & 0.96 & 1.64 & 0.38 \\ 
45 & 11 & 1.01 & 0.4 & 0.35 & 0.21 \\ 
46 & 91 & 5.7 & 1.71 & 2.96 & 0.74 \\ 
47 & 95 & 2.03 & 0.28 & 1.21 & 0.27 \\ 
48 & 3 & 0.09 & 0.02 & 0.05 & 0.01 \\ 
49 & 1 & 0.05 & 0.01 & 0.03 & ---  \\ 
50 & 11 & 0.54 & 0.07 & 0.31 & 0.12 \\ 
51 & 147 & 3.95 & 0.53 & 2.51 & 0.57 \\ 
52 & 51 & 2.92 & 0.59 & 1.73 & 0.46 \\ 
53 & 27 & 0.76 & 0.2 & 0.37 & 0.14 \\ 
54 & 25 & 1.29 & 0.34 & 0.63 & 0.23 \\ 
55 & 548 & 31.89 & 3.65 & 20.84 & 4 \\ 
56 & 31 & 1.58 & 0.35 & 0.86 & 0.3 \\ 
57 & 208 & 11.18 & 2.23 & 6.58 & 1.75 \\ 
58 & 493 & 11.97 & 2.43 & 6.69 & 1.98 \\ 
59 & 338 & 8.63 & 1.53 & 4.93 & 1.39 \\ 
60 & 252 & 11.75 & 2.22 & 6.91 & 2.03 \\ 
61 & 233 & 12.52 & 2.74 & 6.99 & 2.21 \\ 
62 & 5664 & 223.07 & 40 & 133.95 & 34.07 \\ 
63 & 2188 & 53.41 & 15.98 & 24.99 & 7.7 \\ 
64 & 435 & 17.57 & 4.28 & 9.16 & 2.88 \\ 
65 & 372 & 10.18 & 2.81 & 5 & 1.39 \\ 
66 & 51 & 1.81 & 0.61 & 0.64 & 0.39 \\ 
67 & 63 & 0.7 & 0.16 & 0.35 & 0.11 \\ 
68 & 21 & 1.06 & 0.34 & 0.42 & 0.22 \\ 
69 & 75 & 2.77 & 0.87 & 1.27 & 0.43 \\ 
70 & 208 & 9.04 & 3.09 & 3.96 & 1.4 \\ 
71 & 666 & 21.53 & 8.92 & 7.59 & 2.96 \\ 
72 & 240 & 9.68 & 3.05 & 4.63 & 1.4 \\ 
73 & 29 & 2.25 & 1.31 & 0.21 & 0.43 \\ 
74 & 370 & 10.91 & 3.5 & 4.92 & 1.72 \\ 
75 & 5 & 0.33 & 0.05 & 0.24 & 0.03 \\ 
76 & 12 & 0.56 & 0.18 & 0.25 & 0.11 \\ 
77 & 2 & 0.03 & 0.01 & 0.01 & ---  \\ 
78 & 2 & 0.07 & 0.04 & ---  & 0.03 \\ 
79 & 14 & 0.64 & 0.2 & 0.32 & 0.09 \\ 
80 & 3 & 0.43 & 0.22 & 0.13 & 0.07 \\ 
81 & 19 & 0.81 & 0.14 & 0.53 & 0.09 \\ 
82 & 275 & 8.88 & 2.63 & 4.57 & 1.14 \\ 
83 & 11 & 0.46 & 0.08 & 0.27 & 0.06 \\ 
84 & 35 & 1.31 & 0.24 & 0.8 & 0.21 \\ 
85 & 54 & 2.13 & 0.54 & 1.03 & 0.39 \\ 
86 & 71 & 3.63 & 0.59 & 2.34 & 0.48 \\ 
87 & 298 & 7.36 & 2.69 & 2.87 & 1.3 \\ 
88 & 332 & 10.59 & 4.68 & 3.15 & 2.07 \\ 
90 & 5 & 0.19 & 0.08 & 0.03 & 0.06 \\ 
91 & 9 & 0.48 & 0.09 & 0.32 & 0.06 \\ 
92 & 135 & 3.52 & 0.98 & 1.79 & 0.58 \\ 
94 & 1 & 0.02 & ---  & 0.01 & ---  \\ 
95 & 3 & 0.12 & 0.03 & 0.07 & 0.02 \\ 
96 & 37 & 0.93 & 0.24 & 0.51 & 0.13 \\ 
97 & 2 & 0.23 & 0.04 & 0.05 & 0.14 \\ 
98 & 7 & 0.24 & 0.1 & 0.09 & 0.04 \\ 
99 & 195 & 1.23 & 0.26 & 0.62 & 0.18 \\ 
100 & 38 & 1.62 & 0.29 & 1.07 & 0.2 \\ 
102 & 15 & 0.77 & 0.1 & 0.56 & 0.09 \\ 
103 & 9 & 0.13 & 0.03 & 0.08 & 0.02 \\ 
104 & 4 & 0.22 & 0.04 & 0.15 & 0.02 \\ 
105 & 4 & 0.15 & 0.02 & 0.1 & 0.02 \\ 
106 & 1 & 0.06 & 0.01 & 0.04 & 0.01 \\ 
107 & 159 & 2.22 & 0.79 & 0.96 & 0.28 \\ 
108 & 15 & 0.63 & 0.24 & 0.26 & 0.1 \\ 
109 & 54 & 1.37 & 0.62 & 0.48 & 0.2 \\ 
110 & 36 & 2.5 & 0.71 & 1.24 & 0.42 \\ 
111 & 27 & 0.38 & 0.09 & 0.21 & 0.05 \\ 
112 & 4 & 0.07 & 0.02 & 0.03 & 0.01 \\ 
113 & 167 & 5.09 & 1.1 & 3.07 & 0.56 \\ 
114 & 42 & 1.6 & 0.4 & 0.89 & 0.23 \\ 
115 & 24 & 0.34 & 0.07 & 0.17 & 0.05 \\ 
116 & 25 & 1.25 & 0.14 & 0.92 & 0.14 \\ 
117 & 12 & 0.67 & 0.12 & 0.41 & 0.1 \\ 
118 & 12 & 0.57 & 0.11 & 0.36 & 0.07 \\ 
119 & 50 & 1.14 & 0.52 & 0.34 & 0.2 \\ 
120 & 25 & 1.18 & 0.46 & 0.47 & 0.19 \\ 
121 & 216 & 4.82 & 1.86 & 1.84 & 0.7 \\ 
122 & 515 & 10.77 & 4.49 & 3.87 & 1.74 \\ 
123 & 727 & 5.85 & 1.34 & 2.59 & 0.92 \\ 
124 & 34 & 0.62 & 0.2 & 0.24 & 0.13 \\ 
125 & 73 & 1.92 & 0.48 & 1.05 & 0.26 \\ 
126 & 12 & 0.54 & 0.13 & 0.31 & 0.08 \\ 
127 & 574 & 14.47 & 2.42 & 8.93 & 1.73 \\ 
128 & 94 & 4.95 & 1.05 & 2.91 & 0.69 \\ 
129 & 1549 & 13.71 & 3.68 & 6.05 & 2.23 \\ 
130 & 59 & 3 & 0.58 & 1.86 & 0.43 \\ 
131 & 58 & 0.4 & 0.11 & 0.17 & 0.07 \\ 
132 & 10 & 0.22 & 0.07 & 0.1 & 0.03 \\ 
133 & 5 & 0.19 & 0.01 & 0.14 & 0.03 \\ 
134 & 9 & 0.19 & 0.04 & 0.11 & 0.03 \\ 
135 & 91 & 3.28 & 0.7 & 1.92 & 0.47 \\ 
136 & 790 & 11.58 & 2.43 & 6.49 & 1.67 \\ 
137 & 160 & 3.88 & 0.69 & 2.36 & 0.57 \\ 
138 & 26 & 1.41 & 0.25 & 0.89 & 0.13 \\ 
139 & 3533 & 54.9 & 30.93 & 10.47 & 9.08 \\ 
140 & 959 & 21.54 & 8.14 & 8.68 & 3.12 \\ 
141 & 733 & 19.92 & 8.65 & 6.91 & 3.04 \\ 
142 & 104 & 8.49 & 3.69 & 3.04 & 1.33 \\ 
143 & 19 & 1.42 & 0.28 & 0.86 & 0.21 \\ 
144 & 32 & 1.61 & 0.54 & 0.63 & 0.36 \\ 
145 & 5 & 0.38 & 0.08 & 0.24 & 0.05 \\ 
146 & 113 & 5.79 & 1.28 & 3.63 & 0.64 \\ 
147 & 84 & 3.49 & 0.98 & 1.62 & 0.63 \\ 
148 & 822 & 33.51 & 9.81 & 16.67 & 5.36 \\ 
149 & 15 & 0.45 & 0.14 & 0.22 & 0.06 \\ 
150 & 91 & 3.99 & 0.82 & 2.48 & 0.5 \\ 
151 & 5 & 0.27 & 0.07 & 0.15 & 0.04 \\ 
152 & 273 & 4.95 & 1.38 & 2.57 & 0.7 \\ 
153 & 2 & 0.09 & 0.02 & 0.05 & 0.01 \\ 
154 & 140 & 1.81 & 0.58 & 0.81 & 0.28 \\ 
155 & 70 & 2.51 & 0.86 & 1.16 & 0.35 \\ 
156 & 255 & 11.97 & 3.71 & 5.41 & 1.69 \\ 
157 & 29 & 1 & 0.19 & 0.66 & 0.1 \\ 
158 & 3 & 0.14 & 0.04 & 0.07 & 0.02 \\ 
159 & 59 & 2.89 & 0.55 & 1.91 & 0.33 \\ 
160 & 306 & 20.34 & 14.24 & 2.56 & 3.24 \\ 
161 & 252 & 7.86 & 2.48 & 3.82 & 1.21 \\ 
162 & 73 & 1.14 & 0.35 & 0.5 & 0.17 \\ 
163 & 88 & 3.39 & 1.23 & 1.34 & 0.57 \\ 
164 & 1025 & 10.28 & 3.4 & 4.07 & 1.59 \\ 
165 & 68 & 2.87 & 0.69 & 1.63 & 0.43 \\ 
166 & 1795 & 51.92 & 25.8 & 14.99 & 8.24 \\ 
167 & 959 & 25.83 & 10.4 & 10.02 & 4.1 \\ 
169 & 1 & 0.04 & 0.01 & 0.02 & ---  \\ 
173 & 196 & 11.41 & 2.42 & 6.14 & 1.12 \\ 
174 & 72 & 3.58 & 0.7 & 2.29 & 0.39 \\ 
175 & 2 & 0.05 & 0.01 & 0.03 & 0.01 \\ 
176 & 369 & 13.22 & 3.03 & 7.23 & 2.19 \\ 
177 & 3 & 0.12 & 0.05 & 0.05 & 0.02 \\ 
180 & 95 & 0.91 & 0.3 & 0.36 & 0.16 \\ 
181 & 15 & 0.23 & 0.08 & 0.09 & 0.04 \\ 
182 & 51 & 1.25 & 0.38 & 0.62 & 0.18 \\ 
183 & 3 & 0.01 & ---  & ---  & ---  \\ 
185 & 97 & 5.59 & 1.11 & 3.69 & 0.52 \\ 
186 & 877 & 19 & 5.02 & 10.71 & 2.14 \\ 
187 & 251 & 1.52 & 0.54 & 0.54 & 0.24 \\ 
188 & 28 & 1.15 & 0.35 & 0.58 & 0.15 \\ 
189 & 871 & 16.66 & 4.33 & 8.77 & 1.9 \\ 
190 & 73 & 3.06 & 0.73 & 1.76 & 0.38 \\ 
191 & 1401 & 11.26 & 3.08 & 4.84 & 1.72 \\ 
192 & 15 & 0.85 & 0.26 & 0.42 & 0.13 \\ 
193 & 498 & 11.96 & 4.03 & 5.47 & 1.53 \\ 
194 & 3433 & 50.85 & 15.29 & 23.62 & 7.07 \\ 
195 & 1 & 0.04 & 0.02 & 0.02 & 0.01 \\ 
196 & 1 & 0.03 & 0.01 & 0.01 & ---  \\ 
197 & 36 & 1.97 & 0.65 & 0.94 & 0.27 \\ 
198 & 327 & 7.67 & 2.36 & 3.84 & 1.1 \\ 
199 & 41 & 1.64 & 0.54 & 0.75 & 0.23 \\ 
200 & 27 & 0.81 & 0.21 & 0.43 & 0.12 \\ 
201 & 8 & 0.52 & 0.12 & 0.3 & 0.07 \\ 
202 & 26 & 0.69 & 0.33 & 0.19 & 0.12 \\ 
203 & 8 & 0.5 & 0.23 & 0.16 & 0.08 \\ 
204 & 331 & 6.73 & 2.69 & 2.3 & 1.12 \\ 
205 & 368 & 5.8 & 1.95 & 2.58 & 0.93 \\ 
206 & 213 & 13.02 & 4.11 & 6.17 & 1.94 \\ 
208 & 2 & 0.01 & 0.01 & ---  & ---  \\ 
210 & 1 & 0.03 & 0.01 & 0.01 & 0.01 \\ 
211 & 1 & 0.03 & 0.02 & 0.01 & 0.01 \\ 
212 & 29 & 1.02 & 0.32 & 0.5 & 0.16 \\ 
213 & 41 & 1 & 0.3 & 0.49 & 0.15 \\ 
214 & 19 & 0.92 & 0.41 & 0.31 & 0.14 \\ 
215 & 98 & 2.18 & 0.45 & 1.35 & 0.24 \\ 
216 & 1826 & 9.11 & 3.7 & 2.52 & 1.41 \\ 
217 & 185 & 6.53 & 2.14 & 3.04 & 0.82 \\ 
218 & 98 & 4.17 & 1.4 & 2.06 & 0.57 \\ 
219 & 14 & 0.8 & 0.26 & 0.39 & 0.1 \\ 
220 & 299 & 10.02 & 3.84 & 4.06 & 1.43 \\ 
221 & 3559 & 11.29 & 3.52 & 3.15 & 2.19 \\ 
223 & 578 & 11.28 & 3.19 & 5.65 & 1.65 \\ 
224 & 48 & 0.39 & 0.13 & 0.15 & 0.06 \\ 
225 & 5740 & 33.64 & 14.19 & 8.95 & 5.6 \\ 
226 & 110 & 3.17 & 1.21 & 1.23 & 0.46 \\ 
227 & 3108 & 41.14 & 18.06 & 12.18 & 6.54 \\ 
229 & 589 & 4.79 & 1.79 & 1.59 & 0.8 \\ 
230 & 6 & 0.37 & 0.12 & 0.17 & 0.06 \\ 
\hline
Total & 73234 & 2072.48 & 559.38 & 1013.81 & 342.62
\\
\hline
\end{longtable}
}

{\tiny
\begin{longtable}{|c|c|c|c|c|c|}
\caption[Total disk storage space for ab-initio calculations per number of atoms in the primitive cell]{Total disk storage space (in Gb) expended for all of the ab-initio calculations performed in this work, subdivided by the number of atoms in the primitive cell in each material.  In order, the columns in this table list the  number of atoms, the number of ICSDs successfully analyzed with the same number of atoms, the total storage space (in Gb) for the calculations performed with and w/o SOC, the storage space used for the compressed CHGCAR files generated during the SOC self-consistent DFT calculations, the storage used for the compressed PROCAR files generated during the SOC DOS DFT calculations, and the total storage space used for the calculations performed w/o SOC.
\label{tb:statistics_compounds_storage_nbratoms_primitivecell}}\\
\hline
\# atoms & \# ICSDs & \begin{tabular}{c}Total Storage\\SOC+w/o SOC (Gb)\end{tabular} & \begin{tabular}{c}CHGCAR SCC Storage\\(Gb)\end{tabular} & \begin{tabular}{c}PROCAR DOS Storage\\(Gb)\end{tabular} & \begin{tabular}{c}Total w/o SOC Storage\\(Gb)\end{tabular}  \\
\hline 
1 & 1178 & 1.34 & 0.4 & 0.08 & 0.38 \\ 
2 & 5838 & 11.9 & 4.61 & 1.48 & 2.63 \\ 
3 & 2429 & 8.76 & 3.86 & 1.6 & 1.7 \\ 
4 & 4851 & 23.6 & 9.84 & 5.7 & 4.36 \\ 
5 & 3912 & 29.83 & 13.5 & 7.48 & 5.1 \\ 
6 & 6280 & 50.21 & 19.69 & 15.6 & 8.55 \\ 
7 & 1169 & 15.36 & 7.41 & 4.05 & 2.58 \\ 
8 & 4239 & 50.43 & 17.46 & 19.92 & 8.05 \\ 
9 & 1745 & 28.19 & 10.26 & 11.35 & 3.98 \\ 
10 & 3498 & 52.59 & 17.43 & 22.76 & 8.06 \\ 
11 & 488 & 15.25 & 5.89 & 6.03 & 2.21 \\ 
12 & 5023 & 111.11 & 30.79 & 56.24 & 15.91 \\ 
13 & 748 & 20.16 & 6.82 & 9.18 & 2.76 \\ 
14 & 2462 & 62.23 & 20.79 & 28.02 & 8.88 \\ 
15 & 429 & 16.07 & 4.83 & 8.12 & 2.22 \\ 
16 & 2980 & 99.57 & 24.13 & 55.9 & 13.01 \\ 
17 & 336 & 13.85 & 5.66 & 5.38 & 2.06 \\ 
18 & 1607 & 64.96 & 16.25 & 34.47 & 8.18 \\ 
19 & 271 & 13.42 & 3.58 & 7.3 & 1.73 \\ 
20 & 3618 & 152.47 & 29.56 & 91.29 & 20.03 \\ 
21 & 199 & 10.08 & 1.92 & 6.4 & 1.21 \\ 
22 & 1613 & 78.63 & 19.28 & 44.71 & 10.33 \\ 
23 & 148 & 8.11 & 2.55 & 4.1 & 1.05 \\ 
24 & 2635 & 141 & 26.83 & 88.31 & 17.75 \\ 
25 & 65 & 4.93 & 1.03 & 3.02 & 0.62 \\ 
26 & 686 & 45.4 & 10.52 & 27.66 & 5.33 \\ 
27 & 154 & 10.82 & 2.03 & 7.03 & 1.34 \\ 
28 & 2663 & 159.49 & 33.06 & 97.87 & 20.74 \\ 
29 & 441 & 18.63 & 5.6 & 8.85 & 2.69 \\ 
30 & 855 & 60.95 & 11.88 & 38.63 & 7.48 \\ 
31 & 37 & 3.07 & 0.52 & 1.89 & 0.56 \\ 
32 & 1446 & 92.83 & 19.94 & 51.6 & 14.87 \\ 
33 & 76 & 4.35 & 1.42 & 1.9 & 0.74 \\ 
34 & 341 & 21.8 & 5.94 & 9.98 & 4.51 \\ 
35 & 44 & 3.32 & 1.2 & 1.13 & 0.8 \\ 
36 & 1644 & 113.2 & 25.87 & 61.42 & 19.17 \\ 
37 & 34 & 2.13 & 0.86 & 0.48 & 0.66 \\ 
38 & 494 & 34.24 & 8.51 & 17.76 & 5.72 \\ 
39 & 67 & 5.23 & 1.9 & 2.04 & 0.98 \\ 
40 & 1533 & 110.22 & 25.88 & 59.04 & 19.35 \\ 
41 & 46 & 2.93 & 0.76 & 1.37 & 0.6 \\ 
42 & 495 & 37.54 & 6.64 & 19.17 & 6.45 \\ 
43 & 16 & 0.95 & 0.15 & 0.45 & 0.28 \\ 
44 & 791 & 50.33 & 11.77 & 21.13 & 12.38 \\ 
45 & 45 & 2.49 & 0.84 & 0.76 & 0.65 \\ 
46 & 390 & 22.18 & 8.24 & 6.98 & 5.49 \\ 
47 & 10 & 0.45 & 0.21 & ---  & 0.2 \\ 
48 & 846 & 50.64 & 17.71 & 14.2 & 14.46 \\ 
49 & 33 & 2.86 & 0.84 & 1.27 & 0.61 \\ 
50 & 183 & 10.95 & 3.95 & 2.87 & 3.22 \\ 
51 & 12 & 0.51 & 0.21 & 0.05 & 0.17 \\ 
52 & 641 & 37.57 & 14.32 & 6.9 & 11.54 \\ 
53 & 10 & 0.44 & 0.16 & 0.08 & 0.17 \\ 
54 & 257 & 15.03 & 6.1 & 3.5 & 4.36 \\ 
55 & 13 & 0.86 & 0.25 & 0.35 & 0.19 \\ 
56 & 552 & 31.41 & 13 & 4.45 & 10.8 \\ 
57 & 11 & 0.66 & 0.35 & 0.04 & 0.2 \\ 
58 & 161 & 9.55 & 3.8 & 2.12 & 2.73 \\ 
59 & 4 & 0.25 & 0.1 & ---  & 0.12 \\ 
60 & 417 & 23.81 & 9.86 & 2.24 & 9.17 \\ 
62 & 2 & 0.1 & 0.04 & ---  & 0.05 \\ 
64 & 2 & 0.12 & 0.05 & ---  & 0.05 \\ 
68 & 2 & 0.1 & 0.04 & ---  & 0.04 \\ 
70 & 3 & 0.14 & 0.07 & ---  & 0.06 \\ 
72 & 7 & 0.45 & 0.19 & 0.06 & 0.14 \\ 
74 & 2 & 0.1 & 0.04 & ---  & 0.05 \\ 
76 & 6 & 0.29 & 0.12 & ---  & 0.13 \\ 
78 & 1 & 0.06 & 0.02 & ---  & 0.03 \\ 
\hline
Total & 73234 & 2072.48 & 559.38 & 1013.81 & 342.62
\\
\hline
\end{longtable}
}

\afterpage{\clearpage}

\section{Topological Material Classes in the Absence of Spin-Orbit Coupling}\label{App:TopologicalMaterialsNoSOC}

In real materials, spin-orbit interaction (SOC) is always present~\cite{CharlieTI,CharlieZahidReview}; however, depending on the orbitals at the Fermi energy, the dispersion, and the specific atoms in the material, the effects of SOC can be approximated to be negligible.  For each of the materials on~\webNoICSD, we have performed DFT calculations (the details of which are specified in \supappref{App:VASP_appendix}) both incorporating and neglecting the effects of SOC (though as noted in \supappref{App:VASP_appendix}, there exist materials on~\webNoICSD~for which only calculations incorporating SOC were convergent).  As shown in Fig.~\ref{fig:noSOCButton}, users on~\webNoICSD~can toggle between the results of first-principles calculations performed with and w/o SOC for the same material using the switch on the top-right corner of each material page [light blue box in Fig.~\ref{fig:noSOCButton}(A,B)].

In the absence of SOC, the possible stable topological (TQC) classes~\cite{QuantumChemistry,Bandrep1,Bandrep2,Bandrep3,BarryFragile,AndreiMaterials,JenFragile1} for each material -- NLC, SEBR, ES, ESFD, and LCEBR -- are the same as the possible classes in the presence of SOC.  However, as we will detail in the text below, the relative number of materials within each topological class, and the physical meaning of the NLC and SEBR classes, strongly differ with and w/o SOC.  First, in \supappref{App:whySoManyNoSOCES}, we will discuss the relative preponderance of ES-classified materials compared to the relative dearth of NLC- and SEBR-classified compounds in the absence of SOC.  Then, in \supappref{App:physicalMeaningnoSOCNLCSEBR}, we will review the physical meaning and expected topological boundary states of each topological class in the absence of SOC.  In particular, in \supappref{App:physicalMeaningnoSOCNLCSEBR}, we will focus on the cases of NLC- and SEBR-classified nonmagnetic materials, which were shown in Ref.~\onlinecite{ZhidaSemimetals} to be topological semimetals -- as opposed to topological (crystalline) insulators -- when the effects of SOC are neglected.  Throughout this work, we specifically term ICSD entries that are classified as NLC (SEBR) at $E_{F}$ w/o SOC as NLC-SM (SEBR-SM) topological semimetals.  In the language of Refs.~\onlinecite{MTQC,MTQCmaterials}, NLC-SM and SEBR-SM phases w/o SOC are \emph{Smith-index semimetals}, as they are characterized by nontrivial values of the SIs of the nonmagnetic (Type-II Shubnikov) single (spinless) SGs~\cite{ZhidaSemimetals}.

\begin{figure}[h]
\includegraphics[width=0.96\columnwidth]{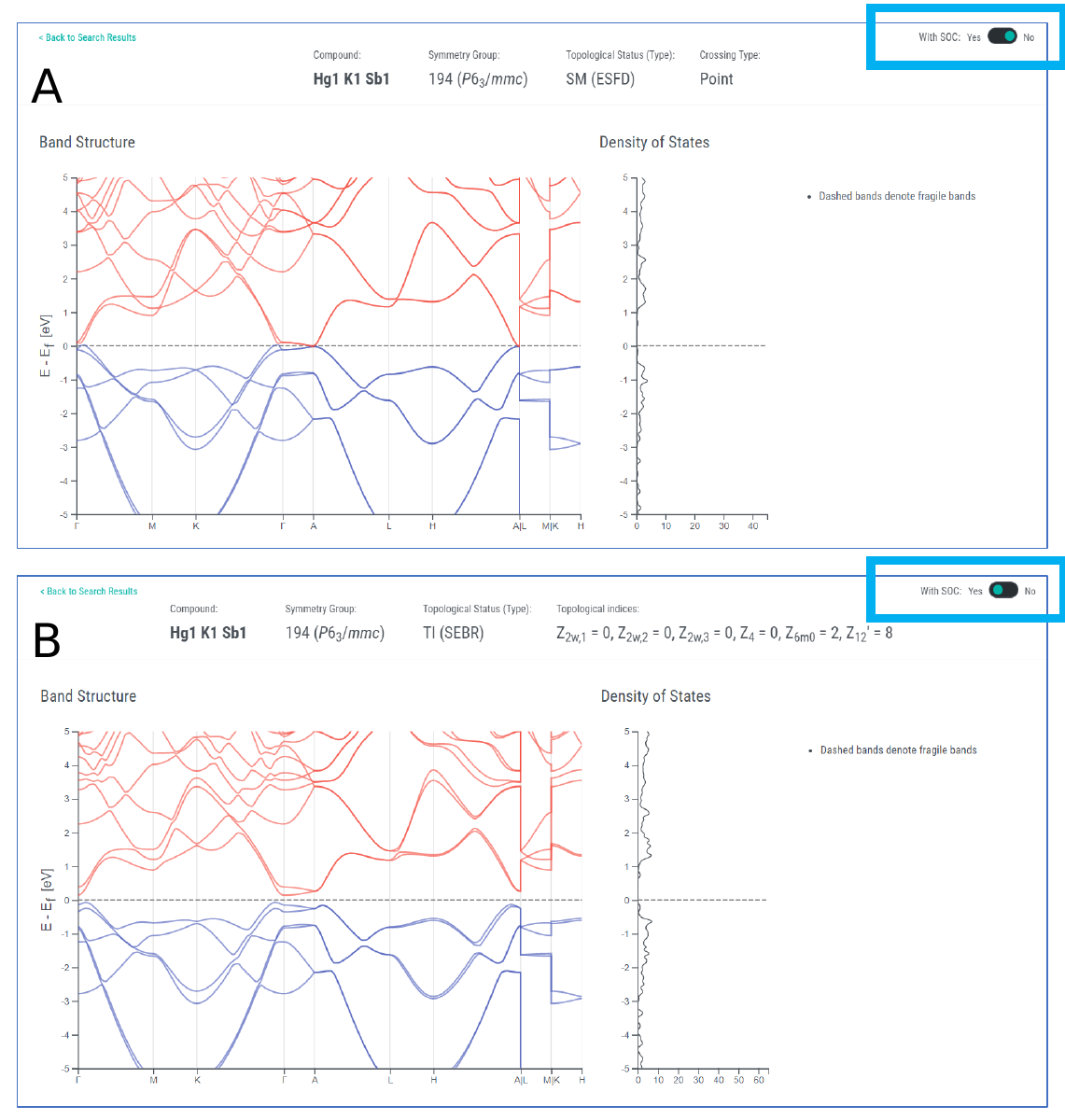}
\caption{The electronic structure of the candidate hourglass TCI KHgSb (\icsdweb{56201}, see Refs.~\cite{HourglassInsulator,Cohomological,DiracInsulator,HourglassExperiment,zeroHallExp} and \supappref{App:z4trivial}) without (A) and with (B) the effects of SOC incorporated.  Users can toggle between the first-principles analysis and topological data for each material on~\webNoICSD~with and without SOC by clicking the switch highlighted in the blue box in the top-right corners of panels A and B.} 
\label{fig:noSOCButton}
\end{figure}

\subsection{Relative Numbers of Materials in Each Topological Class with and without SOC}
\label{App:whySoManyNoSOCES}

\begin{table}[h]
\begin{tabular}{|c|c|c|}
\hline
\multicolumn{3}{|c|}{Number of Unique Materials in Each Stable} \\
\multicolumn{3}{|c|}{Topological Class Among the Materials with} \\
\multicolumn{3}{|c|}{Convergent Calculations in the Absence of SOC} \\ 
\multicolumn{3}{|c|}{(\TQCDBNbrNoSOCUniqueMaterials~Unique Materials)} \\
\hline
Topological Class & Without SOC & With SOC \\
\hline
\hline
NLC (-SM) & \TQCDBNbrNoSOCMaterialsNLCSM~(\TQCDBNbrNoSOCMaterialsNLCSMPercent) & \TQCDBNbrMaterialsWithSOCNoSOCNLC~(\TQCDBNbrMaterialsWithSOCNoSOCNLCPercent) \\
\hline
SEBR (-SM) & \TQCDBNbrNoSOCMaterialsSEBRSM~(\TQCDBNbrNoSOCMaterialsSEBRSMPercent) & \TQCDBNbrMaterialsWithSOCNoSOCSEBR~(\TQCDBNbrMaterialsWithSOCNoSOCSEBRPercent) \\
\hline
ES & \TQCDBNbrNoSOCMaterialsES~(\TQCDBNbrNoSOCMaterialsESPercent) & \TQCDBNbrMaterialsWithSOCNoSOCES~(\TQCDBNbrMaterialsWithSOCNoSOCESPercent) \\
\hline
ESFD & \TQCDBNbrNoSOCMaterialsESFD~(\TQCDBNbrNoSOCMaterialsESFDPercent) & \TQCDBNbrMaterialsWithSOCNoSOCESFD~(\TQCDBNbrMaterialsWithSOCNoSOCESFDPercent) \\
\hline
LCEBR & \TQCDBNbrNoSOCMaterialsLCEBR~(\TQCDBNbrNoSOCMaterialsLCEBRPercent) & \TQCDBNbrMaterialsWithSOCNoSOCLCEBR~(\TQCDBNbrMaterialsWithSOCNoSOCLCEBRPercent) \\
\hline
\end{tabular}
\caption[Number of unique materials in each topological class where both SOC and w/o SOC calculations have converged]{On~\webNoICSD, there are \TQCDBNbrNoSOCUniqueMaterials~unique materials (as defined in Ref.~\onlinecite{AndreiMaterials} and in the main text) for which DFT calculations were successfully performed both with and without incorporating the effects of SOC.  For each of the stable topological classes, we list the number of materials in the class with and without SOC.  As will be discussed in \supappref{App:physicalMeaningnoSOCNLCSEBR}, we have added a parenthetical (-SM) following the NLC and SEBR classes, because NLC- and SEBR-classified nonmagnetic materials are necessarily topological semimetals in the absence of SOC~\cite{AshvinTCI,ZhidaSemimetals,ChenTCI}.  As discussed in this section, the number of LCEBR compounds is relatively similar whether SOC is incorporated or neglected, whereas there are vanishingly few NLC-SM- and SEBR-SM-classified materials in the absence of SOC.  Most specifically, the combined numbers of SEBR- and ESFD-classified materials with SOC ($\TQCDBNbrMaterialsWithSOCNoSOCSEBR + \TQCDBNbrMaterialsWithSOCNoSOCESFD = \TQCDBNbrMaterialsWithSOCNoSOCSEBRESFD$) is close to the number of ESFD-classified materials without SOC (\TQCDBNbrNoSOCMaterialsESFD), the number of NLC- and ES-classified materials with SOC ($ \TQCDBNbrMaterialsWithSOCNoSOCNLC + \TQCDBNbrMaterialsWithSOCNoSOCES = \TQCDBNbrMaterialsWithSOCNoSOCNLCES$) is close to the number of NLC-SM- and ES-classified materials without SOC ($\TQCDBNbrNoSOCMaterialsNLCSM + \TQCDBNbrNoSOCMaterialsES = \TQCDBNbrNoSOCMaterialsNLCSMES$), and the number of SEBR-SM-classified materials without SOC ($\TQCDBNbrNoSOCMaterialsSEBRSM$) is small compared to the overall number of calculated materials (\TQCDBNbrNoSOCMaterialsSEBRSMPercent~of \TQCDBNbrNoSOCUniqueMaterials).  As explained in the text below, the relative absence of NLC-SM and SEBR-SM materials can be understood by recognizing that most NLC- and SEBR-classified materials with SOC become ES-classified when SOC is neglected, because there are significantly many more ways for materials to fail to satisfy the compatibility relations~\cite{QuantumChemistry,Bandrep2,Bandrep3} in the absence of SOC~\cite{Bandrep1}.}
\label{tb:relativeNumberSOC}
\end{table}

In Table~\ref{tb:relativeNumberSOC}, we show the number (and percentage) of materials in each stable topological class at $E_{F}$ in the calculations performed with and w/o SOC for the unique materials (defined in Ref.~\onlinecite{AndreiMaterials} and in the main text) for which DFT calculations (performed as detailed in \supappref{App:VASP_appendix}) converged both with and without incorporating the effects of SOC.  We note that while the number of LCEBR-classified materials are similar with and w/o SOC, there are large differences across the other four stable topological classes.  Most notably, in Table~\ref{tb:relativeNumberSOC}, the number of NLC- and ES-classified materials with SOC ($\TQCDBNbrMaterialsWithSOCNoSOCNLC + \TQCDBNbrMaterialsWithSOCNoSOCES = \TQCDBNbrMaterialsWithSOCNoSOCNLCES$) is close to the number of NLC-SM- and ES-classified materials w/o SOC ($\TQCDBNbrNoSOCMaterialsNLCSM + \TQCDBNbrNoSOCMaterialsES = \TQCDBNbrNoSOCMaterialsNLCSMES$), and the number of SEBR-SM-classified materials w/o SOC ($\TQCDBNbrNoSOCMaterialsSEBRSM$) is small compared to the overall number of calculated materials (\TQCDBNbrNoSOCMaterialsSEBRSMPercent~of \TQCDBNbrNoSOCUniqueMaterials).  In particular, the number of NLC-SM- and SEBR-SM-classified materials w/o SOC in Table~\ref{tb:relativeNumberSOC} ($\TQCDBNbrNoSOCMaterialsNLCSM + \TQCDBNbrNoSOCMaterialsSEBRSM = \TQCDBNbrNoSOCMaterialsNLCSMSEBRSM$) is smaller by roughly a factor of 20 than the number of NLC- and SEBR-classified materials with SOC ($\TQCDBNbrMaterialsWithSOCNoSOCNLC + \TQCDBNbrMaterialsWithSOCNoSOCSEBR = \TQCDBNbrMaterialsWithSOCNoSOCNLCSEBR$).  This difference can be understood by recognizing that the matrix representatives of twofold symmetries, such as mirror reflection and twofold rotation, anticommute in the presence of SOC, but \emph{commute} in the absence of SOC, and that twofold crystal symmetries without translations have real-valued (imaginary-valued) eigenvalues in the absence (presence) of SOC~\cite{BigBook,WiederLayers}.  This implies that many more symmetry operations have simultaneously well-defined eigenvalues at each ${\bf k}$ point in the absence of SOC than in the presence of SOC in nonmagnetic materials.  Consequently, along high-symmetry lines and planes with the same symmetry, there are typically many more small irreducible corepresentation (corep) labels in the absence of SOC than in its presence.  For this reason, when bands are inverted in a material w/o SOC, the material is much more likely to become ES-classified (and thus fail to satisfy the compatibility relations) than it is to become NLC-SM-classified, whereas in the presence of SOC, the material has a significantly greater chance of becoming NLC-classified after a bulk band inversion.

As an example, consider a TRIM point ${\bf k}$ in a material where the little group $G_{\bf k}$ contains twofold rotation $\{C_{2z}|000\}$, inversion $\{\mathcal{I}|000\}$, and time-reversal $\{\mathcal{T}|000\}$ symmetries.  The combination of inversion and twofold rotation also implies the presence of a mirror symmetry $\{M_{z}|000\}=\{\mathcal{I}|000\}\{C_{2z}|000\}$.  Without SOC, the matrix representatives of inversion, twofold rotation, and mirror symmetries commute, and hence all three symmetries have simultaneously well-defined (real-valued) eigenvalues.  There are therefore four one-dimensional (per spin), single-valued coreps at ${\bf k}$, which may be labeled by their $\{C_{2z}|000\}$ and $\{\mathcal{I}|000\}$ eigenvalues.  When bands with distinct parity (inversion) eigenvalues are inverted at ${\bf k}$ in the absence of SOC, protected crossing points appear along the twofold rotation axis if the bands carry distinct $\{C_{2z}|000\}$ eigenvalues and the same $\{M_{z}|000\}$ eigenvalues, or appear in the mirror plane if the inverted bands instead carry distinct $\{M_{z}|000\}$ eigenvalues and the same $\{C_{2z}|000\}$ eigenvalues.  Conversely with SOC, the eigenvalues of $\{C_{2z}|000\}$ are imaginary numbers $\pm i$, and states with distinct $\{C_{2z}|000\}$ eigenvalues are paired by $\{\mathcal{T}|000\}$ symmetry at ${\bf k}$, and by $\{\mathcal{I}\times\mathcal{T}|000\}$ symmetry away from ${\bf k}$.  Therefore in the presence of SOC, $G_{\bf k}$ has only two, two-dimensional, double-valued small coreps that may be labeled by their Kramers pairs of distinct parity (inversion) eigenvalues.  Hence, when bands with distinct parity eigenvalues are inverted at ${\bf k}$ with SOC, protected crossing points do not appear, absent additional symmetries.

Lastly, we note that the general trend of increased band connectivity w/o SOC also extends away from $E_{F}$ across the stoichiometric materials in the ICSD.  We specifically find that $\TQCDBNbrMaterialsBndLowerEfICSDs$ ICSD entries with SOC (corresponding to $\TQCDBNbrMaterialsBndLowerEf$ unique materials) contain a connected grouping of bands with at least $N_{e}$ states at each ${\bf k}$ point, where $N_{e}$ is the number of valence electrons.  Conversely, $\TQCDBNbrNoSOCMaterialsBndLowerEfICSDs$ ICSD entries w/o SOC (corresponding to $\TQCDBNbrNoSOCMaterialsBndLowerEf$ unique materials) contain a connected grouping of bands with at least $N_{e}$ states at each ${\bf k}$ point (taking bands w/o SOC to be spin-degenerate).  In~\supappref{App:TopoBands} we will also discuss the extreme case of \emph{supermetallic} (SMetal) materials, in which \emph{all} of the bands included in our VASP calculations are connected (see~\supappref{App:VASP_appendix} for calculation details).  Further supporting the trend of increased band connectivity in the absence of SOC,~\webNoICSD~contains~$\TQCDBNbrSMetal$~unique SMetal materials with SOC, whereas there are~\emph{$\TQCDBNbrNoSOCSMetal$}~unique SMetal materials w/o SOC, representing a hundred-fold increase compared to calculations performed with SOC in the number of materials with fully-connected valence bands up to the energy (filling) cutoff of $\sim 2N_{e}$ imposed in our VASP calculations (see~\supappref{App:VASP_appendix}).  Further statistics for materials with highly connected bands on~\webNoICSD, as well as material examples, are provided in~\supappref{App:TopoBands}.

\subsection{Physical Interpretation of the NLC and SEBR Materials Classes in the Absence of SOC}
\label{App:physicalMeaningnoSOCNLCSEBR}

To determine the topological class of a material in the absence of SOC, we use the same methods previously employed in Ref.~\onlinecite{AndreiMaterials}.  First, we determine if the coreps of the occupied bands at all high-symmetry points satisfy the compatibility relations~\cite{QuantumChemistry,Bandrep2,Bandrep3} for an insulating gap to exist at $E_{F}$.  If the insulating compatibility relations are not satisfied, then the material is classified as either an ESFD semimetal (in which a band degeneracy at a high-symmetry point is partially occupied) or an ES semimetal (in which the compatibility relations fail along a line or plane in the BZ).

Without SOC, ES and ESFD semimetals exhibit topological boundary states analogous to those of topological semimetals with strong SOC.  For ESFD materials, it has been extensively shown~\cite{RhSiArc,KramersWeyl,BarryMultifold} that if the little group of the ${\bf k}$ point of the nodal degeneracy is isomorphic to a chiral SG (defined in this work as an SG without rotoinversion symmetries, more formally referenced as a \emph{Sohncke} SG, see Ref.~\onlinecite{ChiralSGsRightandWrong}), then the nodal point will carry a topological chiral charge, and will exhibit associated topological surface Fermi arcs.  As demonstrated in several recent theoretical and experimental studies~\cite{RhSiArc,CoSiArc,CoSiObserveJapan,CoSiObserveHasan,CoSiObserveChina,AlPtObserve,PdGaObserve}, the surface Fermi arcs are qualitatively the same in dispersion and length in strong- and weak-SOC ESFD semimetals whose nodal degeneracies lie at high-symmetry points with chiral little groups.  Specifically, large topological Fermi arcs are typically present in ESFD semimetals in chiral SGs both with and without SOC -- in the weak-SOC case, the Fermi arcs are spin-degenerate, whereas in the strong-SOC case, the Fermi arcs are highly-spin-polarized~\cite{RhSiArc,PdGaObserve}, but are typically only weakly split by SOC~\cite{CoSiObserveJapan,CoSiObserveHasan,CoSiObserveChina,AlPtObserve}.  In weak-SOC ES semimetals, the nodal degeneracies are typically nodal lines, which are protected by a combination of coreps with different symmetry eigenvalues and weak-SOC topological invariants (\emph{e.g.}, local $\mathcal{I}\times\mathcal{T}$-symmetry protection, as introduced in Ref.~\onlinecite{YoungkukLineNode}).  When SOC is negligible, all bulk nodal lines carry associated flat-band-like topological ``drumhead'' surface states~\cite{YoungkukLineNode,XiLineNode}, which represent the weak-SOC (nondispersing) limit of the twofold surface Dirac cones of 3D TIs and TCIs.  Additionally, it has recently been shown that more exotic, ``monopole-charged'' variants~\cite{FangWithWithout,YoungkukMonopole,AdrianMonopole,SigristMonopole} of weak-SOC nodal lines also exhibit flat-band-like hinge states that are representative of a \emph{higher-order} bulk-boundary correspondence, and correspond to the weak-SOC (nondispersing) limit of the helical hinge states of higher-order TIs~\cite{TMDHOTI}.

In the cases where a material w/o SOC is classified as NLC or SEBR, the differences with the strong-SOC case are more pronounced.  First, it was shown in Refs.~\onlinecite{AshvinTCI,ZhidaSemimetals,ChenTCI} that 3D, time-reversal- ($\mathcal{T}$-) symmetric (\emph{i.e.} nonmagnetic), symmetry-indicated, topological insulating phases are only stable in the presence of non-negligible SOC.  However, it was also shown in Refs.~\onlinecite{AndreiInversion,ZhidaSemimetals,ChenMaterials,MTQC,MTQCmaterials} that the valence bands of real materials, when SOC is neglected, are still capable of simultaneously satisfying the insulating compatibility relations while not being equivalent to a linear combination of trivial bands.  This presents an apparent contradiction: namely, a set of valence bands in a 3D material without SOC cannot be topologically nontrivial if stable (strong) topological bands do not exist in 3D, $\mathcal{T}$-symmetric systems without SOC.  The resolution -- discovered in Ref.~\onlinecite{ZhidaSemimetals} -- is that in fact \emph{all} NLC- and SEBR-classified materials in the absence of SOC are special cases of topological semimetals that satisfy the compatibility relations for insulators.  Specifically, the compatibility relations are only capable of determining if the symmetry eigenvalues (small coreps) of the occupied bands can be smoothly connected to each other along all ${\bf k}$ paths with crystal symmetry eigenvalues.  However, bulk nodal degeneracies such as Weyl points~\cite{AshvinWeyl1} and nodal lines w/o SOC~\cite{YoungkukMonopole,FangWithWithout} can also be stabilized by topological invariants evaluated on closed manifolds in momentum space (\emph{e.g.} the Chern number evaluated on a sphere surrounding a Weyl point), \emph{even if the crossing bands have the same corep labels}.  In the relatively few materials on~\webNoICSD~that are classified as NLC-SM or SEBR-SM in the calculations performed w/o SOC (Tables~\ref{tb:list_nosoc_uniquematerials_nice_NLC-SM} and~\ref{tb:list_nosoc_uniquematerials_nice_SEBR-SM}, respectively), a bulk gap is permitted along all high-symmetry lines and planes, but there is no gap in the BZ interior in the calculations performed w/o SOC.

To classify the NLC-SM and SEBR-SM phases of the materials on~\webNoICSD~in the calculations performed w/o SOC, we have employed the SI formulas (topological indices and notation) of Ref.~\onlinecite{ZhidaSemimetals}.  We note that in the notation of Ref.~\onlinecite{ZhidaSemimetals}, the $\mathbb{Z}_{2}$-valued strong nodal-line-semimetal parity index introduced in Ref.~\onlinecite{YoungkukLineNode} (which is the vanishing-SOC analog of the Fu-Kane parity index~\cite{FuKaneInversion,FuKaneMele}), does not appear on~\webNoICSD.  Instead, the $\mathbb{Z}_{2}$ nodal-line index from Ref.~\onlinecite{YoungkukLineNode} has been subsumed by the $\mathbb{Z}_{4}$-valued parity index $z_{4}$ that captures both monopole-charged and uncharged nodal lines, and is the vanishing-SOC (spinless) analog of the $\mathbb{Z}_{4}$-valued index $Z_{4}$ for spinful HOTIs discussed in Refs.~\onlinecite{AshvinIndicators,ChenTCI,AshvinTCI,TMDHOTI,MTQC,YoungkukMonopole} and in \supappref{App:z4HOTIs}.  Specifically, in nonmagnetic materials with inversion symmetry, $z_{4}=1,2,3$ phases are all nodal-line semimetals.  If the values of all other spinless SIs (\emph{e.g.} weak indices) are trivial, then the nodal lines in a $z_{4}=1,3$ ($z_{4}=2$) NLC-SM or SEBR-SM carry trivial (nontrivial) monopole charges~\cite{TMDHOTI,YoungkukMonopole}.

To compare the ES nodal-line semimetals that are prevalent throughout~\webNoICSD~(\supappref{App:whySoManyNoSOCES}) to the nodal-line topological indices previously introduced in Refs.~\onlinecite{YoungkukLineNode,ZhidaSemimetals}, we have produced subduction tables at the bottom of each material page on~\webNoICSD.  Specifically, following the procedure employed in Refs~\onlinecite{AndreiMaterials,LuisSubduction,ChenSubduction}, the most familiar~\cite{YoungkukLineNode,ZhidaSemimetals,TMDHOTI} strong $\mathbb{Z}_{4}$ and weak $\mathbb{Z}_{2}$ parity indices for ES-classified, centrosymmetric, nodal line semimetals w/o SOC ($z_{4}$ and $z_{2,i}$ respectively) can be obtained by consulting the subduction table entry for SG 2 ($P\bar{1}$) in calculations performed w/o SOC [noting that, if $z_{4}$ and $z_{2,i}$ are trivial for all $i=1,2,3$, then the subduction table will not display an entry for SG 2 ($P\bar{1}$)].  During the preparation of this work, an alternative method for diagnosing the topology of ES-classified nodal-line semimetals in the absence of SOC was also introduced in Refs.~\onlinecite{NodaLineMap1,NodaLineMap2}, providing additional context for the subduction tables on~\webNoICSD~generated for the present work.

{\tiny
\begin{longtable}{|c|c|c|c|c|c|c|c|c|}
\caption[List of all nonmagnetic NLC-SM unique materials without $f$ electrons]{List of all nonmagnetic unique materials without $f$ electrons that are classified as NLC-SM in calculations performed w/o SOC.  In this table, a chemical formula with a $^{*}$ indicates that some (but not all) of the ICSDs associated to the same unique material (defined using its topological class with SOC) are classified as NLC-SM in calculations for the same ICSD entries performed w/o SOC.  In this table, the topological indices (spinless SIs) at $E_{F}$ of the calculations performed w/o SOC are listed in the notation of Ref.~\onlinecite{ZhidaSemimetals}.  Further details and the physical interpretation of each spinless SI in this table are provided in Ref.~\onlinecite{ZhidaSemimetals}.}
\label{tb:list_nosoc_uniquematerials_nice_NLC-SM}\\
\hline
Chem. formula & SG \# & SG symbol & ICSD & Line/Weyl & no-SOC topological indices \\
\hline  
$\rm{Ca} As_{3}$& 2 & $P\bar{1}$ & \icsdwebshort{193} & Line & $z_{2,1}=1\;\;z_{2,2}=0\;\;z_{2,3}=0\;\;z_4=1$ \\ 
$\rm{Ca} P_{3}$& 2 & $P\bar{1}$ & \icsdwebshort{74479} & Line & $z_{2,1}=0\;\;z_{2,2}=1\;\;z_{2,3}=0\;\;z_4=1$ \\ 
$\rm{Ba} (Mo_{6} S_{8})$& 2 & $P\bar{1}$ & \icsdwebshort{85490} & Line & $z_{2,1}=0\;\;z_{2,2}=0\;\;z_{2,3}=0\;\;z_4=3$ \\ 
$\rm{Bi}$& 2 & $P\bar{1}$ & \icsdwebshort{426929} & Line & $z_{2,1}=0\;\;z_{2,2}=0\;\;z_{2,3}=1\;\;z_4=3$ \\ 
$\rm{Y} As S$& 14 & $P2_1/c$ & \icsdwebshort{611344} & Line & $z_2'=1\;\;z_4=2$ \\ 
$\rm{Fe} H_{4}$& 11 & $P2_1/m$ & \icsdwebshort{187148} & Line & $z_2'=1\;\;z_4=2$ \\ 
$\rm{Al}_{2} Fe_{3} Si_{3}$& 2 & $P\bar{1}$ & \icsdwebshort{422342} & Line & $z_{2,1}=0\;\;z_{2,2}=1\;\;z_{2,3}=0\;\;z_4=3$ \\ 
$\rm{Sr}_{2} Co (Se O_{3})_{3}$& 2 & $P\bar{1}$ & \icsdwebshort{79204} & Line & $z_{2,1}=0\;\;z_{2,2}=1\;\;z_{2,3}=1\;\;z_4=1$ \\ 
$\rm{K} H_{2} P O_{3}$& 14 & $P2_1/c$ & \icsdwebshort{37072} & Line & $z_2'=1\;\;z_4=2$ \\ 
$\rm{Y} As Se$& 14 & $P2_1/c$ & \icsdwebshort{611398} & Line & $z_2'=1\;\;z_4=2$ \\ 
$\rm{Ag} (S O_{4})$& 2 & $P\bar{1}$ & \icsdwebshort{421341} & Line & $z_{2,1}=1\;\;z_{2,2}=1\;\;z_{2,3}=1\;\;z_4=1$ \\ 
$\rm{Sc} H_{3}$& 12 & $C2/m$ & \icsdwebshort{670220} & Line & $z_2'=1\;\;z_{2,2}=1\;\;z_{2,1}=1\;\;z_4=2$ \\ 
$\rm{Cu}_{4} Mo_{6} Se_{8}$& 2 & $P\bar{1}$ & \icsdwebshort{171431} & Line & $z_{2,1}=1\;\;z_{2,2}=1\;\;z_{2,3}=1\;\;z_4=1$ \\ 
$\rm{K} (Os_{2} O_{6})$& 2 & $P\bar{1}$ & \icsdwebshort{419880} & Line & $z_{2,1}=1\;\;z_{2,2}=1\;\;z_{2,3}=0\;\;z_4=3$ \\ 
$\rm{Cu} (N (C N)_{2})_{2} (N H_{3})_{2}$& 2 & $P\bar{1}$ & \icsdwebshort{425593} & Line & $z_{2,1}=0\;\;z_{2,2}=1\;\;z_{2,3}=0\;\;z_4=1$ \\ 
$\rm{Rh}_{3} Ga_{5}$& 2 & $P\bar{1}$ & \icsdwebshort{240179} & Line & $z_{2,1}=1\;\;z_{2,2}=0\;\;z_{2,3}=0\;\;z_4=0$ \\ 
$(\rm{P} Cl_{4})_{2} (Mo_{2} Cl_{10})$& 2 & $P\bar{1}$ & \icsdwebshort{50441} & Line & $z_{2,1}=1\;\;z_{2,2}=0\;\;z_{2,3}=0\;\;z_4=3$ \\ 
$\rm{Pb} Pt_{2} O_{4}$& 2 & $P\bar{1}$ & \icsdwebshort{59657} & Line & $z_{2,1}=0\;\;z_{2,2}=1\;\;z_{2,3}=1\;\;z_4=0$ \\ 
$\rm{Sr} Mo_{6} S_{8}$& 2 & $P\bar{1}$ & \icsdwebshort{65941} & Line & $z_{2,1}=0\;\;z_{2,2}=0\;\;z_{2,3}=0\;\;z_4=3$ \\ 
$\rm{Li}_{5} (W_{2} O_{7})$& 2 & $P\bar{1}$ & \icsdwebshort{194127} & Line & $z_{2,1}=1\;\;z_{2,2}=1\;\;z_{2,3}=1\;\;z_4=2$ \\ 
$\rm{Ta}_{6} S$& 2 & $P\bar{1}$ & \icsdwebshort{202564} & Line & $z_{2,1}=0\;\;z_{2,2}=0\;\;z_{2,3}=1\;\;z_4=2$ \\ 
$\rm{Ca} Mo_{6} S_{8}$& 2 & $P\bar{1}$ & \icsdwebshort{619422} & Line & $z_{2,1}=0\;\;z_{2,2}=0\;\;z_{2,3}=0\;\;z_4=3$ \\ 
$\rm{Ca} Cu (Si O_{4}) (H_{2} O)$& 14 & $P2_1/c$ & \icsdwebshort{30926} & Line & $z_2'=1\;\;z_4=2$ \\ 
$\rm{Li} Cu C_{3} H_{4} P O_{5}$& 15 & $C2/c$ & \icsdwebshort{243640} & Line & $z_2'=1\;\;z_{2,2}=1\;\;z_{2,1}=1\;\;z_4=2$ \\ 
$\rm{Tc}_{2} P_{3}$& 2 & $P\bar{1}$ & \icsdwebshort{41017} & Line & $z_{2,1}=0\;\;z_{2,2}=0\;\;z_{2,3}=0\;\;z_4=1$ \\ 
$\rm{Hg} K$& 2 & $P\bar{1}$ & \icsdwebshort{104302} & Line & $z_{2,1}=0\;\;z_{2,2}=1\;\;z_{2,3}=0\;\;z_4=2$ \\ 
$\rm{Ag} Hg O_{2}$& 2 & $P\bar{1}$ & \icsdwebshort{670059} & Line & $z_{2,1}=0\;\;z_{2,2}=1\;\;z_{2,3}=1\;\;z_4=2$ \\ 
$\rm{Cu} Au O_{2}$& 12 & $C2/m$ & \icsdwebshort{670103} & Line & $z_2'=1\;\;z_{2,2}=1\;\;z_{2,1}=1\;\;z_4=2$ \\ 
$\rm{Cu} (W O_{4})^{*}$& 2 & $P\bar{1}$ & \icsdwebshort{69759} & Line & $z_{2,1}=0\;\;z_{2,2}=1\;\;z_{2,3}=1\;\;z_4=3$ \\ 
$\rm{Cu} (W O_{4})$& 2 & $P\bar{1}$ & \icsdwebshort{84558} & Line & $z_{2,1}=0\;\;z_{2,2}=1\;\;z_{2,3}=1\;\;z_4=1$ \\ 
$\rm{Fe} Se$& 2 & $P\bar{1}$ & \icsdwebshort{196300} & Line & $z_{2,1}=0\;\;z_{2,2}=0\;\;z_{2,3}=1\;\;z_4=1$ \\ 
$\rm{Fe} O_{3}$& 2 & $P\bar{1}$ & \icsdwebshort{671721} & Line & $z_{2,1}=0\;\;z_{2,2}=1\;\;z_{2,3}=1\;\;z_4=2$ \\ 
$\rm{Li}_{4} N_{1}$& 2 & $P\bar{1}$ & \icsdwebshort{675136} & Line & $z_{2,1}=0\;\;z_{2,2}=0\;\;z_{2,3}=0\;\;z_4=3$ \\ 
$\rm{C} Li_{2}$& 12 & $C2/m$ & \icsdwebshort{670915} & Line & $z_2'=1\;\;z_{2,2}=1\;\;z_{2,1}=1\;\;z_4=2$ \\ 
$\rm{Ti}_{11} Ni_{9} Pt_{4}$& 15 & $C2/c$ & \icsdwebshort{168948} & Line & $z_2'=1\;\;z_{2,2}=0\;\;z_{2,1}=0\;\;z_4=2$ \\ 
$\rm{V} O_{2}^{*}$& 2 & $P\bar{1}$ & \icsdwebshort{1502} & Line & $z_{2,1}=0\;\;z_{2,2}=0\;\;z_{2,3}=1\;\;z_4=1$ \\ 
$\rm{Cu} (Mo O_{4})$& 2 & $P\bar{1}$ & \icsdwebshort{39439} & Line & $z_{2,1}=1\;\;z_{2,2}=1\;\;z_{2,3}=0\;\;z_4=0$ \\ 
$\rm{Tc}_{2} As_{3}$& 2 & $P\bar{1}$ & \icsdwebshort{66662} & Line & $z_{2,1}=0\;\;z_{2,2}=0\;\;z_{2,3}=0\;\;z_4=1$ \\ 
$\rm{Mo}_{3} S_{4}$& 2 & $P\bar{1}$ & \icsdwebshort{237587} & Line & $z_{2,1}=0\;\;z_{2,2}=0\;\;z_{2,3}=1\;\;z_4=3$ \\ 
$(\rm{Cu} Cl_{3}) (Cu_{2} Te O_{3})$& 2 & $P\bar{1}$ & \icsdwebshort{431558} & Line & $z_{2,1}=0\;\;z_{2,2}=0\;\;z_{2,3}=0\;\;z_4=3$ \\ 
$\rm{V}_{6} O_{13}^{*}$& 12 & $C2/m$ & \icsdwebshort{50409} & Line & $z_2'=1\;\;z_{2,2}=1\;\;z_{2,1}=1\;\;z_4=2$ \\ 
$\rm{Sc}_{5} Cl_{8} N$& 12 & $C2/m$ & \icsdwebshort{60856} & Line & $z_2'=1\;\;z_{2,2}=1\;\;z_{2,1}=1\;\;z_4=2$ \\ 
$\rm{Li}_{4} N_{1}$& 2 & $P\bar{1}$ & \icsdwebshort{675127} & Line & $z_{2,1}=0\;\;z_{2,2}=0\;\;z_{2,3}=1\;\;z_4=0$ \\ 
$\rm{Cs} Hg$& 2 & $P\bar{1}$ & \icsdwebshort{62000} & Line & $z_{2,1}=0\;\;z_{2,2}=0\;\;z_{2,3}=1\;\;z_4=2$ \\ 
$\rm{Cs} Hg$& 2 & $P\bar{1}$ & \icsdwebshort{150974} & Line & $z_{2,1}=0\;\;z_{2,2}=0\;\;z_{2,3}=1\;\;z_4=0$ \\ 
$\rm{Ni} O O H$& 2 & $P\bar{1}$ & \icsdwebshort{674656} & Line & $z_{2,1}=1\;\;z_{2,2}=0\;\;z_{2,3}=0\;\;z_4=1$ \\ 
$\rm{Ti}_{5} O_{5}$& 12 & $C2/m$ & \icsdwebshort{56694} & Line & $z_2'=1\;\;z_{2,2}=1\;\;z_{2,1}=1\;\;z_4=2$ \\ 
$\rm{Ba} Bi O_{3}$& 12 & $C2/m$ & \icsdwebshort{61499} & Line & $z_2'=1\;\;z_{2,2}=1\;\;z_{2,1}=1\;\;z_4=2$ \\ 
$\rm{Ag} (V_{2} P_{2} O_{10})$& 14 & $P2_1/c$ & \icsdwebshort{73824} & Line & $z_2'=1\;\;z_4=2$ \\ 
$\rm{Hf}_{8} Ni_{21}$& 2 & $P\bar{1}$ & \icsdwebshort{2416} & Line & $z_{2,1}=0\;\;z_{2,2}=0\;\;z_{2,3}=0\;\;z_4=1$ \\ 
$\rm{Ni}_{21} Zr_{8}$& 2 & $P\bar{1}$ & \icsdwebshort{402864} & Line & $z_{2,1}=0\;\;z_{2,2}=1\;\;z_{2,3}=0\;\;z_4=2$ \\ 
$\rm{V} B_{3}$& 12 & $C2/m$ & \icsdwebshort{672902} & Line & $z_2'=0\;\;z_{2,2}=1\;\;z_{2,1}=1\;\;z_4=0$ \\ 
$\rm{Ni}_{21} Zr_{8}$& 2 & $P\bar{1}$ & \icsdwebshort{647148} & Line & $z_{2,1}=0\;\;z_{2,2}=1\;\;z_{2,3}=0\;\;z_4=0$ \\ 
$\rm{Ag} (S O_{4})$& 2 & $P\bar{1}$ & \icsdwebshort{290374} & Line & $z_{2,1}=0\;\;z_{2,2}=0\;\;z_{2,3}=0\;\;z_4=2$ \\ 
$\rm{Ba} Pd_{2} Bi_{2}$& 11 & $P2_1/m$ & \icsdwebshort{416299} & Line & $z_2'=1\;\;z_4=2$ \\ 
$\rm{Ni}_{11} As_{8}$& 15 & $C2/c$ & \icsdwebshort{164878} & Line & $z_2'=1\;\;z_{2,2}=1\;\;z_{2,1}=1\;\;z_4=2$ \\ 
$\rm{B} C_{5}$& 2 & $P\bar{1}$ & \icsdwebshort{166556} & Line & $z_{2,1}=0\;\;z_{2,2}=1\;\;z_{2,3}=1\;\;z_4=3$ \\ 
$\rm{Li}_{6} P$& 2 & $P\bar{1}$ & \icsdwebshort{673928} & Line & $z_{2,1}=0\;\;z_{2,2}=1\;\;z_{2,3}=0\;\;z_4=3$ \\ 
$\rm{Ni} O O H$& 58 & $Pnnm$ & \icsdwebshort{674650} & Line & $z_2'=1\;\;z_4=2\;\;z_{2}^{-}=1\;\;z_{2}^{+}=1$ \\ 
$\rm{Pb}_{2} Sr_{2} Y Cu_{3} O_{8}$& 2 & $P\bar{1}$ & \icsdwebshort{74154} & Line & $z_{2,1}=1\;\;z_{2,2}=1\;\;z_{2,3}=0\;\;z_4=1$ \\ 
$\rm{Cu} (W O_{4})$& 2 & $P\bar{1}$ & \icsdwebshort{169004} & Line & $z_{2,1}=1\;\;z_{2,2}=0\;\;z_{2,3}=0\;\;z_4=3$ \\ 
$\rm{Sc}_{3} Ni Si_{3}$& 12 & $C2/m$ & \icsdwebshort{48004} & Line & $z_2'=1\;\;z_{2,2}=0\;\;z_{2,1}=0\;\;z_4=2$ \\ 
$\rm{C} Li_{8}$& 2 & $P\bar{1}$ & \icsdwebshort{670922} & Line & $z_{2,1}=0\;\;z_{2,2}=0\;\;z_{2,3}=1\;\;z_4=0$ \\ 
$\rm{Sr} Pd_{2} Bi_{2}$& 11 & $P2_1/m$ & \icsdwebshort{416300} & Line & $z_2'=1\;\;z_4=2$ \\ 
$\rm{Na}_{6} Ti_{6} O_{13}$& 12 & $C2/m$ & \icsdwebshort{238302} & Line & $z_2'=1\;\;z_{2,2}=1\;\;z_{2,1}=1\;\;z_4=2$ \\ 
$\rm{Ag} (S O_{4})$& 15 & $C2/c$ & \icsdwebshort{290375} & Line & $z_2'=1\;\;z_{2,2}=0\;\;z_{2,1}=0\;\;z_4=2$ \\ 
$\rm{Au}_{2} Ca_{5}$& 15 & $C2/c$ & \icsdwebshort{58403} & Line & $z_2'=0\;\;z_{2,2}=1\;\;z_{2,1}=1\;\;z_4=0$ \\ 
$\rm{Pt}_{5} P_{2}$& 15 & $C2/c$ & \icsdwebshort{24327} & Line & $z_2'=0\;\;z_{2,2}=1\;\;z_{2,1}=1\;\;z_4=0$ \\ 
$\rm{Sb}_{2} Te_{3}$& 15 & $C2/c$ & \icsdwebshort{187497} & Line & $z_2'=1\;\;z_{2,2}=0\;\;z_{2,1}=0\;\;z_4=2$ \\ 
$\rm{Ni} S_{2}$& 60 & $Pbcn$ & \icsdwebshort{169571} & Line & $z_2'=1\;\;z_4=2$ \\ 
$\rm{Ni}_{10} Sn_{5} P_{3}$& 2 & $P\bar{1}$ & \icsdwebshort{59641} & Line & $z_{2,1}=0\;\;z_{2,2}=1\;\;z_{2,3}=1\;\;z_4=0$ \\ 
$\rm{Ti} Ni$& 2 & $P\bar{1}$ & \icsdwebshort{150629} & Line & $z_{2,1}=1\;\;z_{2,2}=0\;\;z_{2,3}=1\;\;z_4=3$ \\ 
$\rm{Cu}_{7} In_{3}$& 2 & $P\bar{1}$ & \icsdwebshort{429530} & Line & $z_{2,1}=0\;\;z_{2,2}=0\;\;z_{2,3}=0\;\;z_4=2$ \\ 
$\rm{W} Ni_{4} P_{16}$& 15 & $C2/c$ & \icsdwebshort{67920} & Line & $z_2'=1\;\;z_{2,2}=1\;\;z_{2,1}=1\;\;z_4=2$ \\ 
$\rm{Pd}_{13} Pb_{9}$& 15 & $C2/c$ & \icsdwebshort{648362} & Line & $z_2'=1\;\;z_{2,2}=0\;\;z_{2,1}=0\;\;z_4=2$ \\ 
$\rm{Pb}_{9} Pd_{13}$& 15 & $C2/c$ & \icsdwebshort{105593} & Line & $z_2'=1\;\;z_{2,2}=0\;\;z_{2,1}=0\;\;z_4=2$
\\
\hline
\end{longtable}
}

{\tiny
\begin{longtable}{|c|c|c|c|c|c|c|c|c|}
\caption[List of all nonmagnetic SEBR-SM unique materials without $f$ electrons]{List of all nonmagnetic unique materials without $f$ electrons that are classified as SEBR-SM in calculations performed w/o SOC.  In this table, a chemical formula with a $^{*}$ indicates that some (but not all) of the ICSDs associated to the same unique material (defined using its topological class with SOC) are classified as SEBR-SM in calculations for the same ICSD entries performed w/o SOC.  In this table, the topological indices (spinless SIs) at $E_{F}$ of the calculations performed w/o SOC are listed in the notation of Ref.~\onlinecite{ZhidaSemimetals}.  Further details and the physical interpretation of each spinless SI in this table are provided in Ref.~\onlinecite{ZhidaSemimetals}.}
\label{tb:list_nosoc_uniquematerials_nice_SEBR-SM}\\
\hline
Chem. formula & SG \# & SG symbol & ICSD & Line/Weyl & no-SOC topological indices \\
\hline  
$\rm{Tl}_{6} Te O_{12}$& 148 & $R\bar{3}$ & \icsdwebshort{37134} & Line & $z_{2,3}=0\;\;z_4=3\;\;z_{2,1}=0\;\;z_{2,2}=0$ \\ 
$\rm{Mo}_{6} Te_{8}$& 148 & $R\bar{3}$ & \icsdwebshort{59375} & Line & $z_{2,3}=1\;\;z_4=3\;\;z_{2,1}=1\;\;z_{2,2}=1$ \\ 
$\rm{Sc}_{7} Cl_{12} N$& 148 & $R\bar{3}$ & \icsdwebshort{201976} & Line & $z_{2,3}=1\;\;z_4=3\;\;z_{2,1}=1\;\;z_{2,2}=1$ \\ 
$\rm{Mo}_{3} Te_{4}$& 148 & $R\bar{3}$ & \icsdwebshort{644477} & Line & $z_{2,3}=1\;\;z_4=1\;\;z_{2,1}=1\;\;z_{2,2}=1$ \\ 
$\rm{Mo}_{3} S_{4}$& 148 & $R\bar{3}$ & \icsdwebshort{600385} & Line & $z_{2,3}=1\;\;z_4=2\;\;z_{2,1}=1\;\;z_{2,2}=1$ \\ 
$\rm{Ni} Ti_{3} S_{6}$& 148 & $R\bar{3}$ & \icsdwebshort{26312} & Line & $z_{2,3}=1\;\;z_4=3\;\;z_{2,1}=1\;\;z_{2,2}=1$ \\ 
$\rm{Mo}_{3} Se_{4}$& 148 & $R\bar{3}$ & \icsdwebshort{600386} & Line & $z_{2,3}=1\;\;z_4=1\;\;z_{2,1}=1\;\;z_{2,2}=1$ \\ 
$\rm{Mo}_{8} Ga_{40} C$& 148 & $R\bar{3}$ & \icsdwebshort{617918} & Line & $z_{2,3}=1\;\;z_4=3\;\;z_{2,1}=1\;\;z_{2,2}=1$ \\ 
$\rm{Ni}_{3} As In$& 148 & $R\bar{3}$ & \icsdwebshort{671511} & Line & $z_{2,3}=1\;\;z_4=3\;\;z_{2,1}=1\;\;z_{2,2}=1$ \\ 
$\rm{Mo} N_{2}$& 166 & $R\bar{3}m$ & \icsdwebshort{674587} & Line & $z_2'=1\;\;z_4=2$ \\ 
$\rm{Mo}_{6} S_{8}$& 148 & $R\bar{3}$ & \icsdwebshort{252376} & Line & $z_{2,3}=1\;\;z_4=2\;\;z_{2,1}=1\;\;z_{2,2}=1$ \\ 
$\rm{Si} O_{2}$& 162 & $P\bar{3}1m$ & \icsdwebshort{170552} & Line & $z_2'=1\;\;z_4=2$ \\ 
$\rm{Ca}_{3} Au_{4}$& 148 & $R\bar{3}$ & \icsdwebshort{54547} & Line & $z_{2,3}=1\;\;z_4=0\;\;z_{2,1}=1\;\;z_{2,2}=1$ \\ 
$\rm{Na}_{4} Cl_{3}$& 148 & $R\bar{3}$ & \icsdwebshort{672989} & Line & $z_{2,3}=1\;\;z_4=2\;\;z_{2,1}=1\;\;z_{2,2}=1$ \\ 
$\rm{Bi}_{2} Pt$& 147 & $P\bar{3}$ & \icsdwebshort{58847} & Line & $z_{2,3}=1\;\;z_4=2$ \\ 
$\rm{Ni}_{4} Ti_{3}$& 148 & $R\bar{3}$ & \icsdwebshort{105422} & Line & $z_{2,3}=1\;\;z_4=2\;\;z_{2,1}=1\;\;z_{2,2}=1$ \\ 
$\rm{Ge}_{9} Pd_{25}$& 147 & $P\bar{3}$ & \icsdwebshort{637543} & Line & $z_{2,3}=1\;\;z_4=3$
\\
\hline
\end{longtable}
}

\afterpage{\clearpage}

\section{Major Updates to the~\webTQC}\label{App:Website}

In this section, we will briefly overview the most significant changes implemented on the~\webTQC~(\webNoICSD) for this work, following its initial introduction in Ref.~\onlinecite{AndreiMaterials}.  To begin, in one of the most significant changes implemented on~\webNoICSD~for this work, we have processed the complete set of stoichiometric materials in the ICSD, leading to a current total set of~\TQCDBNbrICSDs~ICSD entries on~\webNoICSD~with electronic structures and topological data in the presence of SOC (see \supappref{App:VASP_appendix} for first-principles calculation details).  This represents an increase of roughly $\sim$ 50,000 ICSD entries from the previous content of~\webNoICSD~generated for Ref.~\onlinecite{AndreiMaterials}.  Beyond the large increase in the number of listed ICSD entries, we have also for this work implemented the following major changes on~\webNoICSD:
\begin{itemize}
\item{For the large majority of the ICSD entries accessible on~\webNoICSD~(\TQCDBNbrNoSOCICSDs~of~\TQCDBNbrICSDs~ICSD entries), the electronic structures and topological data are now available with and without the effects of SOC incorporated (see \supappref{App:TopologicalMaterialsNoSOC}).}
\item{The electronic band structure and density of states for each entry on~\webNoICSD~are displayed with dynamical zoom options.  Additionally, by hovering the pointer over a Bloch state in the electronic structure plot or over an energy in the density of states, contextual information can now be accessed, including the energy and the magnitude of the direct gaps above and below a given Bloch state.}
\item{We have for the first time computed the complete set of symmetry-indicated fragile bands above the core shell~\cite{JenFragile1,AshvinFragile,ZhidaFragile,KoreanFragileInversion} for each stoichiometric ICSD entry.  In each band structure plot, the largest sets of bands with cumulative symmetry-indicated fragile topology are labeled with dashed lines.  On each entry on~\webNoICSD~with fragile topological bands, we have also provided a table summarizing the details of the fragile bands.}
\item{We have introduced a ``Topological Data'' table for each entry on~\webNoICSD~that lists the symmetry-indicated stable and fragile topology of each connected grouping of bands, as well as the cumulative stable and fragile topology of the total set of bands occupied up to each insulating electronic filling as determined by band connectivity (see \supappref{App:TQCReview_appendix} and \supappref{App:TopoBands}).}
\item{For each electronic structure calculation on~\webNoICSD, we have made the VASP input POSCAR files and the ${\bf k}$-path files available for download through links provided below the crystallographic lattice data.}
\end{itemize}

In addition to the basic search options based on chemical composition and ICSD accession codes previously implemented for Ref.~\onlinecite{AndreiMaterials}, we have for this work implemented advanced materials search criteria (see Fig.~\ref{fig:websitesearch}), which include:
\begin{itemize}
\item{The SG of the ICSD entry.}
\item{The number of atoms in the unit cell (denoted as ``No. of elements'').}
\item{The number of valence electrons.}
\item{The direct and indirect band gaps at $E_{F}$.}
\item{The dimensionality (point, line, or plane) of the high-symmetry manifold(s) containing the crossing point(s) at $E_{F}$ if the bulk at $E_{F}$ is an enforced topological semimetal (ES- or ESFD-classified).}
\item{The number of times that the bands at $E_{F}$ cross the Fermi level along the plotted ${\bf k}$-path, subdivided by valence and conduction bands.}
\item{The values of the stable SIs~\cite{FuKaneMele,FuKaneInversion,AshvinIndicators,HOTIChen,TMDHOTI,HOTIBismuth,ChenTCI,AshvinTCI,SlagerSymmetry,MTQC} if the bulk is an NLC- or SEBR-classified topological (crystalline) insulator at $E_{F}$ when the effects of SOC are incorporated (see \supappref{App:TQCReview_appendix}).}
\end{itemize}
Importantly, users may simultaneously specify several advanced search criteria.  For example, users may search for weak TIs that contain the element Bi and have $z$-directed weak-index vectors~\cite{FuKaneMele,FuKaneInversion,AdyWeak,MooreBalentsWeak,ChenTCI} by entering Bi into the ``Compound Contains'' bar while specifying the stable SIs $Z_{2w,1}=0$, $Z_{2w,2}=0$, and $Z_{2w,3}=1$ and either $Z_{4}=0$ or $Z_{4}=2$ in the ``Topological Insulator Filters'' in the ``Advanced Search'' drop-down menu (see Fig.~\ref{fig:websitesearch}).  Further discussions regarding the differences between $Z_{4}=0$ and $Z_{4}=2$ weak TIs are provided in~\supappref{App:z4trivial}.

After performing a basic or advanced search, users are provided a list of matching ICSD entries sorted by the symmetry-indicated stable topological class at $E_{F}$ in the presence of SOC [``TI'' (NLC or SEBR), ``SM'' (ES or ESFD), or ``Trivial'' (LCEBR), see Fig.~\ref{fig:websitesearch}].  After choosing an ICSD entry from the search results, users are redirected to the specific page on~\webNoICSD~for the ICSD entry; in Fig.~\ref{fig:websitematerial} we show as an example the material page on~\webNoICSD~for the (obstructed) weak TI Bi$_3$STe$_2$ [\icsdweb{107587}, SG 164 ($P\bar{3}m1$)] (see \supappref{App:z4trivial} and Refs.~\onlinecite{FuKaneMele,FuKaneInversion,AdyWeak,MooreBalentsWeak,ChenTCI,WiederBarryCDW,CDWWeyl,JiabinCDW}).  On the page for each ICSD entry on~\webNoICSD, we provide data including:
\begin{itemize}
\item{The stoichiometric chemical formula, the SG, the crystal lattice parameters, the topological classification at $E_{F}$, a dropdown menu of the other ICSD entries associated to the same unique material (see \supappref{App:VASP_appendix}), the values of the stable SIs~\cite{FuKaneMele,FuKaneInversion,AshvinIndicators,HOTIChen,TMDHOTI,HOTIBismuth,ChenTCI,AshvinTCI,SlagerSymmetry,MTQC} if the ICSD entry is NLC- or SEBR-classified at $E_{F}$, and, if the bulk at $E_{F}$ is an enforced topological semimetal (ES- or ESFD-classified), whether the crossing point lies at a high-symmetry point or along a high-symmetry line or plane.}
\item{Interactive 3D plots of the real-space crystal structure and the first BZ.}
\item{Interactive plots of the electronic band structure and the density of states.}
\item{The direct gaps at $E_{F}$ at the high-symmetry ${\bf k}$ points, defined both by occupation as determined by VASP and by band index.}
\item{The ``Topological Data'' table discussed earlier in this section, which contains the individual and cumulative symmetry-indicated topology of each connected of bands as determined by the compatibility relations (\emph{i.e.} by band connectivity, see \supappref{App:TQCReview_appendix}).}
\item{Information for the largest sets of bands with cumulative symmetry-indicated fragile topology in the electronic structure~\cite{JenFragile1,AshvinFragile,ZhidaFragile,KoreanFragileInversion}.}
\item{If the topological classification at $E_{F}$ is ES, then we provide a table listing the high-symmetry manifolds along which enforced crossing points are required to appear.}  
\item{The stable SIs~\cite{FuKaneMele,FuKaneInversion,AshvinIndicators,HOTIChen,TMDHOTI,HOTIBismuth,ChenTCI,AshvinTCI,SlagerSymmetry,MTQC} of the lowest-symmetry nonmagnetic NLC and SEBR phases that can be realized by breaking crystal symmetries, computed by subduction onto lower-symmetry SGs as detailed in Refs.~\onlinecite{LuisSubduction,MTQC}.  There are two cases in which the ``Transitions upon symmetry lowering'' table is not displayed.  First, the ``Transitions upon symmetry lowering,'' table is not displayed if the bulk cannot be driven by infinitesimal symmetry-breaking into a nonmagnetic NLC or SEBR phase (\emph{i.e.} if the symmetry-indicated stable topology is trivial for all possible group-subgroup subductions).  Second, the ``Transitions upon symmetry lowering'' table is also absent for ESFD materials that are filling-enforced (semi)metals [see Refs.~\onlinecite{WPVZ,WiederLayers,SteveMagnet}] with accidental degeneracies at $E_{F}$ [defined in~\supappref{App:VASP2Trace}], which prevent the unambiguous identification of subduced small coreps.}
\end{itemize}

\begin{figure}[ht]
\centering
\includegraphics[width=0.7\textwidth,angle=0]{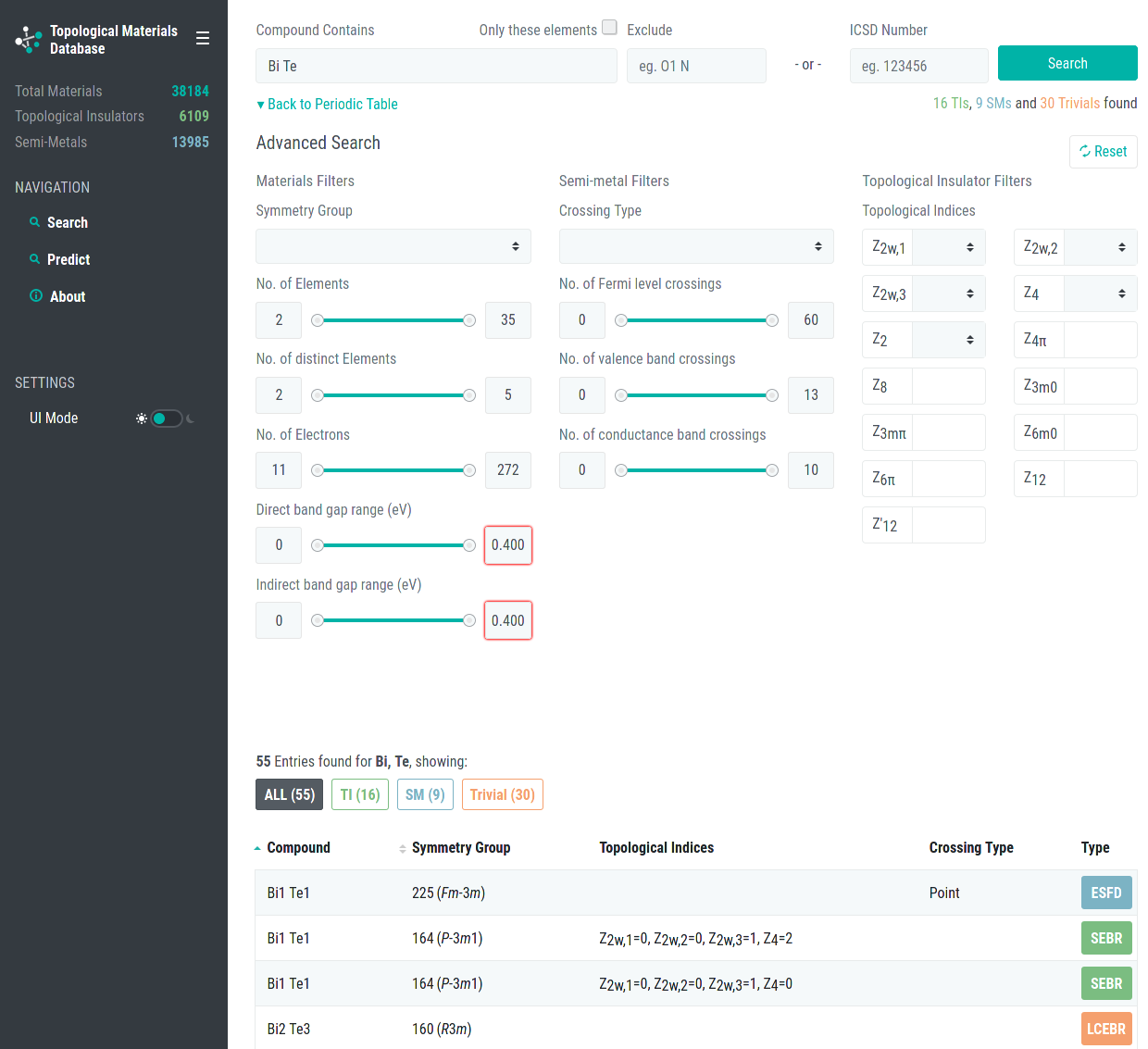}
\caption{A screenshot of the search interface of the updated~\webTQC~(\webNoICSD).  The top three search bars allow basic search options based on chemical composition and ICSD accession codes.  For this work, we have additionally implemented the ``Advanced Search'' features displayed below the ``Back to Periodic Table'' button, which are further detailed in this section.}
\label{fig:websitesearch}
\end{figure}

\begin{figure}[ht]
\centering
\includegraphics[width=0.48\textwidth,angle=0]{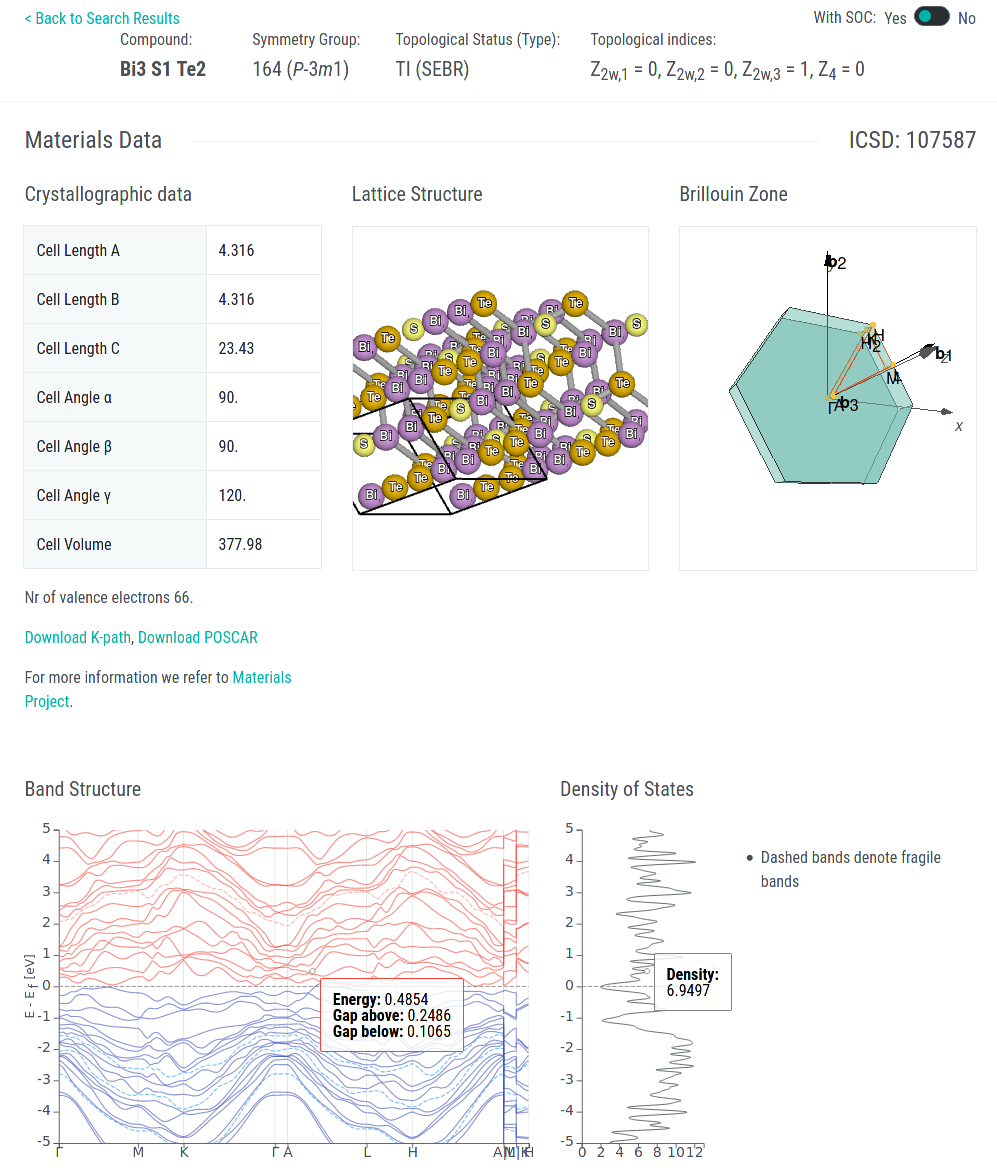}
\includegraphics[width=0.48\textwidth,angle=0]{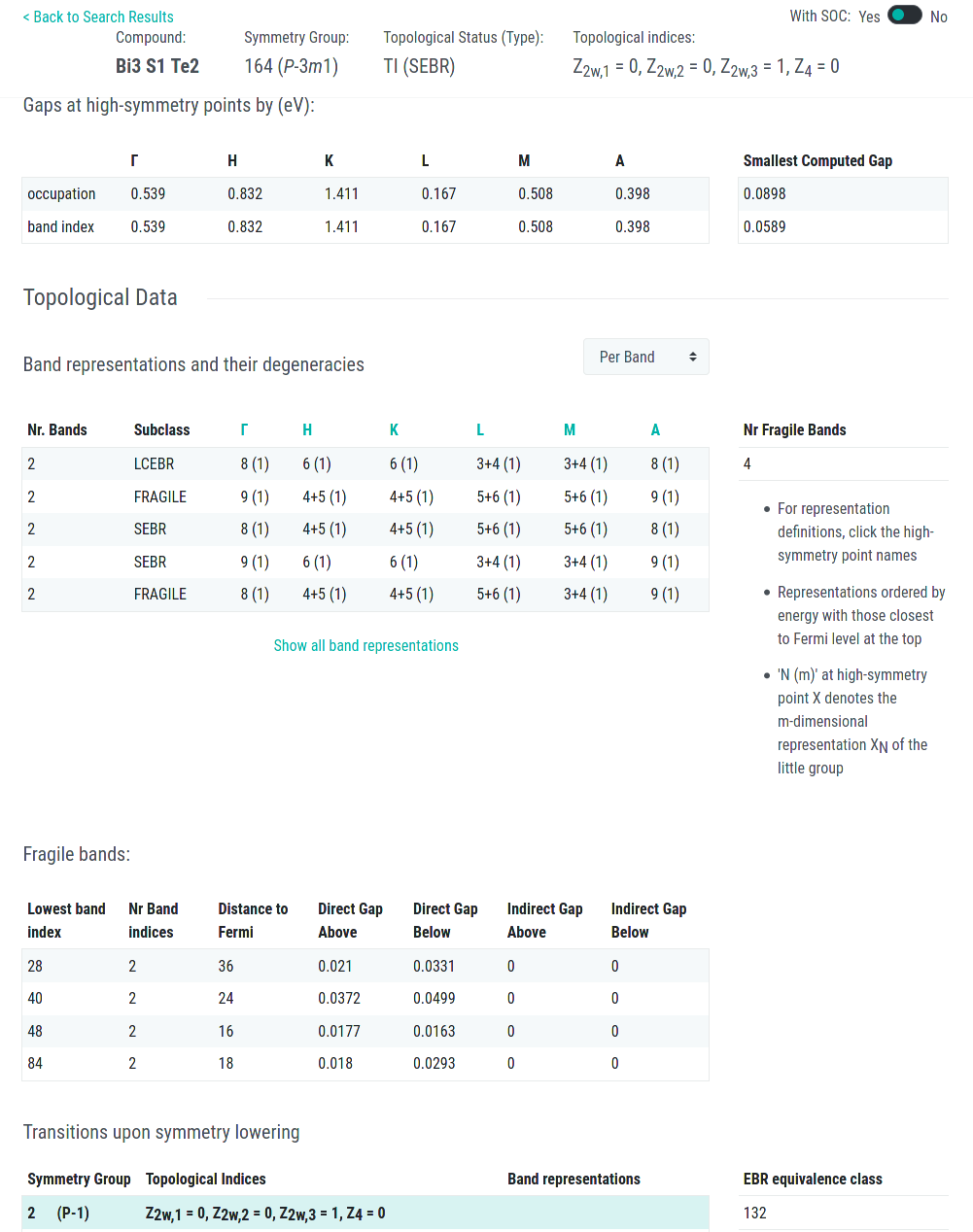}
\caption{An ICSD entry page on~\webNoICSD.  We have chosen as an example the (obstructed) weak TI Bi$_3$STe$_2$ [\icsdweb{107587}, SG 164 ($P\bar{3}m1$)] (see \supappref{App:z4trivial} and Refs.~\onlinecite{FuKaneMele,FuKaneInversion,AdyWeak,MooreBalentsWeak,ChenTCI,WiederBarryCDW,CDWWeyl,JiabinCDW}).  {\it Left panel} We first list the chemical formula, SG, topological classification at $E_{F}$, the stable SIs if the topological classification at $E_{F}$ is NLC (-SM) or SEBR (-SM) [see \supappref{App:TQCReview_appendix} and \supappref{App:TopologicalMaterialsNoSOC}], and, if the bulk at $E_{F}$ is an enforced topological semimetal (ES- or ESFD-classified), whether the crossing point lies at a high-symmetry point or along a high-symmetry line or plane.  On the top-left corner, there is a switch that allows users to toggle between the electronic structures and topological data for the same ICSD entry computed with and without incorporating the effects of SOC.  We then display the ``Materials Data'' for the ICSD entry, which consists of the ICSD accession code, a dropdown list of all other ICSD entries associated to the same unique material (see \supappref{App:VASP_appendix}), the crystallographic lattice data, interactive plots of the real-space unit cell and the first BZ, and the number of valence electrons.  For ICSD entries that are also listed on the~\webmaterialsproject~(MP) \cite{MaterialsProject}, we additionally provide a link to the corresponding page on the MP.  Lastly, below the Materials Data, we provide interactive band structure and density-of-states plots.  {\it Right panel} A continuation of the example page on~\webNoICSD~shown in the left panel.  We first list the direct gaps at $E_{F}$ at the high-symmetry ${\bf k}$ points, defined both by occupation as determined by VASP and by band index.  We then list the ``Topological Data'' table discussed in this section.  Next, we list information for the largest sets of bands with cumulative symmetry-indicated fragile topology in the electronic structure~\cite{JenFragile1,AshvinFragile,ZhidaFragile,KoreanFragileInversion}.  Next, if the topological classification at $E_{F}$ is ES, then we provide a table listing the high-symmetry manifolds along which enforced crossing points are required to appear.  Lastly, using subduction onto lower-symmetry SGs as detailed in Refs.~\onlinecite{LuisSubduction,MTQC}, we compute the stable SIs~\cite{FuKaneMele,FuKaneInversion,AshvinIndicators,HOTIChen,TMDHOTI,HOTIBismuth,ChenTCI,AshvinTCI,SlagerSymmetry,MTQC} of the lowest-symmetry nonmagnetic NLC and SEBR phases that can be realized by breaking crystal symmetries.  If the ``Transitions upon symmetry lowering'' table is not displayed, then the bulk cannot be driven by infinitesimal symmetry-breaking into a nonmagnetic NLC or SEBR phase, or is a filling-enforced, ESFD-classified (semi)metal [see Refs.~\onlinecite{WPVZ,WiederLayers,SteveMagnet}] with at least one accidental degeneracy at $E_{F}$ [defined in~\supappref{App:VASP2Trace}].}
\label{fig:websitematerial}
\end{figure}

Finally, many of the ICSD entries listed on~\webNoICSD~are also analyzed on the~\webmaterialsproject~(MP)~\cite{MaterialsProject}.  While the~\webTQC~introduced in Ref.~\onlinecite{AndreiMaterials} and expanded in this work focuses on the electronic band topology of a given material, the MP provides a general analysis focused on magnetic, structural, and chemical material properties.  For materials that are listed both on~\webNoICSD~and on the MP, we provide on each ICSD entry page on~\webNoICSD~a link to the corresponding MP page.  Additionally, for materials that are listed both on~\webNoICSD~and on the MP, the MP entry page extracts and reproduces the topological classification at $E_{F}$ with SOC from the corresponding ICSD entry page on~\webNoICSD.



\afterpage{\clearpage}

\section{Statistics for all of the Topological Bands and Band Connectivity in the ICSD}\label{App:TopoBands}

The ICSD contains $\TQCDTotICSDs$ entries~\cite{ICSD}.  In this work, we have analyzed the complete set of $\TQCDstoichiometric$ ICSD entries with stoichiometric chemical formulas and processable (non-corrupt) CIF structure files, achieving convergent electronic-structure calculations for $\TQCDBNbrICSDs$ ICSD entries with SOC and $\TQCDBNbrNoSOCICSDs$ ICSD entries w/o SOC, corresponding to $\TQCDBNbrUniqueMaterials$ unique materials (see \supappref{App:VASP_appendix} for calculation details and for the definition of unique materials employed in this work).  In this section, we will provide a detailed statistical breakdown of the symmetry-indicated topological classification of all isolated sets of bands in all of the ICSD entries analyzed in this work as determined by band connectivity through TQC (see~\supappref{App:TQCReview_appendix}).  This includes diagnoses of both symmetry-indicated stable~\cite{FuKaneMele,FuKaneInversion,AshvinIndicators,HOTIChen,TMDHOTI,HOTIBismuth,ChenTCI,AshvinTCI,SlagerSymmetry,MTQC} and fragile~\cite{JenFragile1,ZhidaFragile,KoreanFragileInversion} topology across all of the bands included in our first-principles calculations, as well as statistics for and examples of materials with large band connectivities.  First, in~\supappref{sc:all_mat}, we will provide detailed statistics for the stable topological classification at $E_{F}$ with and w/o SOC of all of the stoichiometric materials in the ICSD.  Then, in~\supappref{sc:all_band}, we will provide detailed statistics for the stable and fragile topological classification of all of the isolated bands (as defined by band connectivity, see~\supappref{App:TQCReview_appendix}) in the ICSD.  We will also in~\supappref{sc:all_band} provide statistics for and examples of materials with extremely large band connectivities.

\subsection{Topological Classification at the Fermi Level for all of the Stoichiometric Unique Materials in the ICSD with and w/o SOC}
\label{sc:all_mat}

In this section, we will provide statistics for the topological classification at the Fermi level for the unique materials studied in this work (defined in the main text and in \supappref{App:VASP_appendix}).  Specifically, in Table~\ref{tb:statistics_soc_uniquematerials} (Table~\ref{tb:statistics_nosoc_uniquematerials}), we list the number of materials whose valence bands are in total classified by one of the four stable topological classes introduced in Ref.~\onlinecite{AndreiMaterials}: NLC (-SM), SEBR (-SM), ES, and ESFD.  We note that, using the symmetry-based indicators for fragile topology introduced in Refs.~\onlinecite{JenFragile1,ZhidaFragile,KoreanFragileInversion}, we did not identify any ICSD entries in which the total set of occupied (valence) bands was fragile topological with or w/o SOC (though we did identify energetically-isolated fragile bands near the Fermi energy in many materials, see \supappref{sc:all_band} and~\supappref{App:fragileBands}).

{\tiny
\begin{longtable}{|c|c|c|c|c|c|c|c|c|c|}
\caption[Number of unique materials in each topological class per SG with SOC]{The number of unique materials in each of the topological classes (as defined in the main text and in Ref.~\onlinecite{AndreiMaterials}) in calculations performed with SOC.  In this table, the topology of a material is taken to be the total (cumulative) topology of all of the bands occupied up to $E_{F}$.  In order, the columns in this table list the SG, the number of unique materials in the SG, the number of TIs and TCIs (defined as either NLC or SEBR) in the SG, the number of semimetals (SMs) in the SG (defined as either ES or ESFD), the number of unique materials with trivial symmetry-indicted topology (LCEBR) in the SG, the number of NLC-classified TIs and TCIs in the SG, the number of SEBR-classified TIs and TCIs in the SG, the number of ES-classified SMs in the SG, and the number of ESFD-classified SMs in the SG.  For all of the topological classification categories in the table, each number of materials is accompanied by the percentage of materials in the SG in the topological class.  When there are no materials in the topological class, the table entry is marked with a dashed horizontal line.
\label{tb:statistics_soc_uniquematerials}}\\
\hline
SG  & \# Mat. & TIs & SMs & Trivial & NLC  &  SEBR & ES & ESFD \\
\hline  
1 & 151 & ---  & 9 {\tiny(6.0\%)} & 142 {\tiny(94.0\%)} & ---  & ---  & ---  & 9 {\tiny(6.0\%)} \\ 
2 & 1574 & 182 {\tiny(11.6\%)} & 116 {\tiny(7.4\%)} & 1276 {\tiny(81.1\%)} & 182 {\tiny(11.6\%)} & ---  & ---  & 116 {\tiny(7.4\%)} \\ 
3 & 13 & ---  & ---  & 13 {\tiny(100\%)} & ---  & ---  & ---  & ---  \\ 
4 & 228 & ---  & 8 {\tiny(3.5\%)} & 220 {\tiny(96.5\%)} & ---  & ---  & 8 {\tiny(3.5\%)} & ---  \\ 
5 & 156 & ---  & 15 {\tiny(9.6\%)} & 141 {\tiny(90.4\%)} & ---  & ---  & ---  & 15 {\tiny(9.6\%)} \\ 
6 & 28 & ---  & 1 {\tiny(3.6\%)} & 27 {\tiny(96.4\%)} & ---  & ---  & ---  & 1 {\tiny(3.6\%)} \\
\hline
\end{longtable}
}

{\tiny
\begin{longtable}{|c|c|c|c|c|c|c|c|c|}
\caption[Number of materials in each topological class per SG w/o SOC]{The number of unique materials in each of the topological classes (as defined in the main text and in \supappref{App:TopologicalMaterialsNoSOC}) in calculations performed w/o SOC.  In this table, the topology of a material is taken to be the total (cumulative) topology of all of the bands occupied up to $E_{F}$.  In order, the columns in this table list the SG, the number of unique materials in the SG (which we emphasize to be defined using the topological classification with SOC, see \supappref{App:VASP_appendix}), the number of semimetals (SMs) in the SG (defined as either NLC-SM, SEBR-SM, ES, or ESFD), the number of unique materials with trivial symmetry-indicted topology (LCEBR) in the SG, the number of NLC-SM-classified SMs in the SG, the number of SEBR-SM-classified SMs in the SG, the number of ES-classified SMs in the SG, and the number of ESFD-classified SMs in the SG.  For all of the topological classification categories in the table, each number of materials is accompanied by the percentage of materials in the SG in the topological class.  When there are no materials in the topological class, the table entry is marked with a dashed horizontal line.
\label{tb:statistics_nosoc_uniquematerials}}\\
\hline
SG  & \# Mat. & SMs & Trivial & NLC-SM  &  SEBR-SM & ES & ESFD \\
\hline  
1 & 151 & 9 {\tiny(6.0\%)} & 142 {\tiny(94.0\%)} & ---  & ---  & ---  & 9 {\tiny(6.0\%)} \\ 
2 & 1466 & 277 {\tiny(18.9\%)} & 1189 {\tiny(81.1\%)} & 165 {\tiny(11.3\%)} & ---  & ---  & 112 {\tiny(7.6\%)} \\ 
3 & 10 & 7 {\tiny(70.0\%)} & 3 {\tiny(30.0\%)} & ---  & ---  & 7 {\tiny(70.0\%)} & ---  \\ 
4 & 219 & 20 {\tiny(9.1\%)} & 199 {\tiny(90.9\%)} & ---  & ---  & 14 {\tiny(6.4\%)} & 6 {\tiny(2.7\%)} \\ 
5 & 148 & 31 {\tiny(20.9\%)} & 117 {\tiny(79.0\%)} & ---  & ---  & 19 {\tiny(12.8\%)} & 12 {\tiny(8.1\%)} \\
\hline
\end{longtable}
}

\clearpage

\subsection{Topological Classification and Connectivity of all of the Bands in all of the Stoichiometric Unique Materials in the ICSD with and w/o SOC}
\label{sc:all_band}

In this section, we will provide statistics on the symmetry-indicated stable~\cite{FuKaneMele,FuKaneInversion,AshvinIndicators,HOTIChen,TMDHOTI,HOTIBismuth,ChenTCI,AshvinTCI,SlagerSymmetry} and fragile~\cite{JenFragile1,ZhidaFragile,KoreanFragileInversion} topology of \emph{all} of the isolated bands in all of the stoichiometric unique materials in the ICSD, where in the calculations performed for this work, we have taken the conduction manifold to contain at least as many states as the valence manifold (\emph{i.e.} there are $\geq 2N_{e}$ total bands, see \supappref{App:VASP_appendix} for further calculation details).  We have also compiled statistics for materials with large or complete band connectivities [defined as \emph{supermetallic} (SMetal) materials in the text below].  For the statistics compiled in this section for symmetry-indicated insulating topology at varying electronic fillings, we emphasize that the isolated bands (as determined by band connectivity through the compatibility relations in TQC, see~\supappref{App:TQCReview_appendix}) can never be classified as ESFD- or ES-classified semimetals, which by definition require a connected group of bands to be partially occupied at a high-symmetry point or along a high-symmetry line or plane~\cite{AndreiMaterials}.  Consequently, all of the isolated topological bands analyzed in this section are labeled as either NLC, SEBR, LCEBR, or FRAGILE.  We emphasize that in the absence of SOC, all symmetry-indicated topological (NLC-SM- and SEBR-SM-classified) bands in nonmagnetic materials correspond to topological semimetal phases, and are thus connected to other bands at higher or lower energies by nodal points in the BZ interior.  However, NLC-SM and SEBR-SM bands are still considered to be isolated by band connectivity, as they satisfy the insulating compatibility relations, which are only defined along high-symmetry BZ lines and planes (see~\supappref{App:TopologicalMaterialsNoSOC}~and Ref.~\onlinecite{ZhidaSemimetals}).

To begin, in this work, we are uniquely able to analyze the symmetry-indicated topology of bands away from the Fermi level by using new features of the~\checktopmat~program, which we have detailed in \supappref{App:Check_Topo_appendix}.  In Table~\ref{table-CTP}, we show the output of~\checktopmat~for the candidate ES compound Ni$_2$SnZr [\icsdweb{105383}, SG 225 ($Fm\bar{3}m$)].  Though Ni$_2$SnZr is an ES-classified SM at $E_{F}$ with SOC, none of the entries in Table~\ref{table-CTP} are labeled as ``ES,'' because the symmetry-indicated topology of energetically isolated groups of bands can only be meaningfully defined by taking the connected bands to be fully occupied (whereas partial occupancy is required to realize an ES-classified semimetal).

Next, in Table~\ref{tb:statistics_soc_bands_uniquematerials}, we list the number of isolated sets of bands in each topological class per SG with SOC as determined by band connectivity.  Remarkably, we find that nearly 2/3 of all of the bands in all of the stoichiometric unique materials in the ICSD exhibit symmetry-indicated stable (NLC or SEBR) topology when incorporating the effects of SOC. Then, in Table~\ref{tb:statistics_soc_band_types_uniquematerials_bandtypes}, we list the number of unique materials in each SG with at least one symmetry-indicated stable or fragile topological set of bands with SOC.  We further find that an overwhelming~\TQCDBNbrTopoBandPercent~of the stoichiometric unique materials in the ICSD have at least one stable or fragile topological band when the effects of SOC are incorporated.  For completeness, we next respectively provide in Tables~\ref{tb:statistics_nosoc_bands_uniquematerials} and~\ref{tb:statistics_nosoc_band_types_uniquematerials_bandtypes} the number of isolated sets of bands in each topological class per SG w/o SOC as determined by band connectivity, and the number of unique materials in each SG with at least one symmetry-indicated stable or fragile topological set of bands w/o SOC.

Lastly, in this work we have discovered the existence of SMetal materials, which realize an extreme limit of band connectivity in which \emph{all} of the bands included our VASP calculations are connected (\emph{i.e.} materials in which there does not exist an LCEBR-, NLC-, or SEBR-classified gap at any integer electronic filling, including filling the complete set of $\sim 2N_{e}$ bands in the VASP calculation, see \supappref{App:VASP_appendix} for calculation details).  As previously discussed in~\supappref{App:TopologicalMaterialsNoSOC},~\webNoICSD~contains~$\TQCDBNbrSMetal$~unique SMetal materials with SOC (Table~\ref{tb:statistics_soc_band_types_uniquematerials_bandtypes}), whereas there are~\emph{$\TQCDBNbrNoSOCSMetal$}~unique SMetal materials w/o SOC (Table~\ref{tb:statistics_nosoc_band_types_uniquematerials_bandtypes}), representing a hundred-fold increase w/o SOC in the number of materials with fully-connected valence bands up to the energy (filling) cutoff of $\sim 2N_{e}$ imposed in our VASP calculations (see~\supappref{App:VASP_appendix}).  We have in this work also discovered the existence of materials in which a very large number of bands ($\geq\TQCDBHighConnectivityNbrBandsThreshold$) are connected, even in the presence of SOC.  In Table~\ref{tb:all_high_connectivity_uniquecompounds_nice}, we list the nonmagnetic unique materials without $f$ electrons that exhibit connected groupings of bands with total dimension greater than or equal to $\TQCDBHighConnectivityNbrBandsThreshold$ in the calculations performed with SOC.

\begin{table}[h]
\begin{tabular}{cccccccccccc}
\hline 
$\Gamma$ & X &L &W & dim & top. type  & ind/band & filling $\nu$ & top. type/all & ind/all  \\
\hline
\hline                                                                                                               
          -GM6(2) &         -X6(2)   &       -L8(2)    &      -W7(2) &   2  & TRIVIAL  &  0 0 0   &    2 & TRIVIAL  &  0 0 0  \\
\hline                                                                                                                        \\
          -GM8(2) &         -X8(2)   &       -L9(2)    &      -W6(2) &   2  & TRIVIAL  &  0 0 0   &    4 & TRIVIAL  &  0 0 0   \\
\hline                                                                                                                         \\
         -GM11(4) &         -X7(2)   &       -L9(2)    &      -W7(2) &   4  &   SEBR   & 2 0 6    &   8  &   SEBR   & 2 0 6   \\
                  &         -X6(2)   &    -L6-L7(2)    &      -W6(2) &      &          &          &      &          &         \\
\hline                                                                                             
         -GM10(4) &         -X8(2)   &       -L9(2)    &      -W6(2) &   4  &   SEBR   & 2 0 2    &  12  & TRIVIAL  &  0 0 0  \\
                  &         -X9(2)   &    -L6-L7(2)    &      -W7(2) &      &          &          &      &          &         \\
\hline                                                                                                                        \\
         -GM10(4) &         -X9(2)   &       -L9(2)    &      -W6(2) &   6  &   SEBR   & 3 1 7    &  18  &   SEBR   & 3 1 7  \\
          -GM7(2) &         -X7(2)   &    -L4-L5(2)    &      -W7(2) &      &          &          &      &          &        \\
                  &         -X6(2)   &       -L8(2)    &      -W6(2) &      &          &          &      &          &         \\
\hline                                                                                                                        \\
         -GM11(4) &         -X9(2)   &       -L8(2)    &      -W7(2) &   4  & TRIVIAL  &  0 0 0   &   22 &    SEBR  &  3 1 7  \\
                  &         -X8(2)   &    -L4-L5(2)    &      -W6(2) &      &          &          &      &          &          \\
\hline                                                                                                                        \\
         -GM10(4) &         -X9(2)   &       -L8(2)    &      -W7(2) &   4  &   SEBR   & 3 1 3    &  26  &   SEBR   & 2 0 2   \\
                  &         -X6(2)   &    -L6-L7(2)    &      -W7(2) &      &          &          &      &          &         \\
\hline                                                                                                                         \\
          -GM7(2) &         -X8(2)   &       -L9(2)    &      -W6(2) &  14  &   SEBR   & 2 0 6    &  40  &  TRIVIAL &   0 0 0  \\
         -GM10(4) &         -X7(2)   &       -L9(2)    &      -W6(2) &      &          &          &      &          &         \\
          -GM8(2) &         -X7(2)   &       -L8(2)    &      -W7(2) &      &          &          &      &          &         \\
         -GM11(4) &         -X6(2)   &       -L9(2)    &      -W6(2) &      &          &          &      &          &         \\
          -GM6(2) &         -X7(2)   &    -L6-L7(2)    &      -W7(2) &      &          &          &      &          &         \\
                  &         -X6(2)   &       -L8(2)    &      -W7(2) &      &          &          &      &          &         \\
                  &         -X6(2)   &    -L4-L5(2)    &      -W7(2) &      &          &          &      &          &         \\
\hline                                                                                                                        \\
          -GM9(2) &         -X7(2)   &       -L9(2)    &      -W6(2) &   2  &   SEBR   & 1 1 1    &  42  &   SEBR   & 1 1 1   \\
\hline                                                                                                                        \\
          -GM8(2) &         -X8(2)   &    -L6-L7(2)    &      -W7(2) &   6  &   SEBR   & 0 0 4    &  48  &   SEBR   & 1 1 5    \\
         -GM11(4) &         -X9(2)   &       -L9(2)    &      -W6(2) &      &          &          &      &          &          \\
                  &         -X8(2)   &       -L8(2)    &      -W6(2) &      &          &          &      &          &          \\
\hline                                                                                                                         \\
         -GM10(4) &         -X6(2)   &    -L4-L5(2)    &      -W7(2) &   4  & TRIVIAL  &  0 0 0   &   52 &    SEBR  &  1 1 5   \\
                  &         -X7(2)   &       -L8(2)    &      -W6(2) &      &          &          &      &          &          \\

\hline
\end{tabular}
\caption[Topology of all of the bands in the ES compound Ni$_2$SnZr ]{A typical table generated by~\checktopmat~(see \supappref{App:Check_Topo_appendix}).  Here, we have used as an example the ES-classified compound Ni$_2$SnZr [\icsdweb{105383}, SG 225 ($Fm\bar{3}m$)].  In this table, the horizontal lines indicate the electronic fillings $\nu$ at which the occupied bands satisfy the compatibility relations.  Columns $\Gamma$, $X$, $L$, and $W$ contain the coreps at each maximal ${\bf k}$ point for each energetically isolated set of bands that satisfy the compatibility relations.  Column ``dim'' shows the dimension of each set of bands, Column ``top. type'' indicates the topology of the isolated bands [which can either be LCEBR, FRAGILE, NLC, or SEBR], and Column ``ind/band'' contains the minimal ($Z_{8}$) and non-minimal ($Z_{4}=Z_{8}\text{ mod }4$, $Z_{2}=Z_{8}\text{ mod }2$) stable SIs~\cite{ChenTCI,MTQC} of SG 225 ($Fm\bar{3}m$), given in the order ($Z_{4},Z_{2},Z_{8}$).  The remaining three columns contain the cumulative information of several combined sets of energetically isolated bands that are filled up to a total number of valence electrons specified by the column ``filling $\nu$.''  Column ``top. type/all'' provides the cumulative topology (LCEBR, FRAGILE, NLC, or SEBR) at the gap specified by the electronic filling in the ``filling $\nu$'' column, and Column ``ind/all'' provides the cumulative stable SIs of all of the filled bands up to the same gap.  Hence, the values of ``ind/all'' at each filling $\nu$ listed in this table can be obtained by summing the values in the ``ind/band'' column up to the same filling $\nu$.  We emphasize that even though Ni$_2$SnZr is ES-classified at the Fermi level (there are 28 valence electrons in Ni$_2$SnZr), none of the entries in this table are labeled as ``ES,'' because the symmetry-indicated topology of energetically isolated groups of bands can only be meaningfully defined by taking the connected bands to be fully occupied (whereas partial occupancy is required to realize an ES-classified semimetal).}
\label{table-CTP}
\end{table}

{\tiny
\begin{longtable}{|c|c|c|c|c|c|c|}
\caption[Number of topological bands per SG among the unique materials with SOC]{In this table, we list the number of isolated sets of bands with SOC as determined by band connectivity (see~\supappref{App:TQCReview_appendix}) in unique materials in each SG in each topological class.  In order, the columns in this table list the SG, the number of unique materials in the SG, the number of NLC-classified sets of bands, the number of SEBR-classified sets of bands, the number of LCEBR-classified sets of bands, the number of sets of connected fragile bands, and the total number of sets of isolated (connected) bands.   For all of the topological classification categories in the table, each number of bands is accompanied by the percentage of bands in the SG in the topological class.  When there are no bands in the topological class, the table entry is marked with a dashed horizontal line.
\label{tb:statistics_soc_bands_uniquematerials}}\\
\hline
SG  & \# Mat. & \# NLC bands & \# SEBR bands & \# LCEBR bands  & \# Fragile bands & Total \# isolated bands \\
\hline  
1& 151 & ---  & ---  & 13263 {\tiny(100\%)} & ---  & 13263 \\ 
2& 1574 & 191942 {\tiny(81.2\%)} & ---  & 44364 {\tiny(18.8\%)} & ---  & 236306 \\ 
3& 13 & ---  & ---  & 1708 {\tiny(100\%)} & ---  & 1708 \\ 
4& 228 & ---  & ---  & 15934 {\tiny(100\%)} & ---  & 15934 \\ 
5& 156 & ---  & ---  & 16121 {\tiny(100\%)} & ---  & 16121 \\ 
6& 28 & ---  & ---  & 2232 {\tiny(100\%)} & ---  & 2232 \\
\hline
\end{longtable}
}

{\tiny
\begin{longtable}{|c|c|c|c|c|c|c|c|c|c|c|}
\caption[Number of unique materials per SG with topological bands at any filling with SOC]{In this table, we list the number of unique materials that contain at least some topological features with SOC in each SG.  In order, the columns in this table list the SG, the number of unique materials in the SG, the number of unique materials with either at least one stable or fragile band in the spectrum or symmetry-enforced crossing points at $E_{F}$ (\emph{i.e.} materials that are either ES- or ESFD-classified semimetals or exhibit stable or fragile topological bands at any filling).  From left to right, in the following columns we list, per SG, the number of unique materials with at least one stable or fragile topological connected set of bands at any filling, at least one fragile topological set of bands at any filling, at least one NLC-classified stable topological band at any filling, at least one SEBR-classified stable topological band at any filling, and at least one stable topological set of bands at any filling.  In the remaining three columns, we list the number of materials that do not contain \emph{any} symmetry-indicated stable or fragile topological bands (STriv), the number of supermetallic (SMetal) materials in which \emph{all} of the bands in at least one of the ICSDs associated to the unique material are \emph{fully} connected, and the number of unique materials in which all bands are stable topological [\emph{i.e.} supertopological (STopo) materials, see \supappref{App:Supertopological}].  For all of the topological classification categories in the table, each number of materials is accompanied by the percentage of materials in the SG in the topological class.  When there are no materials in the topological class, the table entry is marked with a dashed horizontal line.
\label{tb:statistics_soc_band_types_uniquematerials_bandtypes}}\\
\hline
\end{longtable}
}

{\tiny
\begin{longtable}{|c|c|c|c|c|c|c|}
\caption[Number of topological bands per SG w/o SOC among the unique materials]{In this table, we list the number of isolated sets of bands w/o SOC as determined by band connectivity (see~\supappref{App:TQCReview_appendix}) in unique materials in each SG in each topological class.  In order, the columns in this table list the SG, the number of unique materials in the SG, the number of NLC-classified sets of bands, the number of SEBR-classified sets of bands, the number of LCEBR-classified sets of bands, the number of sets of connected fragile bands, and the total number of sets of isolated (connected) bands.   For all of the topological classification categories in the table, each number of bands is accompanied by the percentage of bands in the SG in the topological class.  When there are no bands in the topological class, the table entry is marked with a dashed horizontal line.
\label{tb:statistics_nosoc_bands_uniquematerials}}\\
\hline
SG  & \# Mat. & \# NLC bands & \# SEBR bands & \# LCEBR bands  & \# Fragile bands & Total \# isolated bands \\
\hline  
1& 151 & ---  & ---  & 13271 {\tiny(100\%)} & ---  & 13271 \\ 
2& 1466 & 181126 {\tiny(81.7\%)} & ---  & 40496 {\tiny(18.3\%)} & ---  & 221622 \\ 
3& 10 & 89 {\tiny(29.0\%)} & ---  & 218 {\tiny(71.0\%)} & ---  & 307 \\ 
4& 219 & ---  & ---  & 9057 {\tiny(100\%)} & ---  & 9057 \\ 
5& 148 & ---  & ---  & 7054 {\tiny(100\%)} & ---  & 7054 \\ 
6& 28 & ---  & ---  & 692 {\tiny(100\%)} & ---  & 692 \\
\hline
\end{longtable}
}

{\tiny
\begin{longtable}{|c|c|c|c|c|c|c|c|c|c|c|}
\caption[Number of unique materials per SG with topological bands w/o SOC at any filling]{In this table, we list the number of unique materials that contain at least some topological features w/o SOC in each SG.  In order, the columns in this table list the SG, the number of unique materials in the SG, the number of unique materials with either at least one stable or fragile band in the spectrum or symmetry-enforced crossing points at $E_{F}$ (\emph{i.e.} materials that are either ES-, ESFD-, NLC-SM- or SEBR-SM- classified semimetals at $E_{F}$ or exhibit NLC, SEBR, or fragile topological bands at any filling at which the compatibility relations are satisfied).  From left to right, in the following columns we list, per SG, the number of unique materials with at least one stable or fragile topological connected set of bands at any filling, at least one fragile topological set of bands at any filling, at least one NLC-classified stable topological band at any filling, at least one SEBR-classified stable topological band at any filling, and at least one stable topological set of bands at any filling.  In the remaining three columns, we list the number of materials that do not contain \emph{any} symmetry-indicated stable or fragile topological bands (STriv), the number of supermetallic (SMetal) materials in which \emph{all} of the bands in at least one of the ICSDs associated to the unique material are \emph{fully} connected, and the number of unique materials in which all bands are stable topological [\emph{i.e.} supertopological (STopo) materials, see \supappref{App:Supertopological}].  For all of the topological classification categories in the table, each number of materials is accompanied by the percentage of materials in the SG in the topological class. When there are no materials in the topological class, the table entry is marked with a dashed horizontal line.
\label{tb:statistics_nosoc_band_types_uniquematerials_bandtypes}}\\
\hline
\end{longtable}
}

{\tiny
\begin{longtable}{|c|c|c|c|c|c|c|}
\caption[High-connectivity nonmagnetic unique materials without $f$ electrons]{Nonmagnetic unique materials without $f$ electrons in which one or more associated ICSD entries exhibit a connected grouping of bands with dimension $\geq \TQCDBHighConnectivityNbrBandsThreshold$ in the presence of SOC.  In order, for each unique material containing a grouping of bands with connectivity $\geq \TQCDBHighConnectivityNbrBandsThreshold$ with SOC, the columns in this table list the chemical formula, the SG number, the SG symbol, the ICSD number, the stable topological classification at $E_{F}$, the number of valence electrons (num $e^{-}$), and the largest number of connected bands across the electronic band structures of all of the ICSDs associated to the unique material (noting that the VASP calculations performed in this work have been truncated at fillings of $\sim 2N_{e}$, see~\supappref{App:VASP_appendix} for further calculation details).  In the right-most column of this table, we also list in parentheses a percentage corresponding to the ratio of the maximum band connectivity with SOC in the unique material relative to $2N_{e}$, which is the approximate total number of bands in each VASP calculation performed with SOC.  A chemical formula with a $^{*}$ indicates that only some (but not all) of the ICSDs associated to the same unique material (defined using its topological class with SOC) contain a connected grouping of bands with dimension $\geq \TQCDBHighConnectivityNbrBandsThreshold$ in the presence of SOC.
\label{tb:all_high_connectivity_uniquecompounds_nice}}\\
\hline
Chem. formula & SG \# & SG symbol & ICSD & Top. subclass. & num $e^{-}$ & nbr connected bands\\
\hline  
$\rm{Cu}_{3} \rm{P}^*$& 165 & $P\bar{3}c1$ & \icsdwebshort{628626} & ES & 228 & 144 (31.58\%) \\ 
$\rm{Ag} \rm{Cd}_{3} \rm{Hf}_{3} \rm{F}_{20}$& 176 & $P6_3/m$ & \icsdwebshort{78893} & ESFD & 398 & 136 (17.09\%) \\
\hline
\end{longtable}
}

\afterpage{\clearpage}

\section{Repeat-Topology and Supertopology}\label{App:Supertopological}

In this work, we introduce the concept of a \emph{repeat-topological} (RTopo) material -- a material in which the gap at the Fermi level, as well as the next gap below the Fermi level as measured by band connectivity, exhibit cumulative symmetry-indicated stable topology [see Fig.~\ref{fig:RTopoSTopoSketch}A].  We additionally introduce the concept of a \emph{supertopological} (STopo) material -- a material in which \emph{every} energetically isolated set of bands in the spectrum obtained from our first-principles calculations (see \supappref{App:VASP_appendix}) exhibits symmetry-indicated stable topology [see Fig.~\ref{fig:RTopoSTopoSketch}B].  In this section, we will first rigorously define repeat-topology in \supappref{App:DefRTopo}.  Then, in \supappref{App:DefSTopo}, we will further define supertopology.  Lastly, in \supappref{App:StatForStopo}, we will provide a detailed statistical breakdown of the STopo unique materials with SOC in the ICSD.

\begin{figure}[ht]
\centering
\includegraphics[width=1.0\textwidth,angle=0]{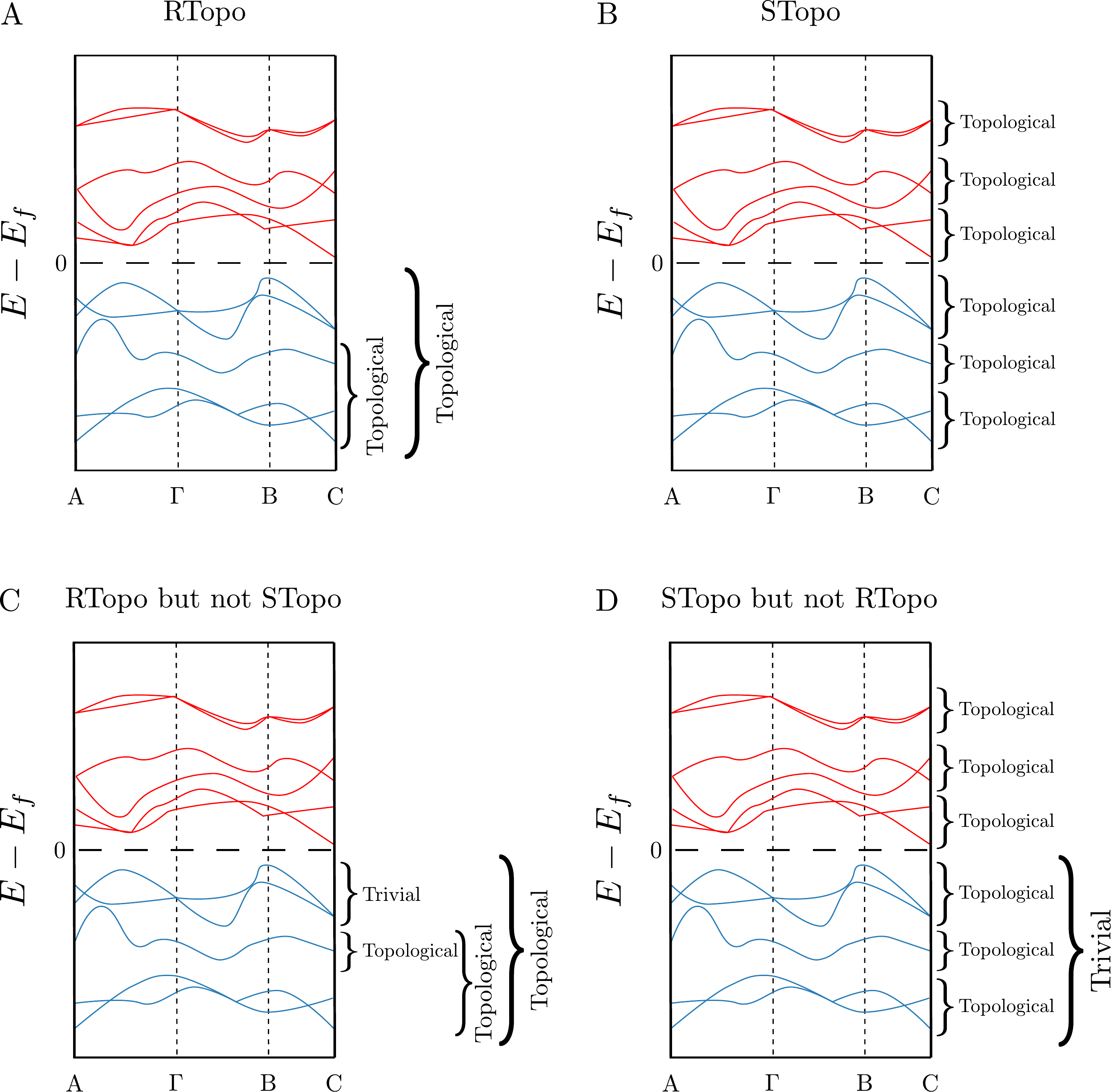}
\caption{Repeat-topological and supertopological materials.  A: A schematic depiction of a repeat-topological (RTopo) band structure.  In RTopo materials, the gap at $E_{F}$, as well as the next gap below $E_{F}$ as measured by band connectivity (see \supappref{App:TQCReview_appendix}), exhibit cumulative symmetry-indicated stable topology (\emph{i.e} the \emph{summed} stable topological indices of all of the bands up to $E_{F}$ and up to the next gap below $E_{F}$ are nontrivial, see \supappref{App:DefRTopo}).  B: A schematic depiction of a supertopological (STopo) band structure.  In STopo materials, \emph{every} energetically isolated set of bands in the spectrum exhibits symmetry-indicated stable topology (see \supappref{App:DefSTopo}).  It is important to emphasize that because repeat-topology is focused on cumulative topology (\emph{i.e.} topological \emph{gaps}), whereas supertopology is defined by the appearance of topological \emph{bands} that may together exhibit cumulative \emph{trivial} topology, then not all RTopo materials are STopo, and not all STopo materials are RTopo.  In panel C, we show a schematic example of a band structure that is RTopo, but not STopo.  Similarly, in panel D, we show a schematic example of a band structure that is STopo, but not RTopo.}
\label{fig:RTopoSTopoSketch}
\end{figure}

\subsection{Definition of Repeat-Topology}
\label{App:DefRTopo}

In this section, we introduce and rigorously define \emph{repeat-topological} (RTopo) materials.  In this work, to define RTopo materials, we first restrict to ICSD entries in which the entire valence manifold exhibits cumulative stable (NLC or SEBR) topology (\emph{i.e.} the summed stable topological indices of all of the occupied bands are nontrivial).  Next, using the compatibility relations (see~\supappref{App:TQCReview_appendix} and Refs.~\cite{QuantumChemistry,Bandrep1,Bandrep2,Bandrep3,JenFragile1}), we determine the next-highest electronic filling at which an insulating gap is permitted.  If the valence manifold in an ICSD entry when filled up to the next-highest gap below $E_{F}$ is also stable topological [see Fig.~\ref{fig:RTopoSTopoSketch}A], then we define the ICSD entry to be RTopo.  We emphasize that by this definition, it is possible for the isolated bands between $E_{F}$ and the next-highest gap below $E_{F}$ to be topologically trivial [see Fig.~\ref{fig:RTopoSTopoSketch}C].  Crucially however, the definition of repeat-topology employed in this work is motivated from an experimental perspective.  Specifically, we have in this work defined RTopo materials as exhibiting consecutive stable topological gaps at $E_{F}$ and just below $E_{F}$ because topological boundary states below $E_{F}$ can straightforwardly be observed through ARPES, whereas states above $E_{F}$ can only be observed by doping or by performing more complicated pump-probe experiments~\cite{ARPESReviewYulin,ARPESReviewHongDing,IlyaPumpProbe1,IlyaPumpProbe2}, and because topological response effects and boundary states in solid-state materials occur as a consequence of \emph{gaps} with nontrivial cumulative band topology, as opposed to isolated topological \emph{bands}.  From a physical perspective, idealized RTopo materials will exhibit two sets of surface or hinge states within consecutive bulk gaps that both lie at energies accessible to ARPES probes without doping ($\sim 1.5$~eV below $E_{F}$).  Hence, the repeated topological surface and hinge states of RTopo ICSD entries in consecutive bulk band gaps are analogous to the Fermi-arc ``quantum ladder'' recently observed in the unconventional chiral semimetal alloy Rh$_x$Ni$_y$Si~\cite{FermiArcQuantumLadder}.  Specifically, while Rh$_x$Ni$_y$Si is gapless at $E_{F}$ and below $E_{F}$, Rh$_x$Ni$_y$Si nevertheless exhibits topological Fermi-arc surface states at $E_{F}$ and at a filling of two electrons below $E_{F}$, representing a closely-related semimetallic analog of the topological (crystalline) insulating RTopo materials discovered in this work.

Next, we extend the definition of RTopo beyond individual ICSD entries to groups of ICSD entries associated to unique materials.  In this work, we define a unique material as a material that exhibits the same topological features (\emph{i.e.} topological class or nodal crossing points) with SOC at the Fermi level (see \supappref{App:VASP_appendix}).  Therefore, away from $E_{F}$, it is possible for two ICSD entries grouped into the same unique material to have isolated sets of bands with different topology.  Because we wish to define RTopo materials in a manner that is robust to numerical uncertainty, we define a unique material to be RTopo if \emph{any} of the ICSD entries associated to the unique material are themselves RTopo.  In \supappref{App:RTopoMaterials}, we list the RTopo TIs and TCIs on~\webNoICSD~with the fewest number of bulk Fermi pockets when the Fermi level is set to $0$, or set to the next highest filling at which an insulating gap is permitted by band connectivity.

Lastly, in the main text, we have highlighted Bi$_2$Mg$_3$ [\icsdweb{659569}, SG 164 ($P\bar{3}m1$)] as a prototypical RTopo (and STopo, see \supappref{App:DefSTopo}) material.  In Table~\ref{tab:BandCharacterizationBiTwoMgThree} of~\supappref{App:Check_Topo_appendix}, we previously computed the symmetry-indicated stable and fragile topology of all of the isolated sets of bands in Bi$_2$Mg$_3$ using the updated version of the~\webchecktopmat~program on the~\webBCSshort~implemented for this work.  Next, in Fig.~\ref{fig:STopoMain}D of the main text, we have plotted the $(0001)$-surface states of Bi$_2$Mg$_3$ obtained from surface Green's functions.  To compute the $(0001)$-surface spectrum shown in Fig.~~\ref{fig:STopoMain}D of the main text, we first used~\textsc{Wannier90}~\cite{Pizzi_2020} to construct a Wannier-based tight-binding model from the $s$ and $p$ orbitals of Mg and the $p$ orbitals of Bi, which we found to accurately reproduce the bulk electronic band structure of Bi$_2$Mg$_3$ obtained from \emph{ab initio} calculations (see~\icsdweb{659569}~on~\webNoICSD).  We next placed the Wannier-based tight-binding model in a $z$- ($c$-axis) directed slab geometry with $100$ layers (slab unit cells) in the stacking ($c$-axis) direction, and then computed the $z$-normal [$(0001)$-] surface states using the iterative Green's function method as implemented in~\textsc{WannierTools}~\cite{WU2018405}.  In the $(0001)$-surface spectrum of Bi$_2$Mg$_3$ shown in Fig.~\ref{fig:STopoMain}D of the main text, we observe topological twofold Dirac-cone surface states in both the projected gap at $E_{F}$ [labeled the ``$0003$ $E_{F}$ gap'' in Fig.~\ref{fig:STopoMain}D of the main text], and just below $E_{F}$ [labeled the ``$0013$ RTopo gap'' in Fig.~\ref{fig:STopoMain}D of the main text].  Most interesting, previous ARPES experiments on Bi$_2$Mg$_3$ samples have also detected surface states below the Fermi level, which were labeled as ``surface resonance bands'' in the earlier works~\cite{Mg3Bi2RTopo1,Mg3Bi2RTopo2,Mg3Bi2RTopo3}.  Hence in this work, we recognize the surface resonance bands detected in Refs.~\cite{Mg3Bi2RTopo1,Mg3Bi2RTopo2,Mg3Bi2RTopo3} to in fact be the RTopo surface states of Bi$_2$Mg$_3$ in the first gap below $E_{F}$ as determined by band connectivity (see~\supappref{App:TQCReview_appendix}).

\subsection{Definition of Supertopology}
\label{App:DefSTopo}

In this section, we introduce and define \emph{supertopological} (STopo) materials.  Similar to the definition of RTopo materials introduced in \supappref{App:DefRTopo}, in this work, we first define an ISCD entry to be STopo if all of the energetically isolated, connected sets of bands with SOC included in our calculations (see \supappref{App:VASP_appendix}) exhibit symmetry-indicated stable (NLC or SEBR) topology [see Fig.~\ref{fig:RTopoSTopoSketch}B].  We note that by this definition, semimetals (ES or ESFD) and trivial insulators (LCEBR) are also capable of being STopo materials, because the definition of supertopology is independent of the Fermi level (\emph{i.e.} supertopology is independent of the electronic filling).  We additionally emphasize that the electronic band structures calculated for this work only contain bands originating from valence atomic orbitals and at least the same number of bands originating from conduction atomic orbitals; we do not consider the topology of bands originating from core-shell atomic orbitals, which are free to be topologically trivial in an STopo material (which is physically motivated, because core-shell atomic orbitals typically lie at energies that are inaccessible in experimental probes).    We additionally emphasize that STopo materials are not necessarily also RTopo.  For example in an STopo material, it is possible for the isolated bands just below $E_{F}$ to exhibit stable topological indices that combine with the stable topological indices of the other sets of bands below $E_{F}$ to sum to trivial values [see Fig.~\ref{fig:RTopoSTopoSketch}D].

As previously with RTopo materials in \supappref{App:DefRTopo}, we next extend the definition of STopo beyond individual ICSD entries to groups of ICSD entries associated to unique materials.  Specifically, in this work, we define a unique material to be STopo if \emph{any} of the ICSD entries associated to the unique material are themselves STopo, thus providing a definition that is less sensitive to numerical uncertainty in materials with small band gaps away from $E_{F}$.  We find that bismuth (Bi) crystals in SG 166 ($R\bar{3}m$) -- recently discovered to be higher-order TIs (HOTIs)~\cite{HOTIBismuth} -- represent a prototypical example of both repeat-topology and supertopology.  In Table~\ref{tableST}, we show the output of the \checktopmat~program (see \supappref{App:Check_Topo_appendix}) applied to two different ICSD entries for Bi [(\icsdweb{64703}) and (\icsdweb{53797})] that are associated to the same unique material.  Notably, we find that, while both Bi (\icsdweb{64703}) and Bi (\icsdweb{53797}) exhibit the cumulative stable indices of a HOTI at the Fermi level [$(Z_{2,1},Z_{2,2},Z_{2,3},Z_{4}) = (0,0,0,2)$ at valence filling $\nu=10$, see \supappref{App:z4HOTIs}], Bi (\icsdweb{64703}) is STopo and RTopo, whereas Bi (\icsdweb{53797}) is RTopo, but not STopo.  Crucially, from a physical perspective, Bi (\icsdweb{64703}) and Bi (\icsdweb{53797}) only differ by weak band inversion away from the Fermi level, and both Bi (\icsdweb{64703}) and Bi (\icsdweb{53797}) exhibit topologically nontrivial (SEBR) gaps with the indices $(Z_{2,1},Z_{2,2},Z_{2,3},Z_{4}) = (0,0,0,3)$ just below the Fermi level (see Fig.~\ref{fig:STopo}).  This indicates that, whether a particular bismuth crystal sample is more closely characterized by the ICSD entry Bi (\icsdweb{64703}) or by Bi (\icsdweb{53797}), ARPES investigations of band gaps below the Fermi level may still reveal fully filled topological surface states (specifically, twofold Dirac cones).  In general, when using \checktopmat, topological surface and hinge states away from $E_{F}$ can be predicted by searching the `` top. type/all'' column (labeled in violet in Table~\ref{tableST}) for NLC and SEBR entries at experimentally accessible gaps away from $E_{F}$.

\begin{table}[h!]
\begin{tabular}{ccccccccccc}
\hline
\hline\\
A\;\;\; Bi & SG 166 ($R\bar{3}m$) & \icsdweb{64703} & &  &  &  &  &  & \\
\\
\hline 
\hline
 $\Gamma$ &T &F &L & dim & \textcolor{red}{top. type}  & ind/band & filling $\nu$ & \textcolor{violet}{top. type/all} & ind/all  \\
\hline
          -GM8(2)   &       -T9(2)  &     -F5-F6(2)   &    -L3-L4(2) &   2  &  SEBR&1 1 1 0&      2&    SEBR&1 1 1 0\\
\hline
          -GM9(2)   &       -T8(2)  &     -F3-F4(2)   &    -L5-L6(2) &   2  &  SEBR&1 1 1 0&      4& LCEBR&0 0 0 0\\
\hline
          -GM8(2)   &       -T9(2)  &     -F3-F4(2)   &    -L3-L4(2) &   2  &  SEBR&1 1 1 1&      6&    SEBR&1 1 1 1\\
\hline
          -GM8(2)   &       -T8(2)  &     -F5-F6(2)   &    -L5-L6(2) &   2  &  SEBR&1 1 1 2&      8&    SEBR&0 0 0 3\\
\hline
      -GM4-GM5(2)   &    -T6-T7(2)  &     -F5-F6(2)   &    -L5-L6(2) &   2  &  SEBR&0 0 0 3&     10&    SEBR&0 0 0 2\\
\hline
          -GM9(2)   &       -T8(2)  &     -F3-F4(2)   &    -L3-L4(2) &   2  &  SEBR&0 0 0 1&     12&    SEBR&0 0 0 3\\
\hline
          -GM9(2)   &       -T9(2)  &     -F3-F4(2)   &    -L3-L4(2) &   2  &  SEBR&1 1 1 2&     14&    SEBR&1 1 1 1\\
\hline
      -GM6-GM7(2)   &    -T4-T5(2)  &     -F3-F4(2)   &    -L5-L6(2) &   2  &  SEBR&1 1 1 0&     16&    SEBR&0 0 0 1\\
\hline
          -GM9(2)   &       -T9(2)  &     -F5-F6(2)   &    -L3-L4(2) &   2  &  SEBR&1 1 1 1&     18&    SEBR&1 1 1 2\\
\hline
\hline
\\
B\;\;\; Bi & SG 166 ($R\bar{3}m$) & \icsdweb{53797} & &  &  &  &  &  & \\
\\
\hline 
\hline
 $\Gamma$ &T &F &L & dim & \textcolor{red}{top. type}  & ind/band & filling $\nu$ & \textcolor{violet}{top. type/all} & ind/all  \\
 \hline
          -GM8(2)    &      -T9(2)   &    -F5-F6(2)    &   -L3-L4(2) &   2 &   SEBR &   1 1 1 0  &    2 &   SEBR   & 1 1 1 0\\
\hline
          -GM9(2)    &      -T8(2)   &    -F3-F4(2)    &   -L5-L6(2) &   2 &   SEBR &   1 1 1 0  &    4 &   LCEBR& 0 0 0 0\\
\hline
          -GM8(2)    &      -T9(2)   &    -F3-F4(2)    &   -L5-L6(2) &   2 &LCEBR &   0 0 0 0  &    6 &   LCEBR& 0 0 0 0\\
\hline
          -GM8(2)    &      -T8(2)   &    -F5-F6(2)    &   -L3-L4(2) &   2 &   SEBR &   0 0 0 3  &    8 &   SEBR   & 0 0 0 3\\
\hline
      -GM4-GM5(2)    &   -T6-T7(2)   &    -F5-F6(2)    &   -L5-L6(2) &   2 &   SEBR &   0 0 0 3  &   10 &   SEBR   & 0 0 0 2\\
\hline
          -GM9(2)    &      -T8(2)   &    -F3-F4(2)    &   -L3-L4(2) &   2 &   SEBR &   0 0 0 1  &   12 &   SEBR   & 0 0 0 3\\
\hline
          -GM9(2)    &      -T9(2)   &    -F3-F4(2)    &   -L3-L4(2) &   2 &   SEBR &   1 1 1 2  &   14 &   SEBR   & 1 1 1 1\\
\hline
      -GM6-GM7(2)    &   -T4-T5(2)   &    -F3-F4(2)    &   -L5-L6(2) &   2 &   SEBR  &  1 1 1 0  &   16 &   SEBR   & 0 0 0 1\\
\hline
          -GM9(2)    &      -T9(2)   &    -F5-F6(2)    &   -L3-L4(2) &   2 &   SEBR  &  1 1 1 1  &   18 &   SEBR   & 1 1 1 2\\
\hline
\end{tabular}
\caption[Topology of all of the bands in the RTopo and STopo compound Bi]{A typical table generated by \checktopmat.  Here, we have used as examples two different ICSD entries A: Bi (\icsdweb{64703}) and B: Bi (\icsdweb{53797}) for the repeat-topological (RTopo) and supertopological (STopo) unique material Bi.  In this table, the horizontal lines indicate the electronic fillings $\nu$ at which the occupied bands satisfy the compatibility relations.  Columns $\Gamma$, $T$, $F$, and $L$ contain the coreps at each maximal ${\bf k}$ point for each energetically isolated set of bands that satisfy the compatibility relations.  Column ``dim'' shows the dimension of each set of bands, Column ``top. type'' indicates the topology of the isolated bands [which can either be LCEBR, FRAGILE, NLC, or SEBR], and Column ``ind/band'' contains the stable SIs ($Z_{2,1}$ $Z_{2,2}$ $Z_{2,3}$ $Z_{4}$) that result from subducing onto SG 2 ($P\bar{1}$).  The remaining three columns contain the cumulative information of several combined sets of energetically isolated bands that are filled up to a total number of valence electrons specified by the column ``filling $\nu$.''  Column ``top. type/all'' provides the cumulative topology (LCEBR, FRAGILE, NLC, or SEBR) at the gap specified by the electronic filling in the ``filling $\nu$'' column, and Column ``ind/all'' provides the cumulative stable SIs of all of the filled bands up to the same gap.  Hence, the values of ``ind/all'' at each filling $\nu$ listed in this table can be obtained by summing the values in the ``ind/band'' column up to the same filling $\nu$.  In both A: Bi (\icsdweb{64703}) and B: Bi (\icsdweb{53797}), the Fermi level lies at the valence filling $\nu=10$, indicating that the gap at the Fermi level exhibits the subduced stable SIs ($0002$).  However away from $E_{F}$, the bands at valence filling $\nu=6$ exhibit trivial symmetry-indicated (LCEBR) topology in B, whereas all of the energetically isolated bands in A are classified as stable topological (SEBR).  Therefore, A: Bi (\icsdweb{64703}) is an STopo ICSD entry, whereas B: Bi (\icsdweb{53797}) is not; however, because A and B are associated to the same unique material, then as a whole, we consider Bi in SG 166 ($R\bar{3}m$) with ($0002$) SEBR topology at the Fermi level to be an STopo unique material.}
\label{tableST}
\end{table}

\begin{figure*}[htb!]
\centering
\begin{tabular}{c c}
A\hspace{0.5cm} {$\rm{Bi}$ - \icsdweb{64703} - $R\bar{3}m$ (SG 166) - SEBR} &B\hspace{0.5cm} {$\rm{Bi}$ - \icsdweb{53797} - $R\bar{3}m$ (SG 166) - SEBR}\\
\small{ $\;Z_{2,1}=0\;Z_{2,2}=0\;Z_{2,3}=0\;Z_4=2$ } & \small{ $\;Z_{2,1}=0\;Z_{2,2}=0\;Z_{2,3}=0\;Z_4=2$ }\\
\includegraphics[height=7cm]{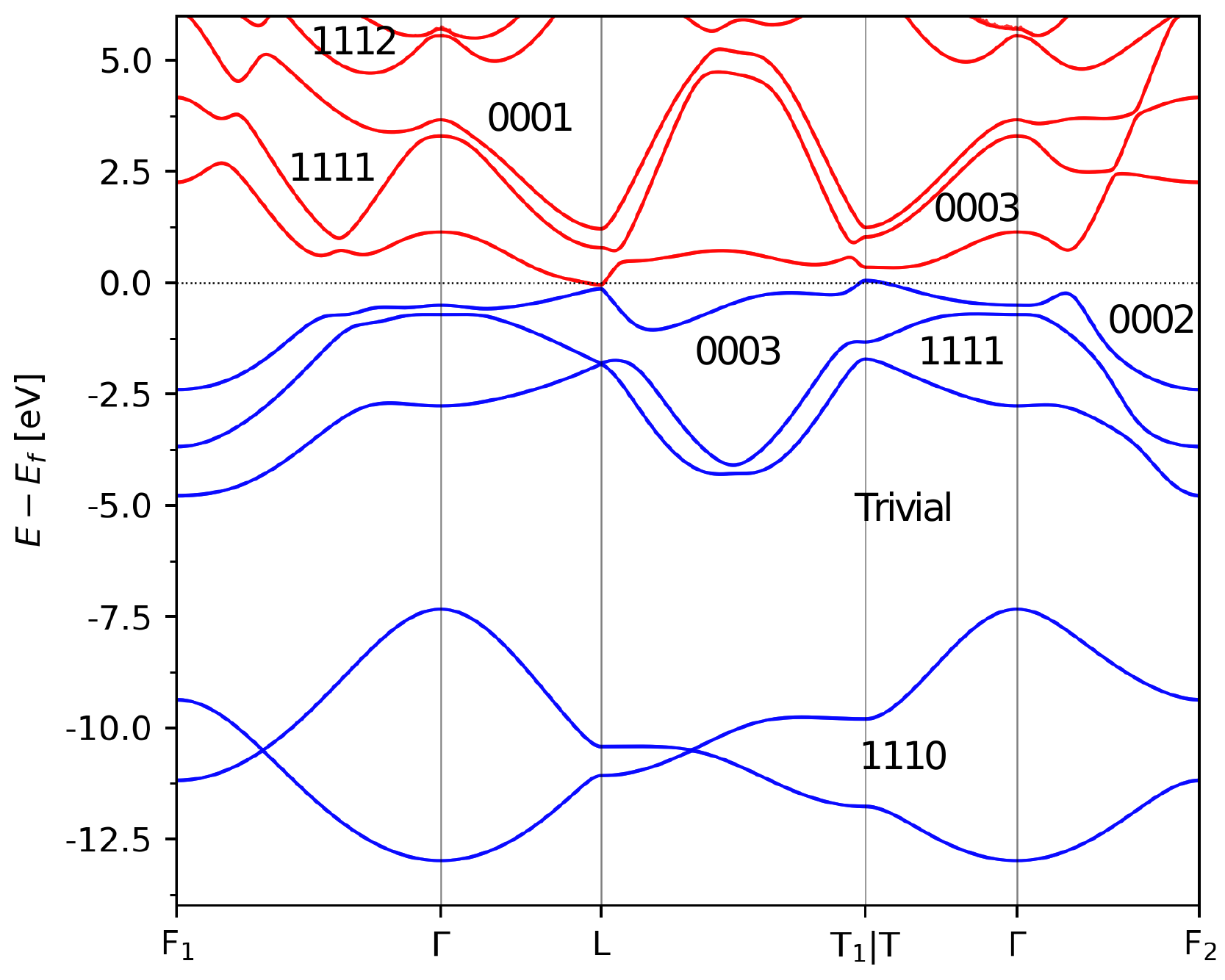} & \includegraphics[height=7cm]{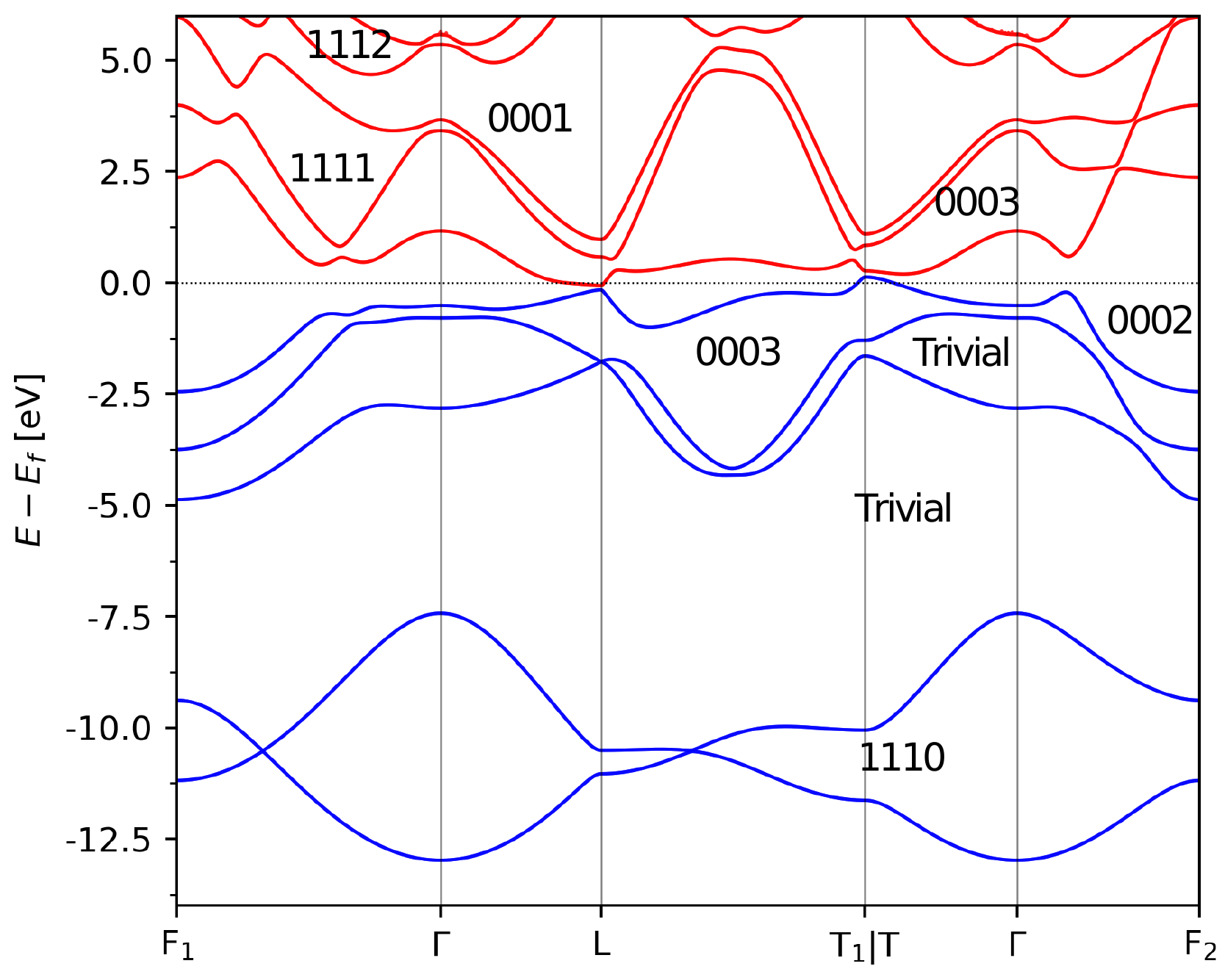}
\end{tabular}
\caption{Bulk bands and topology for two ICSD entries associated to the same unique material Bi.  Blue (red) colors are used to indicate valence (conduction) bands, and the four numbers listed in each gap indicate the cumulative stable SIs that result from subducing onto SG 2 ($P\bar{1}$): $Z_{2,1}$, $Z_{2,2}$, $Z_{2,3}$, $Z_{4}$, reproduced from the `` ind/all'' column in Table~\ref{tableST}.  In both panels A and B, the word ``Trivial'' indicates a gap in which the cumulative values  $(Z_{2,1},Z_{2,2},Z_{2,3},Z_{4}) = (0,0,0,0)$.  In A [B], we show the bulk bands and cumulative stable SIs of A: Bi (\icsdweb{64703}) [B: Bi (\icsdweb{53797})], which is [not] STopo.  Conversely, because the gap at $E_{F}$ [$(0002)$] and the next gap below $E_{F}$ [$(0003)$] exhibit cumulative symmetry-indicated stable topology, then Bi (\icsdweb{64703}) in A and Bi (\icsdweb{53797}) in B are both RTopo ICSD entries (see~\supappref{App:DefRTopo}).  Crucially, though the topological classifications of A and B differ away from $E_{F}$, both A and B exhibit the characteristic stable SIs of a HOTI~\cite{HOTIBismuth} at $E_{F}$ [$(Z_{2,1},Z_{2,2},Z_{2,3},Z_{4}) = (0,0,0,2)$, see \supappref{App:z4HOTIs}].}
\label{fig:STopo}
\end{figure*}

\subsection{Statistics for Supertopological Materials in the ICSD}
\label{App:StatForStopo}

In this section, we provide detailed statistics and lists for all of the STopo unique materials in the ICSD with SOC.  As previously discussed in \supappref{App:DefSTopo}, because the STopo classification is insensitive to the Fermi level, then materials in all of the topological classes (NLC, SEBR, ES, ESFD, and LCEBR) may be STopo.  First, in Table~\ref{tb:statistics_soc_band_types_uniquematerials_bandtypes_supertopo}, we list statistics for all of the STopo unique materials in the ICSD with SOC subdivided by the topology at $E_{F}$ and by SG.  Then, in Tables~\ref{tb:all_supertopological_uniquecompounds_nice_NLC}, \ref{tb:all_supertopological_uniquecompounds_nice_SEBR}, \ref{tb:all_supertopological_uniquecompounds_nice_ES} and~\ref{tb:all_supertopological_uniquecompounds_nice_ESFD}, we respectively list all of the stoichiometric, nonmagnetic unique STopo materials in the ICSD without $f$ electrons that are classified as NLC, SEBR, ES, ESFD, and LCEBR at $E_{F}$.

\begin{longtable}{|c|c|c|c|c|c|c|c|c|c|}
\caption[Statistics for supertopological materials with SOC]{Statistics for supertopological (STopo) materials.  The first and second columns respectively list the SG and the number of unique materials in the SG.  In order, the next columns list the number of unique materials in the SG that are STopo (defined in \supappref{App:DefSTopo}), the number that are STopo in which all of the bands are NLC (STopo-NLC), and the number that are STopo in which all of the bands are SEBR (STopo-SEBR).  In the final five columns, we respectively list the number of unique STopo materials in the SG that are classified as NLC, SEBR, ES, ESFD, and LCEBR at $E_{F}$.  For all of the STopo material subclassifications listed in this table, each number of materials is accompanied by the percentage of unique materials in the SG with the STopo subclassification.  When there are no unique materials with the STopo subclassification, the table entry is marked with a dashed horizontal line.
\label{tb:statistics_soc_band_types_uniquematerials_bandtypes_supertopo}}\\
\hline
SG  & \# Mat. & STopo & STopo-NLC & STopo-SEBR &\begin{tabular}{c}STopo\\ and NLC\end{tabular} &\begin{tabular}{c}STopo\\ and SEBR\end{tabular}&\begin{tabular}{c}STopo\\ and ES\end{tabular}&\begin{tabular}{c}STopo\\ and ESFD\end{tabular}&\begin{tabular}{c}STopo\\ and LCEBR\end{tabular}\\
\hline  
2& 1574 & 7 {\tiny(0.4\%)} & 7 {\tiny(0.4\%)} & ---  & ---  & ---  & ---  & ---  & 7 {\tiny(0.4\%)} \\ 
10& 38 & 1 {\tiny(2.6\%)} & 1 {\tiny(2.6\%)} & ---  & ---  & ---  & ---  & ---  & 1 {\tiny(2.6\%)} \\ 
11& 587 & 1 {\tiny(0.2\%)} & 1 {\tiny(0.2\%)} & ---  & 1 {\tiny(0.2\%)} & ---  & ---  & ---  & ---  \\ 
12& 1345 & 9 {\tiny(0.7\%)} & ---  & ---  & 5 {\tiny(0.4\%)} & 3 {\tiny(0.2\%)} & ---  & ---  & 1 {\tiny(0.1\%)} \\ 
13& 195 & 1 {\tiny(0.5\%)} & 1 {\tiny(0.5\%)} & ---  & ---  & ---  & ---  & ---  & 1 {\tiny(0.5\%)} \\
\hline
\end{longtable}

{\tiny
\begin{longtable}{|c|c|c|c|c|c|c|c|}
\caption[Supertopological nonmagnetic unique materials without $f$ electrons that are NLC at $E_{F}$]{Supertopological (STopo) nonmagnetic unique materials without $f$ electrons that are also NLC-classified at $E_{F}$ with SOC.  In order, the columns list the chemical formula, the SG number, the SG symbol, the ICSD number, whether all of the bands in the STopo material are NLC, whether all of the bands in the STopo material are SEBR, the number of valence electrons, and the cumulative stable SIs at $E_{F}$ (see \supappref{App:MaterialsSelection}).  A chemical formula with a $^{*}$ indicates that only some (but not all) of the ICSDs associated to the same unique material (defined using its topological class with SOC) are STopo.
\label{tb:all_supertopological_uniquecompounds_nice_NLC}}\\
\hline
Chem. formula & SG \# & SG symbol & ICSD & All NLC & All SEBR & num $e^{-}$ & Topological Indices \\
\hline  
$\rm{Ge} \rm{H}_{4}$& 11 & $P2_1/m$ & \icsdwebshort{183082} & Yes & No & 16 & {\tiny $Z_{2,1}=0\;\;Z_{2,2}=0\;\;Z_{2,3}=1\;\;Z_4=3$} \\ 
$\rm{W} \rm{H}_{6}^*$& 12 & $C2/m$ & \icsdwebshort{247613} & No & No & 24 & {\tiny $Z_{2,1}=0\;\;Z_{2,2}=0\;\;Z_{2,3}=0\;\;Z_4=1$} \\ 
$\rm{S} \rm{N}^*$& 14 & $P2_1/c$ & \icsdwebshort{41966} & Yes & No & 44 & {\tiny $Z_{2,1}=1\;\;Z_{2,2}=0\;\;Z_{2,3}=0\;\;Z_4=2$} \\ 
$\rm{Zr} \rm{Ge}_{2}^*$& 63 & $Cmcm$ & \icsdwebshort{56054} & No & No & 24 & {\tiny $Z_{2,1}=1\;\;Z_{2,2}=1\;\;Z_{2,3}=0\;\;Z_4=1$} \\ 
$\rm{Zr} \rm{Si}_{2}^*$& 63 & $Cmcm$ & \icsdwebshort{95179} & Yes & No & 24 & {\tiny $Z_{2,1}=1\;\;Z_{2,2}=1\;\;Z_{2,3}=0\;\;Z_4=1$} \\ 
$\rm{Ga}_{2} \rm{Sc}$& 65 & $Cmmm$ & \icsdwebshort{103956} & No & No & 18 & {\tiny $Z_{2,1}=0\;\;Z_{2,2}=0\;\;Z_{2,3}=1\;\;Z_4=1$} \\
\hline
\end{longtable}
}

{\tiny
\begin{longtable}{|c|c|c|c|c|c|c|c|}
\caption[Supertopological nonmagnetic unique materials without $f$ electrons that are SEBR at $E_{F}$]{Supertopological (STopo) nonmagnetic unique materials without $f$ electrons that are also SEBR-classified at $E_{F}$ with SOC.  In order, the columns list the chemical formula, the SG number, the SG symbol, the ICSD number, whether all of the bands in the STopo material are NLC, whether all of the bands in the STopo material are SEBR, the number of valence electrons, and the cumulative stable SIs at $E_{F}$ (see \supappref{App:MaterialsSelection}).  A chemical formula with a $^{*}$ indicates that only some (but not all) of the ICSDs associated to the same unique material (defined using its topological class with SOC) are STopo.
\label{tb:all_supertopological_uniquecompounds_nice_SEBR}}\\
\hline
Chem. formula & SG \# & SG symbol & ICSD & All NLC & All SEBR & num $e^{-}$ & Topological Indices \\
\hline  
$\rm{Bi}$& 12 & $C2/m$ & \icsdwebshort{409752} & No & No & 10 & {\tiny $Z_{2,1}=0\;\;Z_{2,2}=0\;\;Z_{2,3}=1\;\;Z_4=2$} \\ 
$\rm{Ta} \rm{Sb}_{2}^*$& 12 & $C2/m$ & \icsdwebshort{651600} & No & No & 30 & {\tiny $Z_{2,1}=1\;\;Z_{2,2}=1\;\;Z_{2,3}=1\;\;Z_4=2$} \\ 
$\rm{Si}_{2} \rm{Ti}^*$& 63 & $Cmcm$ & \icsdwebshort{168418} & No & No & 24 & {\tiny $Z_{2,1}=1\;\;Z_{2,2}=1\;\;Z_{2,3}=0\;\;Z_4=0$} \\ 
$\rm{Ca}$& 64 & $Cmce$ & \icsdwebshort{107510} & No & Yes & 8 & {\tiny $Z_{2,1}=0\;\;Z_{2,2}=0\;\;Z_{2,3}=0\;\;Z_4=3$} \\ 
$\rm{Ga}$& 64 & $Cmce$ & \icsdwebshort{43388} & No & Yes & 12 & {\tiny $Z_{2,1}=1\;\;Z_{2,2}=1\;\;Z_{2,3}=0\;\;Z_4=1$} \\ 
$\rm{Mo} \rm{Ni}_{4}$& 87 & $I4/m$ & \icsdwebshort{105047} & No & No & 46 & {\tiny $Z_{2,1}=0\;\;Z_{2,2}=0\;\;Z_{2,3}=0\;\;Z_4=3\;\;Z_2=1\;\;Z_8=7$} \\
\hline
\end{longtable}
}

{\tiny
\begin{longtable}{|c|c|c|c|c|c|c|c|}
\caption[Supertopological nonmagnetic unique materials without $f$ electrons that are ES at $E_{F}$]{Supertopological (STopo) nonmagnetic unique materials without $f$ electrons that are also ES-classified at $E_{F}$ with SOC.  In order, the columns list the chemical formula, the SG number, the SG symbol, the ICSD number, whether all of the bands in the STopo material are NLC, whether all of the bands in the STopo material are SEBR, and the number of valence electrons.  A chemical formula with a $^{*}$ indicates that only some (but not all) of the ICSDs associated to the same unique material (defined using its topological class with SOC) are STopo.
\label{tb:all_supertopological_uniquecompounds_nice_ES}}\\
\hline
Chem. formula & SG \# & SG symbol & ICSD & All NLC & All SEBR & num $e^{-}$ \\
\hline  
$\rm{Ir} \rm{V}$& 123 & $P4/mmm$ & \icsdwebshort{104589} & Yes & No & 14 \\ 
$\rm{Rh} \rm{V}^*$& 123 & $P4/mmm$ & \icsdwebshort{169385} & Yes & No & 14 \\ 
$\rm{Mg} \rm{Ni} \rm{H}_{2}$& 123 & $P4/mmm$ & \icsdwebshort{187258} & No & No & 14 \\ 
$\rm{Ru} \rm{Si}_{2}$& 123 & $P4/mmm$ & \icsdwebshort{154013} & No & No & 16 \\ 
$\rm{Au} \rm{Cu}$& 123 & $P4/mmm$ & \icsdwebshort{42574} & Yes & No & 22 \\ 
$\rm{Ni} \rm{Zn}$& 123 & $P4/mmm$ & \icsdwebshort{105469} & No & No & 22 \\
\hline
\end{longtable}
}

{\tiny
\begin{longtable}{|c|c|c|c|c|c|c|c|}
\caption[Supertopological nonmagnetic unique materials without $f$ electrons that are ESFD at $E_{F}$]{Supertopological (STopo) nonmagnetic unique materials without $f$ electrons that are also ESFD-classified at $E_{F}$ with SOC.  In order, the columns list the chemical formula, the SG number, the SG symbol, the ICSD number, whether all of the bands in the STopo material are NLC, whether all of the bands in the STopo material are SEBR, and the number of valence electrons.  A chemical formula with a $^{*}$ indicates that only some (but not all) of the ICSDs associated to the same unique material (defined using its topological class with SOC) are STopo.
\label{tb:all_supertopological_uniquecompounds_nice_ESFD}}\\
\hline
Chem. formula & SG \# & SG symbol & ICSD & All NLC & All SEBR & num $e^{-}$ \\
\hline  
$\rm{Mo} \rm{Rh}$& 51 & $Pmma$ & \icsdwebshort{108608} & Yes & No & 30 \\ 
$\rm{Ca} \rm{Rh} \rm{In}_{4}$& 51 & $Pmma$ & \icsdwebshort{410891} & Yes & No & 46 \\ 
$\rm{Ga}$& 63 & $Cmcm$ & \icsdwebshort{43539} & Yes & No & 6 \\ 
$\rm{Al} \rm{Ti}^*$& 123 & $P4/mmm$ & \icsdwebshort{58187} & Yes & No & 7 \\ 
$\rm{Ga} \rm{Ti}$& 123 & $P4/mmm$ & \icsdwebshort{103989} & No & No & 7 \\ 
$\rm{Rh} \rm{Ti}$& 123 & $P4/mmm$ & \icsdwebshort{105951} & No & No & 13 \\
\hline
\end{longtable}
}

{\tiny
\begin{longtable}{|c|c|c|c|c|c|c|c|}
\caption[Supertopological nonmagnetic unique materials without $f$ electrons that are LCEBR at $E_{F}$]{Supertopological (STopo) nonmagnetic unique materials without $f$ electrons that are LCEBR-classified at $E_{F}$ with SOC.  In order, the columns list the chemical formula, the SG number, the SG symbol, the ICSD number, whether all of the bands in the STopo material are NLC, whether all of the bands in the STopo material are SEBR, and the number of valence electrons.  A chemical formula with a $^{*}$ indicates that only some (but not all) of the ICSDs associated to the same unique material (defined using its topological class with SOC) are STopo.
\label{tb:all_supertopological_uniquecompounds_nice_LCEBR}}\\
\hline
Chem. formula & SG \# & SG symbol & ICSD & All NLC & All SEBR & num $e^{-}$ \\
\hline  
$\rm{Na} \rm{Ca} \rm{H}_{3}$& 2 & $P\bar{1}$ & \icsdwebshort{168714} & Yes & No & 12 \\ 
$\rm{Li} \rm{Be} \rm{H}_{3}$& 2 & $P\bar{1}$ & \icsdwebshort{162771} & Yes & No & 36 \\ 
$\rm{Na} \rm{Sb} \rm{S}_{2}$& 2 & $P\bar{1}$ & \icsdwebshort{200597} & Yes & No & 36 \\ 
$\rm{Mn} \rm{P}_{4}^*$& 2 & $P\bar{1}$ & \icsdwebshort{100786} & Yes & No & 54 \\ 
$\rm{Mg} (\rm{Si} \rm{O}_{4} \rm{H}_{2})^*$& 10 & $P2/m$ & \icsdwebshort{190936} & Yes & No & 32 \\ 
$\rm{Ba} \rm{Pb} \rm{O}_{3}^*$& 12 & $C2/m$ & \icsdwebshort{67811} & No & No & 64 \\
\hline
\end{longtable}
}

\afterpage{\clearpage}

\section{SOC-Driven Topological Phase Transitions}\label{App:PhaseTransitionsNoSOCSOC}

In this section, we provide statistics for the SOC-driven topological phase transitions in the materials studied in this work.  To begin, there are two broad mechanisms by which a material can be driven into a topological (crystalline) insulating phase.  The first mechanism occurs when a semimetal w/o SOC gaps directly into a TI or TCI under the introduction of SOC; in this case, the TI or TCI phase originates from \emph{splitting} bands that were previously connected in the absence of SOC (and possibly in the presence of additional crystal symmetries that are subsequently broken by Rashba- or Dresselhaus-like SOC).  The prototypical example of SOC-driven topological splitting occurs in the Kane-Mele model of graphene, which is an ESFD semimetal in the absence of SOC~\cite{QuantumChemistry,YoungkukLineNode,GrapheneReview}, and an SEBR 2D TI when the (weak) effects of SOC are taken into consideration~\cite{CharlieTI,KaneMeleZ2}.  The second mechanism for realizing a TI or TCI is SOC-driven \emph{band inversion}.  In the case of band inversion, a narrow-gap, topologically trivial semiconductor w/o SOC becomes a TI or TCI when introducing SOC changes the band ordering at a small number of ${\bf k}$ points in the BZ.  In real materials, band inversion arises due to the interplay of SOC and other interactions, which are typically controlled by doping and chemical substitution.  A prototypical of example of SOC- and dopant-driven band inversion occurs in the family of (higher-order) topological crystalline and trivial insulators Pb$_{1-x}$Sn$_x$Te (see Refs.~\onlinecite{FuKaneInversion,TeoFuKaneTCI,HsiehTCI,TanakaSnTeExp,HOTIBernevig,ChenRotation,ChenTCI,FradkinPbTe,BarryPbTe} and \supappref{App:rotationAnomaly}).

In \supappref{App:TopologicalMaterialsNoSOC}, we reviewed the previously established result that all 3D, symmetry-indicated topological phases w/o SOC are topological semimetals (either NLC-SM, SEBR-SM, ES, or ESFD).  In the case of ES and ESFD semimetals w/o SOC, introducing SOC may split bands and open a topological gap, as occurs in the hourglass insulator KHgSb (\icsdweb{56201}, see Refs.~\onlinecite{HourglassInsulator,Cohomological,DiracInsulator,HourglassExperiment,zeroHallExp} and \supappref{App:z4trivial}), or the bulk may remain gapless, as occurs in B20 compounds in the CoSi family of chiral crystals (\icsdweb{189221}, see Refs.~\onlinecite{RhSiArc,CoSiArc,CoSiObserveJapan,CoSiObserveHasan,CoSiObserveChina,AlPtObserve,PdGaObserve} and \supappref{App:ESFD}).  However, in the case of NLC-SM- and SEBR-SM-classified semimetals w/o SOC, infinitesimal SOC will always open a direct gap if the bulk is a centrosymmetric nodal-line semimetal~\cite{YoungkukLineNode,ZhidaSemimetals,TMDHOTI}, but will conversely not open a gap if the bulk is a noncentrosymmetric spinless (spin-degenerate) Weyl semimetal in the absence of SOC~\cite{SpinlessWeylGuoqing}.  Specifically, if the bulk is a spinless Weyl semimetal when the effects of SOC are neglected, then introducing infinitesimal SOC will \emph{not} generically open a bulk gap -- whether or not the spinless SIs in the absence of SOC are nontrivial -- as the two Weyl points within each spin-degenerate pair carry the same chiral charges (Chern numbers), and hence cannot pairwise annihilate~\cite{ZhidaSemimetals,S4Weyl,SpinlessWeylGuoqing}.

In the tables below, we provide detailed statistics comparing the topology of all of the stoichiometric unique materials in the ISCD at the Fermi level with and w/o SOC (we emphasize that in this work, a unique material is defined by the topological classification or crossing points at $E_{F}$ with SOC, see \supappref{App:VASP_appendix} for further details).  Specifically, in Tables~\ref{tb:statistics_nosoc_to_soc_uniquematerials_ES},~\ref{tb:statistics_nosoc_to_soc_uniquematerials_ESFD},~\ref{tb:statistics_nosoc_to_soc_uniquematerials_LCEBR},~\ref{tb:statistics_nosoc_to_soc_uniquematerials_NLC-SM}, and~\ref{tb:statistics_nosoc_to_soc_uniquematerials_SEBR-SM}, we respectively list the number of unique materials in each SG that are classified w/o SOC as ES, ESFD, LCEBR, NLC-SM, and SEBR-SM, which we then further divide by the topological classification of the same material in the presence of SOC.  Notably, there are relatively few materials in the ICSD that are classified as NLC-SM (Table~\ref{tb:statistics_nosoc_to_soc_uniquematerials_NLC-SM}) or SEBR-SM (Table~\ref{tb:statistics_nosoc_to_soc_uniquematerials_SEBR-SM} [see \supappref{App:whySoManyNoSOCES} for a more detailed discussion of NLC-SM and SEBR-SM materials in the ICSD].

{\tiny
\begin{longtable}{|c|c|c|c|c|c|c|c|c|c|}
\caption[SOC-driven topological phase transition statistics for ES materials w/o SOC]{SOC-driven topological phase transition statistics for ES-classified semimetals w/o SOC.  In order, the first two columns in this table list the SG and the number of ES-classified unique materials in the SG w/o SOC.  If an SG is not listed in the first column, then there are no ES-classified unique materials in the SG w/o SOC.  In the remaining five columns, restricting to the same unique materials listed in the second column, we respectively list the number of unique materials that become classified as NLC, SEBR, ES, ESFD, and LCEBR when the effects of SOC are incorporated.  For all of the topological classification categories in the table, each number of materials is accompanied by the percentage of materials in the SG in the topological class.  When there are no materials in the topological class, the table entry is marked with a dashed horizontal line.  
 \label{tb:statistics_nosoc_to_soc_uniquematerials_ES}}\\
\hline
SG  & \# Mat. & NLC & SEBR & ES & ESFD & LCEBR\\
\hline  
3& 7 & ---  & ---  & ---  & ---  & 7 (100.00\%) \\ 
4& 14 & ---  & ---  & ---  & ---  & 14 (100.00\%) \\ 
5& 19 & ---  & ---  & ---  & ---  & 19 (100.00\%) \\ 
6& 4 & ---  & ---  & ---  & ---  & 4 (100.00\%) \\ 
7& 6 & ---  & ---  & ---  & ---  & 6 (100.00\%) \\ 
8& 14 & ---  & ---  & ---  & ---  & 14 (100.00\%) \\ 
9& 4 & ---  & ---  & ---  & ---  & 4 (100.00\%) \\ 
10& 16 & 15 (93.75\%) & ---  & ---  & ---  & 1 (6.25\%) \\ 
11& 97 & 74 (76.29\%) & ---  & ---  & ---  & 23 (23.71\%) \\ 
12& 406 & 177 (43.60\%) & 160 (39.41\%) & ---  & ---  & 69 (17.00\%) \\ 
13& 21 & 12 (57.14\%) & ---  & ---  & ---  & 9 (42.86\%) \\ 
14& 330 & 227 (68.79\%) & ---  & 41 (12.42\%) & ---  & 62 (18.79\%) \\ 
15& 137 & 78 (56.93\%) & ---  & ---  & ---  & 59 (43.07\%) \\ 
16& 1 & ---  & ---  & ---  & ---  & 1 (100.00\%) \\ 
17& 3 & ---  & ---  & ---  & ---  & 3 (100.00\%) \\ 
18& 4 & ---  & ---  & ---  & ---  & 4 (100.00\%) \\ 
19& 9 & ---  & ---  & ---  & ---  & 9 (100.00\%) \\ 
20& 3 & ---  & ---  & ---  & ---  & 3 (100.00\%) \\ 
21& 3 & ---  & ---  & ---  & ---  & 3 (100.00\%) \\ 
22& 11 & ---  & ---  & ---  & ---  & 11 (100.00\%) \\ 
23& 3 & ---  & ---  & ---  & ---  & 3 (100.00\%) \\ 
25& 12 & ---  & ---  & ---  & ---  & 12 (100.00\%) \\ 
26& 8 & ---  & ---  & ---  & ---  & 8 (100.00\%) \\ 
28& 5 & ---  & ---  & ---  & ---  & 5 (100.00\%) \\ 
29& 1 & ---  & ---  & ---  & ---  & 1 (100.00\%) \\ 
30& 1 & ---  & ---  & ---  & ---  & 1 (100.00\%) \\ 
31& 19 & ---  & ---  & ---  & ---  & 19 (100.00\%) \\ 
33& 20 & ---  & ---  & ---  & ---  & 20 (100.00\%) \\ 
34& 2 & ---  & ---  & ---  & ---  & 2 (100.00\%) \\ 
35& 2 & ---  & ---  & ---  & ---  & 2 (100.00\%) \\ 
36& 21 & ---  & ---  & ---  & ---  & 21 (100.00\%) \\ 
38& 38 & ---  & ---  & ---  & ---  & 38 (100.00\%) \\ 
39& 3 & ---  & ---  & ---  & ---  & 3 (100.00\%) \\ 
40& 3 & ---  & ---  & ---  & ---  & 3 (100.00\%) \\ 
41& 9 & ---  & ---  & ---  & ---  & 9 (100.00\%) \\ 
43& 8 & ---  & ---  & ---  & ---  & 8 (100.00\%) \\ 
44& 24 & ---  & ---  & ---  & ---  & 24 (100.00\%) \\ 
45& 2 & ---  & ---  & ---  & ---  & 2 (100.00\%) \\ 
46& 23 & ---  & ---  & ---  & ---  & 23 (100.00\%) \\ 
47& 29 & 25 (86.21\%) & ---  & ---  & ---  & 4 (13.79\%) \\ 
48& 1 & ---  & ---  & ---  & ---  & 1 (100.00\%) \\ 
50& 5 & 3 (60.00\%) & 1 (20.00\%) & ---  & ---  & 1 (20.00\%) \\ 
51& 42 & 40 (95.24\%) & ---  & ---  & ---  & 2 (4.76\%) \\ 
53& 3 & ---  & 2 (66.67\%) & ---  & ---  & 1 (33.33\%) \\ 
54& 4 & 3 (75.00\%) & ---  & ---  & ---  & 1 (25.00\%) \\ 
55& 211 & 173 (81.99\%) & ---  & ---  & ---  & 38 (18.01\%) \\ 
56& 1 & ---  & ---  & ---  & ---  & 1 (100.00\%) \\ 
57& 31 & 25 (80.65\%) & ---  & ---  & ---  & 6 (19.35\%) \\ 
58& 62 & 37 (59.68\%) & ---  & 8 (12.90\%) & ---  & 17 (27.42\%) \\ 
59& 78 & 27 (34.62\%) & 33 (42.31\%) & ---  & ---  & 18 (23.08\%) \\ 
60& 20 & 13 (65.00\%) & ---  & ---  & ---  & 7 (35.00\%) \\ 
61& 16 & 8 (50.00\%) & ---  & 4 (25.00\%) & ---  & 4 (25.00\%) \\ 
62& 777 & 572 (73.62\%) & ---  & ---  & ---  & 205 (26.38\%) \\ 
63& 454 & 236 (51.98\%) & 153 (33.70\%) & ---  & ---  & 65 (14.32\%) \\ 
64& 72 & ---  & 50 (69.44\%) & 6 (8.33\%) & ---  & 16 (22.22\%) \\ 
65& 123 & 68 (55.28\%) & 44 (35.77\%) & ---  & ---  & 11 (8.94\%) \\ 
66& 7 & 4 (57.14\%) & ---  & ---  & ---  & 3 (42.86\%) \\ 
67& 11 & 9 (81.82\%) & ---  & ---  & ---  & 2 (18.18\%) \\ 
68& 2 & ---  & ---  & ---  & ---  & 2 (100.00\%) \\ 
69& 25 & ---  & 22 (88.00\%) & ---  & ---  & 3 (12.00\%) \\ 
70& 36 & ---  & 28 (77.78\%) & ---  & ---  & 8 (22.22\%) \\ 
71& 162 & ---  & 134 (82.72\%) & ---  & ---  & 28 (17.28\%) \\ 
72& 51 & 25 (49.02\%) & 13 (25.49\%) & ---  & ---  & 13 (25.49\%) \\ 
74& 75 & 51 (68.00\%) & ---  & ---  & ---  & 24 (32.00\%) \\ 
76& 1 & ---  & ---  & 1 (100.00\%) & ---  & ---  \\ 
82& 10 & 8 (80.00\%) & ---  & ---  & ---  & 2 (20.00\%) \\ 
83& 2 & ---  & 1 (50.00\%) & 1 (50.00\%) & ---  & ---  \\ 
84& 4 & 2 (50.00\%) & ---  & 2 (50.00\%) & ---  & ---  \\ 
85& 3 & ---  & 3 (100.00\%) & ---  & ---  & ---  \\ 
86& 9 & ---  & 1 (11.11\%) & 6 (66.67\%) & ---  & 2 (22.22\%) \\ 
87& 15 & ---  & 6 (40.00\%) & 9 (60.00\%) & ---  & ---  \\ 
88& 12 & ---  & 2 (16.67\%) & 4 (33.33\%) & ---  & 6 (50.00\%) \\ 
90& 2 & ---  & ---  & ---  & ---  & 2 (100.00\%) \\ 
91& 1 & ---  & ---  & ---  & ---  & 1 (100.00\%) \\ 
92& 6 & ---  & ---  & 4 (66.67\%) & ---  & 2 (33.33\%) \\ 
94& 1 & ---  & ---  & ---  & ---  & 1 (100.00\%) \\ 
95& 1 & ---  & ---  & 1 (100.00\%) & ---  & ---  \\ 
99& 1 & ---  & ---  & ---  & ---  & 1 (100.00\%) \\ 
100& 1 & ---  & ---  & ---  & ---  & 1 (100.00\%) \\ 
102& 3 & ---  & ---  & 1 (33.33\%) & ---  & 2 (66.67\%) \\ 
105& 1 & ---  & ---  & ---  & ---  & 1 (100.00\%) \\ 
107& 13 & ---  & ---  & 10 (76.92\%) & ---  & 3 (23.08\%) \\ 
108& 2 & ---  & ---  & 1 (50.00\%) & ---  & 1 (50.00\%) \\ 
109& 7 & ---  & ---  & ---  & ---  & 7 (100.00\%) \\ 
110& 2 & ---  & ---  & 1 (50.00\%) & ---  & 1 (50.00\%) \\ 
111& 1 & ---  & ---  & ---  & ---  & 1 (100.00\%) \\ 
113& 5 & 2 (40.00\%) & ---  & ---  & ---  & 3 (60.00\%) \\ 
114& 3 & ---  & 2 (66.67\%) & 1 (33.33\%) & ---  & ---  \\ 
115& 1 & ---  & ---  & ---  & ---  & 1 (100.00\%) \\ 
116& 3 & ---  & 1 (33.33\%) & ---  & ---  & 2 (66.67\%) \\ 
117& 3 & ---  & ---  & ---  & ---  & 3 (100.00\%) \\ 
118& 2 & 1 (50.00\%) & ---  & ---  & ---  & 1 (50.00\%) \\ 
119& 8 & ---  & ---  & ---  & ---  & 8 (100.00\%) \\ 
120& 3 & 2 (66.67\%) & ---  & ---  & ---  & 1 (33.33\%) \\ 
121& 17 & ---  & 14 (82.35\%) & ---  & ---  & 3 (17.65\%) \\ 
122& 11 & ---  & 10 (90.91\%) & ---  & ---  & 1 (9.09\%) \\ 
123& 66 & 22 (33.33\%) & 17 (25.76\%) & 25 (37.88\%) & ---  & 2 (3.03\%) \\ 
124& 4 & ---  & ---  & 4 (100.00\%) & ---  & ---  \\ 
125& 7 & 1 (14.29\%) & 3 (42.86\%) & 3 (42.86\%) & ---  & ---  \\ 
126& 2 & ---  & ---  & 2 (100.00\%) & ---  & ---  \\ 
127& 78 & 30 (38.46\%) & ---  & 42 (53.85\%) & ---  & 6 (7.69\%) \\ 
128& 12 & ---  & 5 (41.67\%) & 7 (58.33\%) & ---  & ---  \\ 
129& 213 & 50 (23.47\%) & 83 (38.97\%) & 43 (20.19\%) & ---  & 37 (17.37\%) \\ 
130& 10 & 2 (20.00\%) & 1 (10.00\%) & 1 (10.00\%) & ---  & 6 (60.00\%) \\ 
131& 11 & 4 (36.36\%) & ---  & 4 (36.36\%) & ---  & 3 (27.27\%) \\ 
132& 1 & ---  & ---  & 1 (100.00\%) & ---  & ---  \\ 
135& 9 & 2 (22.22\%) & ---  & 6 (66.67\%) & ---  & 1 (11.11\%) \\ 
136& 95 & ---  & 24 (25.26\%) & 49 (51.58\%) & ---  & 22 (23.16\%) \\ 
137& 10 & ---  & 6 (60.00\%) & 3 (30.00\%) & ---  & 1 (10.00\%) \\ 
138& 4 & ---  & ---  & 1 (25.00\%) & ---  & 3 (75.00\%) \\ 
139& 332 & ---  & 203 (61.14\%) & 100 (30.12\%) & ---  & 29 (8.73\%) \\ 
140& 100 & 32 (32.00\%) & 16 (16.00\%) & 39 (39.00\%) & ---  & 13 (13.00\%) \\ 
141& 77 & ---  & 43 (55.84\%) & 19 (24.68\%) & ---  & 15 (19.48\%) \\ 
142& 11 & ---  & 5 (45.45\%) & 2 (18.18\%) & ---  & 4 (36.36\%) \\ 
147& 1 & ---  & ---  & ---  & ---  & 1 (100.00\%) \\ 
148& 5 & ---  & ---  & 5 (100.00\%) & ---  & ---  \\ 
150& 4 & ---  & ---  & 1 (25.00\%) & ---  & 3 (75.00\%) \\ 
152& 2 & ---  & ---  & 1 (50.00\%) & ---  & 1 (50.00\%) \\ 
153& 1 & ---  & ---  & ---  & ---  & 1 (100.00\%) \\ 
154& 1 & ---  & ---  & ---  & ---  & 1 (100.00\%) \\ 
155& 2 & ---  & ---  & 1 (50.00\%) & ---  & 1 (50.00\%) \\ 
156& 7 & ---  & ---  & 4 (57.14\%) & ---  & 3 (42.86\%) \\ 
157& 2 & ---  & ---  & ---  & ---  & 2 (100.00\%) \\ 
158& 1 & ---  & ---  & ---  & ---  & 1 (100.00\%) \\ 
160& 6 & ---  & ---  & ---  & ---  & 6 (100.00\%) \\ 
161& 3 & ---  & ---  & 2 (66.67\%) & ---  & 1 (33.33\%) \\ 
162& 5 & ---  & 3 (60.00\%) & 2 (40.00\%) & ---  & ---  \\ 
163& 3 & ---  & 3 (100.00\%) & ---  & ---  & ---  \\ 
164& 47 & ---  & 33 (70.21\%) & 7 (14.89\%) & ---  & 7 (14.89\%) \\ 
165& 6 & ---  & 3 (50.00\%) & 1 (16.67\%) & ---  & 2 (33.33\%) \\ 
166& 153 & ---  & 96 (62.75\%) & 30 (19.61\%) & ---  & 27 (17.65\%) \\ 
167& 21 & ---  & 6 (28.57\%) & 11 (52.38\%) & ---  & 4 (19.05\%) \\ 
173& 11 & ---  & ---  & 10 (90.91\%) & ---  & 1 (9.09\%) \\ 
176& 5 & ---  & 3 (60.00\%) & 2 (40.00\%) & ---  & ---  \\ 
180& 2 & ---  & ---  & ---  & 1 (50.00\%) & 1 (50.00\%) \\ 
181& 1 & ---  & ---  & ---  & ---  & 1 (100.00\%) \\ 
182& 5 & ---  & ---  & 4 (80.00\%) & ---  & 1 (20.00\%) \\ 
185& 10 & ---  & ---  & 6 (60.00\%) & ---  & 4 (40.00\%) \\ 
186& 40 & ---  & ---  & 30 (75.00\%) & 2 (5.00\%) & 8 (20.00\%) \\ 
187& 31 & 9 (29.03\%) & ---  & 19 (61.29\%) & ---  & 3 (9.68\%) \\ 
189& 94 & ---  & 53 (56.38\%) & 40 (42.55\%) & ---  & 1 (1.06\%) \\ 
190& 3 & 2 (66.67\%) & ---  & 1 (33.33\%) & ---  & ---  \\ 
191& 63 & ---  & 22 (34.92\%) & 41 (65.08\%) & ---  & ---  \\ 
193& 24 & ---  & 10 (41.67\%) & 14 (58.33\%) & ---  & ---  \\ 
194& 235 & ---  & 98 (41.70\%) & 124 (52.77\%) & ---  & 13 (5.53\%) \\ 
198& 1 & ---  & ---  & ---  & ---  & 1 (100.00\%) \\ 
199& 1 & ---  & ---  & ---  & ---  & 1 (100.00\%) \\ 
200& 2 & ---  & 2 (100.00\%) & ---  & ---  & ---  \\ 
204& 4 & ---  & 4 (100.00\%) & ---  & ---  & ---  \\ 
205& 9 & 6 (66.67\%) & ---  & 1 (11.11\%) & ---  & 2 (22.22\%) \\ 
206& 1 & ---  & 1 (100.00\%) & ---  & ---  & ---  \\ 
208& 2 & ---  & ---  & ---  & ---  & 2 (100.00\%) \\ 
212& 2 & ---  & ---  & 2 (100.00\%) & ---  & ---  \\ 
213& 1 & ---  & ---  & 1 (100.00\%) & ---  & ---  \\ 
214& 1 & ---  & ---  & ---  & 1 (100.00\%) & ---  \\ 
216& 21 & 13 (61.90\%) & ---  & 2 (9.52\%) & 1 (4.76\%) & 5 (23.81\%) \\ 
217& 6 & ---  & 1 (16.67\%) & ---  & 2 (33.33\%) & 3 (50.00\%) \\ 
219& 2 & ---  & ---  & ---  & ---  & 2 (100.00\%) \\ 
220& 15 & 10 (66.67\%) & ---  & ---  & 4 (26.67\%) & 1 (6.67\%) \\ 
221& 107 & 4 (3.74\%) & 50 (46.73\%) & 46 (42.99\%) & 6 (5.61\%) & 1 (0.93\%) \\ 
223& 13 & ---  & 3 (23.08\%) & 9 (69.23\%) & 1 (7.69\%) & ---  \\ 
225& 74 & ---  & 34 (45.95\%) & 33 (44.59\%) & 3 (4.05\%) & 4 (5.41\%) \\ 
226& 3 & ---  & 1 (33.33\%) & 1 (33.33\%) & 1 (33.33\%) & ---  \\ 
227& 131 & ---  & 71 (54.20\%) & 37 (28.24\%) & 13 (9.92\%) & 10 (7.63\%) \\ 
229& 8 & ---  & 2 (25.00\%) & 3 (37.50\%) & 3 (37.50\%) & ---  \\ 
230& 1 & ---  & ---  & 1 (100.00\%) & ---  & ---  \\ 
\hline
Total& 6006 & 2,104 (35.03\%) & 1,585 (26.39\%) & 944 (15.72\%) & 38 (0.63\%) & 1,335 (22.23\%)
\\
\hline
\end{longtable}
}

{\tiny
\begin{longtable}{|c|c|c|c|c|c|c|c|c|c|}
\caption[SOC-driven topological phase transition statistics for ESFD materials w/o SOC]{SOC-driven topological phase transition statistics for ESFD-classified semimetals w/o SOC.  In order, the first two columns in this table list the SG and the number of ESFD-classified unique materials in the SG w/o SOC.  If an SG is not listed in the first column, then there are no ESFD-classified unique materials in the SG w/o SOC.  In the remaining five columns, restricting to the same unique materials listed in the second column, we respectively list the number of unique materials that become classified as NLC, SEBR, ES, ESFD, and LCEBR when the effects of SOC are incorporated.  For all of the topological classification categories in the table, each number of materials is accompanied by the percentage of materials in the SG in the topological class.  When there are no materials in the topological class, the table entry is marked with a dashed horizontal line.
\label{tb:statistics_nosoc_to_soc_uniquematerials_ESFD}}\\
\hline
SG  & \# Mat. & NLC & SEBR & ES & ESFD & LCEBR\\
\hline  
1& 9 & ---  & ---  & ---  & 9 (100.00\%) & ---  \\ 
2& 112 & 1 (0.89\%) & ---  & ---  & 111 (99.11\%) & ---  \\ 
4& 6 & ---  & ---  & 6 (100.00\%) & ---  & ---  \\ 
5& 12 & ---  & ---  & ---  & 12 (100.00\%) & ---  \\ 
6& 1 & ---  & ---  & ---  & 1 (100.00\%) & ---  \\ 
7& 3 & ---  & ---  & 3 (100.00\%) & ---  & ---  \\ 
8& 12 & ---  & ---  & ---  & 12 (100.00\%) & ---  \\ 
9& 11 & ---  & ---  & 11 (100.00\%) & ---  & ---  \\ 
10& 5 & ---  & ---  & ---  & 5 (100.00\%) & ---  \\ 
11& 96 & ---  & ---  & ---  & 95 (98.96\%) & 1 (1.04\%) \\ 
12& 191 & ---  & ---  & ---  & 191 (100.00\%) & ---  \\ 
13& 20 & ---  & ---  & ---  & 20 (100.00\%) & ---  \\ 
14& 137 & ---  & ---  & ---  & 137 (100.00\%) & ---  \\ 
15& 173 & ---  & ---  & ---  & 173 (100.00\%) & ---  \\ 
18& 2 & ---  & ---  & ---  & 2 (100.00\%) & ---  \\ 
19& 19 & ---  & ---  & 19 (100.00\%) & ---  & ---  \\ 
20& 2 & ---  & ---  & 2 (100.00\%) & ---  & ---  \\ 
21& 6 & ---  & ---  & ---  & 6 (100.00\%) & ---  \\ 
23& 2 & ---  & ---  & ---  & 2 (100.00\%) & ---  \\ 
25& 2 & ---  & ---  & ---  & 2 (100.00\%) & ---  \\ 
26& 4 & ---  & ---  & ---  & 4 (100.00\%) & ---  \\ 
28& 2 & ---  & ---  & 2 (100.00\%) & ---  & ---  \\ 
29& 1 & ---  & ---  & 1 (100.00\%) & ---  & ---  \\ 
30& 1 & ---  & ---  & ---  & 1 (100.00\%) & ---  \\ 
31& 14 & ---  & ---  & ---  & 14 (100.00\%) & ---  \\ 
33& 23 & ---  & ---  & 23 (100.00\%) & ---  & ---  \\ 
34& 3 & ---  & ---  & ---  & 3 (100.00\%) & ---  \\ 
35& 2 & ---  & ---  & ---  & 2 (100.00\%) & ---  \\ 
36& 32 & ---  & ---  & ---  & 32 (100.00\%) & ---  \\ 
38& 31 & ---  & ---  & ---  & 31 (100.00\%) & ---  \\ 
40& 4 & ---  & ---  & 4 (100.00\%) & ---  & ---  \\ 
41& 1 & ---  & ---  & 1 (100.00\%) & ---  & ---  \\ 
43& 13 & ---  & ---  & ---  & 13 (100.00\%) & ---  \\ 
44& 11 & ---  & ---  & ---  & 11 (100.00\%) & ---  \\ 
46& 7 & ---  & ---  & 7 (100.00\%) & ---  & ---  \\ 
47& 13 & ---  & ---  & ---  & 13 (100.00\%) & ---  \\ 
51& 35 & ---  & ---  & ---  & 35 (100.00\%) & ---  \\ 
52& 4 & ---  & ---  & 4 (100.00\%) & ---  & ---  \\ 
53& 3 & ---  & ---  & ---  & 3 (100.00\%) & ---  \\ 
54& 2 & ---  & ---  & 2 (100.00\%) & ---  & ---  \\ 
55& 41 & ---  & ---  & ---  & 41 (100.00\%) & ---  \\ 
56& 4 & ---  & ---  & 4 (100.00\%) & ---  & ---  \\ 
57& 17 & ---  & ---  & 17 (100.00\%) & ---  & ---  \\ 
58& 35 & ---  & ---  & ---  & 35 (100.00\%) & ---  \\ 
59& 65 & ---  & ---  & ---  & 65 (100.00\%) & ---  \\ 
60& 9 & ---  & ---  & 9 (100.00\%) & ---  & ---  \\ 
61& 2 & ---  & ---  & 2 (100.00\%) & ---  & ---  \\ 
62& 674 & ---  & ---  & 674 (100.00\%) & ---  & ---  \\ 
63& 470 & ---  & ---  & ---  & 470 (100.00\%) & ---  \\ 
64& 39 & ---  & ---  & ---  & 39 (100.00\%) & ---  \\ 
65& 70 & ---  & ---  & ---  & 70 (100.00\%) & ---  \\ 
66& 9 & ---  & ---  & ---  & 9 (100.00\%) & ---  \\ 
67& 6 & ---  & ---  & ---  & 6 (100.00\%) & ---  \\ 
68& 4 & ---  & ---  & ---  & 4 (100.00\%) & ---  \\ 
69& 18 & ---  & ---  & ---  & 18 (100.00\%) & ---  \\ 
70& 11 & ---  & ---  & ---  & 11 (100.00\%) & ---  \\ 
71& 153 & ---  & ---  & ---  & 153 (100.00\%) & ---  \\ 
72& 58 & ---  & ---  & ---  & 58 (100.00\%) & ---  \\ 
73& 1 & ---  & ---  & 1 (100.00\%) & ---  & ---  \\ 
74& 77 & ---  & ---  & ---  & 77 (100.00\%) & ---  \\ 
75& 2 & ---  & ---  & ---  & 2 (100.00\%) & ---  \\ 
76& 1 & ---  & ---  & 1 (100.00\%) & ---  & ---  \\ 
79& 1 & ---  & ---  & 1 (100.00\%) & ---  & ---  \\ 
81& 3 & ---  & ---  & ---  & ---  & 3 (100.00\%) \\ 
82& 25 & 7 (28.00\%) & ---  & ---  & 10 (40.00\%) & 8 (32.00\%) \\ 
83& 6 & 2 (33.33\%) & 2 (33.33\%) & 2 (33.33\%) & ---  & ---  \\ 
84& 3 & ---  & ---  & 1 (33.33\%) & 2 (66.67\%) & ---  \\ 
85& 9 & ---  & ---  & 1 (11.11\%) & 8 (88.89\%) & ---  \\ 
86& 29 & ---  & 8 (27.59\%) & 10 (34.48\%) & 11 (37.93\%) & ---  \\ 
87& 89 & ---  & 14 (15.73\%) & 13 (14.61\%) & 59 (66.29\%) & 3 (3.37\%) \\ 
88& 15 & ---  & 2 (13.33\%) & ---  & 12 (80.00\%) & 1 (6.67\%) \\ 
90& 2 & ---  & ---  & ---  & 2 (100.00\%) & ---  \\ 
91& 2 & ---  & ---  & 1 (50.00\%) & ---  & 1 (50.00\%) \\ 
92& 12 & ---  & ---  & 7 (58.33\%) & 5 (41.67\%) & ---  \\ 
96& 1 & ---  & ---  & 1 (100.00\%) & ---  & ---  \\ 
97& 1 & ---  & ---  & 1 (100.00\%) & ---  & ---  \\ 
98& 1 & ---  & ---  & ---  & ---  & 1 (100.00\%) \\ 
99& 26 & ---  & ---  & 7 (26.92\%) & 15 (57.69\%) & 4 (15.38\%) \\ 
100& 13 & ---  & ---  & 1 (7.69\%) & 11 (84.62\%) & 1 (7.69\%) \\ 
102& 3 & ---  & ---  & 1 (33.33\%) & 1 (33.33\%) & 1 (33.33\%) \\ 
103& 3 & ---  & ---  & ---  & 3 (100.00\%) & ---  \\ 
104& 1 & ---  & ---  & ---  & 1 (100.00\%) & ---  \\ 
106& 1 & ---  & ---  & 1 (100.00\%) & ---  & ---  \\ 
107& 88 & ---  & ---  & 6 (6.82\%) & 51 (57.95\%) & 31 (35.23\%) \\ 
108& 5 & ---  & ---  & 1 (20.00\%) & 3 (60.00\%) & 1 (20.00\%) \\ 
109& 16 & ---  & ---  & ---  & 13 (81.25\%) & 3 (18.75\%) \\ 
110& 1 & ---  & ---  & 1 (100.00\%) & ---  & ---  \\ 
111& 6 & ---  & 1 (16.67\%) & ---  & 4 (66.67\%) & 1 (16.67\%) \\ 
113& 21 & ---  & ---  & ---  & 19 (90.48\%) & 2 (9.52\%) \\ 
114& 3 & ---  & ---  & ---  & ---  & 3 (100.00\%) \\ 
115& 14 & ---  & 4 (28.57\%) & ---  & 6 (42.86\%) & 4 (28.57\%) \\ 
117& 3 & 1 (33.33\%) & ---  & ---  & 1 (33.33\%) & 1 (33.33\%) \\ 
118& 1 & ---  & ---  & ---  & 1 (100.00\%) & ---  \\ 
119& 9 & 2 (22.22\%) & ---  & ---  & 6 (66.67\%) & 1 (11.11\%) \\ 
120& 3 & ---  & ---  & 2 (66.67\%) & ---  & 1 (33.33\%) \\ 
121& 39 & ---  & 4 (10.26\%) & ---  & 20 (51.28\%) & 15 (38.46\%) \\ 
122& 12 & ---  & 3 (25.00\%) & 1 (8.33\%) & 8 (66.67\%) & ---  \\ 
123& 325 & 28 (8.62\%) & 18 (5.54\%) & 64 (19.69\%) & 210 (64.62\%) & 5 (1.54\%) \\ 
124& 15 & 1 (6.67\%) & ---  & 3 (20.00\%) & 11 (73.33\%) & ---  \\ 
125& 25 & 3 (12.00\%) & 3 (12.00\%) & 5 (20.00\%) & 14 (56.00\%) & ---  \\ 
126& 2 & ---  & ---  & ---  & 2 (100.00\%) & ---  \\ 
127& 248 & 32 (12.90\%) & ---  & 46 (18.55\%) & 167 (67.34\%) & 3 (1.21\%) \\ 
128& 33 & ---  & 8 (24.24\%) & 14 (42.42\%) & 10 (30.30\%) & 1 (3.03\%) \\ 
129& 478 & 51 (10.67\%) & 34 (7.11\%) & 48 (10.04\%) & 330 (69.04\%) & 15 (3.14\%) \\ 
130& 13 & 1 (7.69\%) & ---  & 2 (15.38\%) & 9 (69.23\%) & 1 (7.69\%) \\ 
131& 25 & 3 (12.00\%) & ---  & 5 (20.00\%) & 17 (68.00\%) & ---  \\ 
133& 2 & 1 (50.00\%) & ---  & 1 (50.00\%) & ---  & ---  \\ 
134& 3 & ---  & ---  & ---  & 3 (100.00\%) & ---  \\ 
135& 6 & 2 (33.33\%) & ---  & 1 (16.67\%) & 3 (50.00\%) & ---  \\ 
136& 93 & ---  & 9 (9.68\%) & 21 (22.58\%) & 58 (62.37\%) & 5 (5.38\%) \\ 
137& 27 & ---  & 2 (7.41\%) & 7 (25.93\%) & 17 (62.96\%) & 1 (3.70\%) \\ 
138& 3 & 1 (33.33\%) & ---  & 1 (33.33\%) & ---  & 1 (33.33\%) \\ 
139& 1052 & ---  & 251 (23.86\%) & 180 (17.11\%) & 582 (55.32\%) & 39 (3.71\%) \\ 
140& 219 & 18 (8.22\%) & 8 (3.65\%) & 22 (10.05\%) & 167 (76.26\%) & 4 (1.83\%) \\ 
141& 92 & ---  & 8 (8.70\%) & 4 (4.35\%) & 73 (79.35\%) & 7 (7.61\%) \\ 
142& 24 & ---  & 4 (16.67\%) & 10 (41.67\%) & 10 (41.67\%) & ---  \\ 
143& 5 & ---  & ---  & 1 (20.00\%) & 3 (60.00\%) & 1 (20.00\%) \\ 
144& 2 & ---  & ---  & ---  & 1 (50.00\%) & 1 (50.00\%) \\ 
146& 20 & ---  & ---  & 5 (25.00\%) & 9 (45.00\%) & 6 (30.00\%) \\ 
147& 15 & ---  & 5 (33.33\%) & 5 (33.33\%) & 3 (20.00\%) & 2 (13.33\%) \\ 
148& 184 & ---  & 37 (20.11\%) & 22 (11.96\%) & 102 (55.43\%) & 23 (12.50\%) \\ 
149& 5 & ---  & ---  & 1 (20.00\%) & 2 (40.00\%) & 2 (40.00\%) \\ 
150& 11 & ---  & ---  & 2 (18.18\%) & 8 (72.73\%) & 1 (9.09\%) \\ 
152& 7 & ---  & ---  & 1 (14.29\%) & 6 (85.71\%) & ---  \\ 
154& 1 & ---  & ---  & 1 (100.00\%) & ---  & ---  \\ 
155& 17 & ---  & ---  & 2 (11.76\%) & 7 (41.18\%) & 8 (47.06\%) \\ 
156& 9 & ---  & ---  & 2 (22.22\%) & 6 (66.67\%) & 1 (11.11\%) \\ 
157& 3 & ---  & ---  & 1 (33.33\%) & 2 (66.67\%) & ---  \\ 
158& 2 & ---  & ---  & 1 (50.00\%) & 1 (50.00\%) & ---  \\ 
159& 6 & ---  & ---  & 3 (50.00\%) & 2 (33.33\%) & 1 (16.67\%) \\ 
160& 52 & ---  & ---  & 7 (13.46\%) & 33 (63.46\%) & 12 (23.08\%) \\ 
161& 15 & ---  & ---  & 6 (40.00\%) & 7 (46.67\%) & 2 (13.33\%) \\ 
162& 21 & ---  & 2 (9.52\%) & 6 (28.57\%) & 11 (52.38\%) & 2 (9.52\%) \\ 
163& 22 & ---  & 1 (4.55\%) & 8 (36.36\%) & 10 (45.45\%) & 3 (13.64\%) \\ 
164& 271 & ---  & 58 (21.40\%) & 37 (13.65\%) & 150 (55.35\%) & 26 (9.59\%) \\ 
165& 18 & ---  & 1 (5.56\%) & 6 (33.33\%) & 8 (44.44\%) & 3 (16.67\%) \\ 
166& 463 & ---  & 84 (18.14\%) & 68 (14.69\%) & 281 (60.69\%) & 30 (6.48\%) \\ 
167& 100 & ---  & 5 (5.00\%) & 16 (16.00\%) & 66 (66.00\%) & 13 (13.00\%) \\ 
173& 28 & ---  & ---  & 24 (85.71\%) & 2 (7.14\%) & 2 (7.14\%) \\ 
174& 33 & ---  & ---  & 2 (6.06\%) & 31 (93.94\%) & ---  \\ 
175& 2 & ---  & ---  & ---  & 2 (100.00\%) & ---  \\ 
176& 77 & ---  & 3 (3.90\%) & 16 (20.78\%) & 52 (67.53\%) & 6 (7.79\%) \\ 
180& 16 & ---  & ---  & 6 (37.50\%) & 9 (56.25\%) & 1 (6.25\%) \\ 
181& 4 & ---  & ---  & 1 (25.00\%) & 3 (75.00\%) & ---  \\ 
182& 19 & ---  & ---  & 16 (84.21\%) & 2 (10.53\%) & 1 (5.26\%) \\ 
185& 19 & ---  & ---  & 4 (21.05\%) & 15 (78.95\%) & ---  \\ 
186& 163 & ---  & ---  & 38 (23.31\%) & 116 (71.17\%) & 9 (5.52\%) \\ 
187& 92 & 3 (3.26\%) & ---  & 31 (33.70\%) & 58 (63.04\%) & ---  \\ 
188& 4 & ---  & ---  & 2 (50.00\%) & 2 (50.00\%) & ---  \\ 
189& 389 & ---  & 47 (12.08\%) & 81 (20.82\%) & 260 (66.84\%) & 1 (0.26\%) \\ 
190& 22 & 2 (9.09\%) & ---  & 4 (18.18\%) & 16 (72.73\%) & ---  \\ 
191& 384 & ---  & 51 (13.28\%) & 130 (33.85\%) & 203 (52.86\%) & ---  \\ 
193& 233 & ---  & 5 (2.15\%) & 84 (36.05\%) & 140 (60.09\%) & 4 (1.72\%) \\ 
194& 1043 & ---  & 107 (10.26\%) & 414 (39.69\%) & 505 (48.42\%) & 17 (1.63\%) \\ 
196& 1 & ---  & ---  & ---  & 1 (100.00\%) & ---  \\ 
197& 10 & ---  & ---  & ---  & 9 (90.00\%) & 1 (10.00\%) \\ 
198& 76 & ---  & ---  & 2 (2.63\%) & 74 (97.37\%) & ---  \\ 
199& 8 & ---  & ---  & 1 (12.50\%) & 7 (87.50\%) & ---  \\ 
200& 15 & ---  & 1 (6.67\%) & ---  & 13 (86.67\%) & 1 (6.67\%) \\ 
201& 4 & ---  & 1 (25.00\%) & ---  & 3 (75.00\%) & ---  \\ 
202& 10 & ---  & ---  & ---  & 10 (100.00\%) & ---  \\ 
203& 1 & ---  & ---  & 1 (100.00\%) & ---  & ---  \\ 
204& 119 & ---  & 7 (5.88\%) & ---  & 104 (87.39\%) & 8 (6.72\%) \\ 
205& 40 & 2 (5.00\%) & ---  & 2 (5.00\%) & 34 (85.00\%) & 2 (5.00\%) \\ 
206& 29 & ---  & 9 (31.03\%) & 2 (6.90\%) & 13 (44.83\%) & 5 (17.24\%) \\ 
210& 1 & ---  & ---  & ---  & 1 (100.00\%) & ---  \\ 
211& 1 & ---  & ---  & ---  & 1 (100.00\%) & ---  \\ 
212& 6 & ---  & ---  & 1 (16.67\%) & 5 (83.33\%) & ---  \\ 
213& 16 & ---  & ---  & 2 (12.50\%) & 14 (87.50\%) & ---  \\ 
214& 6 & ---  & ---  & 1 (16.67\%) & 5 (83.33\%) & ---  \\ 
215& 23 & ---  & 3 (13.04\%) & ---  & 13 (56.52\%) & 7 (30.43\%) \\ 
216& 496 & 45 (9.07\%) & ---  & 15 (3.02\%) & 412 (83.06\%) & 24 (4.84\%) \\ 
217& 60 & ---  & 5 (8.33\%) & 1 (1.67\%) & 50 (83.33\%) & 4 (6.67\%) \\ 
218& 12 & ---  & 1 (8.33\%) & 6 (50.00\%) & 5 (41.67\%) & ---  \\ 
219& 3 & ---  & ---  & 2 (66.67\%) & 1 (33.33\%) & ---  \\ 
220& 100 & 1 (1.00\%) & ---  & 2 (2.00\%) & 95 (95.00\%) & 2 (2.00\%) \\ 
221& 1063 & ---  & 63 (5.93\%) & 114 (10.72\%) & 878 (82.60\%) & 8 (0.75\%) \\ 
223& 201 & ---  & 6 (2.99\%) & 37 (18.41\%) & 158 (78.61\%) & ---  \\ 
224& 9 & ---  & ---  & 1 (11.11\%) & 8 (88.89\%) & ---  \\ 
225& 1203 & ---  & 128 (10.64\%) & 135 (11.22\%) & 904 (75.15\%) & 36 (2.99\%) \\ 
226& 39 & ---  & 1 (2.56\%) & 4 (10.26\%) & 34 (87.18\%) & ---  \\ 
227& 542 & ---  & 69 (12.73\%) & 108 (19.93\%) & 325 (59.96\%) & 40 (7.38\%) \\ 
229& 177 & ---  & 5 (2.82\%) & 15 (8.47\%) & 156 (88.14\%) & 1 (0.56\%) \\ 
230& 4 & ---  & 1 (25.00\%) & ---  & 3 (75.00\%) & ---  \\ 
\hline
Total& 13997 & 207 (1.48\%) & 1,089 (7.78\%) & 2,792 (19.95\%) & 9,423 (67.32\%) & 486 (3.47\%)
\\
\hline
\end{longtable}
}

{\tiny
\begin{longtable}{|c|c|c|c|c|c|c|c|c|c|}
\caption[SOC-driven topological phase transition statistics for LCEBR materials w/o SOC]{SOC-driven topological phase transition statistics for LCEBR-classified materials w/o SOC.  In order, the first two columns in this table list the SG and the number of LCEBR-classified unique materials in the SG w/o SOC.  If an SG is not listed in the first column, then there are no LCEBR-classified unique materials in the SG w/o SOC.  In the remaining five columns, restricting to the same unique materials listed in the second column, we respectively list the number of unique materials that become classified as NLC, SEBR, ES, ESFD, and LCEBR when the effects of SOC are incorporated.  For all of the topological classification categories in the table, each number of materials is accompanied by the percentage of materials in the SG in the topological class.  When there are no materials in the topological class, the table entry is marked with a dashed horizontal line.
\label{tb:statistics_nosoc_to_soc_uniquematerials_LCEBR}}\\
\hline
SG  & \# Mat. & NLC & SEBR & ES & ESFD & LCEBR\\
\hline  
1& 142 & ---  & ---  & ---  & ---  & 142 (100.00\%) \\ 
2& 1189 & ---  & ---  & ---  & ---  & 1,189 (100.00\%) \\ 
3& 3 & ---  & ---  & ---  & ---  & 3 (100.00\%) \\ 
4& 199 & ---  & ---  & ---  & ---  & 199 (100.00\%) \\ 
5& 117 & ---  & ---  & ---  & ---  & 117 (100.00\%) \\ 
6& 23 & ---  & ---  & ---  & ---  & 23 (100.00\%) \\ 
7& 76 & ---  & ---  & ---  & ---  & 76 (100.00\%) \\ 
8& 115 & ---  & ---  & ---  & ---  & 115 (100.00\%) \\ 
9& 182 & ---  & ---  & ---  & ---  & 182 (100.00\%) \\ 
10& 12 & ---  & ---  & ---  & ---  & 12 (100.00\%) \\ 
11& 350 & 2 (0.57\%) & ---  & ---  & ---  & 348 (99.43\%) \\ 
12& 629 & 7 (1.11\%) & 3 (0.48\%) & ---  & ---  & 619 (98.41\%) \\ 
13& 122 & ---  & ---  & ---  & ---  & 122 (100.00\%) \\ 
14& 2029 & 6 (0.30\%) & ---  & 1 (0.05\%) & ---  & 2,022 (99.66\%) \\ 
15& 1070 & 2 (0.19\%) & ---  & ---  & ---  & 1,068 (99.81\%) \\ 
16& 1 & ---  & ---  & ---  & ---  & 1 (100.00\%) \\ 
17& 5 & ---  & ---  & ---  & ---  & 5 (100.00\%) \\ 
18& 24 & ---  & ---  & ---  & ---  & 24 (100.00\%) \\ 
19& 233 & ---  & ---  & ---  & ---  & 233 (100.00\%) \\ 
20& 53 & ---  & ---  & ---  & ---  & 53 (100.00\%) \\ 
21& 6 & ---  & ---  & ---  & ---  & 6 (100.00\%) \\ 
22& 3 & ---  & ---  & ---  & ---  & 3 (100.00\%) \\ 
23& 10 & ---  & ---  & ---  & ---  & 10 (100.00\%) \\ 
24& 2 & ---  & ---  & ---  & ---  & 2 (100.00\%) \\ 
25& 31 & ---  & ---  & ---  & ---  & 31 (100.00\%) \\ 
26& 47 & ---  & ---  & ---  & ---  & 47 (100.00\%) \\ 
27& 1 & ---  & ---  & ---  & ---  & 1 (100.00\%) \\ 
28& 4 & ---  & ---  & ---  & ---  & 4 (100.00\%) \\ 
29& 57 & ---  & ---  & ---  & ---  & 57 (100.00\%) \\ 
30& 3 & ---  & ---  & ---  & ---  & 3 (100.00\%) \\ 
31& 125 & ---  & ---  & ---  & ---  & 125 (100.00\%) \\ 
32& 10 & ---  & ---  & ---  & ---  & 10 (100.00\%) \\ 
33& 208 & ---  & ---  & ---  & ---  & 208 (100.00\%) \\ 
34& 16 & ---  & ---  & ---  & ---  & 16 (100.00\%) \\ 
35& 4 & ---  & ---  & ---  & ---  & 4 (100.00\%) \\ 
36& 188 & ---  & ---  & ---  & ---  & 188 (100.00\%) \\ 
37& 5 & ---  & ---  & ---  & ---  & 5 (100.00\%) \\ 
38& 26 & ---  & ---  & ---  & ---  & 26 (100.00\%) \\ 
39& 9 & ---  & ---  & ---  & ---  & 9 (100.00\%) \\ 
40& 47 & ---  & ---  & ---  & ---  & 47 (100.00\%) \\ 
41& 21 & ---  & ---  & ---  & ---  & 21 (100.00\%) \\ 
42& 4 & ---  & ---  & ---  & ---  & 4 (100.00\%) \\ 
43& 62 & ---  & ---  & ---  & ---  & 62 (100.00\%) \\ 
44& 27 & ---  & ---  & ---  & ---  & 27 (100.00\%) \\ 
45& 6 & ---  & ---  & ---  & ---  & 6 (100.00\%) \\ 
46& 25 & ---  & ---  & ---  & ---  & 25 (100.00\%) \\ 
47& 3 & ---  & ---  & ---  & ---  & 3 (100.00\%) \\ 
48& 2 & ---  & ---  & ---  & ---  & 2 (100.00\%) \\ 
50& 2 & ---  & ---  & ---  & ---  & 2 (100.00\%) \\ 
51& 23 & 1 (4.35\%) & ---  & ---  & ---  & 22 (95.65\%) \\ 
52& 32 & ---  & ---  & ---  & ---  & 32 (100.00\%) \\ 
53& 12 & ---  & ---  & ---  & ---  & 12 (100.00\%) \\ 
54& 12 & ---  & ---  & ---  & ---  & 12 (100.00\%) \\ 
55& 84 & ---  & ---  & ---  & ---  & 84 (100.00\%) \\ 
56& 14 & ---  & ---  & ---  & ---  & 14 (100.00\%) \\ 
57& 80 & ---  & ---  & ---  & ---  & 80 (100.00\%) \\ 
58& 104 & 1 (0.96\%) & ---  & ---  & ---  & 103 (99.04\%) \\ 
59& 65 & ---  & ---  & ---  & ---  & 65 (100.00\%) \\ 
60& 109 & ---  & ---  & ---  & ---  & 109 (100.00\%) \\ 
61& 101 & ---  & ---  & ---  & ---  & 101 (100.00\%) \\ 
62& 1349 & 17 (1.26\%) & ---  & ---  & ---  & 1,332 (98.74\%) \\ 
63& 353 & 2 (0.57\%) & 3 (0.85\%) & ---  & ---  & 348 (98.58\%) \\ 
64& 133 & ---  & ---  & ---  & ---  & 133 (100.00\%) \\ 
65& 20 & ---  & ---  & ---  & ---  & 20 (100.00\%) \\ 
66& 18 & 1 (5.56\%) & ---  & ---  & ---  & 17 (94.44\%) \\ 
67& 12 & ---  & ---  & ---  & ---  & 12 (100.00\%) \\ 
68& 13 & ---  & ---  & ---  & ---  & 13 (100.00\%) \\ 
69& 22 & ---  & ---  & ---  & ---  & 22 (100.00\%) \\ 
70& 83 & ---  & ---  & ---  & ---  & 83 (100.00\%) \\ 
71& 70 & ---  & ---  & ---  & ---  & 70 (100.00\%) \\ 
72& 73 & ---  & ---  & ---  & ---  & 73 (100.00\%) \\ 
73& 15 & ---  & ---  & ---  & ---  & 15 (100.00\%) \\ 
74& 50 & 1 (2.00\%) & ---  & ---  & ---  & 49 (98.00\%) \\ 
75& 3 & ---  & ---  & ---  & ---  & 3 (100.00\%) \\ 
76& 10 & ---  & ---  & ---  & ---  & 10 (100.00\%) \\ 
77& 2 & ---  & ---  & ---  & ---  & 2 (100.00\%) \\ 
78& 2 & ---  & ---  & ---  & ---  & 2 (100.00\%) \\ 
79& 10 & ---  & ---  & ---  & ---  & 10 (100.00\%) \\ 
80& 3 & ---  & ---  & ---  & ---  & 3 (100.00\%) \\ 
81& 9 & ---  & ---  & ---  & ---  & 9 (100.00\%) \\ 
82& 93 & 2 (2.15\%) & ---  & ---  & ---  & 91 (97.85\%) \\ 
83& 1 & ---  & ---  & ---  & ---  & 1 (100.00\%) \\ 
84& 21 & ---  & ---  & ---  & ---  & 21 (100.00\%) \\ 
85& 22 & ---  & ---  & ---  & ---  & 22 (100.00\%) \\ 
86& 11 & ---  & 1 (9.09\%) & ---  & ---  & 10 (90.91\%) \\ 
87& 69 & ---  & ---  & ---  & ---  & 69 (100.00\%) \\ 
88& 102 & ---  & ---  & ---  & ---  & 102 (100.00\%) \\ 
90& 1 & ---  & ---  & ---  & ---  & 1 (100.00\%) \\ 
91& 4 & ---  & ---  & ---  & ---  & 4 (100.00\%) \\ 
92& 29 & ---  & ---  & ---  & ---  & 29 (100.00\%) \\ 
95& 2 & ---  & ---  & ---  & ---  & 2 (100.00\%) \\ 
96& 15 & ---  & ---  & ---  & ---  & 15 (100.00\%) \\ 
97& 1 & ---  & ---  & ---  & ---  & 1 (100.00\%) \\ 
98& 3 & ---  & ---  & ---  & ---  & 3 (100.00\%) \\ 
99& 71 & ---  & ---  & ---  & ---  & 71 (100.00\%) \\ 
100& 6 & ---  & ---  & ---  & ---  & 6 (100.00\%) \\ 
102& 6 & ---  & ---  & ---  & ---  & 6 (100.00\%) \\ 
104& 2 & ---  & ---  & ---  & ---  & 2 (100.00\%) \\ 
105& 3 & ---  & ---  & ---  & ---  & 3 (100.00\%) \\ 
107& 11 & ---  & ---  & ---  & ---  & 11 (100.00\%) \\ 
108& 7 & ---  & ---  & ---  & ---  & 7 (100.00\%) \\ 
109& 14 & ---  & ---  & 1 (7.14\%) & ---  & 13 (92.86\%) \\ 
110& 5 & ---  & ---  & ---  & ---  & 5 (100.00\%) \\ 
111& 13 & ---  & ---  & ---  & ---  & 13 (100.00\%) \\ 
112& 4 & ---  & ---  & ---  & ---  & 4 (100.00\%) \\ 
113& 45 & ---  & ---  & ---  & ---  & 45 (100.00\%) \\ 
114& 25 & ---  & ---  & ---  & ---  & 25 (100.00\%) \\ 
115& 6 & ---  & ---  & ---  & ---  & 6 (100.00\%) \\ 
116& 14 & ---  & ---  & ---  & ---  & 14 (100.00\%) \\ 
117& 5 & ---  & ---  & ---  & ---  & 5 (100.00\%) \\ 
118& 8 & ---  & ---  & ---  & ---  & 8 (100.00\%) \\ 
119& 18 & ---  & ---  & ---  & ---  & 18 (100.00\%) \\ 
120& 9 & ---  & ---  & ---  & ---  & 9 (100.00\%) \\ 
121& 67 & ---  & ---  & ---  & ---  & 67 (100.00\%) \\ 
122& 101 & ---  & 3 (2.97\%) & 1 (0.99\%) & ---  & 97 (96.04\%) \\ 
123& 59 & ---  & ---  & ---  & ---  & 59 (100.00\%) \\ 
124& 3 & ---  & ---  & ---  & ---  & 3 (100.00\%) \\ 
125& 17 & ---  & ---  & ---  & ---  & 17 (100.00\%) \\ 
126& 7 & ---  & ---  & 1 (14.29\%) & ---  & 6 (85.71\%) \\ 
127& 42 & 1 (2.38\%) & ---  & ---  & ---  & 41 (97.62\%) \\ 
128& 26 & ---  & ---  & ---  & ---  & 26 (100.00\%) \\ 
129& 170 & 1 (0.59\%) & ---  & 5 (2.94\%) & ---  & 164 (96.47\%) \\ 
130& 14 & ---  & ---  & ---  & ---  & 14 (100.00\%) \\ 
131& 7 & ---  & ---  & ---  & ---  & 7 (100.00\%) \\ 
132& 9 & ---  & ---  & ---  & ---  & 9 (100.00\%) \\ 
133& 1 & ---  & ---  & ---  & ---  & 1 (100.00\%) \\ 
134& 1 & ---  & ---  & ---  & ---  & 1 (100.00\%) \\ 
135& 11 & ---  & ---  & ---  & ---  & 11 (100.00\%) \\ 
136& 63 & ---  & 1 (1.59\%) & ---  & ---  & 62 (98.41\%) \\ 
137& 31 & ---  & ---  & ---  & ---  & 31 (100.00\%) \\ 
138& 9 & ---  & ---  & ---  & ---  & 9 (100.00\%) \\ 
139& 159 & ---  & 2 (1.26\%) & 3 (1.89\%) & ---  & 154 (96.86\%) \\ 
140& 112 & 1 (0.89\%) & ---  & ---  & ---  & 111 (99.11\%) \\ 
141& 77 & ---  & 1 (1.30\%) & 1 (1.30\%) & ---  & 75 (97.40\%) \\ 
142& 38 & ---  & ---  & ---  & ---  & 38 (100.00\%) \\ 
143& 12 & ---  & ---  & 1 (8.33\%) & ---  & 11 (91.67\%) \\ 
144& 11 & ---  & ---  & ---  & ---  & 11 (100.00\%) \\ 
145& 5 & ---  & ---  & ---  & ---  & 5 (100.00\%) \\ 
146& 63 & ---  & ---  & ---  & 1 (1.59\%) & 62 (98.41\%) \\ 
147& 52 & ---  & ---  & 1 (1.92\%) & ---  & 51 (98.08\%) \\ 
148& 255 & ---  & 1 (0.39\%) & ---  & ---  & 254 (99.61\%) \\ 
149& 5 & ---  & ---  & ---  & ---  & 5 (100.00\%) \\ 
150& 49 & ---  & ---  & ---  & ---  & 49 (100.00\%) \\ 
151& 4 & ---  & ---  & ---  & ---  & 4 (100.00\%) \\ 
152& 31 & ---  & ---  & 1 (3.23\%) & 2 (6.45\%) & 28 (90.32\%) \\ 
154& 15 & ---  & ---  & ---  & 1 (6.67\%) & 14 (93.33\%) \\ 
155& 31 & ---  & ---  & ---  & ---  & 31 (100.00\%) \\ 
156& 41 & ---  & ---  & 1 (2.44\%) & ---  & 40 (97.56\%) \\ 
157& 9 & ---  & ---  & ---  & ---  & 9 (100.00\%) \\ 
159& 18 & ---  & ---  & ---  & ---  & 18 (100.00\%) \\ 
160& 100 & ---  & ---  & ---  & ---  & 100 (100.00\%) \\ 
161& 69 & ---  & ---  & ---  & ---  & 69 (100.00\%) \\ 
162& 25 & ---  & ---  & ---  & ---  & 25 (100.00\%) \\ 
163& 32 & ---  & 1 (3.12\%) & ---  & ---  & 31 (96.88\%) \\ 
164& 254 & ---  & 13 (5.12\%) & 2 (0.79\%) & ---  & 239 (94.09\%) \\ 
165& 9 & ---  & ---  & ---  & ---  & 9 (100.00\%) \\ 
166& 337 & ---  & 15 (4.45\%) & 3 (0.89\%) & ---  & 319 (94.66\%) \\ 
167& 179 & ---  & ---  & ---  & ---  & 179 (100.00\%) \\ 
169& 1 & ---  & ---  & ---  & ---  & 1 (100.00\%) \\ 
173& 74 & ---  & ---  & ---  & ---  & 74 (100.00\%) \\ 
174& 17 & ---  & ---  & ---  & ---  & 17 (100.00\%) \\ 
176& 107 & ---  & ---  & ---  & ---  & 107 (100.00\%) \\ 
177& 1 & ---  & ---  & ---  & ---  & 1 (100.00\%) \\ 
180& 10 & ---  & ---  & ---  & ---  & 10 (100.00\%) \\ 
181& 3 & ---  & ---  & ---  & ---  & 3 (100.00\%) \\ 
182& 9 & ---  & ---  & ---  & ---  & 9 (100.00\%) \\ 
183& 1 & ---  & ---  & ---  & ---  & 1 (100.00\%) \\ 
185& 20 & ---  & ---  & 2 (10.00\%) & ---  & 18 (90.00\%) \\ 
186& 167 & ---  & ---  & 3 (1.80\%) & ---  & 164 (98.20\%) \\ 
187& 31 & ---  & ---  & ---  & ---  & 31 (100.00\%) \\ 
188& 16 & ---  & ---  & ---  & ---  & 16 (100.00\%) \\ 
189& 54 & ---  & 2 (3.70\%) & ---  & ---  & 52 (96.30\%) \\ 
190& 16 & ---  & ---  & ---  & ---  & 16 (100.00\%) \\ 
191& 5 & ---  & ---  & ---  & ---  & 5 (100.00\%) \\ 
192& 2 & ---  & ---  & ---  & ---  & 2 (100.00\%) \\ 
193& 13 & ---  & ---  & ---  & ---  & 13 (100.00\%) \\ 
194& 252 & ---  & 4 (1.59\%) & 6 (2.38\%) & ---  & 242 (96.03\%) \\ 
195& 1 & ---  & ---  & ---  & ---  & 1 (100.00\%) \\ 
197& 10 & ---  & ---  & ---  & ---  & 10 (100.00\%) \\ 
198& 93 & ---  & ---  & ---  & 1 (1.08\%) & 92 (98.92\%) \\ 
199& 23 & ---  & ---  & ---  & ---  & 23 (100.00\%) \\ 
200& 5 & ---  & ---  & ---  & 1 (20.00\%) & 4 (80.00\%) \\ 
201& 2 & ---  & ---  & ---  & ---  & 2 (100.00\%) \\ 
202& 8 & ---  & ---  & ---  & ---  & 8 (100.00\%) \\ 
203& 5 & ---  & ---  & ---  & ---  & 5 (100.00\%) \\ 
204& 22 & ---  & ---  & ---  & 1 (4.55\%) & 21 (95.45\%) \\ 
205& 81 & 1 (1.23\%) & ---  & 1 (1.23\%) & ---  & 79 (97.53\%) \\ 
206& 30 & ---  & 1 (3.33\%) & ---  & 1 (3.33\%) & 28 (93.33\%) \\ 
212& 6 & ---  & ---  & ---  & ---  & 6 (100.00\%) \\ 
213& 5 & ---  & ---  & ---  & ---  & 5 (100.00\%) \\ 
214& 6 & ---  & ---  & ---  & ---  & 6 (100.00\%) \\ 
215& 30 & ---  & ---  & ---  & ---  & 30 (100.00\%) \\ 
216& 234 & 3 (1.28\%) & ---  & 6 (2.56\%) & 8 (3.42\%) & 217 (92.74\%) \\ 
217& 42 & ---  & ---  & ---  & ---  & 42 (100.00\%) \\ 
218& 30 & ---  & ---  & ---  & ---  & 30 (100.00\%) \\ 
219& 3 & ---  & ---  & ---  & ---  & 3 (100.00\%) \\ 
220& 27 & ---  & ---  & ---  & ---  & 27 (100.00\%) \\ 
221& 164 & ---  & 12 (7.32\%) & 1 (0.61\%) & ---  & 151 (92.07\%) \\ 
223& 3 & ---  & ---  & ---  & ---  & 3 (100.00\%) \\ 
224& 7 & ---  & ---  & 1 (14.29\%) & ---  & 6 (85.71\%) \\ 
225& 374 & ---  & 14 (3.74\%) & ---  & 6 (1.60\%) & 354 (94.65\%) \\ 
226& 1 & ---  & ---  & ---  & ---  & 1 (100.00\%) \\ 
227& 143 & ---  & 1 (0.70\%) & ---  & 2 (1.40\%) & 140 (97.90\%) \\ 
229& 20 & ---  & ---  & ---  & ---  & 20 (100.00\%) \\ 
230& 1 & ---  & ---  & ---  & ---  & 1 (100.00\%) \\ 
\hline
Total& 15865 & 49 (0.31\%) & 78 (0.49\%) & 42 (0.26\%) & 24 (0.15\%) & 15,672 (98.78\%)
\\
\hline
\end{longtable}
}

{\tiny
\begin{longtable}{|c|c|c|c|c|c|c|c|c|c|}
\caption[SOC-driven topological phase transition statistics for NLC-SM materials w/o SOC]{SOC-driven topological phase transition statistics for NLC-SM-classified semimetals w/o SOC.  In order, the first two columns in this table list the SG and the number of NLC-SM-classified unique materials in the SG w/o SOC.  If an SG is not listed in the first column, then there are no NLC-SM-classified unique materials in the SG w/o SOC.  In the remaining five columns, restricting to the same unique materials listed in the second column, we respectively list the number of unique materials that become classified as NLC, SEBR, ES, ESFD, and LCEBR when the effects of SOC are incorporated.  For all of the topological classification categories in the table, each number of materials is accompanied by the percentage of materials in the SG in the topological class.  When there are no materials in the topological class, the table entry is marked with a dashed horizontal line.
\label{tb:statistics_nosoc_to_soc_uniquematerials_NLC-SM}}\\
\hline
SG  & \# Mat. & NLC & SEBR & ES & ESFD & LCEBR\\
\hline  
2& 165 & 158 (95.76\%) & ---  & ---  & ---  & 7 (4.24\%) \\ 
11& 4 & 4 (100.00\%) & ---  & ---  & ---  & ---  \\ 
12& 24 & 2 (8.33\%) & 20 (83.33\%) & ---  & ---  & 2 (8.33\%) \\ 
13& 1 & 1 (100.00\%) & ---  & ---  & ---  & ---  \\ 
14& 24 & 23 (95.83\%) & ---  & ---  & ---  & 1 (4.17\%) \\ 
15& 22 & 18 (81.82\%) & ---  & ---  & ---  & 4 (18.18\%) \\ 
58& 1 & ---  & ---  & 1 (100.00\%) & ---  & ---  \\ 
60& 1 & 1 (100.00\%) & ---  & ---  & ---  & ---  \\ 
81& 1 & ---  & ---  & ---  & ---  & 1 (100.00\%) \\ 
82& 2 & 1 (50.00\%) & ---  & 1 (50.00\%) & ---  & ---  \\ 
85& 2 & ---  & ---  & ---  & ---  & 2 (100.00\%) \\ 
87& 2 & ---  & 2 (100.00\%) & ---  & ---  & ---  \\ 
166& 2 & ---  & 2 (100.00\%) & ---  & ---  & ---  \\ 
\hline
Total& 251 & 208 (82.87\%) & 24 (9.56\%) & 2 (0.8\%) & ---  & 17 (6.77\%)
\\
\hline
\end{longtable}
}

{\tiny
\begin{longtable}{|c|c|c|c|c|c|c|c|c|c|}
\caption[SOC-driven topological phase transition statistics for SEBR-SM materials w/o SOC]{SOC-driven topological phase transition statistics for SEBR-SM-classified semimetals w/o SOC.  In order, the first two columns in this table list the SG and the number of SEBR-SM-classified unique materials in the SG w/o SOC.  If an SG is not listed in the first column, then there are no SEBR-SM-classified unique materials in the SG w/o SOC.  In the remaining five columns, restricting to the same unique materials listed in the second column, we respectively list the number of unique materials that become classified as NLC, SEBR, ES, ESFD, and LCEBR when the effects of SOC are incorporated.  For all of the topological classification categories in the table, each number of materials is accompanied by the percentage of materials in the SG in the topological class.  When there are no materials in the topological class, the table entry is marked with a dashed horizontal line.
\label{tb:statistics_nosoc_to_soc_uniquematerials_SEBR-SM}}\\
\hline
SG  & \# Mat. & NLC & SEBR & ES & ESFD & LCEBR\\
\hline
147& 3 & ---  & 3 (100.00\%) & ---  & ---  & ---  \\ 
148& 35 & ---  & 29 (82.86\%) & 5 (14.29\%) & ---  & 1 (2.86\%) \\ 
162& 1 & ---  & 1 (100.00\%) & ---  & ---  & ---  \\ 
164& 2 & ---  & 2 (100.00\%) & ---  & ---  & ---  \\ 
166& 2 & ---  & 2 (100.00\%) & ---  & ---  & ---  \\ 
167& 1 & ---  & 1 (100.00\%) & ---  & ---  & ---  \\ 
\hline
Total& 44 & ---  & 38 (86.36\%) & 5 (11.36\%) & ---  & 1 (2.27\%)
\\
\hline
\end{longtable}
}

\afterpage{\clearpage}

\section{Lists of Representative Topological Materials}\label{App:MaterialsSelection}

\newcommand{\ztwoTINLC}[1]{The NLC-classified 3D strong TIs with the largest band gaps or the fewest and smallest bulk Fermi pockets. (part #1/5)}
\newcommand{\ztwoTISEBR}[1]{The SEBR-classified 3D strong TIs with the largest band gaps or the fewest and smallest bulk Fermi pockets. (part #1/5)}
\newcommand{\zfourHOTIsNLC}[1]{The NLC-classified, inversion-symmetry-indicated HOTIs with the largest band gaps or the fewest and smallest bulk Fermi pockets. (part #1/6)}
\newcommand{\zfourHOTIsSEBR}[1]{The SEBR-classified, inversion-symmetry-indicated HOTIs with the largest band gaps or the fewest and smallest bulk Fermi pockets. (part #1/2)}
\newcommand{\zfourTCIsandWTIsNLC}[1]{The NLC-classified, inversion-symmetry-indicated weak TIs and TCIs with the largest band gaps or the fewest and smallest bulk Fermi pockets. (part #1/2)}
\newcommand{\zfourTCIsandWTIsSEBR}[1]{The SEBR-classified, inversion-symmetry-indicated weak TIs and TCIs with the largest band gaps or the fewest and smallest bulk Fermi pockets. (part #1/14)}
\newcommand{\zeightHOTIsTCISEBR}[1]{The SEBR-classified, fourfold-rotation-anomaly and mirror TCIs characterized by $Z_{8}=4$ with the largest band gaps or the fewest and smallest bulk Fermi pockets. (part #1/4)}
\newcommand{\zfourTCIweakSEBR}[1]{The SEBR-classified TCIs with the largest band gaps or the fewest and smallest bulk Fermi pockets whose topology is characterized by $Z_{4}=0$ and undefined or trivial values of $Z_{8}$. (part #1/4)}
\newcommand{\strongSOCNOTCHIRALfoldNewFermionESFD}[1]{The ESFD-classified topological semimetals with the simplest bulk Fermi surfaces and fourfold-degenerate, achiral spin-3/2 fermions close to $E_{F}$. (part #1/10)}
\newcommand{\strongSOCchiralNewFermionESFD}[1]{The ESFD-classified topological semimetals in chiral space groups with the simplest bulk Fermi surfaces and enforced chiral fermions close to $E_{F}$. (part #1/6)}
\newcommand{\strongSOCeightfoldESFD}[1]{The ESFD-classified topological semimetals with the simplest bulk Fermi surfaces and eightfold-degenerate double Dirac fermions close to $E_{F}$. (part #1/3)}
\newcommand{\strongSOCES}[1]{The ES-classified topological semimetals with the simplest bulk Fermi surfaces. (part #1/6)}
\newcommand{\ESFDnoSOCtoTI}[1]{The spinful topological (crystalline) insulators with the largest bulk gaps or the fewest Fermi pockets that are classified as ESFD topological semimetals when the effects of SOC are neglected. (part #1/2)}
\newcommand{\ESnoSOCtoTI}[1]{The spinful topological (crystalline) insulators with the largest bulk gaps or the fewest Fermi pockets that are classified as ES topological semimetals when the effects of SOC are neglected. (part #1/2)}
\newcommand{\fragileSEBR}[1]{The SEBR-classified topological (crystalline) insulators with the largest bulk gaps or the fewest and smallest bulk Fermi pockets that host well-isolated fragile bands at or close to $E_{F}$. (part #1/2)}
\newcommand{\fragileESFD}[1]{The ESFD-classified topological semimetals with the simplest bulk Fermi surfaces and well-isolated fragile bands at or close to $E_{F}$. (part #1/2)}
\newcommand{\strongSOCquadraticotherESFD}[1]{The ESFD-classified topological semimetals with the simplest bulk Fermi surfaces beyond those shown in Figs.~\ref{fig:strongSOC_3fold_ESFD} through~\ref{fig:strongSOC_8fold_ESFD3}. (part #1/2)}

\newcommand{\rTopoNLC}[1]{The RTopo TIs and TCIs that are NLC-classified at the Fermi level and have the fewest and smallest bulk Fermi pockets when the Fermi level is set to $0$, or set to the next highest filling at which an insulating gap is permitted by band connectivity. (part #1/2)}
\newcommand{\rTopoSEBR}[1]{The RTopo TIs and TCIs that are SEBR-classified at the Fermi level and have the fewest and smallest bulk Fermi pockets when the Fermi level is set to $0$, or set to the next highest filling at which an insulating gap is permitted by band connectivity. (part #1/6)}

In this section, we will provide representative examples of materials on~\webNoICSD~with nontrivial bulk topology that can be identified from symmetry eigenvalues and band connectivity (TQC)~\cite{QuantumChemistry,Bandrep1,Bandrep2,Bandrep3,JenFragile1}.  Though many of the materials listed in this section have been identified in previous theoretical and experimental works, most of the materials listed in this section were not previously recognized as hosting nontrivial electronic band topology.  In each part of this section, we will highlight the most well-studied examples of known topological materials.  First, in \supappref{App:insulators}, we will enumerate the 3D topological insulators (TIs)~\cite{FuKaneMele,FuKaneInversion,AndreiInversion}, topological crystalline insulators (TCIs)~\cite{LiangTCIOriginal,TeoFuKaneTCI,HsiehTCI,HourglassInsulator,Cohomological,DiracInsulator,MobiusInsulator}, and higher-order TIs (HOTIs)~\cite{multipole,WladPump,HOTIBernevig,HOTIBismuth,ChenRotation,HOTIChen,HigherOrderTIPiet,AshvinIndicators,ChenTCI,AshvinTCI,TMDHOTI,WiederAxion} with the fewest electron and hole pockets at the Fermi energy ($E_{F}$).  Next, in \supappref{App:semimetals}, we will list the topological semimetals with the simplest bulk Fermi surfaces, including examples of both conventional Dirac~\cite{ZJDirac,ZJDirac2,SteveDirac,JuliaDirac,CavaDirac1,CavaDirac2,YulinCadmiumExp,NagaosaDirac,NaDirac,SYDiracSurface,HingeSM} as well as unconventional fermion~\cite{DDP,NewFermions,KramersWeyl,RhSiArc,CoSiArc,CoSiObserveJapan,CoSiObserveHasan,CoSiObserveChina,AlPtObserve,PdGaObserve} semimetals.  Then, in \supappref{App:transitions}, we will provide representative examples of spin-orbit-coupling- (SOC-) driven semimetal-to-insulator transitions that result in topological (crystalline) insulators with large gaps at $E_{F}$.  Finally, in \supappref{App:fragileBands}, we will show the materials with the cleanest insulating or semimetallic Fermi surfaces that host fragile topological bands~\cite{AshvinFragile,JenFragile1,AshvinFragile2,BarryFragile,AdrianFragile,KoreanFragile,ZhidaFragile,FragileFlowMeta,ZhidaBLG,ZhidaFragile2,KoreanFragileInversion,DelicateAris,MBP_fragile} at or near $E_{F}$.  We note that in this section, we only display electronic structures and topological data calculated in the presence of SOC; for each material, the electronic structure and topological data calculated in the absence of SOC can be accessed by clicking on the ICSD number listed above each plot.  In Fig.~\ref{fig:BSTypicalPlot}, we provide an example detailing the labeling scheme and information contained within each of the band-structure plots shown in this section.

\begin{figure}[ht]
\centering
\includegraphics[width=0.8\textwidth,angle=0]{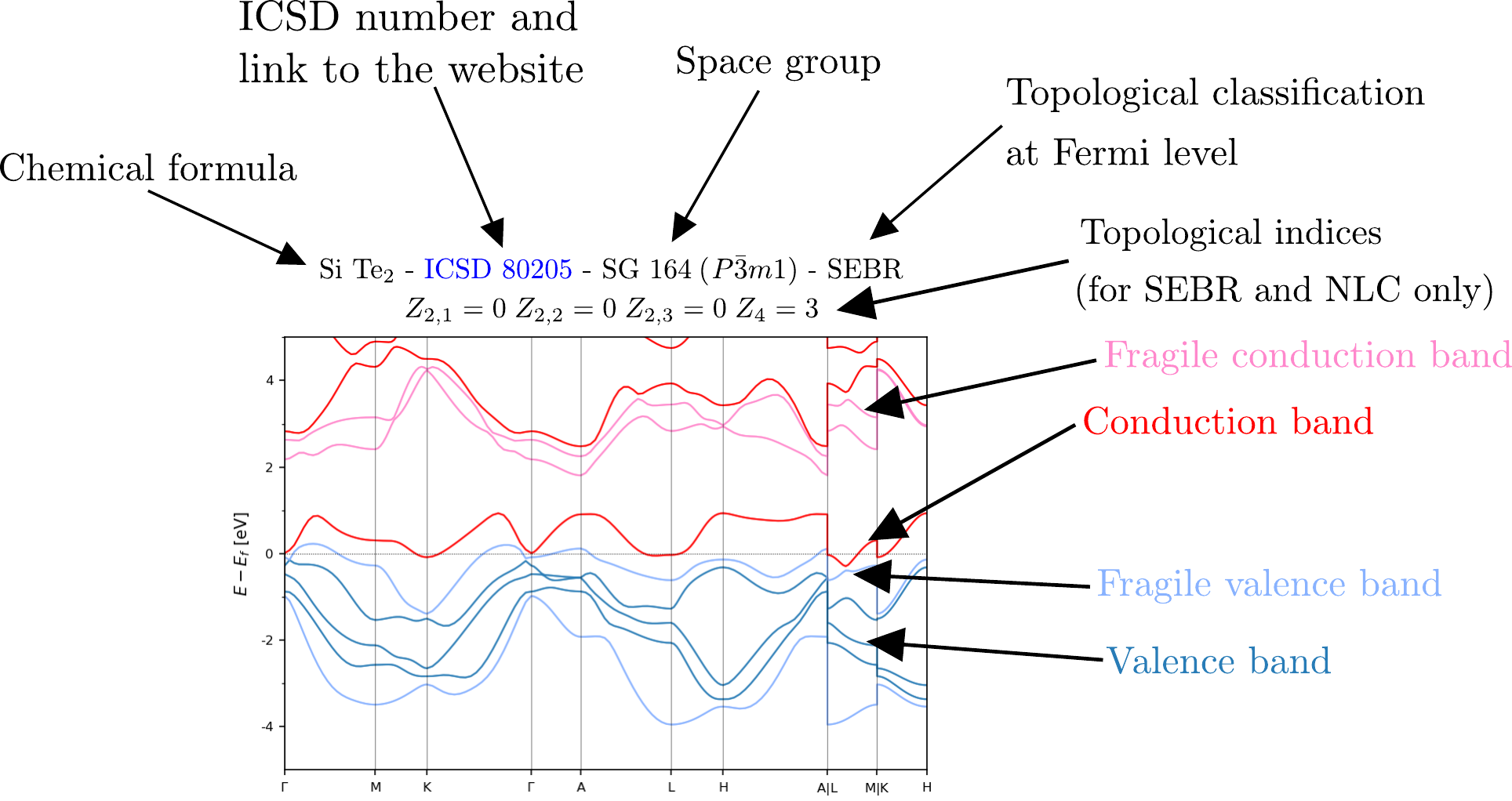}
\caption{Example band-structure plot.  In this figure, we show the band structure and topological indices of SiTe$_2$ [\icsdweb{80205}, SG 164 ($P\bar{3}m1$)] as an example of the labeling scheme and information contained within each of the band-structure plots shown in this section.  First, at the top of each band-structure plot, we provide the chemical formula, the ICSD number with a hyperlink to~\webNoICSD, the space group symbol and number, and the topological classification at the Fermi level (\emph{e.g.} NLC, SEBR, ES, or ESFD).  If the ICSD entry is classified as SEBR or NLC at $E_{F}$, then we further list the cumulative symmetry-based topological indices (stable SIs) at $E_{F}$ in the notation established in Ref.~\onlinecite{ChenTCI}.  In the band-structure plot, trivial and stable topological conduction (valence) bands are labeled in red (blue), and fragile conduction (valence) bands are labeled in pink (light blue).}
\label{fig:BSTypicalPlot}
\end{figure}

We additionally emphasize that for the materials highlighted in this section, we cannot preclude the existence of non-symmetry-indicated topological (crystalline) insulating (\emph{e.g.} hourglass insulating~\cite{HourglassInsulator,Cohomological,DiracInsulator,MobiusInsulator} or high-fold mirror-Chern TCI~\cite{TeoFuKaneTCI,ChenBernevigTCI}) or Weyl semimetal~\cite{AshvinWeyl1,ZahidWeyl,AndreiWeyl} phases.  These non-symmetry-indicated phases cannot be identified through TQC alone, and must instead be diagnosed by performing additional (nested) Wilson-loop calculations~\cite{Fidkowski2011,AndreiXiZ2,ArisInversion,HourglassInsulator,Cohomological,DiracInsulator,multipole,TMDHOTI,WiederAxion,S4Weyl}.

\subsection{Symmetry-Indicated TIs, TCIs, and HOTIs}
\label{App:insulators}

In this section, we will list the 3D TIs, TCIs, and HOTIs that can be diagnosed from their symmetry eigenvalues through the topological-index formulas established in Refs.~\onlinecite{FuKaneMele,FuKaneInversion,HOTIChen,ChenTCI,AshvinTCI}.  First, in \supappref{App:strongTIs}, we will list the 3D strong TIs indicated through the Fu-Kane parity criterion~\cite{FuKaneMele,FuKaneInversion}.  Next, in \supappref{App:s4weylTI}, we will list the $S_{4z}$ rotoinversion-indicated 3D TIs and Weyl semimetals~\cite{ChenTCI,AshvinTCI,S4Weyl}.  Then, in \supappref{App:z4HOTIs}, \supappref{App:weakTIs}, \supappref{App:rotationAnomaly}, \supappref{App:z4trivial}, and \supappref{App:z3TCIs}, we will list the TCIs and HOTIs diagnosed through the symmetry-based indicators (stable SIs) more recently introduced in Refs.~\onlinecite{AshvinIndicators,HOTIChen,TMDHOTI,HOTIBismuth,ChenTCI,AshvinTCI,HOTIBernevig}.  Lastly, in \supappref{App:RTopoMaterials}, we will list the repeat-topological (RTopo) TIs and TCIs (rigorously defined in~\supappref{App:DefRTopo}), which exhibit symmetry-indicated stable topology at the Fermi level and at the next gap below $E_{F}$ as determined by band connectivity (see~\supappref{App:TQCReview_appendix}).

\subsubsection{3D Strong TIs}
\label{App:strongTIs}

In this section, we list the symmetry-indicated 3D strong TIs with the largest band gaps or the fewest and smallest bulk Fermi pockets.  First, in Figs.~\ref{fig:z2TI_NLC1},~\ref{fig:z2TI_NLC2},~\ref{fig:z2TI_NLC3},~\ref{fig:z2TI_NLC4}, and~\ref{fig:z2TI_NLC5}, we list the 3D TIs classified as NLC, and then, in Figs.~\ref{fig:z2TI_SEBR1},~\ref{fig:z2TI_SEBR2},~\ref{fig:z2TI_SEBR3},~\ref{fig:z2TI_SEBR4}, and~\ref{fig:z2TI_SEBR5}, we list the 3D TIs classified as SEBR.  3D TIs characteristically exhibit odd numbers of twofold Dirac-cone surface states~\cite{FuKaneInversion,FuKaneMele,HsiehBismuthSelenide,HsiehDiracInsulator} that are protected by time-reversal ($\mathcal{T}$) symmetry.  Though it was originally recognized that 3D TIs could be identified through a $\mathbb{Z}_{2}$-valued (Fu-Kane) index based on parity [inversion ($\mathcal{I}$)] eigenvalues~\cite{FuKaneMele,FuKaneInversion}, a more exhaustive consideration of eigenvalue indicators (stable SIs, see Refs.~\onlinecite{AshvinIndicators,HOTIChen,TMDHOTI,HOTIBismuth,ChenTCI,AshvinTCI,HOTIBernevig} and \supappref{App:TQCReview_appendix}) has revealed that the Fu-Kane parity index is in fact subsumed by a $\mathbb{Z}_{4}$-valued index $Z_{4}$ that includes both HOTIs and strong TIs.  In the notation of Ref.~\onlinecite{ChenTCI}, which we use throughout this work, 3D TIs are indicated either by $Z_{4}=1,3$ (which coincide with the Fu-Kane parity criterion).  The most well known example of a 3D TI~\cite{HsiehBismuthSelenide} -- Bi$_2$Se$_3$ [\icsdweb{617079}, SG 166 ($R\bar{3}m$)] -- is classified as SEBR, and accordingly appears in Fig.~\ref{fig:z2TI_SEBR3}.


\begin{figure}[ht]
\centering
\begin{tabular}{c c}
\scriptsize{$\rm{Ca} \rm{As}_{3}$ - \icsdweb{193} - SG 2 ($P\bar{1}$) - NLC} & \scriptsize{$\rm{Ca} \rm{P}_{3}$ - \icsdweb{74479} - SG 2 ($P\bar{1}$) - NLC}\\
\tiny{ $\;Z_{2,1}=1\;Z_{2,2}=0\;Z_{2,3}=0\;Z_4=1$ } & \tiny{ $\;Z_{2,1}=0\;Z_{2,2}=1\;Z_{2,3}=0\;Z_4=1$ }\\
\includegraphics[width=0.38\textwidth,angle=0]{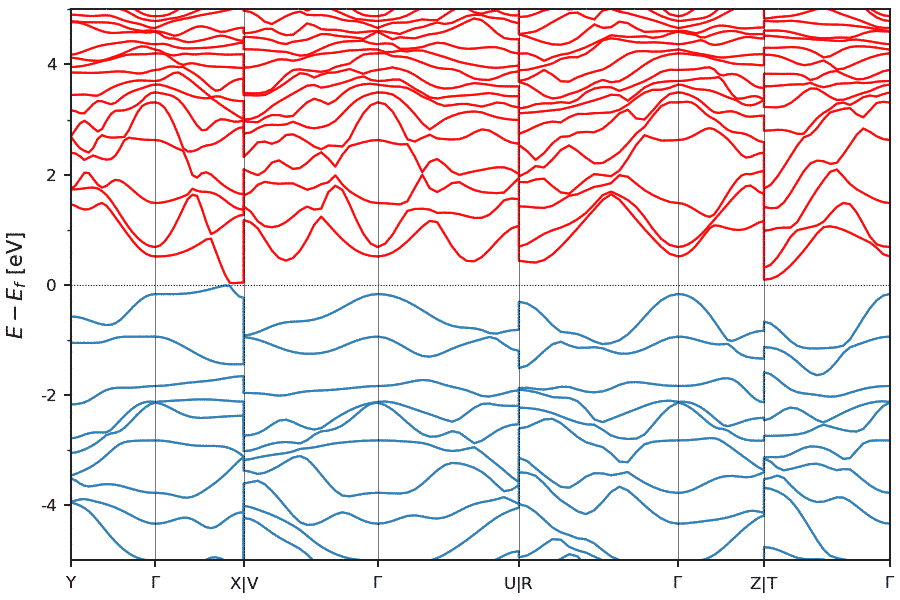} & \includegraphics[width=0.38\textwidth,angle=0]{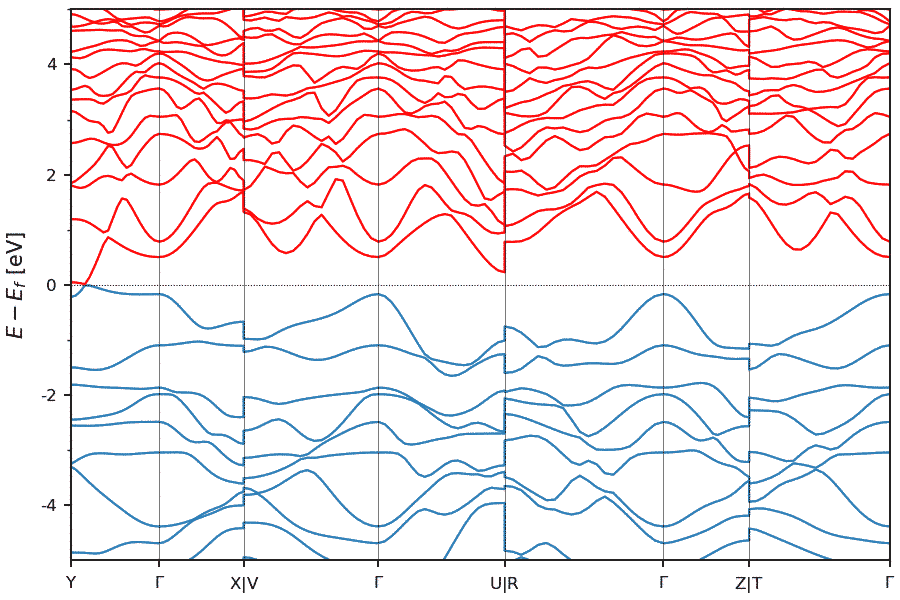}\\
\end{tabular}
\begin{tabular}{c c}
\scriptsize{$\rm{Ni} (\rm{Ta}_{2} \rm{Se}_{7})$ - \icsdweb{61352} - SG 12 ($C2/m$) - NLC} & \scriptsize{$\rm{Pb}_{3} \rm{Bi}_{2} \rm{S}_{6}$ - \icsdweb{92981} - SG 12 ($C2/m$) - NLC}\\
\tiny{ $\;Z_{2,1}=0\;Z_{2,2}=0\;Z_{2,3}=0\;Z_4=3$ } & \tiny{ $\;Z_{2,1}=1\;Z_{2,2}=1\;Z_{2,3}=0\;Z_4=3$ }\\
\includegraphics[width=0.38\textwidth,angle=0]{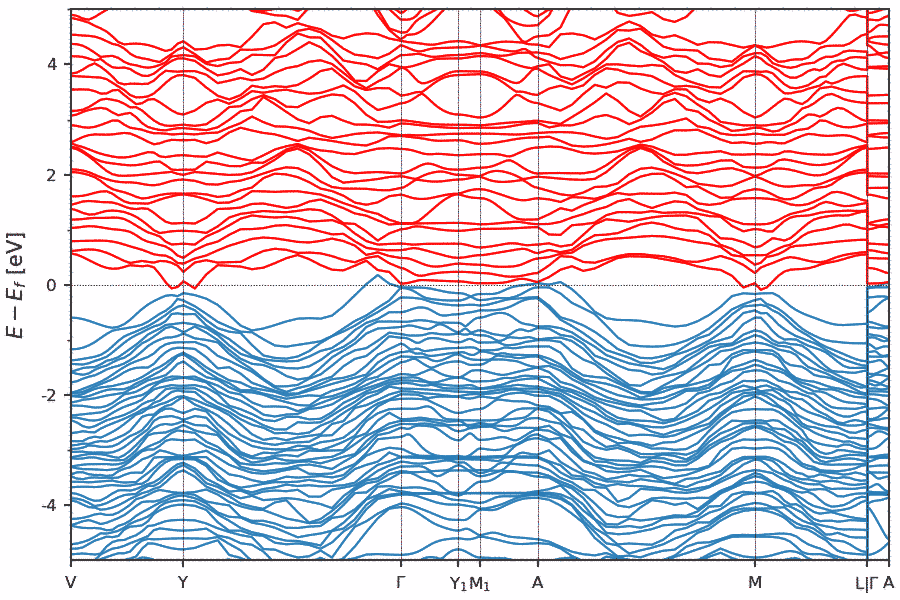} & \includegraphics[width=0.38\textwidth,angle=0]{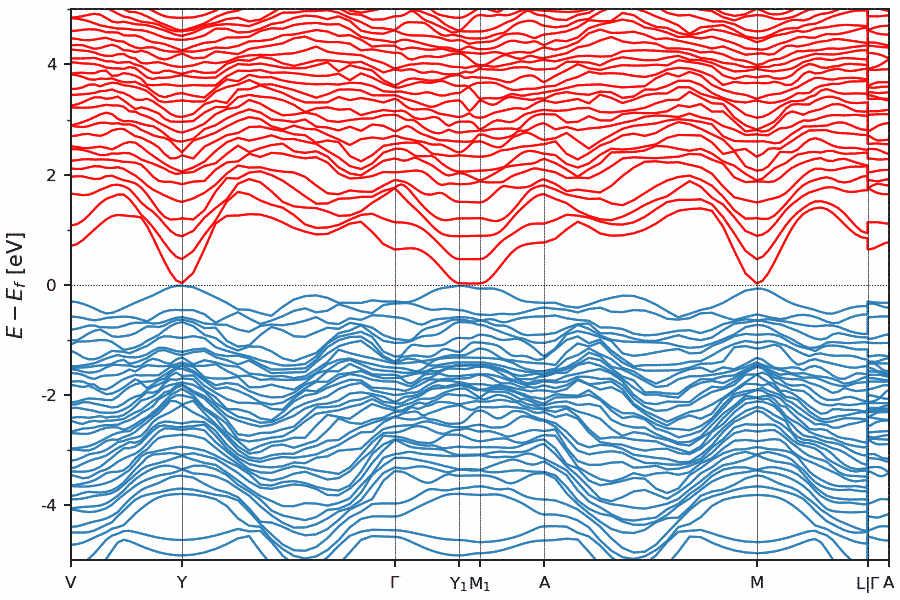}\\
\end{tabular}
\begin{tabular}{c c}
\scriptsize{$\rm{Nb}_{2} \rm{Pd} \rm{S}_{5}$ - \icsdweb{192215} - SG 12 ($C2/m$) - NLC} & \scriptsize{$\rm{Sr}_{3} \rm{Al}_{2} \rm{Ge}_{4}$ - \icsdweb{417968} - SG 12 ($C2/m$) - NLC}\\
\tiny{ $\;Z_{2,1}=0\;Z_{2,2}=0\;Z_{2,3}=1\;Z_4=1$ } & \tiny{ $\;Z_{2,1}=1\;Z_{2,2}=1\;Z_{2,3}=0\;Z_4=3$ }\\
\includegraphics[width=0.38\textwidth,angle=0]{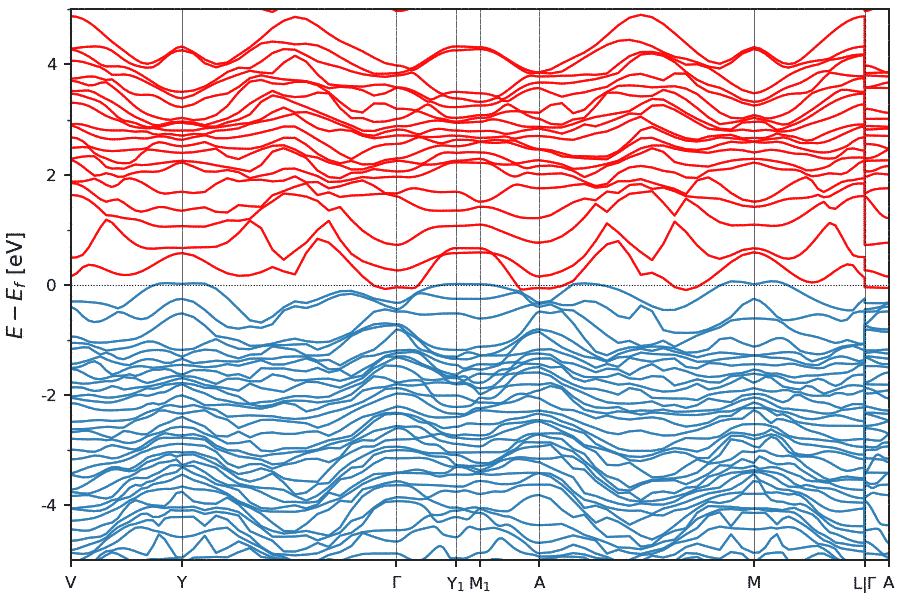} & \includegraphics[width=0.38\textwidth,angle=0]{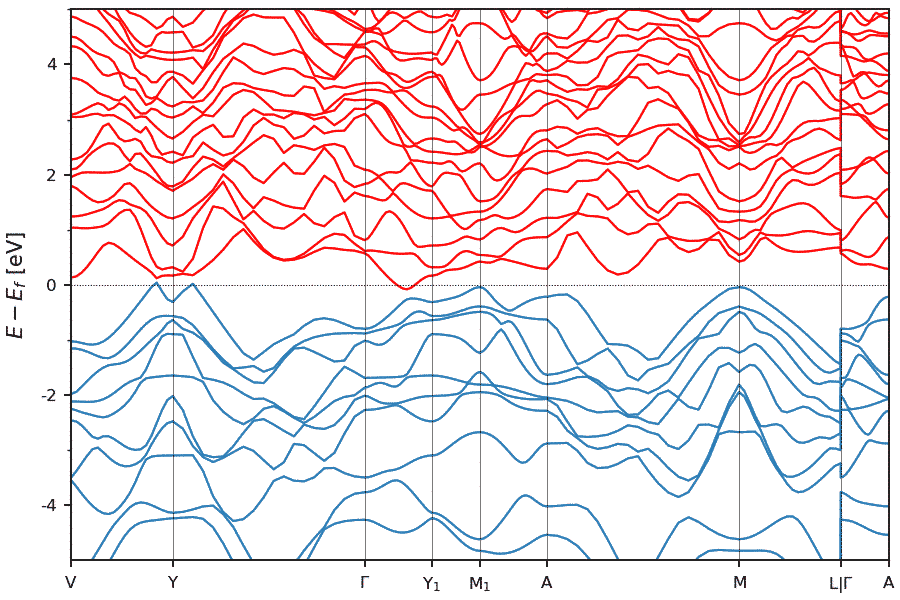}\\
\end{tabular}
\begin{tabular}{c c}
\scriptsize{$\rm{Ba}_{11} \rm{Cd}_{6} \rm{Sb}_{12}$ - \icsdweb{418886} - SG 12 ($C2/m$) - NLC} & \scriptsize{$\rm{Ba}_{2} \rm{Cd}_{2} \rm{Sb}_{3}$ - \icsdweb{420834} - SG 12 ($C2/m$) - NLC}\\
\tiny{ $\;Z_{2,1}=0\;Z_{2,2}=0\;Z_{2,3}=0\;Z_4=3$ } & \tiny{ $\;Z_{2,1}=1\;Z_{2,2}=1\;Z_{2,3}=0\;Z_4=1$ }\\
\includegraphics[width=0.38\textwidth,angle=0]{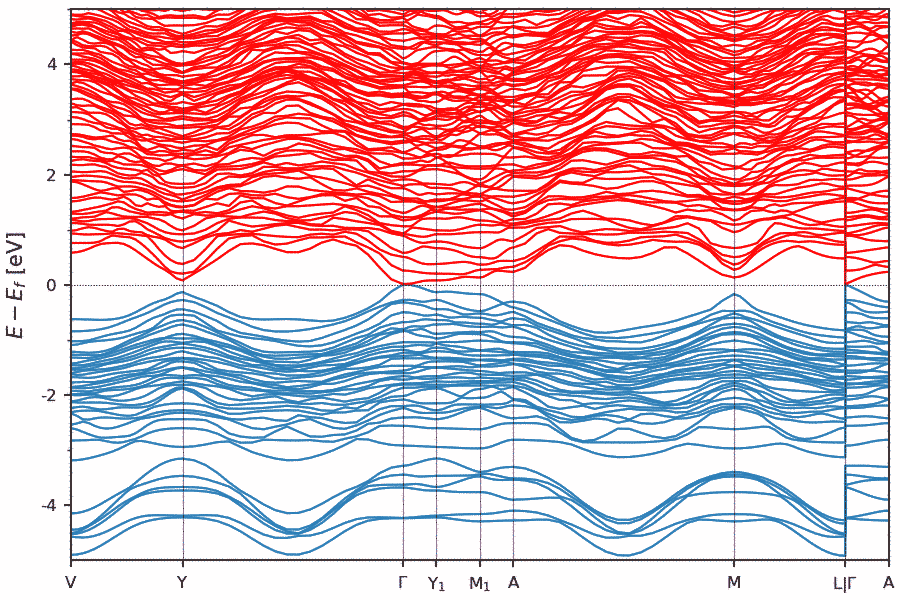} & \includegraphics[width=0.38\textwidth,angle=0]{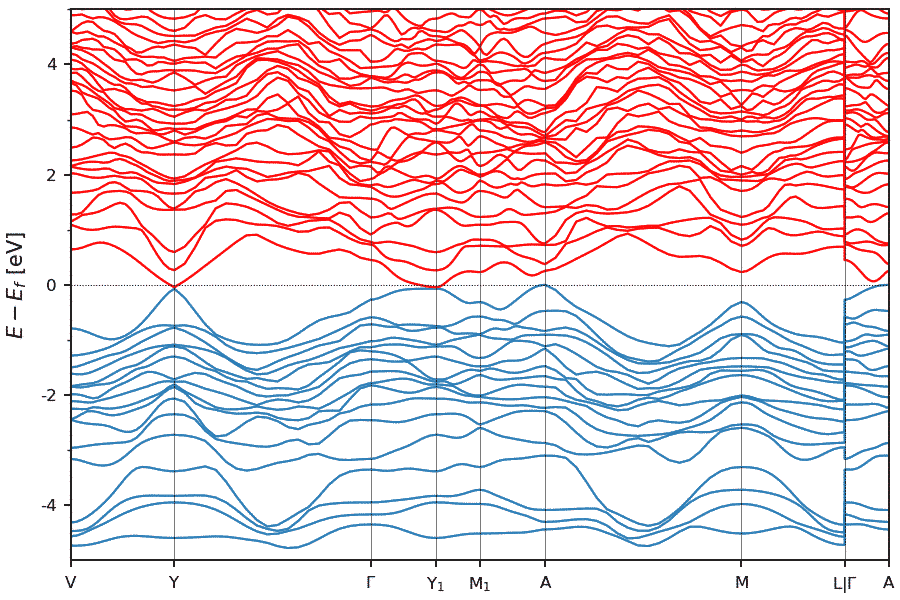}\\
\end{tabular}

\caption{\ztwoTINLC{1}}
\label{fig:z2TI_NLC1}
\end{figure}

\begin{figure}[ht]
\centering
\begin{tabular}{c c}
\scriptsize{$\rm{Ba}_{5} \rm{Cd}_{2} \rm{Sb}_{4} \rm{O}_{2}$ - \icsdweb{423458} - SG 12 ($C2/m$) - NLC} & \scriptsize{$\rm{Pb}_{7} \rm{Bi}_{4} \rm{Se}_{13}$ - \icsdweb{427410} - SG 12 ($C2/m$) - NLC}\\
\tiny{ $\;Z_{2,1}=1\;Z_{2,2}=1\;Z_{2,3}=1\;Z_4=3$ } & \tiny{ $\;Z_{2,1}=1\;Z_{2,2}=1\;Z_{2,3}=1\;Z_4=1$ }\\
\includegraphics[width=0.38\textwidth,angle=0]{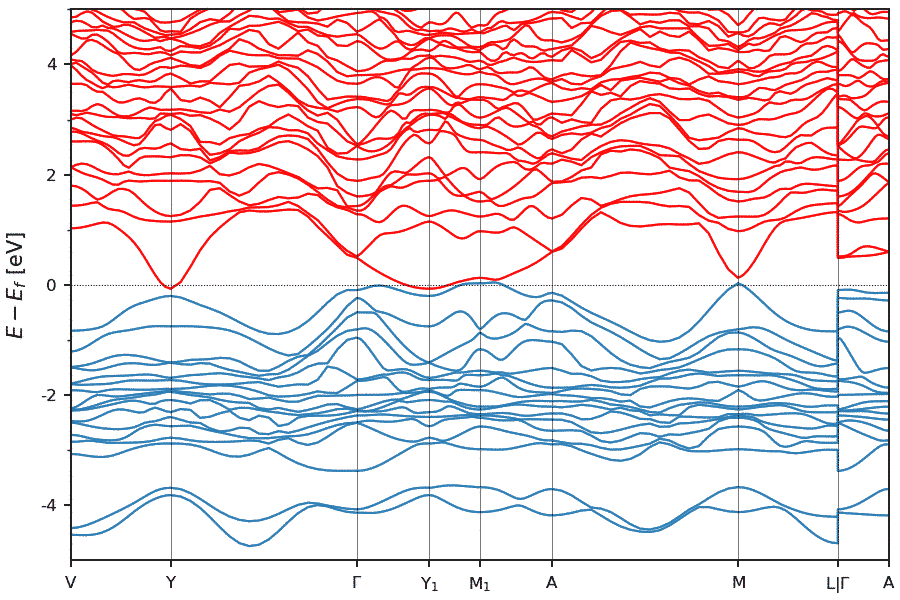} & \includegraphics[width=0.38\textwidth,angle=0]{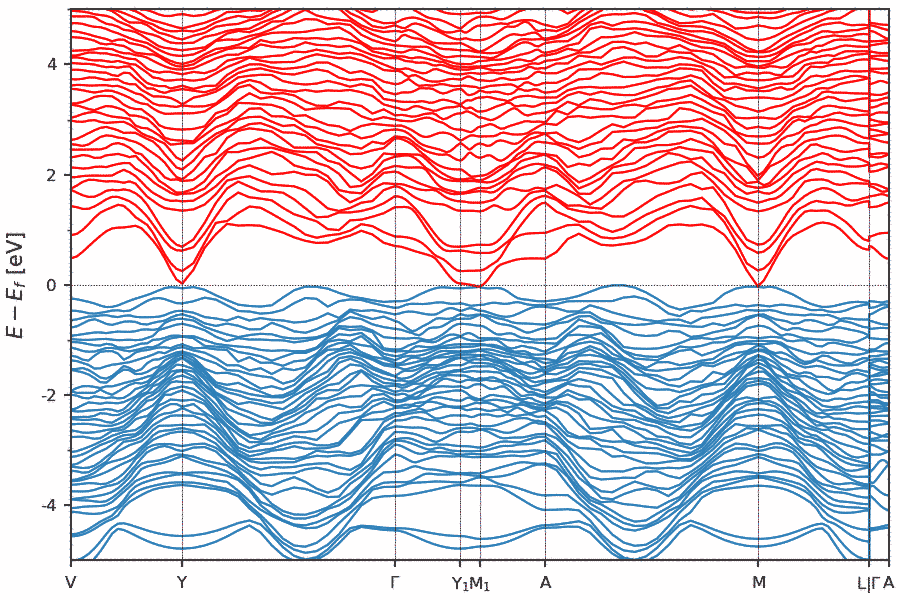}\\
\end{tabular}
\begin{tabular}{c c}
\scriptsize{$\rm{Co} \rm{Sb}_{2}$ - \icsdweb{63553} - SG 14 ($P2_1/c$) - NLC} & \scriptsize{$\rm{Ag}_{2} \rm{Te}$ - \icsdweb{73230} - SG 14 ($P2_1/c$) - NLC}\\
\tiny{ $\;Z_{2,1}=1\;Z_{2,2}=0\;Z_{2,3}=0\;Z_4=3$ } & \tiny{ $\;Z_{2,1}=0\;Z_{2,2}=0\;Z_{2,3}=0\;Z_4=3$ }\\
\includegraphics[width=0.38\textwidth,angle=0]{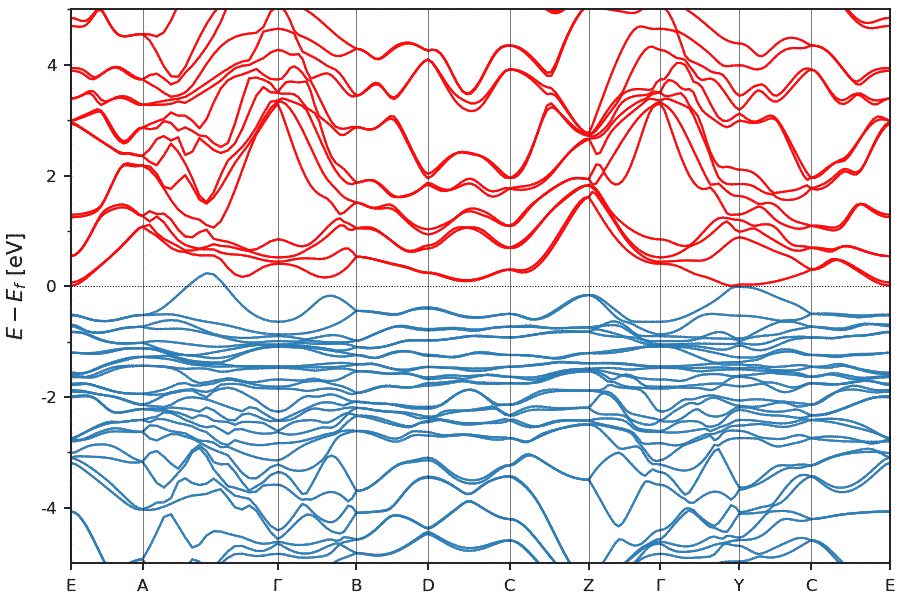} & \includegraphics[width=0.38\textwidth,angle=0]{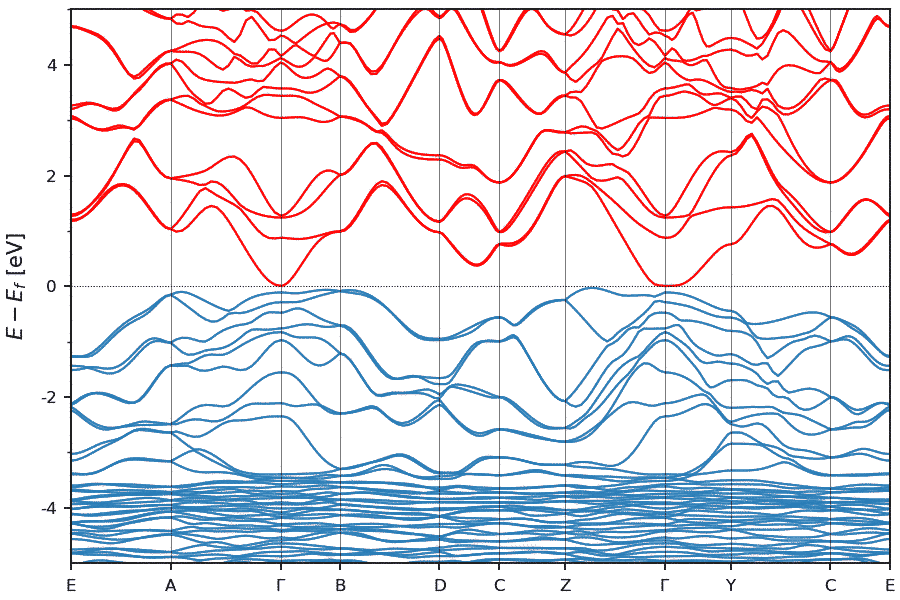}\\
\end{tabular}
\begin{tabular}{c c}
\scriptsize{$\rm{K}_{10} (\rm{Sn}_{2} \rm{Bi}_{6})$ - \icsdweb{300261} - SG 14 ($P2_1/c$) - NLC} & \scriptsize{$\rm{Ta}_{2} \rm{Ni} \rm{Se}_{5}$ - \icsdweb{61148} - SG 15 ($C2/c$) - NLC}\\
\tiny{ $\;Z_{2,1}=0\;Z_{2,2}=0\;Z_{2,3}=0\;Z_4=3$ } & \tiny{ $\;Z_{2,1}=0\;Z_{2,2}=0\;Z_{2,3}=0\;Z_4=3$ }\\
\includegraphics[width=0.38\textwidth,angle=0]{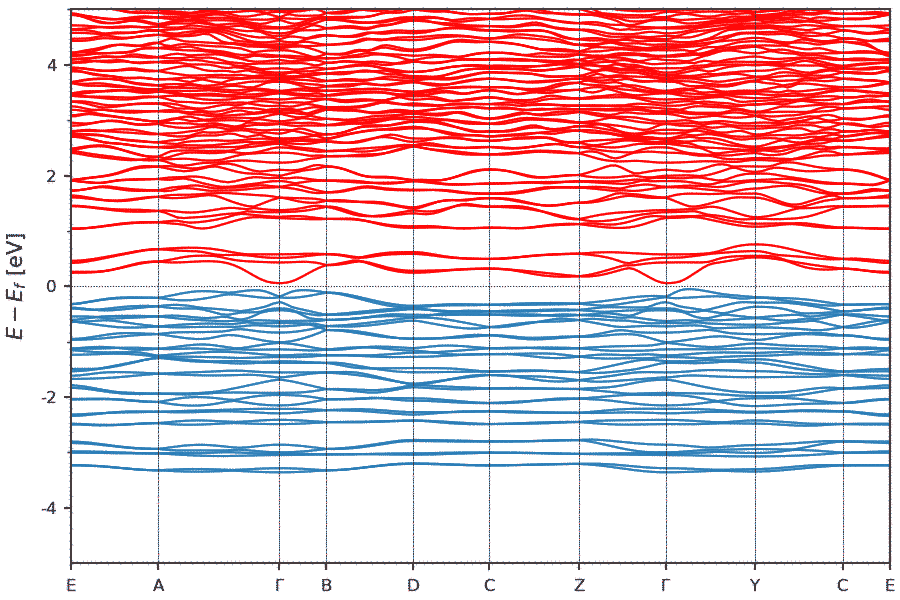} & \includegraphics[width=0.38\textwidth,angle=0]{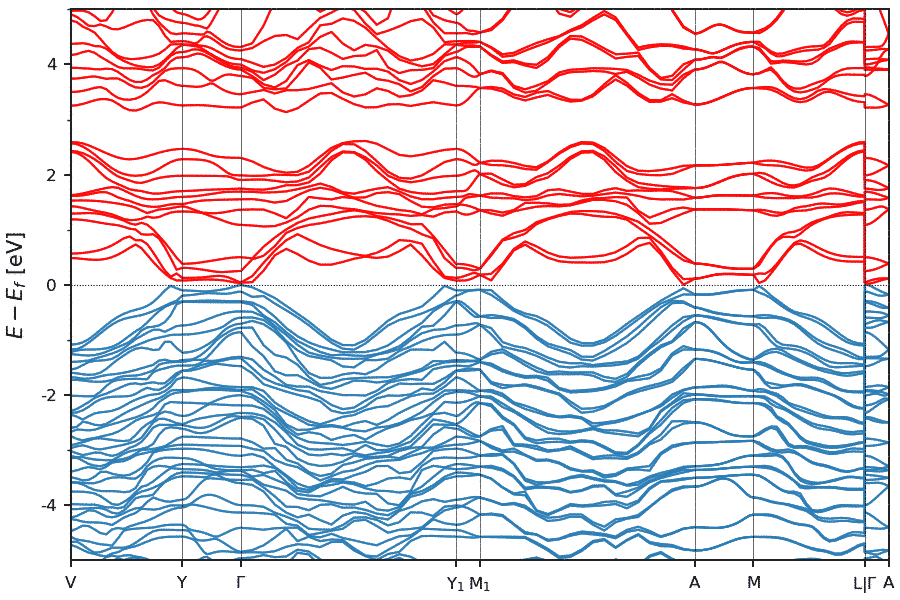}\\
\end{tabular}
\begin{tabular}{c c}
\scriptsize{$\rm{Y} \rm{V} \rm{B}_{4}$ - \icsdweb{615679} - SG 55 ($Pbam$) - NLC} & \scriptsize{$\rm{In}_{2} \rm{O}_{3}$ - \icsdweb{236900} - SG 60 ($Pbcn$) - NLC}\\
\tiny{ $\;Z_{2,1}=0\;Z_{2,2}=0\;Z_{2,3}=0\;Z_4=1$ } & \tiny{ $\;Z_{2,1}=0\;Z_{2,2}=0\;Z_{2,3}=0\;Z_4=3$ }\\
\includegraphics[width=0.38\textwidth,angle=0]{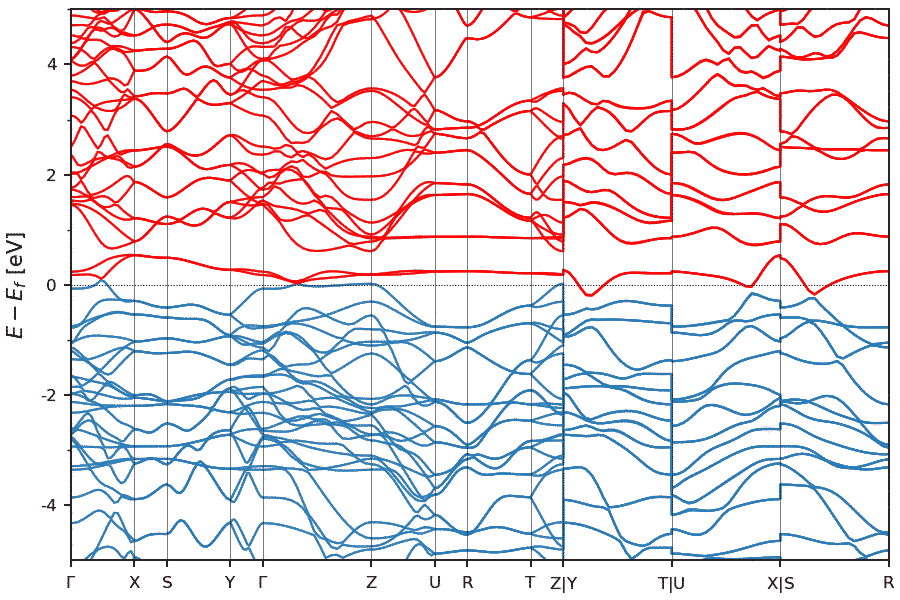} & \includegraphics[width=0.38\textwidth,angle=0]{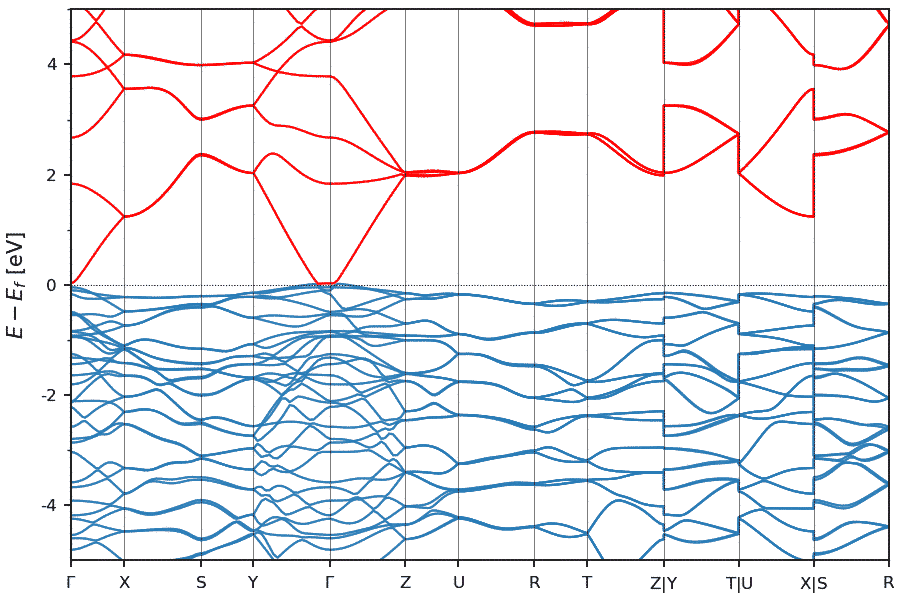}\\
\end{tabular}

\caption{\ztwoTINLC{2}}
\label{fig:z2TI_NLC2}
\end{figure}

\begin{figure}[ht]
\centering
\begin{tabular}{c c}
\scriptsize{$\rm{Cu}_{4} (\rm{Sn} \rm{S}_{4})$ - \icsdweb{833} - SG 62 ($Pnma$) - NLC} & \scriptsize{$\rm{Na} \rm{Cd} \rm{Sb}$ - \icsdweb{12159} - SG 62 ($Pnma$) - NLC}\\
\tiny{ $\;Z_{2,1}=0\;Z_{2,2}=0\;Z_{2,3}=0\;Z_4=3$ } & \tiny{ $\;Z_{2,1}=0\;Z_{2,2}=0\;Z_{2,3}=0\;Z_4=3$ }\\
\includegraphics[width=0.38\textwidth,angle=0]{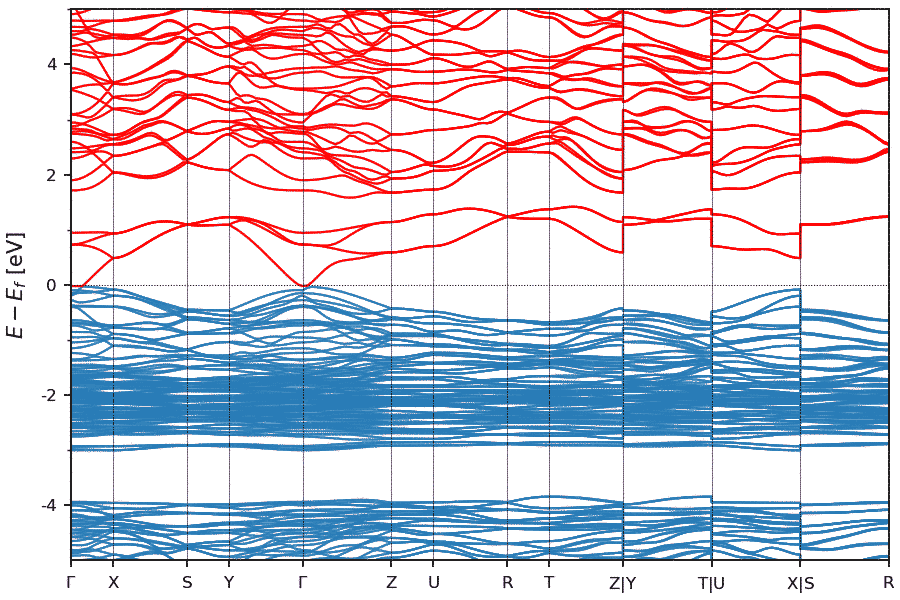} & \includegraphics[width=0.38\textwidth,angle=0]{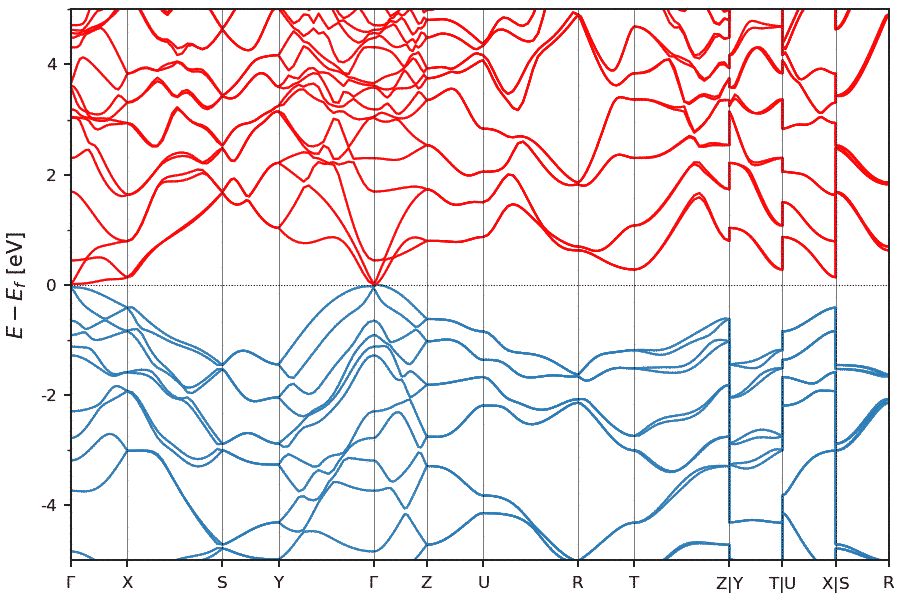}\\
\end{tabular}
\begin{tabular}{c c}
\scriptsize{$\rm{Ba}_{2} \rm{Ge}$ - \icsdweb{25351} - SG 62 ($Pnma$) - NLC} & \scriptsize{$\rm{Ba}_{2} \rm{Si}$ - \icsdweb{52698} - SG 62 ($Pnma$) - NLC}\\
\tiny{ $\;Z_{2,1}=0\;Z_{2,2}=0\;Z_{2,3}=0\;Z_4=1$ } & \tiny{ $\;Z_{2,1}=0\;Z_{2,2}=0\;Z_{2,3}=0\;Z_4=1$ }\\
\includegraphics[width=0.38\textwidth,angle=0]{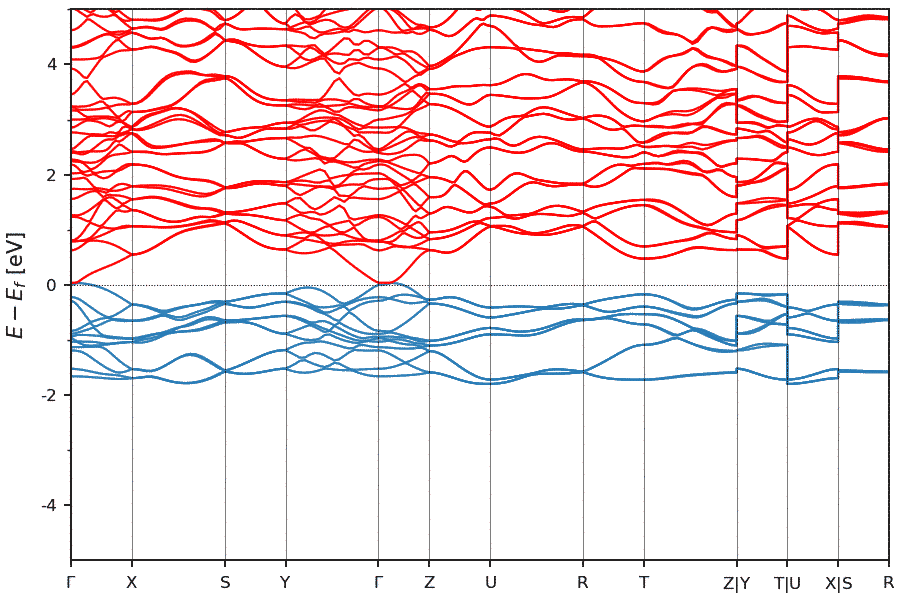} & \includegraphics[width=0.38\textwidth,angle=0]{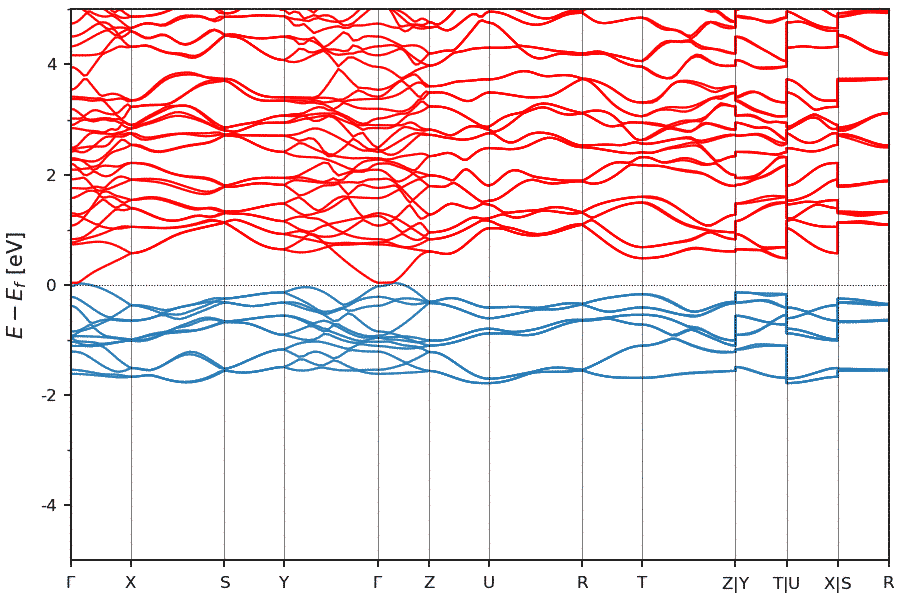}\\
\end{tabular}
\begin{tabular}{c c}
\scriptsize{$\rm{Pb} \rm{Sr}_{2}$ - \icsdweb{105624} - SG 62 ($Pnma$) - NLC} & \scriptsize{$\rm{Pb} \rm{O}_{2}$ - \icsdweb{189979} - SG 62 ($Pnma$) - NLC}\\
\tiny{ $\;Z_{2,1}=0\;Z_{2,2}=0\;Z_{2,3}=0\;Z_4=3$ } & \tiny{ $\;Z_{2,1}=0\;Z_{2,2}=0\;Z_{2,3}=0\;Z_4=3$ }\\
\includegraphics[width=0.38\textwidth,angle=0]{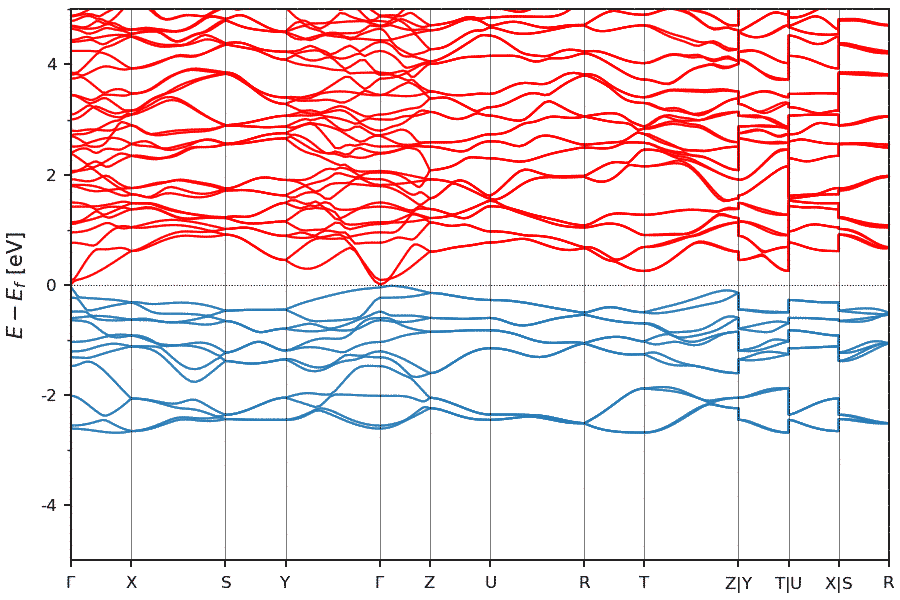} & \includegraphics[width=0.38\textwidth,angle=0]{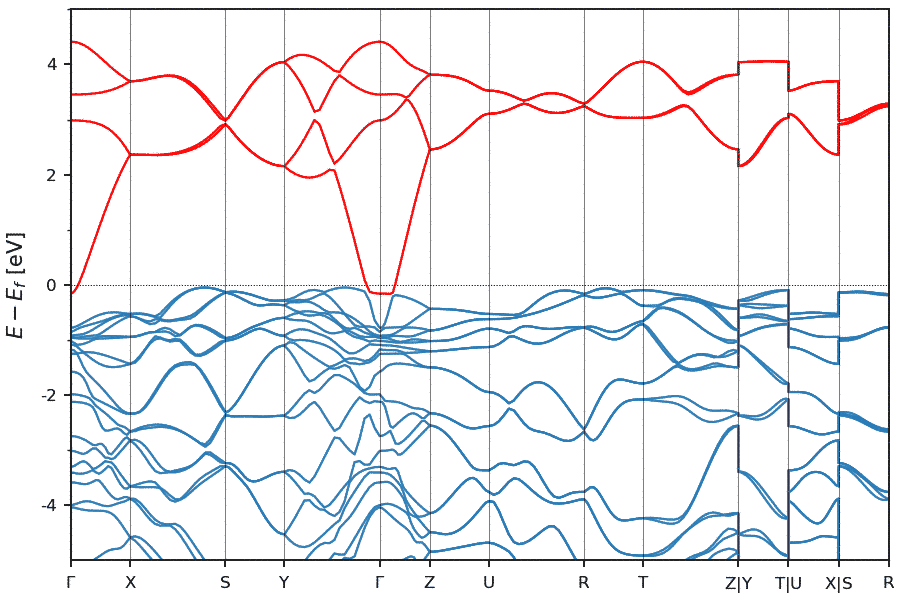}\\
\end{tabular}
\begin{tabular}{c c}
\scriptsize{$\rm{Zr} \rm{As}_{2}$ - \icsdweb{196782} - SG 62 ($Pnma$) - NLC} & \scriptsize{$\rm{Pt} \rm{Si} \rm{Ti}$ - \icsdweb{260794} - SG 62 ($Pnma$) - NLC}\\
\tiny{ $\;Z_{2,1}=0\;Z_{2,2}=0\;Z_{2,3}=0\;Z_4=3$ } & \tiny{ $\;Z_{2,1}=0\;Z_{2,2}=0\;Z_{2,3}=0\;Z_4=3$ }\\
\includegraphics[width=0.38\textwidth,angle=0]{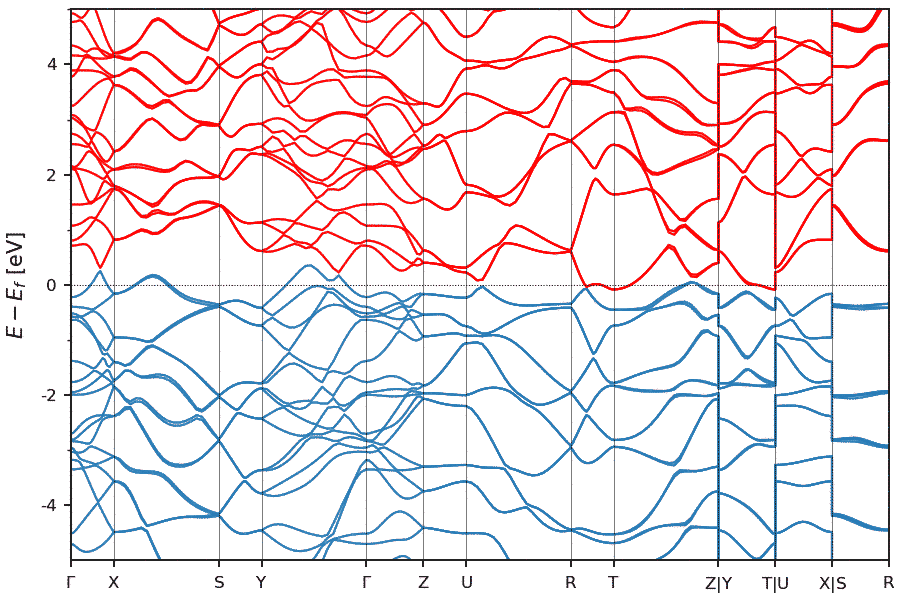} & \includegraphics[width=0.38\textwidth,angle=0]{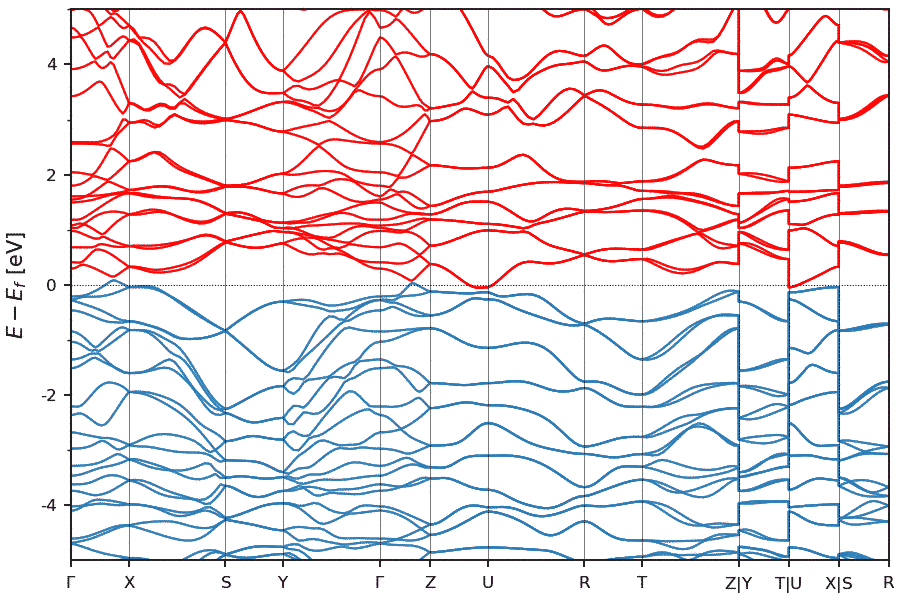}\\
\end{tabular}

\caption{\ztwoTINLC{3}}
\label{fig:z2TI_NLC3}
\end{figure}

\begin{figure}[ht]
\centering
\begin{tabular}{c c}
\scriptsize{$\rm{Ba}_{2} \rm{Pb}$ - \icsdweb{416561} - SG 62 ($Pnma$) - NLC} & \scriptsize{$\rm{Cu}_{4} \rm{Bi} \rm{Se}_{4} \rm{I}$ - \icsdweb{422827} - SG 62 ($Pnma$) - NLC}\\
\tiny{ $\;Z_{2,1}=0\;Z_{2,2}=0\;Z_{2,3}=0\;Z_4=1$ } & \tiny{ $\;Z_{2,1}=0\;Z_{2,2}=0\;Z_{2,3}=0\;Z_4=3$ }\\
\includegraphics[width=0.38\textwidth,angle=0]{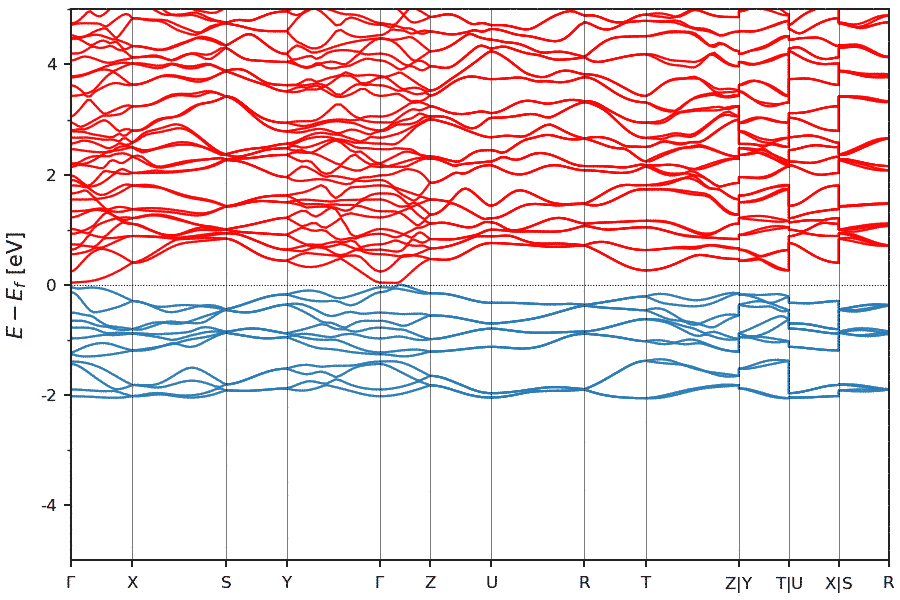} & \includegraphics[width=0.38\textwidth,angle=0]{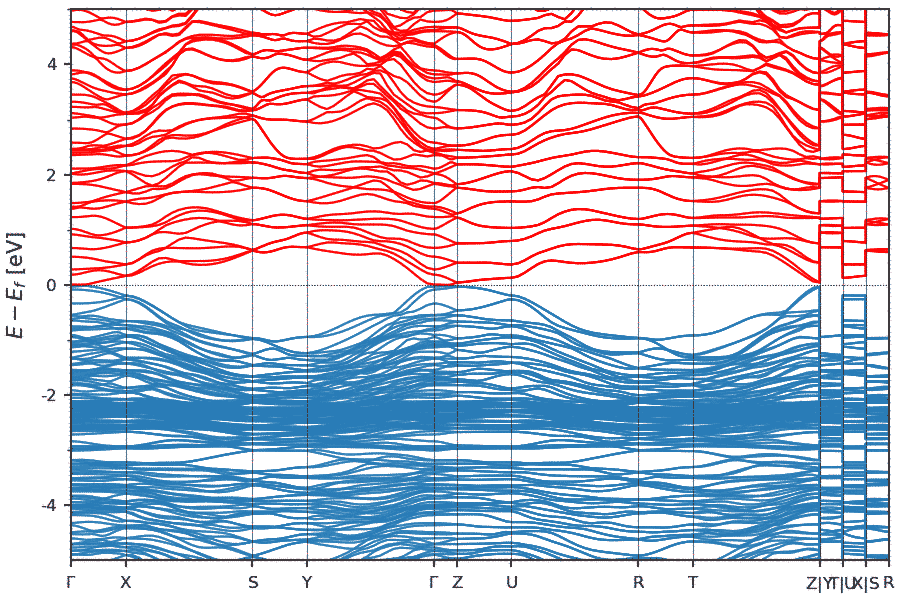}\\
\end{tabular}
\begin{tabular}{c c}
\scriptsize{$\rm{Bi} \rm{Ca} \rm{Li}$ - \icsdweb{616539} - SG 62 ($Pnma$) - NLC} & \scriptsize{$\rm{Sr} \rm{Cd} \rm{Ge}$ - \icsdweb{619960} - SG 62 ($Pnma$) - NLC}\\
\tiny{ $\;Z_{2,1}=0\;Z_{2,2}=0\;Z_{2,3}=0\;Z_4=3$ } & \tiny{ $\;Z_{2,1}=0\;Z_{2,2}=0\;Z_{2,3}=0\;Z_4=3$ }\\
\includegraphics[width=0.38\textwidth,angle=0]{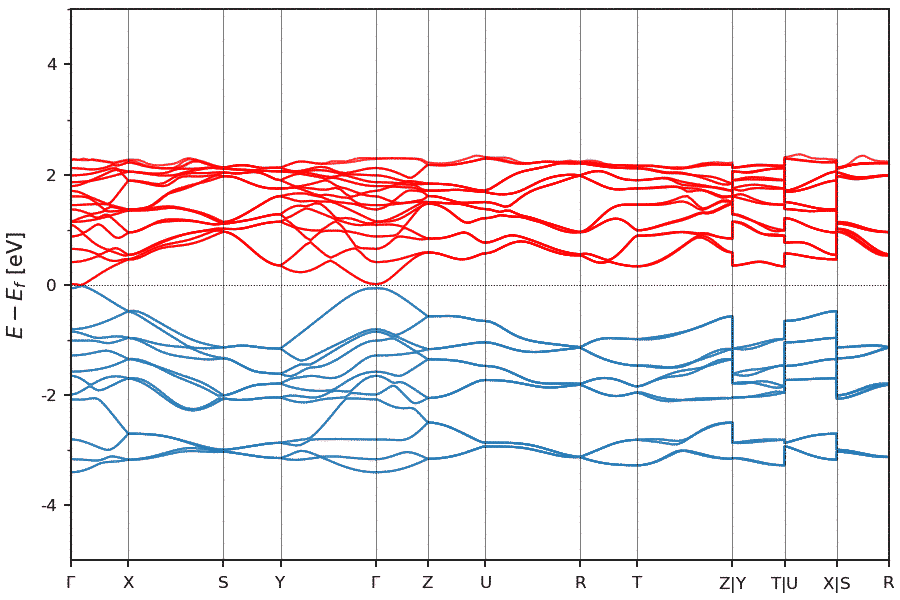} & \includegraphics[width=0.38\textwidth,angle=0]{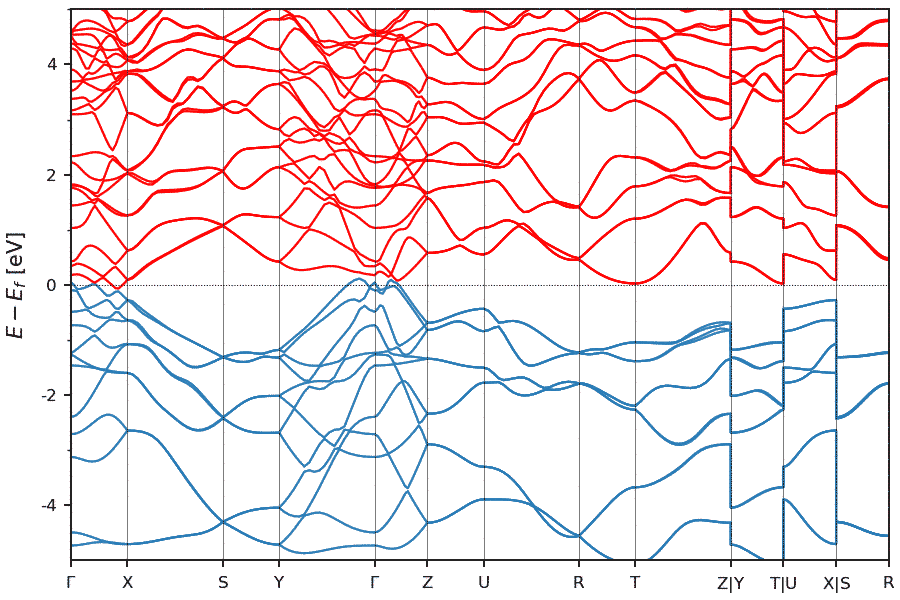}\\
\end{tabular}
\begin{tabular}{c c}
\scriptsize{$\rm{Ta} \rm{Co} \rm{Si}$ - \icsdweb{625050} - SG 62 ($Pnma$) - NLC} & \scriptsize{$\rm{Hf} \rm{Fe} \rm{Si}$ - \icsdweb{632263} - SG 62 ($Pnma$) - NLC}\\
\tiny{ $\;Z_{2,1}=0\;Z_{2,2}=0\;Z_{2,3}=0\;Z_4=3$ } & \tiny{ $\;Z_{2,1}=0\;Z_{2,2}=0\;Z_{2,3}=0\;Z_4=1$ }\\
\includegraphics[width=0.38\textwidth,angle=0]{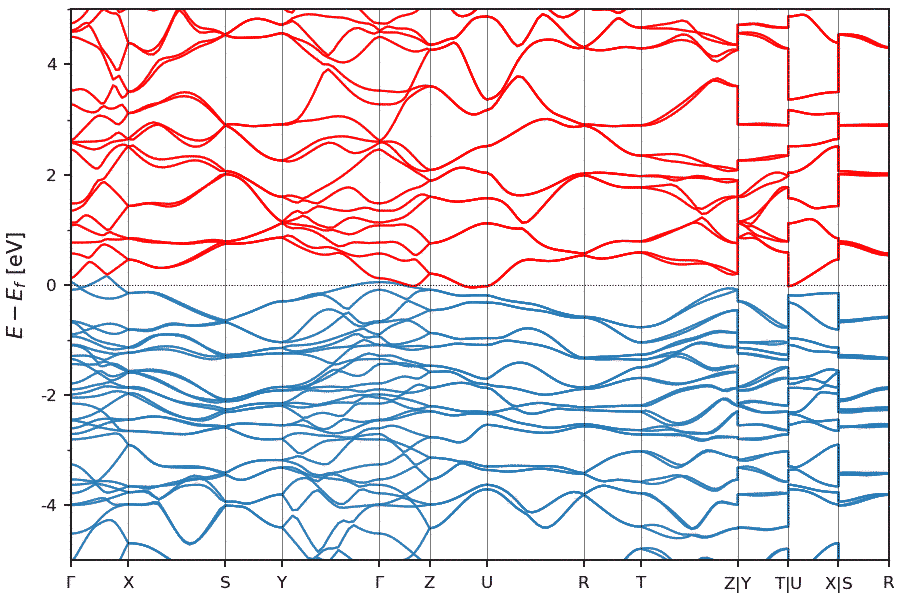} & \includegraphics[width=0.38\textwidth,angle=0]{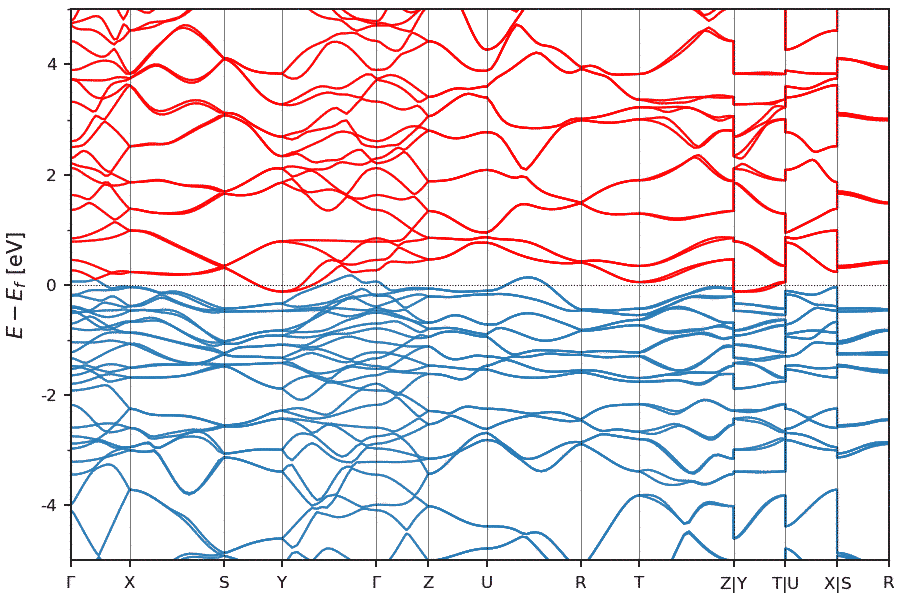}\\
\end{tabular}
\begin{tabular}{c c}
\scriptsize{$\rm{Ir} \rm{Sc} \rm{Ge}$ - \icsdweb{636730} - SG 62 ($Pnma$) - NLC} & \scriptsize{$\rm{Rh} \rm{Zr} \rm{P}$ - \icsdweb{648009} - SG 62 ($Pnma$) - NLC}\\
\tiny{ $\;Z_{2,1}=0\;Z_{2,2}=0\;Z_{2,3}=0\;Z_4=1$ } & \tiny{ $\;Z_{2,1}=0\;Z_{2,2}=0\;Z_{2,3}=0\;Z_4=3$ }\\
\includegraphics[width=0.38\textwidth,angle=0]{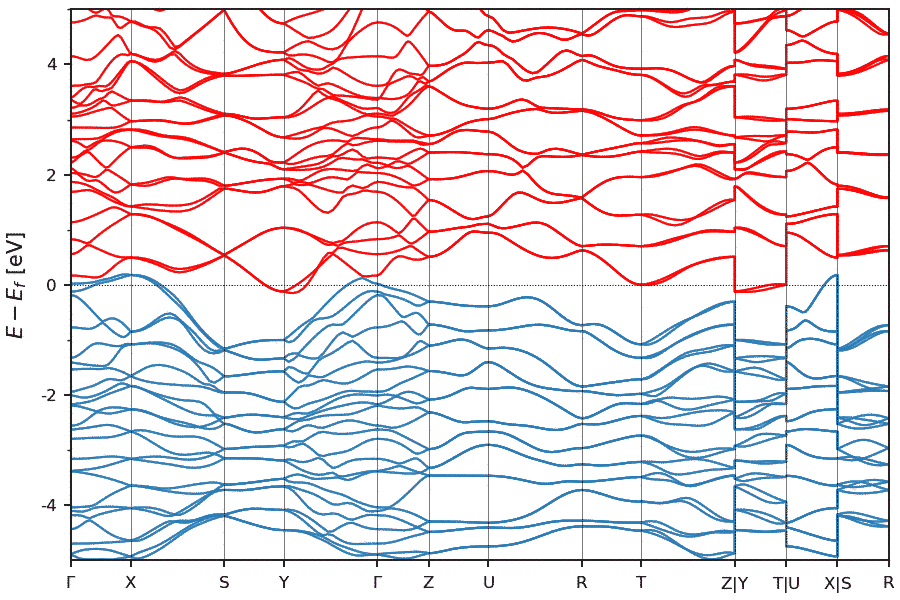} & \includegraphics[width=0.38\textwidth,angle=0]{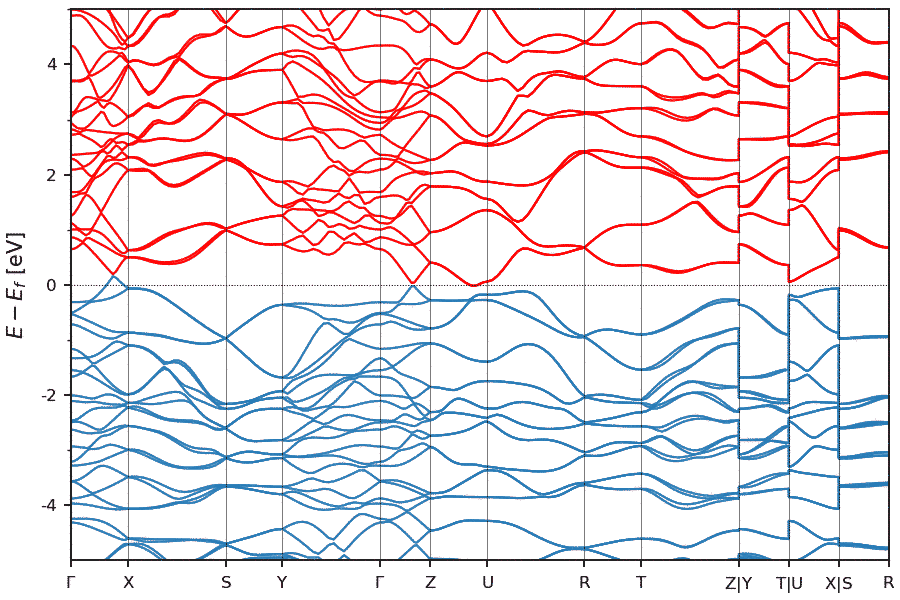}\\
\end{tabular}

\caption{\ztwoTINLC{4}}
\label{fig:z2TI_NLC4}
\end{figure}

\begin{figure}[ht]
\centering
\begin{tabular}{c c}
\scriptsize{$\rm{Rb}_{2} \rm{Cd}_{5} \rm{As}_{4}$ - \icsdweb{290262} - SG 63 ($Cmcm$) - NLC} & \scriptsize{$\rm{Na}_{15} \rm{Ge}_{8} \rm{Sn} \rm{P}$ - \icsdweb{417884} - SG 63 ($Cmcm$) - NLC}\\
\tiny{ $\;Z_{2,1}=0\;Z_{2,2}=0\;Z_{2,3}=0\;Z_4=3$ } & \tiny{ $\;Z_{2,1}=1\;Z_{2,2}=1\;Z_{2,3}=0\;Z_4=3$ }\\
\includegraphics[width=0.38\textwidth,angle=0]{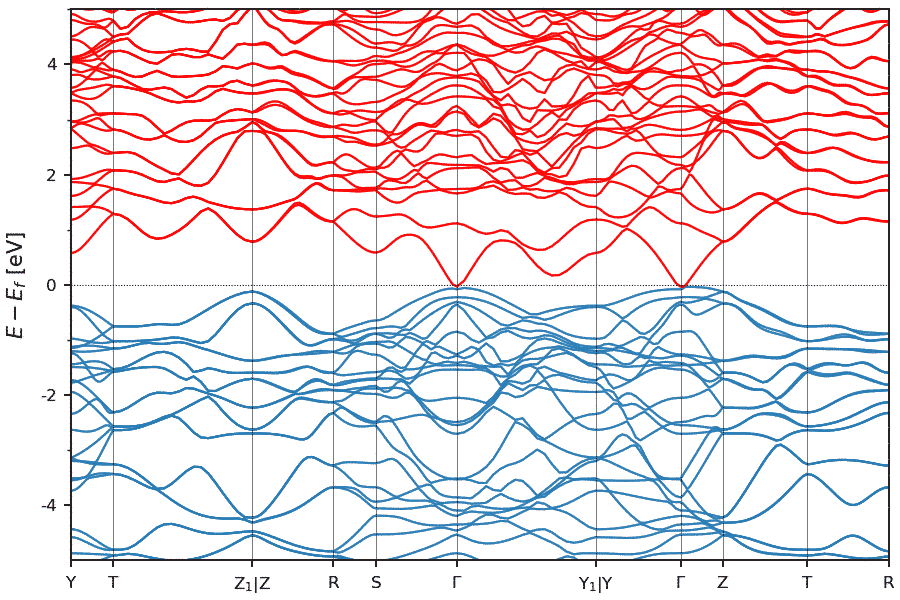} & \includegraphics[width=0.38\textwidth,angle=0]{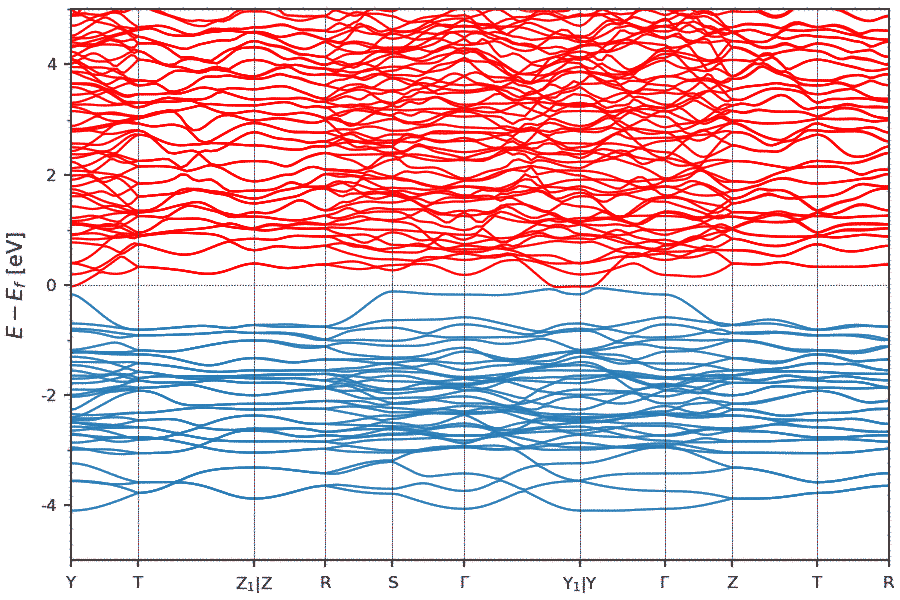}\\
\end{tabular}
\begin{tabular}{c c}
\scriptsize{$\rm{Ba}_{2} \rm{Mg} \rm{Ir}_{4} \rm{Ge}_{4}$ - \icsdweb{429723} - SG 63 ($Cmcm$) - NLC} & \scriptsize{$\rm{Zr} \rm{Te}_{5}$ - \icsdweb{657473} - SG 63 ($Cmcm$) - NLC}\\
\tiny{ $\;Z_{2,1}=1\;Z_{2,2}=1\;Z_{2,3}=0\;Z_4=3$ } & \tiny{ $\;Z_{2,1}=1\;Z_{2,2}=1\;Z_{2,3}=0\;Z_4=3$ }\\
\includegraphics[width=0.38\textwidth,angle=0]{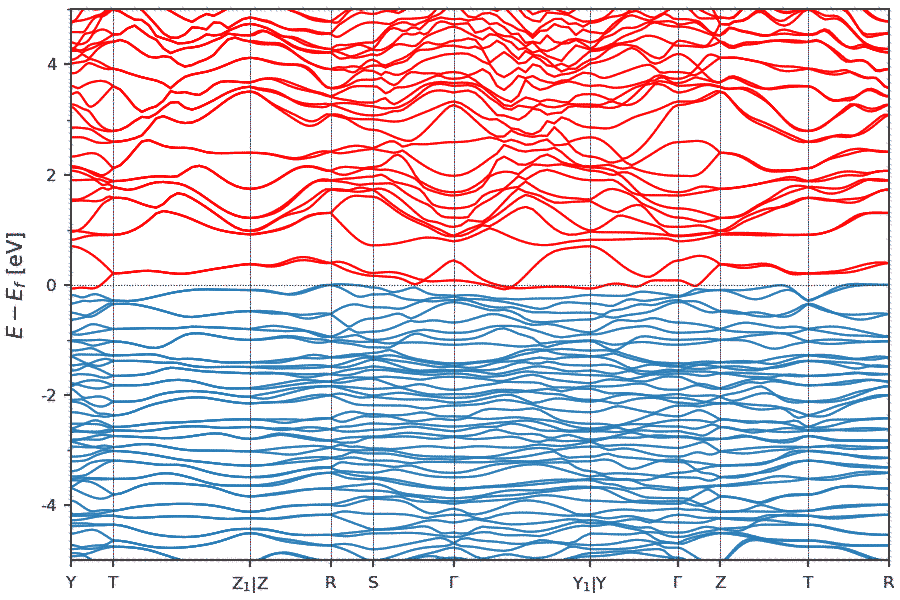} & \includegraphics[width=0.38\textwidth,angle=0]{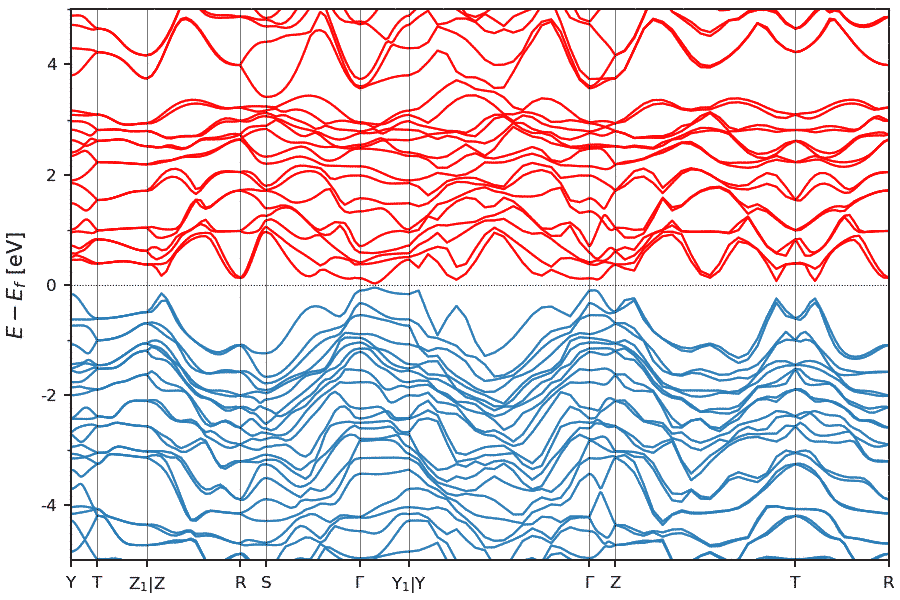}\\
\end{tabular}
\begin{tabular}{c c}
\scriptsize{$\rm{Na}_{2} \rm{K}_{3} (\rm{Sn} \rm{Bi}_{3})$ - \icsdweb{50534} - SG 73 ($Ibca$) - NLC} & \scriptsize{$\rm{Cs} \rm{Sn} \rm{I}_{3}$ - \icsdweb{262925} - SG 127 ($P4/mbm$) - NLC}\\
\tiny{ $\;Z_{2,1}=0\;Z_{2,2}=0\;Z_{2,3}=0\;Z_4=3$ } & \tiny{ $\;Z_{2,1}=0\;Z_{2,2}=0\;Z_{2,3}=1\;Z_4=3\;Z_{4m,\pi}=3\;Z_2=1\;Z_8=7$ }\\
\includegraphics[width=0.38\textwidth,angle=0]{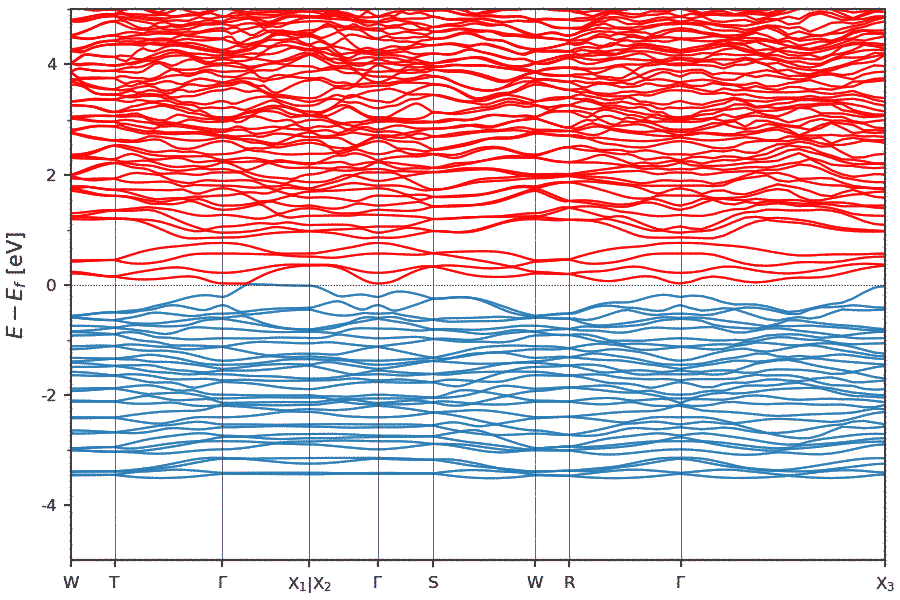} & \includegraphics[width=0.38\textwidth,angle=0]{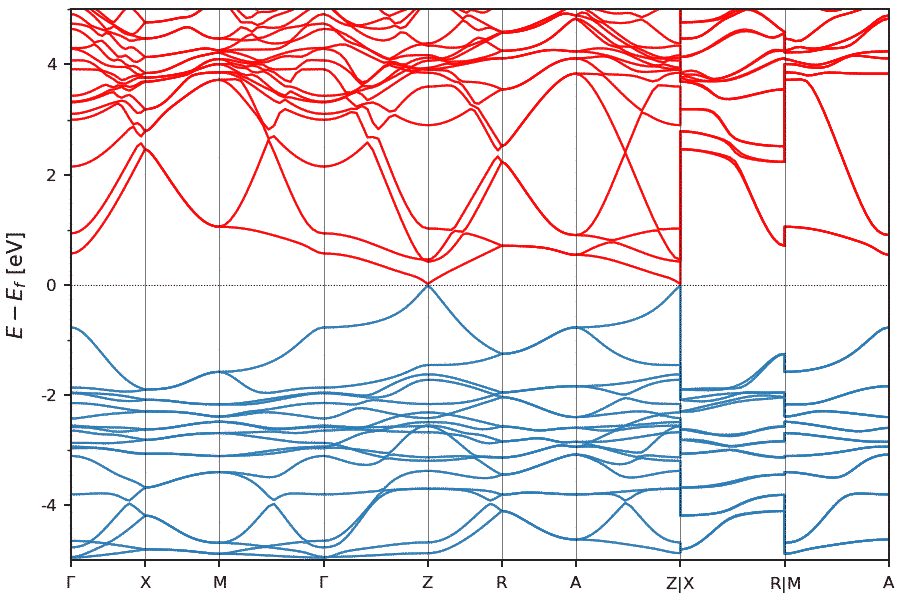}\\
\end{tabular}
\begin{tabular}{c c}
\scriptsize{$\rm{In} \rm{Pb}_{2} \rm{I}_{5}$ - \icsdweb{152028} - SG 140 ($I4/mcm$) - NLC}\\
\tiny{ $\;Z_{2,1}=0\;Z_{2,2}=0\;Z_{2,3}=0\;Z_4=3\;Z_2=1\;Z_8=7$ }\\
\includegraphics[width=0.38\textwidth,angle=0]{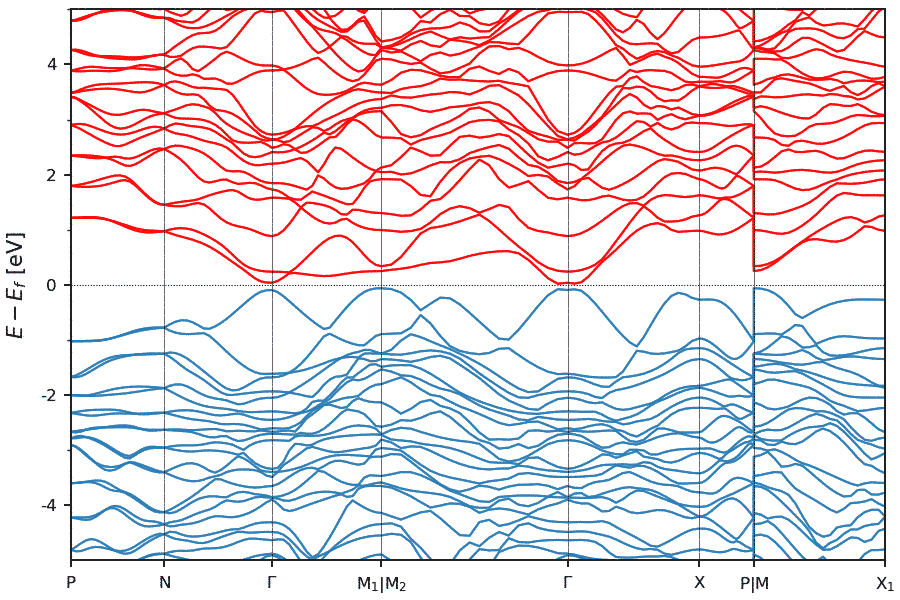}\\
\end{tabular}

\caption{\ztwoTINLC{5}}
\label{fig:z2TI_NLC5}
\end{figure}


\begin{figure}[ht]
\centering
\begin{tabular}{c c}
\scriptsize{$\rm{Cu} \rm{In} \rm{Se}_{2}$ - \icsdweb{252780} - SG 122 ($I\bar{4}2d$) - SEBR} & \scriptsize{$\rm{Cd} \rm{Ge} \rm{As}_{2}$ - \icsdweb{253229} - SG 122 ($I\bar{4}2d$) - SEBR}\\
\tiny{ $\;Z_2=1$ } & \tiny{ $\;Z_2=1$ }\\
\includegraphics[width=0.38\textwidth,angle=0]{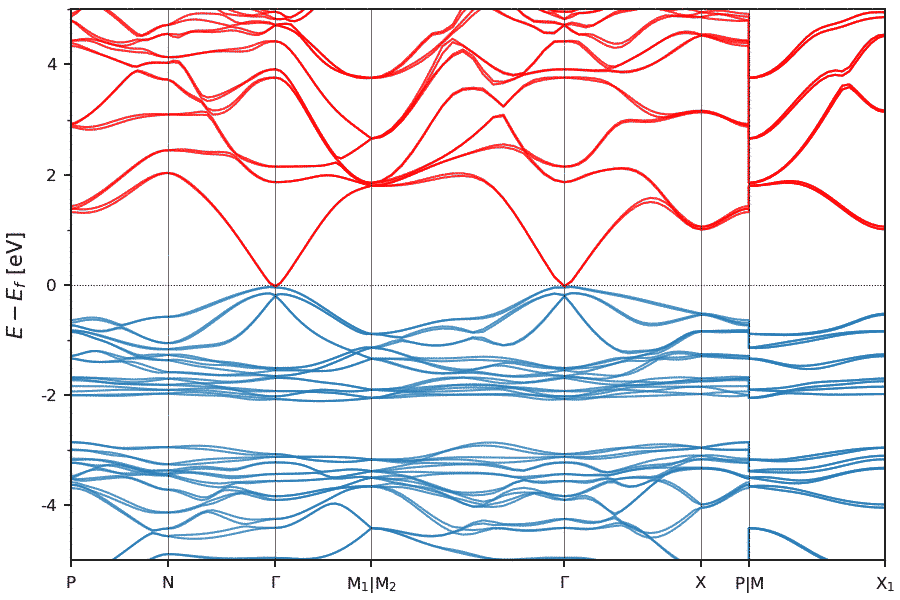} & \includegraphics[width=0.38\textwidth,angle=0]{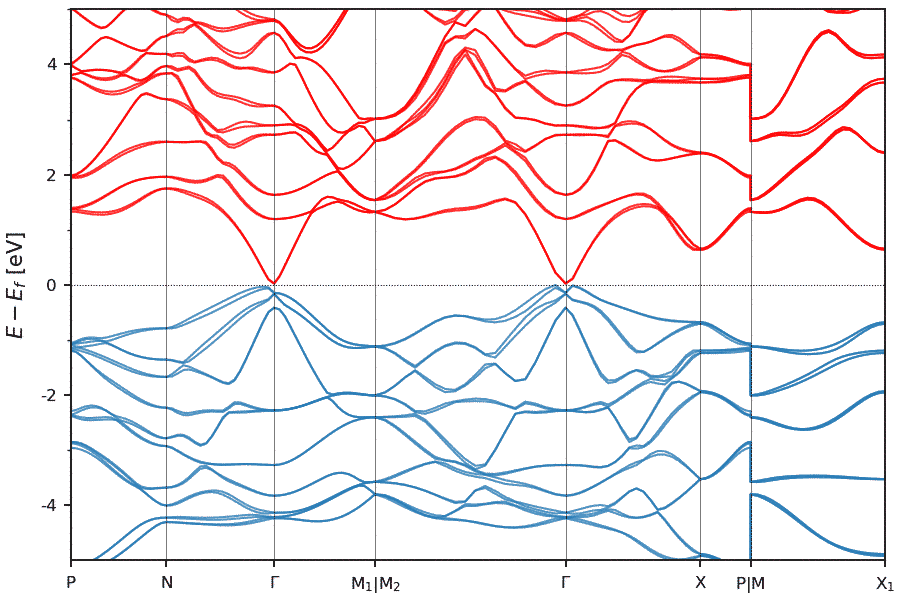}\\
\end{tabular}
\begin{tabular}{c c}
\scriptsize{$\rm{Cs} \rm{Ag}_{5} \rm{Te}_{3}$ - \icsdweb{79215} - SG 136 ($P4_2/mnm$) - SEBR} & \scriptsize{$\rm{Ba}_{8} (\rm{Bi}_{4}) \rm{O} \rm{H}_{2}$ - \icsdweb{415033} - SG 139 ($I4/mmm$) - SEBR}\\
\tiny{ $\;Z_{2,1}=0\;Z_{2,2}=0\;Z_{2,3}=0\;Z_4=3\;Z_2=1$ } & \tiny{ $\;Z_{2,1}=1\;Z_{2,2}=1\;Z_{2,3}=1\;Z_4=3\;Z_2=1\;Z_8=3$ }\\
\includegraphics[width=0.38\textwidth,angle=0]{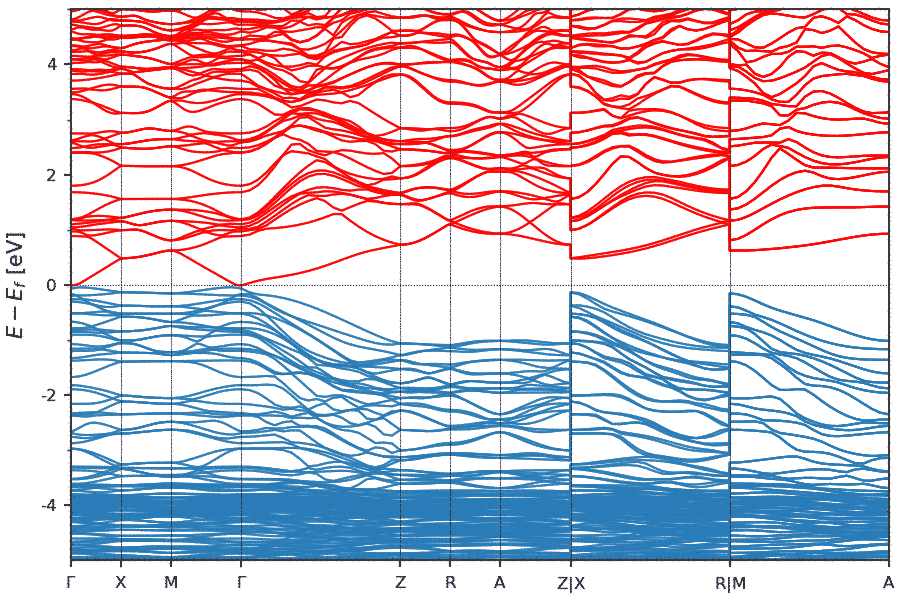} & \includegraphics[width=0.38\textwidth,angle=0]{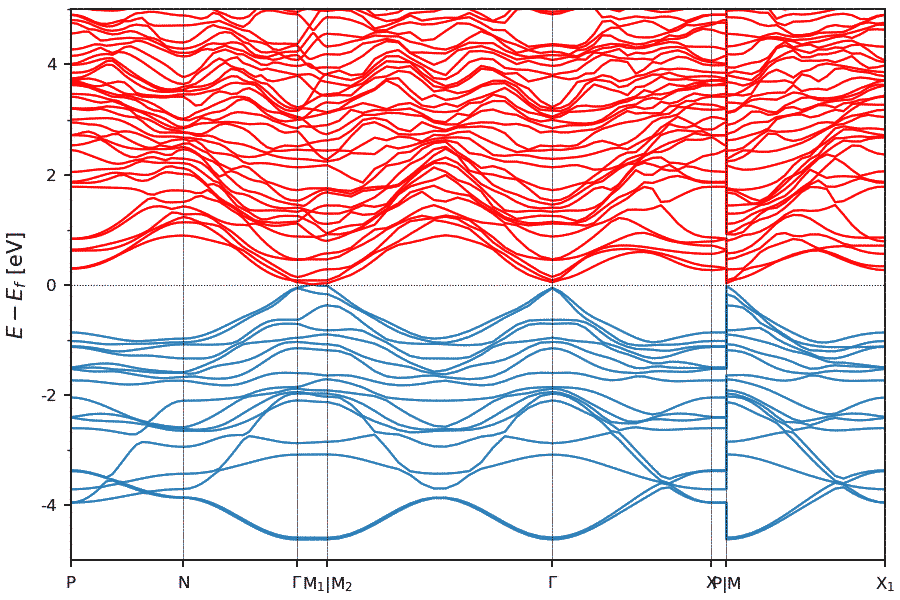}\\
\end{tabular}
\begin{tabular}{c c}
\scriptsize{$\rm{Tl}_{6} \rm{Te} \rm{O}_{12}$ - \icsdweb{37134} - SG 148 ($R\bar{3}$) - SEBR} & \scriptsize{$\rm{Mo}_{3} \rm{Te}_{4}$ - \icsdweb{644477} - SG 148 ($R\bar{3}$) - SEBR}\\
\tiny{ $\;Z_{2,1}=0\;Z_{2,2}=0\;Z_{2,3}=0\;Z_4=3$ } & \tiny{ $\;Z_{2,1}=1\;Z_{2,2}=1\;Z_{2,3}=1\;Z_4=1$ }\\
\includegraphics[width=0.38\textwidth,angle=0]{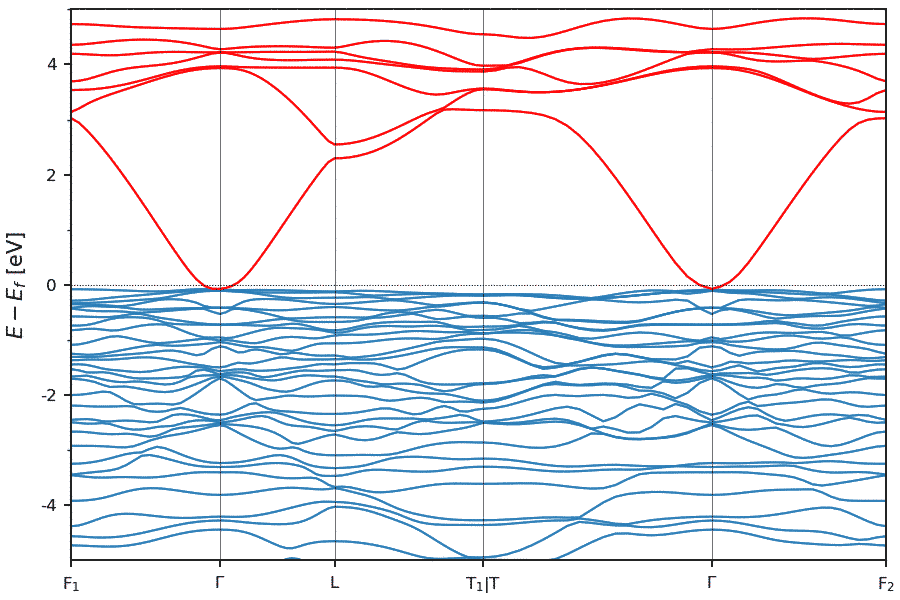} & \includegraphics[width=0.38\textwidth,angle=0]{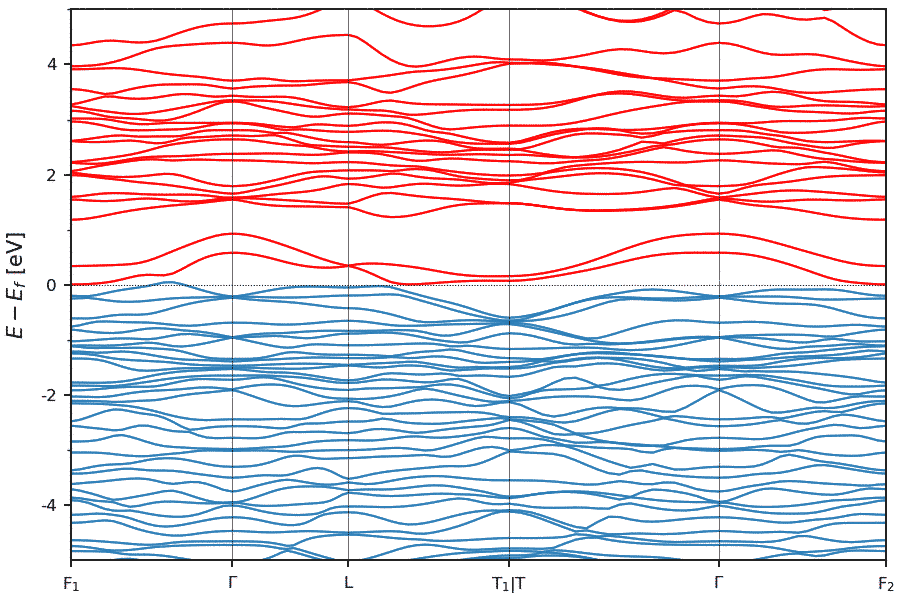}\\
\end{tabular}
\begin{tabular}{c c}
\scriptsize{$\rm{Pb}_{2} \rm{Bi}_{2} \rm{Se}_{5}$ - \icsdweb{30372} - SG 164 ($P\bar{3}m1$) - SEBR} & \scriptsize{$\rm{Ge} \rm{Sb}_{4} \rm{Te}_{7}$ - \icsdweb{42875} - SG 164 ($P\bar{3}m1$) - SEBR}\\
\tiny{ $\;Z_{2,1}=0\;Z_{2,2}=0\;Z_{2,3}=0\;Z_4=3$ } & \tiny{ $\;Z_{2,1}=0\;Z_{2,2}=0\;Z_{2,3}=1\;Z_4=3$ }\\
\includegraphics[width=0.38\textwidth,angle=0]{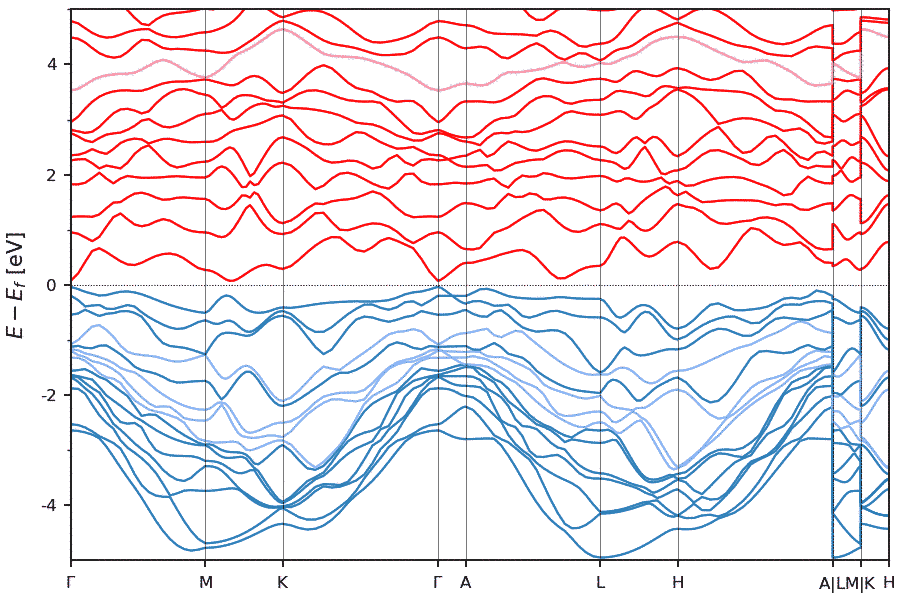} & \includegraphics[width=0.38\textwidth,angle=0]{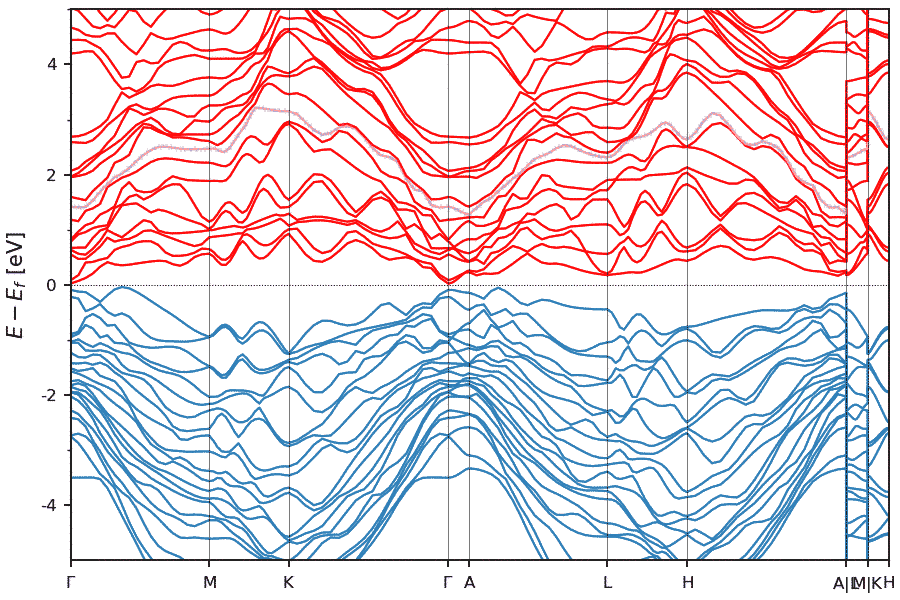}\\
\end{tabular}

\caption{\ztwoTISEBR{1}}
\label{fig:z2TI_SEBR1}
\end{figure}

\begin{figure}[ht]
\centering
\begin{tabular}{c c}
\scriptsize{$\rm{Sn} \rm{Bi}_{4} \rm{Te}_{7}$ - \icsdweb{236253} - SG 164 ($P\bar{3}m1$) - SEBR} & \scriptsize{$\rm{Pb} \rm{Bi}_{4} \rm{Te}_{7}$ - \icsdweb{250249} - SG 164 ($P\bar{3}m1$) - SEBR}\\
\tiny{ $\;Z_{2,1}=0\;Z_{2,2}=0\;Z_{2,3}=1\;Z_4=1$ } & \tiny{ $\;Z_{2,1}=0\;Z_{2,2}=0\;Z_{2,3}=1\;Z_4=1$ }\\
\includegraphics[width=0.38\textwidth,angle=0]{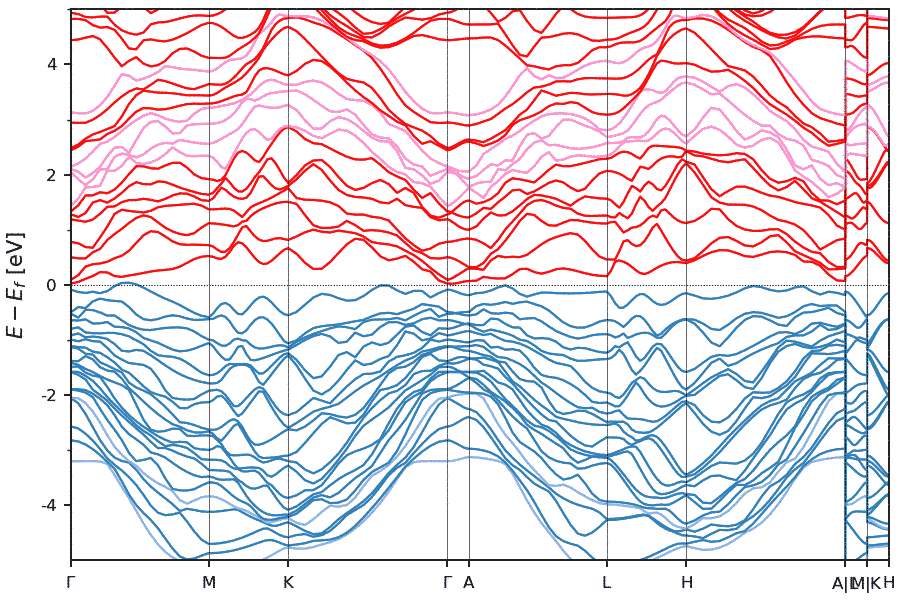} & \includegraphics[width=0.38\textwidth,angle=0]{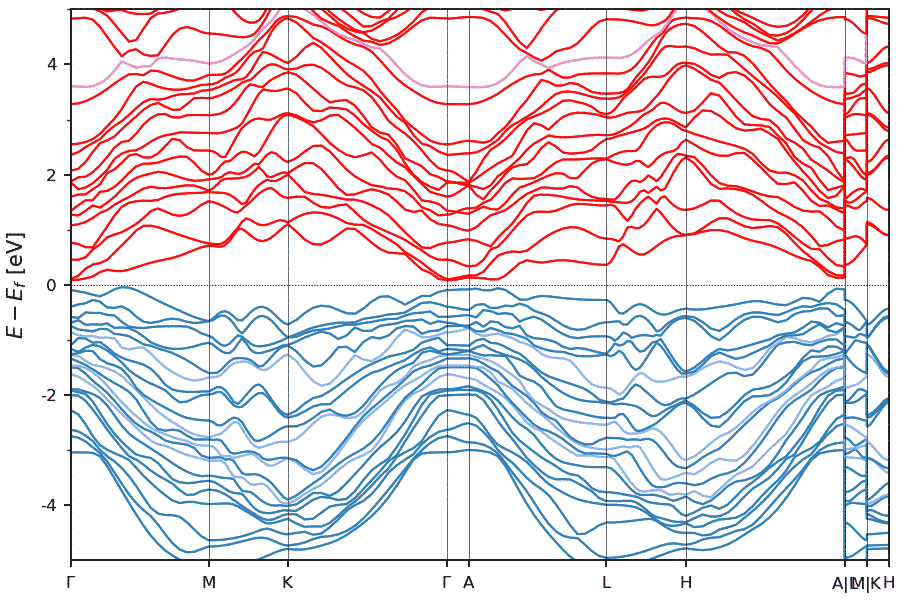}\\
\end{tabular}
\begin{tabular}{c c}
\scriptsize{$\rm{Ba} \rm{Sn}_{2}$ - \icsdweb{409958} - SG 164 ($P\bar{3}m1$) - SEBR} & \scriptsize{$\rm{Ti}_{8} \rm{C}_{5}$ - \icsdweb{20822} - SG 166 ($R\bar{3}m$) - SEBR}\\
\tiny{ $\;Z_{2,1}=0\;Z_{2,2}=0\;Z_{2,3}=1\;Z_4=1$ } & \tiny{ $\;Z_{2,1}=0\;Z_{2,2}=0\;Z_{2,3}=0\;Z_4=1$ }\\
\includegraphics[width=0.38\textwidth,angle=0]{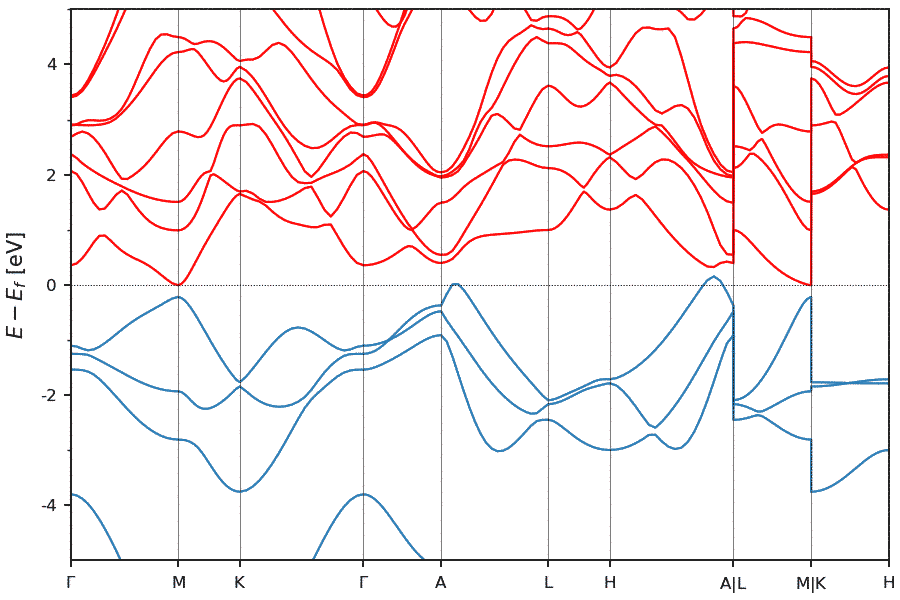} & \includegraphics[width=0.38\textwidth,angle=0]{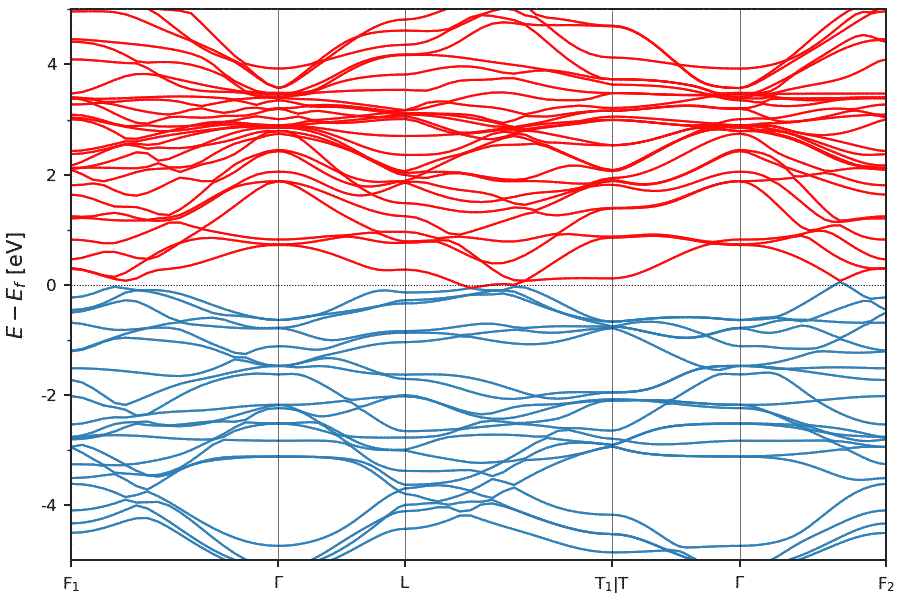}\\
\end{tabular}
\begin{tabular}{c c}
\scriptsize{$\rm{Ge} \rm{Bi}_{2} \rm{Te}_{4}$ - \icsdweb{30394} - SG 166 ($R\bar{3}m$) - SEBR} & \scriptsize{$\rm{Bi}_{2} \rm{Se} \rm{Te}_{2}$ - \icsdweb{43512} - SG 166 ($R\bar{3}m$) - SEBR}\\
\tiny{ $\;Z_{2,1}=1\;Z_{2,2}=1\;Z_{2,3}=1\;Z_4=1$ } & \tiny{ $\;Z_{2,1}=0\;Z_{2,2}=0\;Z_{2,3}=0\;Z_4=3$ }\\
\includegraphics[width=0.38\textwidth,angle=0]{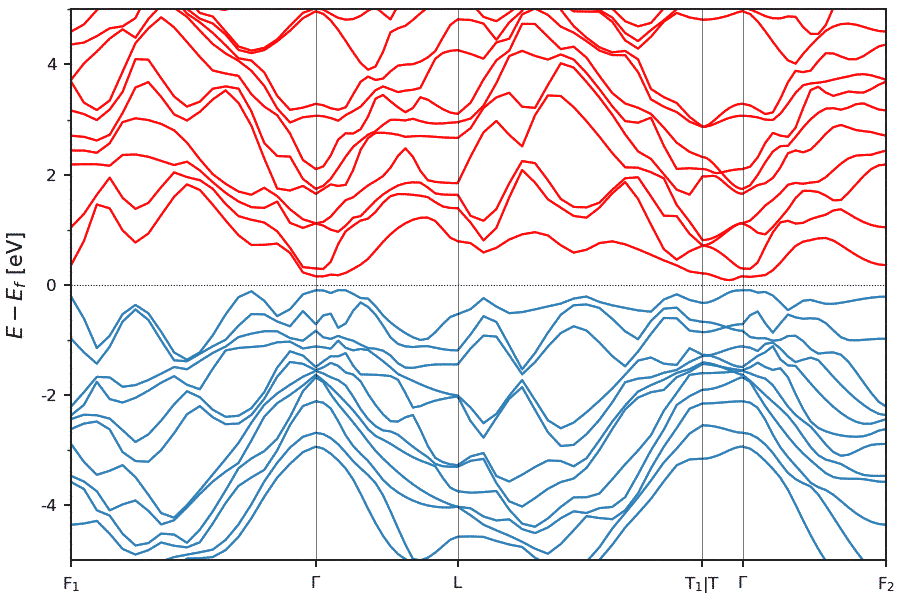} & \includegraphics[width=0.38\textwidth,angle=0]{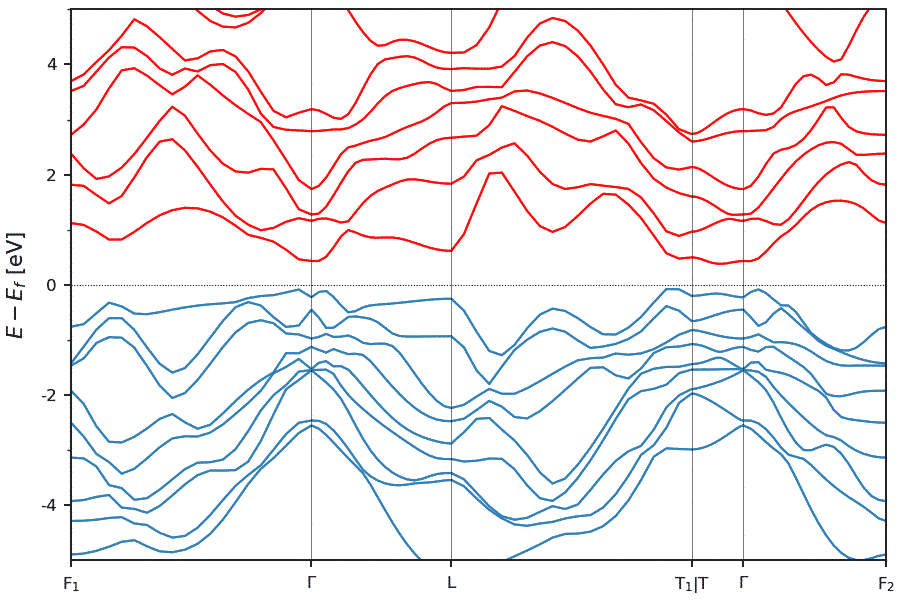}\\
\end{tabular}
\begin{tabular}{c c}
\scriptsize{$\rm{Sb}_{2} \rm{Se}_{2} \rm{Te}$ - \icsdweb{52295} - SG 166 ($R\bar{3}m$) - SEBR} & \scriptsize{$\rm{As}_{2} \rm{Te}_{3}$ - \icsdweb{68110} - SG 166 ($R\bar{3}m$) - SEBR}\\
\tiny{ $\;Z_{2,1}=0\;Z_{2,2}=0\;Z_{2,3}=0\;Z_4=3$ } & \tiny{ $\;Z_{2,1}=0\;Z_{2,2}=0\;Z_{2,3}=0\;Z_4=3$ }\\
\includegraphics[width=0.38\textwidth,angle=0]{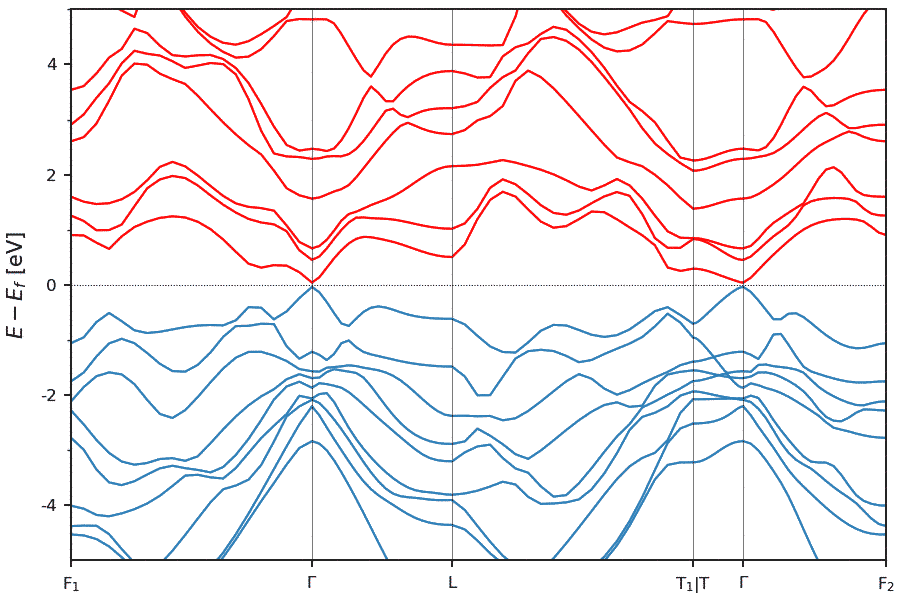} & \includegraphics[width=0.38\textwidth,angle=0]{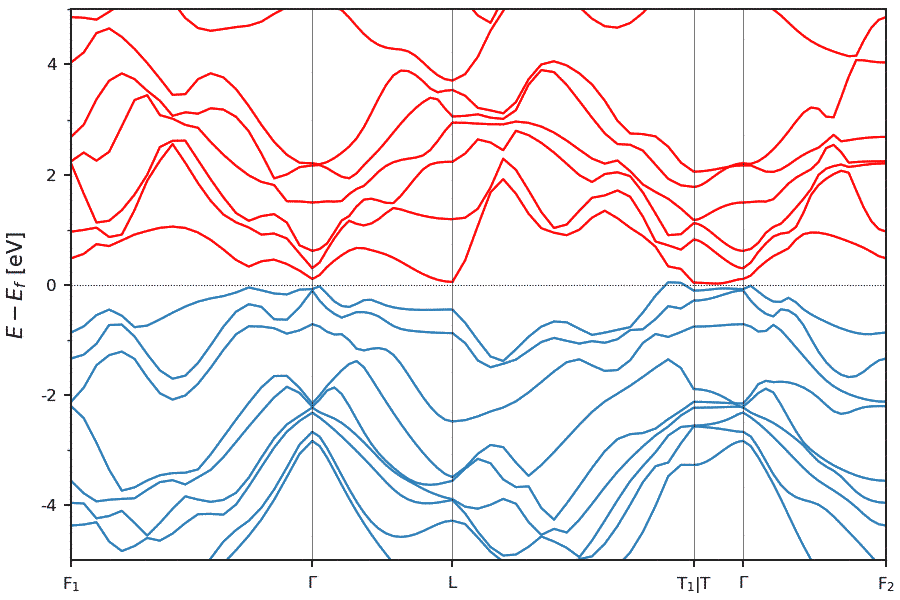}\\
\end{tabular}

\caption{\ztwoTISEBR{2}}
\label{fig:z2TI_SEBR2}
\end{figure}

\begin{figure}[ht]
\centering
\begin{tabular}{c c}
\scriptsize{$\rm{As}_{2} \rm{Ge}_{4} \rm{Te}_{7}$ - \icsdweb{68114} - SG 166 ($R\bar{3}m$) - SEBR} & \scriptsize{$\rm{Cu} \rm{I}$ - \icsdweb{80259} - SG 166 ($R\bar{3}m$) - SEBR}\\
\tiny{ $\;Z_{2,1}=0\;Z_{2,2}=0\;Z_{2,3}=0\;Z_4=3$ } & \tiny{ $\;Z_{2,1}=0\;Z_{2,2}=0\;Z_{2,3}=0\;Z_4=3$ }\\
\includegraphics[width=0.38\textwidth,angle=0]{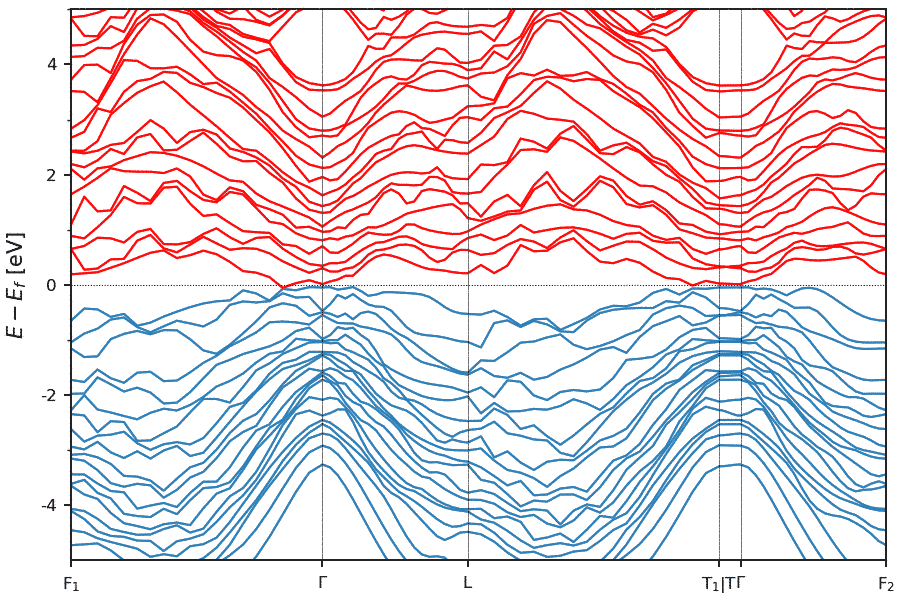} & \includegraphics[width=0.38\textwidth,angle=0]{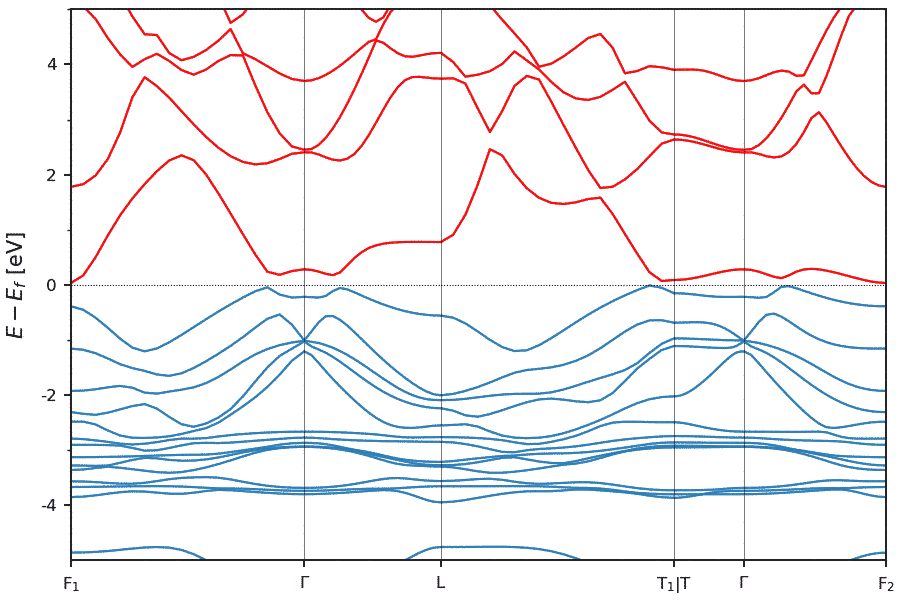}\\
\end{tabular}
\begin{tabular}{c c}
\scriptsize{$\rm{Sr} (\rm{Sn}_{2} \rm{As}_{2})$ - \icsdweb{82371} - SG 166 ($R\bar{3}m$) - SEBR} & \scriptsize{$\rm{Pb} \rm{Bi}_{6} \rm{Te}_{10}$ - \icsdweb{95551} - SG 166 ($R\bar{3}m$) - SEBR}\\
\tiny{ $\;Z_{2,1}=0\;Z_{2,2}=0\;Z_{2,3}=0\;Z_4=3$ } & \tiny{ $\;Z_{2,1}=1\;Z_{2,2}=1\;Z_{2,3}=1\;Z_4=1$ }\\
\includegraphics[width=0.38\textwidth,angle=0]{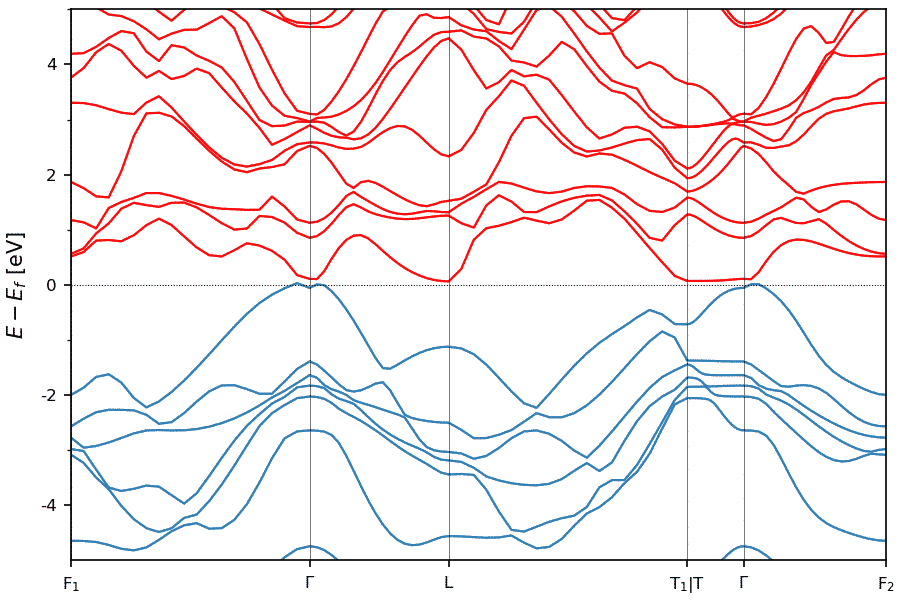} & \includegraphics[width=0.38\textwidth,angle=0]{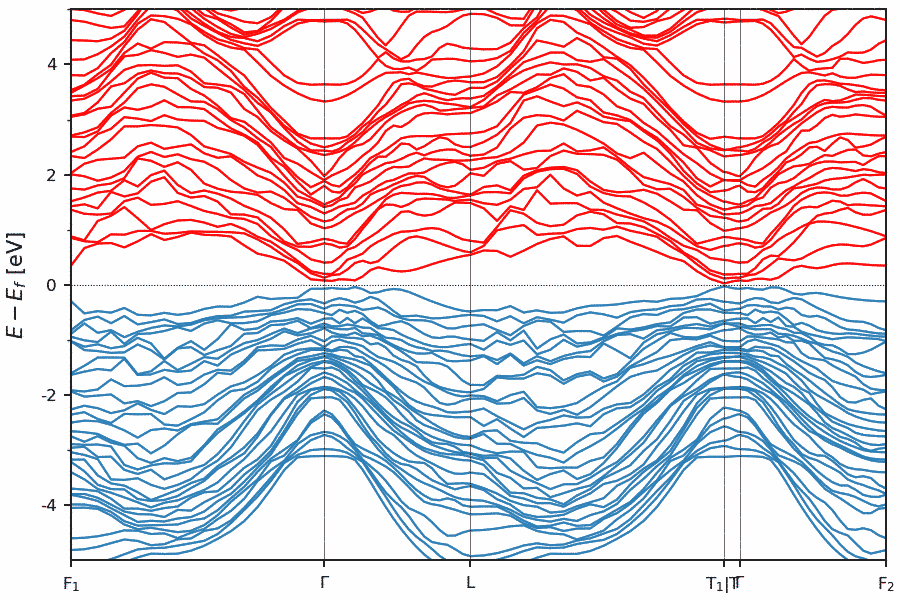}\\
\end{tabular}
\begin{tabular}{c c}
\scriptsize{$\rm{Sb}_{8} \rm{Te}_{3}$ - \icsdweb{152188} - SG 166 ($R\bar{3}m$) - SEBR} & \scriptsize{$\rm{Sb}_{48} \rm{Te}_{9}$ - \icsdweb{152189} - SG 166 ($R\bar{3}m$) - SEBR}\\
\tiny{ $\;Z_{2,1}=1\;Z_{2,2}=1\;Z_{2,3}=1\;Z_4=3$ } & \tiny{ $\;Z_{2,1}=1\;Z_{2,2}=1\;Z_{2,3}=1\;Z_4=3$ }\\
\includegraphics[width=0.38\textwidth,angle=0]{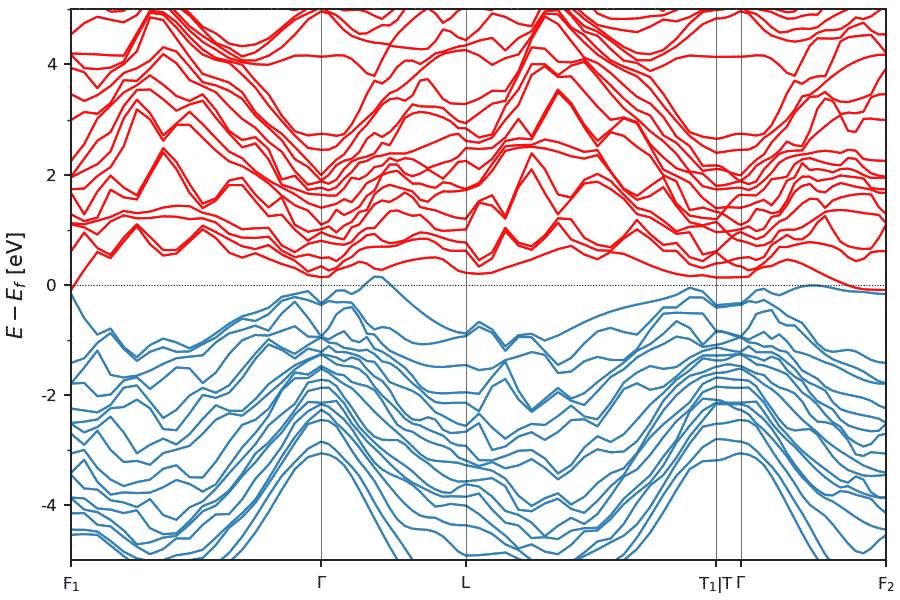} & \includegraphics[width=0.38\textwidth,angle=0]{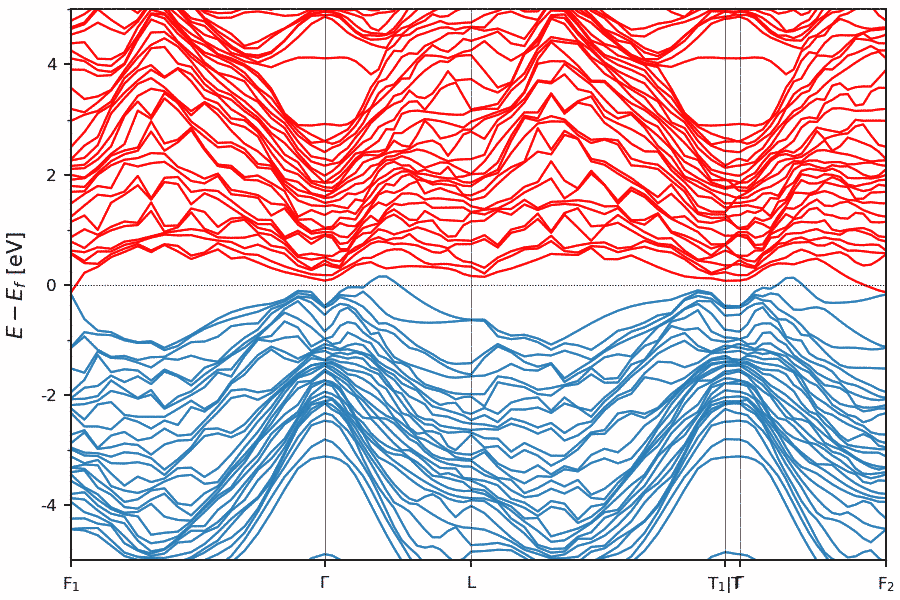}\\
\end{tabular}
\begin{tabular}{c c}
\scriptsize{$\rm{As}$ - \icsdweb{162837} - SG 166 ($R\bar{3}m$) - SEBR} & \scriptsize{$\rm{Bi}_{2} \rm{Te}_{3}$ - \icsdweb{193330} - SG 166 ($R\bar{3}m$) - SEBR}\\
\tiny{ $\;Z_{2,1}=1\;Z_{2,2}=1\;Z_{2,3}=1\;Z_4=3$ } & \tiny{ $\;Z_{2,1}=0\;Z_{2,2}=0\;Z_{2,3}=0\;Z_4=3$ }\\
\includegraphics[width=0.38\textwidth,angle=0]{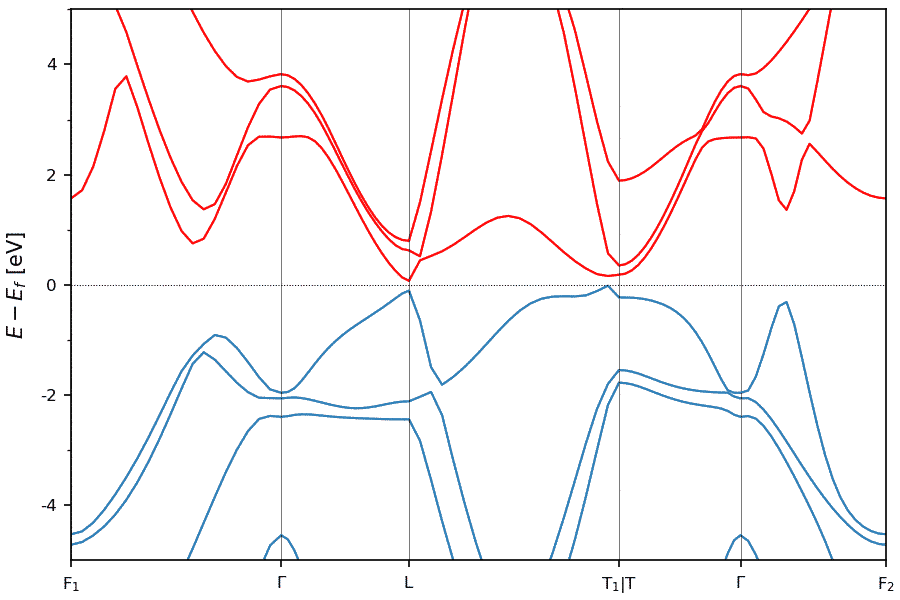} & \includegraphics[width=0.38\textwidth,angle=0]{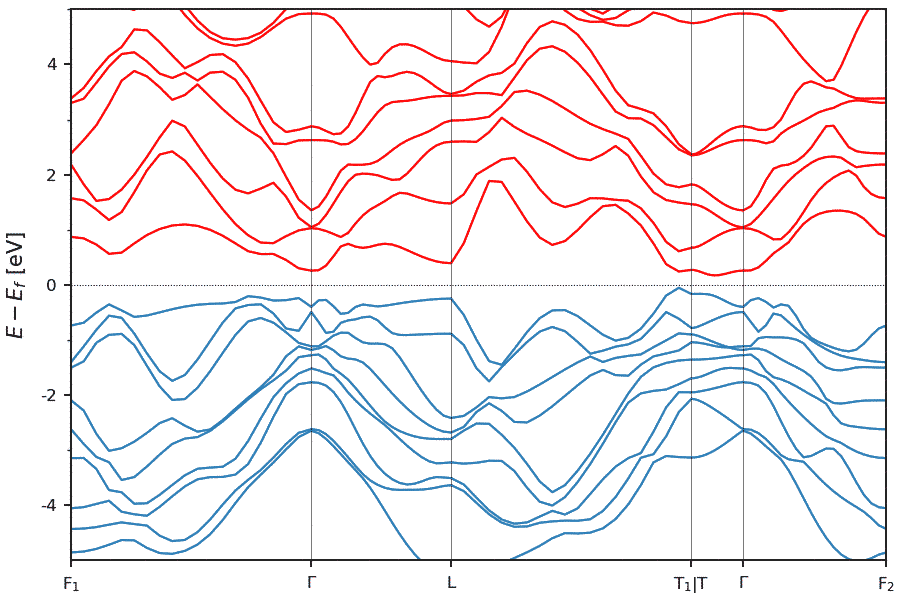}\\
\end{tabular}

\caption{\ztwoTISEBR{3}}
\label{fig:z2TI_SEBR3}
\end{figure}

\begin{figure}[ht]
\centering
\begin{tabular}{c c}
\scriptsize{$\rm{Ca} \rm{Ga}_{2} \rm{As}_{2}$ - \icsdweb{422526} - SG 166 ($R\bar{3}m$) - SEBR} & \scriptsize{$\rm{Pb} \rm{Bi}_{2} \rm{Te}_{4}$ - \icsdweb{616936} - SG 166 ($R\bar{3}m$) - SEBR}\\
\tiny{ $\;Z_{2,1}=1\;Z_{2,2}=1\;Z_{2,3}=1\;Z_4=3$ } & \tiny{ $\;Z_{2,1}=1\;Z_{2,2}=1\;Z_{2,3}=1\;Z_4=1$ }\\
\includegraphics[width=0.38\textwidth,angle=0]{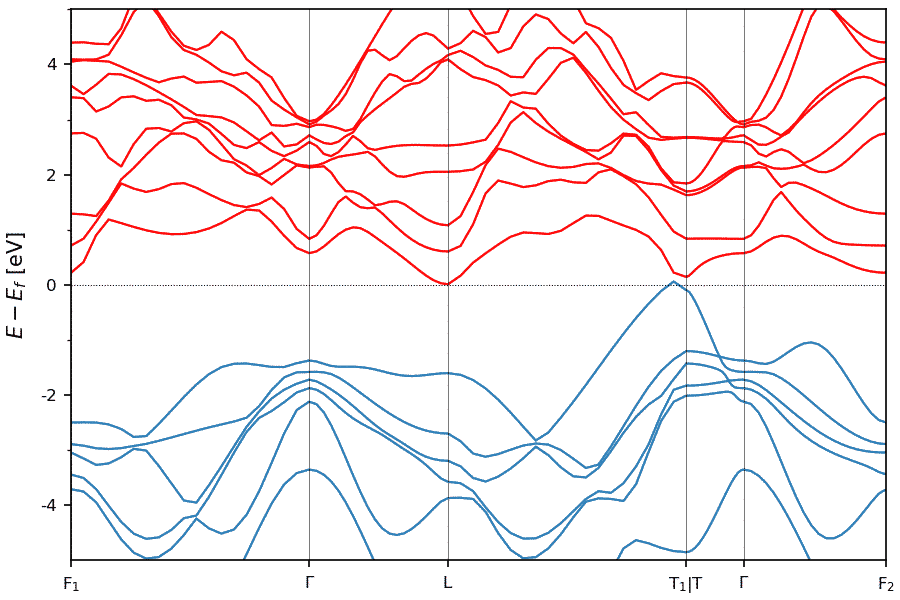} & \includegraphics[width=0.38\textwidth,angle=0]{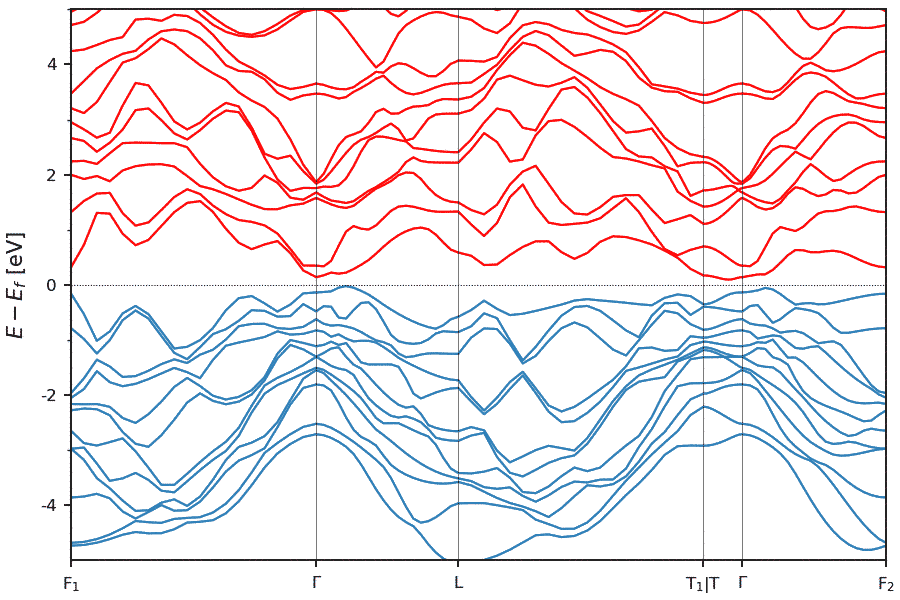}\\
\end{tabular}
\begin{tabular}{c c}
\scriptsize{$\rm{Bi}_{2} \rm{Se}_{3}$ - \icsdweb{617079} - SG 166 ($R\bar{3}m$) - SEBR} & \scriptsize{$\rm{Tl} \rm{Bi} \rm{Se}_{2}$ - \icsdweb{617113} - SG 166 ($R\bar{3}m$) - SEBR}\\
\tiny{ $\;Z_{2,1}=0\;Z_{2,2}=0\;Z_{2,3}=0\;Z_4=3$ } & \tiny{ $\;Z_{2,1}=0\;Z_{2,2}=0\;Z_{2,3}=0\;Z_4=1$ }\\
\includegraphics[width=0.38\textwidth,angle=0]{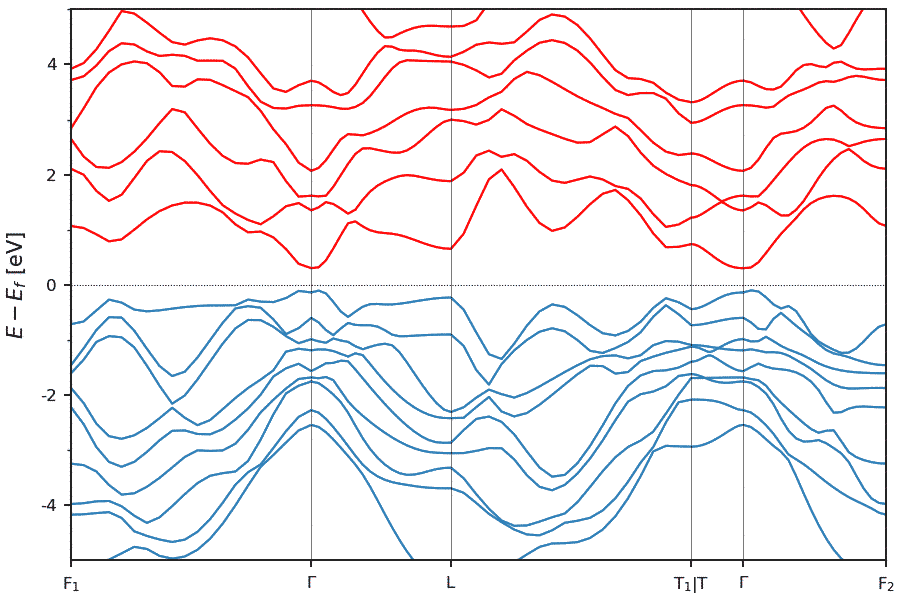} & \includegraphics[width=0.38\textwidth,angle=0]{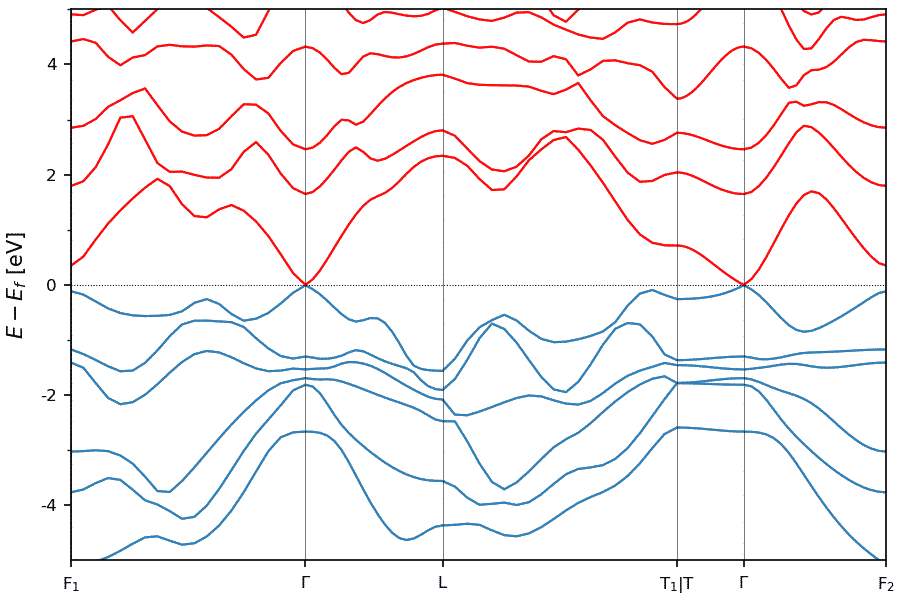}\\
\end{tabular}
\begin{tabular}{c c}
\scriptsize{$\rm{Sb}$ - \icsdweb{651490} - SG 166 ($R\bar{3}m$) - SEBR} & \scriptsize{$\rm{Sb} \rm{Tl} \rm{Te}_{2}$ - \icsdweb{651646} - SG 166 ($R\bar{3}m$) - SEBR}\\
\tiny{ $\;Z_{2,1}=1\;Z_{2,2}=1\;Z_{2,3}=1\;Z_4=3$ } & \tiny{ $\;Z_{2,1}=0\;Z_{2,2}=0\;Z_{2,3}=0\;Z_4=1$ }\\
\includegraphics[width=0.38\textwidth,angle=0]{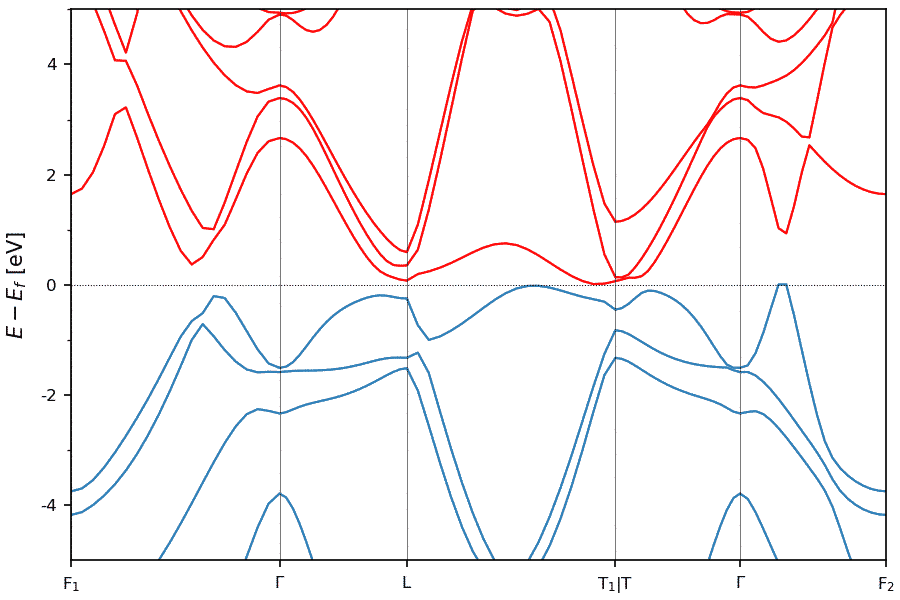} & \includegraphics[width=0.38\textwidth,angle=0]{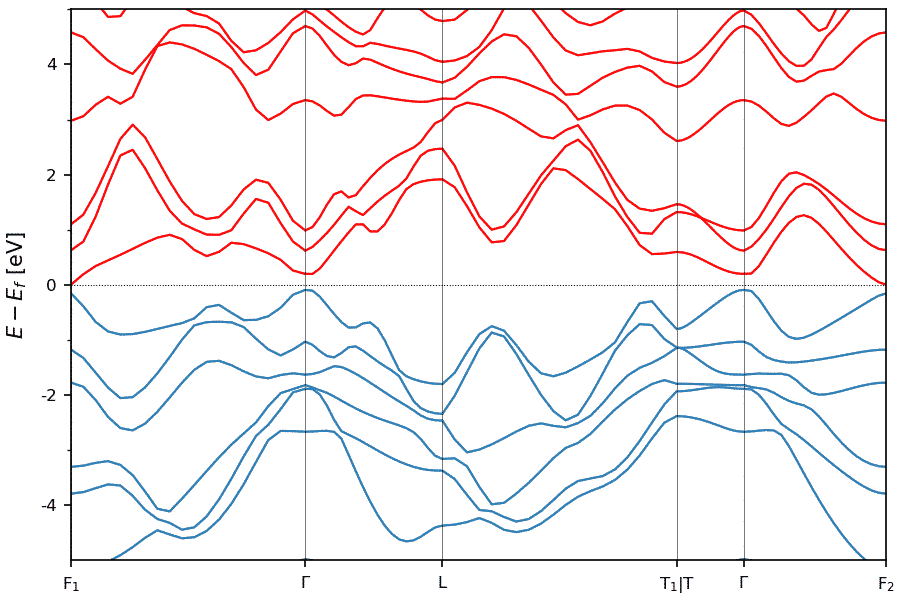}\\
\end{tabular}
\begin{tabular}{c c}
\scriptsize{$\rm{Sn} \rm{Sb}_{2} \rm{Te}_{4}$ - \icsdweb{657316} - SG 166 ($R\bar{3}m$) - SEBR} & \scriptsize{$\rm{Mg} \rm{Po}$ - \icsdweb{150664} - SG 194 ($P6_3/mmc$) - SEBR}\\
\tiny{ $\;Z_{2,1}=1\;Z_{2,2}=1\;Z_{2,3}=1\;Z_4=1$ } & \tiny{ $\;Z_{2,1}=0\;Z_{2,2}=0\;Z_{2,3}=0\;Z_4=3\;Z_{6m,0}=5\;Z_{12}'=11$ }\\
\includegraphics[width=0.38\textwidth,angle=0]{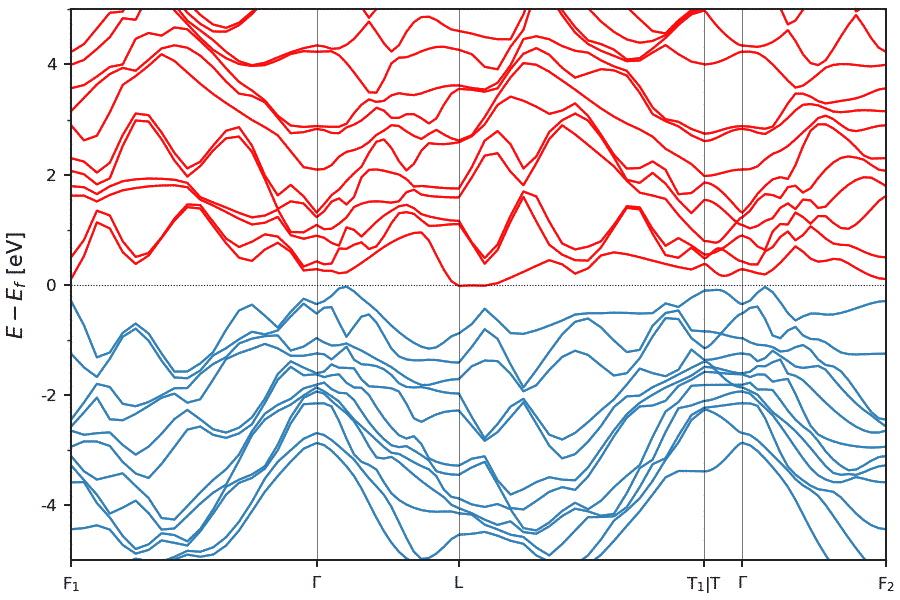} & \includegraphics[width=0.38\textwidth,angle=0]{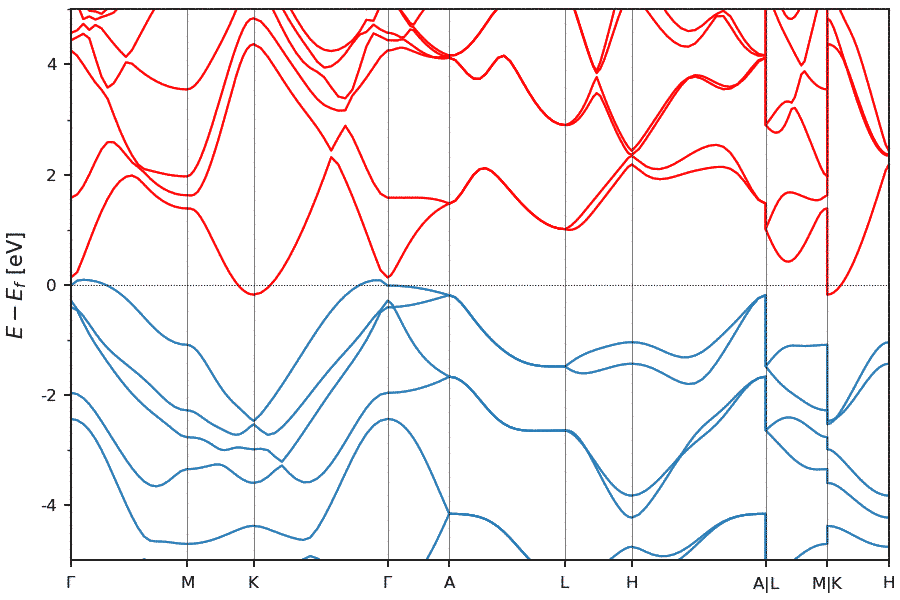}\\
\end{tabular}

\caption{\ztwoTISEBR{4}}
\label{fig:z2TI_SEBR4}
\end{figure}

\begin{figure}[ht]
\centering
\begin{tabular}{c c}
\scriptsize{$\rm{Pb} \rm{S}$ - \icsdweb{183244} - SG 194 ($P6_3/mmc$) - SEBR} & \scriptsize{$\rm{Zn}_{3} \rm{N}_{2}$ - \icsdweb{84918} - SG 206 ($Ia\bar{3}$) - SEBR}\\
\tiny{ $\;Z_{2,1}=0\;Z_{2,2}=0\;Z_{2,3}=0\;Z_4=1\;Z_{6m,0}=5\;Z_{12}'=5$ } & \tiny{ $\;Z_{2,1}=0\;Z_{2,2}=0\;Z_{2,3}=0\;Z_4=3$ }\\
\includegraphics[width=0.38\textwidth,angle=0]{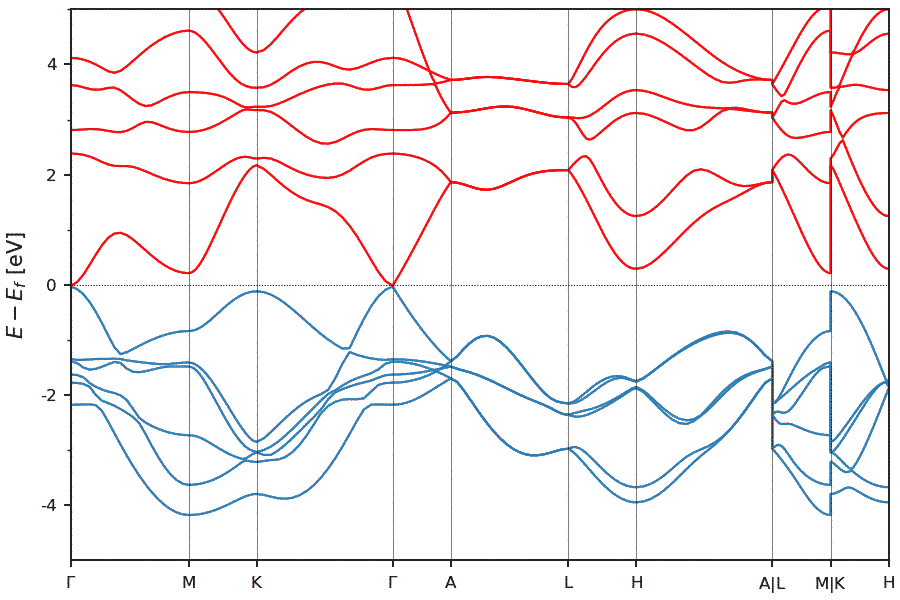} & \includegraphics[width=0.38\textwidth,angle=0]{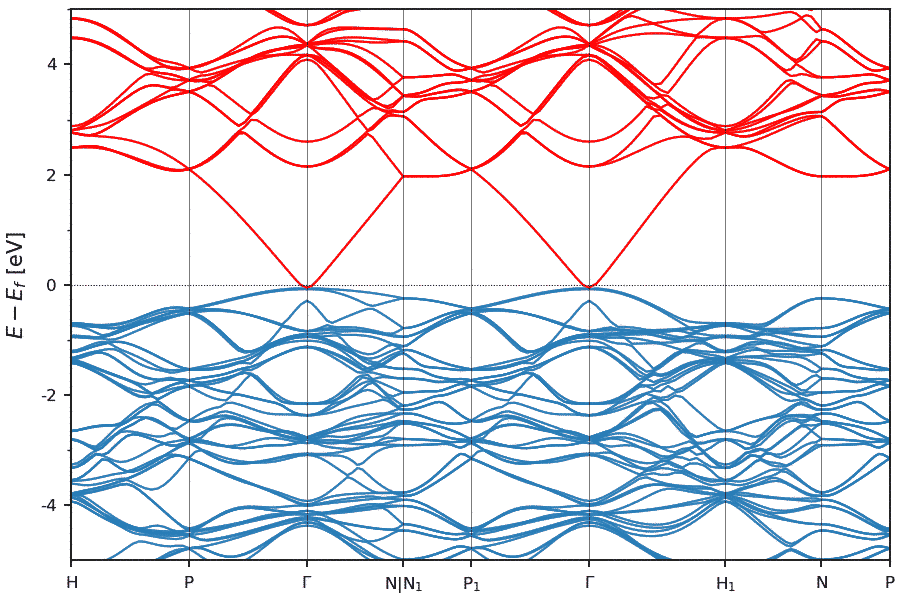}\\
\end{tabular}
\begin{tabular}{c c}
\scriptsize{$\rm{Cd}_{3} \rm{N}_{2}$ - \icsdweb{416908} - SG 206 ($Ia\bar{3}$) - SEBR} & \scriptsize{$\rm{Cs} \rm{Ge} \rm{Br}_{3}$ - \icsdweb{80320} - SG 221 ($Pm\bar{3}m$) - SEBR}\\
\tiny{ $\;Z_{2,1}=0\;Z_{2,2}=0\;Z_{2,3}=0\;Z_4=3$ } & \tiny{ $\;Z_{2,1}=1\;Z_{2,2}=1\;Z_{2,3}=1\;Z_4=3\;Z_{4m,\pi}=3\;Z_2=1\;Z_8=3$ }\\
\includegraphics[width=0.38\textwidth,angle=0]{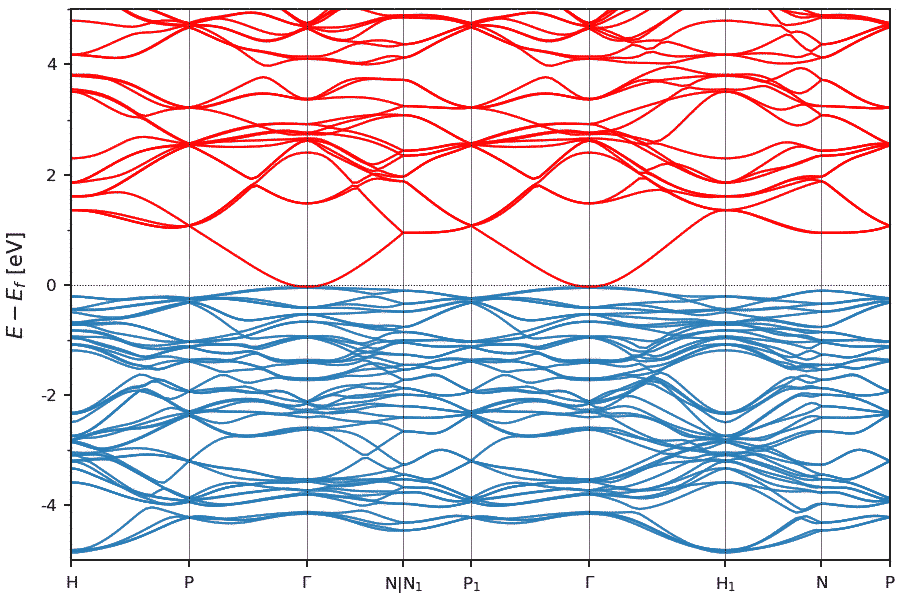} & \includegraphics[width=0.38\textwidth,angle=0]{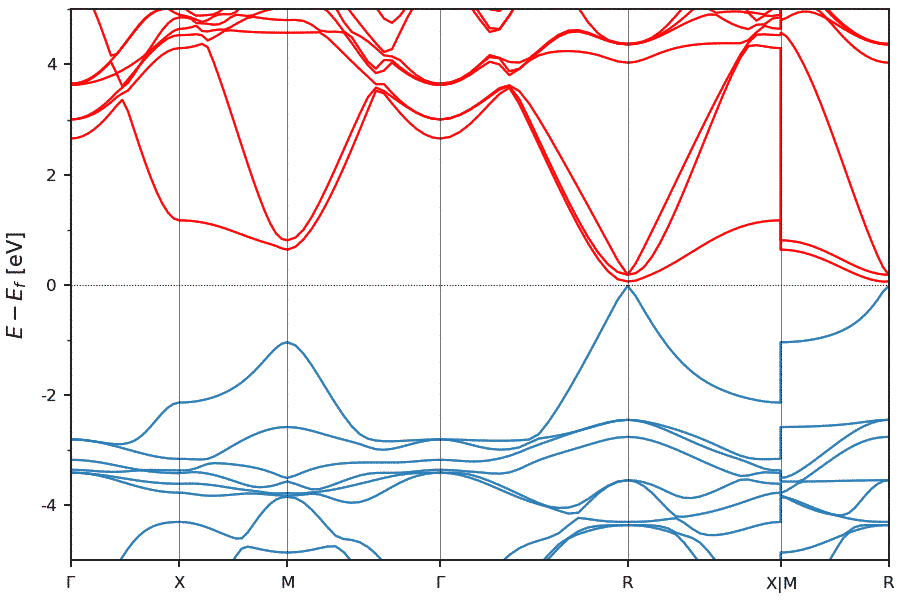}\\
\end{tabular}
\begin{tabular}{c c}
\scriptsize{$\rm{Cs} \rm{Pb} \rm{Cl}_{3}$ - \icsdweb{181781} - SG 221 ($Pm\bar{3}m$) - SEBR} & \scriptsize{$\rm{Cs} \rm{Pb} \rm{Br}_{3}$ - \icsdweb{201285} - SG 221 ($Pm\bar{3}m$) - SEBR}\\
\tiny{ $\;Z_{2,1}=1\;Z_{2,2}=1\;Z_{2,3}=1\;Z_4=3\;Z_{4m,\pi}=3\;Z_2=1\;Z_8=3$ } & \tiny{ $\;Z_{2,1}=1\;Z_{2,2}=1\;Z_{2,3}=1\;Z_4=3\;Z_{4m,\pi}=3\;Z_2=1\;Z_8=3$ }\\
\includegraphics[width=0.38\textwidth,angle=0]{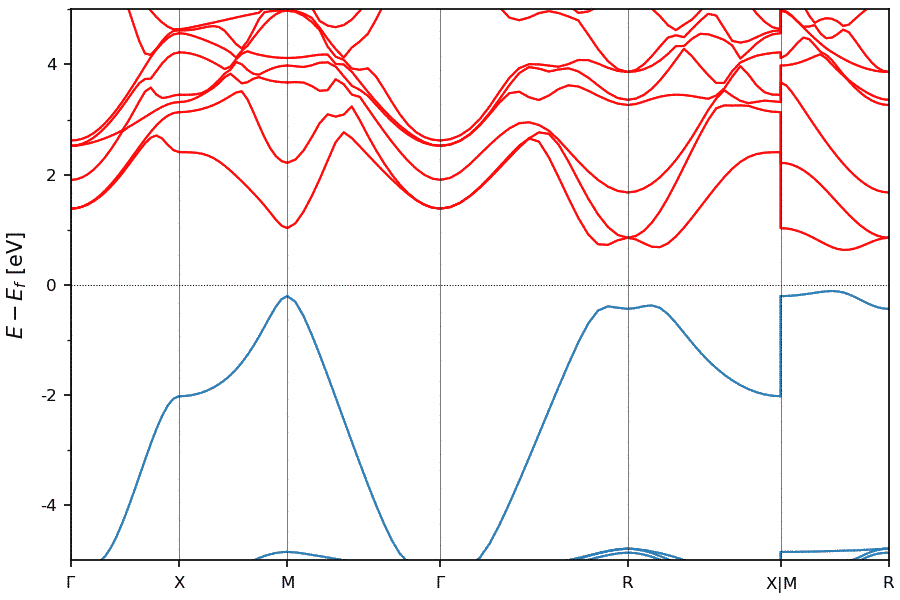} & \includegraphics[width=0.38\textwidth,angle=0]{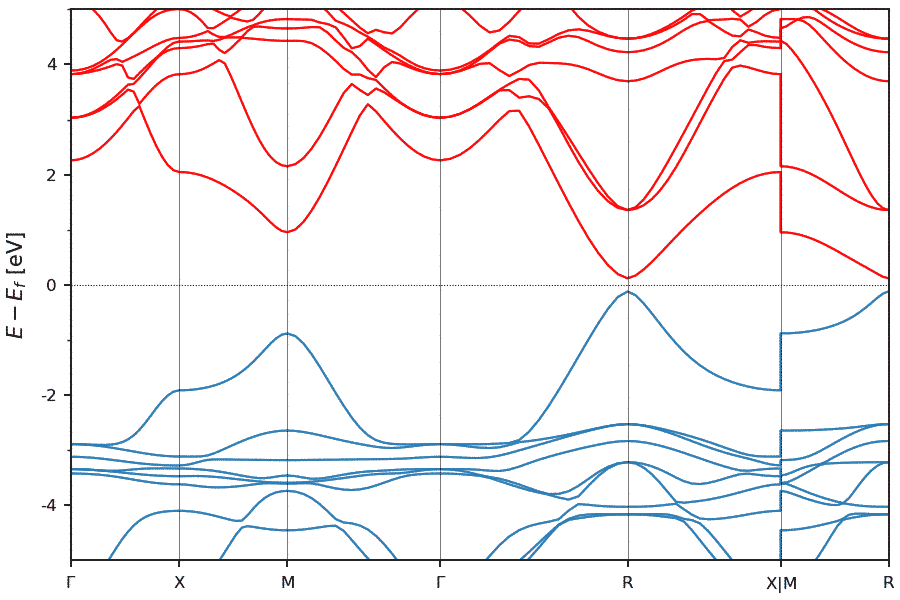}\\
\end{tabular}
\begin{tabular}{c c}
\scriptsize{$\rm{Cs} \rm{Sn} \rm{I}_{3}$ - \icsdweb{262924} - SG 221 ($Pm\bar{3}m$) - SEBR} & \scriptsize{$\rm{Cu}_{2} \rm{S}$ - \icsdweb{628818} - SG 225 ($Fm\bar{3}m$) - SEBR}\\
\tiny{ $\;Z_{2,1}=1\;Z_{2,2}=1\;Z_{2,3}=1\;Z_4=3\;Z_{4m,\pi}=3\;Z_2=1\;Z_8=3$ } & \tiny{ $\;Z_{2,1}=0\;Z_{2,2}=0\;Z_{2,3}=0\;Z_4=3\;Z_2=1\;Z_8=7$ }\\
\includegraphics[width=0.38\textwidth,angle=0]{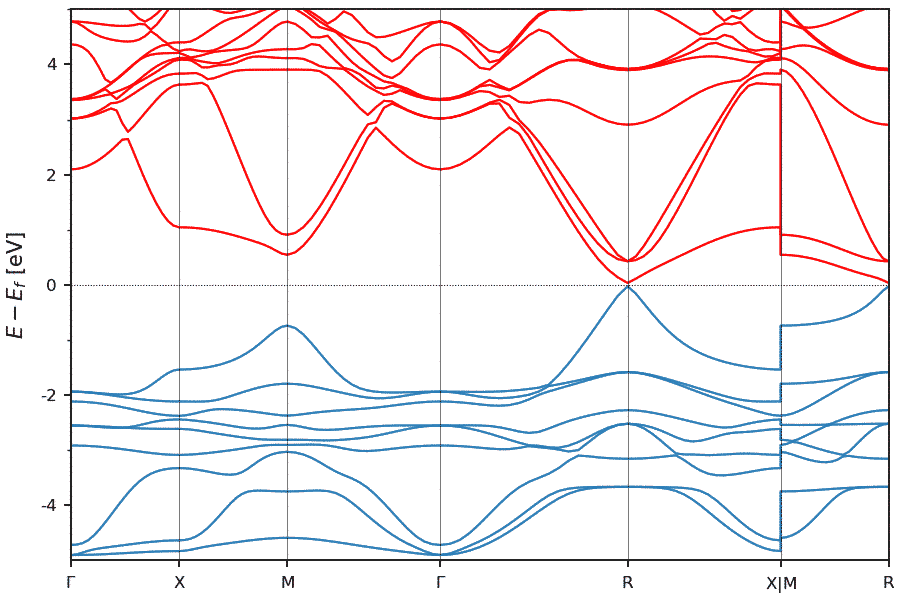} & \includegraphics[width=0.38\textwidth,angle=0]{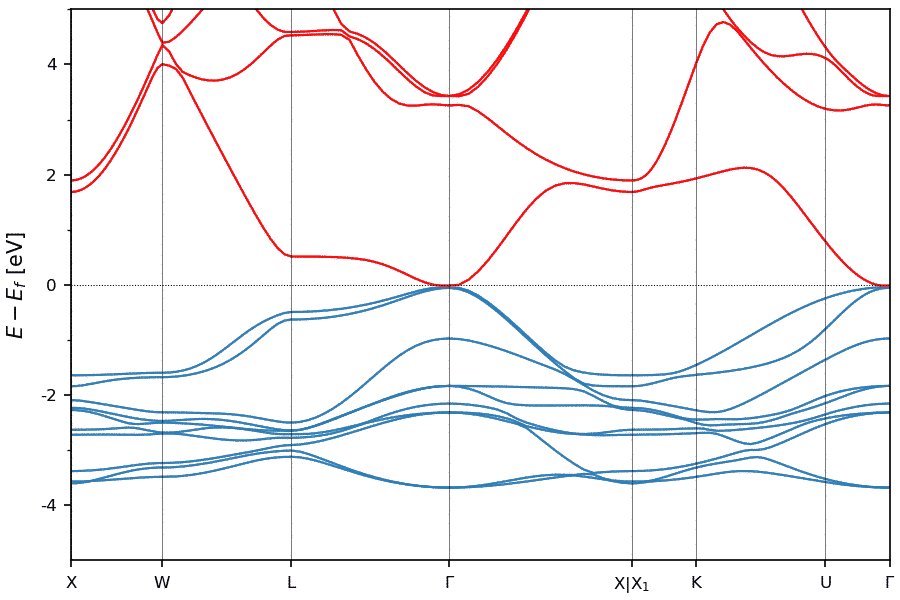}\\
\end{tabular}

\caption{\ztwoTISEBR{5}}
\label{fig:z2TI_SEBR5}
\end{figure}

\clearpage

\subsubsection{$S_{4}$-Rotoinversion-Indicated Strong TIs and Weyl Semimetals}
\label{App:s4weylTI}

In this section, we list the noncentrosymmetric stable topological materials with fourfold rotoinversion $S_{4} = C_{4}\times\mathcal{I}$ symmetry and with the largest gaps or the fewest Fermi pockets along all high-symmetry BZ lines.  As shown in Ref.~\onlinecite{ChenTCI}, these materials can either be strong 3D TIs or Weyl semimetals, and can only be further distinguished by performing an additional calculation beyond symmetry-based indicators~\cite{S4Weyl}.  In terms of the indices introduced in Ref.~\onlinecite{ChenTCI}, the materials listed in this section exhibit $Z_{2}=1$ (which is not to be confused with the Fu-Kane parity criterion~\cite{FuKaneMele,FuKaneInversion}, see \supappref{App:strongTIs} for further details), and do not have a well-defined $Z_{4}$ index, because they are noncentrosymmetric.  First, in Fig.~\ref{fig:z2rotoWeyl_TI_NLC}, we list the $S_{4}$ TIs and Weyl semimetals classified as NLC, and then, in Fig.~\ref{fig:z2rotoWeyl_TI_SEBR}, we list the $S_{4}$ TIs and Weyl semimetals classified as SEBR.  Among the stoichiometric materials in the ICSD, we find that there are relatively few examples of noncentrosymmetric $S_{4}$-symmetry-indicated TIs and Weyl semimetals that either have SGs in which the only stable SI is $Z_{2}$, or which have trivial values for all stable SIs aside from $Z_{2}$.


\begin{figure}[ht]
\centering
\begin{tabular}{c c}
\scriptsize{$\rm{Cu}_{2} \rm{Zn} \rm{Sn} \rm{Se}_{4}$ - \icsdweb{189278} - SG 82 ($I\bar{4}$) - NLC} & \scriptsize{$\rm{Y}_{2} \rm{C}_{3}$ - \icsdweb{77572} - SG 220 ($I\bar{4}3d$) - NLC}\\
\tiny{ $\;Z_2=1$ } & \tiny{ $\;Z_2=1$ }\\
\includegraphics[width=0.38\textwidth,angle=0]{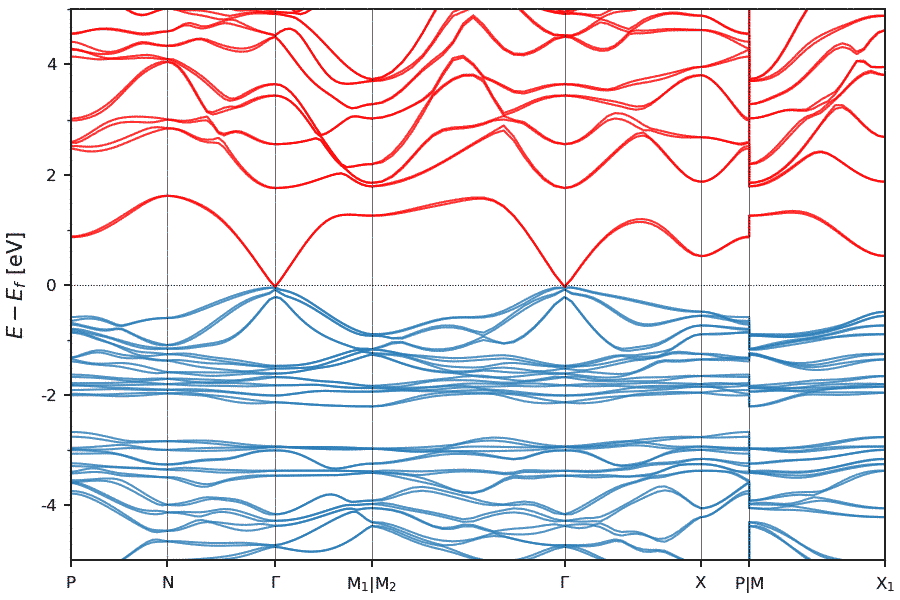} & \includegraphics[width=0.38\textwidth,angle=0]{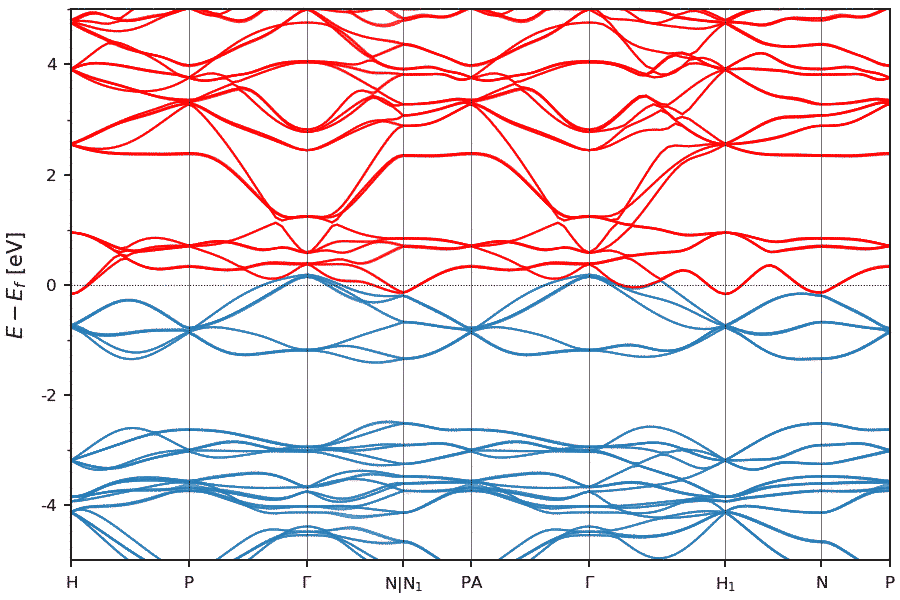}\\
\end{tabular}

\caption{The NLC-classified, $S_{4}$-rotoinversion-indicated, noncentrosymmetric 3D strong TIs and Weyl semimetals with the largest band gaps along high-symmetry lines or the fewest and smallest bulk Fermi pockets.}
\label{fig:z2rotoWeyl_TI_NLC}
\end{figure}


\begin{figure}[ht]
\centering
\begin{tabular}{c c}
\scriptsize{$\rm{Cu}_{2} \rm{Hg} \rm{Sn} \rm{Se}_{4}$ - \icsdweb{95119} - SG 121 ($I\bar{4}2m$) - SEBR} & \scriptsize{$\rm{Cu}_{2} \rm{Hg} \rm{Ge} \rm{Se}_{4}$ - \icsdweb{152761} - SG 121 ($I\bar{4}2m$) - SEBR}\\
\tiny{ $\;Z_2=1$ } & \tiny{ $\;Z_2=1$ }\\
\includegraphics[width=0.38\textwidth,angle=0]{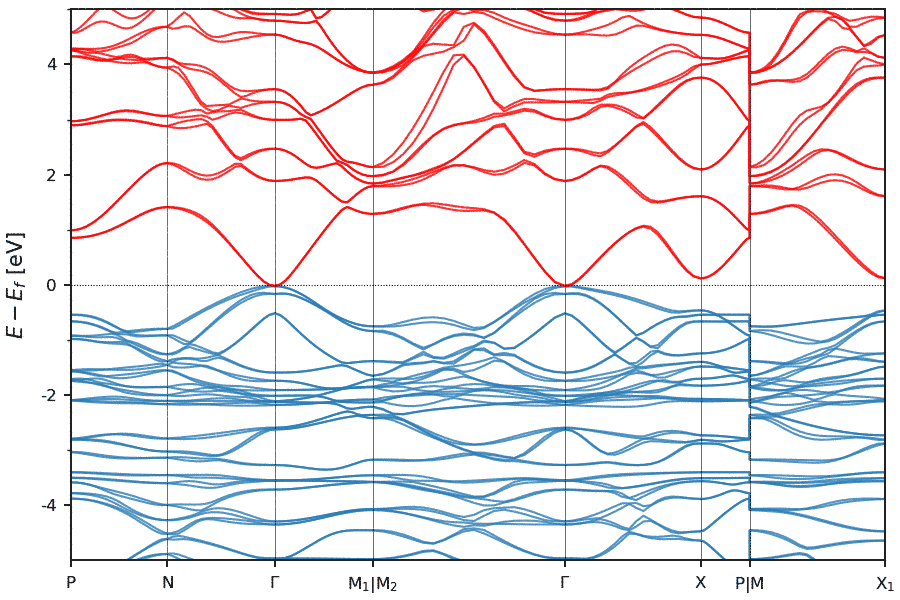} & \includegraphics[width=0.38\textwidth,angle=0]{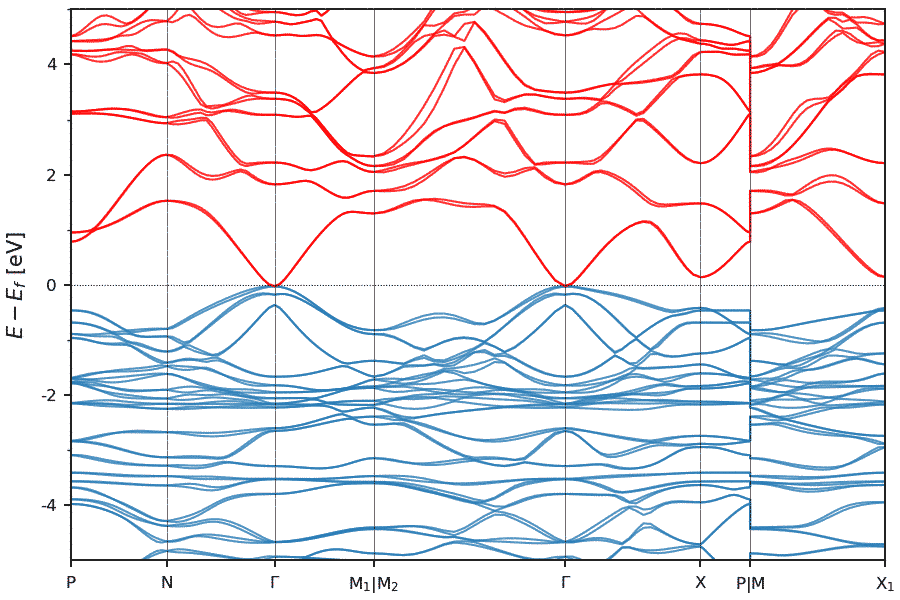}\\
\end{tabular}
\begin{tabular}{c c}
\scriptsize{$\rm{Cd} \rm{Ge} \rm{As}_{2}$ - \icsdweb{16736} - SG 122 ($I\bar{4}2d$) - SEBR} & \scriptsize{$\rm{Zn} \rm{Sn} \rm{As}_{2}$ - \icsdweb{22178} - SG 122 ($I\bar{4}2d$) - SEBR}\\
\tiny{ $\;Z_2=1$ } & \tiny{ $\;Z_2=1$ }\\
\includegraphics[width=0.38\textwidth,angle=0]{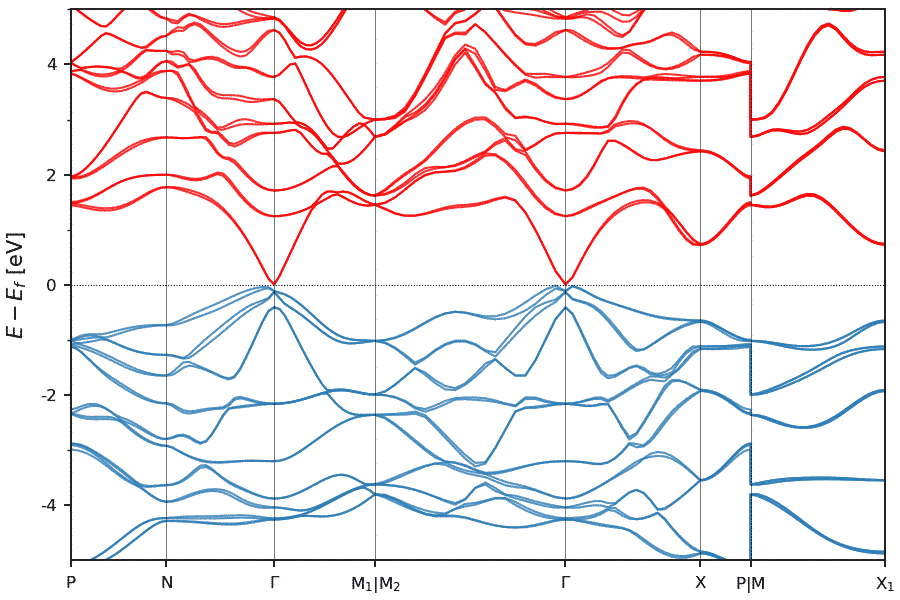} & \includegraphics[width=0.38\textwidth,angle=0]{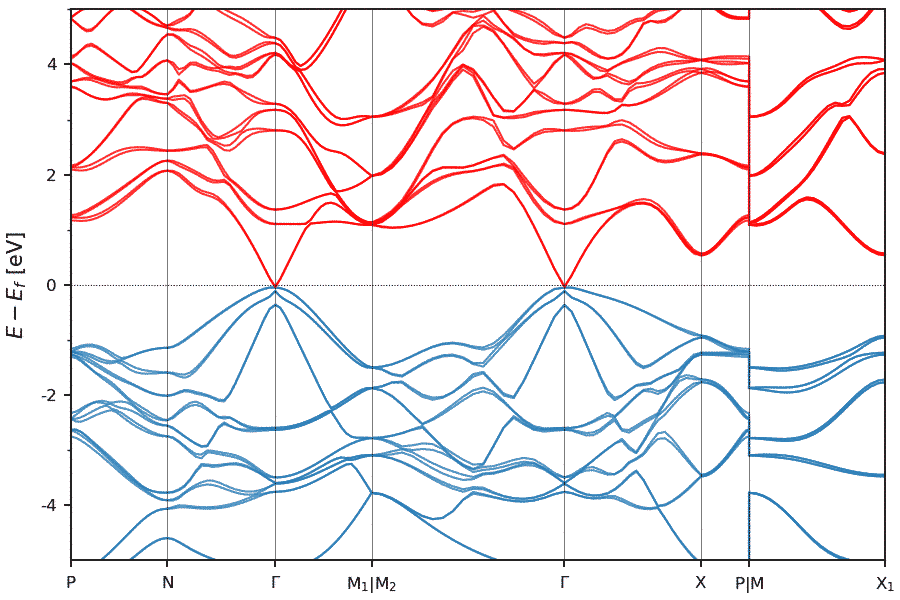}\\
\end{tabular}
\begin{tabular}{c c}
\scriptsize{$\rm{Cu} \rm{Tl} \rm{Se}_{2}$ - \icsdweb{28743} - SG 122 ($I\bar{4}2d$) - SEBR} & \scriptsize{$\rm{Cu} \rm{In} \rm{S}_{2}$ - \icsdweb{600582} - SG 122 ($I\bar{4}2d$) - SEBR}\\
\tiny{ $\;Z_2=1$ } & \tiny{ $\;Z_2=1$ }\\
\includegraphics[width=0.38\textwidth,angle=0]{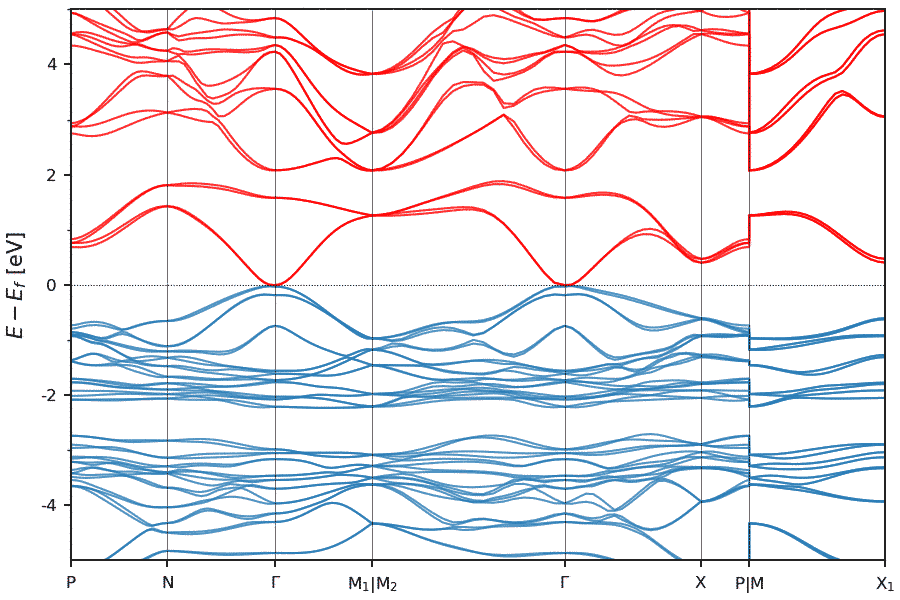} & \includegraphics[width=0.38\textwidth,angle=0]{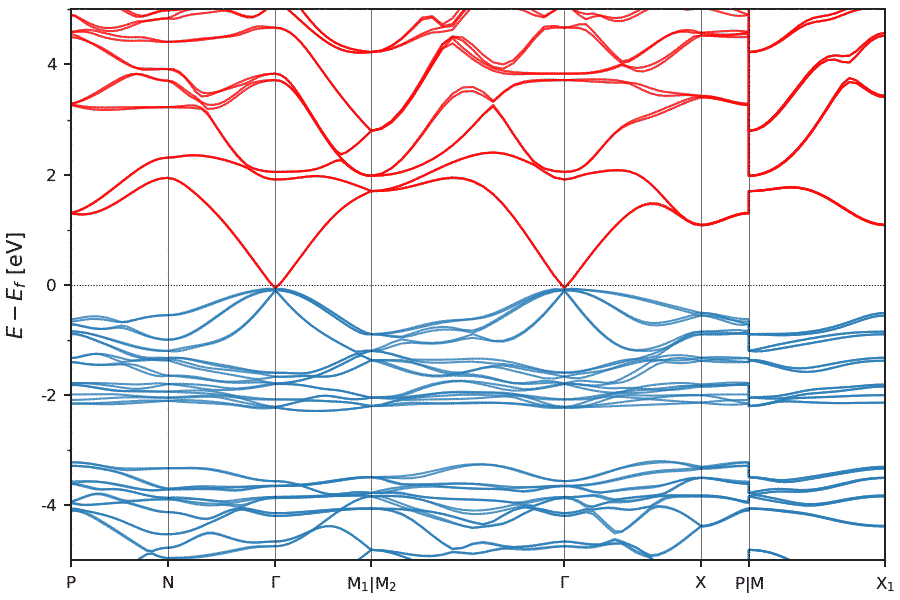}\\
\end{tabular}
\begin{tabular}{c c}
\scriptsize{$\rm{Cd} \rm{Sn} \rm{As}_{2}$ - \icsdweb{609982} - SG 122 ($I\bar{4}2d$) - SEBR} & \scriptsize{$\rm{Cu} \rm{Tl} \rm{S}_{2}$ - \icsdweb{628931} - SG 122 ($I\bar{4}2d$) - SEBR}\\
\tiny{ $\;Z_2=1$ } & \tiny{ $\;Z_2=1$ }\\
\includegraphics[width=0.38\textwidth,angle=0]{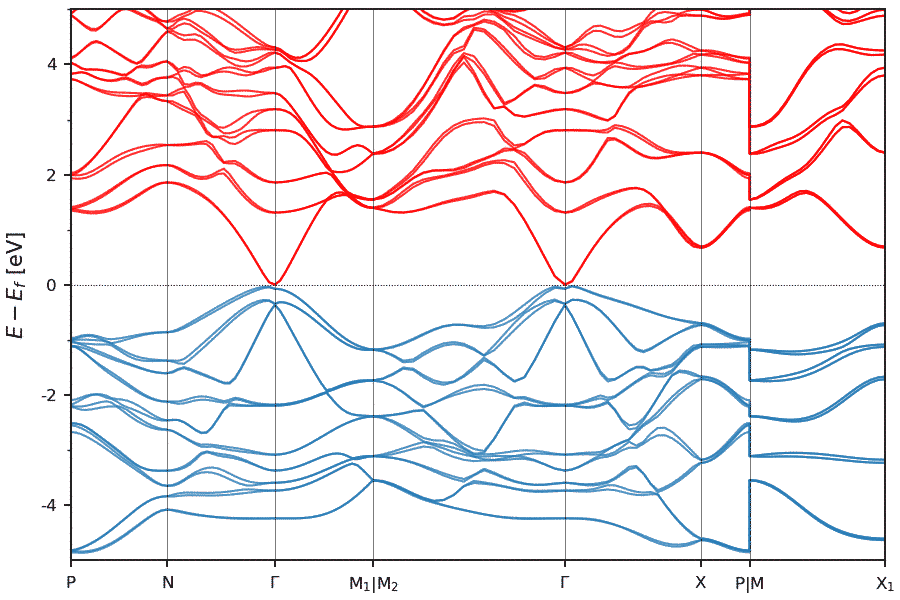} & \includegraphics[width=0.38\textwidth,angle=0]{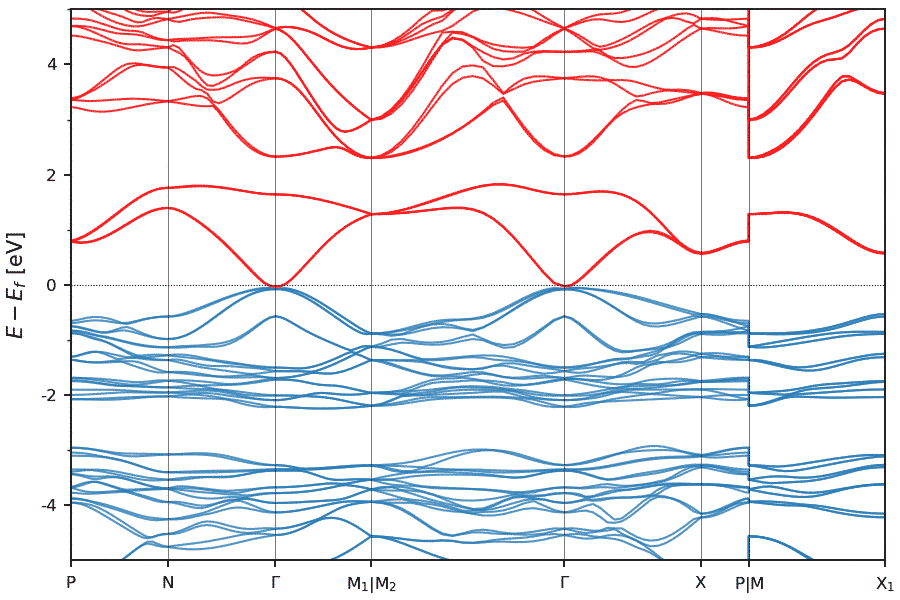}\\
\end{tabular}

\caption{The SEBR-classified, $S_{4}$-rotoinversion-indicated, noncentrosymmetric 3D strong TIs and Weyl semimetals with the largest band gaps along high-symmetry lines or the fewest and smallest bulk Fermi pockets.}
\label{fig:z2rotoWeyl_TI_SEBR}
\end{figure}

\clearpage

\subsubsection{Inversion-Symmetry-Protected HOTIs}
\label{App:z4HOTIs}

In this section, we list the inversion-symmetry-indicated HOTIs with the largest bulk gaps or the fewest and smallest bulk Fermi pockets.  In ideal models of inversion-symmetry-indicated HOTIs, when symmetries other than $\mathcal{I}$ and $\mathcal{T}$ are relaxed, finite samples exhibit gapped 2D surfaces and gapless 1D hinges with anomalous helical modes~\cite{ChenRotation,HOTIBismuth,TMDHOTI,AshvinIndicators,AshvinTCI,ChenTCI}.  In the presence of additional symmetries, such as mirror reflection and twofold rotation ($C_{2}$), inversion-symmetry-indicated HOTIs may also exhibit $2 + 4n$ twofold Dirac cones on specific, high-symmetry surfaces~\cite{ChenRotation,AshvinTCI,ChenTCI,DiracInsulator}.  In terms of the indices introduced in Ref.~\cite{ChenTCI}, the materials listed in this section are characterized by $Z_{4}=2$, and exhibit trivial values for all other independent stable SIs.  First, in Figs.~\ref{fig:z4_2HOTIs_NLC1},~\ref{fig:z4_2HOTIs_NLC2},~\ref{fig:z4_2HOTIs_NLC3},~\ref{fig:z4_2HOTIs_NLC4},~\ref{fig:z4_2HOTIs_NLC5}, and~\ref{fig:z4_2HOTIs_NLC6}, we list the HOTIs classified as NLC, and then, in Figs.~\ref{fig:z4_2HOTIs_SEBR1} and~\ref{fig:z4_2HOTIs_SEBR2}, we list the HOTIs classified as SEBR.  The materials listed in this section include the recently identified HOTIs bismuth [\icsdweb{64705}, SG 166 ($R\bar{3}m$)]~\cite{HOTIBismuth,BismuthFacet} and BiBr [\icsdweb{1560}, SG 12 ($C2/m$)]~\cite{BiBrFan,AshvinFirstMaterials,BiBrSuyang,BiBrFanHOTI,BiBrNatMater}, which are classified as SEBR, and MoTe$_2$ [\icsdweb{14349}, SG 11 ($P2_{1}/m$)]~\cite{TMDHOTI,AshvinFirstMaterials}, which is classified as NLC.  Bismuth, BiBr, and MoTe$_2$, as well as members of closely-related material families have been the subject of recent intense theoretical and experimental investigations seeking to distinguish $\mathcal{I}$- and $\mathcal{T}$-symmetric HOTIs from weak TIs and other TI and TCI phases~\cite{BismuthHaimDefect,RaquelPartial,DavidMoTe2Exp,MazWTe2Exp,WTe2HingeStep,PhuanOngMoTe2Hinge,MTQC}.  Notably, $\mathcal{I}$- and $\mathcal{T}$-symmetric HOTIs exhibit trivial axion angles $\theta\text{ mod }2\pi=0$, and it remains an open and urgent question whether there exist quantized electromagnetic response effects beyond the axionic magnetoelectric effect that can distinguish $\mathcal{I}$- and $\mathcal{T}$-symmetric HOTI phases from other TCIs and from trivial insulators~\cite{WiederAxion,MTQC,WiederBarryCDW}.


\begin{figure}[ht]
\centering
\begin{tabular}{c c}
\scriptsize{$\rm{Bi}$ - \icsdweb{426929} - SG 2 ($P\bar{1}$) - NLC} & \scriptsize{$\rm{Mo} \rm{Te}_{2}$ - \icsdweb{14349} - SG 11 ($P2_1/m$) - NLC}\\
\tiny{ $\;Z_{2,1}=0\;Z_{2,2}=0\;Z_{2,3}=0\;Z_4=2$ } & \tiny{ $\;Z_{2,1}=0\;Z_{2,2}=0\;Z_{2,3}=0\;Z_4=2$ }\\
\includegraphics[width=0.38\textwidth,angle=0]{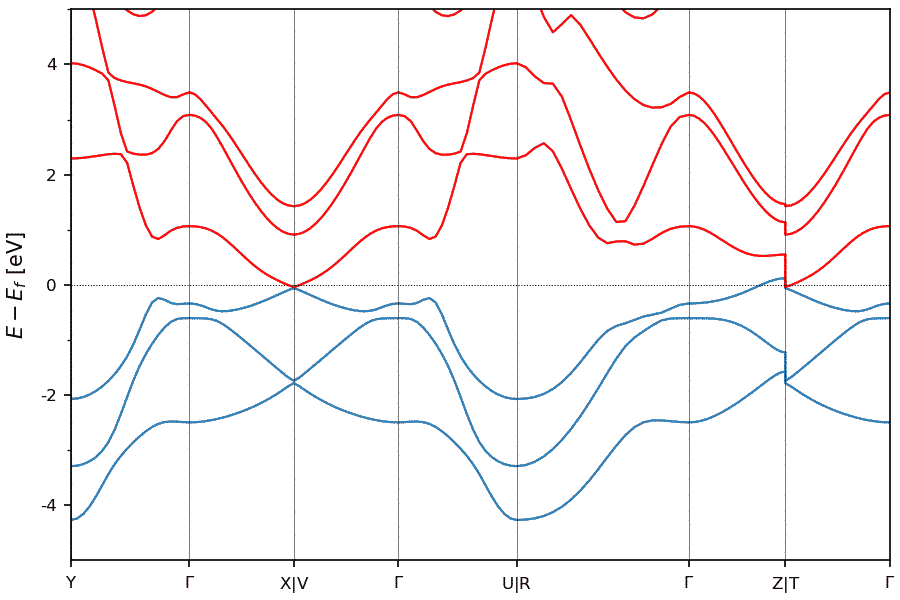} & \includegraphics[width=0.38\textwidth,angle=0]{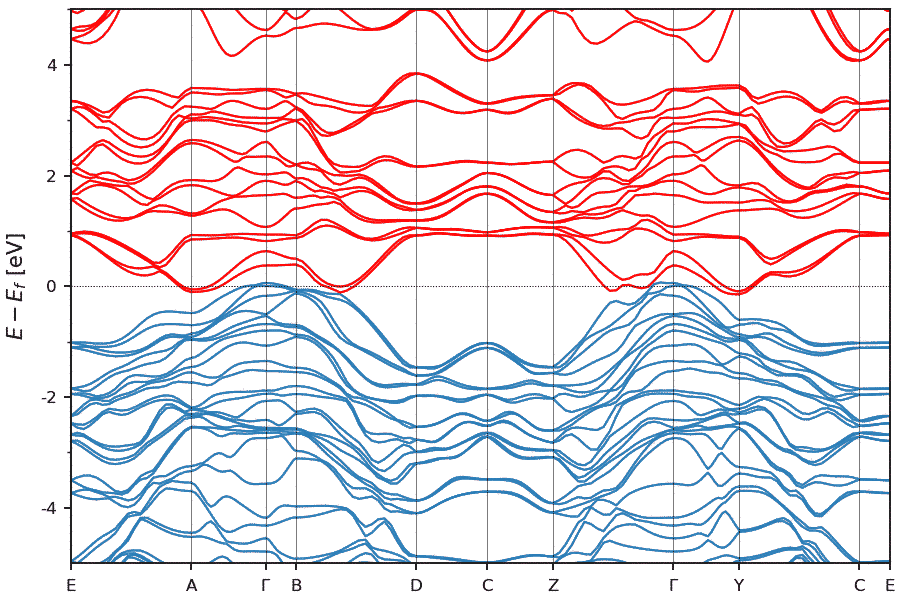}\\
\end{tabular}
\begin{tabular}{c c}
\scriptsize{$\rm{Fe} \rm{H}_{4}$ - \icsdweb{187148} - SG 11 ($P2_1/m$) - NLC} & \scriptsize{$\rm{Ba} \rm{Pd}_{2} \rm{Bi}_{2}$ - \icsdweb{416299} - SG 11 ($P2_1/m$) - NLC}\\
\tiny{ $\;Z_{2,1}=0\;Z_{2,2}=0\;Z_{2,3}=0\;Z_4=2$ } & \tiny{ $\;Z_{2,1}=0\;Z_{2,2}=0\;Z_{2,3}=0\;Z_4=2$ }\\
\includegraphics[width=0.38\textwidth,angle=0]{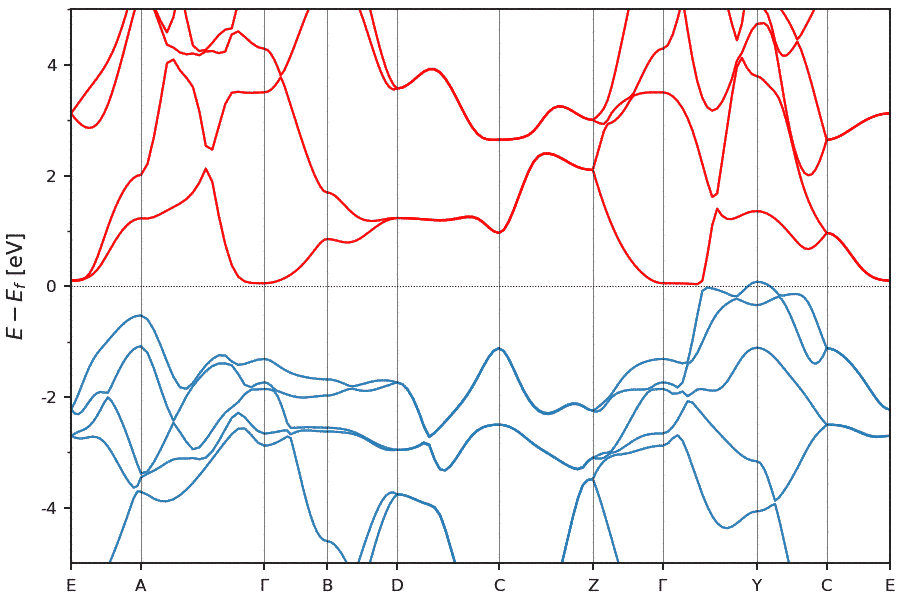} & \includegraphics[width=0.38\textwidth,angle=0]{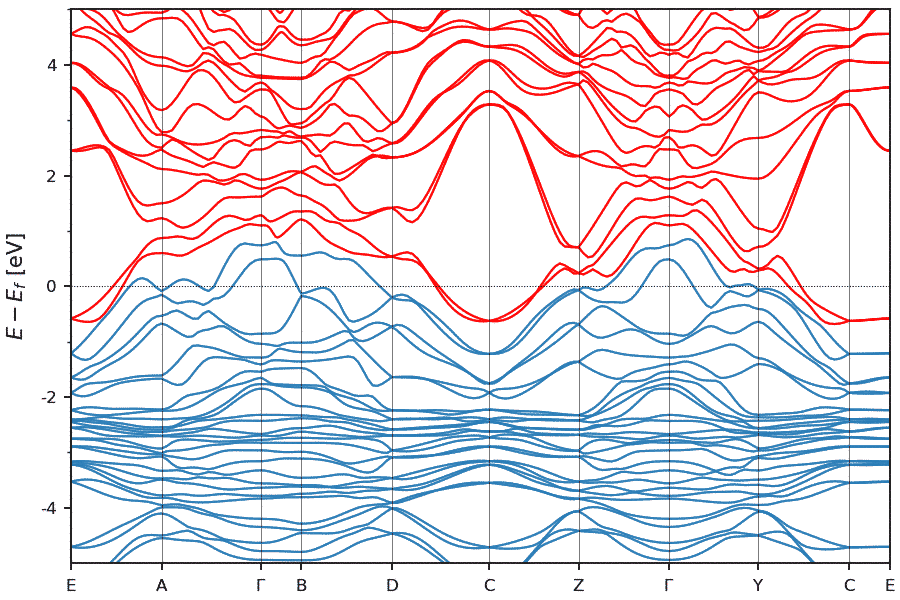}\\
\end{tabular}
\begin{tabular}{c c}
\scriptsize{$\rm{Sr} \rm{Pd}_{2} \rm{Bi}_{2}$ - \icsdweb{416300} - SG 11 ($P2_1/m$) - NLC} & \scriptsize{$\rm{Cu} (\rm{Te} \rm{O}_{4})$ - \icsdweb{1671} - SG 14 ($P2_1/c$) - NLC}\\
\tiny{ $\;Z_{2,1}=0\;Z_{2,2}=0\;Z_{2,3}=0\;Z_4=2$ } & \tiny{ $\;Z_{2,1}=0\;Z_{2,2}=0\;Z_{2,3}=0\;Z_4=2$ }\\
\includegraphics[width=0.38\textwidth,angle=0]{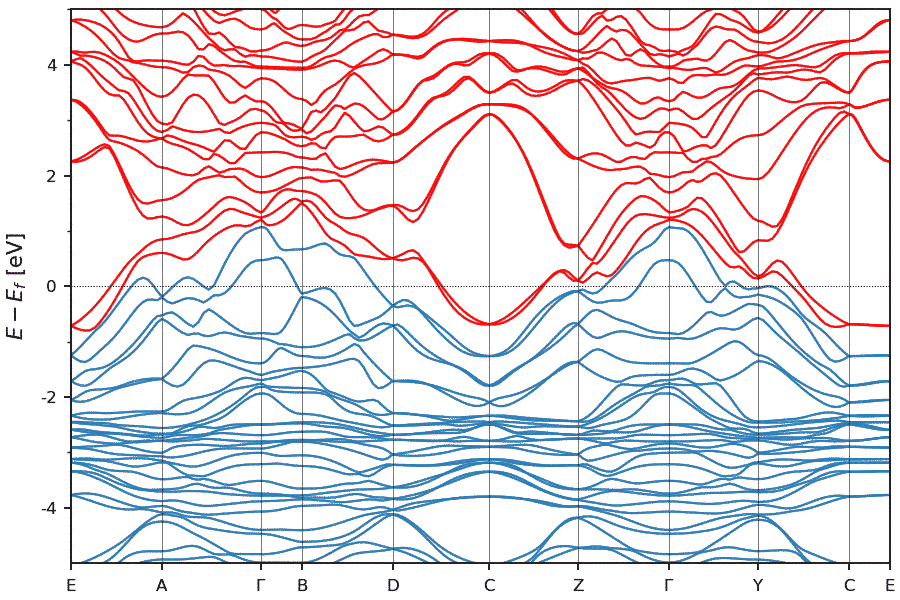} & \includegraphics[width=0.38\textwidth,angle=0]{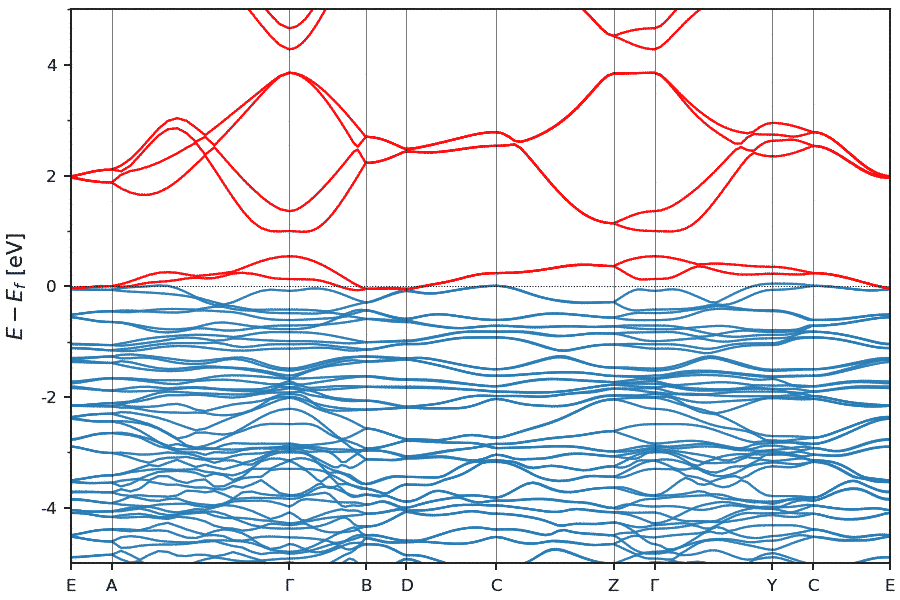}\\
\end{tabular}
\begin{tabular}{c c}
\scriptsize{$\rm{Re}_{2} \rm{Si}$ - \icsdweb{57480} - SG 14 ($P2_1/c$) - NLC} & \scriptsize{$\rm{Os} \rm{N}_{2}$ - \icsdweb{240759} - SG 14 ($P2_1/c$) - NLC}\\
\tiny{ $\;Z_{2,1}=0\;Z_{2,2}=0\;Z_{2,3}=0\;Z_4=2$ } & \tiny{ $\;Z_{2,1}=0\;Z_{2,2}=0\;Z_{2,3}=0\;Z_4=2$ }\\
\includegraphics[width=0.38\textwidth,angle=0]{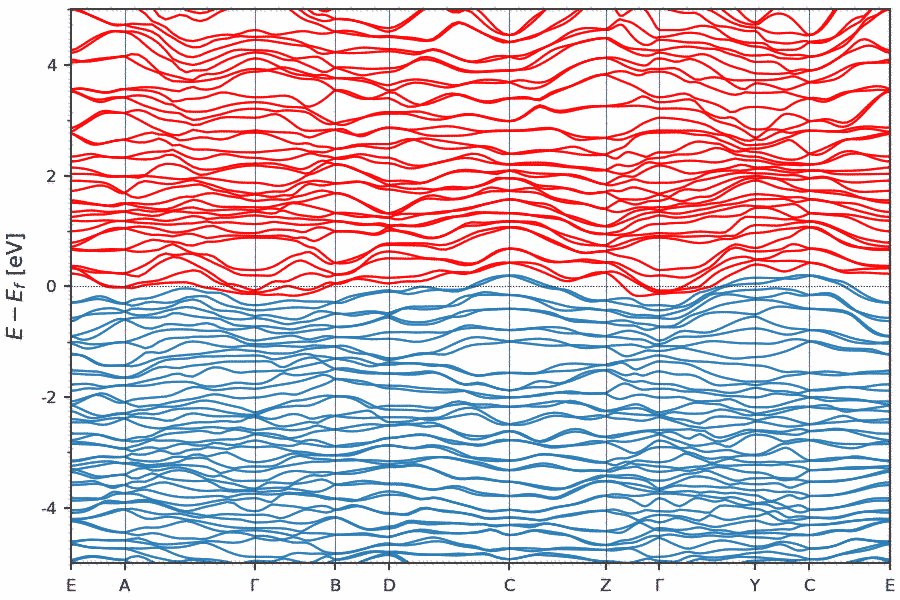} & \includegraphics[width=0.38\textwidth,angle=0]{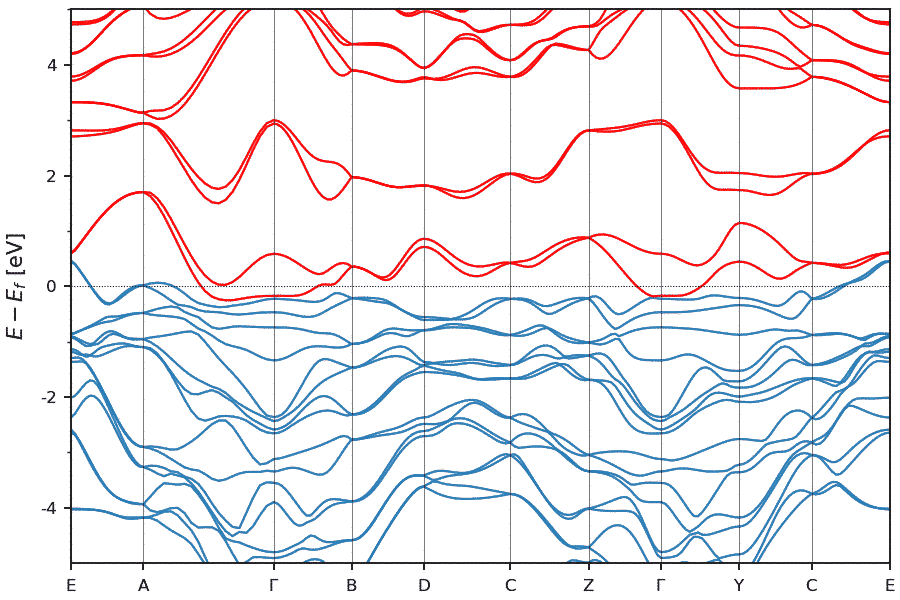}\\
\end{tabular}

\caption{\zfourHOTIsNLC{1}}
\label{fig:z4_2HOTIs_NLC1}
\end{figure}

\begin{figure}[ht]
\centering
\begin{tabular}{c c}
\scriptsize{$\rm{Y} \rm{As} \rm{S}$ - \icsdweb{611344} - SG 14 ($P2_1/c$) - NLC} & \scriptsize{$\rm{Y} \rm{As} \rm{Se}$ - \icsdweb{611398} - SG 14 ($P2_1/c$) - NLC}\\
\tiny{ $\;Z_{2,1}=0\;Z_{2,2}=0\;Z_{2,3}=0\;Z_4=2$ } & \tiny{ $\;Z_{2,1}=0\;Z_{2,2}=0\;Z_{2,3}=0\;Z_4=2$ }\\
\includegraphics[width=0.38\textwidth,angle=0]{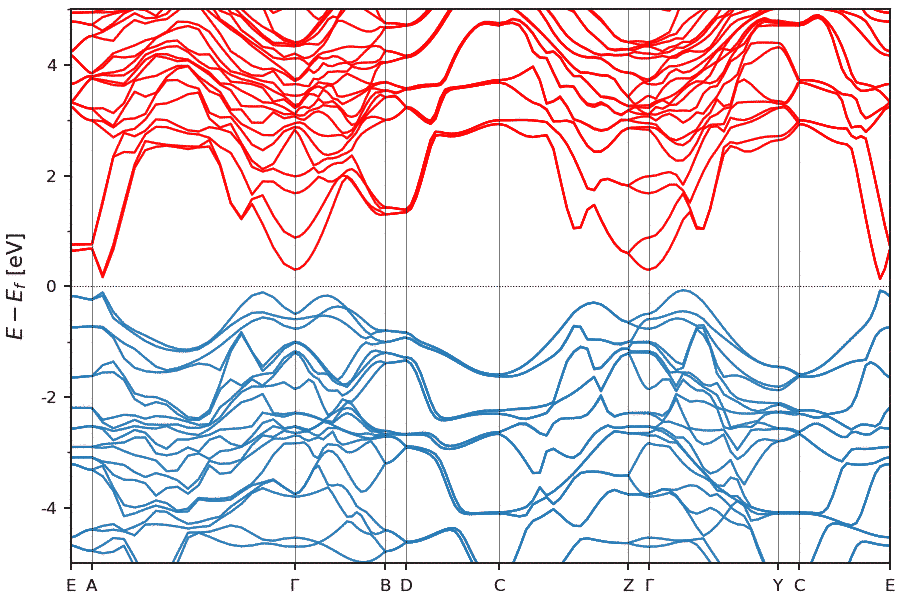} & \includegraphics[width=0.38\textwidth,angle=0]{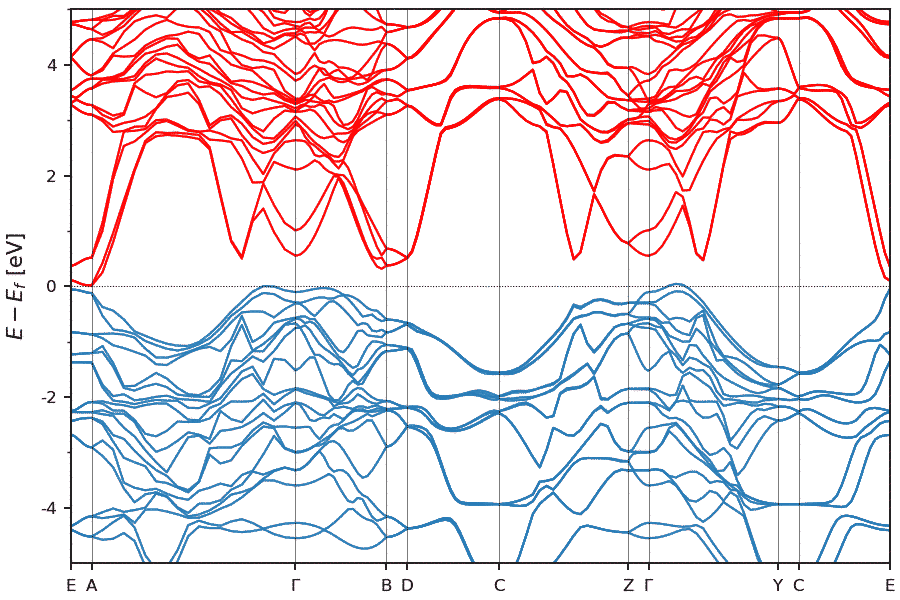}\\
\end{tabular}
\begin{tabular}{c c}
\scriptsize{$\rm{Cu}_{6} \rm{Sn}_{5}$ - \icsdweb{106530} - SG 15 ($C2/c$) - NLC} & \scriptsize{$\rm{Ca}_{5} \rm{Ga}_{2} \rm{Sb}_{6}$ - \icsdweb{36466} - SG 55 ($Pbam$) - NLC}\\
\tiny{ $\;Z_{2,1}=0\;Z_{2,2}=0\;Z_{2,3}=0\;Z_4=2$ } & \tiny{ $\;Z_{2,1}=0\;Z_{2,2}=0\;Z_{2,3}=0\;Z_4=2$ }\\
\includegraphics[width=0.38\textwidth,angle=0]{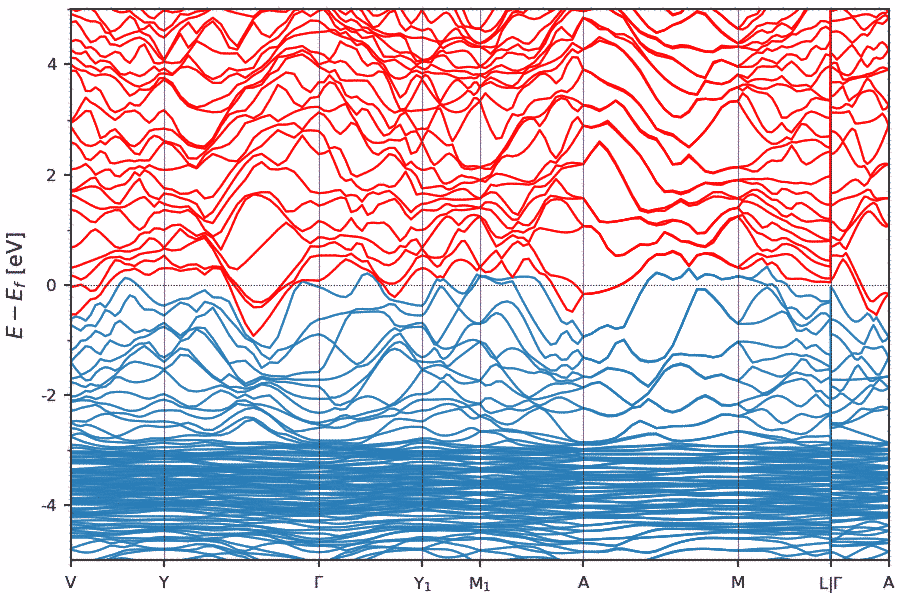} & \includegraphics[width=0.38\textwidth,angle=0]{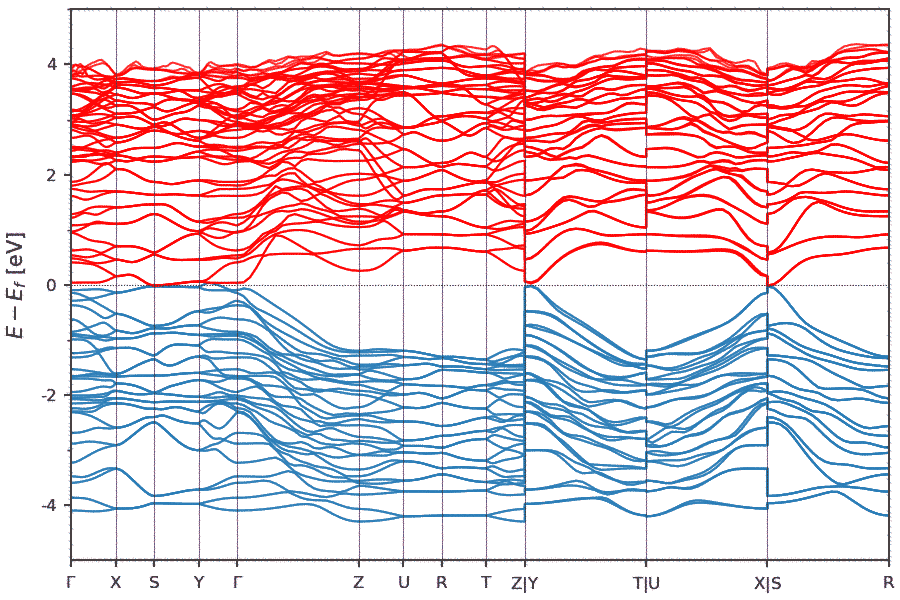}\\
\end{tabular}
\begin{tabular}{c c}
\scriptsize{$\rm{Pt} \rm{Ge}_{2}$ - \icsdweb{43684} - SG 58 ($Pnnm$) - NLC} & \scriptsize{$\rm{Ru} \rm{N}_{2}$ - \icsdweb{240754} - SG 58 ($Pnnm$) - NLC}\\
\tiny{ $\;Z_{2,1}=0\;Z_{2,2}=0\;Z_{2,3}=0\;Z_4=2$ } & \tiny{ $\;Z_{2,1}=0\;Z_{2,2}=0\;Z_{2,3}=0\;Z_4=2$ }\\
\includegraphics[width=0.38\textwidth,angle=0]{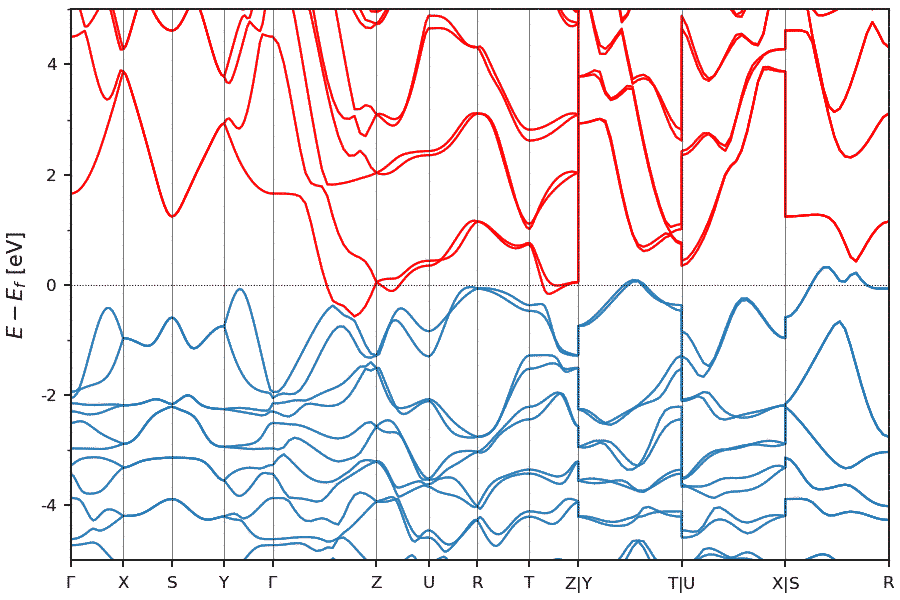} & \includegraphics[width=0.38\textwidth,angle=0]{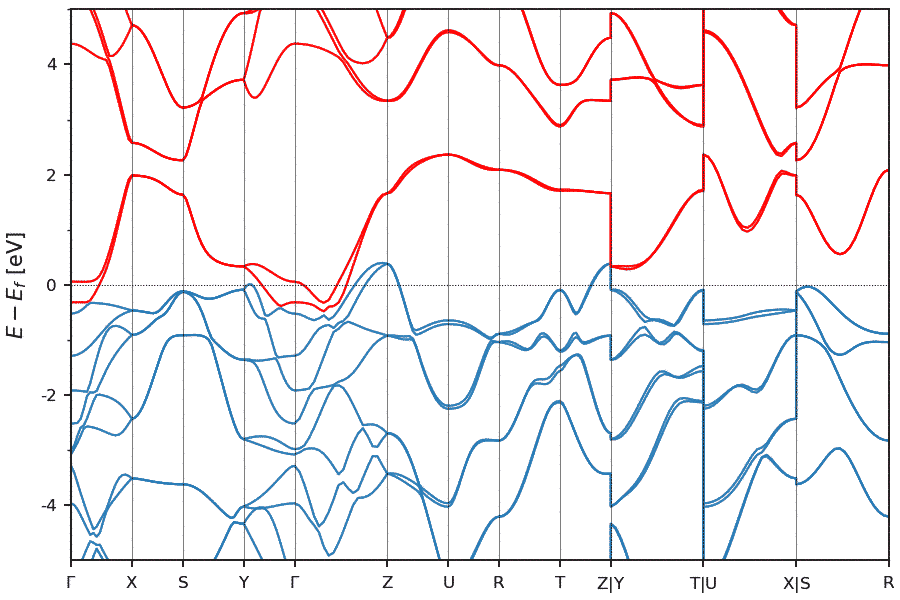}\\
\end{tabular}
\begin{tabular}{c c}
\scriptsize{$\rm{Co}_{2} \rm{C}$ - \icsdweb{617391} - SG 58 ($Pnnm$) - NLC} & \scriptsize{$\rm{Ni} \rm{S}_{2}$ - \icsdweb{169571} - SG 60 ($Pbcn$) - NLC}\\
\tiny{ $\;Z_{2,1}=0\;Z_{2,2}=0\;Z_{2,3}=0\;Z_4=2$ } & \tiny{ $\;Z_{2,1}=0\;Z_{2,2}=0\;Z_{2,3}=0\;Z_4=2$ }\\
\includegraphics[width=0.38\textwidth,angle=0]{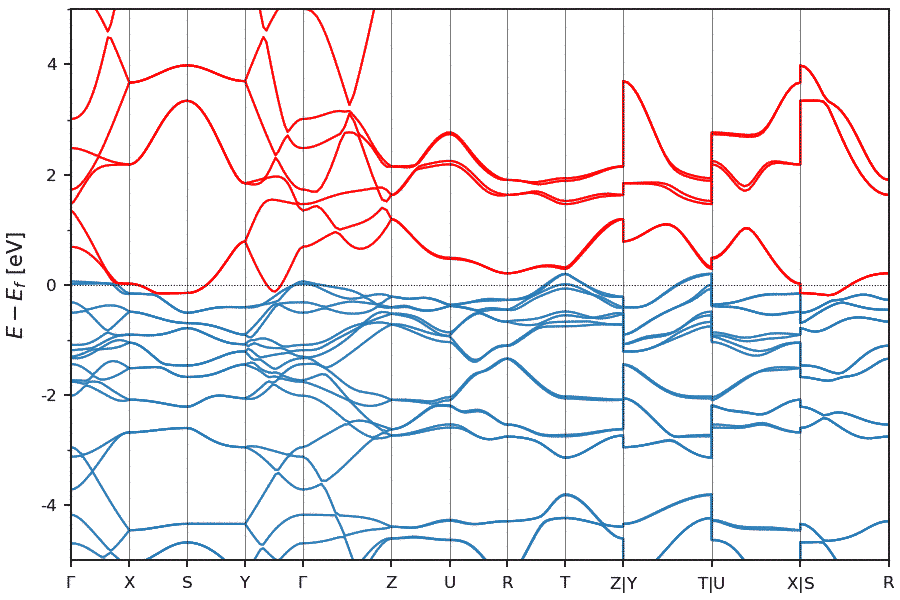} & \includegraphics[width=0.38\textwidth,angle=0]{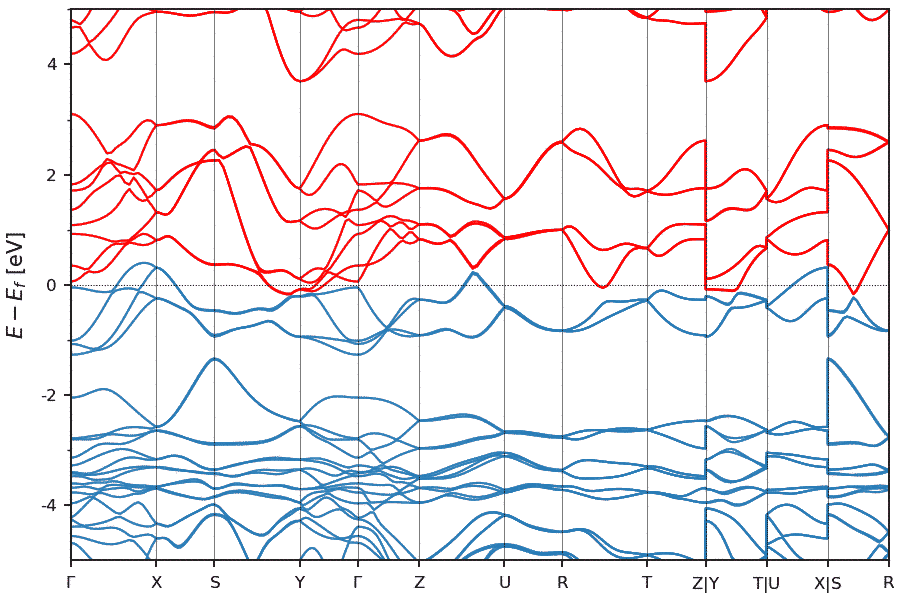}\\
\end{tabular}

\caption{\zfourHOTIsNLC{2}}
\label{fig:z4_2HOTIs_NLC2}
\end{figure}

\begin{figure}[ht]
\centering
\begin{tabular}{c c}
\scriptsize{$\rm{Pt} \rm{Ge}$ - \icsdweb{2624} - SG 62 ($Pnma$) - NLC} & \scriptsize{$\rm{Ni} \rm{Si}$ - \icsdweb{30626} - SG 62 ($Pnma$) - NLC}\\
\tiny{ $\;Z_{2,1}=0\;Z_{2,2}=0\;Z_{2,3}=0\;Z_4=2$ } & \tiny{ $\;Z_{2,1}=0\;Z_{2,2}=0\;Z_{2,3}=0\;Z_4=2$ }\\
\includegraphics[width=0.38\textwidth,angle=0]{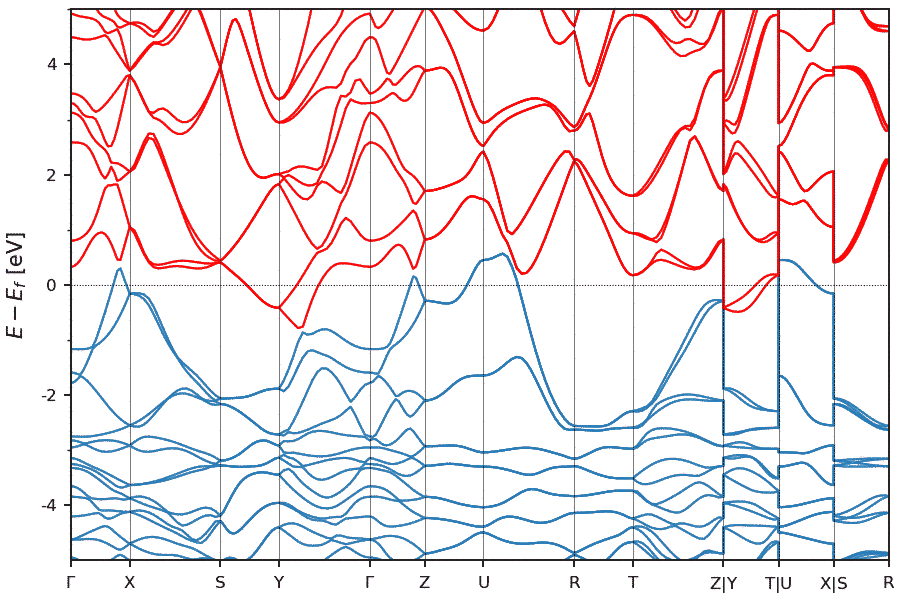} & \includegraphics[width=0.38\textwidth,angle=0]{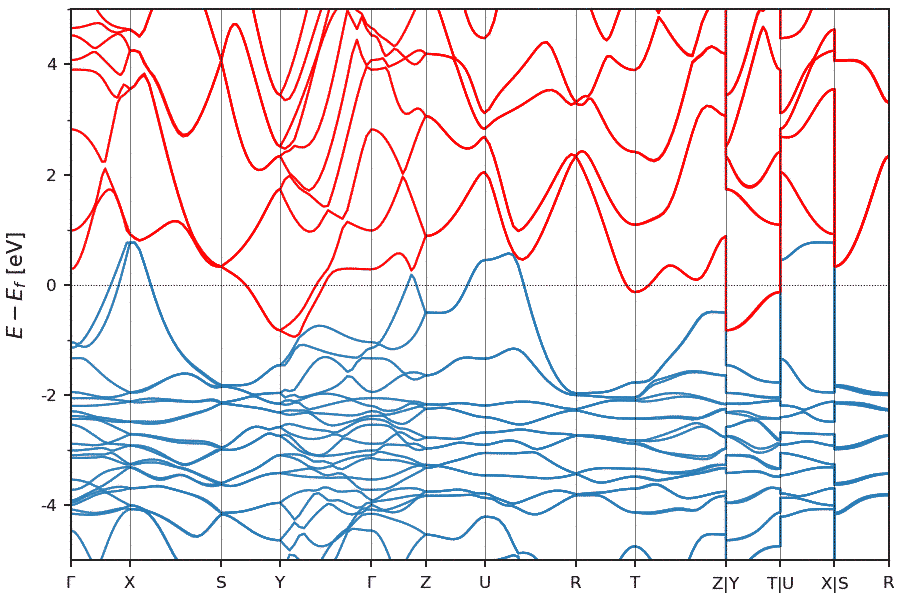}\\
\end{tabular}
\begin{tabular}{c c}
\scriptsize{$\rm{Ca} \rm{Te}$ - \icsdweb{41959} - SG 62 ($Pnma$) - NLC} & \scriptsize{$\rm{V} \rm{As}$ - \icsdweb{42446} - SG 62 ($Pnma$) - NLC}\\
\tiny{ $\;Z_{2,1}=0\;Z_{2,2}=0\;Z_{2,3}=0\;Z_4=2$ } & \tiny{ $\;Z_{2,1}=0\;Z_{2,2}=0\;Z_{2,3}=0\;Z_4=2$ }\\
\includegraphics[width=0.38\textwidth,angle=0]{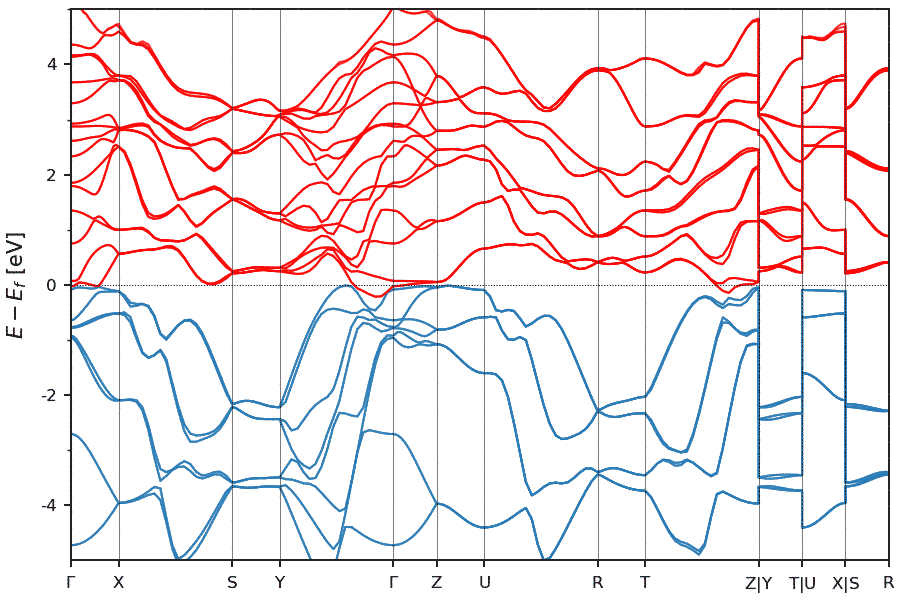} & \includegraphics[width=0.38\textwidth,angle=0]{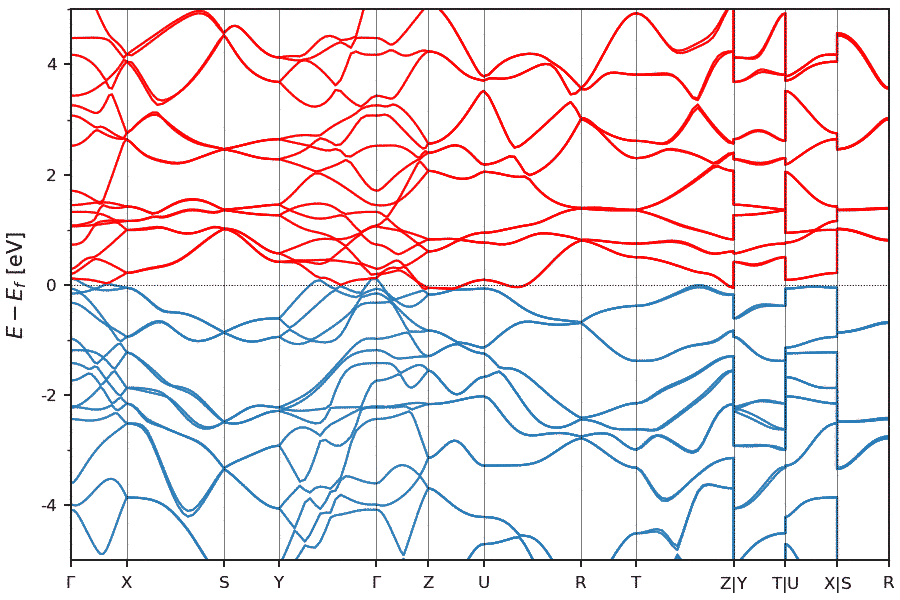}\\
\end{tabular}
\begin{tabular}{c c}
\scriptsize{$\rm{Co}_{2} \rm{Si}$ - \icsdweb{44858} - SG 62 ($Pnma$) - NLC} & \scriptsize{$\rm{Rh}_{2} \rm{Ge}$ - \icsdweb{44860} - SG 62 ($Pnma$) - NLC}\\
\tiny{ $\;Z_{2,1}=0\;Z_{2,2}=0\;Z_{2,3}=0\;Z_4=2$ } & \tiny{ $\;Z_{2,1}=0\;Z_{2,2}=0\;Z_{2,3}=0\;Z_4=2$ }\\
\includegraphics[width=0.38\textwidth,angle=0]{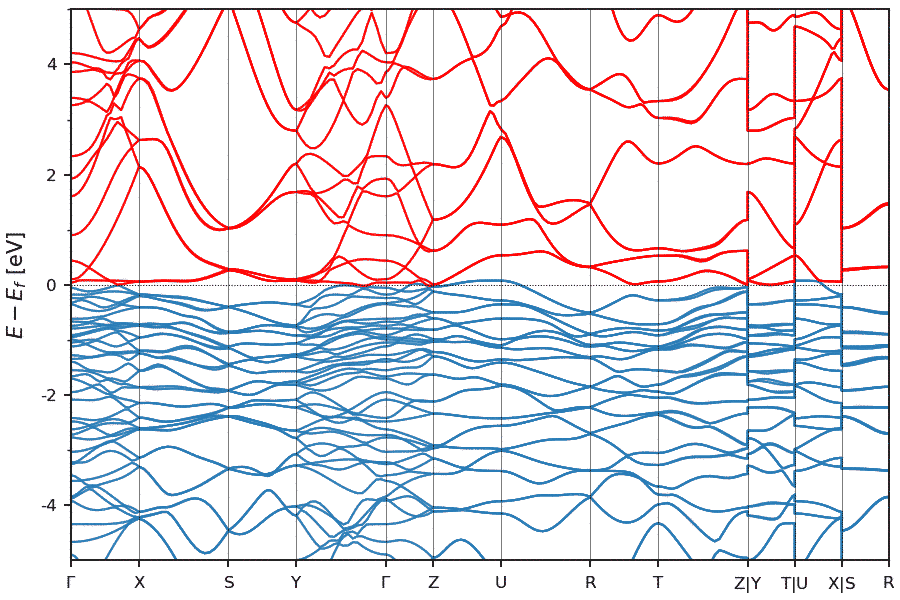} & \includegraphics[width=0.38\textwidth,angle=0]{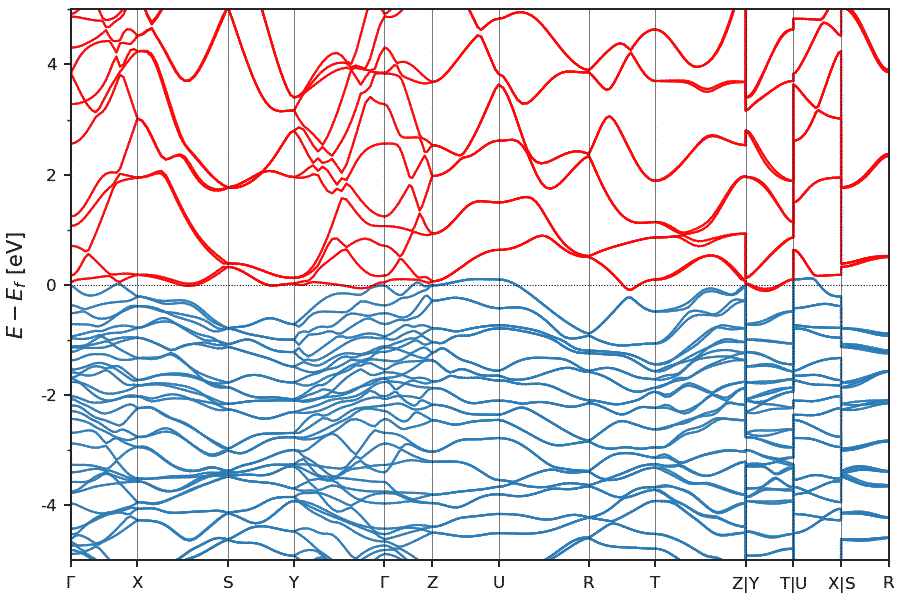}\\
\end{tabular}
\begin{tabular}{c c}
\scriptsize{$\rm{Ni} \rm{Ge}$ - \icsdweb{52124} - SG 62 ($Pnma$) - NLC} & \scriptsize{$\rm{Rh} \rm{Sb}$ - \icsdweb{76621} - SG 62 ($Pnma$) - NLC}\\
\tiny{ $\;Z_{2,1}=0\;Z_{2,2}=0\;Z_{2,3}=0\;Z_4=2$ } & \tiny{ $\;Z_{2,1}=0\;Z_{2,2}=0\;Z_{2,3}=0\;Z_4=2$ }\\
\includegraphics[width=0.38\textwidth,angle=0]{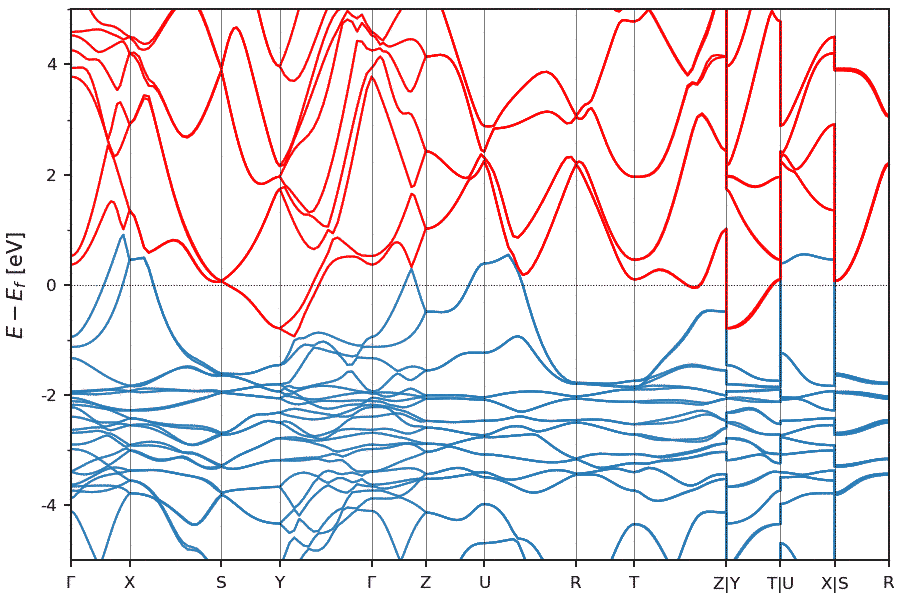} & \includegraphics[width=0.38\textwidth,angle=0]{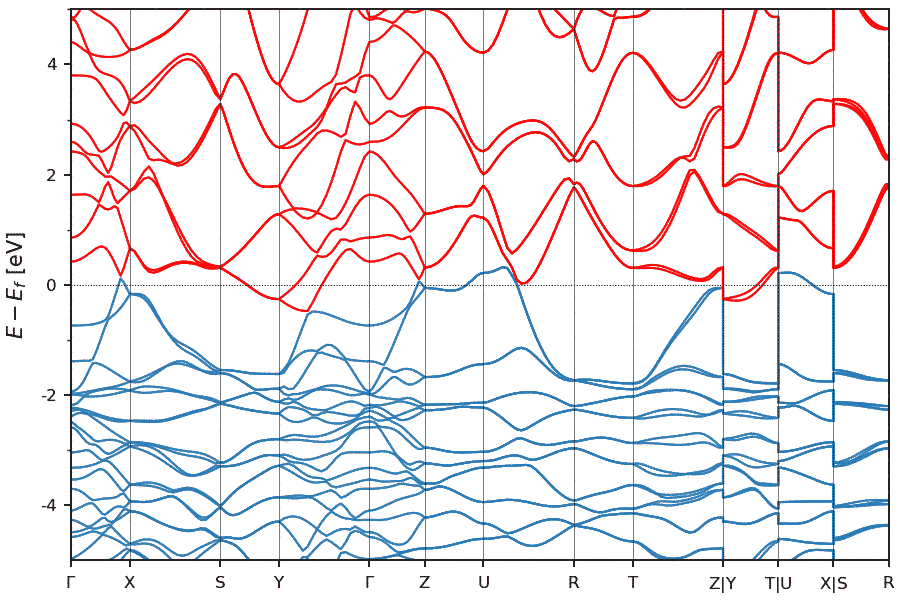}\\
\end{tabular}

\caption{\zfourHOTIsNLC{3}}
\label{fig:z4_2HOTIs_NLC3}
\end{figure}

\begin{figure}[ht]
\centering
\begin{tabular}{c c}
\scriptsize{$\rm{Pd} \rm{Ge}$ - \icsdweb{76624} - SG 62 ($Pnma$) - NLC} & \scriptsize{$\rm{Pd} \rm{Si}$ - \icsdweb{76626} - SG 62 ($Pnma$) - NLC}\\
\tiny{ $\;Z_{2,1}=0\;Z_{2,2}=0\;Z_{2,3}=0\;Z_4=2$ } & \tiny{ $\;Z_{2,1}=0\;Z_{2,2}=0\;Z_{2,3}=0\;Z_4=2$ }\\
\includegraphics[width=0.38\textwidth,angle=0]{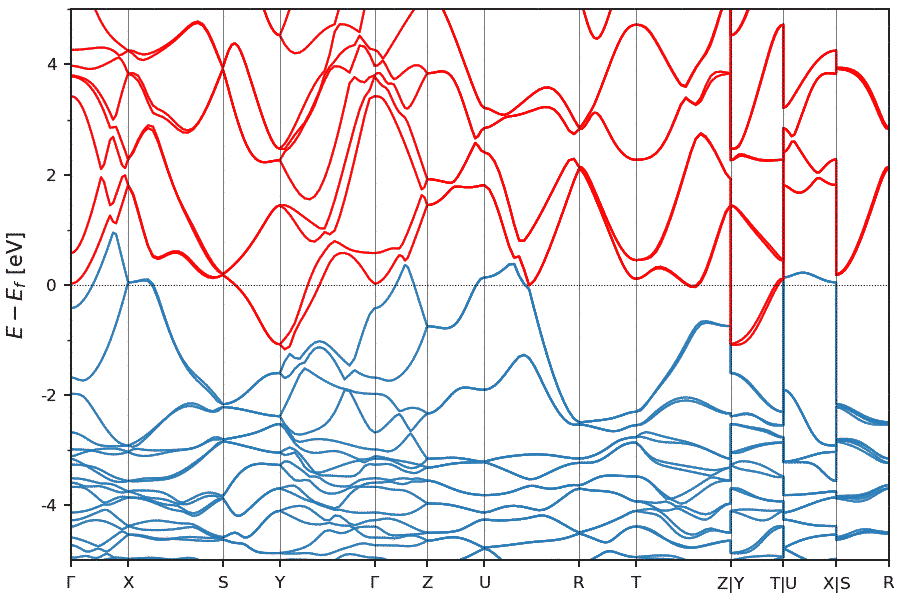} & \includegraphics[width=0.38\textwidth,angle=0]{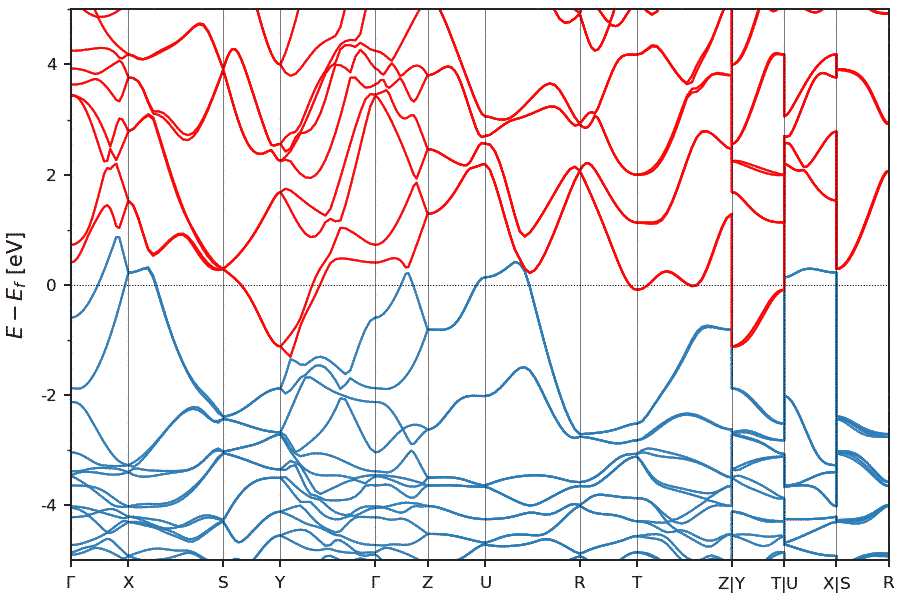}\\
\end{tabular}
\begin{tabular}{c c}
\scriptsize{$\rm{Pt} \rm{Sc}_{2}$ - \icsdweb{105785} - SG 62 ($Pnma$) - NLC} & \scriptsize{$\rm{Rh}_{2} \rm{Sn}$ - \icsdweb{105931} - SG 62 ($Pnma$) - NLC}\\
\tiny{ $\;Z_{2,1}=0\;Z_{2,2}=0\;Z_{2,3}=0\;Z_4=2$ } & \tiny{ $\;Z_{2,1}=0\;Z_{2,2}=0\;Z_{2,3}=0\;Z_4=2$ }\\
\includegraphics[width=0.38\textwidth,angle=0]{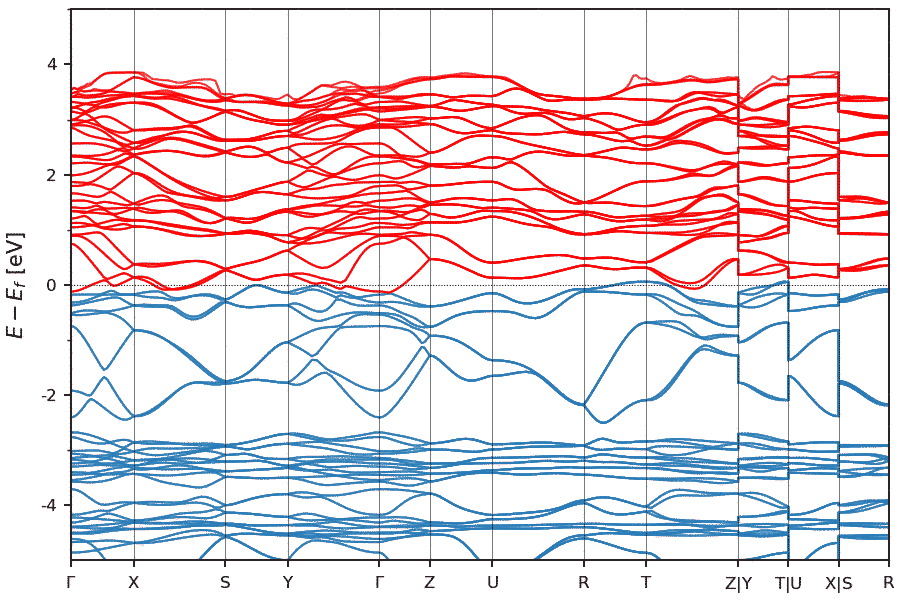} & \includegraphics[width=0.38\textwidth,angle=0]{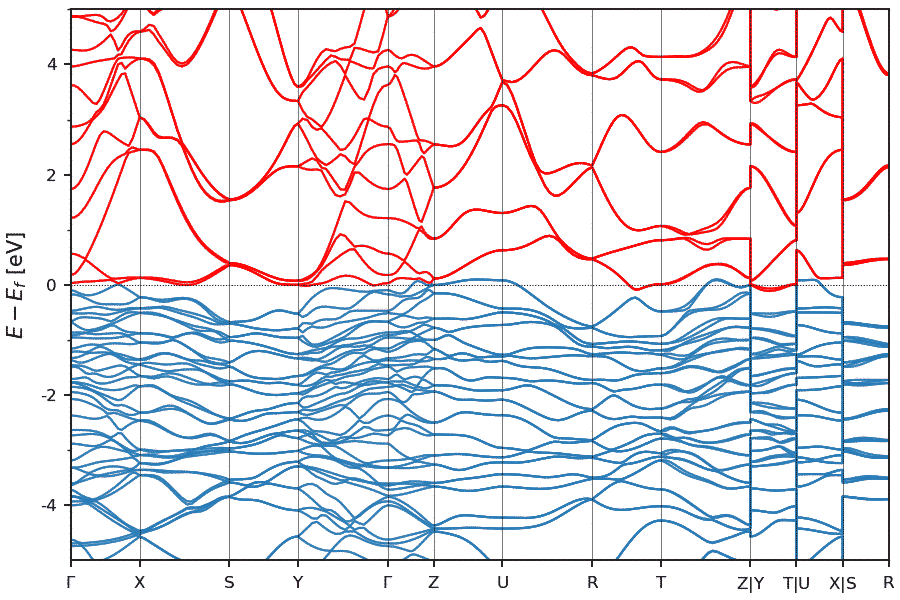}\\
\end{tabular}
\begin{tabular}{c c}
\scriptsize{$\rm{Pd}_{2} \rm{Sn}$ - \icsdweb{158365} - SG 62 ($Pnma$) - NLC} & \scriptsize{$\rm{W} \rm{H}_{2}$ - \icsdweb{247595} - SG 62 ($Pnma$) - NLC}\\
\tiny{ $\;Z_{2,1}=0\;Z_{2,2}=0\;Z_{2,3}=0\;Z_4=2$ } & \tiny{ $\;Z_{2,1}=0\;Z_{2,2}=0\;Z_{2,3}=0\;Z_4=2$ }\\
\includegraphics[width=0.38\textwidth,angle=0]{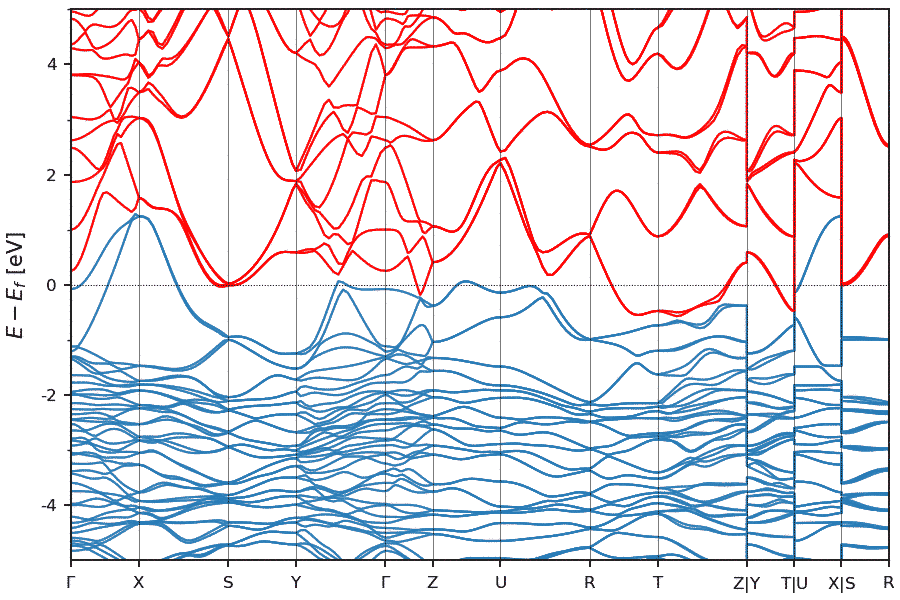} & \includegraphics[width=0.38\textwidth,angle=0]{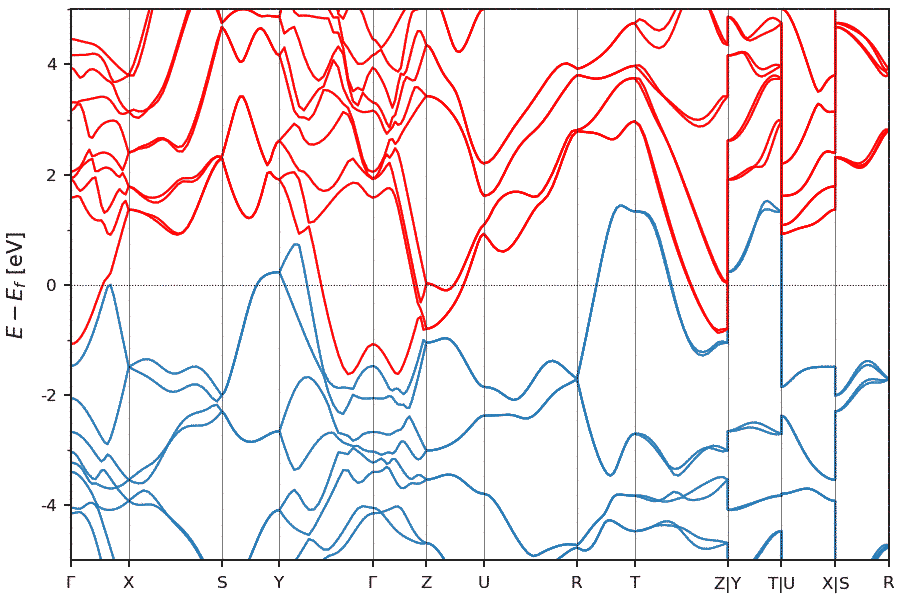}\\
\end{tabular}
\begin{tabular}{c c}
\scriptsize{$\rm{V} \rm{Co} \rm{Si}$ - \icsdweb{409847} - SG 62 ($Pnma$) - NLC} & \scriptsize{$\rm{Ca} \rm{Ir} \rm{In}_{2}$ - \icsdweb{410839} - SG 62 ($Pnma$) - NLC}\\
\tiny{ $\;Z_{2,1}=0\;Z_{2,2}=0\;Z_{2,3}=0\;Z_4=2$ } & \tiny{ $\;Z_{2,1}=0\;Z_{2,2}=0\;Z_{2,3}=0\;Z_4=2$ }\\
\includegraphics[width=0.38\textwidth,angle=0]{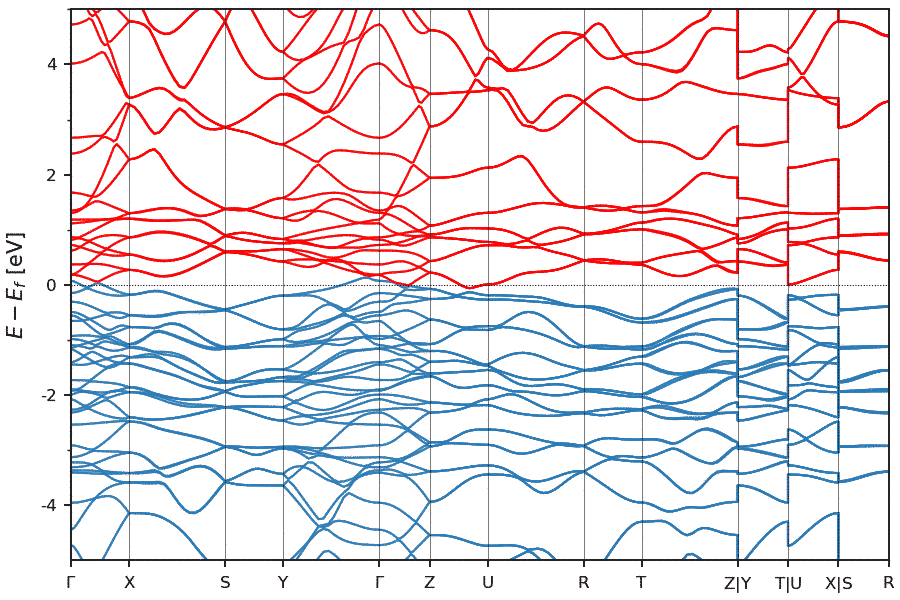} & \includegraphics[width=0.38\textwidth,angle=0]{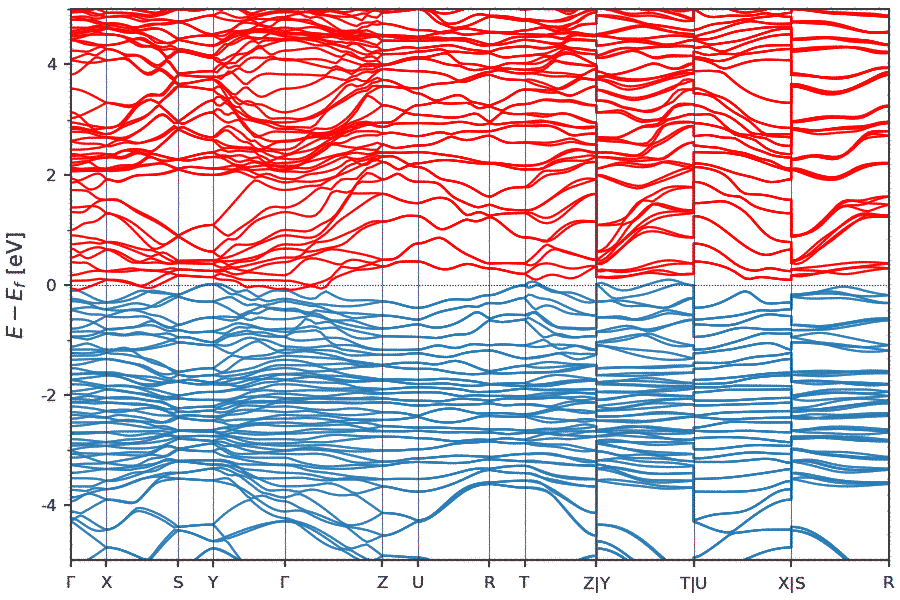}\\
\end{tabular}

\caption{\zfourHOTIsNLC{4}}
\label{fig:z4_2HOTIs_NLC4}
\end{figure}

\begin{figure}[ht]
\centering
\begin{tabular}{c c}
\scriptsize{$\rm{Nb} \rm{Ir} \rm{Si}$ - \icsdweb{411882} - SG 62 ($Pnma$) - NLC} & \scriptsize{$\rm{Nb} \rm{Ir} \rm{Ge}$ - \icsdweb{411883} - SG 62 ($Pnma$) - NLC}\\
\tiny{ $\;Z_{2,1}=0\;Z_{2,2}=0\;Z_{2,3}=0\;Z_4=2$ } & \tiny{ $\;Z_{2,1}=0\;Z_{2,2}=0\;Z_{2,3}=0\;Z_4=2$ }\\
\includegraphics[width=0.38\textwidth,angle=0]{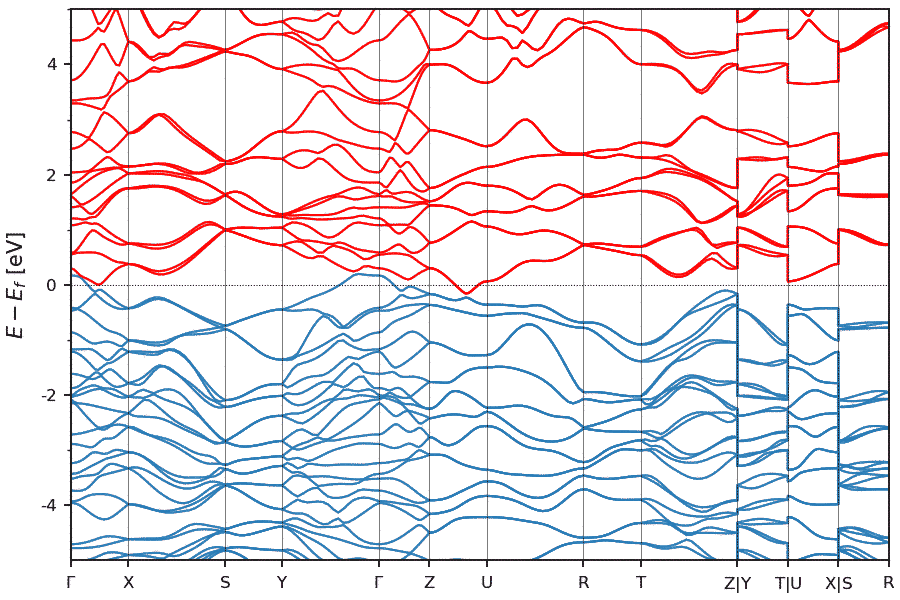} & \includegraphics[width=0.38\textwidth,angle=0]{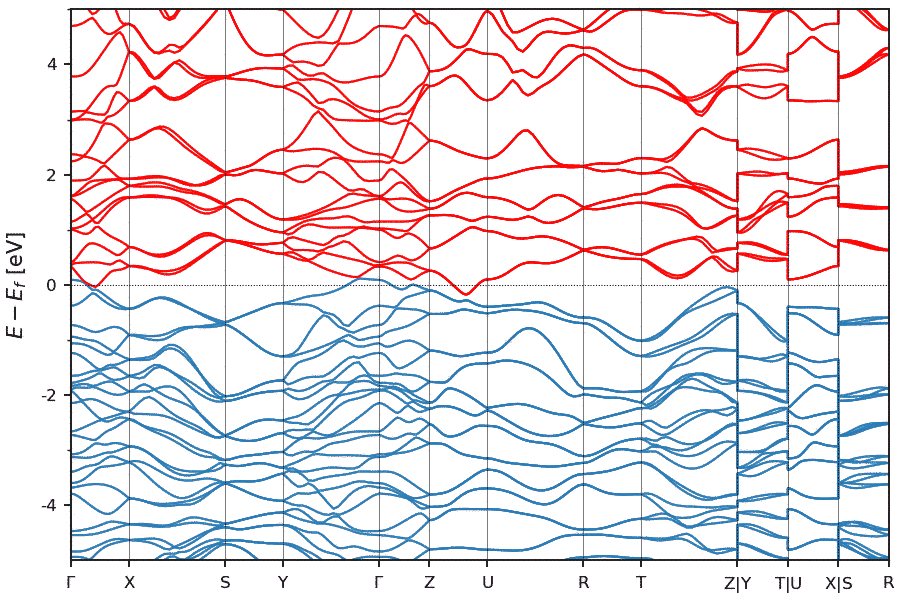}\\
\end{tabular}
\begin{tabular}{c c}
\scriptsize{$\rm{Ta} \rm{Ir} \rm{Si}$ - \icsdweb{411884} - SG 62 ($Pnma$) - NLC} & \scriptsize{$\rm{Rh} \rm{Bi}_{3}$ - \icsdweb{600489} - SG 62 ($Pnma$) - NLC}\\
\tiny{ $\;Z_{2,1}=0\;Z_{2,2}=0\;Z_{2,3}=0\;Z_4=2$ } & \tiny{ $\;Z_{2,1}=0\;Z_{2,2}=0\;Z_{2,3}=0\;Z_4=2$ }\\
\includegraphics[width=0.38\textwidth,angle=0]{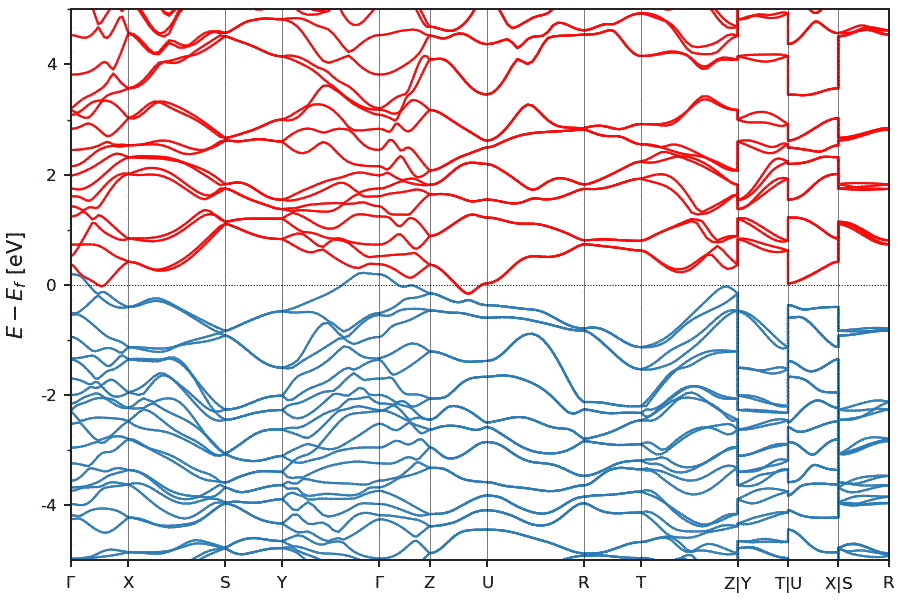} & \includegraphics[width=0.38\textwidth,angle=0]{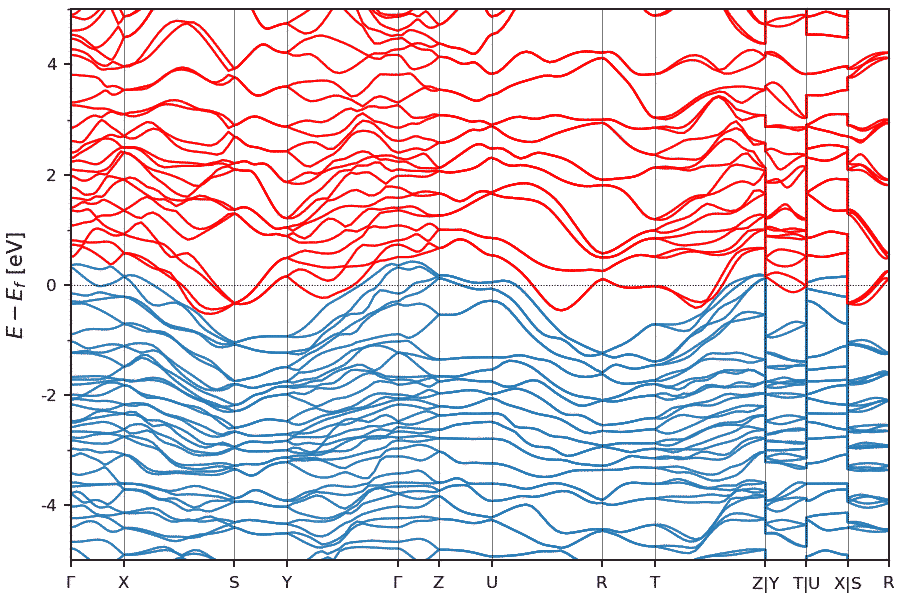}\\
\end{tabular}
\begin{tabular}{c c}
\scriptsize{$\rm{Co} \rm{As}$ - \icsdweb{610038} - SG 62 ($Pnma$) - NLC} & \scriptsize{$\rm{Ta} \rm{Fe} \rm{As}$ - \icsdweb{610528} - SG 62 ($Pnma$) - NLC}\\
\tiny{ $\;Z_{2,1}=0\;Z_{2,2}=0\;Z_{2,3}=0\;Z_4=2$ } & \tiny{ $\;Z_{2,1}=0\;Z_{2,2}=0\;Z_{2,3}=0\;Z_4=2$ }\\
\includegraphics[width=0.38\textwidth,angle=0]{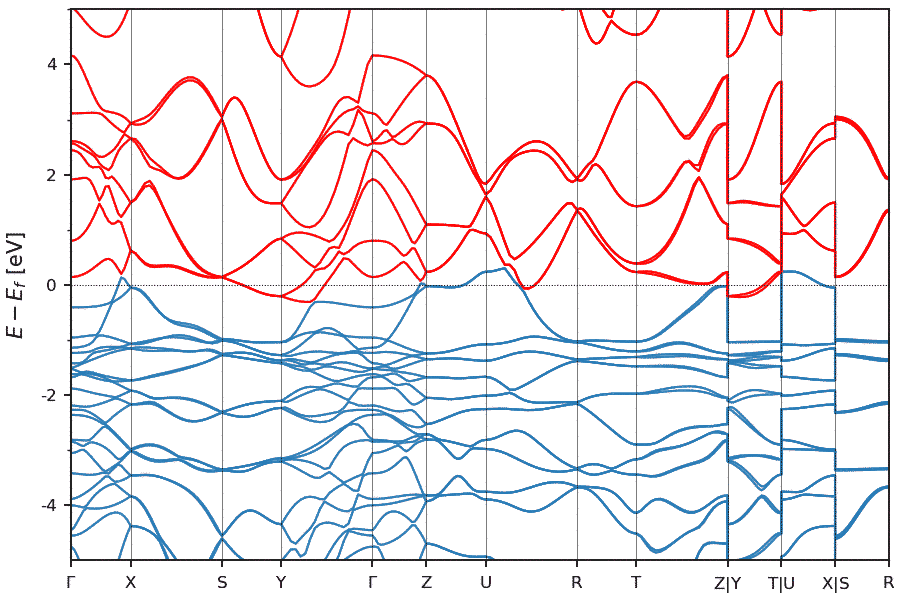} & \includegraphics[width=0.38\textwidth,angle=0]{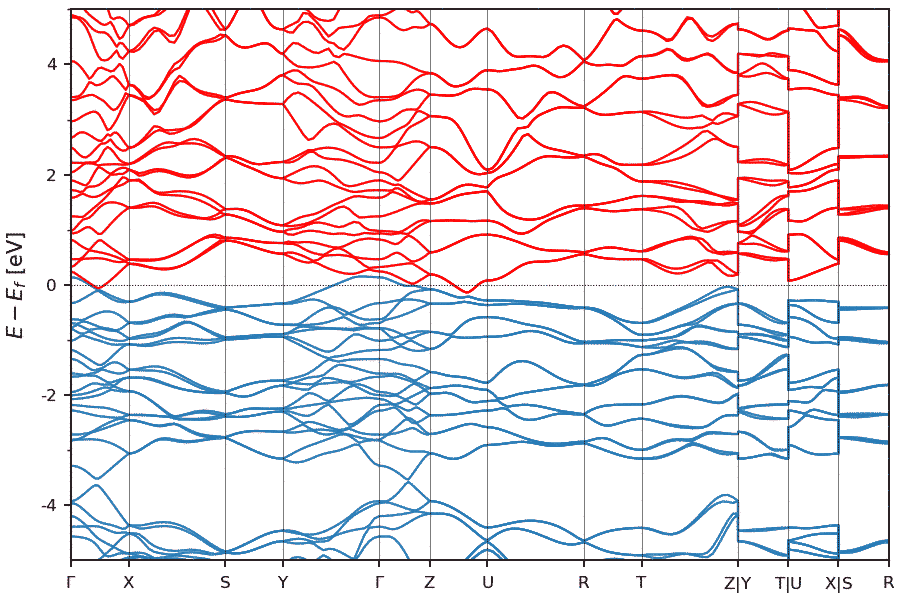}\\
\end{tabular}
\begin{tabular}{c c}
\scriptsize{$\rm{Mn} \rm{Co} \rm{Ge}$ - \icsdweb{623484} - SG 62 ($Pnma$) - NLC} & \scriptsize{$\rm{Co} \rm{Mn} \rm{Si}$ - \icsdweb{624142} - SG 62 ($Pnma$) - NLC}\\
\tiny{ $\;Z_{2,1}=0\;Z_{2,2}=0\;Z_{2,3}=0\;Z_4=2$ } & \tiny{ $\;Z_{2,1}=0\;Z_{2,2}=0\;Z_{2,3}=0\;Z_4=2$ }\\
\includegraphics[width=0.38\textwidth,angle=0]{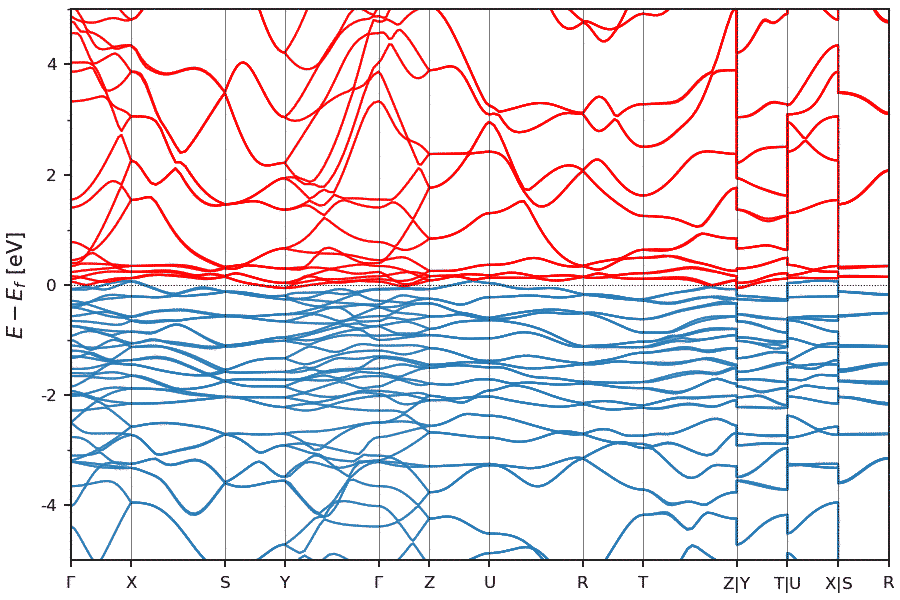} & \includegraphics[width=0.38\textwidth,angle=0]{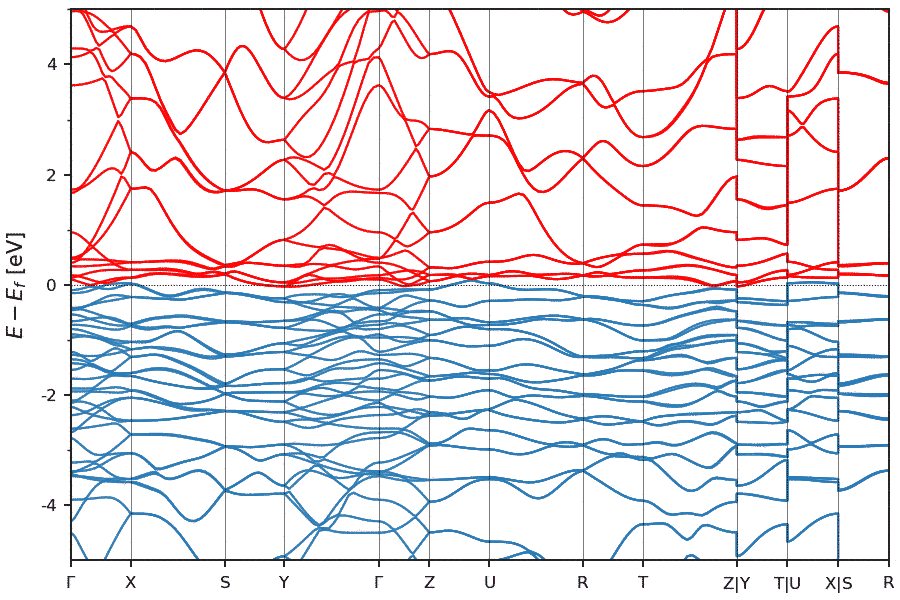}\\
\end{tabular}

\caption{\zfourHOTIsNLC{5}}
\label{fig:z4_2HOTIs_NLC5}
\end{figure}

\begin{figure}[ht]
\centering
\begin{tabular}{c c}
\scriptsize{$\rm{Co} \rm{P}$ - \icsdweb{624588} - SG 62 ($Pnma$) - NLC} & \scriptsize{$\rm{Nb} \rm{Fe} \rm{P}$ - \icsdweb{632794} - SG 62 ($Pnma$) - NLC}\\
\tiny{ $\;Z_{2,1}=0\;Z_{2,2}=0\;Z_{2,3}=0\;Z_4=2$ } & \tiny{ $\;Z_{2,1}=0\;Z_{2,2}=0\;Z_{2,3}=0\;Z_4=2$ }\\
\includegraphics[width=0.38\textwidth,angle=0]{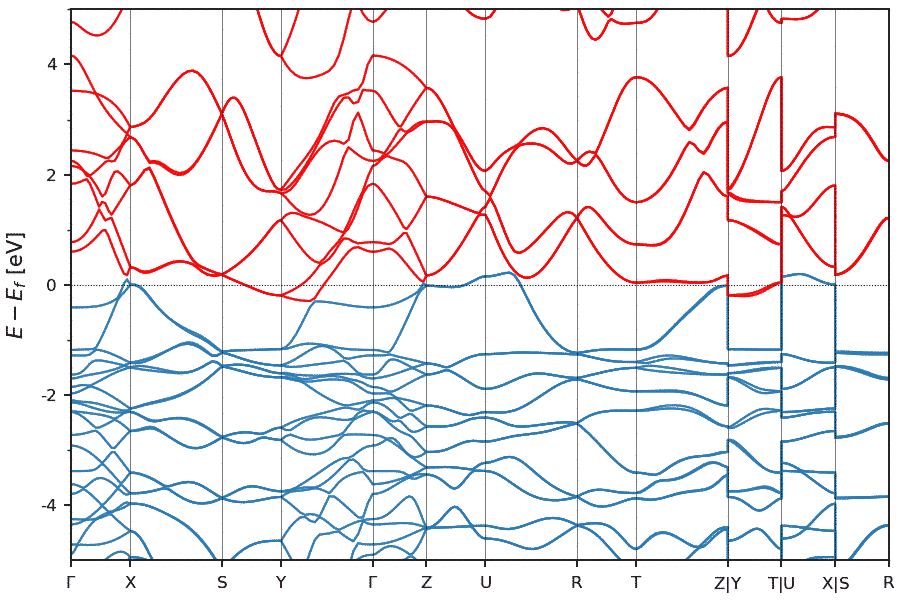} & \includegraphics[width=0.38\textwidth,angle=0]{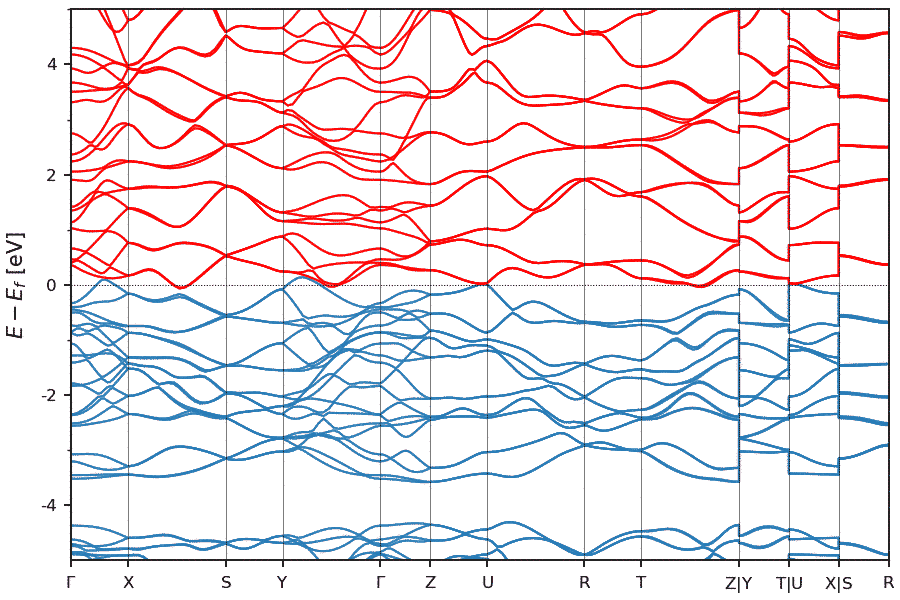}\\
\end{tabular}
\begin{tabular}{c c}
\scriptsize{$\rm{Nb} \rm{Ni} \rm{P}$ - \icsdweb{645088} - SG 62 ($Pnma$) - NLC} & \scriptsize{$\rm{Nb} \rm{Ru} \rm{P}$ - \icsdweb{645175} - SG 62 ($Pnma$) - NLC}\\
\tiny{ $\;Z_{2,1}=0\;Z_{2,2}=0\;Z_{2,3}=0\;Z_4=2$ } & \tiny{ $\;Z_{2,1}=0\;Z_{2,2}=0\;Z_{2,3}=0\;Z_4=2$ }\\
\includegraphics[width=0.38\textwidth,angle=0]{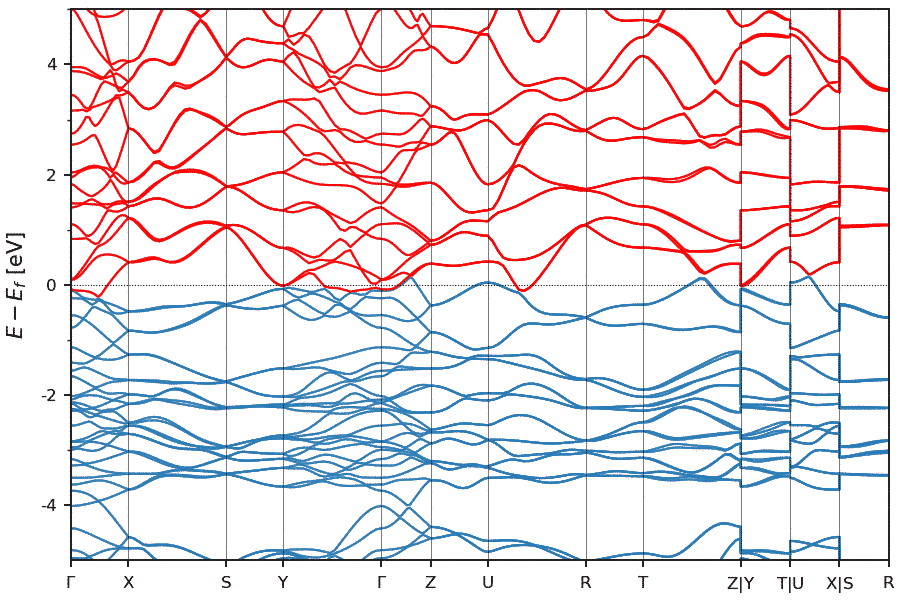} & \includegraphics[width=0.38\textwidth,angle=0]{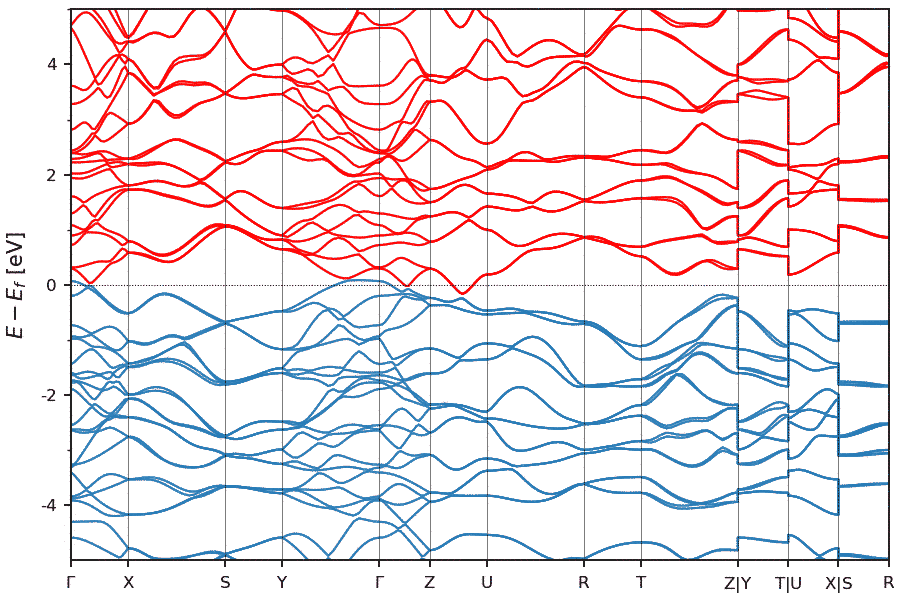}\\
\end{tabular}
\begin{tabular}{c c}
\scriptsize{$\rm{Rh}_{2} \rm{Si}$ - \icsdweb{650300} - SG 62 ($Pnma$) - NLC}\\
\tiny{ $\;Z_{2,1}=0\;Z_{2,2}=0\;Z_{2,3}=0\;Z_4=2$ }\\
\includegraphics[width=0.38\textwidth,angle=0]{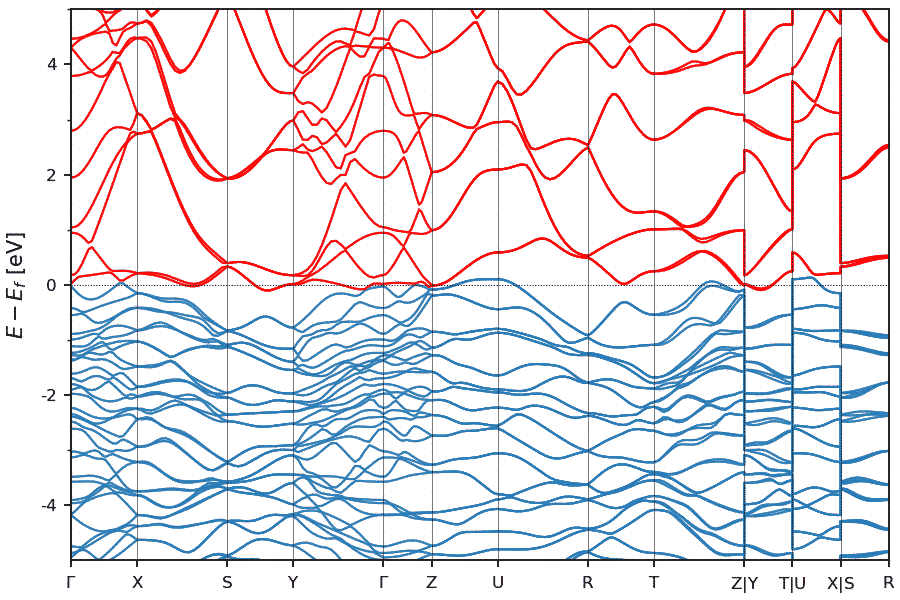}\\
\end{tabular}

\caption{\zfourHOTIsNLC{6}}
\label{fig:z4_2HOTIs_NLC6}
\end{figure}


\begin{figure}[ht]
\centering
\begin{tabular}{c c}
\scriptsize{$\rm{Bi} \rm{Br}$ - \icsdweb{1560} - SG 12 ($C2/m$) - SEBR} & \scriptsize{$\rm{Sr} \rm{In}_{4}$ - \icsdweb{240135} - SG 12 ($C2/m$) - SEBR}\\
\tiny{ $\;Z_{2,1}=0\;Z_{2,2}=0\;Z_{2,3}=0\;Z_4=2$ } & \tiny{ $\;Z_{2,1}=0\;Z_{2,2}=0\;Z_{2,3}=0\;Z_4=2$ }\\
\includegraphics[width=0.38\textwidth,angle=0]{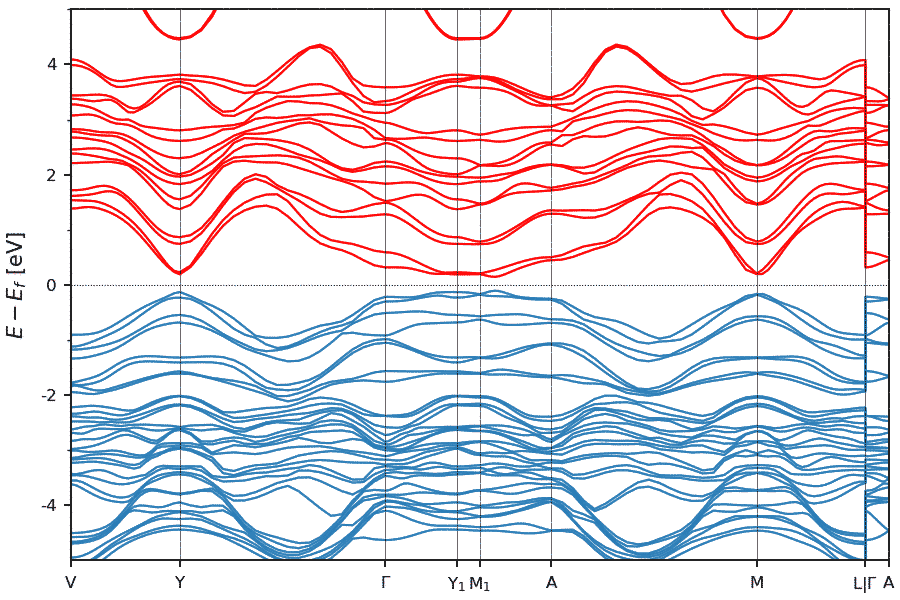} & \includegraphics[width=0.38\textwidth,angle=0]{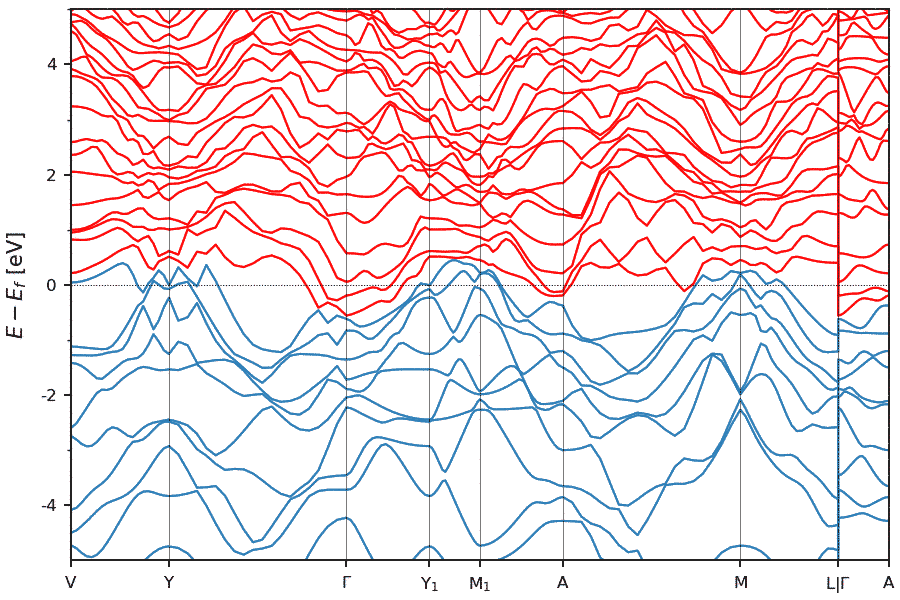}\\
\end{tabular}
\begin{tabular}{c c}
\scriptsize{$\rm{Pt}_{3} \rm{Ge}_{2}$ - \icsdweb{43682} - SG 63 ($Cmcm$) - SEBR} & \scriptsize{$\rm{V}_{3} \rm{Ge} \rm{N}$ - \icsdweb{637164} - SG 63 ($Cmcm$) - SEBR}\\
\tiny{ $\;Z_{2,1}=0\;Z_{2,2}=0\;Z_{2,3}=0\;Z_4=2$ } & \tiny{ $\;Z_{2,1}=0\;Z_{2,2}=0\;Z_{2,3}=0\;Z_4=2$ }\\
\includegraphics[width=0.38\textwidth,angle=0]{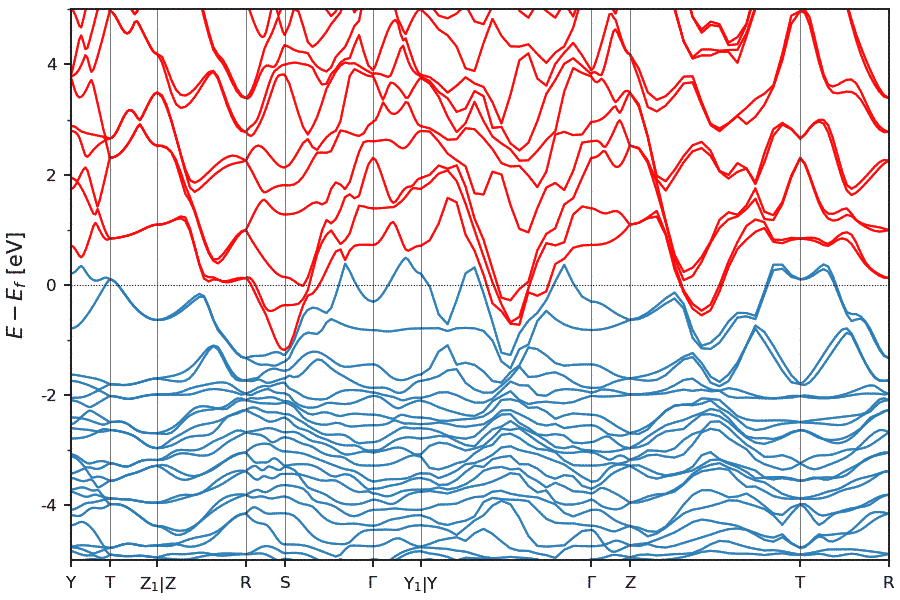} & \includegraphics[width=0.38\textwidth,angle=0]{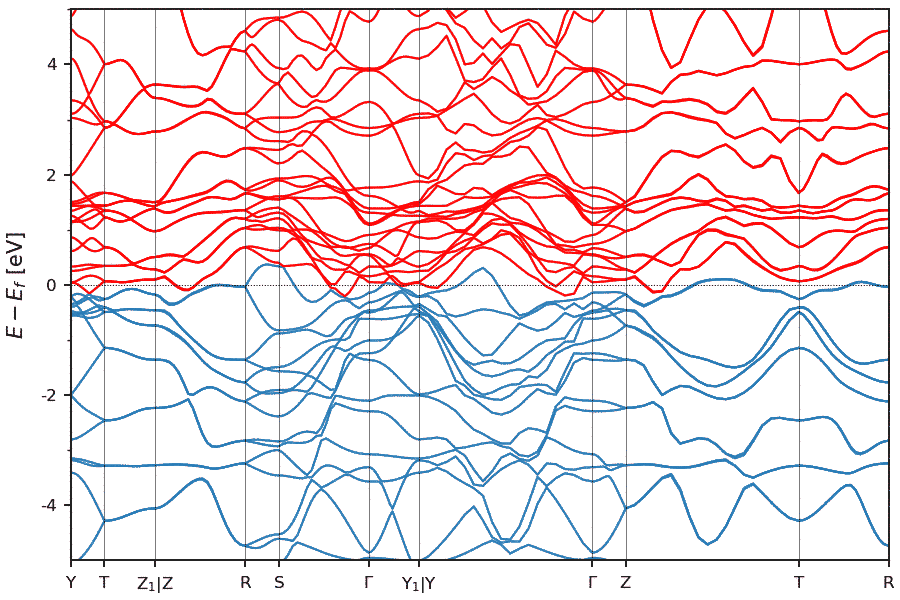}\\
\end{tabular}
\begin{tabular}{c c}
\scriptsize{$\rm{Sr}_{3} \rm{Rh}_{8} \rm{B}_{6}$ - \icsdweb{66771} - SG 69 ($Fmmm$) - SEBR} & \scriptsize{$\rm{Zr} \rm{Te}_{2}$ - \icsdweb{653213} - SG 164 ($P\bar{3}m1$) - SEBR}\\
\tiny{ $\;Z_{2,1}=0\;Z_{2,2}=0\;Z_{2,3}=0\;Z_4=2$ } & \tiny{ $\;Z_{2,1}=0\;Z_{2,2}=0\;Z_{2,3}=0\;Z_4=2$ }\\
\includegraphics[width=0.38\textwidth,angle=0]{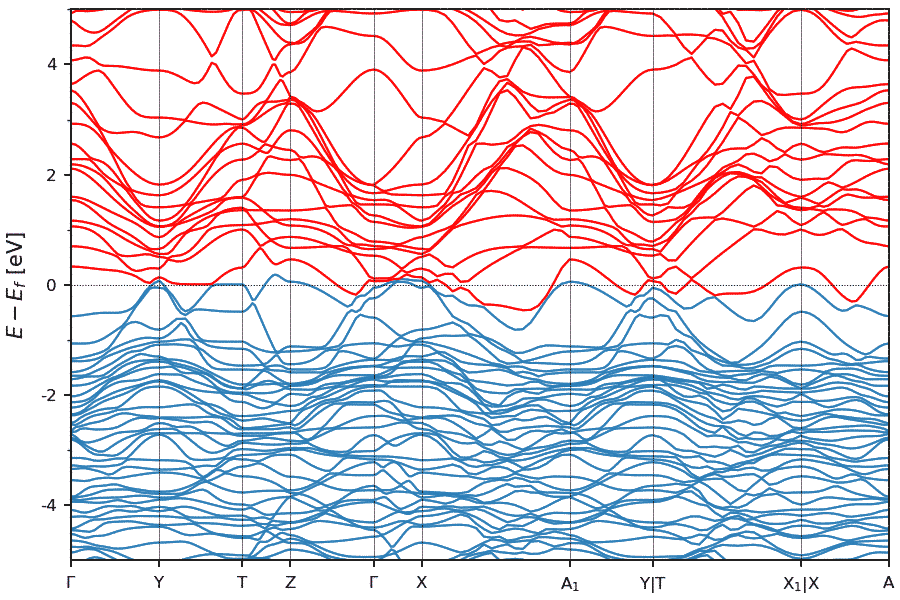} & \includegraphics[width=0.38\textwidth,angle=0]{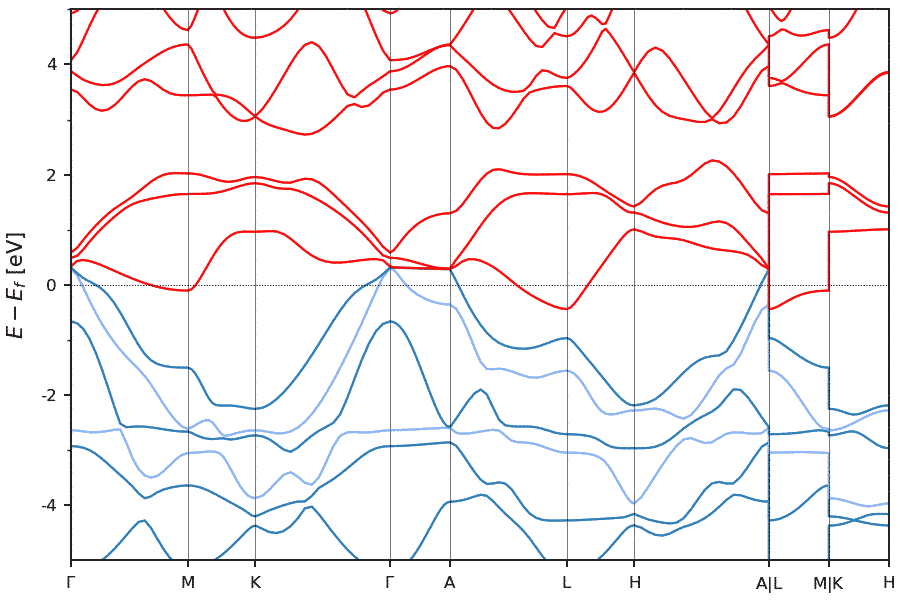}\\
\end{tabular}
\begin{tabular}{c c}
\scriptsize{$\rm{V}_{3} \rm{Sb}_{2}$ - \icsdweb{41814} - SG 166 ($R\bar{3}m$) - SEBR} & \scriptsize{$\rm{P}$ - \icsdweb{53301} - SG 166 ($R\bar{3}m$) - SEBR}\\
\tiny{ $\;Z_{2,1}=0\;Z_{2,2}=0\;Z_{2,3}=0\;Z_4=2$ } & \tiny{ $\;Z_{2,1}=0\;Z_{2,2}=0\;Z_{2,3}=0\;Z_4=2$ }\\
\includegraphics[width=0.38\textwidth,angle=0]{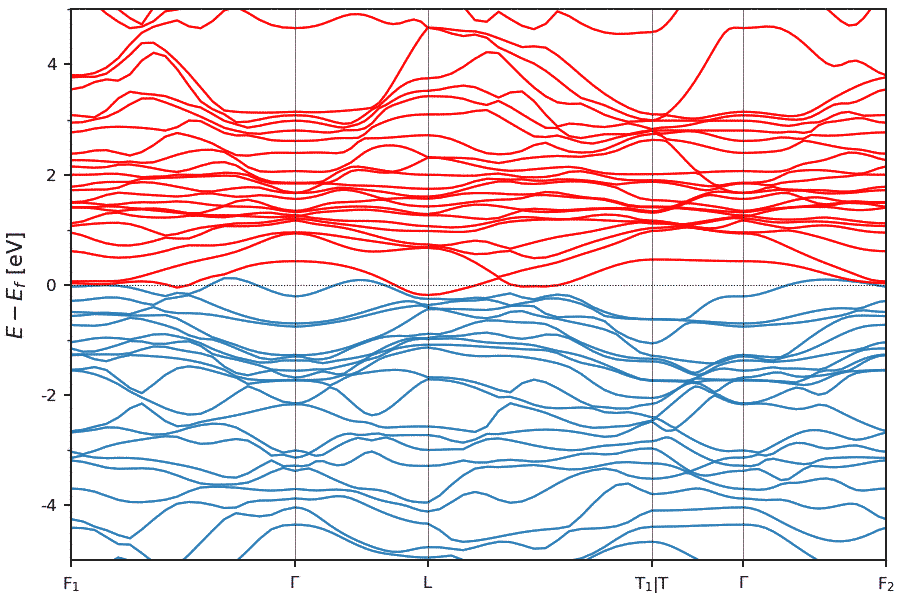} & \includegraphics[width=0.38\textwidth,angle=0]{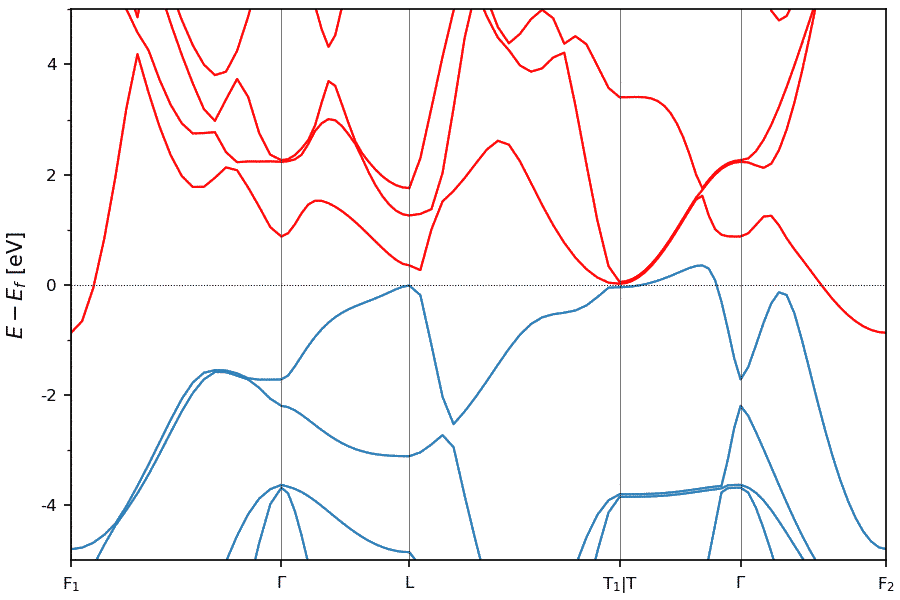}\\
\end{tabular}

\caption{\zfourHOTIsSEBR{1}}
\label{fig:z4_2HOTIs_SEBR1}
\end{figure}

\begin{figure}[ht]
\centering
\begin{tabular}{c c}
\scriptsize{$\rm{Bi}$ - \icsdweb{64705} - SG 166 ($R\bar{3}m$) - SEBR} & \scriptsize{$\rm{Co} \rm{N}_{3}$ - \icsdweb{162105} - SG 204 ($Im\bar{3}$) - SEBR}\\
\tiny{ $\;Z_{2,1}=0\;Z_{2,2}=0\;Z_{2,3}=0\;Z_4=2$ } & \tiny{ $\;Z_{2,1}=0\;Z_{2,2}=0\;Z_{2,3}=0\;Z_4=2$ }\\
\includegraphics[width=0.38\textwidth,angle=0]{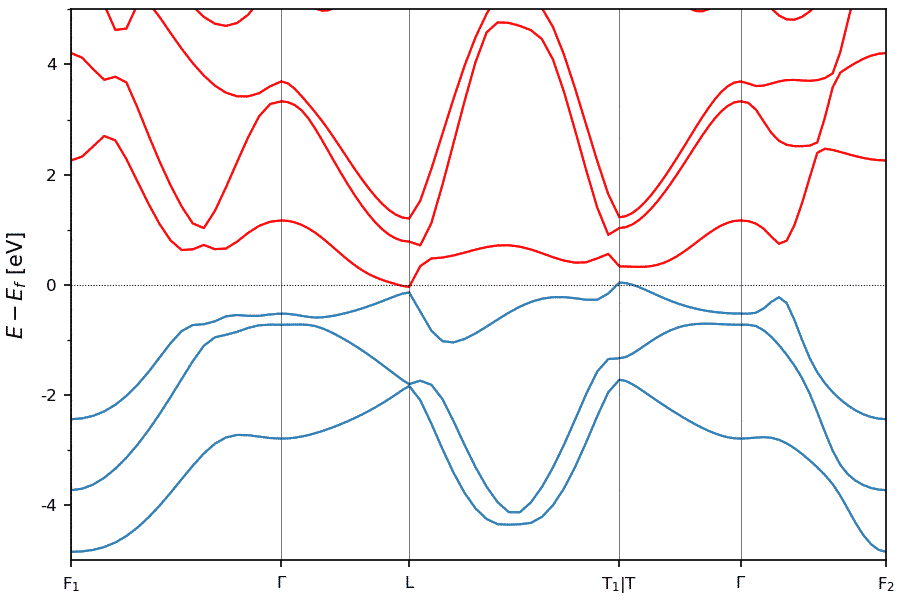} & \includegraphics[width=0.38\textwidth,angle=0]{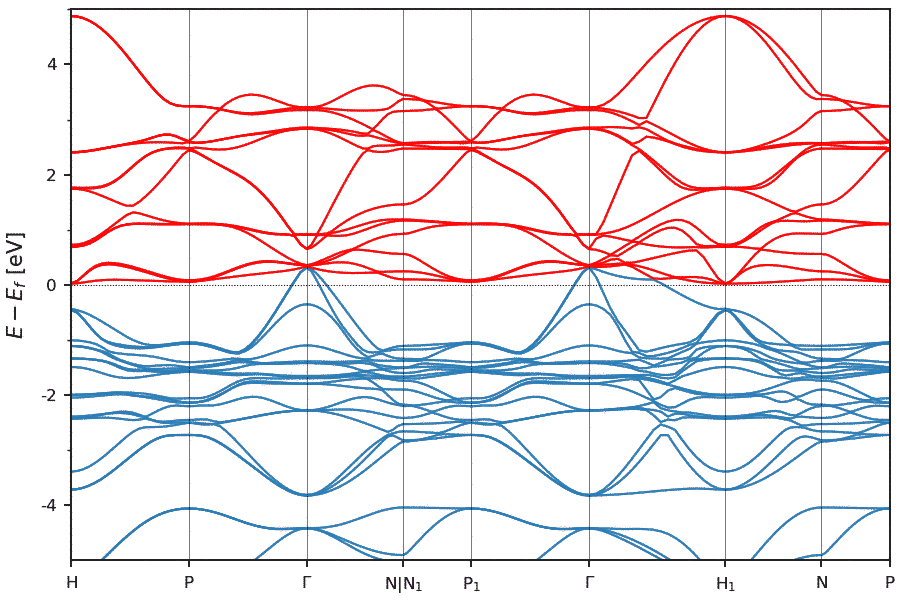}\\
\end{tabular}
\begin{tabular}{c c}
\scriptsize{$\rm{Pt}_{2} \rm{Sr}$ - \icsdweb{108713} - SG 227 ($Fd\bar{3}m$) - SEBR}\\
\tiny{ $\;Z_{2,1}=0\;Z_{2,2}=0\;Z_{2,3}=0\;Z_4=2\;Z_2=0$ }\\
\includegraphics[width=0.38\textwidth,angle=0]{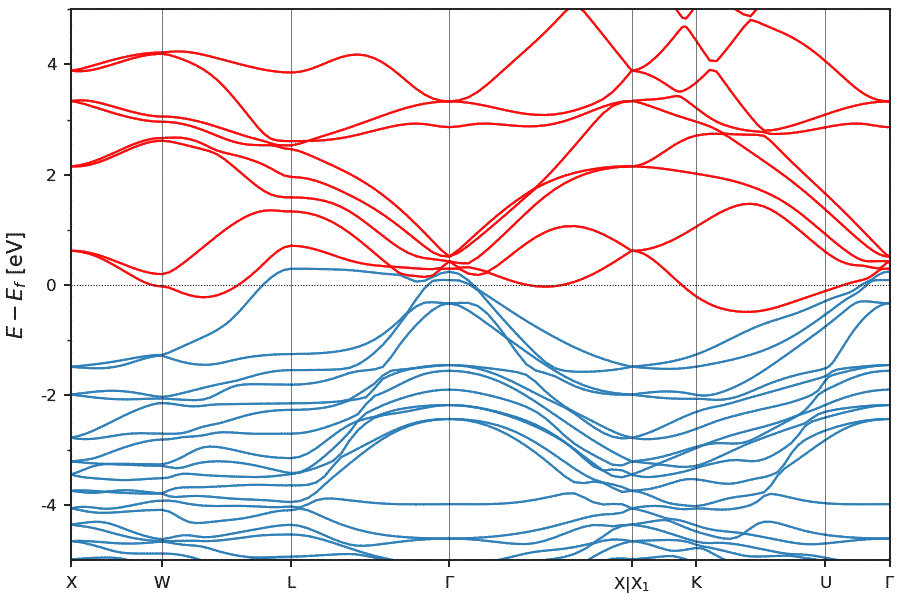}\\
\end{tabular}

\caption{\zfourHOTIsSEBR{2}}
\label{fig:z4_2HOTIs_SEBR2}
\end{figure}

\clearpage

\subsubsection{Inversion-Symmetry-Indicated Weak TIs and TCIs}
\label{App:weakTIs}

In this section, we list the inversion-symmetry-indicated weak TIs and TCIs with the largest bulk gaps or the fewest and smallest bulk Fermi pockets.  Like the HOTIs previously listed in \supappref{App:z4HOTIs}, the weak TIs and TCIs listed in this section have 2D surfaces with even numbers of massless or massive twofold Dirac cones, depending on the Miller indices of the surfaces~\cite{FuKaneMele,FuKaneInversion,AdyWeak,MooreBalentsWeak}.  In terms of the indices introduced in Ref.~\onlinecite{ChenTCI}, the materials in this section are characterized by $Z_{4}=2$, and exhibit nontrivial values for some or all other independent stable SIs.  First, in Figs.~\ref{fig:z4_2TCIs_andWTIs_NLC1} and~\ref{fig:z4_2TCIs_andWTIs_NLC2}, we list the weak TIs and TCIs classified as NLC, and then, in Figs.~\ref{fig:z4_2TCIs_andWTIs_SEBR1},~\ref{fig:z4_2TCIs_andWTIs_SEBR2},~\ref{fig:z4_2TCIs_andWTIs_SEBR3},~\ref{fig:z4_2TCIs_andWTIs_SEBR4},~\ref{fig:z4_2TCIs_andWTIs_SEBR5},~\ref{fig:z4_2TCIs_andWTIs_SEBR6},~\ref{fig:z4_2TCIs_andWTIs_SEBR7},~\ref{fig:z4_2TCIs_andWTIs_SEBR8},~\ref{fig:z4_2TCIs_andWTIs_SEBR9},~\ref{fig:z4_2TCIs_andWTIs_SEBR10},~\ref{fig:z4_2TCIs_andWTIs_SEBR11},~\ref{fig:z4_2TCIs_andWTIs_SEBR12},~\ref{fig:z4_2TCIs_andWTIs_SEBR13}, and~\ref{fig:z4_2TCIs_andWTIs_SEBR14}, we list the weak TIs and TCIs classified as SEBR.  The materials listed in this section include the experimentally confirmed weak TI BiI [\icsdweb{1559}, SG 12 ($C2/m$)]~\cite{BiBrFan,BiIWTIExp,BiBrFanHOTI,BiBrNatMater}, which is classified as SEBR.


\begin{figure}[ht]
\centering
\begin{tabular}{c c}
\scriptsize{$\rm{Hg} \rm{K}$ - \icsdweb{104302} - SG 2 ($P\bar{1}$) - NLC} & \scriptsize{$\rm{Ta}_{6} \rm{S}$ - \icsdweb{202564} - SG 2 ($P\bar{1}$) - NLC}\\
\tiny{ $\;Z_{2,1}=0\;Z_{2,2}=1\;Z_{2,3}=0\;Z_4=2$ } & \tiny{ $\;Z_{2,1}=0\;Z_{2,2}=0\;Z_{2,3}=1\;Z_4=2$ }\\
\includegraphics[width=0.38\textwidth,angle=0]{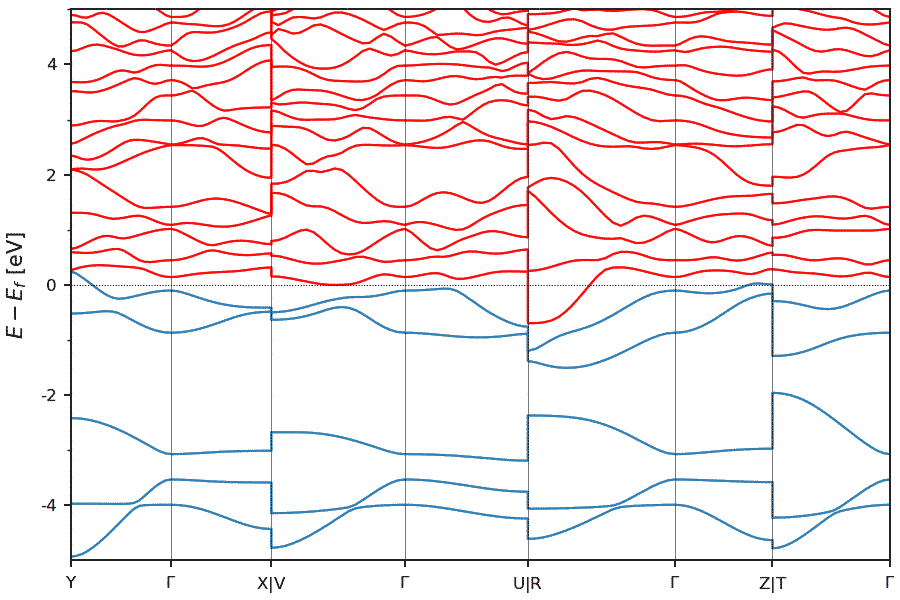} & \includegraphics[width=0.38\textwidth,angle=0]{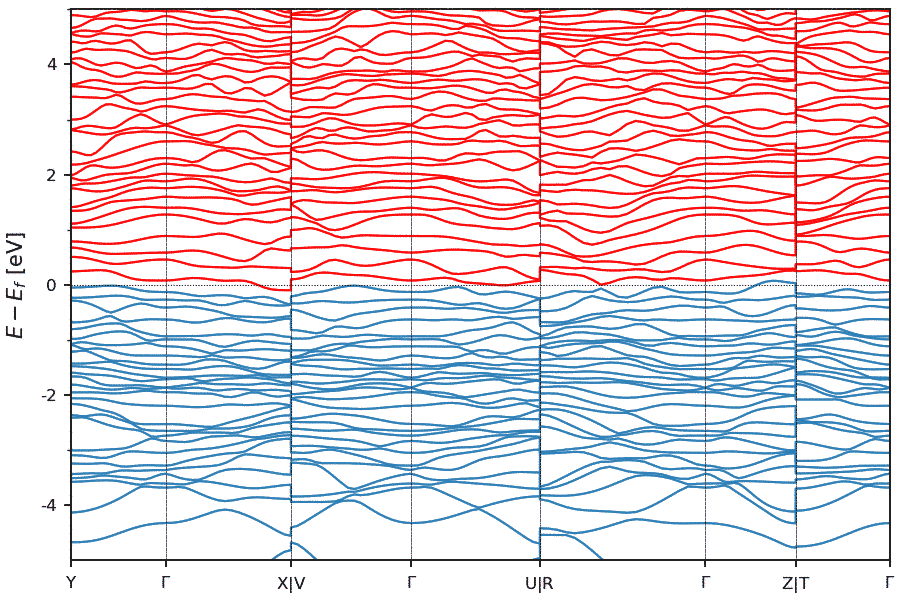}\\
\end{tabular}
\begin{tabular}{c c}
\scriptsize{$\rm{Ti} \rm{Zr} \rm{Se}_{4}$ - \icsdweb{104188} - SG 10 ($P2/m$) - NLC} & \scriptsize{$\rm{Sr} \rm{Sb}_{2}$ - \icsdweb{52307} - SG 11 ($P2_1/m$) - NLC}\\
\tiny{ $\;Z_{2,1}=1\;Z_{2,2}=0\;Z_{2,3}=0\;Z_4=2$ } & \tiny{ $\;Z_{2,1}=1\;Z_{2,2}=0\;Z_{2,3}=0\;Z_4=2$ }\\
\includegraphics[width=0.38\textwidth,angle=0]{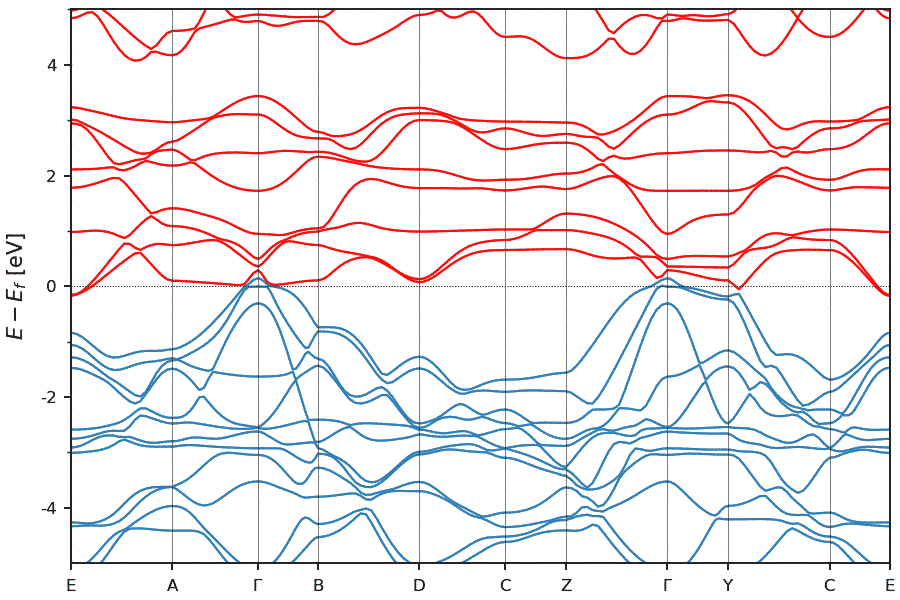} & \includegraphics[width=0.38\textwidth,angle=0]{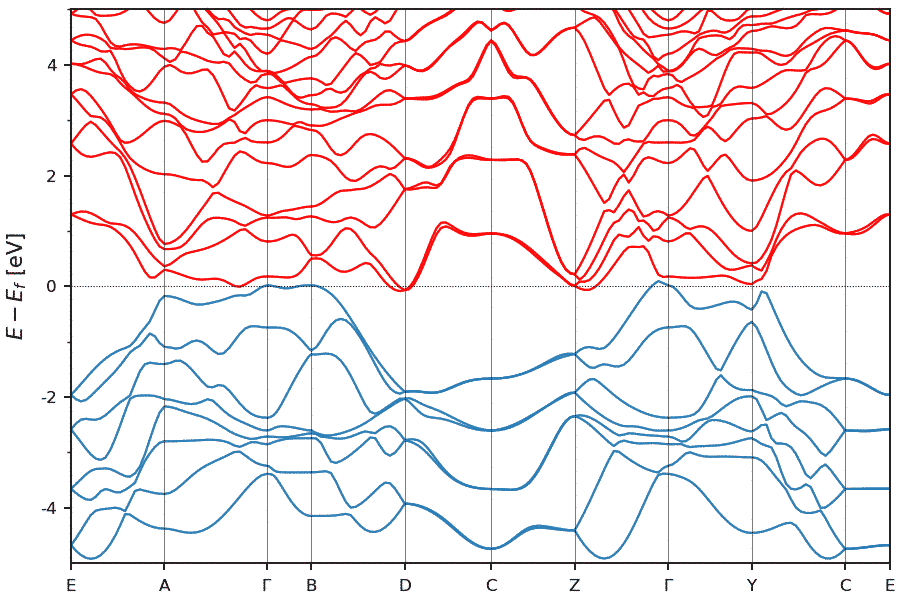}\\
\end{tabular}
\begin{tabular}{c c}
\scriptsize{$\rm{Ti} \rm{Ni}$ - \icsdweb{657957} - SG 11 ($P2_1/m$) - NLC} & \scriptsize{$\rm{S} \rm{N}$ - \icsdweb{41966} - SG 14 ($P2_1/c$) - NLC}\\
\tiny{ $\;Z_{2,1}=0\;Z_{2,2}=0\;Z_{2,3}=1\;Z_4=2$ } & \tiny{ $\;Z_{2,1}=1\;Z_{2,2}=0\;Z_{2,3}=0\;Z_4=2$ }\\
\includegraphics[width=0.38\textwidth,angle=0]{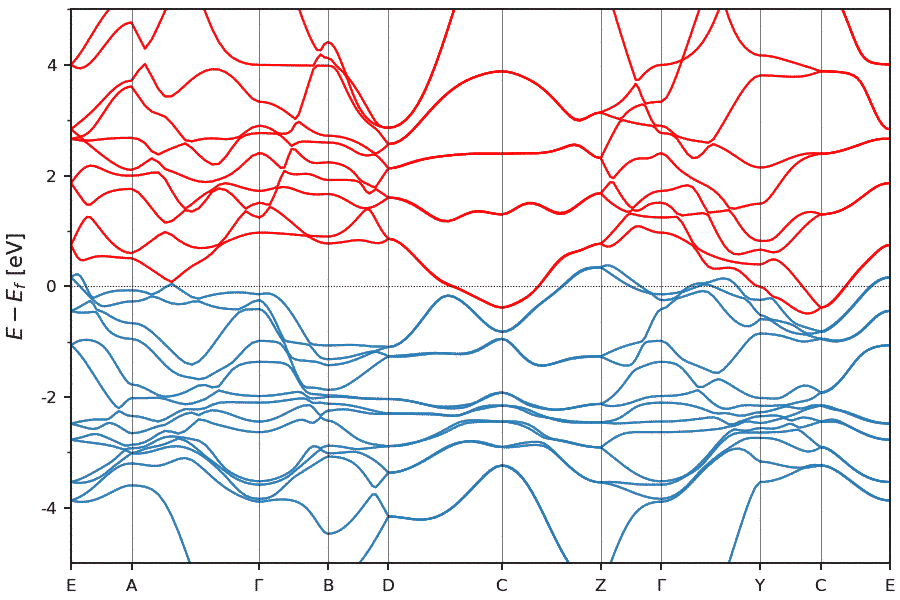} & \includegraphics[width=0.38\textwidth,angle=0]{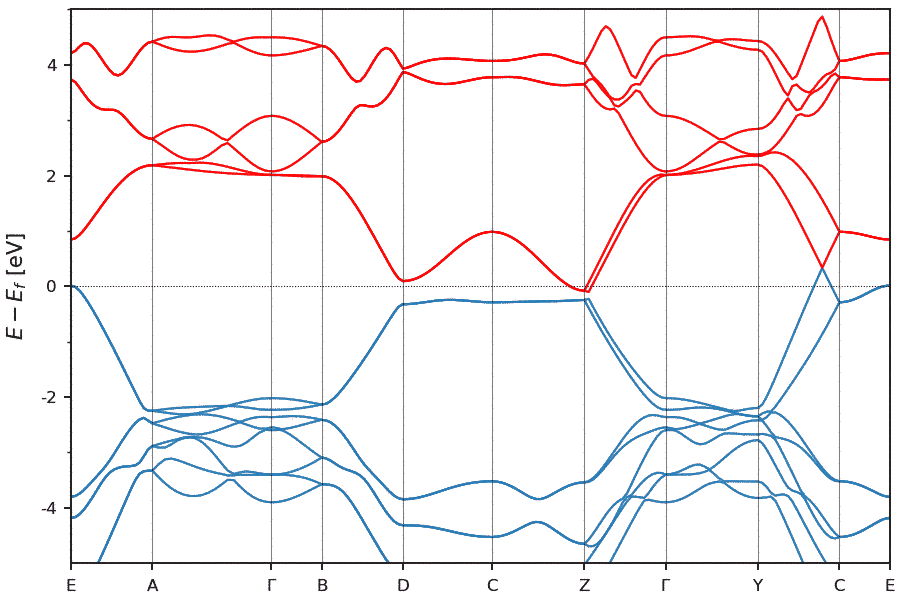}\\
\end{tabular}
\begin{tabular}{c c}
\scriptsize{$\rm{Ca} \rm{Cu} (\rm{As}_{2} \rm{O}_{7})$ - \icsdweb{408693} - SG 14 ($P2_1/c$) - NLC} & \scriptsize{$\rm{Ca}_{2} \rm{Ir}_{2} \rm{Si}$ - \icsdweb{95789} - SG 15 ($C2/c$) - NLC}\\
\tiny{ $\;Z_{2,1}=1\;Z_{2,2}=0\;Z_{2,3}=0\;Z_4=2$ } & \tiny{ $\;Z_{2,1}=1\;Z_{2,2}=1\;Z_{2,3}=0\;Z_4=2$ }\\
\includegraphics[width=0.38\textwidth,angle=0]{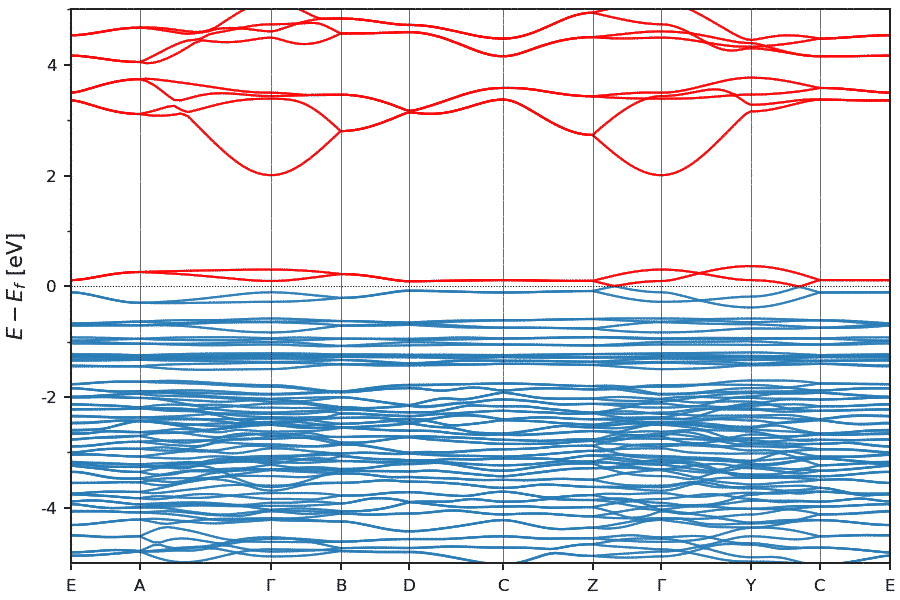} & \includegraphics[width=0.38\textwidth,angle=0]{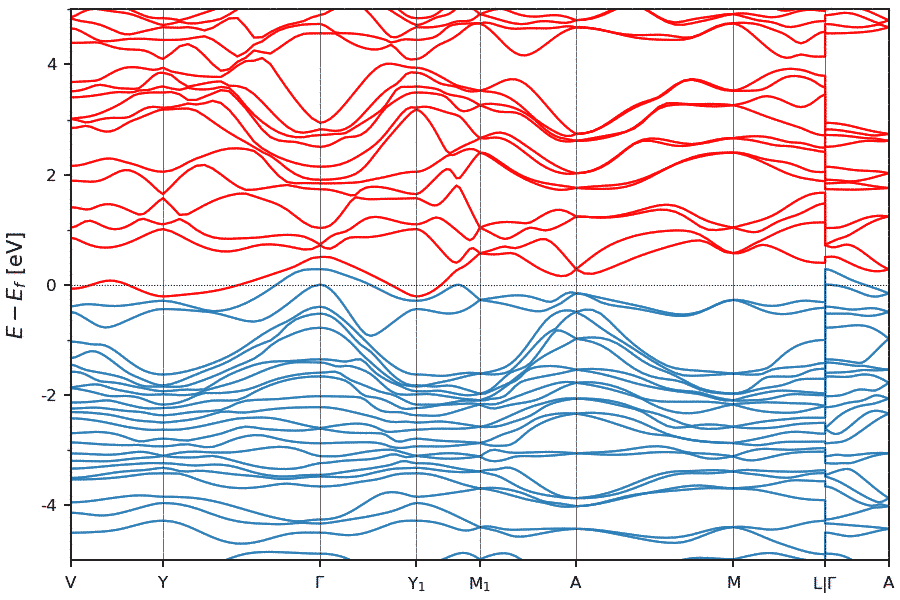}\\
\end{tabular}

\caption{\zfourTCIsandWTIsNLC{1}}
\label{fig:z4_2TCIs_andWTIs_NLC1}
\end{figure}

\begin{figure}[ht]
\centering
\begin{tabular}{c c}
\scriptsize{$\rm{Ta}_{4} \rm{O}$ - \icsdweb{76022} - SG 47 ($Pmmm$) - NLC} & \scriptsize{$\rm{Ga} \rm{Pt}_{2}$ - \icsdweb{103918} - SG 51 ($Pmma$) - NLC}\\
\tiny{ $\;Z_{2,1}=1\;Z_{2,2}=1\;Z_{2,3}=1\;Z_4=2$ } & \tiny{ $\;Z_{2,1}=0\;Z_{2,2}=1\;Z_{2,3}=1\;Z_4=2$ }\\
\includegraphics[width=0.38\textwidth,angle=0]{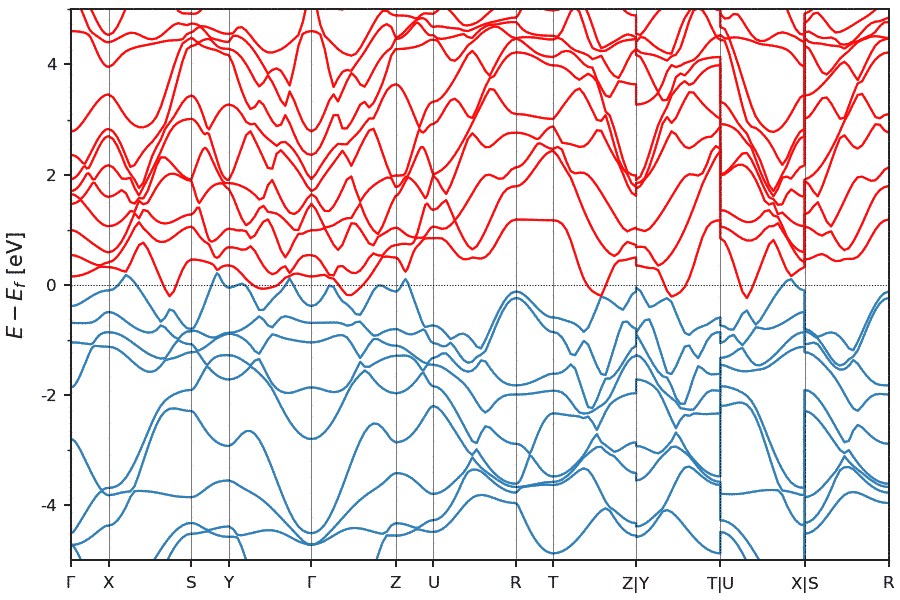} & \includegraphics[width=0.38\textwidth,angle=0]{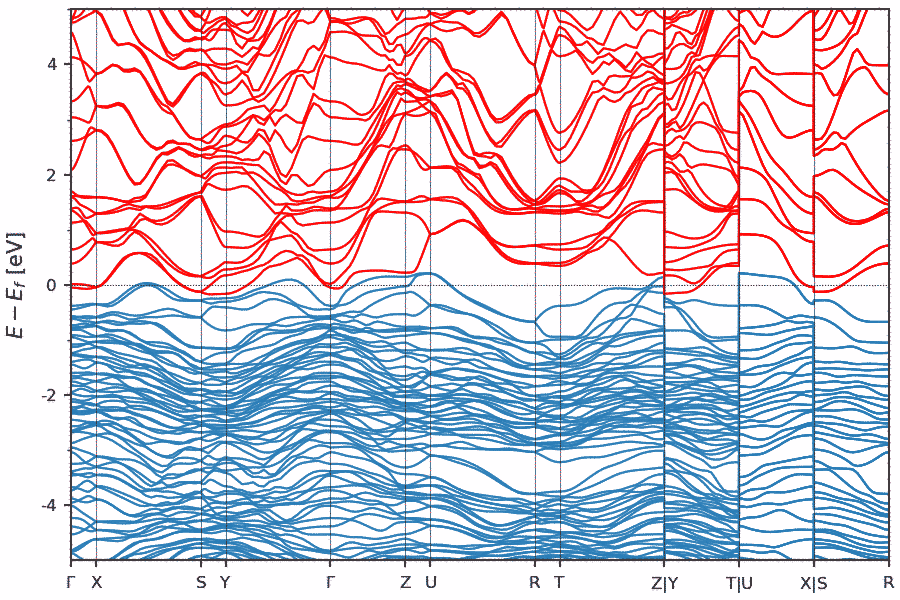}\\
\end{tabular}
\begin{tabular}{c c}
\scriptsize{$\rm{Fe} \rm{Se}$ - \icsdweb{169315} - SG 67 ($Cmme$) - NLC} & \scriptsize{$\rm{Pb}_{3} \rm{Sr}_{2}$ - \icsdweb{105627} - SG 127 ($P4/mbm$) - NLC}\\
\tiny{ $\;Z_{2,1}=0\;Z_{2,2}=0\;Z_{2,3}=1\;Z_4=2$ } & \tiny{ $\;Z_{2,1}=0\;Z_{2,2}=0\;Z_{2,3}=0\;Z_4=2\;Z_{4m,\pi}=2\;Z_2=0\;Z_8=2$ }\\
\includegraphics[width=0.38\textwidth,angle=0]{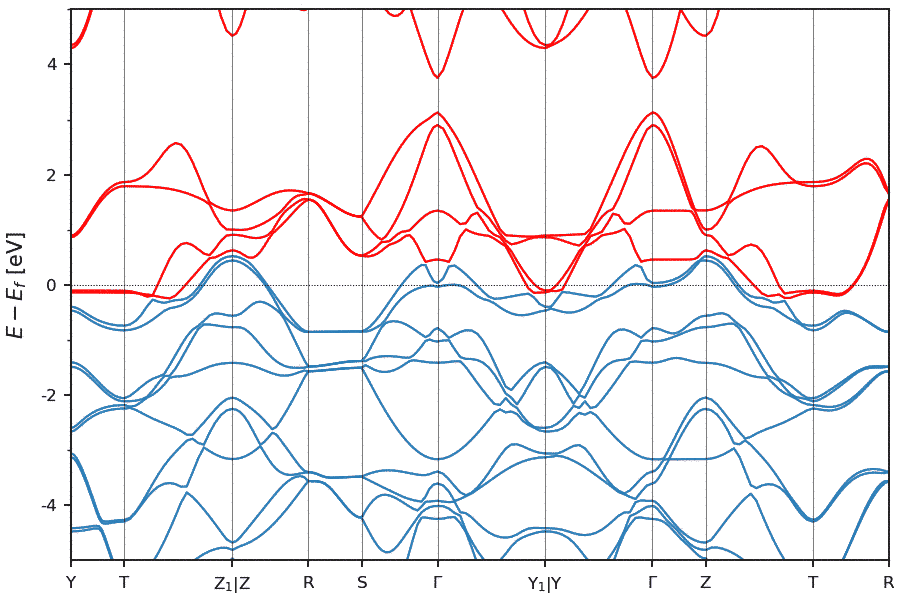} & \includegraphics[width=0.38\textwidth,angle=0]{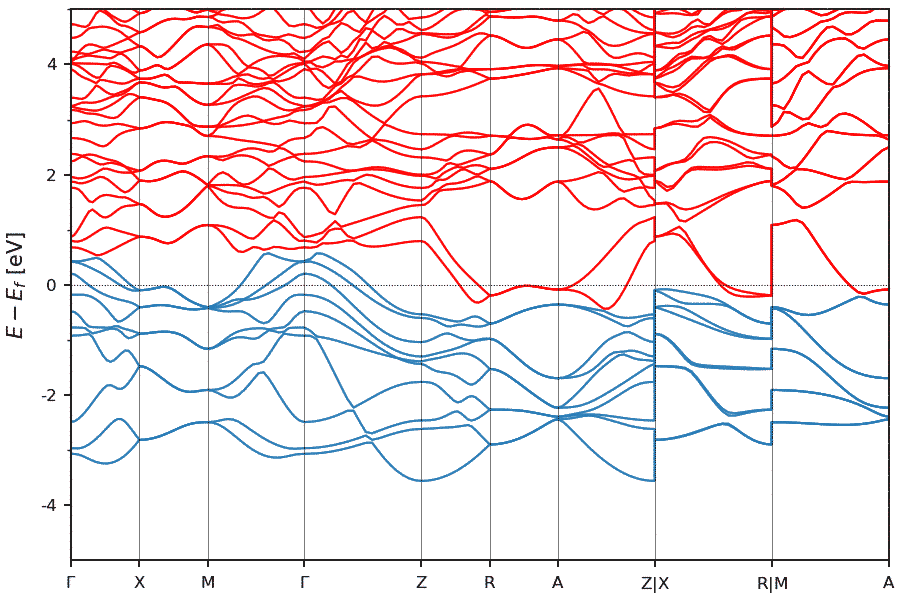}\\
\end{tabular}
\begin{tabular}{c c}
\scriptsize{$\rm{Re} \rm{N}$ - \icsdweb{187713} - SG 131 ($P4_2/mmc$) - NLC} & \scriptsize{$\rm{Ir} \rm{Sc}$ - \icsdweb{640974} - SG 221 ($Pm\bar{3}m$) - NLC}\\
\tiny{ $\;Z_{2,1}=1\;Z_{2,2}=1\;Z_{2,3}=0\;Z_4=2\;Z_2=0$ } & \tiny{ $\;Z_{2,1}=0\;Z_{2,2}=0\;Z_{2,3}=0\;Z_4=2\;Z_{4m,\pi}=0\;Z_2=0\;Z_8=6$ }\\
\includegraphics[width=0.38\textwidth,angle=0]{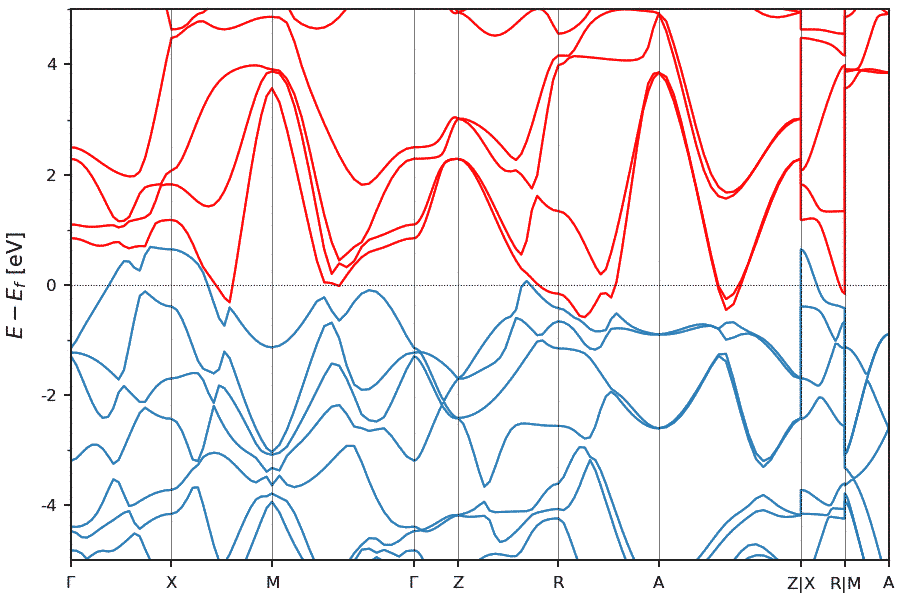} & \includegraphics[width=0.38\textwidth,angle=0]{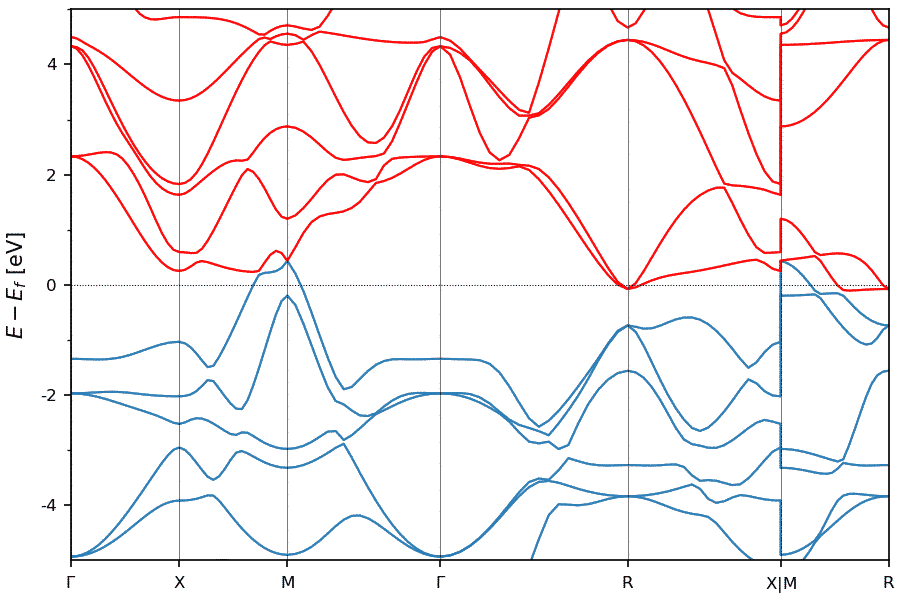}\\
\end{tabular}

\caption{\zfourTCIsandWTIsNLC{2}}
\label{fig:z4_2TCIs_andWTIs_NLC2}
\end{figure}


\begin{figure}[ht]
\centering
\begin{tabular}{c c}
\scriptsize{$\rm{Zr} \rm{Cl}$ - \icsdweb{868} - SG 12 ($C2/m$) - SEBR} & \scriptsize{$\rm{Bi} \rm{I}$ - \icsdweb{1559} - SG 12 ($C2/m$) - SEBR}\\
\tiny{ $\;Z_{2,1}=0\;Z_{2,2}=0\;Z_{2,3}=1\;Z_4=2$ } & \tiny{ $\;Z_{2,1}=0\;Z_{2,2}=0\;Z_{2,3}=1\;Z_4=2$ }\\
\includegraphics[width=0.38\textwidth,angle=0]{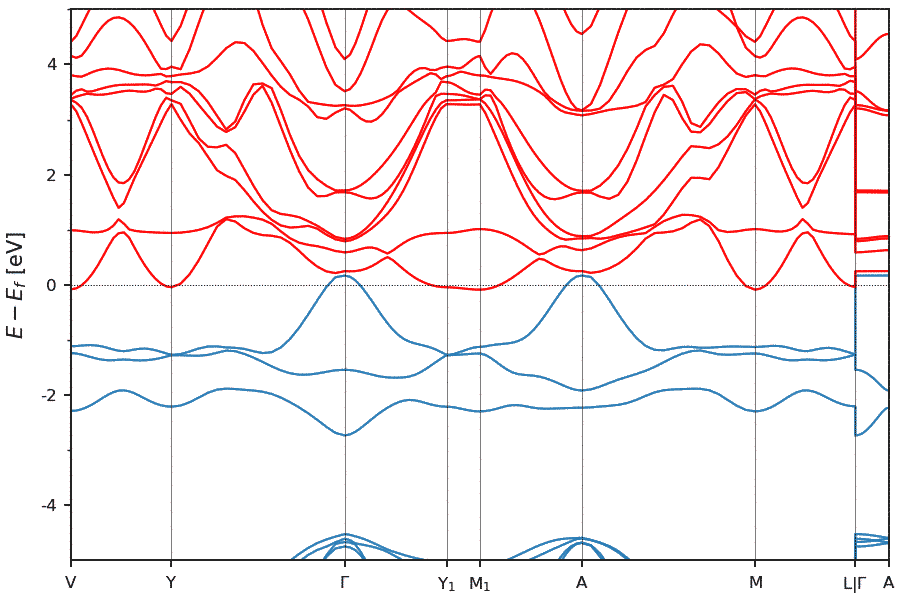} & \includegraphics[width=0.38\textwidth,angle=0]{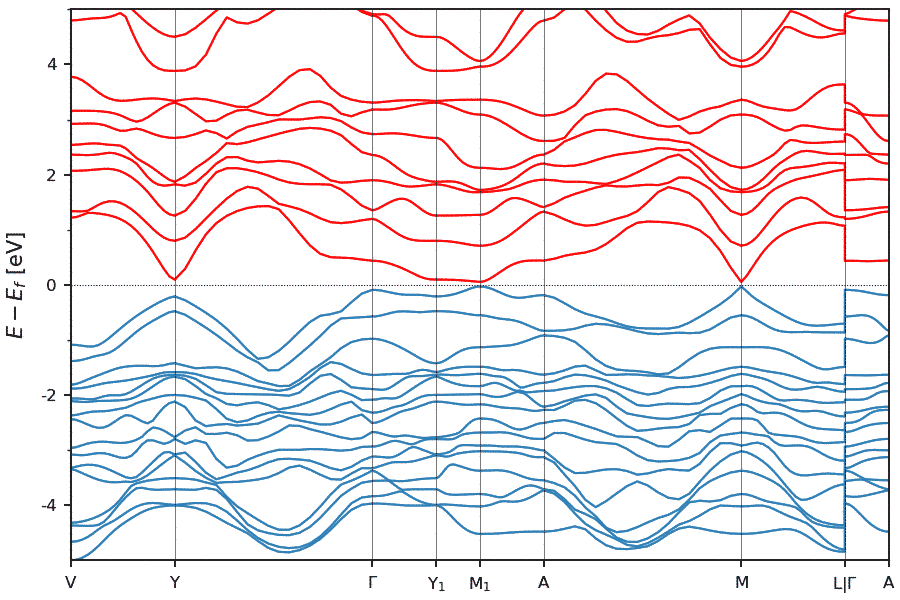}\\
\end{tabular}
\begin{tabular}{c c}
\scriptsize{$\rm{Na} \rm{N}_{3}$ - \icsdweb{29383} - SG 12 ($C2/m$) - SEBR} & \scriptsize{$\rm{Ta} \rm{As}_{2}$ - \icsdweb{44069} - SG 12 ($C2/m$) - SEBR}\\
\tiny{ $\;Z_{2,1}=1\;Z_{2,2}=1\;Z_{2,3}=0\;Z_4=2$ } & \tiny{ $\;Z_{2,1}=1\;Z_{2,2}=1\;Z_{2,3}=1\;Z_4=2$ }\\
\includegraphics[width=0.38\textwidth,angle=0]{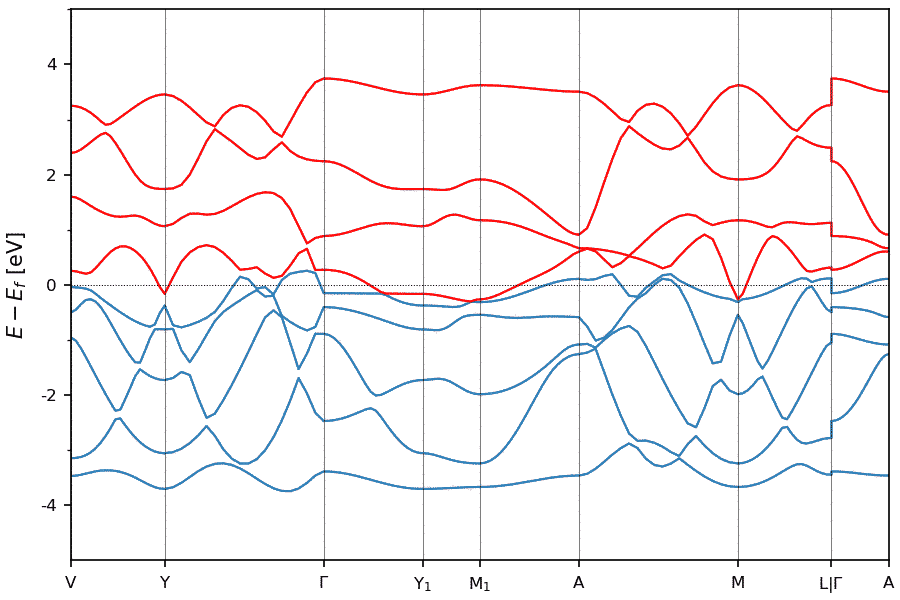} & \includegraphics[width=0.38\textwidth,angle=0]{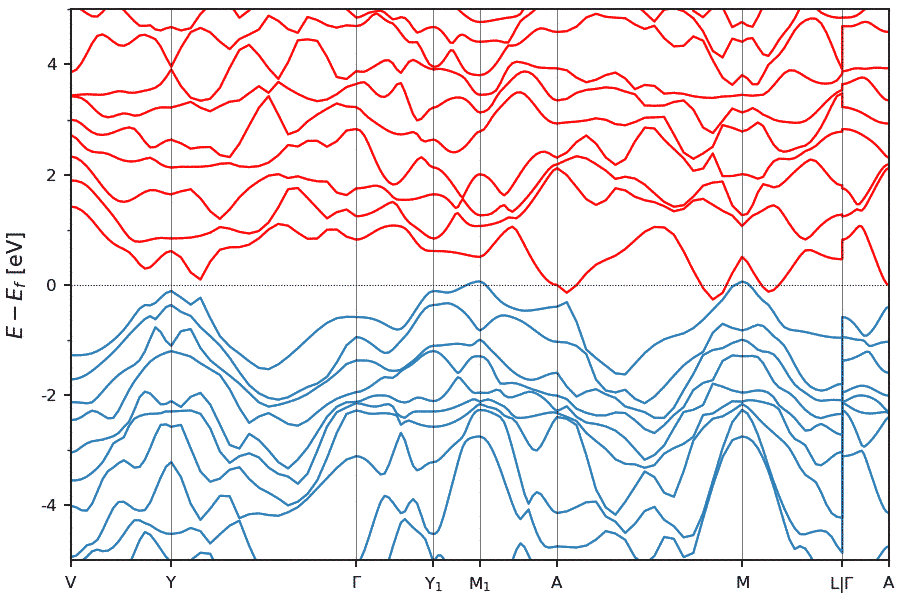}\\
\end{tabular}
\begin{tabular}{c c}
\scriptsize{$\rm{Nb} \rm{Sb}_{2}$ - \icsdweb{57430} - SG 12 ($C2/m$) - SEBR} & \scriptsize{$\rm{Nb} \rm{As}_{2}$ - \icsdweb{81218} - SG 12 ($C2/m$) - SEBR}\\
\tiny{ $\;Z_{2,1}=1\;Z_{2,2}=1\;Z_{2,3}=1\;Z_4=2$ } & \tiny{ $\;Z_{2,1}=1\;Z_{2,2}=1\;Z_{2,3}=1\;Z_4=2$ }\\
\includegraphics[width=0.38\textwidth,angle=0]{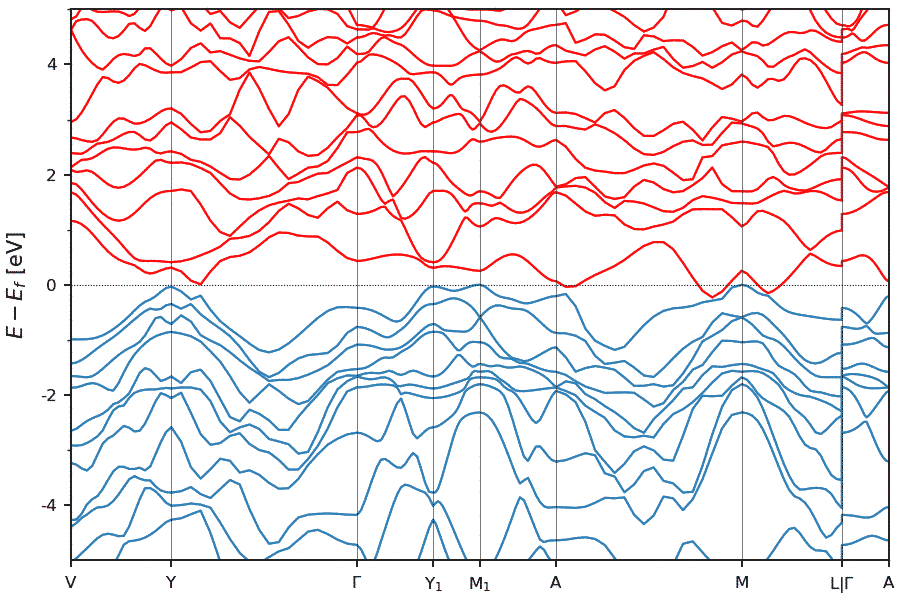} & \includegraphics[width=0.38\textwidth,angle=0]{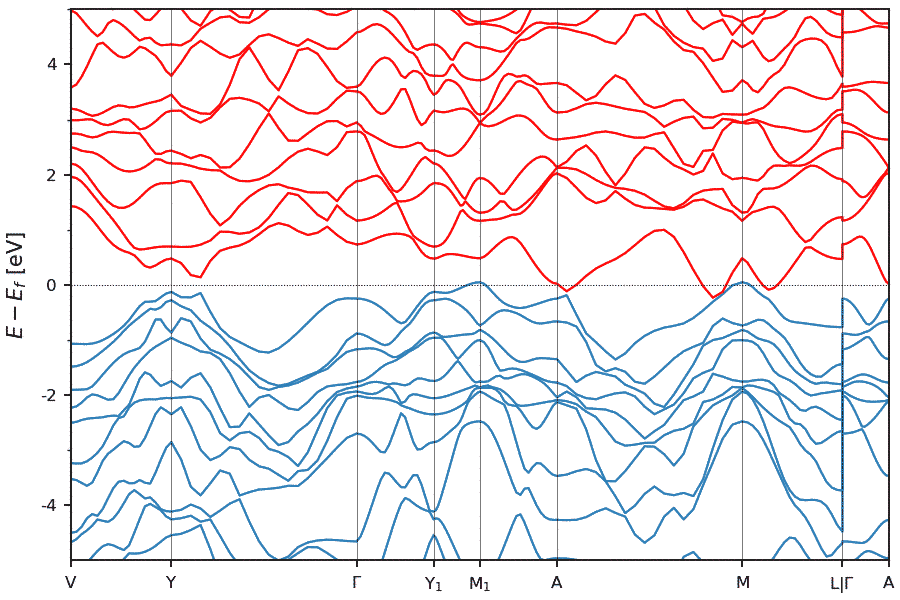}\\
\end{tabular}
\begin{tabular}{c c}
\scriptsize{$\rm{Ba}_{3} \rm{Cd}_{2} \rm{Sb}_{4}$ - \icsdweb{173685} - SG 12 ($C2/m$) - SEBR} & \scriptsize{$\rm{Ba}_{3} \rm{Zn}_{2} \rm{As}_{4}$ - \icsdweb{424760} - SG 12 ($C2/m$) - SEBR}\\
\tiny{ $\;Z_{2,1}=1\;Z_{2,2}=1\;Z_{2,3}=0\;Z_4=2$ } & \tiny{ $\;Z_{2,1}=1\;Z_{2,2}=1\;Z_{2,3}=0\;Z_4=2$ }\\
\includegraphics[width=0.38\textwidth,angle=0]{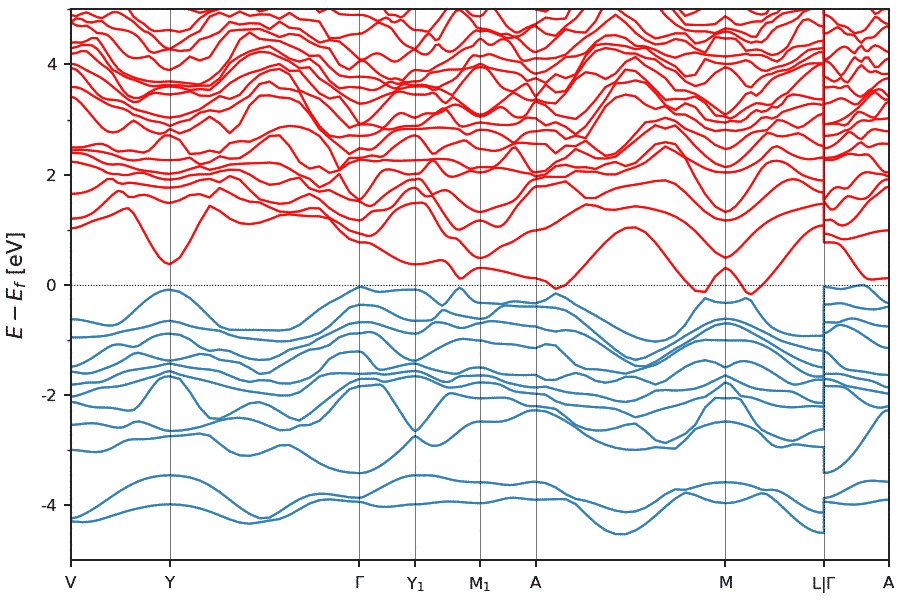} & \includegraphics[width=0.38\textwidth,angle=0]{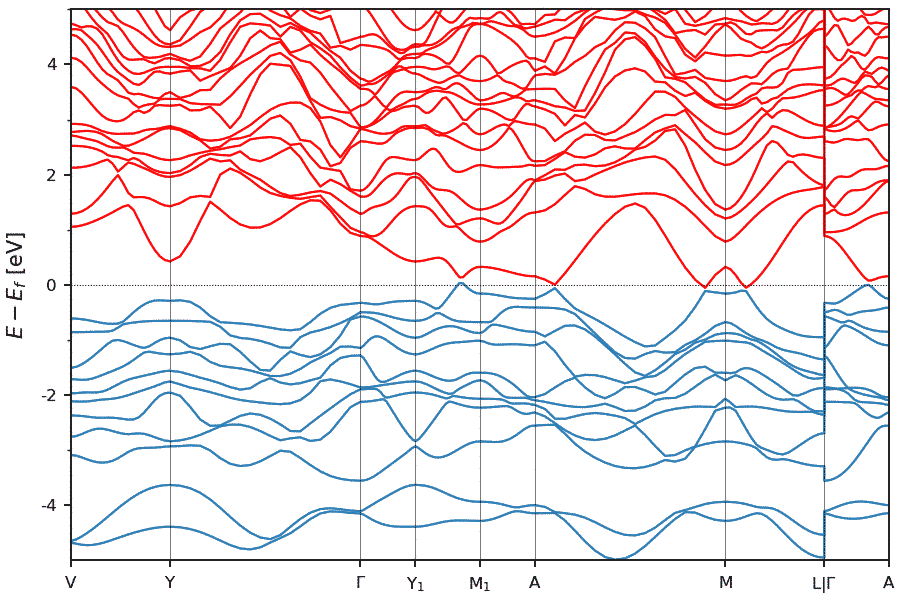}\\
\end{tabular}

\caption{\zfourTCIsandWTIsSEBR{1}}
\label{fig:z4_2TCIs_andWTIs_SEBR1}
\end{figure}

\begin{figure}[ht]
\centering
\begin{tabular}{c c}
\scriptsize{$\rm{Ba}_{3} \rm{Cd}_{2} \rm{As}_{4}$ - \icsdweb{424761} - SG 12 ($C2/m$) - SEBR} & \scriptsize{$\rm{V} \rm{As}_{2}$ - \icsdweb{611574} - SG 12 ($C2/m$) - SEBR}\\
\tiny{ $\;Z_{2,1}=1\;Z_{2,2}=1\;Z_{2,3}=0\;Z_4=2$ } & \tiny{ $\;Z_{2,1}=1\;Z_{2,2}=1\;Z_{2,3}=1\;Z_4=2$ }\\
\includegraphics[width=0.38\textwidth,angle=0]{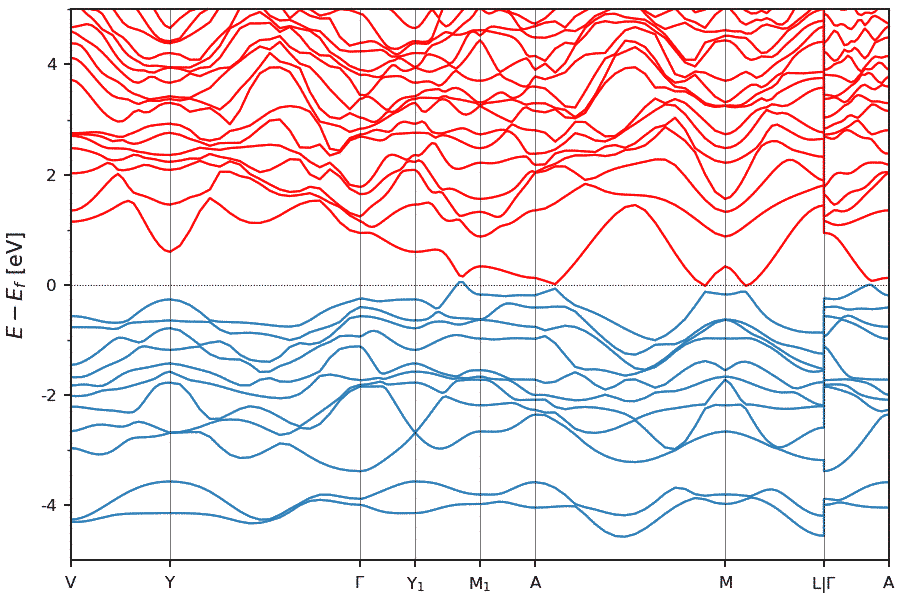} & \includegraphics[width=0.38\textwidth,angle=0]{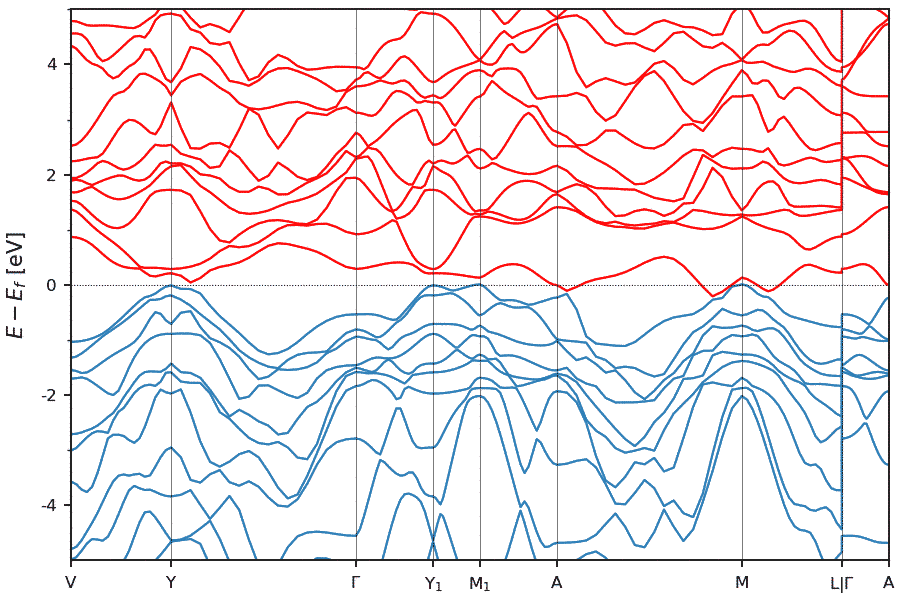}\\
\end{tabular}%

\caption{\zfourTCIsandWTIsSEBR{2}}
\label{fig:z4_2TCIs_andWTIs_SEBR2}
\end{figure}

\begin{figure}[ht]
\centering
%

\caption{\zfourTCIsandWTIsSEBR{3}}
\label{fig:z4_2TCIs_andWTIs_SEBR3}
\end{figure}

\begin{figure}[ht]
\centering
%

\caption{\zfourTCIsandWTIsSEBR{14}}
\label{fig:z4_2TCIs_andWTIs_SEBR14}
\end{figure}

\clearpage

\subsubsection{Rotation-Anomaly TCIs and High-Fold-Rotation Mirror TCIs}
\label{App:rotationAnomaly}

In this section, we list the symmetry-indicated rotation-anomaly and high-fold-rotation mirror TCIs~\cite{ChenRotation,HOTIChen,ChenTCI,AshvinTCI,WiederAxion,ChenBernevigTCI} with the largest bulk gaps or the fewest and smallest bulk Fermi pockets.  The TCIs listed in this section exhibit varying, even numbers of massless or massive twofold surface Dirac cones, depending on the Miller indices of the surfaces~\cite{ChenRotation,HOTIChen,ChenTCI,AshvinTCI,WiederAxion}.  On the edges between gapped surfaces, helical hinge modes may also be present, depending on the crystallographic (Miller) indices of the hinges.  In terms of the stable SIs introduced in Ref.~\onlinecite{ChenTCI}, the materials in this section either exhibit $Z_{8}=4$ or $Z_{12}'=6$; we do not find any examples of band insulators with $Z_{12}=6$ and few bulk Fermi pockets.  First, in Fig.~\ref{fig:z12prime6HOTIs_SEBR}, we list the few sixfold-rotation-anomaly TCIs with $Z_{12}'=6$, all of which are classified as SEBR.  In Fig.~\ref{fig:z8_4HOTIs_NLC}, we then show the only TCI with $Z_{8}=4$ and a relatively clean Fermi surface that is classified as NLC.  Finally, in Figs.~\ref{fig:z8_4HOTIs_TCI_SEBR1},~\ref{fig:z8_4HOTIs_TCI_SEBR2},~\ref{fig:z8_4HOTIs_TCI_SEBR3}, and~\ref{fig:z8_4HOTIs_TCI_SEBR4}, we list the TCIs with $Z_{8}=4$ that are classified as SEBR.  The materials identified in this section include members of the well-studied SnTe family, including SnTe [\icsdweb{652759}, SG 225 ($Fm\bar{3}m$)]~\cite{TeoFuKaneTCI,HsiehTCI,TanakaSnTeExp,HOTIBernevig,ChenRotation,ChenTCI} -- a prototypical fourfold rotation-anomaly TCI -- and PbTe [\icsdweb{648615}, SG 225 ($Fm\bar{3}m$)]~\cite{FradkinPbTe,BarryPbTe}, which also appears as a fourfold-rotation anomaly TCI, depending on the choice of lattice parameters and DFT functionals.  SnTe and PbTe are both classified as SEBR and thus appear in Figs.~\ref{fig:z8_4HOTIs_TCI_SEBR3} and~\ref{fig:z8_4HOTIs_TCI_SEBR4}, respectively.


\begin{figure}[ht]
\centering
\begin{tabular}{c c}
\scriptsize{$\rm{Co}_{2} \rm{Sn}$ - \icsdweb{102673} - SG 194 ($P6_3/mmc$) - SEBR} & \scriptsize{$\rm{Sn} \rm{Ti}_{2}$ - \icsdweb{169007} - SG 194 ($P6_3/mmc$) - SEBR}\\
\tiny{ $\;Z_{2,1}=0\;Z_{2,2}=0\;Z_{2,3}=0\;Z_4=2\;Z_{6m,0}=0\;Z_{12}'=6$ } & \tiny{ $\;Z_{2,1}=0\;Z_{2,2}=0\;Z_{2,3}=0\;Z_4=2\;Z_{6m,0}=0\;Z_{12}'=6$ }\\
\includegraphics[width=0.38\textwidth,angle=0]{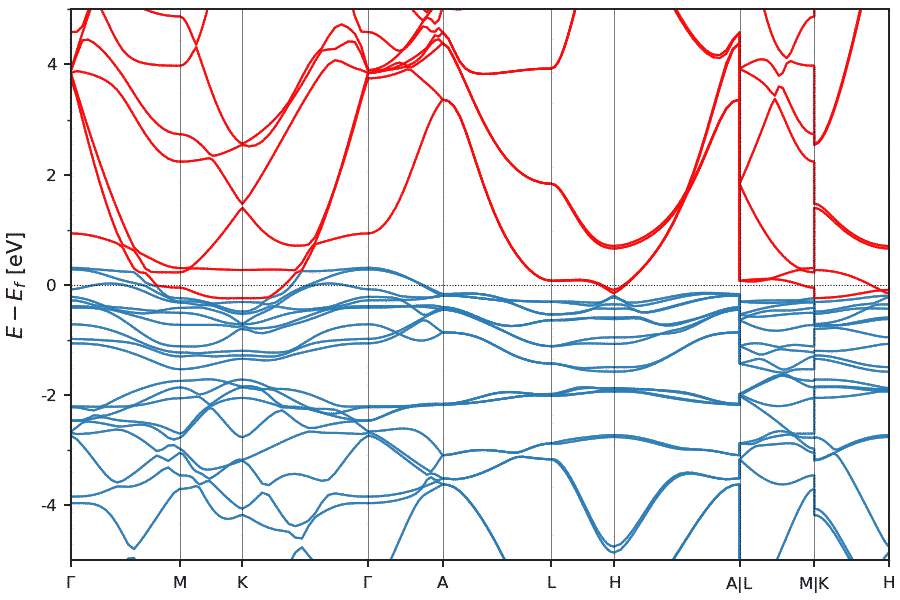} & \includegraphics[width=0.38\textwidth,angle=0]{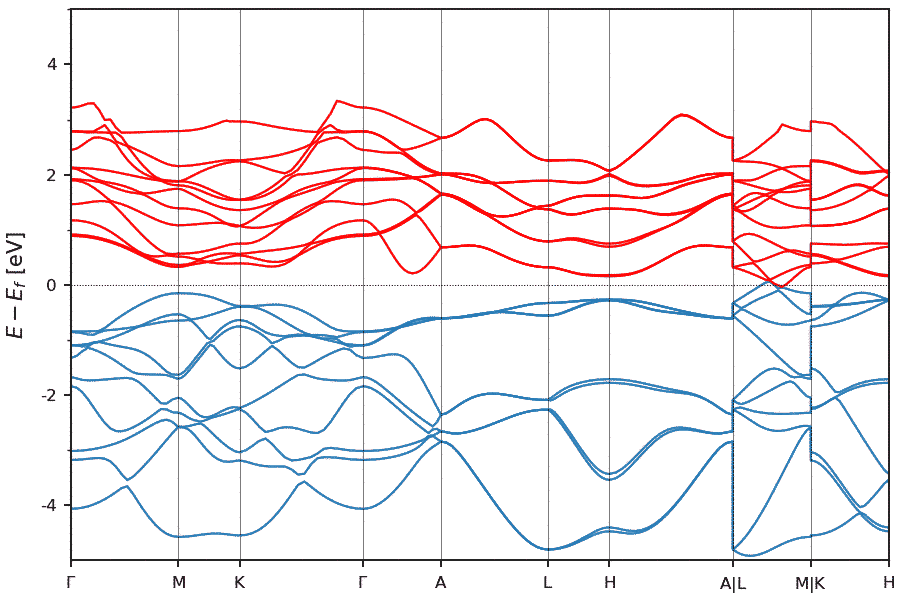}\\
\end{tabular}
\begin{tabular}{c c}
\scriptsize{$\rm{Co}_{2} \rm{Ge}$ - \icsdweb{623418} - SG 194 ($P6_3/mmc$) - SEBR}\\
\tiny{ $\;Z_{2,1}=0\;Z_{2,2}=0\;Z_{2,3}=0\;Z_4=2\;Z_{6m,0}=0\;Z_{12}'=6$ }\\
\includegraphics[width=0.38\textwidth,angle=0]{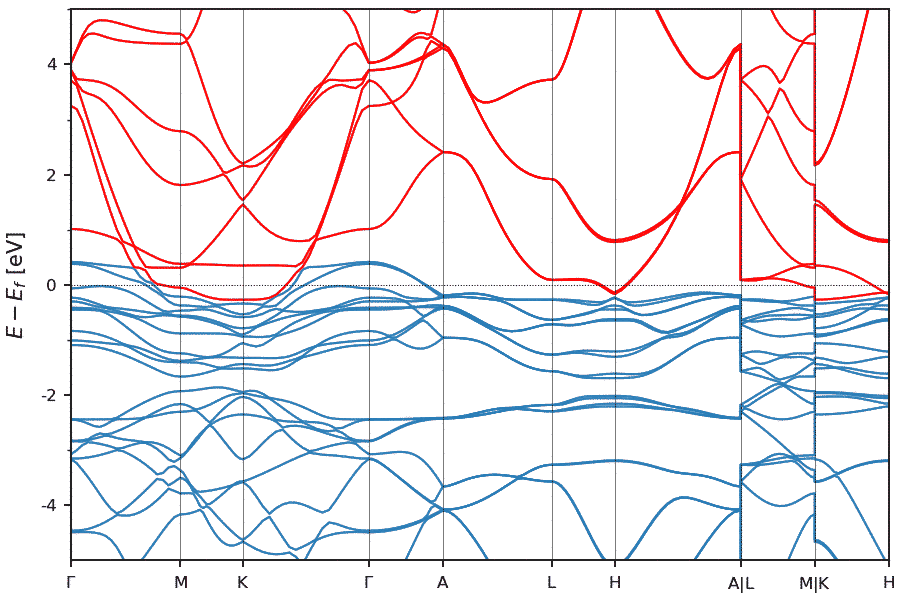}\\
\end{tabular}

\caption{The sixfold-rotation-anomaly TCIs with the largest band gaps or the fewest and smallest bulk Fermi pockets.  All of these materials exhibit $Z_{12}'=6$, and are classified as SEBR.}
\label{fig:z12prime6HOTIs_SEBR}
\end{figure}


\begin{figure}[ht]
\centering
\begin{tabular}{c c}
\scriptsize{$\rm{Ga}_{2} \rm{Zr}_{3}$ - \icsdweb{104042} - SG 127 ($P4/mbm$) - NLC}\\
\tiny{ $\;Z_{2,1}=0\;Z_{2,2}=0\;Z_{2,3}=1\;Z_4=0\;Z_{4m,\pi}=3\;Z_2=0\;Z_8=4$ }\\
\includegraphics[width=0.38\textwidth,angle=0]{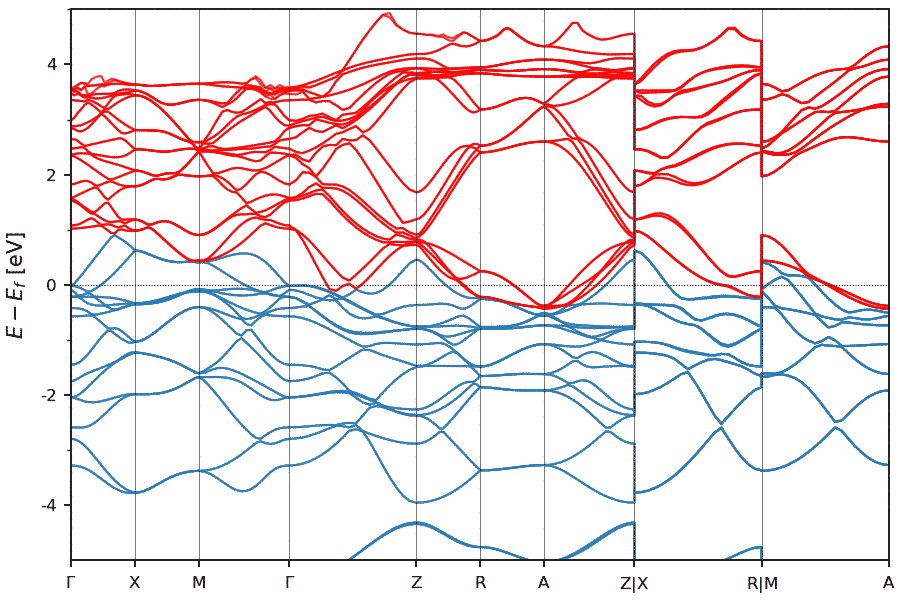}\\
\end{tabular}

\caption{The NLC-classified, fourfold-rotation-anomaly weak TI characterized by $Z_{8}=4$ with the fewest and smallest bulk Fermi pockets.}
\label{fig:z8_4HOTIs_NLC}
\end{figure}


\begin{figure}[ht]
\centering
\begin{tabular}{c c}
\scriptsize{$\rm{Au}_{4} \rm{Ti}$ - \icsdweb{109132} - SG 87 ($I4/m$) - SEBR} & \scriptsize{$\rm{Ba} \rm{Mg}_{4} \rm{Ge}_{3}$ - \icsdweb{165631} - SG 123 ($P4/mmm$) - SEBR}\\
\tiny{ $\;Z_{2,1}=1\;Z_{2,2}=1\;Z_{2,3}=1\;Z_4=0\;Z_2=0\;Z_8=4$ } & \tiny{ $\;Z_{2,1}=0\;Z_{2,2}=0\;Z_{2,3}=1\;Z_4=0\;Z_{4m,\pi}=1\;Z_2=0\;Z_8=4$ }\\
\includegraphics[width=0.38\textwidth,angle=0]{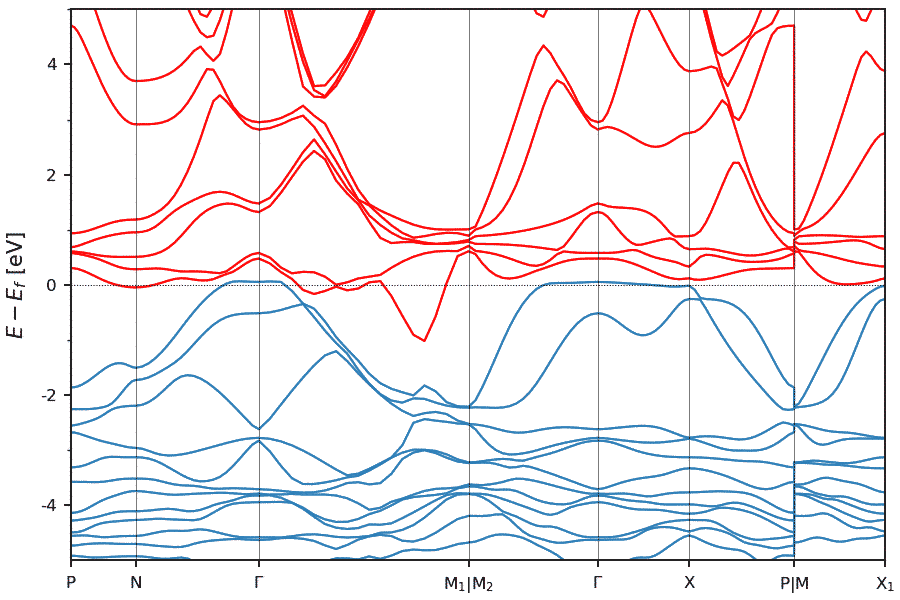} & \includegraphics[width=0.38\textwidth,angle=0]{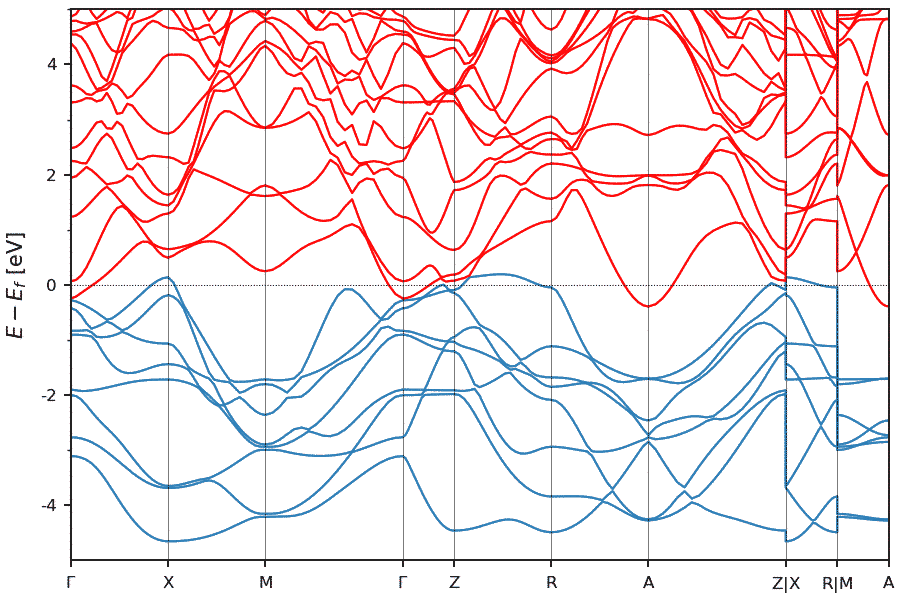}\\
\end{tabular}
\begin{tabular}{c c}
\scriptsize{$\rm{Ca} \rm{Rh}_{2} \rm{P}_{2}$ - \icsdweb{50185} - SG 139 ($I4/mmm$) - SEBR} & \scriptsize{$\rm{Ca} \rm{Zn}_{2} \rm{Si}_{2}$ - \icsdweb{59649} - SG 139 ($I4/mmm$) - SEBR}\\
\tiny{ $\;Z_{2,1}=1\;Z_{2,2}=1\;Z_{2,3}=1\;Z_4=0\;Z_2=0\;Z_8=4$ } & \tiny{ $\;Z_{2,1}=0\;Z_{2,2}=0\;Z_{2,3}=0\;Z_4=0\;Z_2=0\;Z_8=4$ }\\
\includegraphics[width=0.38\textwidth,angle=0]{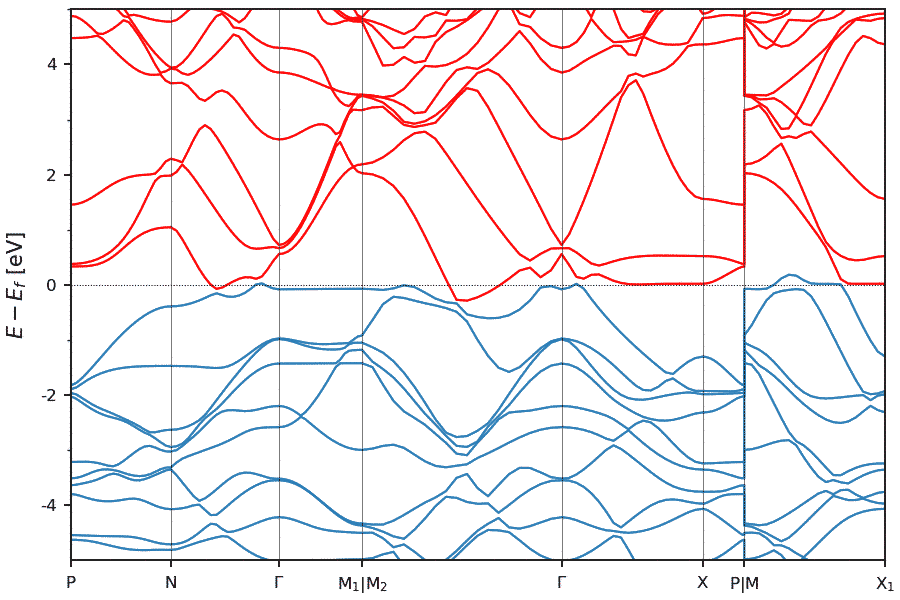} & \includegraphics[width=0.38\textwidth,angle=0]{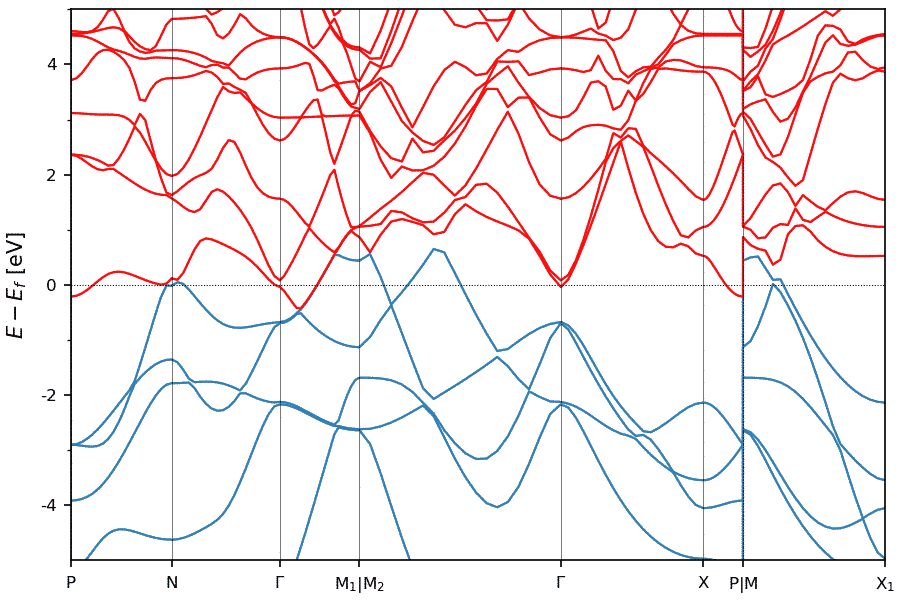}\\
\end{tabular}
\begin{tabular}{c c}
\scriptsize{$\rm{Na}_{2} \rm{Ti}_{2} \rm{Sb}_{2} \rm{O}$ - \icsdweb{91202} - SG 139 ($I4/mmm$) - SEBR} & \scriptsize{$\rm{Zr} \rm{H}_{2}$ - \icsdweb{108539} - SG 139 ($I4/mmm$) - SEBR}\\
\tiny{ $\;Z_{2,1}=0\;Z_{2,2}=0\;Z_{2,3}=0\;Z_4=0\;Z_2=0\;Z_8=4$ } & \tiny{ $\;Z_{2,1}=0\;Z_{2,2}=0\;Z_{2,3}=0\;Z_4=0\;Z_2=0\;Z_8=4$ }\\
\includegraphics[width=0.38\textwidth,angle=0]{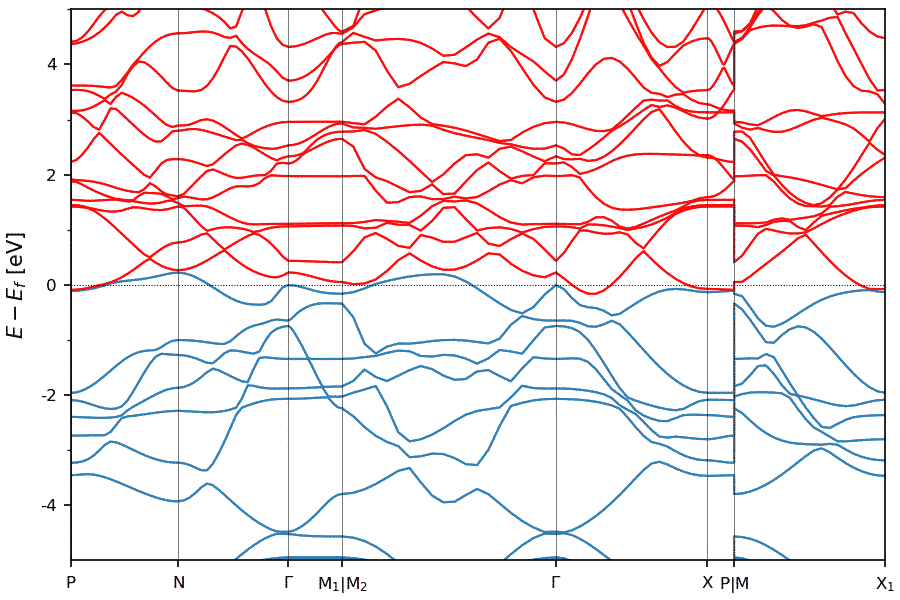} & \includegraphics[width=0.38\textwidth,angle=0]{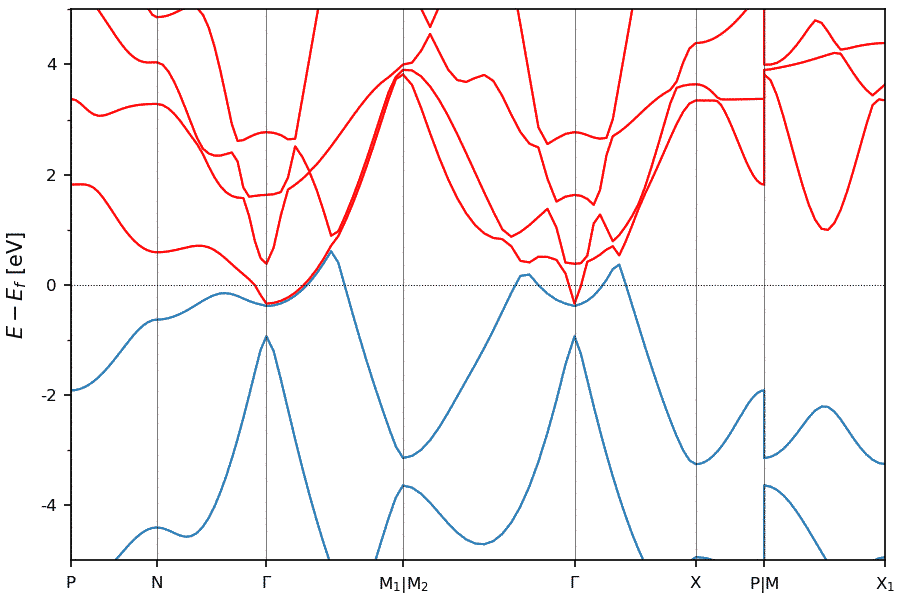}\\
\end{tabular}
\begin{tabular}{c c}
\scriptsize{$(\rm{Sr} \rm{F})_{2} \rm{Ti}_{2} \rm{Sb}_{2} \rm{O}$ - \icsdweb{167014} - SG 139 ($I4/mmm$) - SEBR} & \scriptsize{$\rm{K}_{2} \rm{Mg}_{5} \rm{Sn}_{3}$ - \icsdweb{421342} - SG 139 ($I4/mmm$) - SEBR}\\
\tiny{ $\;Z_{2,1}=0\;Z_{2,2}=0\;Z_{2,3}=0\;Z_4=0\;Z_2=0\;Z_8=4$ } & \tiny{ $\;Z_{2,1}=0\;Z_{2,2}=0\;Z_{2,3}=0\;Z_4=0\;Z_2=0\;Z_8=4$ }\\
\includegraphics[width=0.38\textwidth,angle=0]{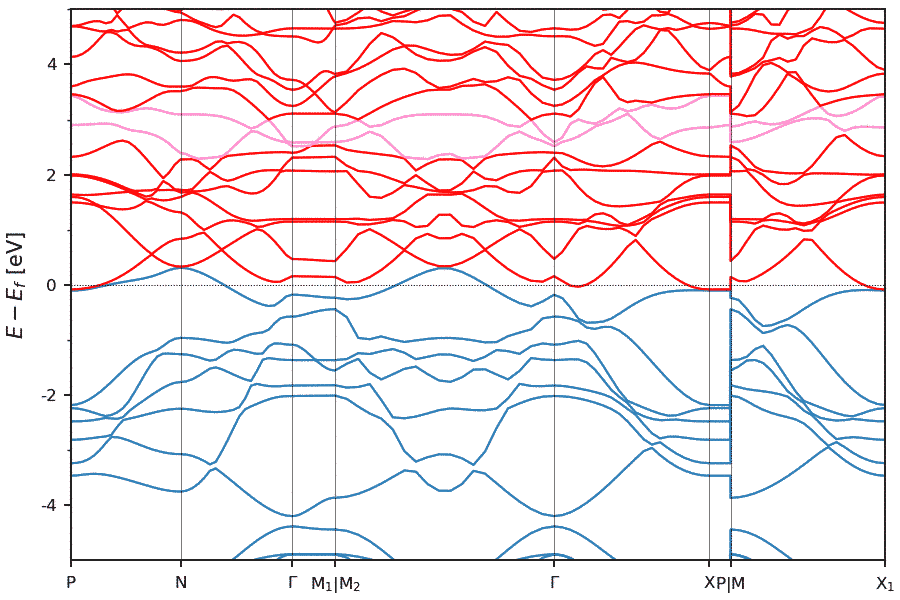} & \includegraphics[width=0.38\textwidth,angle=0]{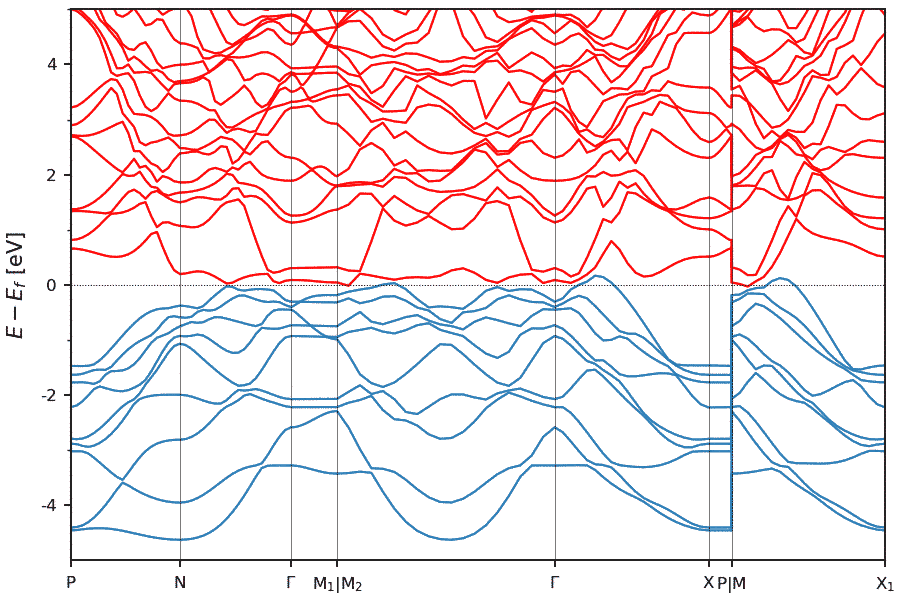}\\
\end{tabular}

\caption{\zeightHOTIsTCISEBR{1}}
\label{fig:z8_4HOTIs_TCI_SEBR1}
\end{figure}

\begin{figure}[ht]
\centering
\begin{tabular}{c c}
\scriptsize{$(\rm{Sr} \rm{F})_{2} \rm{Ti}_{2} \rm{Bi}_{2} \rm{O}$ - \icsdweb{430064} - SG 139 ($I4/mmm$) - SEBR} & \scriptsize{$\rm{Ca} \rm{Co}_{2} \rm{As}_{2}$ - \icsdweb{609899} - SG 139 ($I4/mmm$) - SEBR}\\
\tiny{ $\;Z_{2,1}=0\;Z_{2,2}=0\;Z_{2,3}=0\;Z_4=0\;Z_2=0\;Z_8=4$ } & \tiny{ $\;Z_{2,1}=1\;Z_{2,2}=1\;Z_{2,3}=1\;Z_4=0\;Z_2=0\;Z_8=4$ }\\
\includegraphics[width=0.38\textwidth,angle=0]{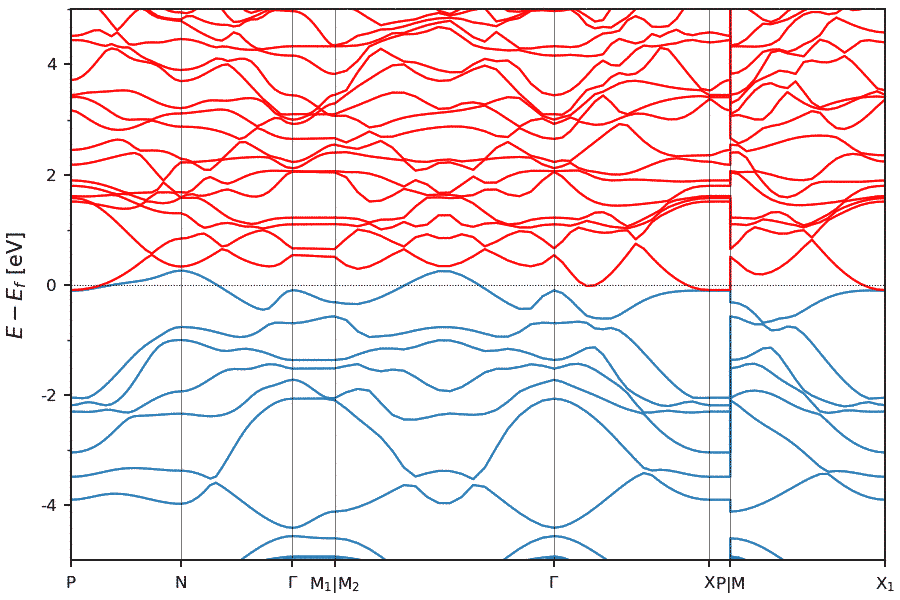} & \includegraphics[width=0.38\textwidth,angle=0]{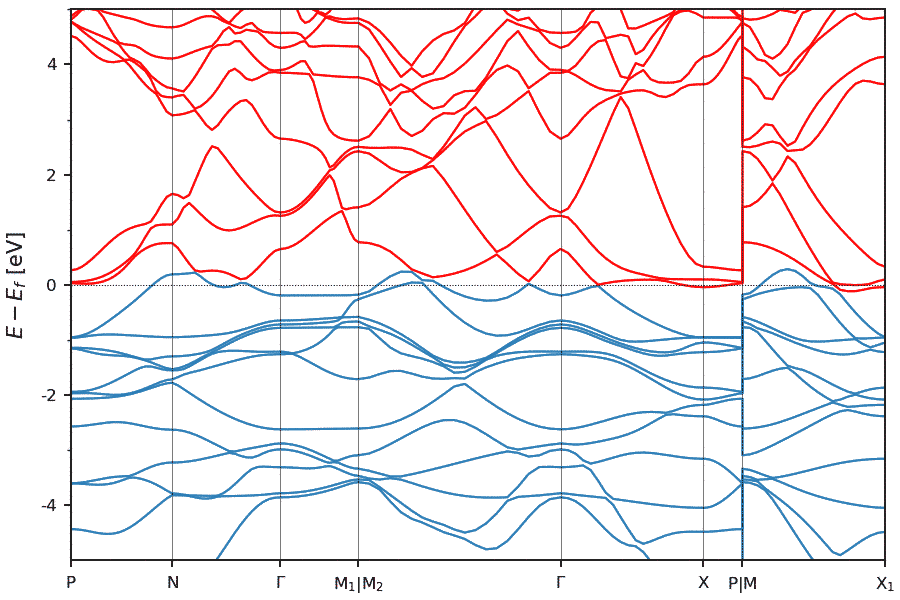}\\
\end{tabular}
\begin{tabular}{c c}
\scriptsize{$\rm{Ti} \rm{Sb}_{2}$ - \icsdweb{52322} - SG 140 ($I4/mcm$) - SEBR} & \scriptsize{$\rm{Pt}_{3} \rm{Ge}$ - \icsdweb{77962} - SG 140 ($I4/mcm$) - SEBR}\\
\tiny{ $\;Z_{2,1}=1\;Z_{2,2}=1\;Z_{2,3}=1\;Z_4=0\;Z_2=0\;Z_8=4$ } & \tiny{ $\;Z_{2,1}=0\;Z_{2,2}=0\;Z_{2,3}=0\;Z_4=0\;Z_2=0\;Z_8=4$ }\\
\includegraphics[width=0.38\textwidth,angle=0]{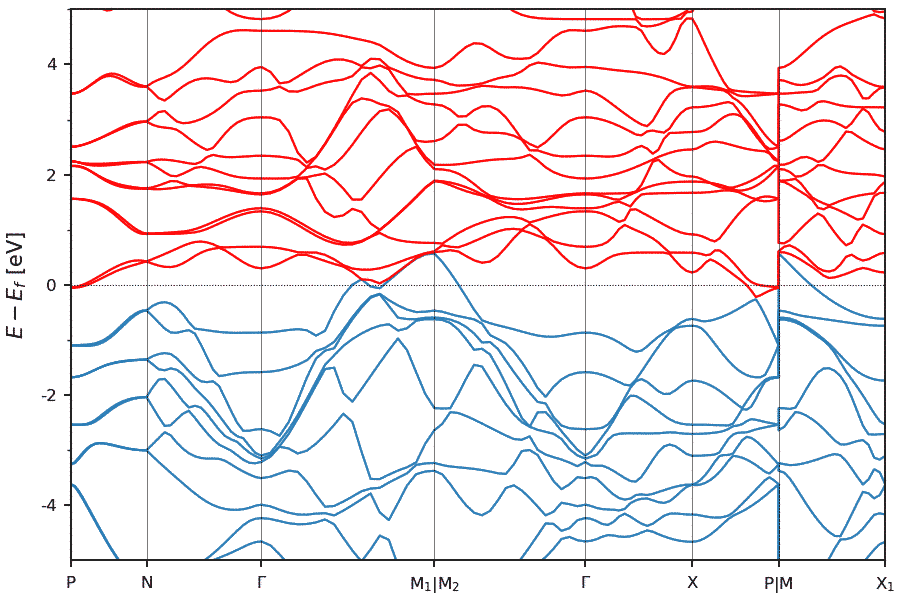} & \includegraphics[width=0.38\textwidth,angle=0]{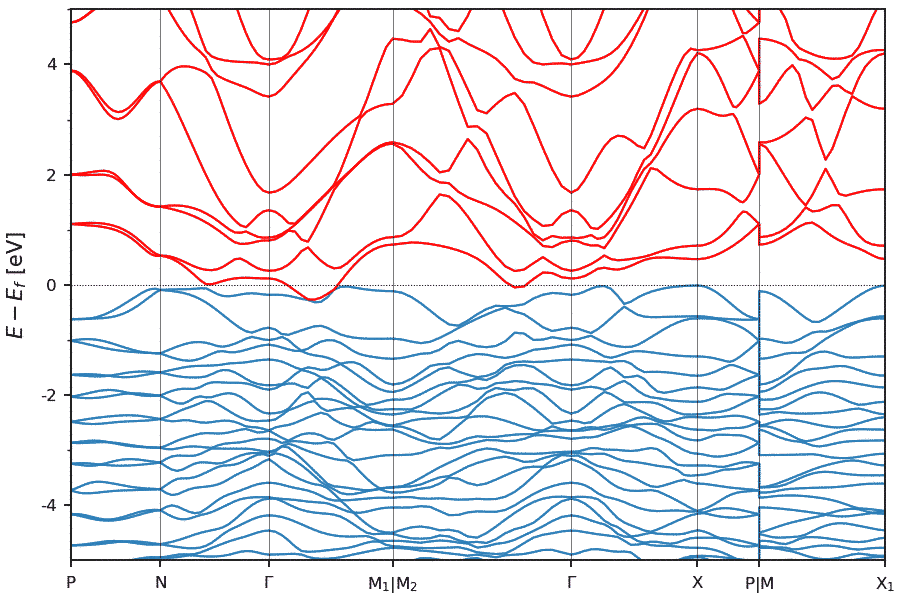}\\
\end{tabular}
\begin{tabular}{c c}
\scriptsize{$\rm{Ca} \rm{Pd}$ - \icsdweb{619498} - SG 221 ($Pm\bar{3}m$) - SEBR} & \scriptsize{$\rm{Sn} \rm{Se}$ - \icsdweb{52424} - SG 225 ($Fm\bar{3}m$) - SEBR}\\
\tiny{ $\;Z_{2,1}=0\;Z_{2,2}=0\;Z_{2,3}=0\;Z_4=0\;Z_{4m,\pi}=2\;Z_2=0\;Z_8=4$ } & \tiny{ $\;Z_{2,1}=0\;Z_{2,2}=0\;Z_{2,3}=0\;Z_4=0\;Z_2=0\;Z_8=4$ }\\
\includegraphics[width=0.38\textwidth,angle=0]{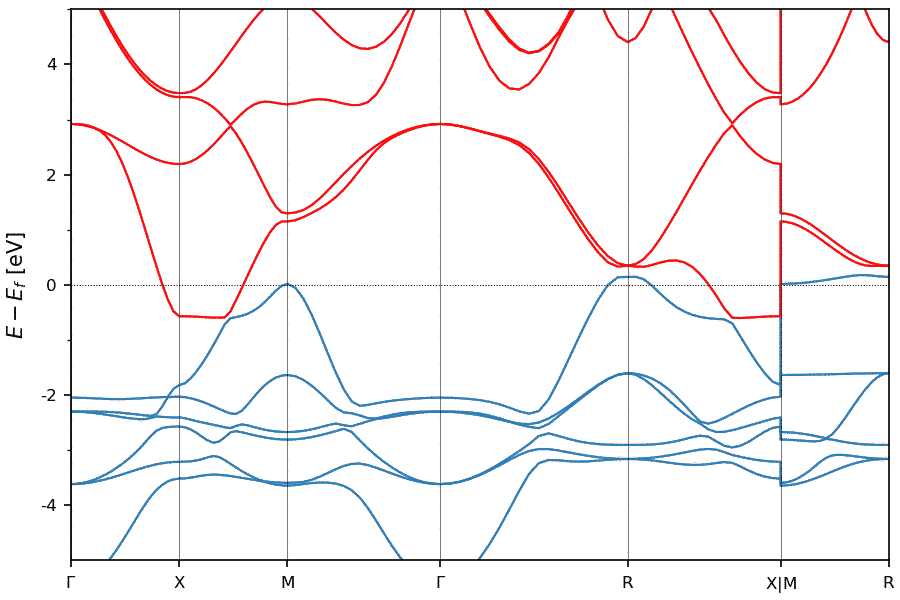} & \includegraphics[width=0.38\textwidth,angle=0]{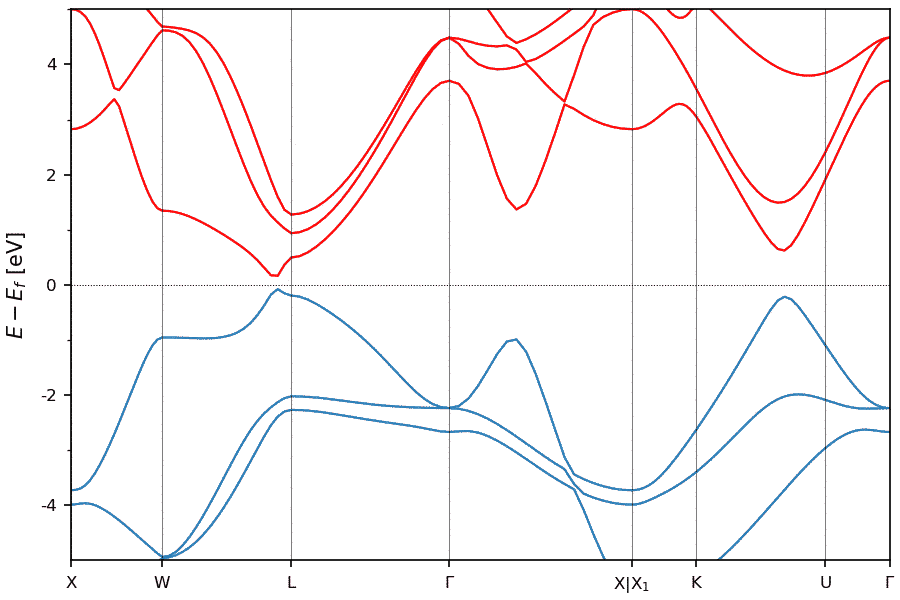}\\
\end{tabular}
\begin{tabular}{c c}
\scriptsize{$\rm{Ba}$ - \icsdweb{52679} - SG 225 ($Fm\bar{3}m$) - SEBR} & \scriptsize{$\rm{Pb} \rm{Po}$ - \icsdweb{105598} - SG 225 ($Fm\bar{3}m$) - SEBR}\\
\tiny{ $\;Z_{2,1}=0\;Z_{2,2}=0\;Z_{2,3}=0\;Z_4=0\;Z_2=0\;Z_8=4$ } & \tiny{ $\;Z_{2,1}=0\;Z_{2,2}=0\;Z_{2,3}=0\;Z_4=0\;Z_2=0\;Z_8=4$ }\\
\includegraphics[width=0.38\textwidth,angle=0]{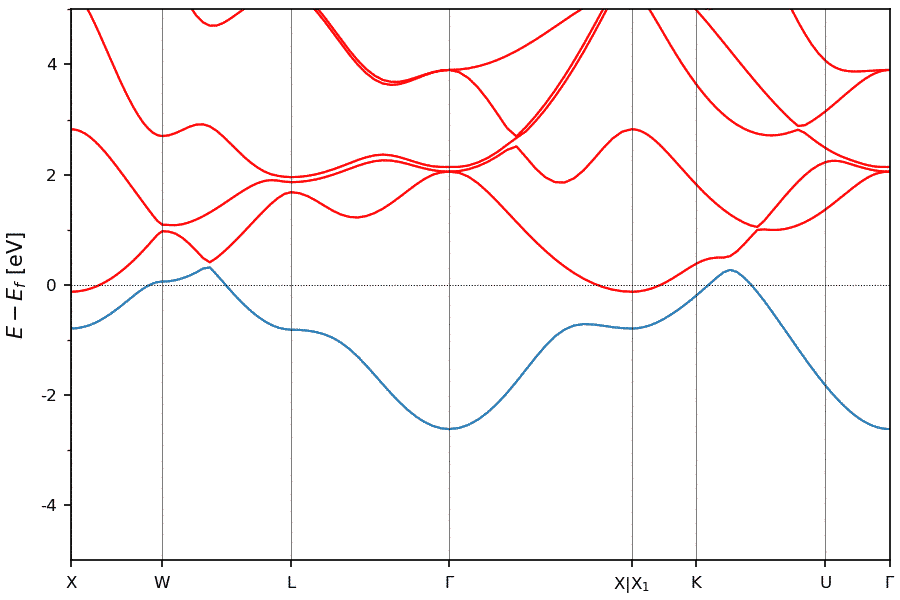} & \includegraphics[width=0.38\textwidth,angle=0]{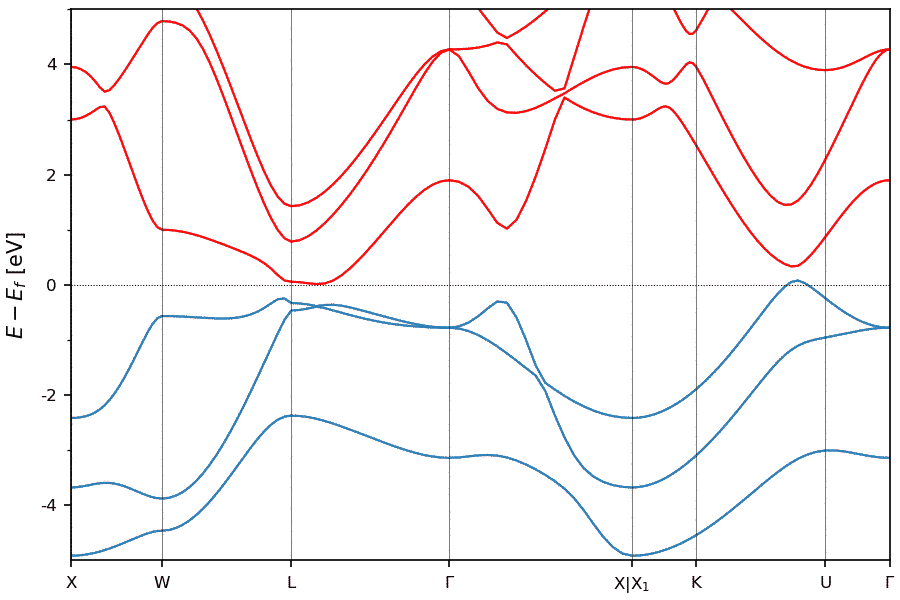}\\
\end{tabular}

\caption{\zeightHOTIsTCISEBR{2}}
\label{fig:z8_4HOTIs_TCI_SEBR2}
\end{figure}

\begin{figure}[ht]
\centering
\begin{tabular}{c c}
\scriptsize{$\rm{Mg}$ - \icsdweb{180453} - SG 225 ($Fm\bar{3}m$) - SEBR} & \scriptsize{$\rm{Bi} \rm{I}_{3}$ - \icsdweb{187609} - SG 225 ($Fm\bar{3}m$) - SEBR}\\
\tiny{ $\;Z_{2,1}=0\;Z_{2,2}=0\;Z_{2,3}=0\;Z_4=0\;Z_2=0\;Z_8=4$ } & \tiny{ $\;Z_{2,1}=0\;Z_{2,2}=0\;Z_{2,3}=0\;Z_4=0\;Z_2=0\;Z_8=4$ }\\
\includegraphics[width=0.38\textwidth,angle=0]{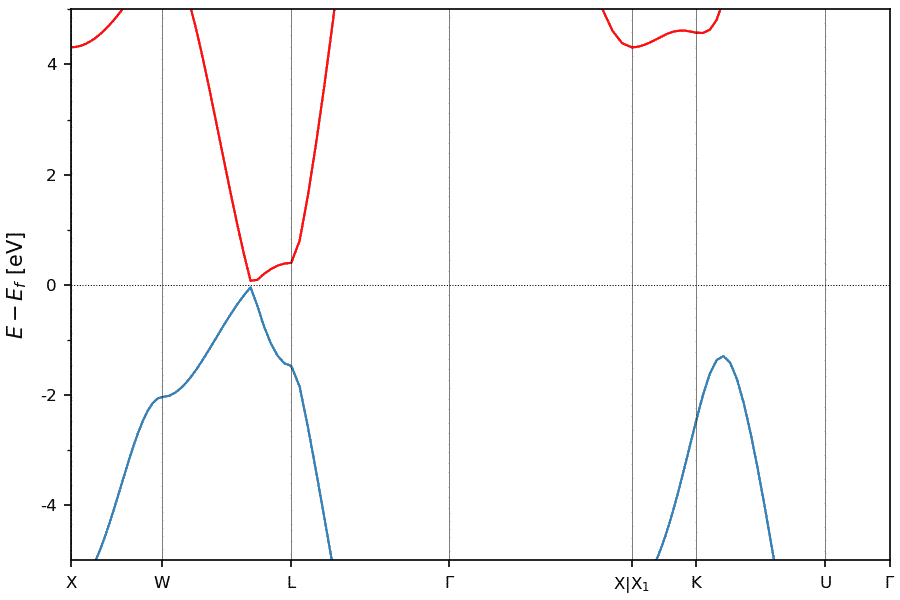} & \includegraphics[width=0.38\textwidth,angle=0]{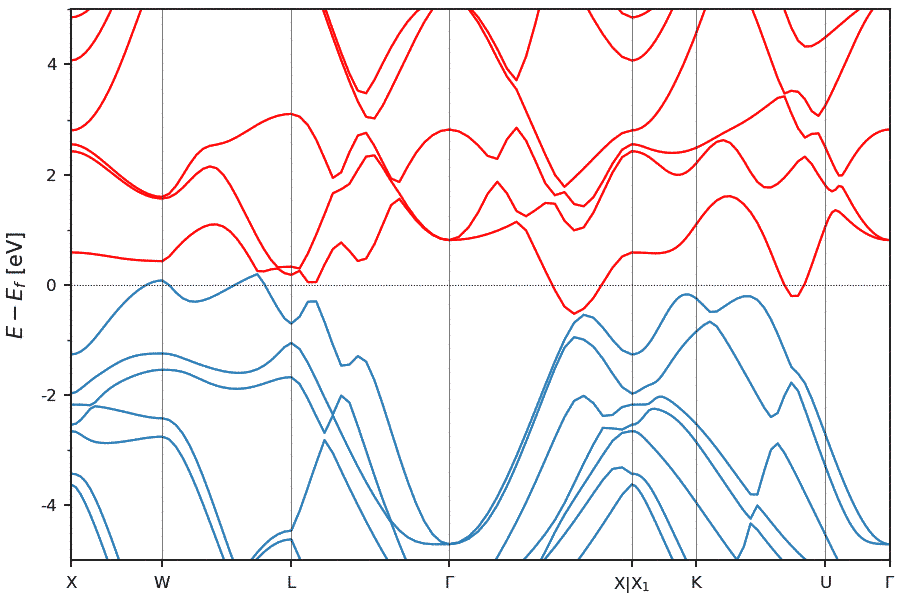}\\
\end{tabular}
\begin{tabular}{c c}
\scriptsize{$\rm{Ca}$ - \icsdweb{188390} - SG 225 ($Fm\bar{3}m$) - SEBR} & \scriptsize{$\rm{Pb}_{4} \rm{Se}_{4}$ - \icsdweb{238502} - SG 225 ($Fm\bar{3}m$) - SEBR}\\
\tiny{ $\;Z_{2,1}=0\;Z_{2,2}=0\;Z_{2,3}=0\;Z_4=0\;Z_2=0\;Z_8=4$ } & \tiny{ $\;Z_{2,1}=0\;Z_{2,2}=0\;Z_{2,3}=0\;Z_4=0\;Z_2=0\;Z_8=4$ }\\
\includegraphics[width=0.38\textwidth,angle=0]{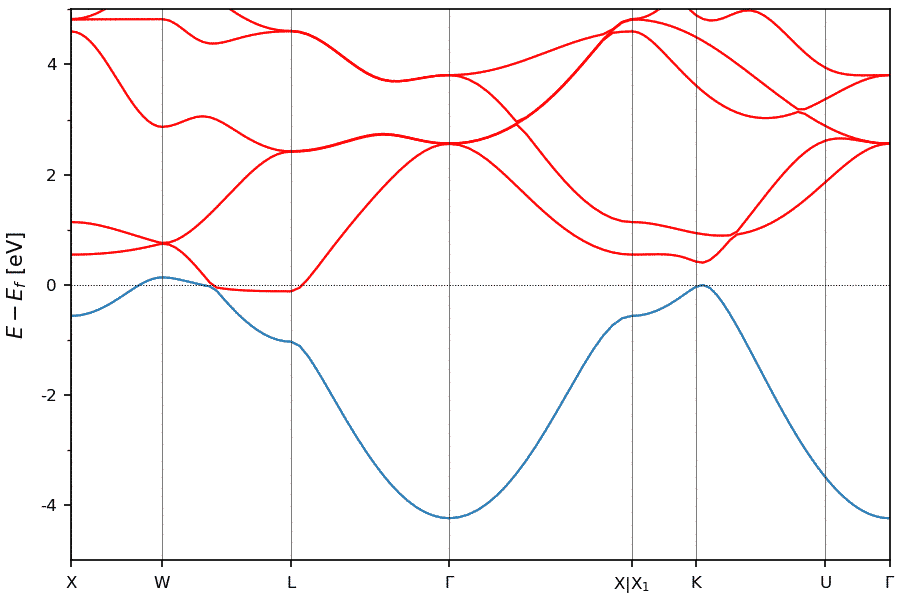} & \includegraphics[width=0.38\textwidth,angle=0]{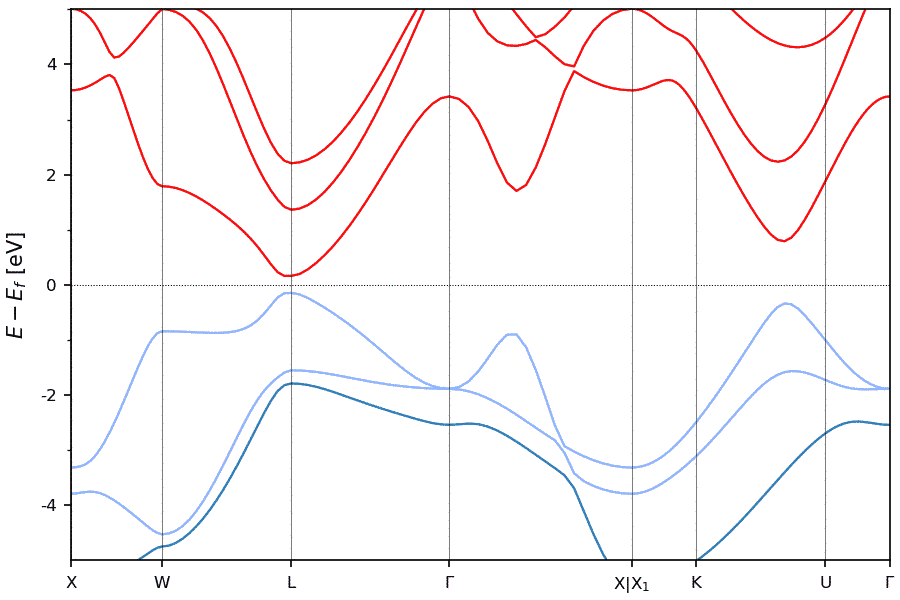}\\
\end{tabular}
\begin{tabular}{c c}
\scriptsize{$\rm{Pb} \rm{S}$ - \icsdweb{250762} - SG 225 ($Fm\bar{3}m$) - SEBR} & \scriptsize{$\rm{Pb} \rm{Se}$ - \icsdweb{291926} - SG 225 ($Fm\bar{3}m$) - SEBR}\\
\tiny{ $\;Z_{2,1}=0\;Z_{2,2}=0\;Z_{2,3}=0\;Z_4=0\;Z_2=0\;Z_8=4$ } & \tiny{ $\;Z_{2,1}=0\;Z_{2,2}=0\;Z_{2,3}=0\;Z_4=0\;Z_2=0\;Z_8=4$ }\\
\includegraphics[width=0.38\textwidth,angle=0]{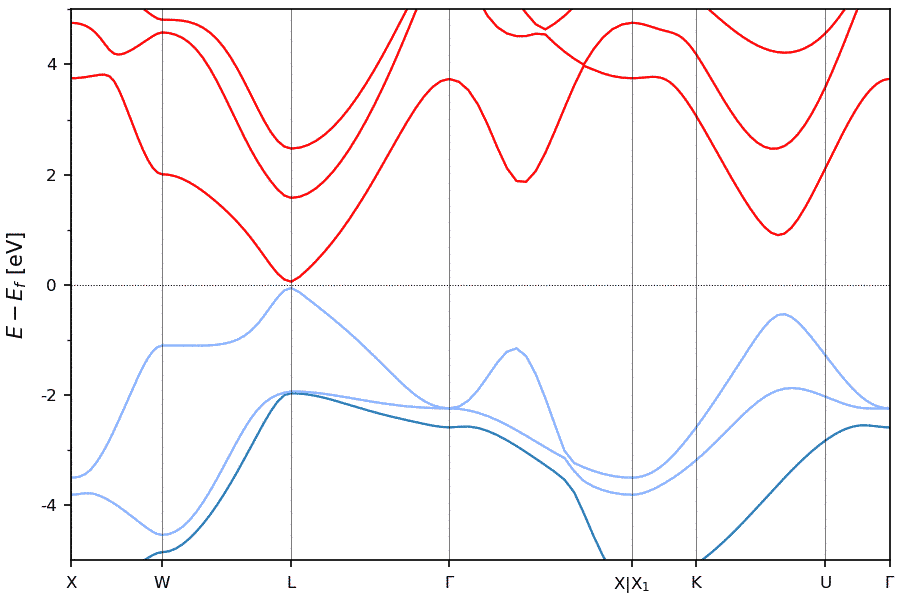} & \includegraphics[width=0.38\textwidth,angle=0]{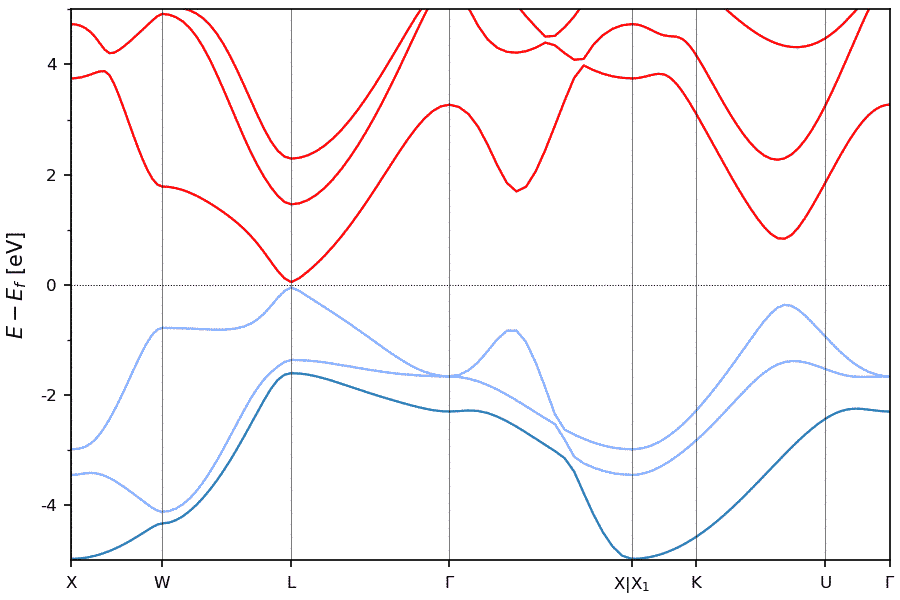}\\
\end{tabular}
\begin{tabular}{c c}
\scriptsize{$\rm{Pb} \rm{Te}$ - \icsdweb{648615} - SG 225 ($Fm\bar{3}m$) - SEBR} & \scriptsize{$\rm{Sn} \rm{S}$ - \icsdweb{651015} - SG 225 ($Fm\bar{3}m$) - SEBR}\\
\tiny{ $\;Z_{2,1}=0\;Z_{2,2}=0\;Z_{2,3}=0\;Z_4=0\;Z_2=0\;Z_8=4$ } & \tiny{ $\;Z_{2,1}=0\;Z_{2,2}=0\;Z_{2,3}=0\;Z_4=0\;Z_2=0\;Z_8=4$ }\\
\includegraphics[width=0.38\textwidth,angle=0]{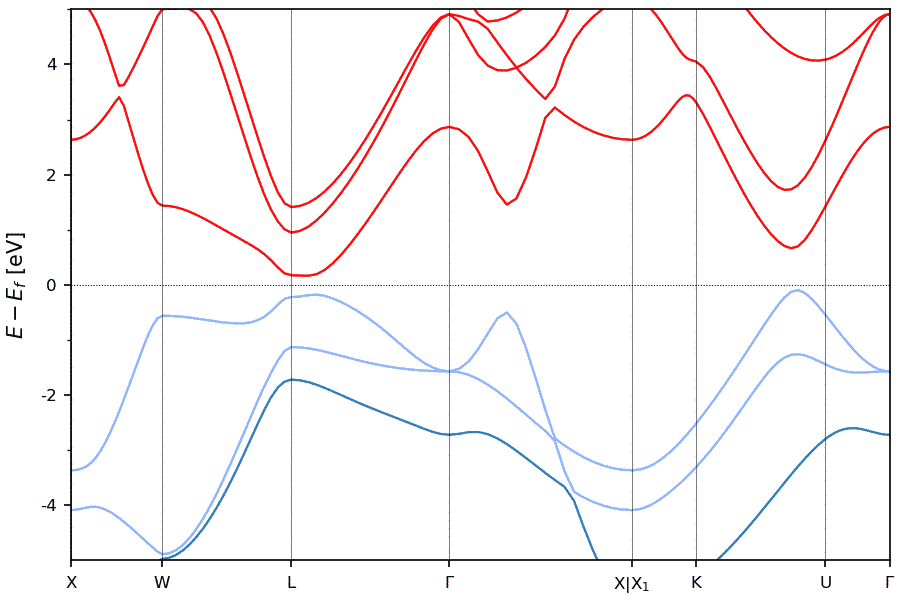} & \includegraphics[width=0.38\textwidth,angle=0]{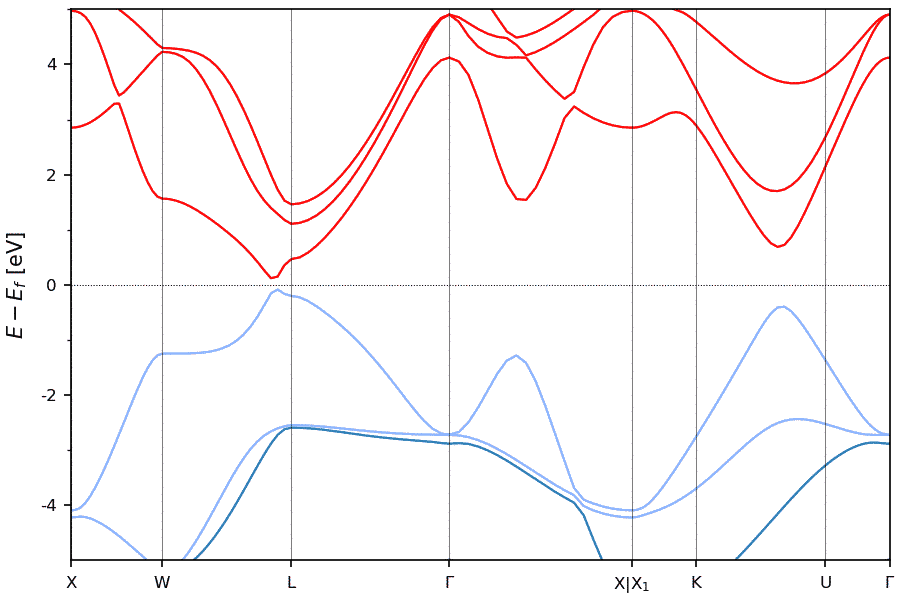}\\
\end{tabular}

\caption{\zeightHOTIsTCISEBR{3}}
\label{fig:z8_4HOTIs_TCI_SEBR3}
\end{figure}

\begin{figure}[ht]
\centering
\begin{tabular}{c c}
\scriptsize{$\rm{Sn} \rm{Te}$ - \icsdweb{652759} - SG 225 ($Fm\bar{3}m$) - SEBR} & \scriptsize{$\rm{Sr}$ - \icsdweb{652875} - SG 225 ($Fm\bar{3}m$) - SEBR}\\
\tiny{ $\;Z_{2,1}=0\;Z_{2,2}=0\;Z_{2,3}=0\;Z_4=0\;Z_2=0\;Z_8=4$ } & \tiny{ $\;Z_{2,1}=0\;Z_{2,2}=0\;Z_{2,3}=0\;Z_4=0\;Z_2=0\;Z_8=4$ }\\
\includegraphics[width=0.38\textwidth,angle=0]{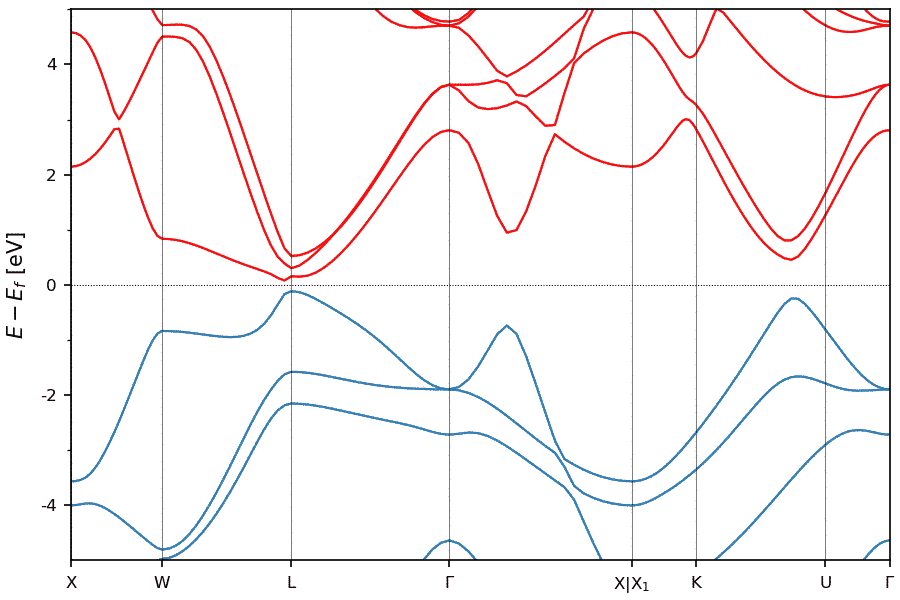} & \includegraphics[width=0.38\textwidth,angle=0]{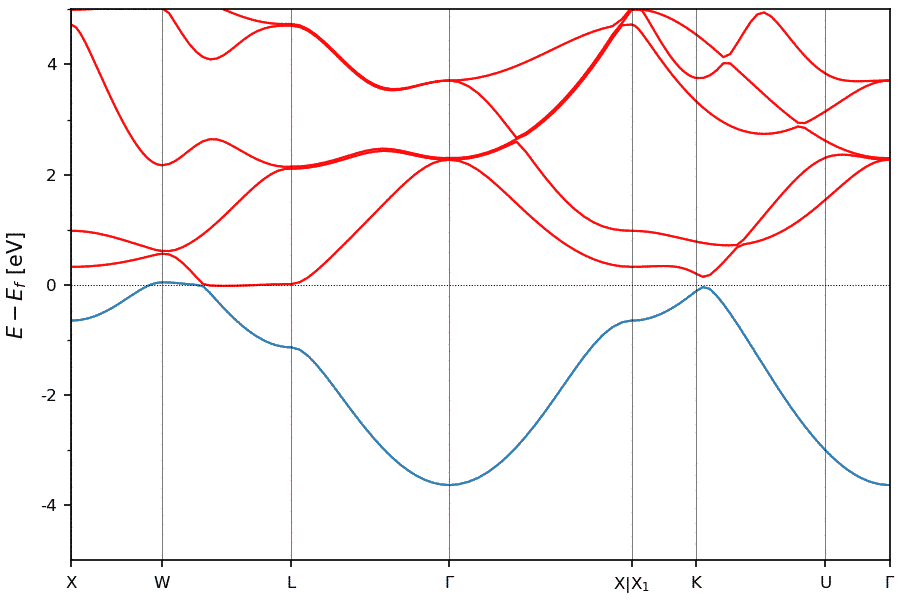}\\
\end{tabular}

\caption{\zeightHOTIsTCISEBR{4}}
\label{fig:z8_4HOTIs_TCI_SEBR4}
\end{figure}

\clearpage

\subsubsection{Other TCIs with Trivial $Z_{4}$ Indices}
\label{App:z4trivial}

In this section, we list the remaining TCIs with trivial strong $Z_{4}$ indices and nontrivial values for at least one other stable SI, and which have simple bulk Fermi surfaces or insulating gaps.  In terms of the stable SIs introduced in Ref.~\onlinecite{ChenTCI}, the materials listed in this section specifically exhibit $Z_{4}=0$ and nontrivial values for at least one other stable SI, excluding the cases of $Z_{8}=4$ or $Z_{12}'=6$, which were previously addressed in~\supappref{App:rotationAnomaly}.   The TCIs listed in this section exhibit differing numbers of twofold surface Dirac cones, whose presence is indicated through a combination of weak indices and mirror and rotation-anomaly TCI indices.

Notably, a subset of the TCIs listed in this section exhibit completely trivial strong indices (\emph{i.e.} $Z_{4}=0$ without any other nontrivial strong indices).  As shown in Refs.~\cite{ChenTCI,WiederBarryCDW,CDWWeyl,JiabinCDW}, materials with $Z_{4}=0$ and nontrivial weak indices realize ``obstructed'' weak-TI phases that differ from weak-TI phases by fractional lattice translation.  Specifically, when the weak indices $Z_{2,i}$ are nontrivial, the value of the strong index $Z_{4}$ can be advanced by $2$ ($\text{mod }4$) by shifting the origin of the unit cell by half of a lattice translation along the direction of the weak-index vector.  In an obstructed weak-TI phase, the bulk is topologically equivalent to a stack of 2D TIs oriented along the weak-index vector, but crucially a stack in which the 2D TI layers in each stack unit cell are displaced from the origin by half of a lattice translation along the stacking direction, analogous to the obstructed-atomic-limit phase of the Su-Schrieffer-Heeger model of polyacetylene~\cite{SSH,SSHExp,WiederBarryCDW,JiabinCDW,JenOAL,JeanNoelSSH}.  Like weak TIs, obstructed weak TIs exhibit twofold Dirac-cone surface states and topological defect bound states~\cite{AshvinScrewTI,RaquelPartial,WiederBarryCDW,JiabinCDW}.

Below, in Fig.~\ref{fig:z4_0_TCI_weak_NLC}, we first list the $Z_{4}=0$ TCIs that are classified as NLC, and then, in Figs.~\ref{fig:z4_0_TCI_weak_SEBR1},~\ref{fig:z4_0_TCI_weak_SEBR2},~\ref{fig:z4_0_TCI_weak_SEBR3}, and~\ref{fig:z4_0_TCI_weak_SEBR4}, we list the $Z_{4}=0$ TCIs that are classified as SEBR.  In addition to obstructed weak-TI phases, the materials listed in this section include members of the KHgSb family [\icsdweb{56201}, SG 194 ($P6_{3}/mmc$)], which have been shown theoretically and experimentally~\cite{HourglassInsulator,Cohomological,DiracInsulator,HourglassExperiment,zeroHallExp} to exhibit a combination of symmetry-indicated mirror-TCI surface states and non-symmetry-indicated glide-protected surface states.  In particular, KHgSb itself is classified as SEBR, and thus appears in Fig.~\ref{fig:z4_0_TCI_weak_SEBR3} under the alternative chemical formula HgKSb.


\begin{figure}[ht]
\centering
\begin{tabular}{c c}
\scriptsize{$\rm{Pb} \rm{Pt}_{2} \rm{O}_{4}$ - \icsdweb{82204} - SG 2 ($P\bar{1}$) - NLC} & \scriptsize{$\rm{Cs} \rm{Hg}$ - \icsdweb{150974} - SG 2 ($P\bar{1}$) - NLC}\\
\tiny{ $\;Z_{2,1}=0\;Z_{2,2}=1\;Z_{2,3}=1\;Z_4=0$ } & \tiny{ $\;Z_{2,1}=0\;Z_{2,2}=0\;Z_{2,3}=1\;Z_4=0$ }\\
\includegraphics[width=0.38\textwidth,angle=0]{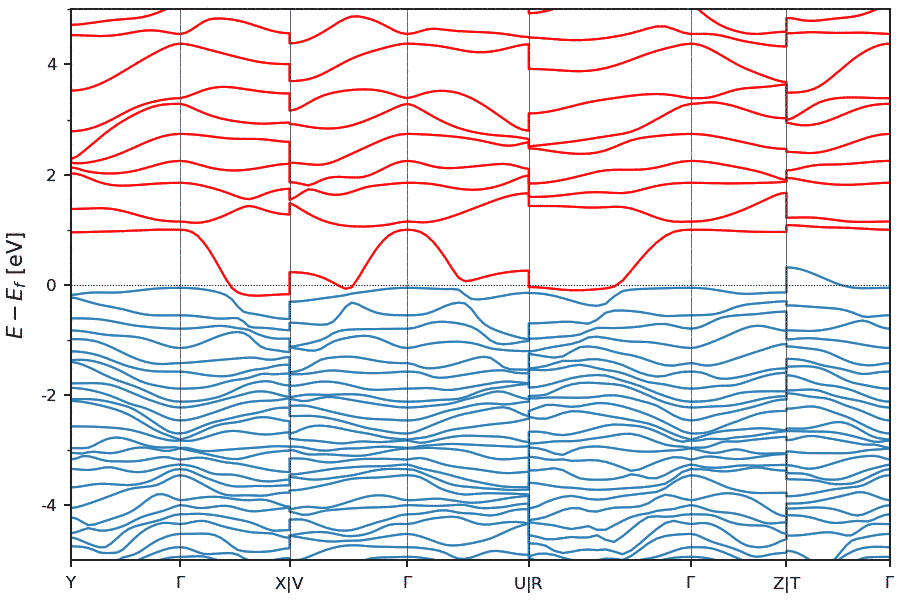} & \includegraphics[width=0.38\textwidth,angle=0]{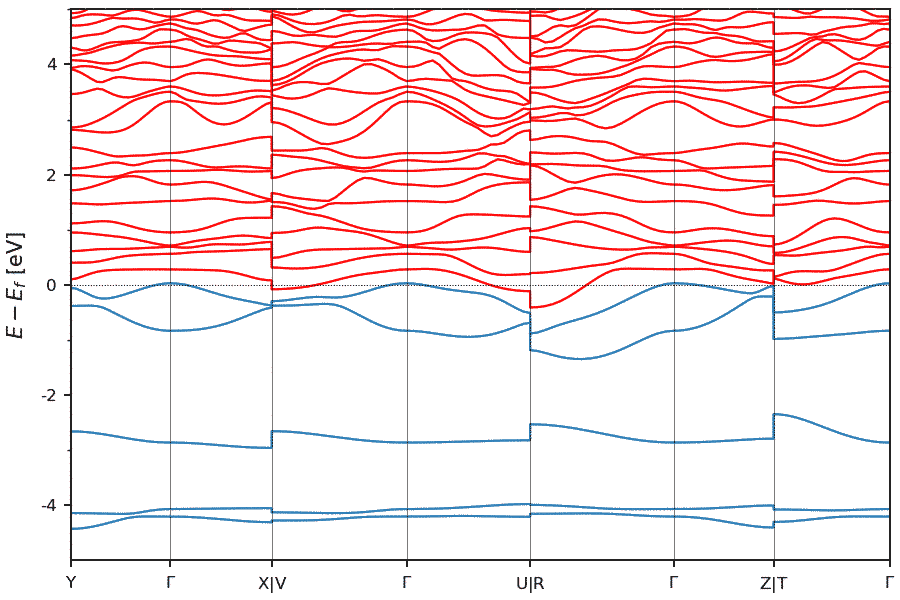}\\
\end{tabular}
\begin{tabular}{c c}
\scriptsize{$\rm{Rh}_{3} \rm{Ga}_{5}$ - \icsdweb{240179} - SG 2 ($P\bar{1}$) - NLC} & \scriptsize{$\rm{Ba} \rm{Sb}_{2}$ - \icsdweb{409517} - SG 11 ($P2_1/m$) - NLC}\\
\tiny{ $\;Z_{2,1}=1\;Z_{2,2}=0\;Z_{2,3}=0\;Z_4=0$ } & \tiny{ $\;Z_{2,1}=1\;Z_{2,2}=0\;Z_{2,3}=0\;Z_4=0$ }\\
\includegraphics[width=0.38\textwidth,angle=0]{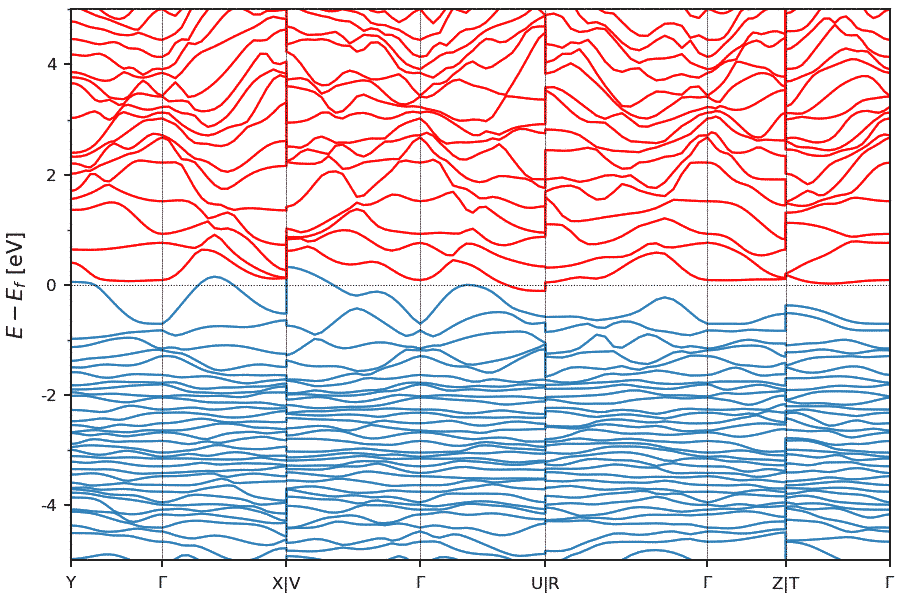} & \includegraphics[width=0.38\textwidth,angle=0]{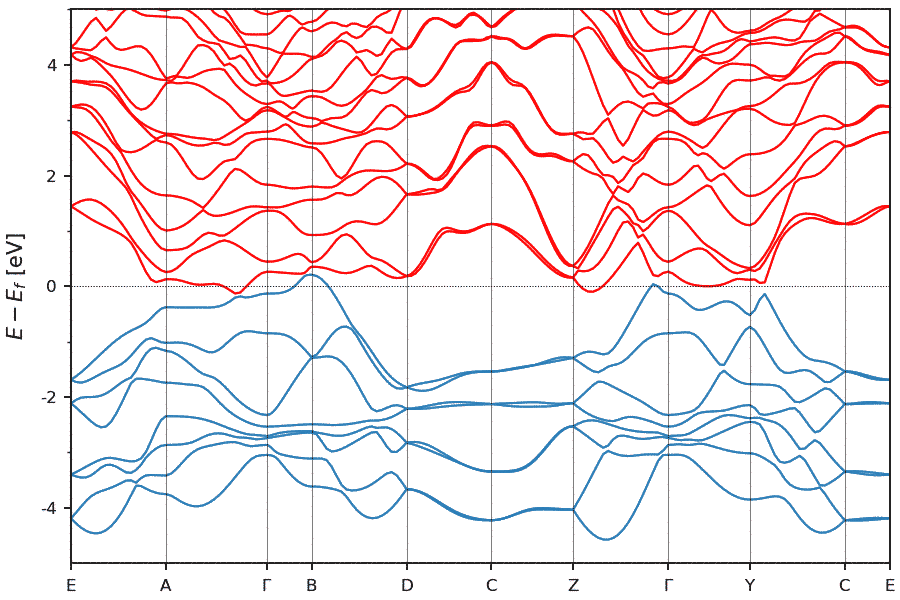}\\
\end{tabular}
\begin{tabular}{c c}
\scriptsize{$\rm{Pt}_{5} \rm{P}_{2}$ - \icsdweb{24327} - SG 15 ($C2/c$) - NLC} & \scriptsize{$\rm{Ba} \rm{Hg}_{2}$ - \icsdweb{58655} - SG 74 ($Imma$) - NLC}\\
\tiny{ $\;Z_{2,1}=1\;Z_{2,2}=1\;Z_{2,3}=0\;Z_4=0$ } & \tiny{ $\;Z_{2,1}=1\;Z_{2,2}=1\;Z_{2,3}=1\;Z_4=0$ }\\
\includegraphics[width=0.38\textwidth,angle=0]{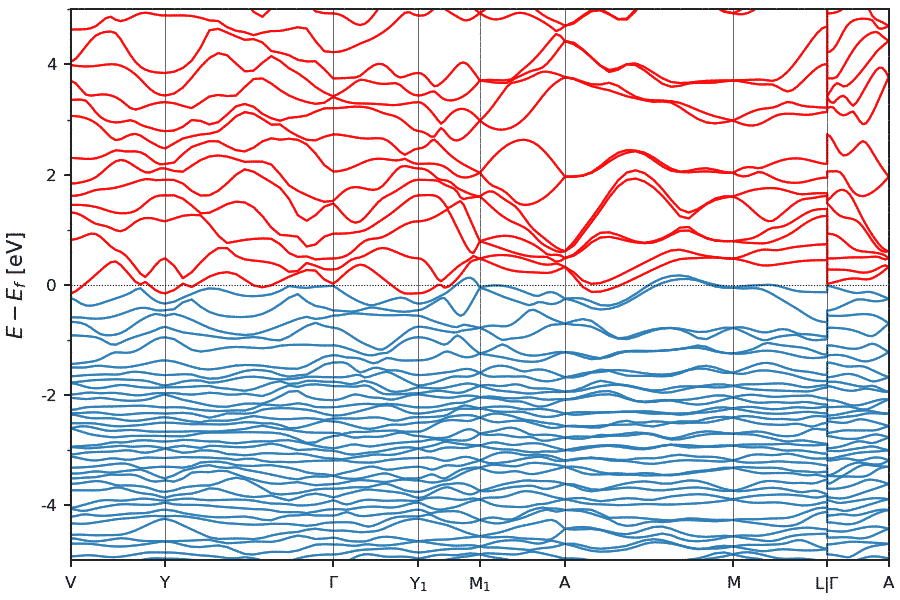} & \includegraphics[width=0.38\textwidth,angle=0]{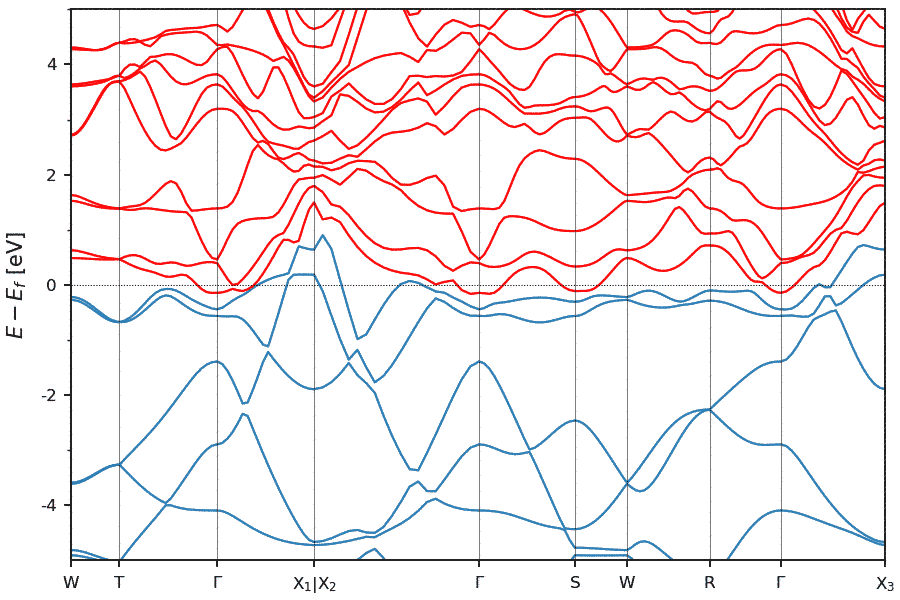}\\
\end{tabular}
\begin{tabular}{c c}
\scriptsize{$\rm{Hg}_{2} \rm{Sr}$ - \icsdweb{104347} - SG 74 ($Imma$) - NLC}\\
\tiny{ $\;Z_{2,1}=1\;Z_{2,2}=1\;Z_{2,3}=1\;Z_4=0$ }\\
\includegraphics[width=0.38\textwidth,angle=0]{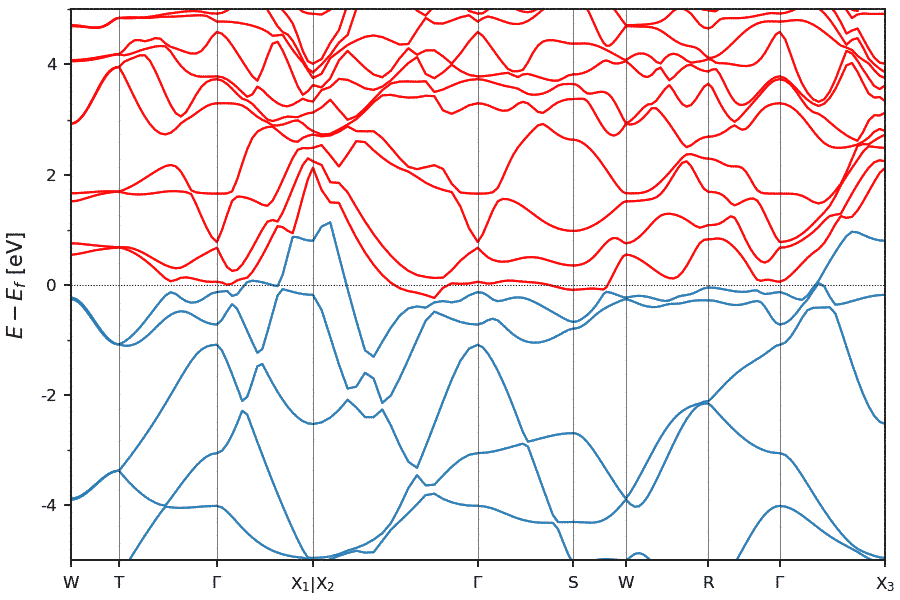}\\
\end{tabular}

\caption{The NLC-classified TCIs with the largest band gaps or the fewest and smallest bulk Fermi pockets whose topology is characterized by $Z_{4}=0$ and undefined or trivial values of $Z_{8}$.}
\label{fig:z4_0_TCI_weak_NLC}
\end{figure}


\begin{figure}[ht]
\centering
\begin{tabular}{c c}
\scriptsize{$\rm{Nb} \rm{Sb}_{2}$ - \icsdweb{18144} - SG 12 ($C2/m$) - SEBR} & \scriptsize{$\rm{V} \rm{P}_{2}$ - \icsdweb{42077} - SG 12 ($C2/m$) - SEBR}\\
\tiny{ $\;Z_{2,1}=1\;Z_{2,2}=1\;Z_{2,3}=1\;Z_4=0$ } & \tiny{ $\;Z_{2,1}=1\;Z_{2,2}=1\;Z_{2,3}=1\;Z_4=0$ }\\
\includegraphics[width=0.38\textwidth,angle=0]{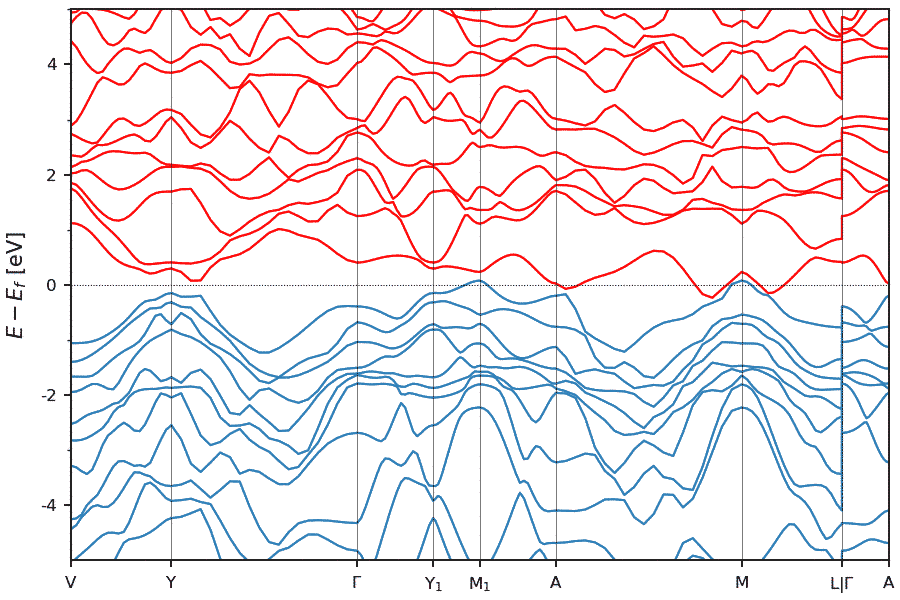} & \includegraphics[width=0.38\textwidth,angle=0]{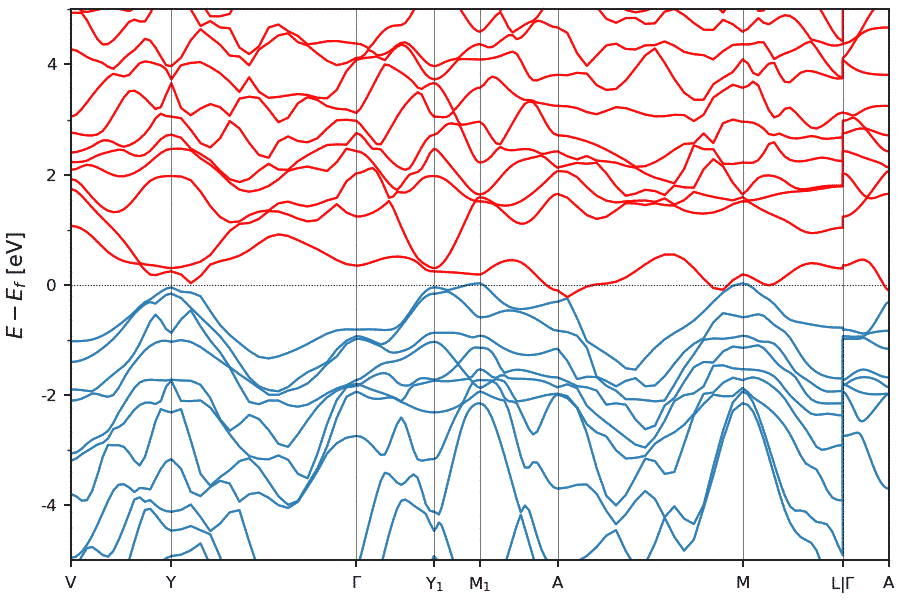}\\
\end{tabular}
\begin{tabular}{c c}
\scriptsize{$\rm{Ni} \rm{Si}_{2} \rm{Se}_{4}$ - \icsdweb{646533} - SG 12 ($C2/m$) - SEBR} & \scriptsize{$\rm{W}_{3} \rm{Co} \rm{B}_{3}$ - \icsdweb{25753} - SG 63 ($Cmcm$) - SEBR}\\
\tiny{ $\;Z_{2,1}=0\;Z_{2,2}=0\;Z_{2,3}=1\;Z_4=0$ } & \tiny{ $\;Z_{2,1}=1\;Z_{2,2}=1\;Z_{2,3}=0\;Z_4=0$ }\\
\includegraphics[width=0.38\textwidth,angle=0]{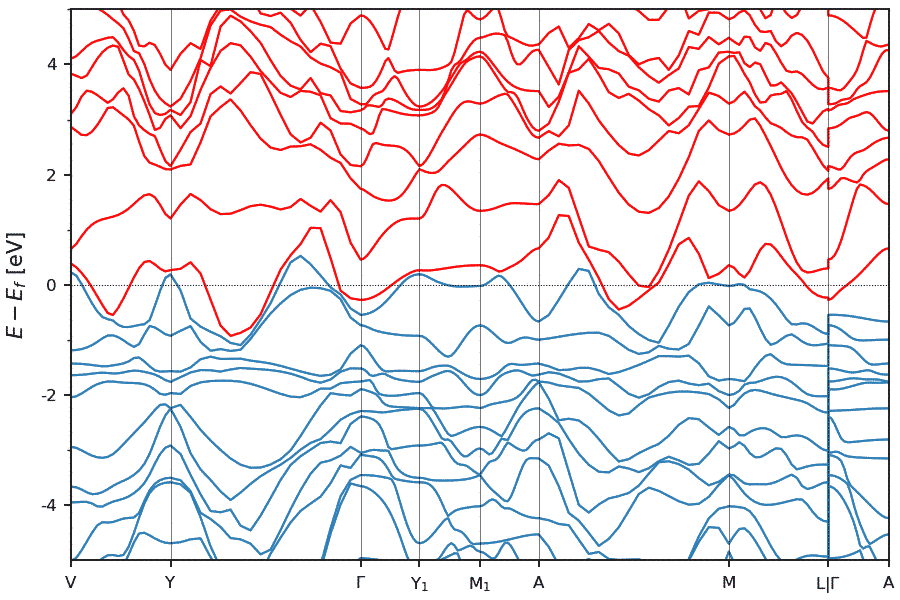} & \includegraphics[width=0.38\textwidth,angle=0]{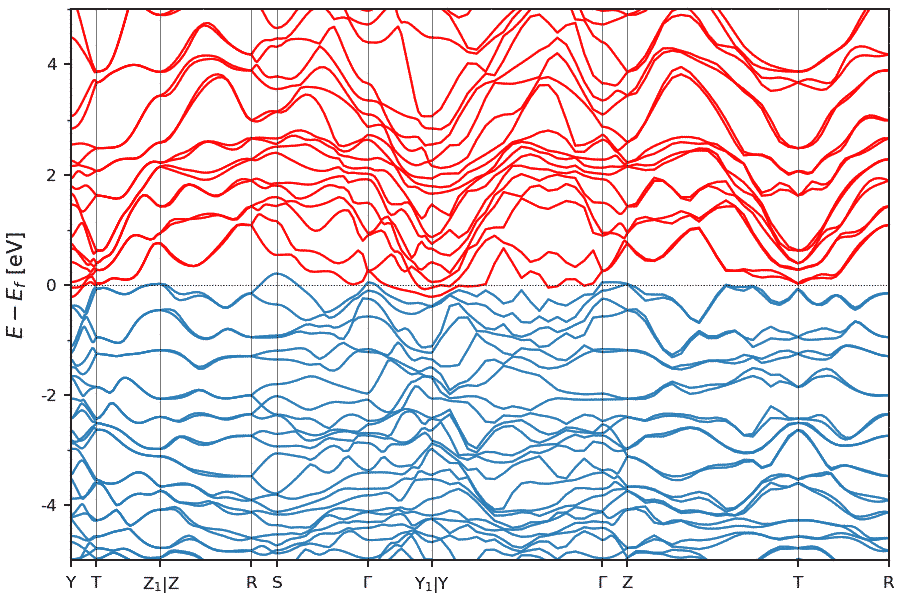}\\
\end{tabular}
\begin{tabular}{c c}
\scriptsize{$\rm{Mg} \rm{Os}_{3} \rm{B}_{4}$ - \icsdweb{91243} - SG 63 ($Cmcm$) - SEBR} & \scriptsize{$\rm{Pd}_{3} \rm{Sn} \rm{C}$ - \icsdweb{618636} - SG 123 ($P4/mmm$) - SEBR}\\
\tiny{ $\;Z_{2,1}=1\;Z_{2,2}=1\;Z_{2,3}=0\;Z_4=0$ } & \tiny{ $\;Z_{2,1}=1\;Z_{2,2}=1\;Z_{2,3}=1\;Z_4=0\;Z_{4m,\pi}=3\;Z_2=0\;Z_8=0$ }\\
\includegraphics[width=0.38\textwidth,angle=0]{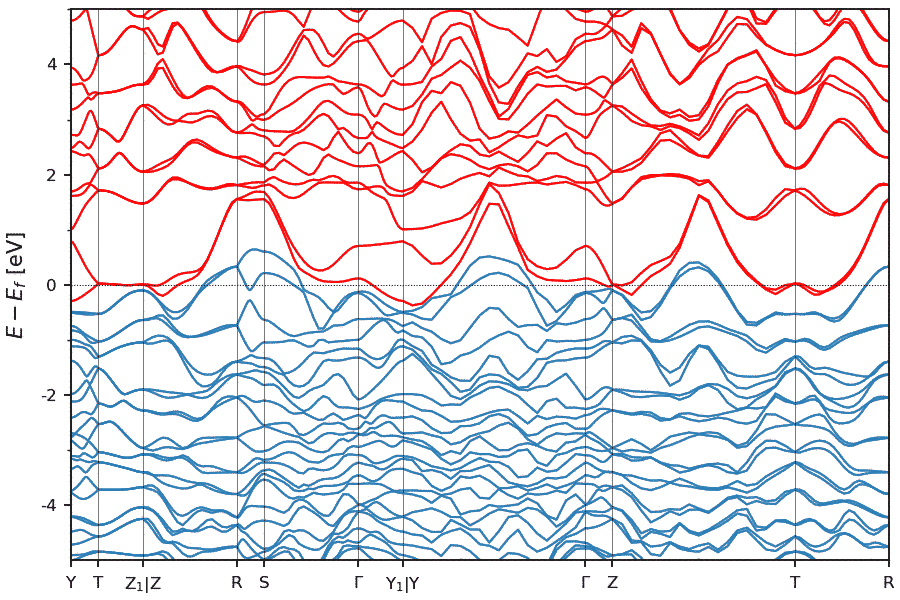} & \includegraphics[width=0.38\textwidth,angle=0]{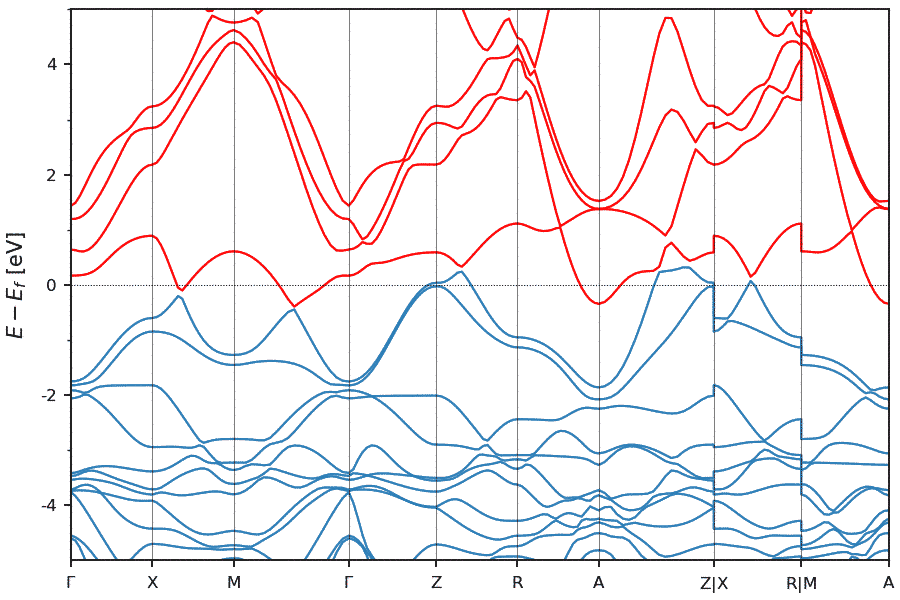}\\
\end{tabular}
\begin{tabular}{c c}
\scriptsize{$\rm{Zr} \rm{Si} \rm{Te}$ - \icsdweb{74522} - SG 129 ($P4/nmm$) - SEBR} & \scriptsize{$\rm{Bi}$ - \icsdweb{51674} - SG 140 ($I4/mcm$) - SEBR}\\
\tiny{ $\;Z_{2,1}=0\;Z_{2,2}=0\;Z_{2,3}=1\;Z_4=0\;Z_2=0$ } & \tiny{ $\;Z_{2,1}=1\;Z_{2,2}=1\;Z_{2,3}=1\;Z_4=0\;Z_2=0\;Z_8=0$ }\\
\includegraphics[width=0.38\textwidth,angle=0]{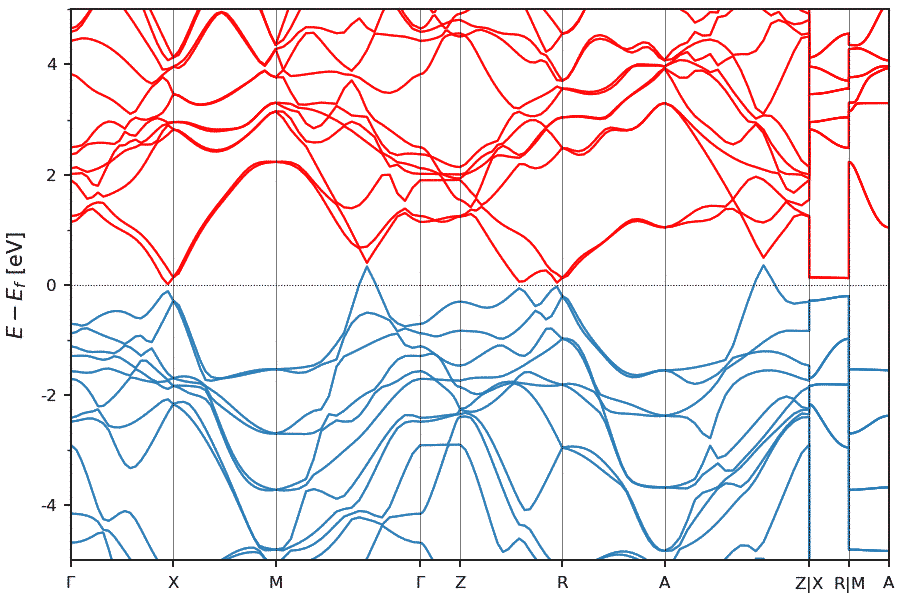} & \includegraphics[width=0.38\textwidth,angle=0]{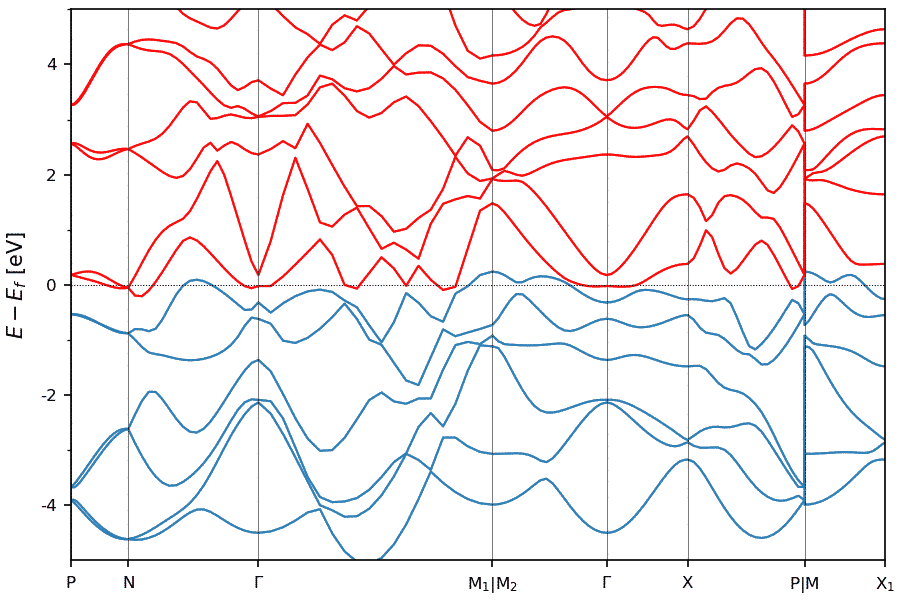}\\
\end{tabular}

\caption{\zfourTCIweakSEBR{1}}
\label{fig:z4_0_TCI_weak_SEBR1}
\end{figure}

\begin{figure}[ht]
\centering
\begin{tabular}{c c}
\scriptsize{$\rm{Ba} \rm{Si}_{2}$ - \icsdweb{1237} - SG 164 ($P\bar{3}m1$) - SEBR} & \scriptsize{$\rm{Bi}_{2} \rm{Pb}_{2} \rm{Te}_{5}$ - \icsdweb{42708} - SG 164 ($P\bar{3}m1$) - SEBR}\\
\tiny{ $\;Z_{2,1}=0\;Z_{2,2}=0\;Z_{2,3}=1\;Z_4=0$ } & \tiny{ $\;Z_{2,1}=0\;Z_{2,2}=0\;Z_{2,3}=1\;Z_4=0$ }\\
\includegraphics[width=0.38\textwidth,angle=0]{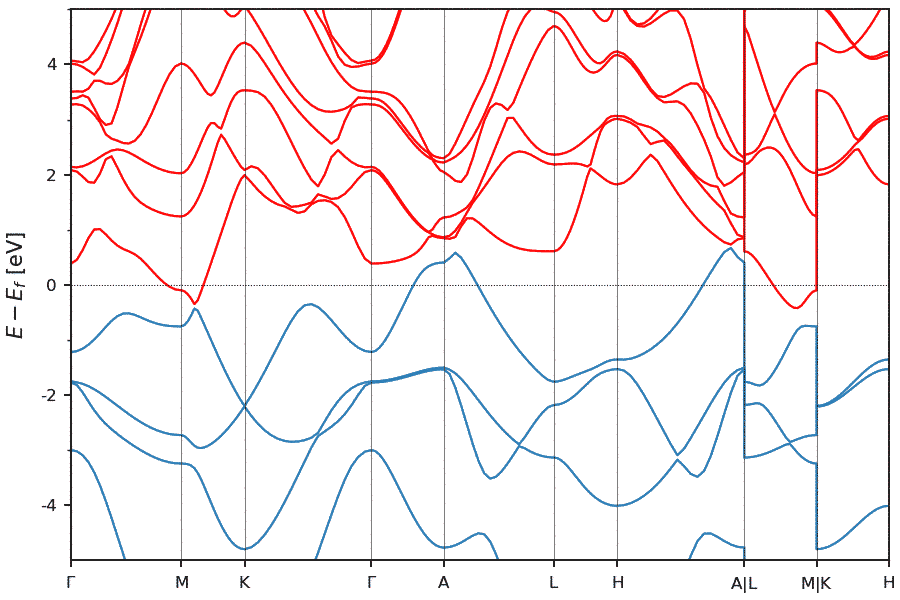} & \includegraphics[width=0.38\textwidth,angle=0]{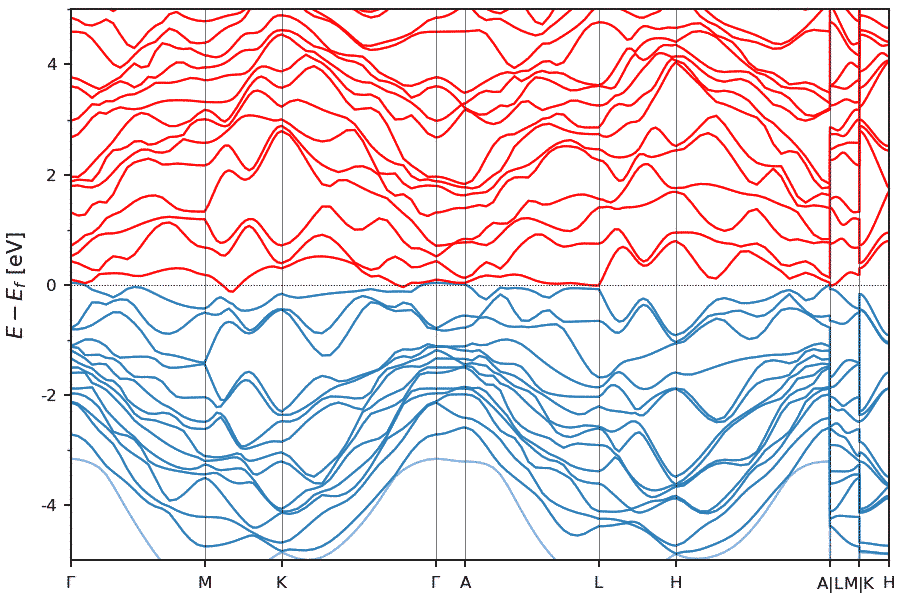}\\
\end{tabular}
\begin{tabular}{c c}
\scriptsize{$\rm{Bi}_{3} \rm{Te}_{2} \rm{S}$ - \icsdweb{107587} - SG 164 ($P\bar{3}m1$) - SEBR} & \scriptsize{$\rm{Bi} \rm{Se}$ - \icsdweb{617073} - SG 164 ($P\bar{3}m1$) - SEBR}\\
\tiny{ $\;Z_{2,1}=0\;Z_{2,2}=0\;Z_{2,3}=1\;Z_4=0$ } & \tiny{ $\;Z_{2,1}=0\;Z_{2,2}=0\;Z_{2,3}=1\;Z_4=0$ }\\
\includegraphics[width=0.38\textwidth,angle=0]{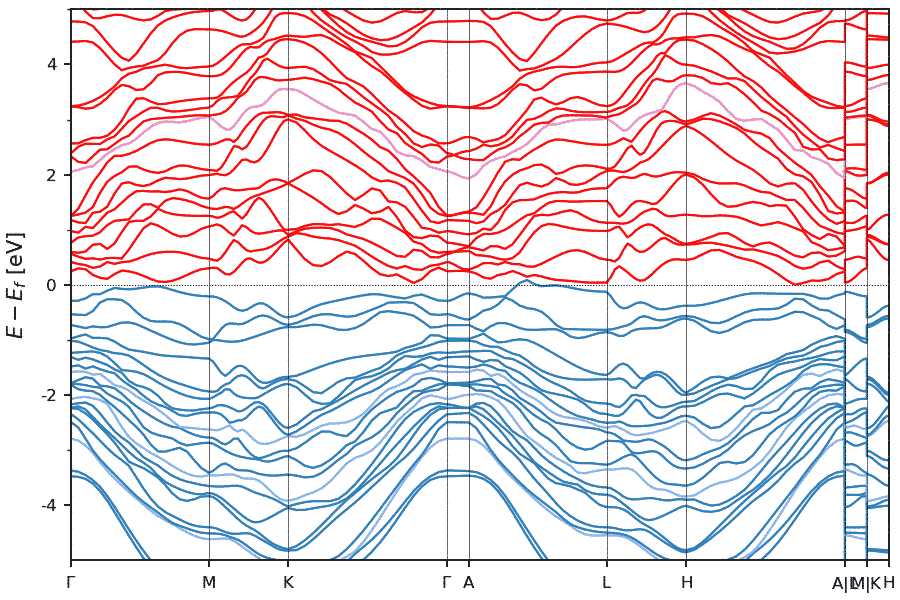} & \includegraphics[width=0.38\textwidth,angle=0]{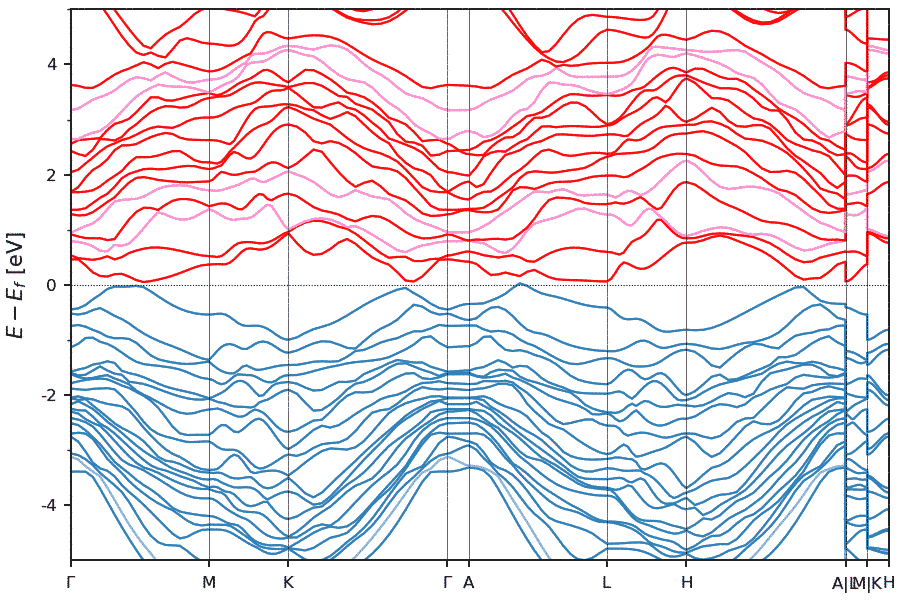}\\
\end{tabular}
\begin{tabular}{c c}
\scriptsize{$\rm{Bi} \rm{Te}$ - \icsdweb{617181} - SG 164 ($P\bar{3}m1$) - SEBR} & \scriptsize{$\rm{Zr} \rm{Cl}$ - \icsdweb{869} - SG 166 ($R\bar{3}m$) - SEBR}\\
\tiny{ $\;Z_{2,1}=0\;Z_{2,2}=0\;Z_{2,3}=1\;Z_4=0$ } & \tiny{ $\;Z_{2,1}=1\;Z_{2,2}=1\;Z_{2,3}=1\;Z_4=0$ }\\
\includegraphics[width=0.38\textwidth,angle=0]{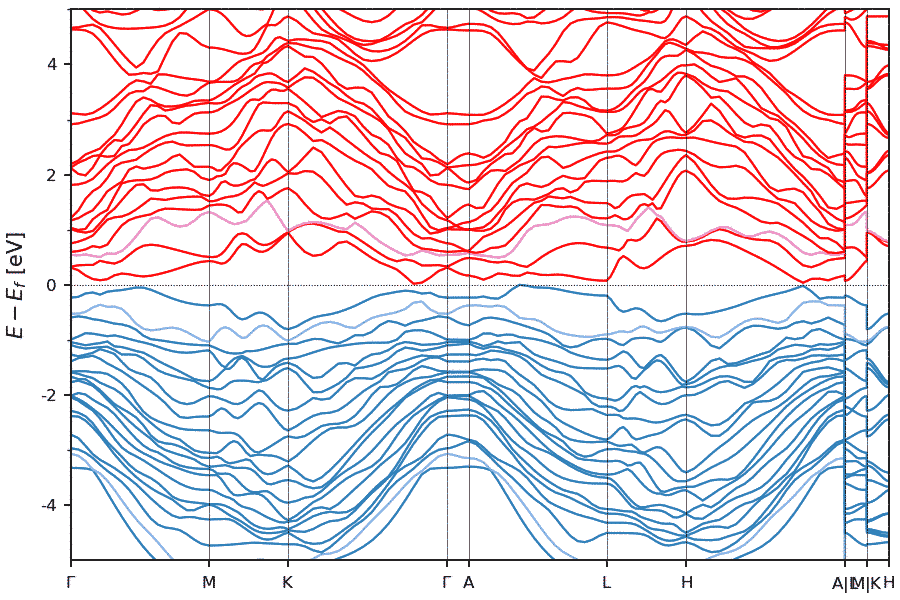} & \includegraphics[width=0.38\textwidth,angle=0]{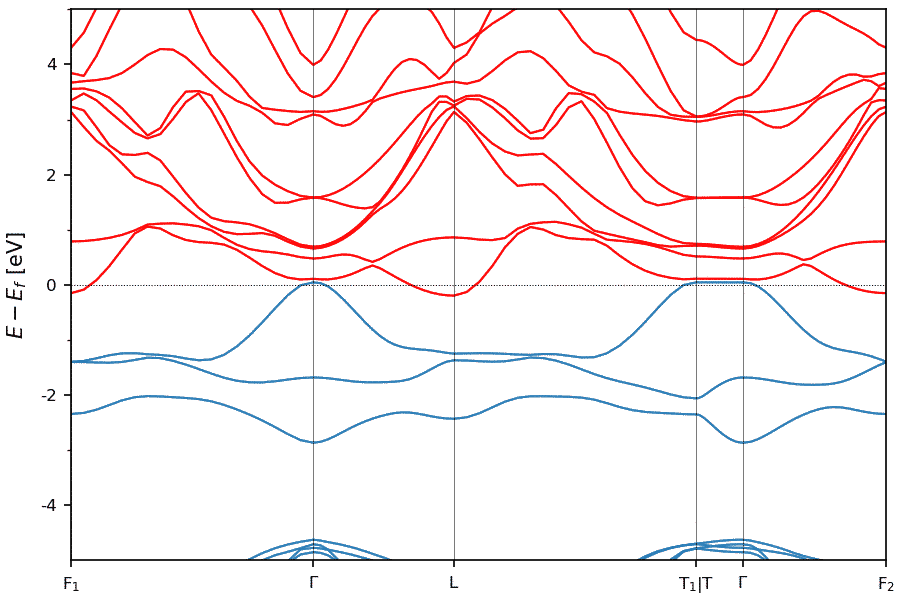}\\
\end{tabular}
\begin{tabular}{c c}
\scriptsize{$\rm{Zr} \rm{Br}$ - \icsdweb{1168} - SG 166 ($R\bar{3}m$) - SEBR} & \scriptsize{$\rm{Bi}_{8} \rm{Se}_{9}$ - \icsdweb{42665} - SG 166 ($R\bar{3}m$) - SEBR}\\
\tiny{ $\;Z_{2,1}=1\;Z_{2,2}=1\;Z_{2,3}=1\;Z_4=0$ } & \tiny{ $\;Z_{2,1}=1\;Z_{2,2}=1\;Z_{2,3}=1\;Z_4=0$ }\\
\includegraphics[width=0.38\textwidth,angle=0]{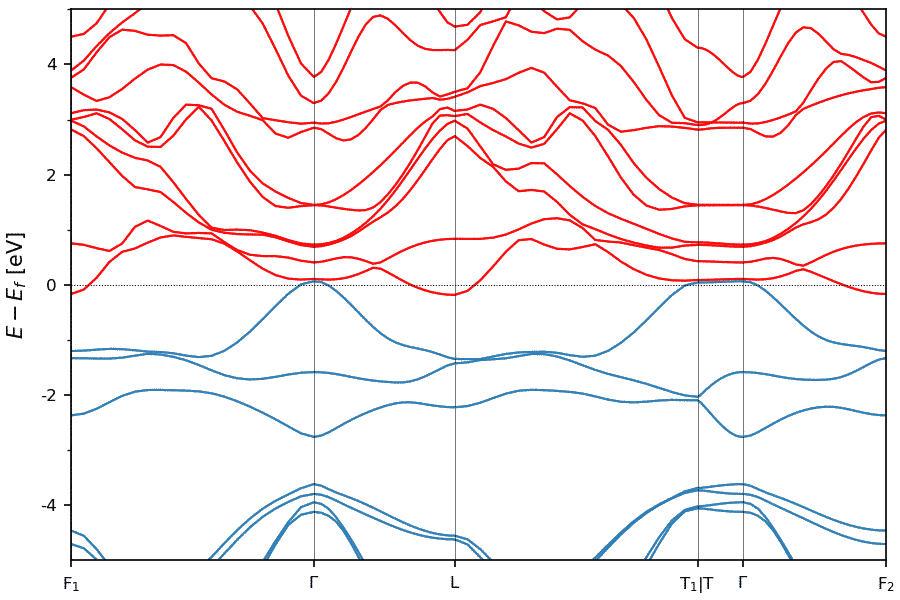} & \includegraphics[width=0.38\textwidth,angle=0]{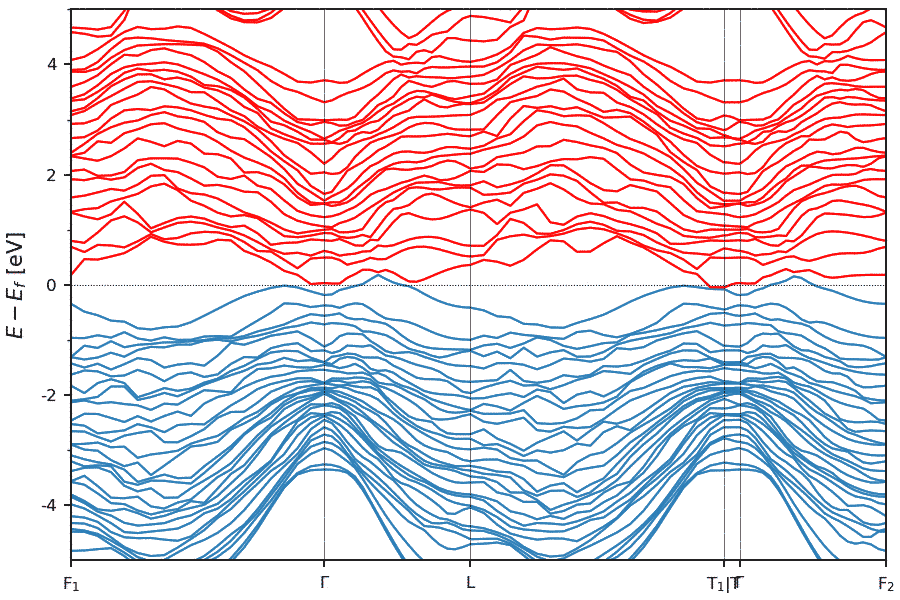}\\
\end{tabular}

\caption{\zfourTCIweakSEBR{2}}
\label{fig:z4_0_TCI_weak_SEBR2}
\end{figure}

\begin{figure}[ht]
\centering
\begin{tabular}{c c}
\scriptsize{$\rm{Bi}_{4} \rm{Se}_{3}$ - \icsdweb{617074} - SG 166 ($R\bar{3}m$) - SEBR} & \scriptsize{$\rm{Ni}_{3} \rm{In}_{2} \rm{Se}_{2}$ - \icsdweb{640140} - SG 166 ($R\bar{3}m$) - SEBR}\\
\tiny{ $\;Z_{2,1}=1\;Z_{2,2}=1\;Z_{2,3}=1\;Z_4=0$ } & \tiny{ $\;Z_{2,1}=1\;Z_{2,2}=1\;Z_{2,3}=1\;Z_4=0$ }\\
\includegraphics[width=0.38\textwidth,angle=0]{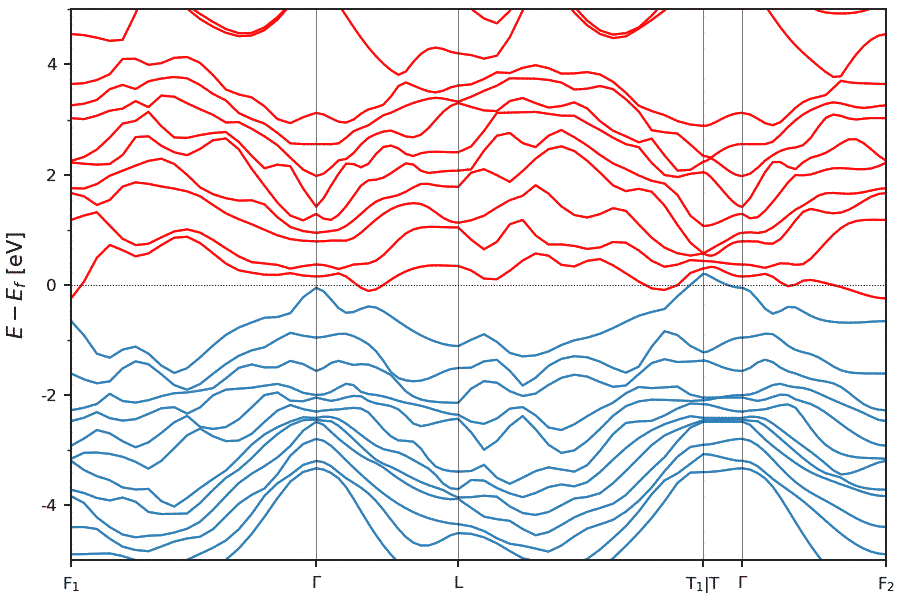} & \includegraphics[width=0.38\textwidth,angle=0]{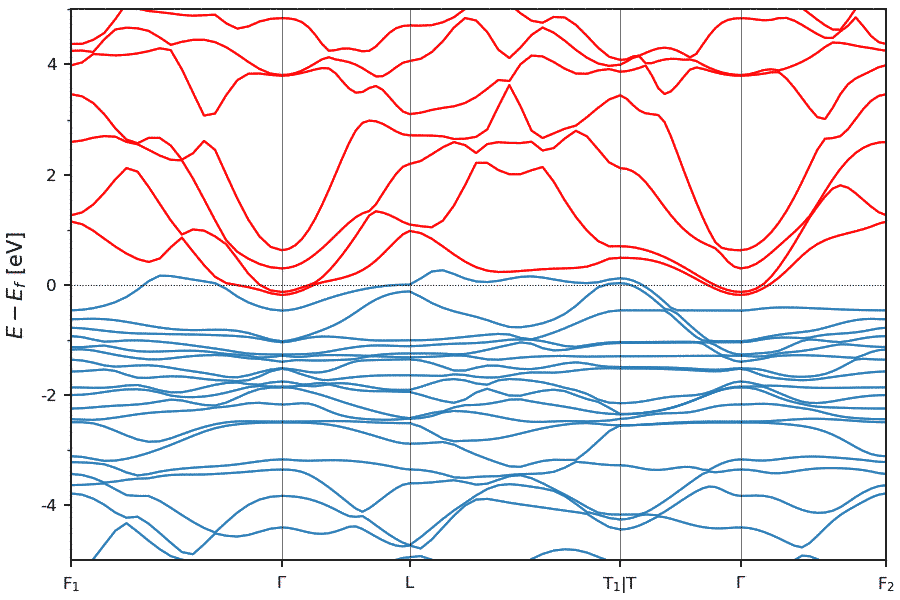}\\
\end{tabular}
\begin{tabular}{c c}
\scriptsize{$\rm{Ni}_{3} \rm{Tl}_{2} \rm{S}_{2}$ - \icsdweb{646398} - SG 166 ($R\bar{3}m$) - SEBR} & \scriptsize{$\rm{Pd}_{3} \rm{Tl}_{2} \rm{S}_{2}$ - \icsdweb{648762} - SG 166 ($R\bar{3}m$) - SEBR}\\
\tiny{ $\;Z_{2,1}=1\;Z_{2,2}=1\;Z_{2,3}=1\;Z_4=0$ } & \tiny{ $\;Z_{2,1}=1\;Z_{2,2}=1\;Z_{2,3}=1\;Z_4=0$ }\\
\includegraphics[width=0.38\textwidth,angle=0]{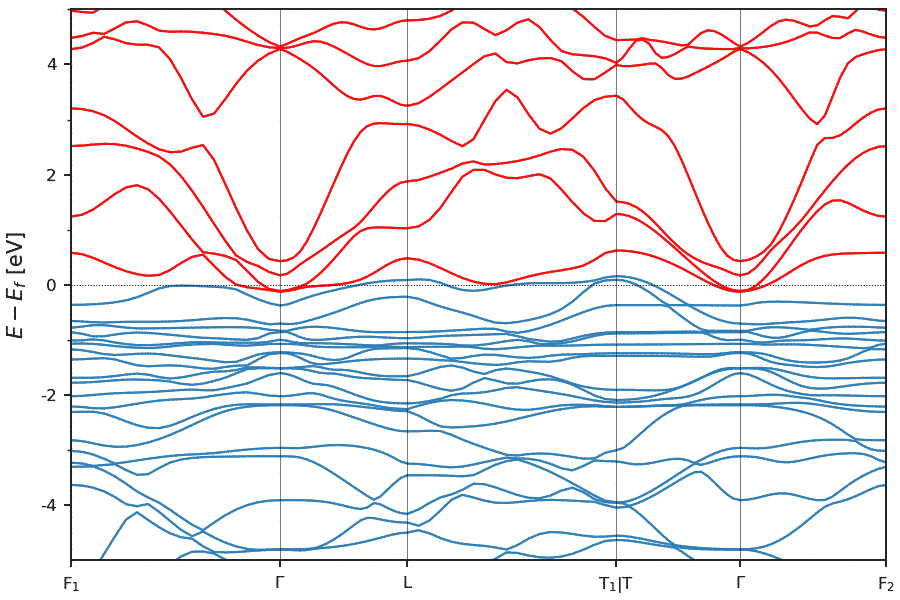} & \includegraphics[width=0.38\textwidth,angle=0]{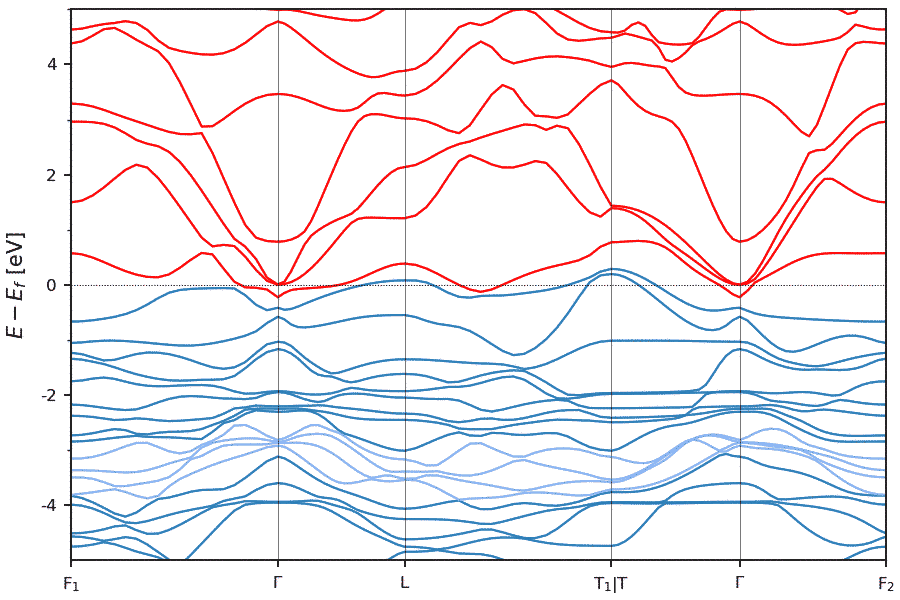}\\
\end{tabular}
\begin{tabular}{c c}
\scriptsize{$\rm{Zr} \rm{B}_{2}$ - \icsdweb{169458} - SG 191 ($P6/mmm$) - SEBR} & \scriptsize{$\rm{K} \rm{Hg} \rm{As}$ - \icsdweb{10458} - SG 194 ($P6_3/mmc$) - SEBR}\\
\tiny{ $\;Z_{2,1}=0\;Z_{2,2}=0\;Z_{2,3}=1\;Z_4=0\;Z_{6m,0}=5\;Z_{6m,\pi}=3\;Z_{12}=8$ } & \tiny{ $\;Z_{2,1}=0\;Z_{2,2}=0\;Z_{2,3}=0\;Z_4=0\;Z_{6m,0}=2\;Z_{12}'=8$ }\\
\includegraphics[width=0.38\textwidth,angle=0]{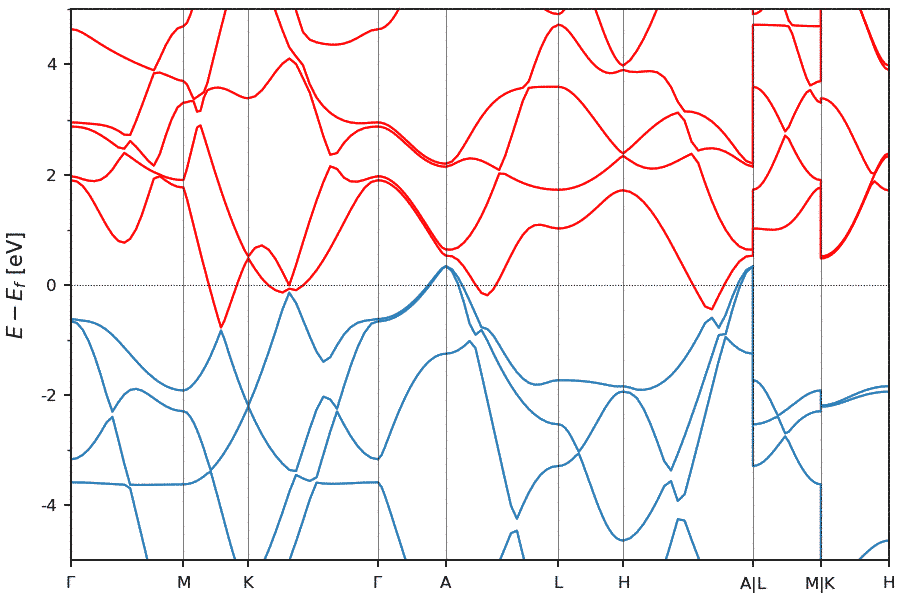} & \includegraphics[width=0.38\textwidth,angle=0]{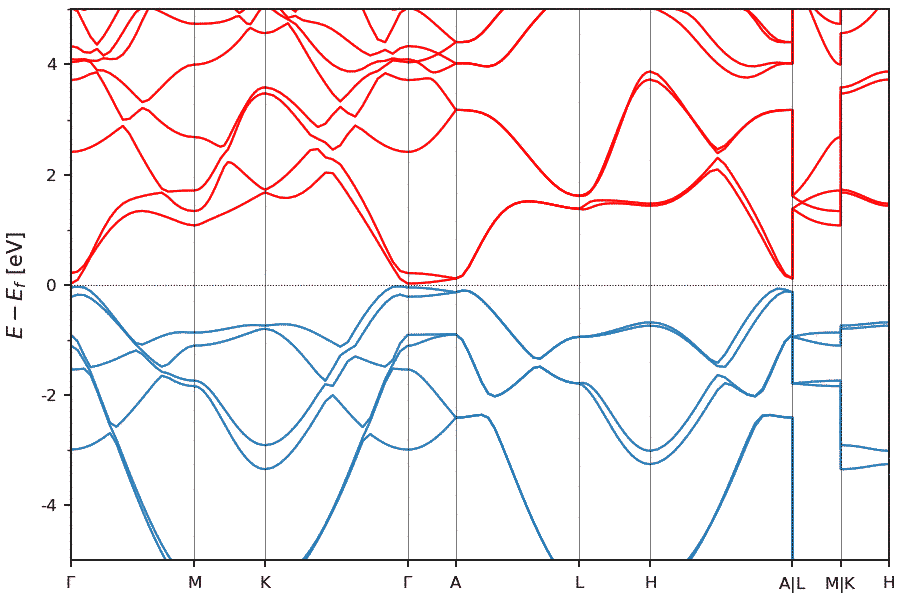}\\
\end{tabular}
\begin{tabular}{c c}
\scriptsize{$\rm{Hg} \rm{K} \rm{Sb}$ - \icsdweb{56201} - SG 194 ($P6_3/mmc$) - SEBR} & \scriptsize{$\rm{Cd} \rm{Na}_{2} \rm{Sn}$ - \icsdweb{102034} - SG 194 ($P6_3/mmc$) - SEBR}\\
\tiny{ $\;Z_{2,1}=0\;Z_{2,2}=0\;Z_{2,3}=0\;Z_4=0\;Z_{6m,0}=2\;Z_{12}'=8$ } & \tiny{ $\;Z_{2,1}=0\;Z_{2,2}=0\;Z_{2,3}=0\;Z_4=0\;Z_{6m,0}=2\;Z_{12}'=8$ }\\
\includegraphics[width=0.38\textwidth,angle=0]{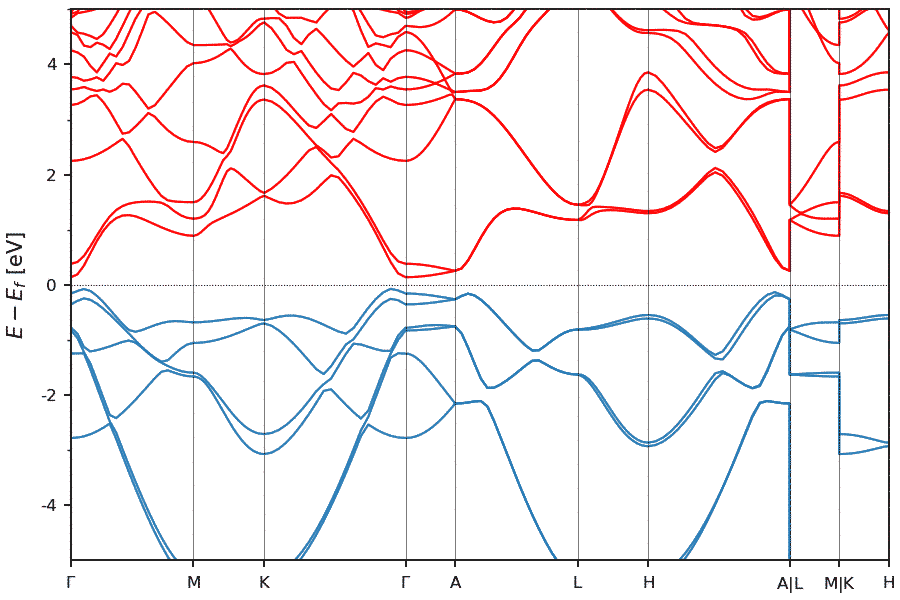} & \includegraphics[width=0.38\textwidth,angle=0]{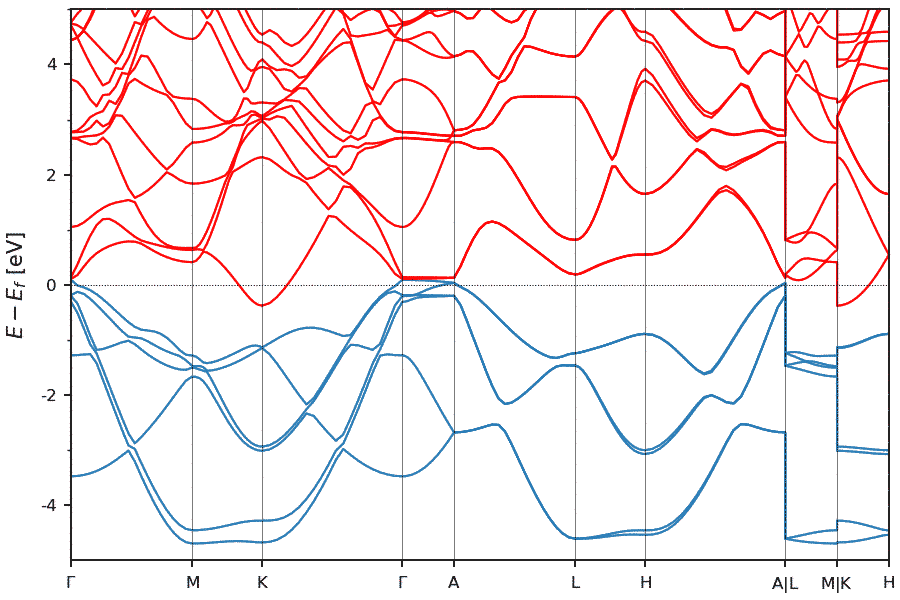}\\
\end{tabular}

\caption{\zfourTCIweakSEBR{3}}
\label{fig:z4_0_TCI_weak_SEBR3}
\end{figure}

\begin{figure}[ht]
\centering
\begin{tabular}{c c}
\scriptsize{$\rm{Cu}_{2} \rm{S}$ - \icsdweb{166578} - SG 194 ($P6_3/mmc$) - SEBR} & \scriptsize{$\rm{Ge} \rm{Ni}_{3} \rm{C}$ - \icsdweb{77001} - SG 221 ($Pm\bar{3}m$) - SEBR}\\
\tiny{ $\;Z_{2,1}=0\;Z_{2,2}=0\;Z_{2,3}=0\;Z_4=0\;Z_{6m,0}=4\;Z_{12}'=4$ } & \tiny{ $\;Z_{2,1}=1\;Z_{2,2}=1\;Z_{2,3}=1\;Z_4=0\;Z_{4m,\pi}=3\;Z_2=0\;Z_8=0$ }\\
\includegraphics[width=0.38\textwidth,angle=0]{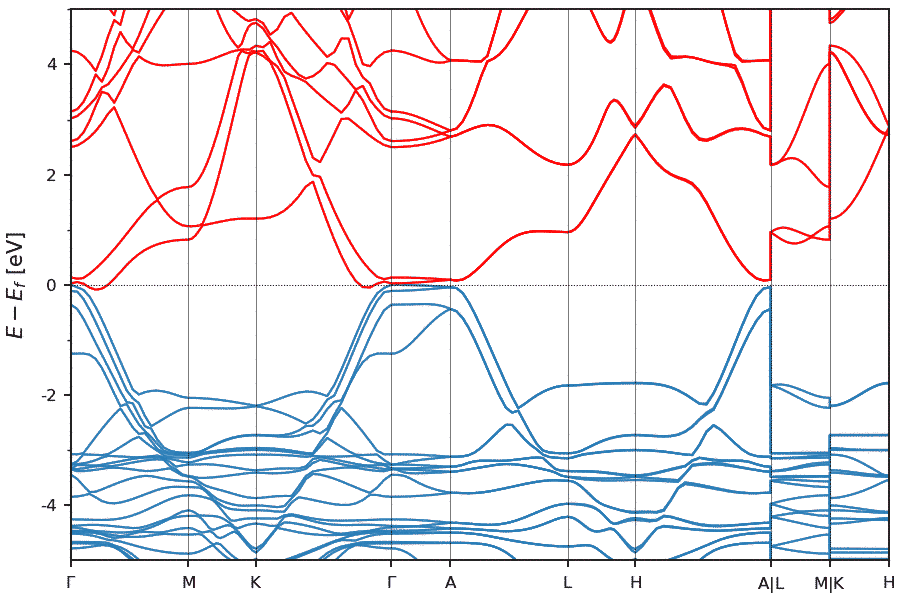} & \includegraphics[width=0.38\textwidth,angle=0]{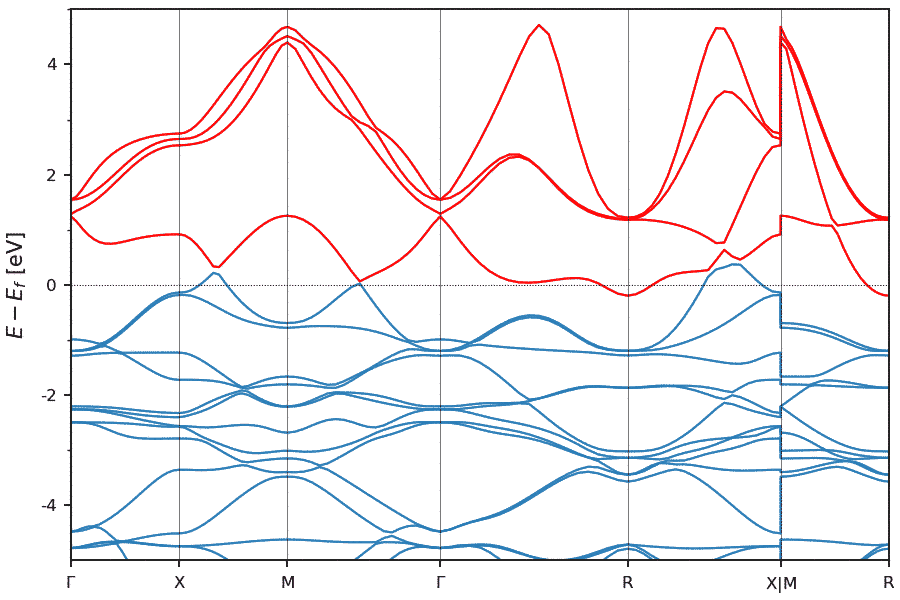}\\
\end{tabular}
\begin{tabular}{c c}
\scriptsize{$\rm{Pd}_{3} \rm{Pb} \rm{C}$ - \icsdweb{108178} - SG 221 ($Pm\bar{3}m$) - SEBR} & \scriptsize{$\rm{In} \rm{N} \rm{Ni}_{3}$ - \icsdweb{247065} - SG 221 ($Pm\bar{3}m$) - SEBR}\\
\tiny{ $\;Z_{2,1}=1\;Z_{2,2}=1\;Z_{2,3}=1\;Z_4=0\;Z_{4m,\pi}=3\;Z_2=0\;Z_8=0$ } & \tiny{ $\;Z_{2,1}=1\;Z_{2,2}=1\;Z_{2,3}=1\;Z_4=0\;Z_{4m,\pi}=3\;Z_2=0\;Z_8=0$ }\\
\includegraphics[width=0.38\textwidth,angle=0]{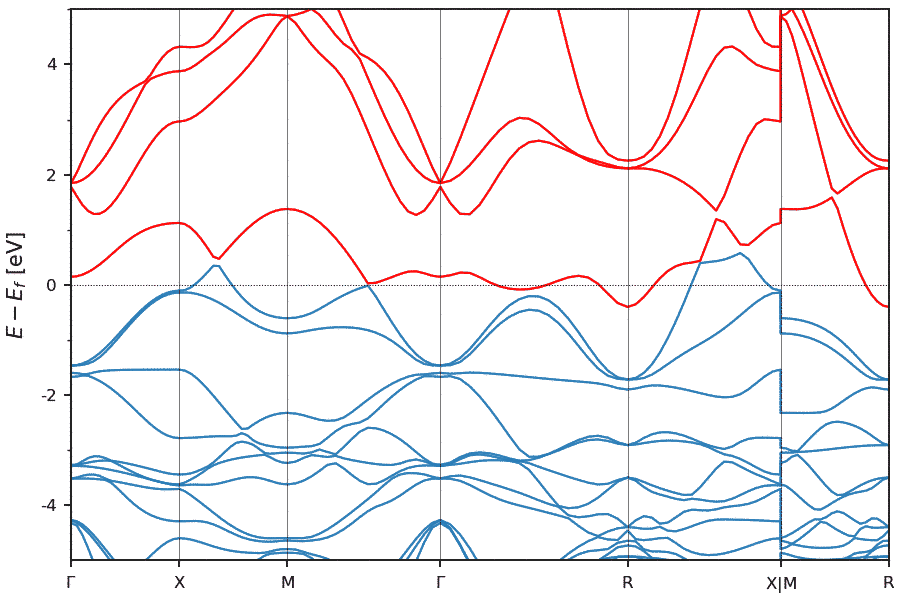} & \includegraphics[width=0.38\textwidth,angle=0]{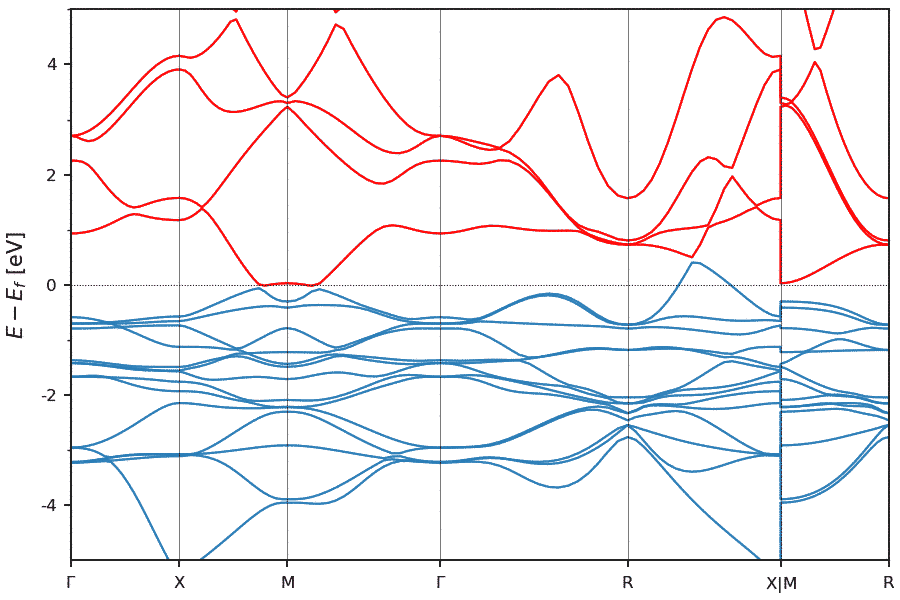}\\
\end{tabular}
\begin{tabular}{c c}
\scriptsize{$\rm{Tl} \rm{Pd}_{3} \rm{H}$ - \icsdweb{247273} - SG 221 ($Pm\bar{3}m$) - SEBR} & \scriptsize{$\rm{Hf} \rm{Rh}_{3} \rm{B}$ - \icsdweb{614447} - SG 221 ($Pm\bar{3}m$) - SEBR}\\
\tiny{ $\;Z_{2,1}=1\;Z_{2,2}=1\;Z_{2,3}=1\;Z_4=0\;Z_{4m,\pi}=3\;Z_2=0\;Z_8=0$ } & \tiny{ $\;Z_{2,1}=1\;Z_{2,2}=1\;Z_{2,3}=1\;Z_4=0\;Z_{4m,\pi}=1\;Z_2=0\;Z_8=0$ }\\
\includegraphics[width=0.38\textwidth,angle=0]{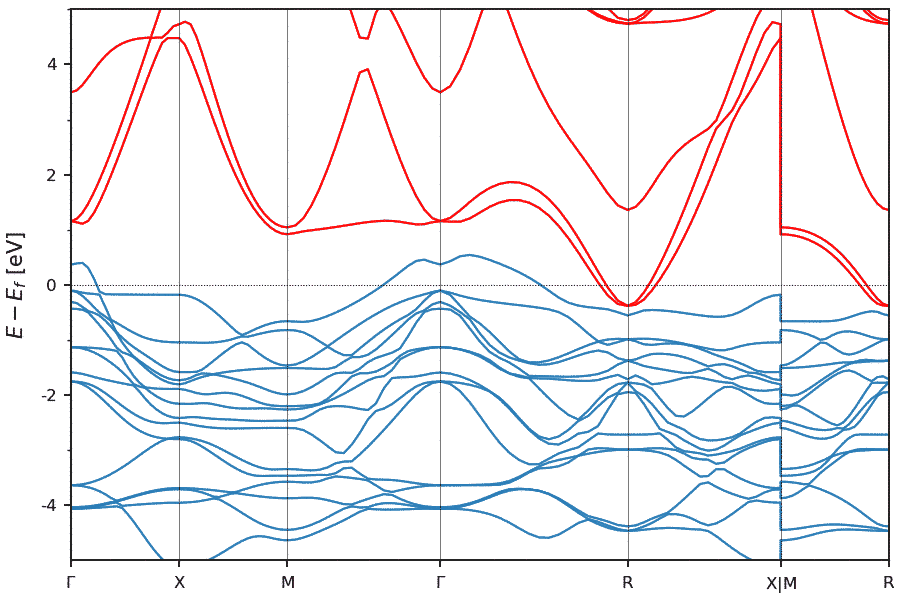} & \includegraphics[width=0.38\textwidth,angle=0]{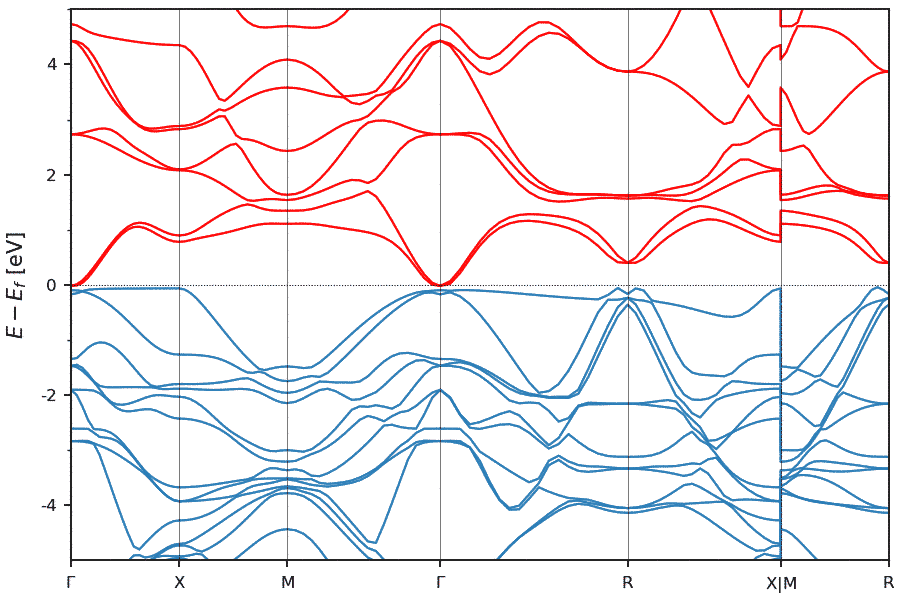}\\
\end{tabular}

\caption{\zfourTCIweakSEBR{4}}
\label{fig:z4_0_TCI_weak_SEBR4}
\end{figure}

\clearpage

\subsubsection{TCIs with Threefold Rotation Symmetry}
\label{App:z3TCIs}

In this section, we list the TCIs with threefold rotation symmetry and nontrivial $Z_{3m0,\pi}$ indices.  As discussed in Refs.~\onlinecite{ChenTCI,AshvinTCI,MTQC}, TCIs with nontrivial $Z_{3m0,\pi}$ indices may also be 3D TIs.  Specifically, $Z_{3m0,\pi}$ respectively indicate the mirror Chern numbers in the $k_{z}=0,\pi$ planes modulo $3$.  Insulators with net-odd momentum-space mirror Chern numbers $C_{M_{z}}$ are 3D TIs -- however, because insulators with $C_{M_{z}}=2$ and $C_{M_{z}}=-1$ exhibit the same mirror Chern numbers modulo $3$, then further Wilson-loop~\cite{Fidkowski2011,AndreiXiZ2,ArisInversion,HourglassInsulator,Cohomological,DiracInsulator} or surface-state calculations must be performed to determine whether insulators with nontrivial $Z_{3m0,\pi}$ are additionally 3D TIs.

\begin{figure}[ht]
\centering
\begin{tabular}{c c}
\scriptsize{$\rm{Y} \rm{N}$ - \icsdweb{185567} - SG 187 ($P\bar{6}m2$) - NLC}\\
\tiny{ $\;Z_{3m,0}=1\;Z_{3m,\pi}=0$ }\\
\includegraphics[width=0.38\textwidth,angle=0]{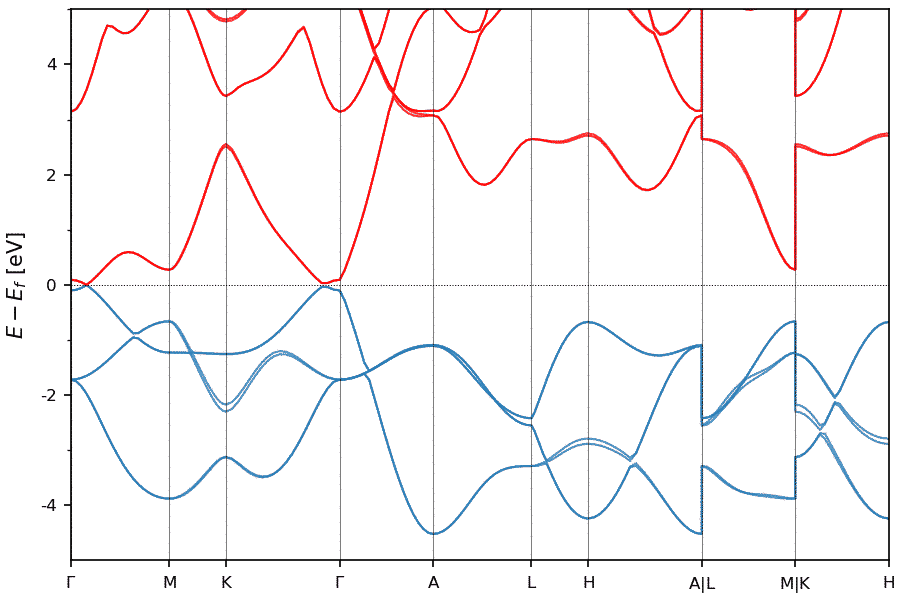}\\
\end{tabular}

\caption{The NLC-classified, threefold-rotation-indicated TCI with the fewest and smallest bulk Fermi pockets.}
\label{fig:z3TCI_NLC}
\end{figure}

\begin{figure}[ht]
\centering
\begin{tabular}{c c}
\scriptsize{$\rm{Ca} \rm{Ag} \rm{P}$ - \icsdweb{10016} - SG 189 ($P\bar{6}2m$) - SEBR} & \scriptsize{$\rm{Na} \rm{Ba} \rm{Bi}$ - \icsdweb{413810} - SG 189 ($P\bar{6}2m$) - SEBR}\\
\tiny{ $\;Z_{3m,0}=1\;Z_{3m,\pi}=0$ } & \tiny{ $\;Z_{3m,0}=1\;Z_{3m,\pi}=0$ }\\
\includegraphics[width=0.38\textwidth,angle=0]{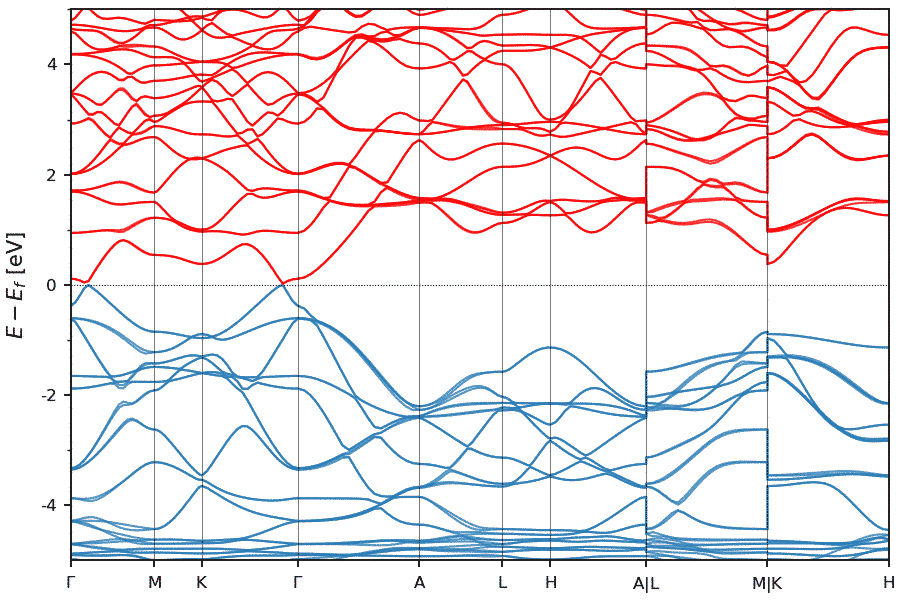} & \includegraphics[width=0.38\textwidth,angle=0]{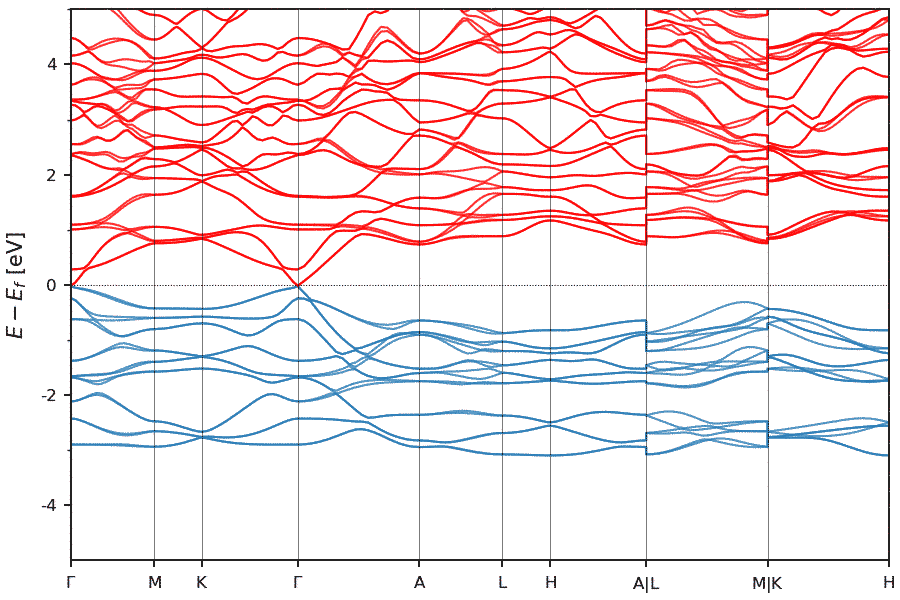}\\
\end{tabular}
\begin{tabular}{c c}
\scriptsize{$\rm{Ag} \rm{Ca} \rm{As}$ - \icsdweb{604730} - SG 189 ($P\bar{6}2m$) - SEBR}\\
\tiny{ $\;Z_{3m,0}=1\;Z_{3m,\pi}=0$ }\\
\includegraphics[width=0.38\textwidth,angle=0]{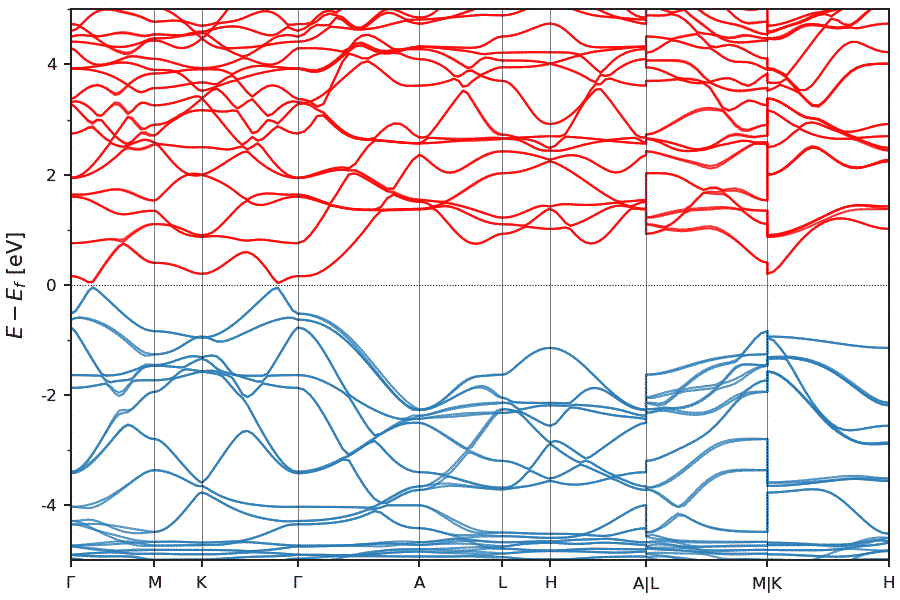}\\
\end{tabular}

\caption{The SEBR-classified, threefold-rotation-indicated TIs and TCIs with the largest band gaps or the fewest and smallest bulk Fermi pockets.}
\label{fig:z3TCI_SEBR}
\end{figure}

\clearpage

\subsection{Repeat-Topological Materials}
\label{App:RTopoMaterials}

In this section, we list the RTopo TIs and TCIs with the largest band gaps or the fewest and smallest bulk Fermi pockets when the Fermi level is set to $0$, and when the Fermi level is set to the next highest filling below $E_{F}$ at which an insulating gap is permitted by band connectivity (see \supappref{App:DefRTopo} for further details and a rigorous definition of RTopo materials).  As introduced in this work, RTopo materials specifically exhibit stable topological gaps at $E_{F}$, and at the next-highest gap below $E_{F}$ as determined by band connectivity through TQC (see~\supappref{App:TQCReview_appendix} and~\supappref{App:DefRTopo}).  First in Figs.~\ref{fig:rTopo_NLC1} and~\ref{fig:rTopo_NLC2}, we list the RTopo materials that are classified as NLC at $E_{F}$, and then, in Figs.~\ref{fig:rTopo_SEBR1},~\ref{fig:rTopo_SEBR2},~\ref{fig:rTopo_SEBR3},~\ref{fig:rTopo_SEBR4},~\ref{fig:rTopo_SEBR5}, and~\ref{fig:rTopo_SEBR6}, we list the RTopo materials that are classified as SEBR at $E_{F}$.

\begin{figure}[ht]
\centering
\begin{tabular}{c c}
\scriptsize{$\rm{Ca} \rm{As}_{3}$ - \icsdweb{193} - SG 2 ($P\bar{1}$) - NLC} & \scriptsize{$\rm{Ca} \rm{P}_{3}$ - \icsdweb{74479} - SG 2 ($P\bar{1}$) - NLC}\\
\tiny{ $\;Z_{2,1}=1\;Z_{2,2}=0\;Z_{2,3}=0\;Z_4=1$ } & \tiny{ $\;Z_{2,1}=0\;Z_{2,2}=1\;Z_{2,3}=0\;Z_4=1$ }\\
\includegraphics[width=0.38\textwidth,angle=0]{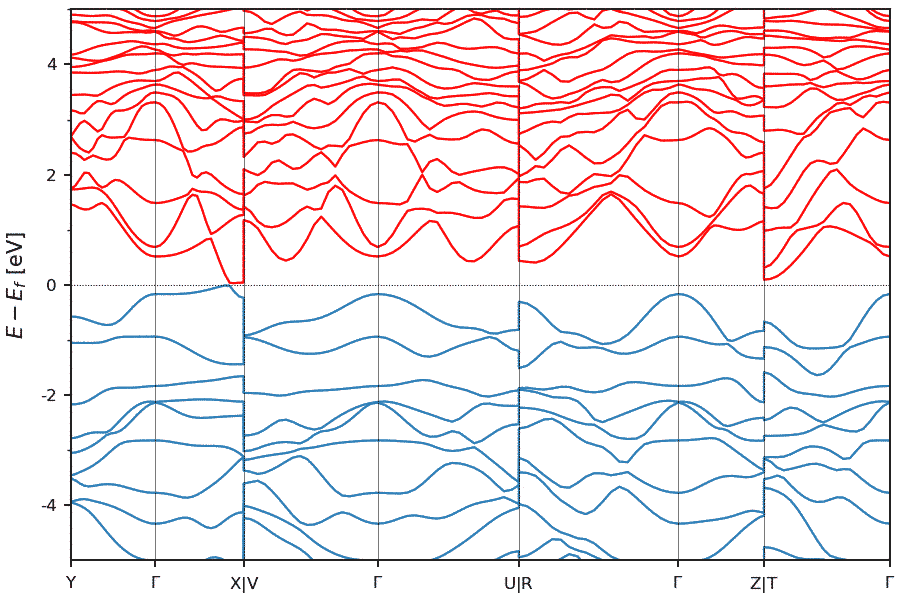} & \includegraphics[width=0.38\textwidth,angle=0]{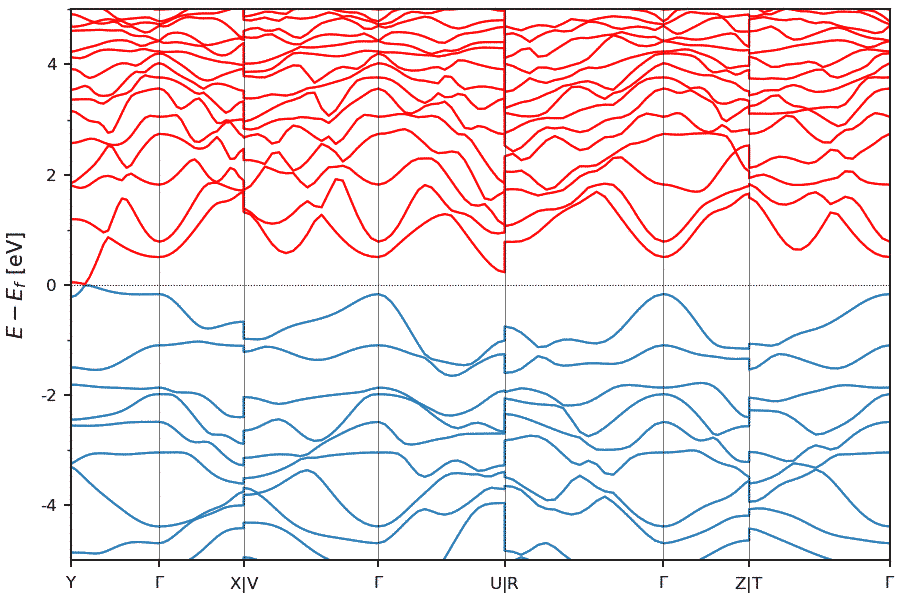}\\
\end{tabular}
\begin{tabular}{c c}
\scriptsize{$\rm{Sr} \rm{As}_{3}$ - \icsdweb{611440} - SG 2 ($P\bar{1}$) - NLC} & \scriptsize{$\rm{Ca} \rm{Sb}_{2}$ - \icsdweb{862} - SG 11 ($P2_1/m$) - NLC}\\
\tiny{ $\;Z_{2,1}=1\;Z_{2,2}=0\;Z_{2,3}=1\;Z_4=1$ } & \tiny{ $\;Z_{2,1}=1\;Z_{2,2}=0\;Z_{2,3}=1\;Z_4=1$ }\\
\includegraphics[width=0.38\textwidth,angle=0]{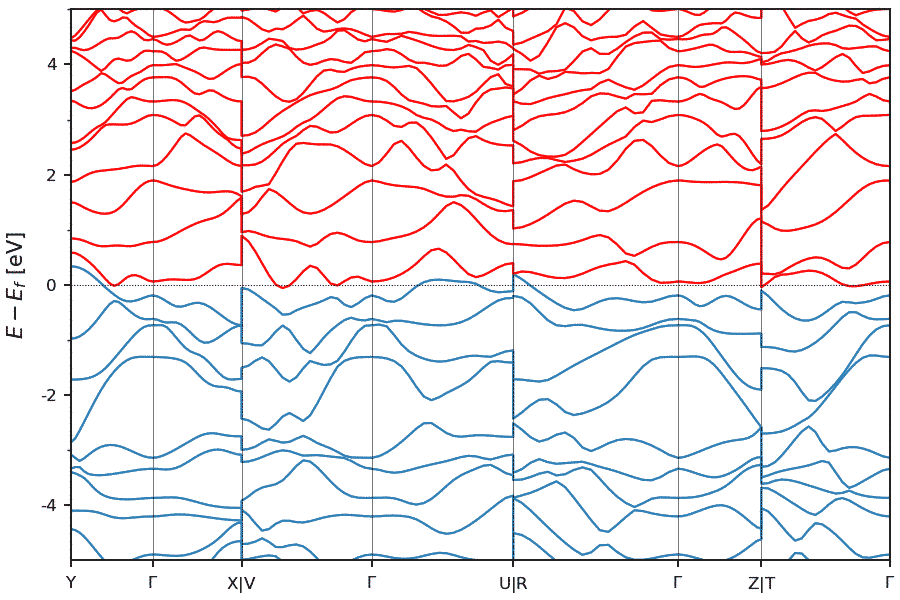} & \includegraphics[width=0.38\textwidth,angle=0]{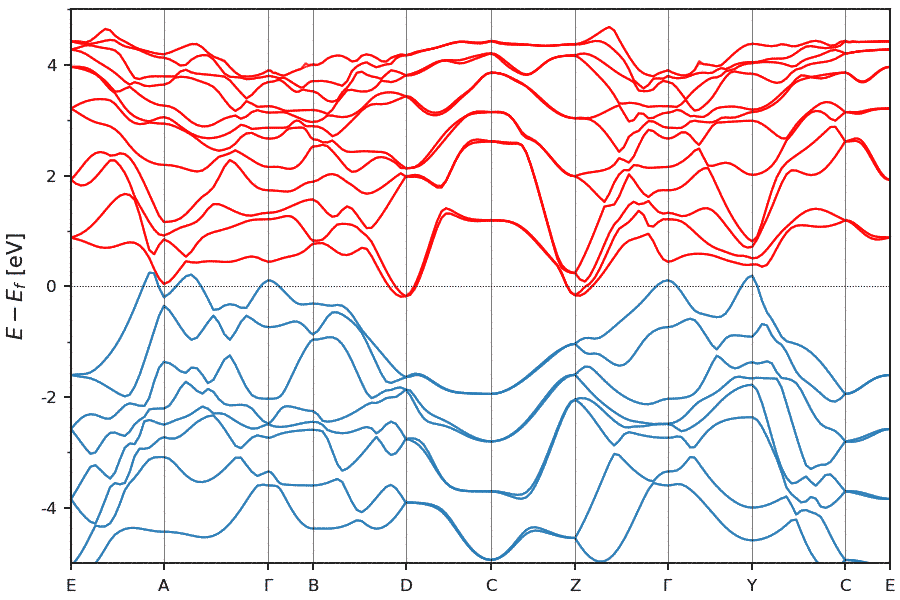}\\
\end{tabular}
\begin{tabular}{c c}
\scriptsize{$\rm{Bi}$ - \icsdweb{42679} - SG 11 ($P2_1/m$) - NLC} & \scriptsize{$\rm{Ta}_{2} \rm{Se}_{3}$ - \icsdweb{42982} - SG 11 ($P2_1/m$) - NLC}\\
\tiny{ $\;Z_{2,1}=1\;Z_{2,2}=0\;Z_{2,3}=1\;Z_4=3$ } & \tiny{ $\;Z_{2,1}=1\;Z_{2,2}=0\;Z_{2,3}=0\;Z_4=3$ }\\
\includegraphics[width=0.38\textwidth,angle=0]{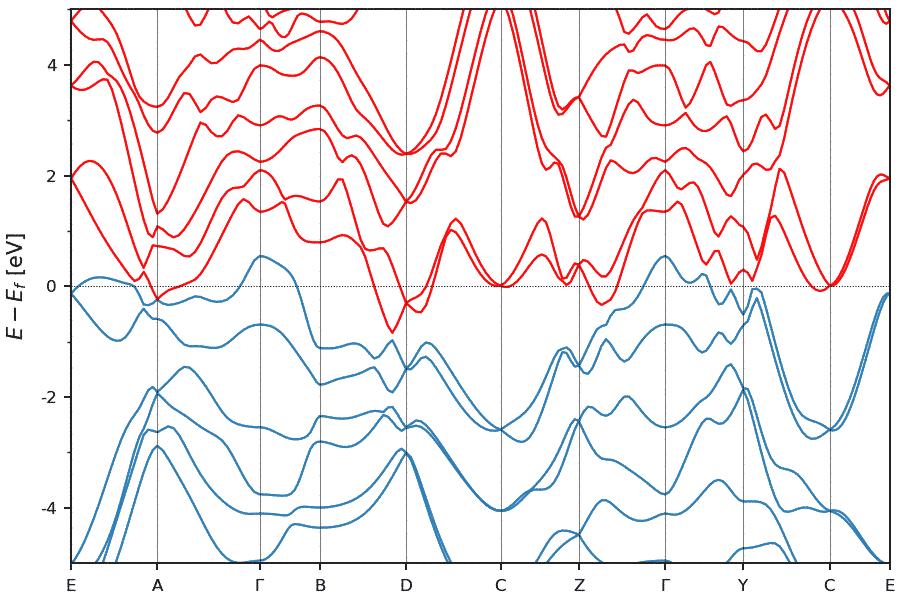} & \includegraphics[width=0.38\textwidth,angle=0]{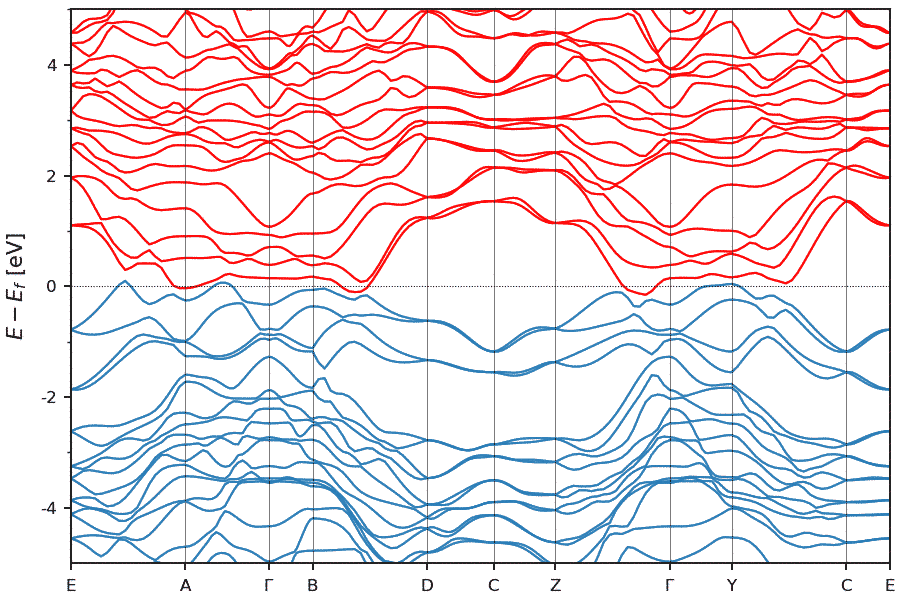}\\
\end{tabular}
\begin{tabular}{c c}
\scriptsize{$\rm{Sr} \rm{Sb}_{2}$ - \icsdweb{52307} - SG 11 ($P2_1/m$) - NLC} & \scriptsize{$\rm{Au}_{2} \rm{P}_{3}$ - \icsdweb{8058} - SG 12 ($C2/m$) - NLC}\\
\tiny{ $\;Z_{2,1}=1\;Z_{2,2}=0\;Z_{2,3}=0\;Z_4=2$ } & \tiny{ $\;Z_{2,1}=0\;Z_{2,2}=0\;Z_{2,3}=1\;Z_4=3$ }\\
\includegraphics[width=0.38\textwidth,angle=0]{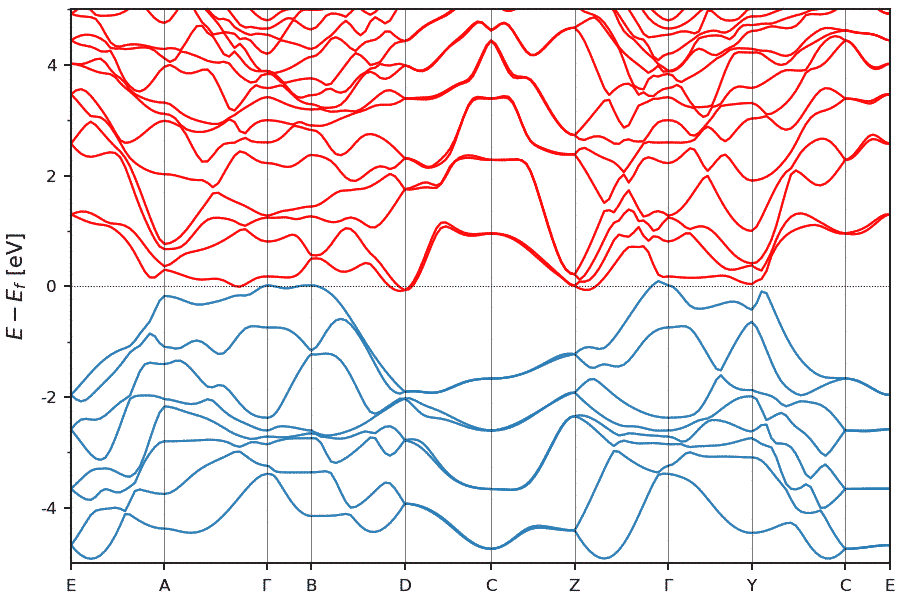} & \includegraphics[width=0.38\textwidth,angle=0]{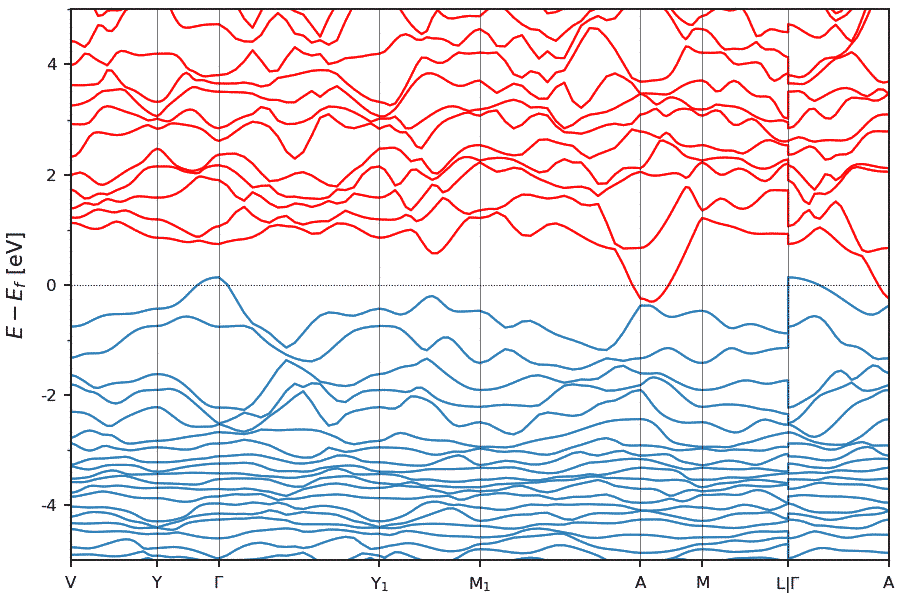}\\
\end{tabular}

\caption{\rTopoNLC{1}}
\label{fig:rTopo_NLC1}
\end{figure}

\begin{figure}[ht]
\centering
\begin{tabular}{c c}
\scriptsize{$\rm{Ba} \rm{Sb}_{3}$ - \icsdweb{49000} - SG 12 ($C2/m$) - NLC} & \scriptsize{$\rm{Ta}_{2} \rm{Pd} \rm{Se}_{6}$ - \icsdweb{61005} - SG 12 ($C2/m$) - NLC}\\
\tiny{ $\;Z_{2,1}=0\;Z_{2,2}=0\;Z_{2,3}=1\;Z_4=3$ } & \tiny{ $\;Z_{2,1}=1\;Z_{2,2}=1\;Z_{2,3}=0\;Z_4=3$ }\\
\includegraphics[width=0.38\textwidth,angle=0]{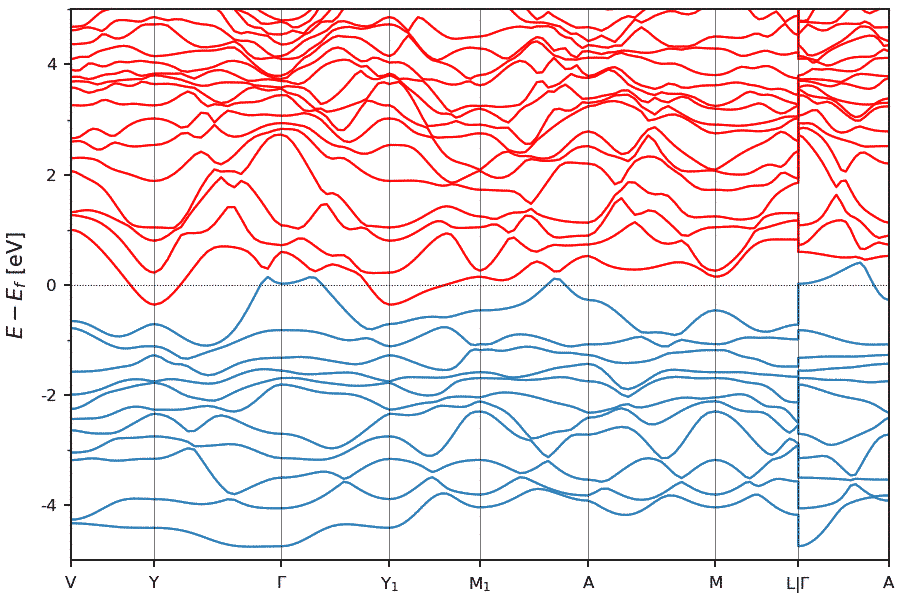} & \includegraphics[width=0.38\textwidth,angle=0]{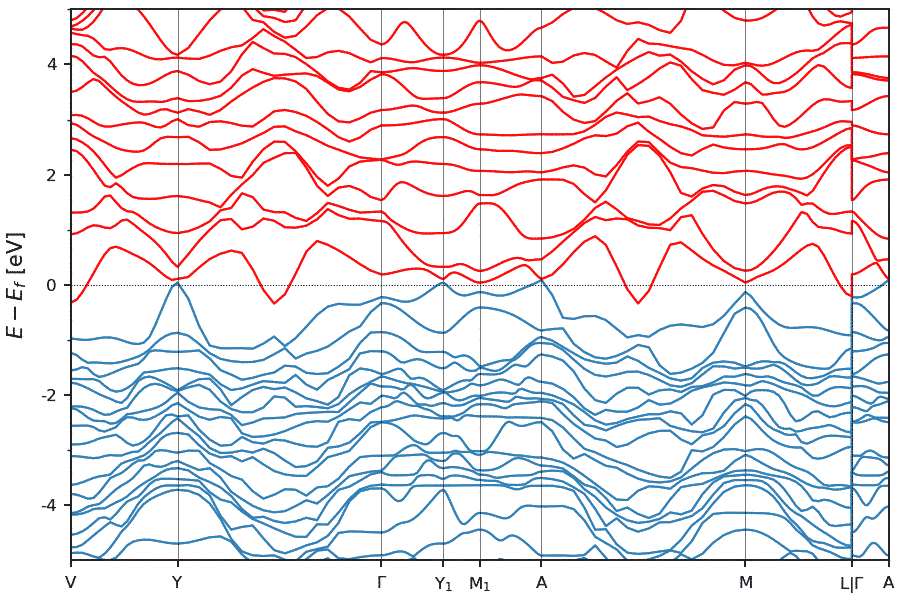}\\
\end{tabular}
\begin{tabular}{c c}
\scriptsize{$\rm{Ba} \rm{Tl}_{4}$ - \icsdweb{261206} - SG 12 ($C2/m$) - NLC} & \scriptsize{$\rm{Ba} \rm{Ti}_{2} \rm{As}_{2} \rm{O}$ - \icsdweb{169074} - SG 123 ($P4/mmm$) - NLC}\\
\tiny{ $\;Z_{2,1}=0\;Z_{2,2}=0\;Z_{2,3}=0\;Z_4=3$ } & \tiny{ $\;Z_{2,1}=0\;Z_{2,2}=0\;Z_{2,3}=1\;Z_4=1\;Z_{4m,\pi}=3\;Z_2=1\;Z_8=5$ }\\
\includegraphics[width=0.38\textwidth,angle=0]{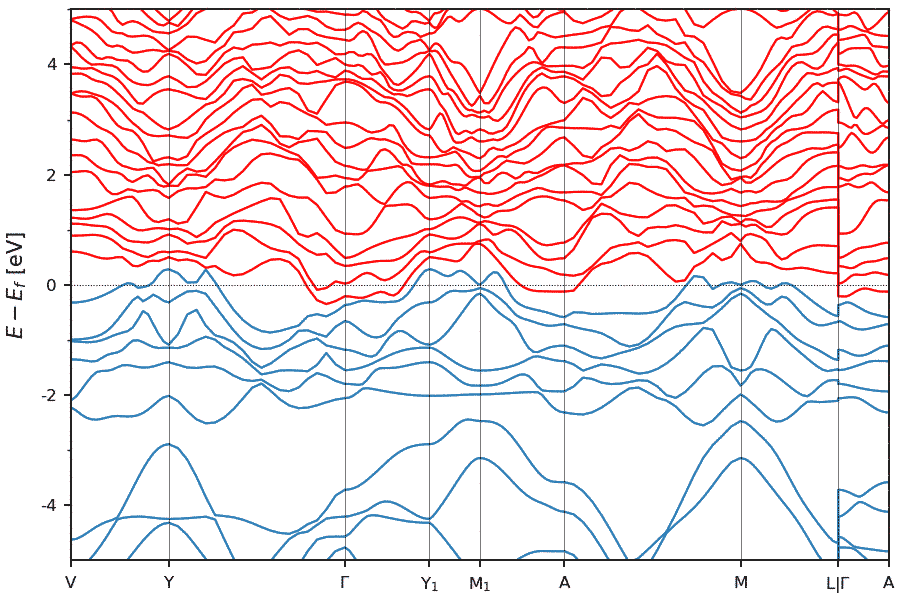} & \includegraphics[width=0.38\textwidth,angle=0]{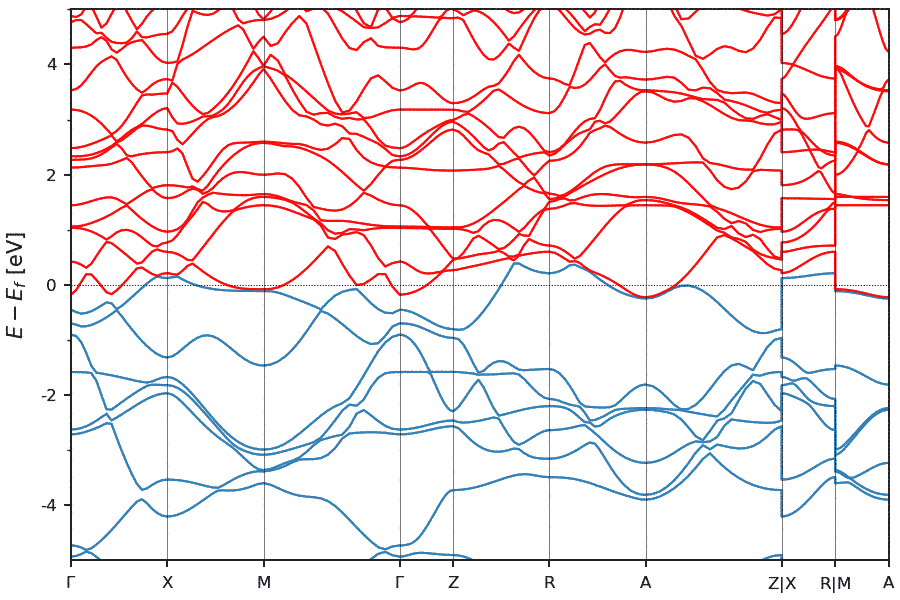}\\
\end{tabular}

\caption{\rTopoNLC{2}}
\label{fig:rTopo_NLC2}
\end{figure}

\begin{figure}[ht]
\centering
\begin{tabular}{c c}
\scriptsize{$\rm{Nb}_{6} \rm{C}_{5}$ - \icsdweb{39297} - SG 12 ($C2/m$) - SEBR} & \scriptsize{$\rm{W}_{2} \rm{As}_{3}$ - \icsdweb{43185} - SG 12 ($C2/m$) - SEBR}\\
\tiny{ $\;Z_{2,1}=0\;Z_{2,2}=0\;Z_{2,3}=1\;Z_4=0$ } & \tiny{ $\;Z_{2,1}=1\;Z_{2,2}=1\;Z_{2,3}=1\;Z_4=2$ }\\
\includegraphics[width=0.38\textwidth,angle=0]{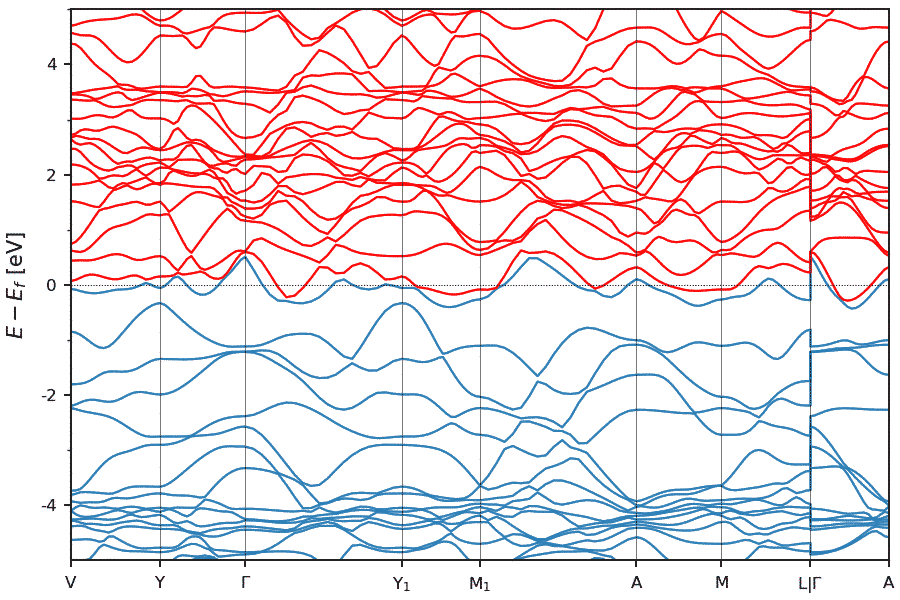} & \includegraphics[width=0.38\textwidth,angle=0]{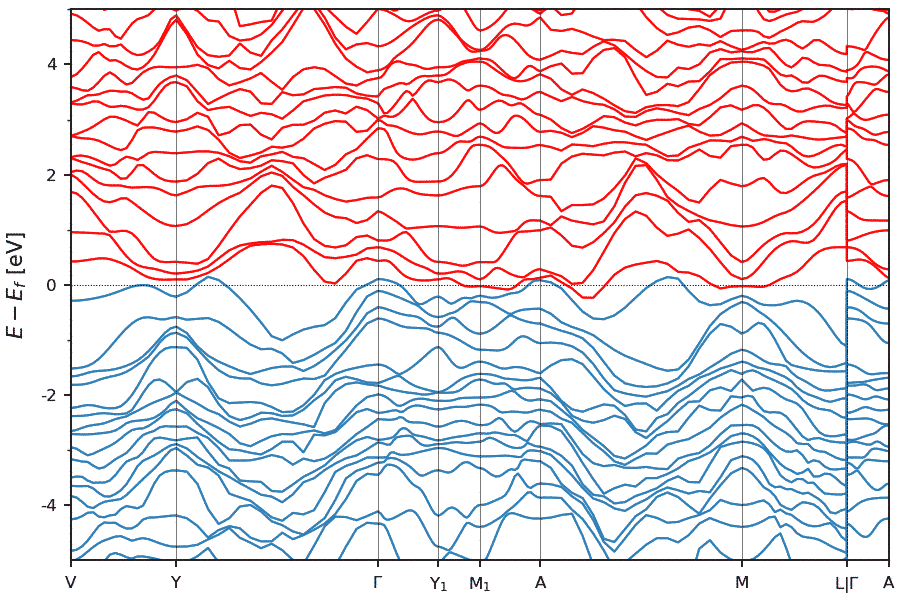}\\
\end{tabular}
\begin{tabular}{c c}
\scriptsize{$\rm{Ba}_{3} \rm{Cd}_{2} \rm{Sb}_{4}$ - \icsdweb{173685} - SG 12 ($C2/m$) - SEBR} & \scriptsize{$\rm{Ta} \rm{Sb}_{2}$ - \icsdweb{651600} - SG 12 ($C2/m$) - SEBR}\\
\tiny{ $\;Z_{2,1}=1\;Z_{2,2}=1\;Z_{2,3}=0\;Z_4=2$ } & \tiny{ $\;Z_{2,1}=1\;Z_{2,2}=1\;Z_{2,3}=1\;Z_4=2$ }\\
\includegraphics[width=0.38\textwidth,angle=0]{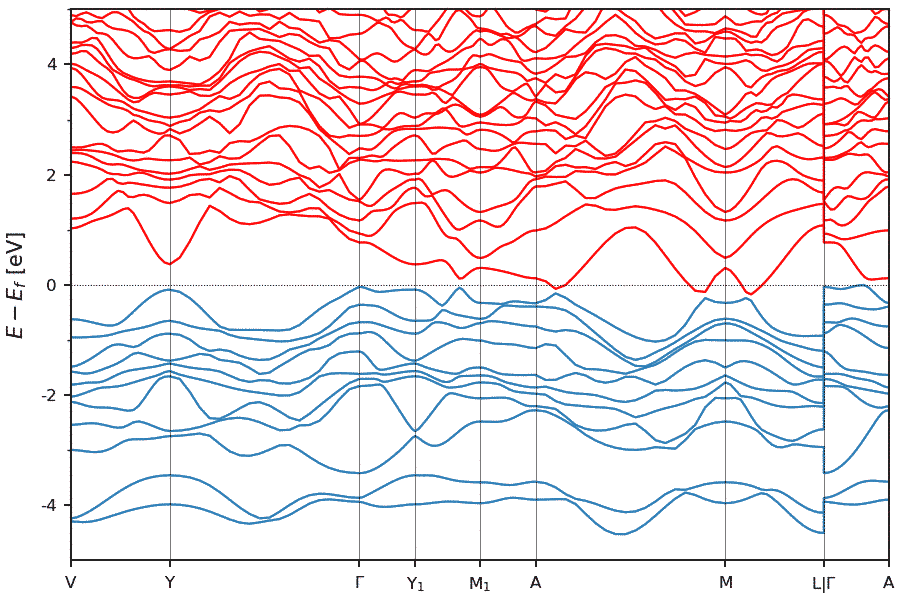} & \includegraphics[width=0.38\textwidth,angle=0]{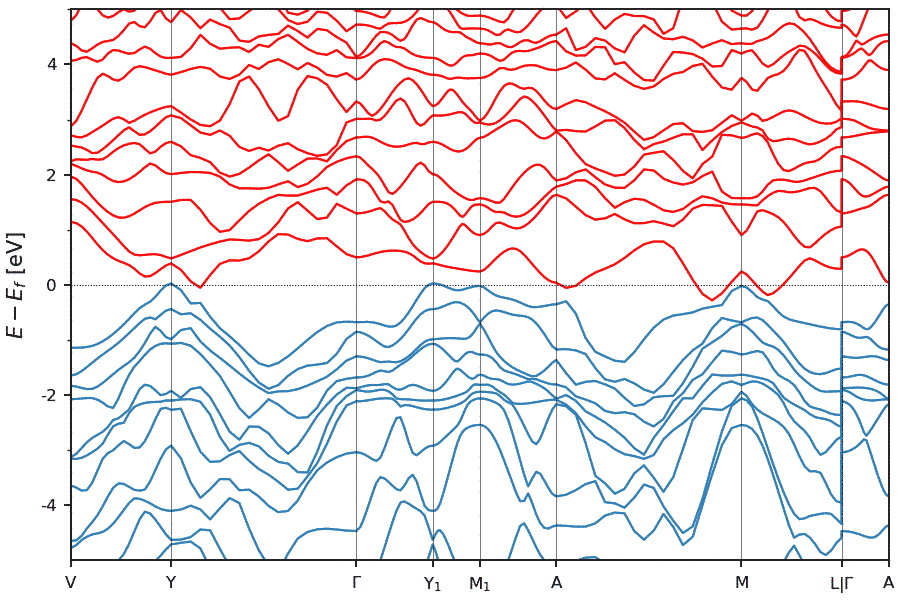}\\
\end{tabular}
\begin{tabular}{c c}
\scriptsize{$\rm{Bi}$ - \icsdweb{653719} - SG 12 ($C2/m$) - SEBR} & \scriptsize{$\rm{Sr}_{3} \rm{Li}_{4} \rm{Sb}_{4}$ - \icsdweb{25307} - SG 71 ($Immm$) - SEBR}\\
\tiny{ $\;Z_{2,1}=0\;Z_{2,2}=0\;Z_{2,3}=1\;Z_4=2$ } & \tiny{ $\;Z_{2,1}=1\;Z_{2,2}=1\;Z_{2,3}=1\;Z_4=2$ }\\
\includegraphics[width=0.38\textwidth,angle=0]{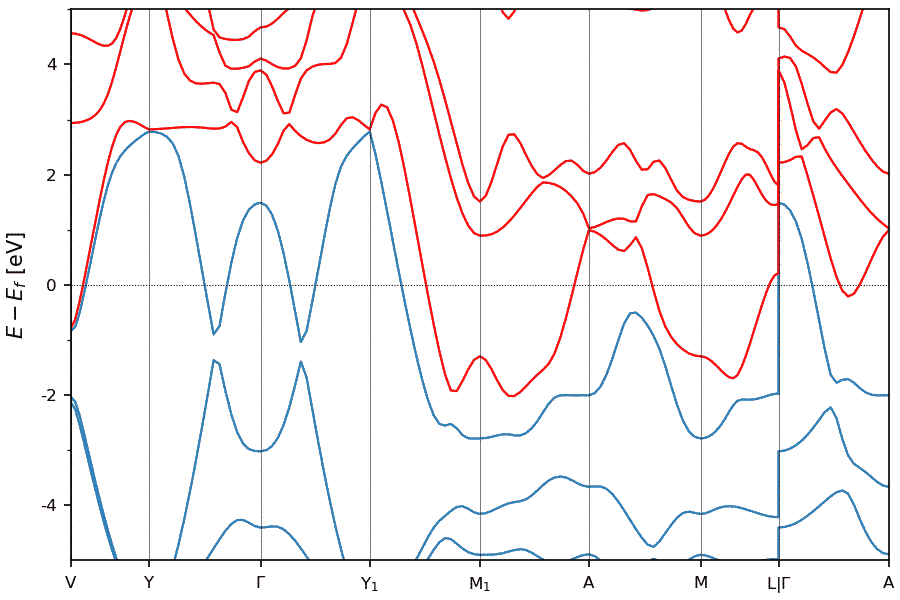} & \includegraphics[width=0.38\textwidth,angle=0]{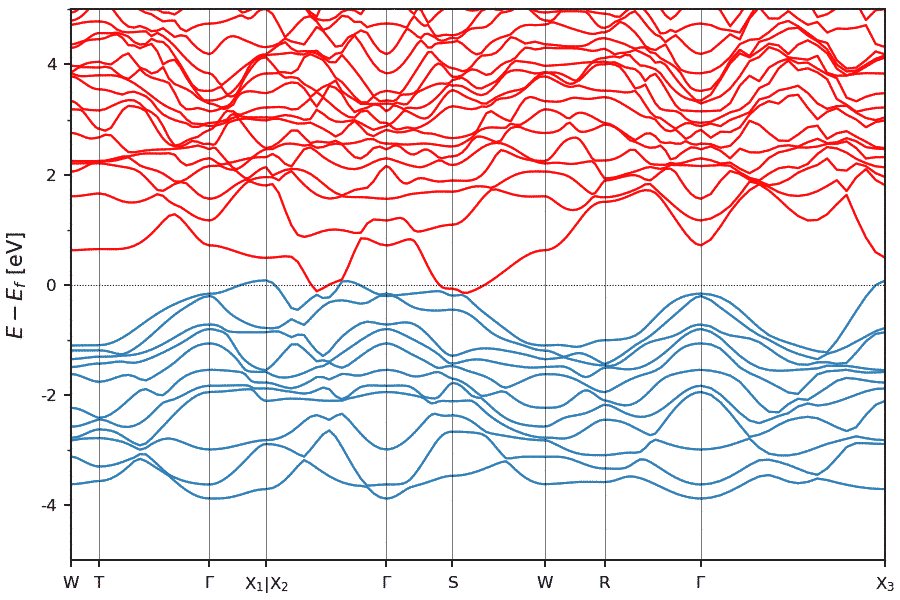}\\
\end{tabular}
\begin{tabular}{c c}
\scriptsize{$\rm{Al}_{2} \rm{Ba}_{3} \rm{Ge}_{2}$ - \icsdweb{52612} - SG 71 ($Immm$) - SEBR} & \scriptsize{$\rm{Sr}_{2} \rm{Cd} \rm{Pt}_{2}$ - \icsdweb{252178} - SG 71 ($Immm$) - SEBR}\\
\tiny{ $\;Z_{2,1}=0\;Z_{2,2}=0\;Z_{2,3}=0\;Z_4=1$ } & \tiny{ $\;Z_{2,1}=1\;Z_{2,2}=1\;Z_{2,3}=1\;Z_4=2$ }\\
\includegraphics[width=0.38\textwidth,angle=0]{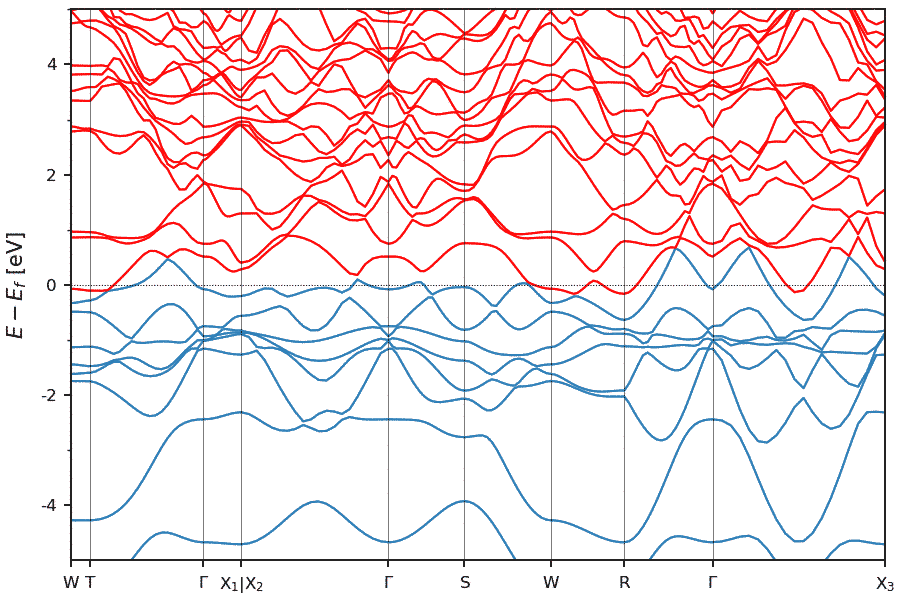} & \includegraphics[width=0.38\textwidth,angle=0]{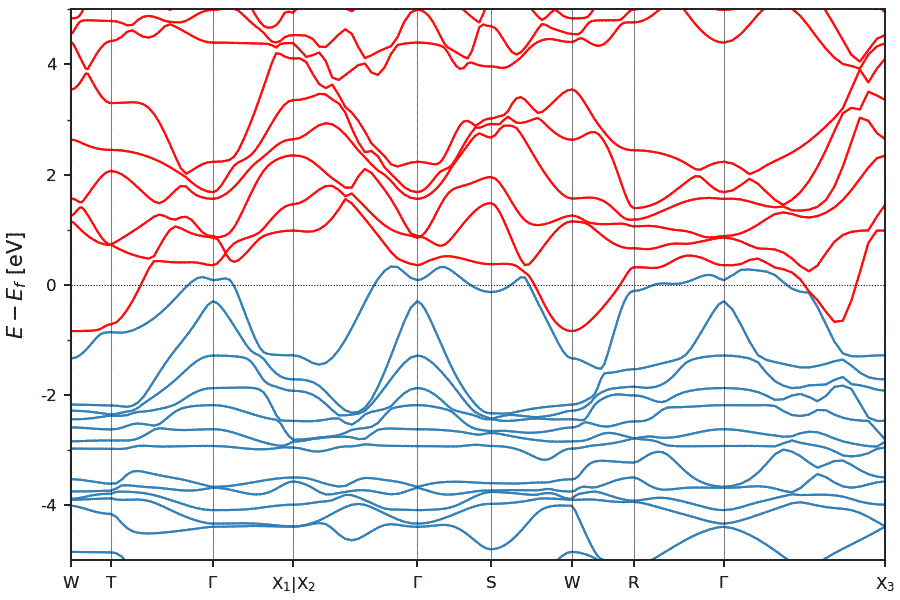}\\
\end{tabular}

\caption{\rTopoSEBR{1}}
\label{fig:rTopo_SEBR1}
\end{figure}

\begin{figure}[ht]
\centering
\begin{tabular}{c c}
\scriptsize{$\rm{W}_{2} \rm{Ni} \rm{B}_{2}$ - \icsdweb{615069} - SG 71 ($Immm$) - SEBR} & \scriptsize{$\rm{Hf}_{5} \rm{Te}_{4}$ - \icsdweb{154358} - SG 87 ($I4/m$) - SEBR}\\
\tiny{ $\;Z_{2,1}=0\;Z_{2,2}=0\;Z_{2,3}=0\;Z_4=3$ } & \tiny{ $\;Z_{2,1}=1\;Z_{2,2}=1\;Z_{2,3}=1\;Z_4=1\;Z_2=1\;Z_8=5$ }\\
\includegraphics[width=0.38\textwidth,angle=0]{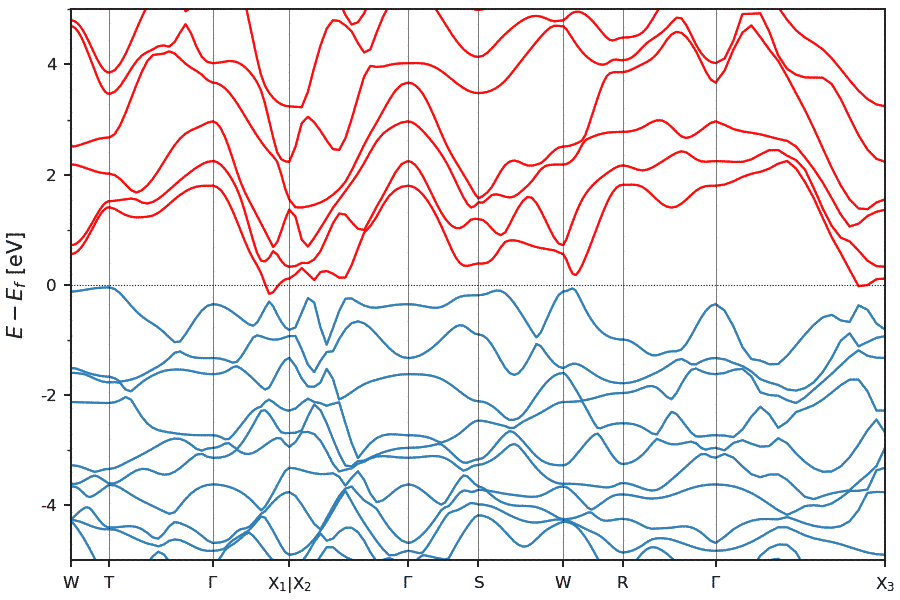} & \includegraphics[width=0.38\textwidth,angle=0]{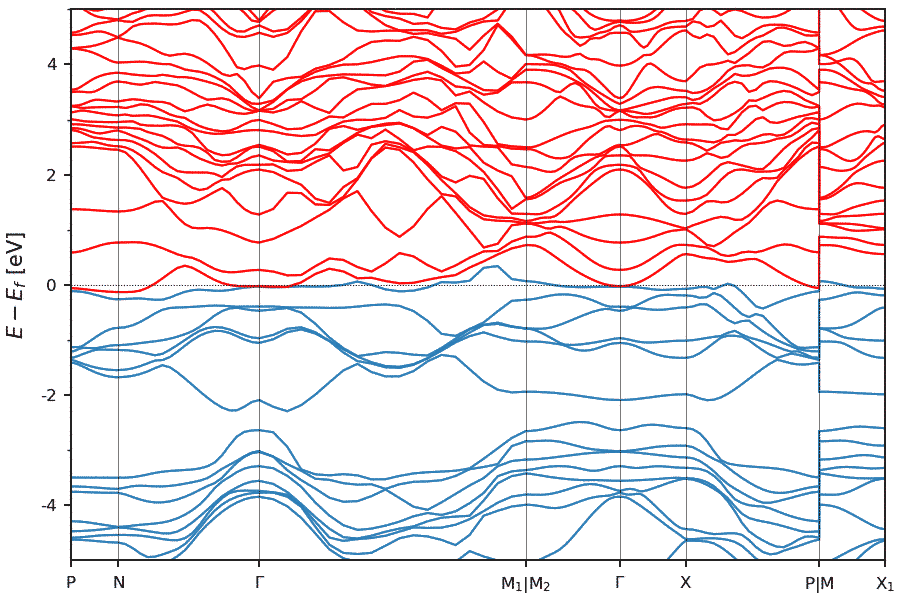}\\
\end{tabular}
\begin{tabular}{c c}
\scriptsize{$\rm{Ba} \rm{Ti}_{2} \rm{Sb}_{2} \rm{O}$ - \icsdweb{430061} - SG 123 ($P4/mmm$) - SEBR} & \scriptsize{$\rm{Sr}_{2} \rm{Bi}$ - \icsdweb{41836} - SG 139 ($I4/mmm$) - SEBR}\\
\tiny{ $\;Z_{2,1}=1\;Z_{2,2}=1\;Z_{2,3}=1\;Z_4=2\;Z_{4m,\pi}=3\;Z_2=0\;Z_8=2$ } & \tiny{ $\;Z_{2,1}=1\;Z_{2,2}=1\;Z_{2,3}=1\;Z_4=2\;Z_2=0\;Z_8=2$ }\\
\includegraphics[width=0.38\textwidth,angle=0]{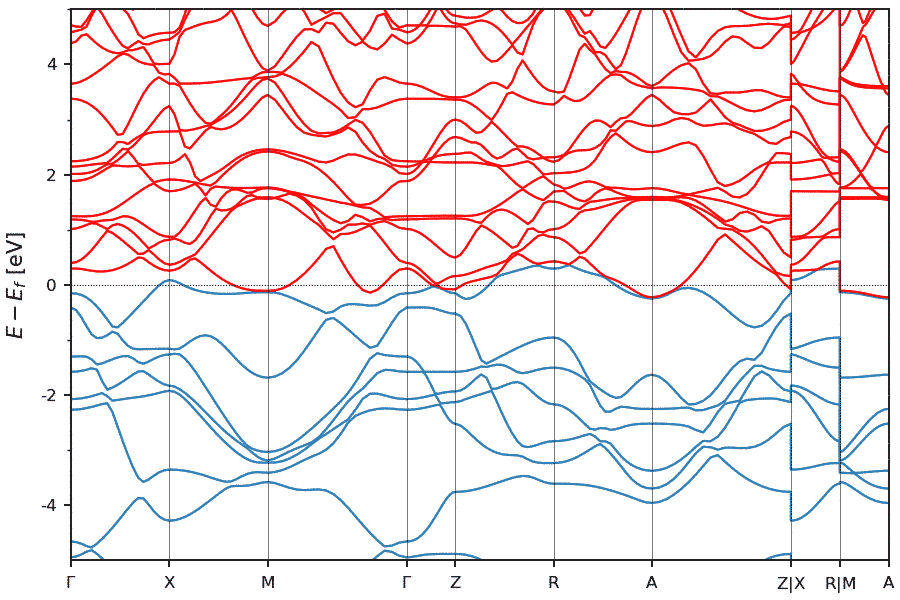} & \includegraphics[width=0.38\textwidth,angle=0]{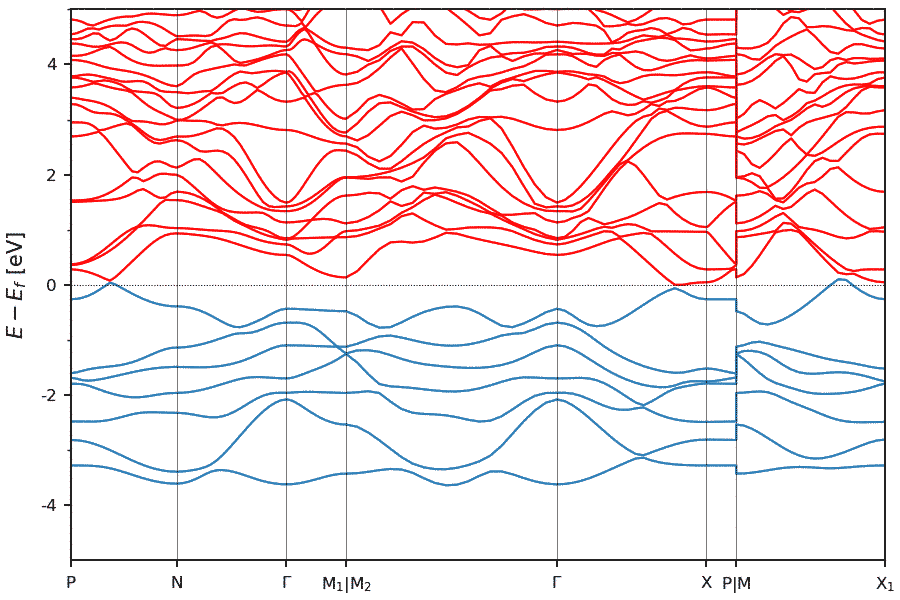}\\
\end{tabular}
\begin{tabular}{c c}
\scriptsize{$\rm{Ba}_{2} \rm{Sb}$ - \icsdweb{41837} - SG 139 ($I4/mmm$) - SEBR} & \scriptsize{$\rm{Sr}_{2} \rm{Sb}$ - \icsdweb{42119} - SG 139 ($I4/mmm$) - SEBR}\\
\tiny{ $\;Z_{2,1}=1\;Z_{2,2}=1\;Z_{2,3}=1\;Z_4=2\;Z_2=0\;Z_8=2$ } & \tiny{ $\;Z_{2,1}=1\;Z_{2,2}=1\;Z_{2,3}=1\;Z_4=2\;Z_2=0\;Z_8=2$ }\\
\includegraphics[width=0.38\textwidth,angle=0]{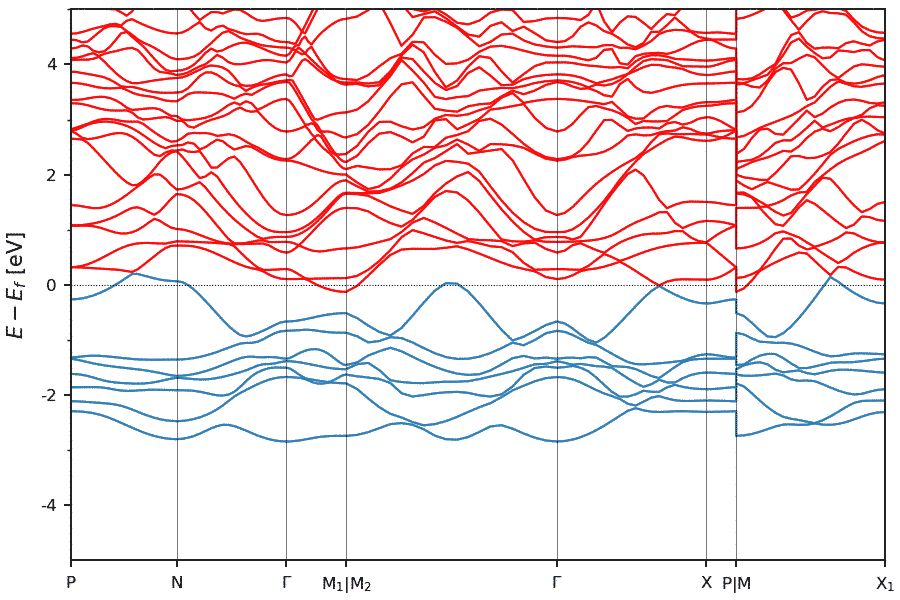} & \includegraphics[width=0.38\textwidth,angle=0]{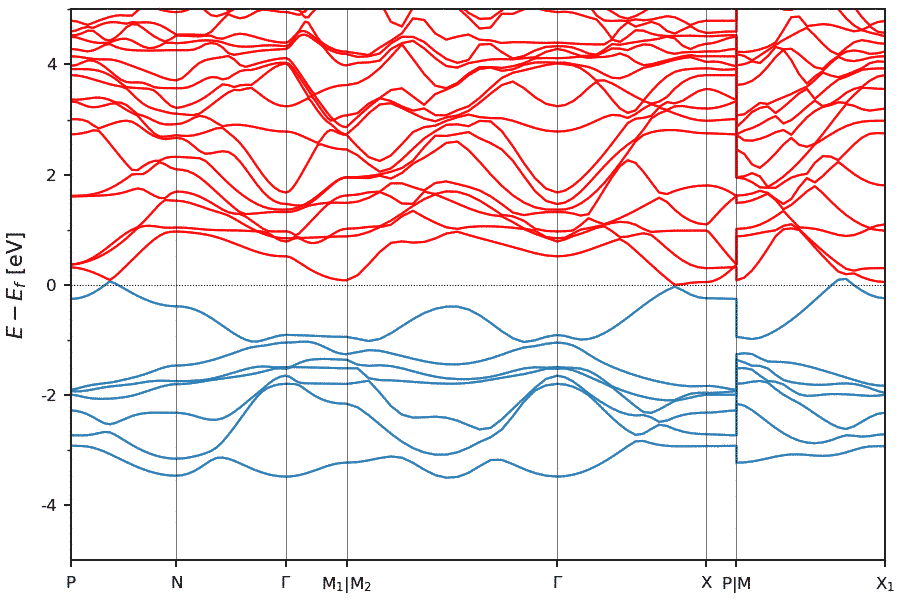}\\
\end{tabular}
\begin{tabular}{c c}
\scriptsize{$\rm{Hf}_{2} \rm{Pd}$ - \icsdweb{104255} - SG 139 ($I4/mmm$) - SEBR} & \scriptsize{$\rm{Hf} \rm{Pd}_{2}$ - \icsdweb{241367} - SG 139 ($I4/mmm$) - SEBR}\\
\tiny{ $\;Z_{2,1}=1\;Z_{2,2}=1\;Z_{2,3}=1\;Z_4=3\;Z_2=1\;Z_8=7$ } & \tiny{ $\;Z_{2,1}=1\;Z_{2,2}=1\;Z_{2,3}=1\;Z_4=1\;Z_2=1\;Z_8=5$ }\\
\includegraphics[width=0.38\textwidth,angle=0]{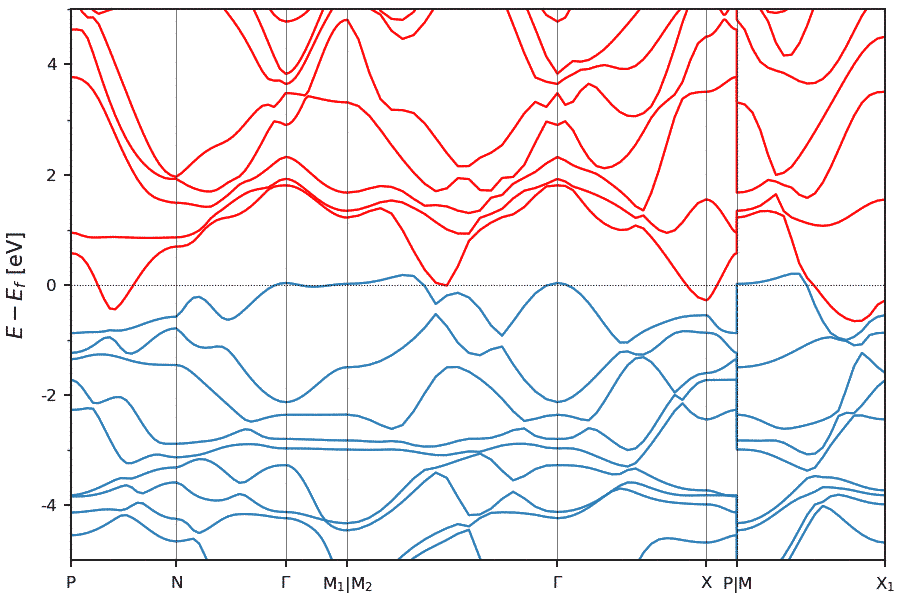} & \includegraphics[width=0.38\textwidth,angle=0]{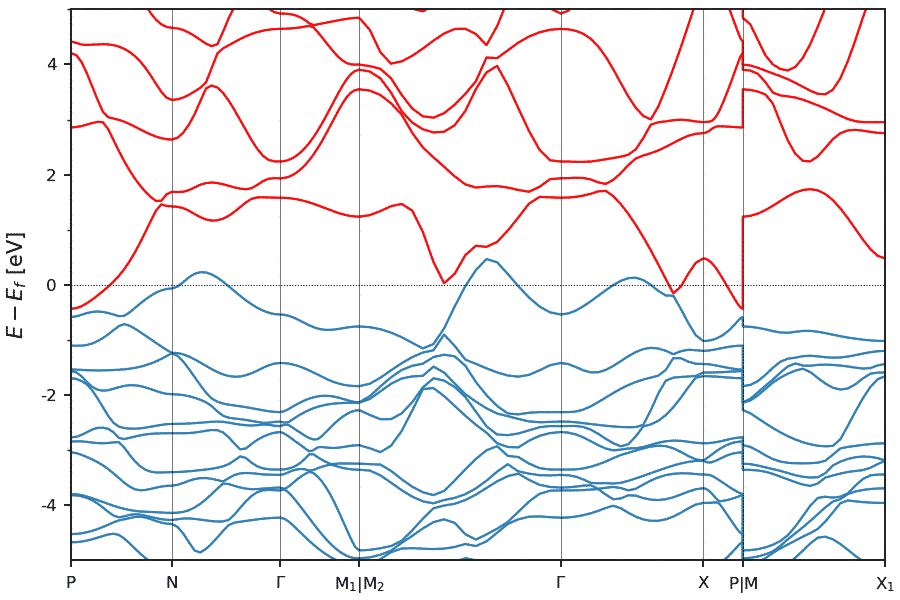}\\
\end{tabular}

\caption{\rTopoSEBR{2}}
\label{fig:rTopo_SEBR2}
\end{figure}

\begin{figure}[ht]
\centering
\begin{tabular}{c c}
\scriptsize{$(\rm{Sr} \rm{F})_{2} \rm{Ti}_{2} \rm{Bi}_{2} \rm{O}$ - \icsdweb{430064} - SG 139 ($I4/mmm$) - SEBR} & \scriptsize{$\rm{Au}_{2} \rm{Zr}$ - \icsdweb{612510} - SG 139 ($I4/mmm$) - SEBR}\\
\tiny{ $\;Z_{2,1}=0\;Z_{2,2}=0\;Z_{2,3}=0\;Z_4=0\;Z_2=0\;Z_8=4$ } & \tiny{ $\;Z_{2,1}=0\;Z_{2,2}=0\;Z_{2,3}=0\;Z_4=3\;Z_2=1\;Z_8=7$ }\\
\includegraphics[width=0.38\textwidth,angle=0]{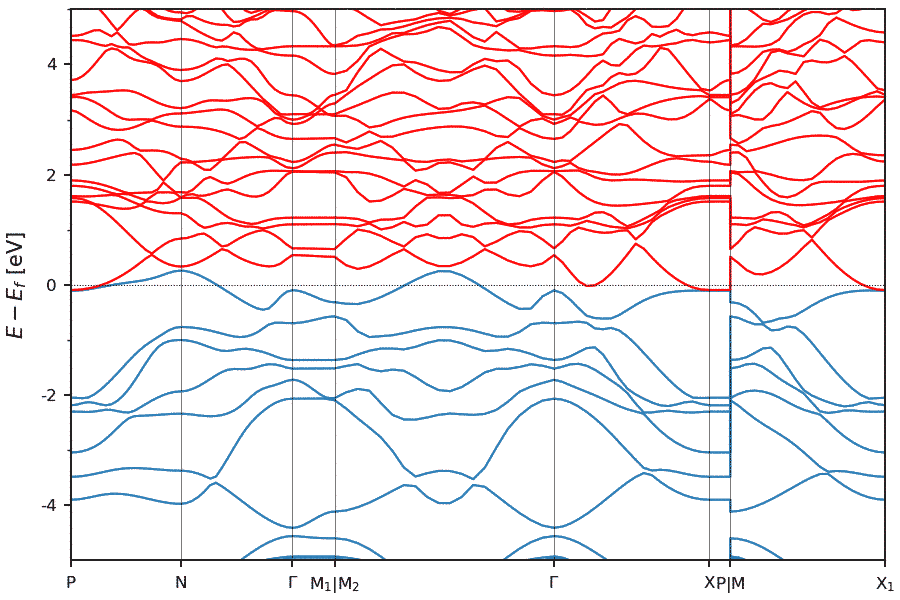} & \includegraphics[width=0.38\textwidth,angle=0]{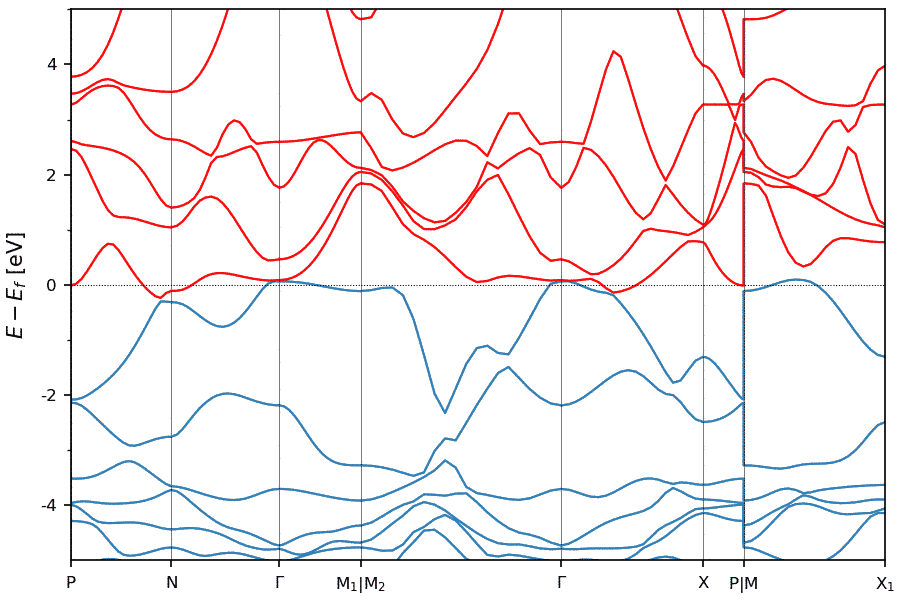}\\
\end{tabular}
\begin{tabular}{c c}
\scriptsize{$\rm{Cd} \rm{Hf}_{2}$ - \icsdweb{619965} - SG 139 ($I4/mmm$) - SEBR} & \scriptsize{$\rm{Hf}_{2} \rm{Pd} \rm{H}_{2}$ - \icsdweb{638175} - SG 139 ($I4/mmm$) - SEBR}\\
\tiny{ $\;Z_{2,1}=1\;Z_{2,2}=1\;Z_{2,3}=1\;Z_4=2\;Z_2=0\;Z_8=2$ } & \tiny{ $\;Z_{2,1}=1\;Z_{2,2}=1\;Z_{2,3}=1\;Z_4=2\;Z_2=0\;Z_8=2$ }\\
\includegraphics[width=0.38\textwidth,angle=0]{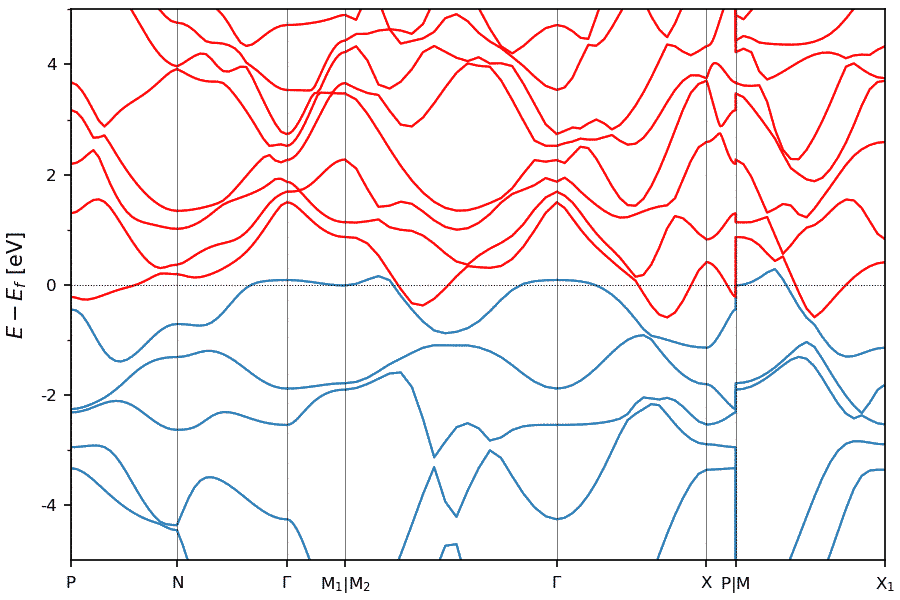} & \includegraphics[width=0.38\textwidth,angle=0]{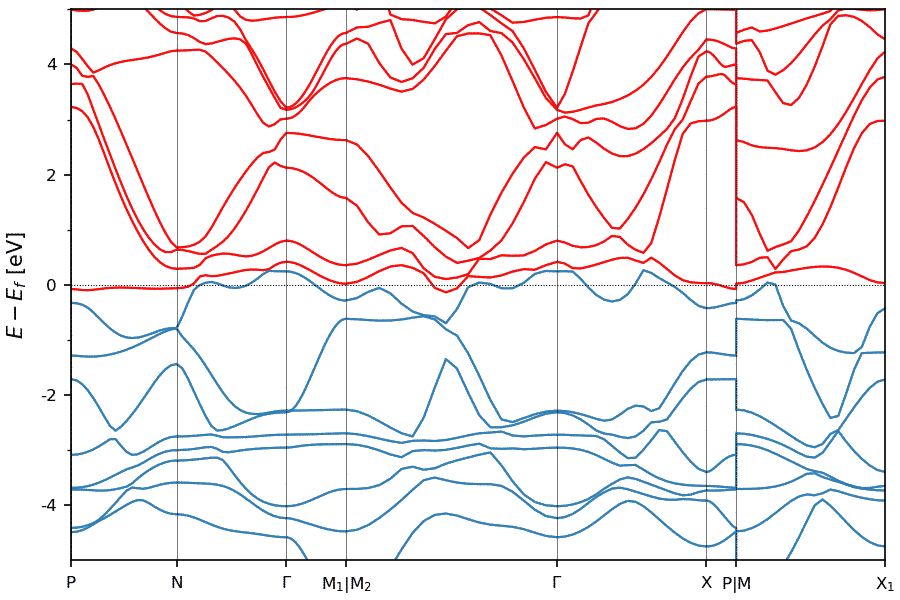}\\
\end{tabular}
\begin{tabular}{c c}
\scriptsize{$\rm{Pd} \rm{Zr}_{2} \rm{H}_{2}$ - \icsdweb{638425} - SG 139 ($I4/mmm$) - SEBR} & \scriptsize{$\rm{Hf}_{2} \rm{Hg}$ - \icsdweb{638564} - SG 139 ($I4/mmm$) - SEBR}\\
\tiny{ $\;Z_{2,1}=1\;Z_{2,2}=1\;Z_{2,3}=1\;Z_4=2\;Z_2=0\;Z_8=2$ } & \tiny{ $\;Z_{2,1}=1\;Z_{2,2}=1\;Z_{2,3}=1\;Z_4=2\;Z_2=0\;Z_8=2$ }\\
\includegraphics[width=0.38\textwidth,angle=0]{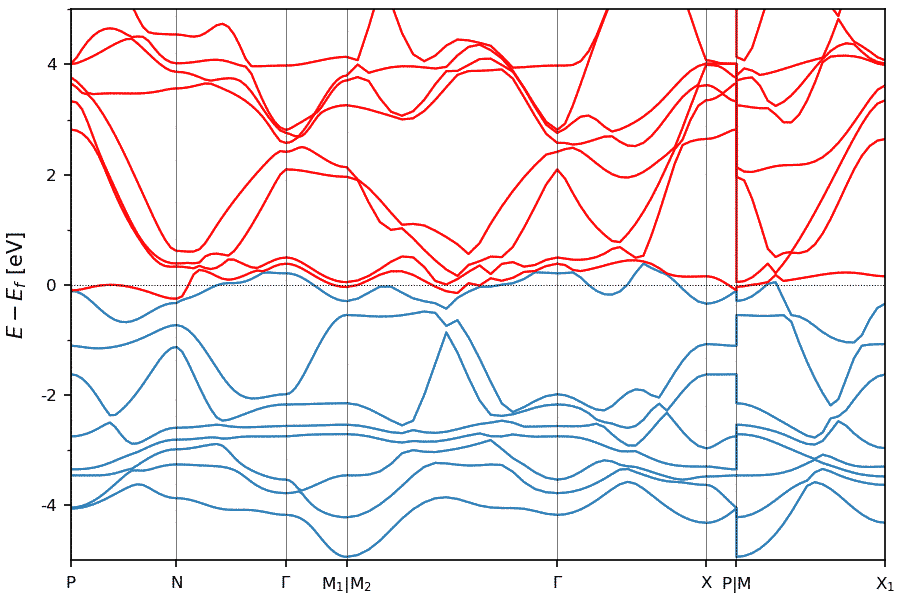} & \includegraphics[width=0.38\textwidth,angle=0]{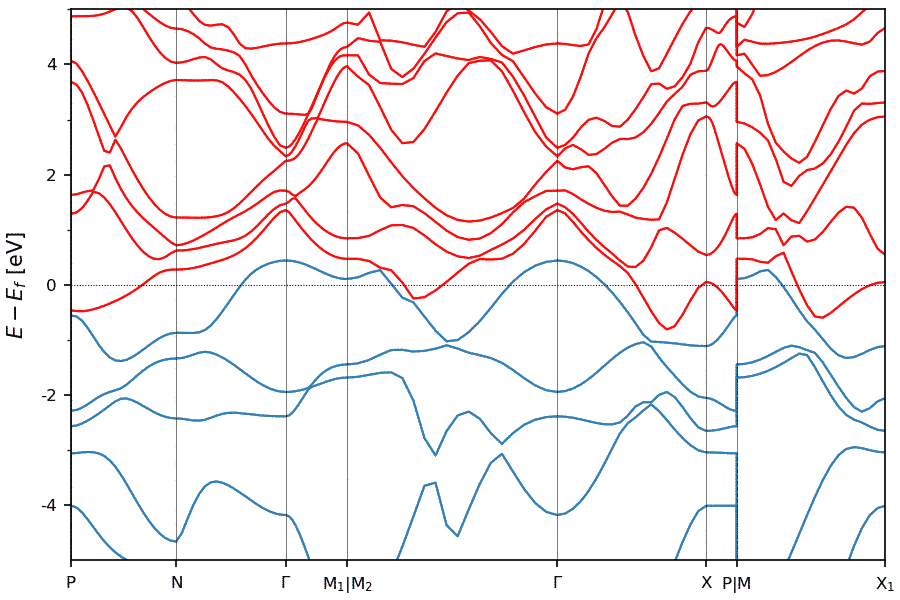}\\
\end{tabular}
\begin{tabular}{c c}
\scriptsize{$\rm{Hf}_{2} \rm{Zn}$ - \icsdweb{639016} - SG 139 ($I4/mmm$) - SEBR} & \scriptsize{$\rm{Pd}_{2} \rm{Zr}$ - \icsdweb{649147} - SG 139 ($I4/mmm$) - SEBR}\\
\tiny{ $\;Z_{2,1}=1\;Z_{2,2}=1\;Z_{2,3}=1\;Z_4=2\;Z_2=0\;Z_8=2$ } & \tiny{ $\;Z_{2,1}=1\;Z_{2,2}=1\;Z_{2,3}=1\;Z_4=1\;Z_2=1\;Z_8=5$ }\\
\includegraphics[width=0.38\textwidth,angle=0]{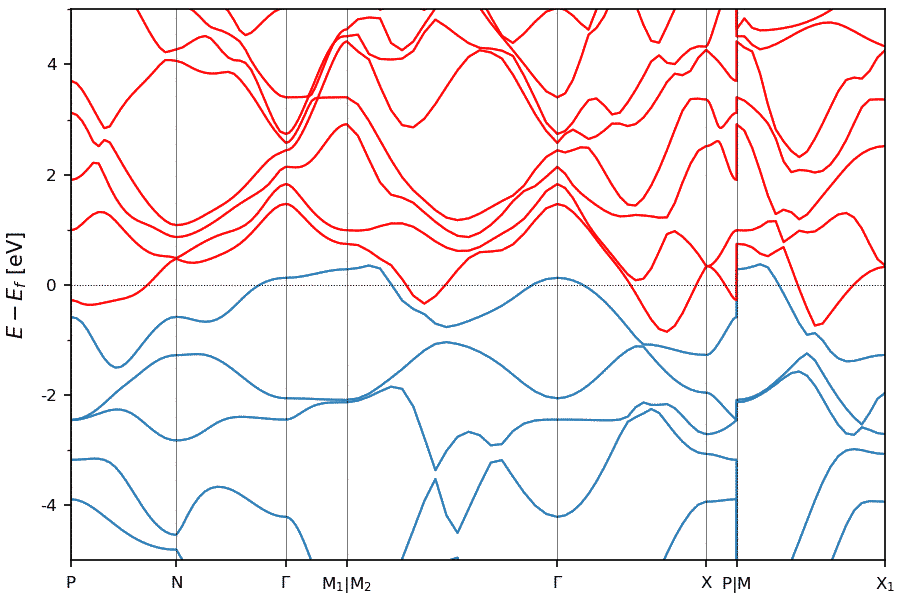} & \includegraphics[width=0.38\textwidth,angle=0]{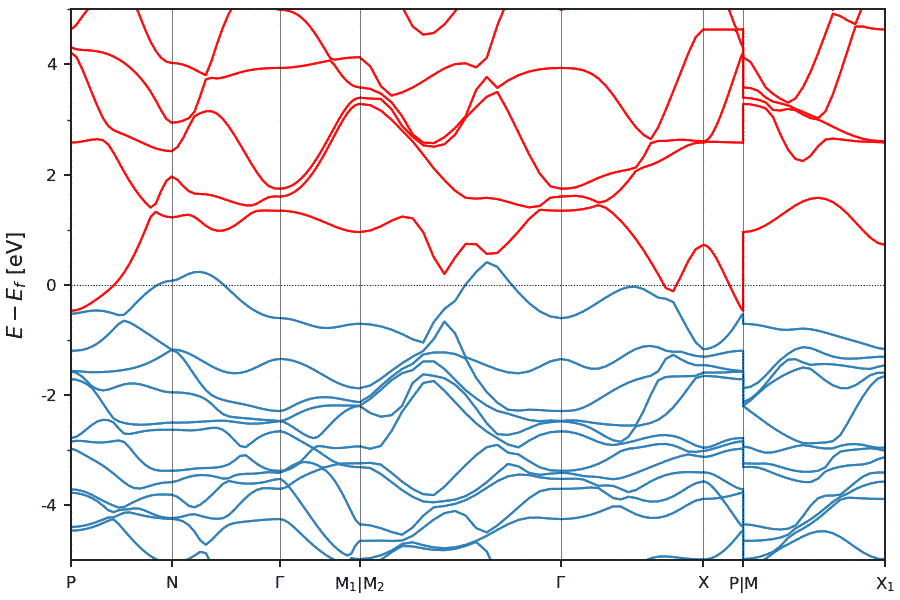}\\
\end{tabular}

\caption{\rTopoSEBR{3}}
\label{fig:rTopo_SEBR3}
\end{figure}

\begin{figure}[ht]
\centering
\begin{tabular}{c c}
\scriptsize{$\rm{Pt}_{3} \rm{Sb}$ - \icsdweb{649555} - SG 139 ($I4/mmm$) - SEBR} & \scriptsize{$\rm{Zn} \rm{Zr}_{2}$ - \icsdweb{653511} - SG 139 ($I4/mmm$) - SEBR}\\
\tiny{ $\;Z_{2,1}=1\;Z_{2,2}=1\;Z_{2,3}=1\;Z_4=1\;Z_2=1\;Z_8=1$ } & \tiny{ $\;Z_{2,1}=1\;Z_{2,2}=1\;Z_{2,3}=1\;Z_4=2\;Z_2=0\;Z_8=2$ }\\
\includegraphics[width=0.38\textwidth,angle=0]{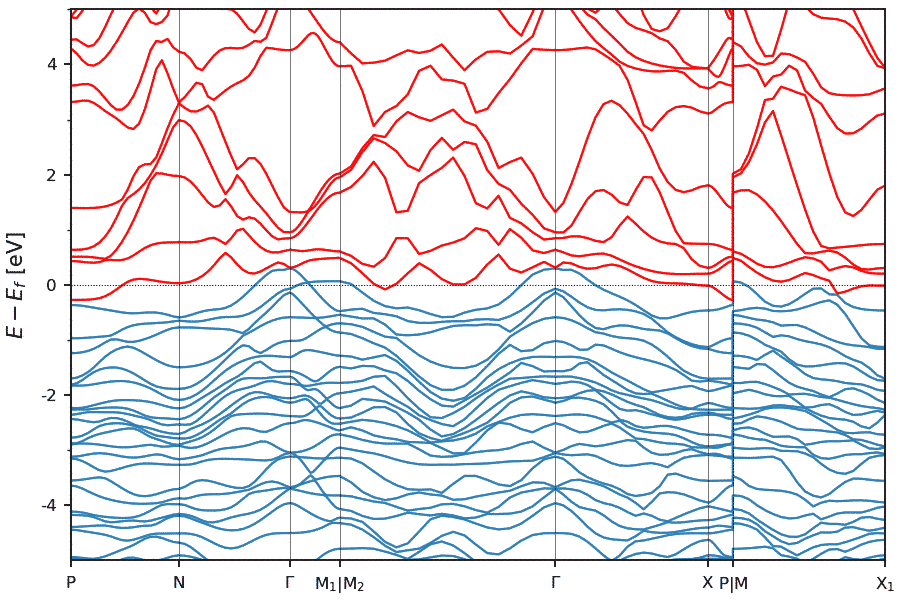} & \includegraphics[width=0.38\textwidth,angle=0]{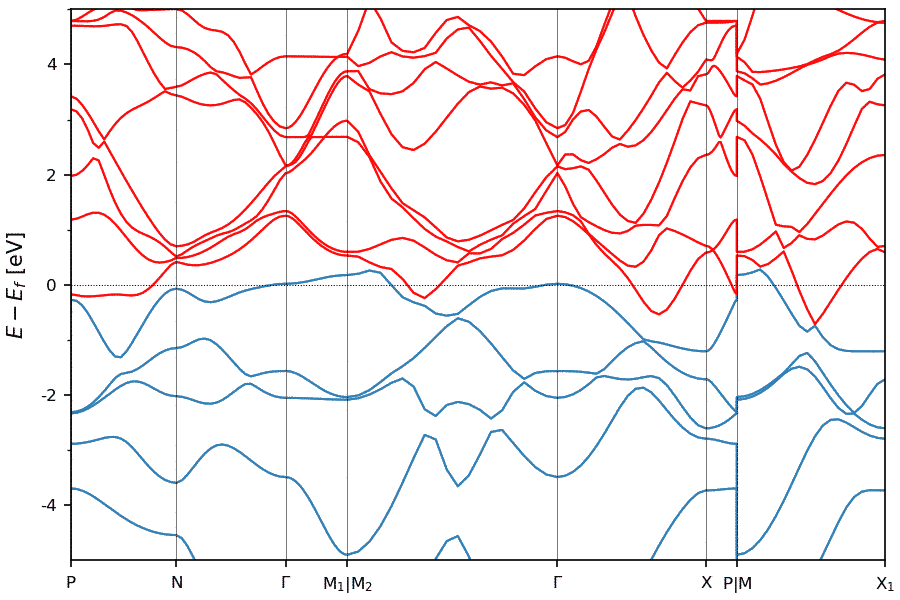}\\
\end{tabular}
\begin{tabular}{c c}
\scriptsize{$\rm{Ag}_{2} \rm{O}$ - \icsdweb{20368} - SG 164 ($P\bar{3}m1$) - SEBR} & \scriptsize{$\rm{Bi}_{2} \rm{Se}_{2}$ - \icsdweb{20458} - SG 164 ($P\bar{3}m1$) - SEBR}\\
\tiny{ $\;Z_{2,1}=0\;Z_{2,2}=0\;Z_{2,3}=0\;Z_4=3$ } & \tiny{ $\;Z_{2,1}=0\;Z_{2,2}=0\;Z_{2,3}=1\;Z_4=0$ }\\
\includegraphics[width=0.38\textwidth,angle=0]{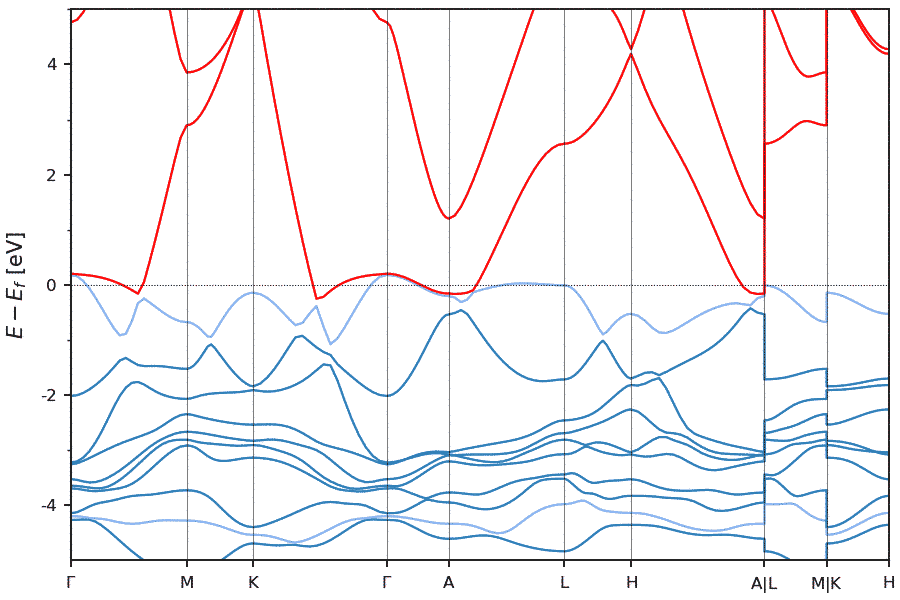} & \includegraphics[width=0.38\textwidth,angle=0]{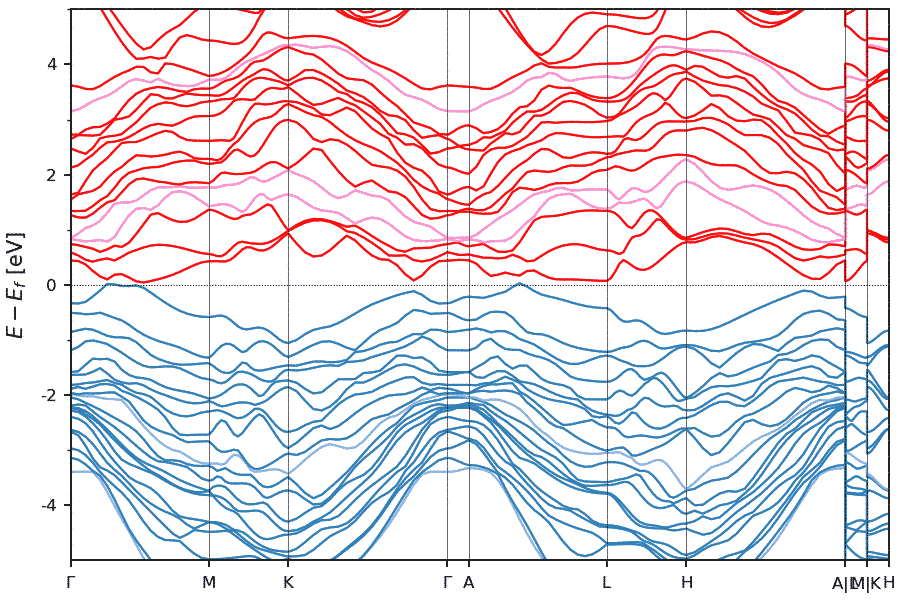}\\
\end{tabular}
\begin{tabular}{c c}
\scriptsize{$\rm{Sb}_{2} \rm{Te}_{2}$ - \icsdweb{20459} - SG 164 ($P\bar{3}m1$) - SEBR} & \scriptsize{$\rm{Ta}_{2} \rm{S}_{2} \rm{C}$ - \icsdweb{23790} - SG 164 ($P\bar{3}m1$) - SEBR}\\
\tiny{ $\;Z_{2,1}=0\;Z_{2,2}=0\;Z_{2,3}=1\;Z_4=3$ } & \tiny{ $\;Z_{2,1}=0\;Z_{2,2}=0\;Z_{2,3}=1\;Z_4=1$ }\\
\includegraphics[width=0.38\textwidth,angle=0]{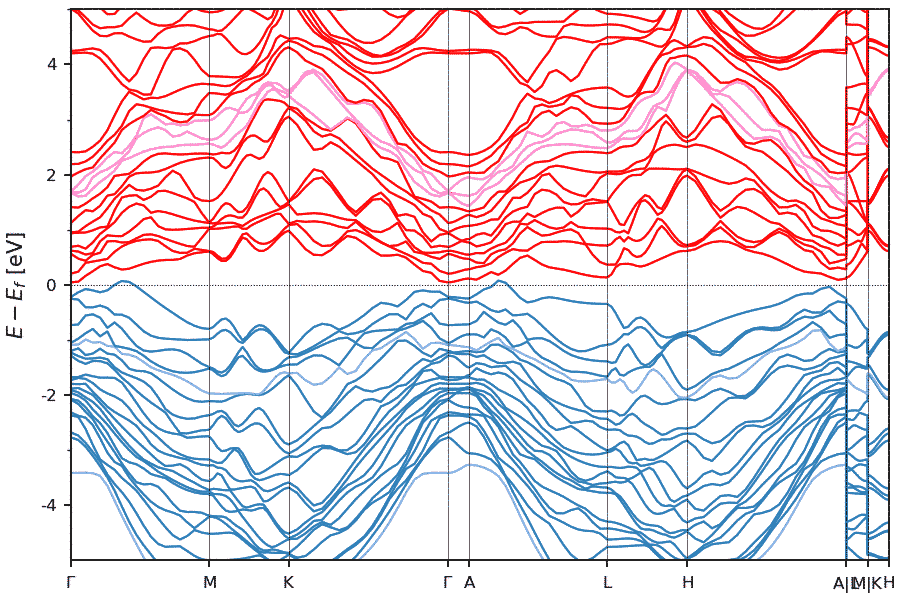} & \includegraphics[width=0.38\textwidth,angle=0]{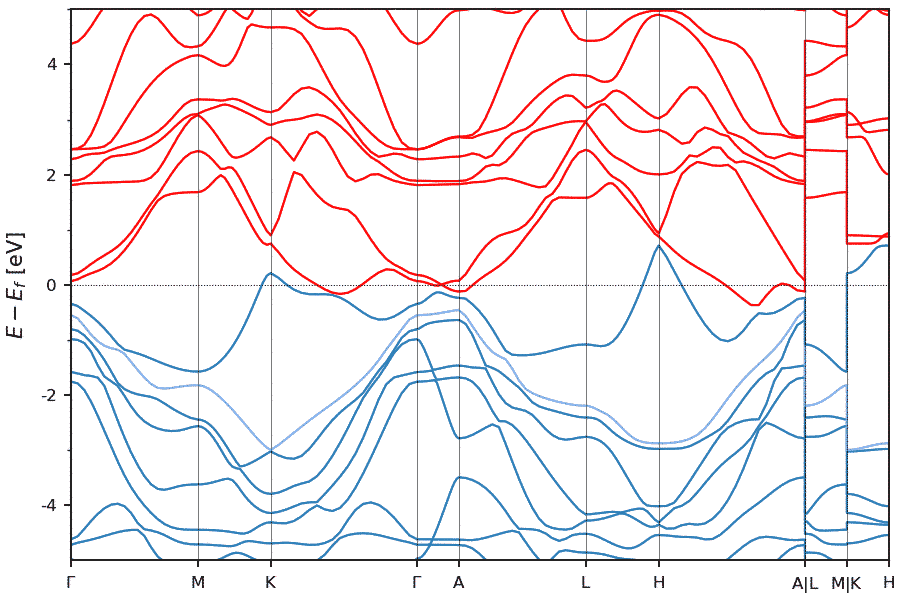}\\
\end{tabular}
\begin{tabular}{c c}
\scriptsize{$\rm{Bi}_{2} \rm{Pb}_{2} \rm{Te}_{5}$ - \icsdweb{42708} - SG 164 ($P\bar{3}m1$) - SEBR} & \scriptsize{$\rm{Bi} \rm{Te}$ - \icsdweb{100654} - SG 164 ($P\bar{3}m1$) - SEBR}\\
\tiny{ $\;Z_{2,1}=0\;Z_{2,2}=0\;Z_{2,3}=1\;Z_4=0$ } & \tiny{ $\;Z_{2,1}=0\;Z_{2,2}=0\;Z_{2,3}=1\;Z_4=2$ }\\
\includegraphics[width=0.38\textwidth,angle=0]{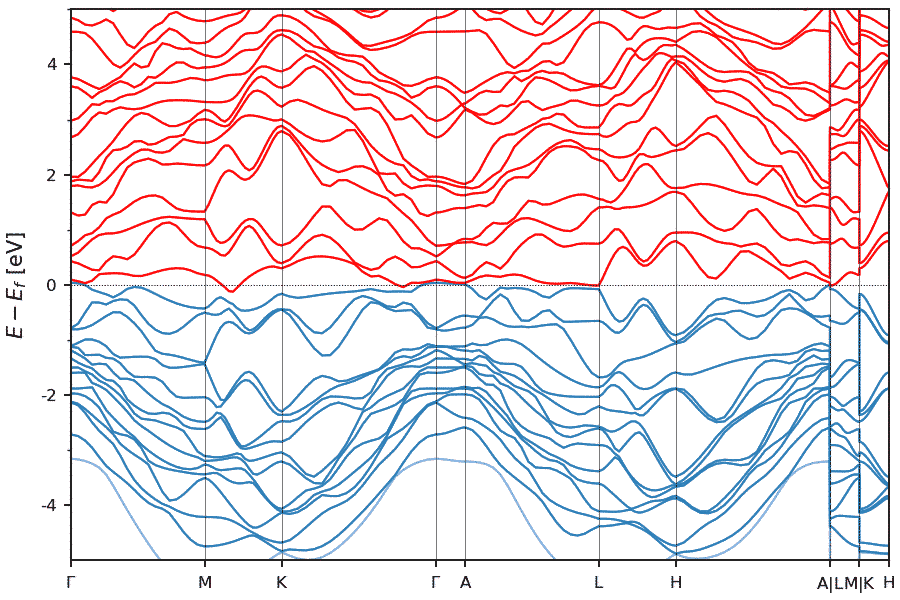} & \includegraphics[width=0.38\textwidth,angle=0]{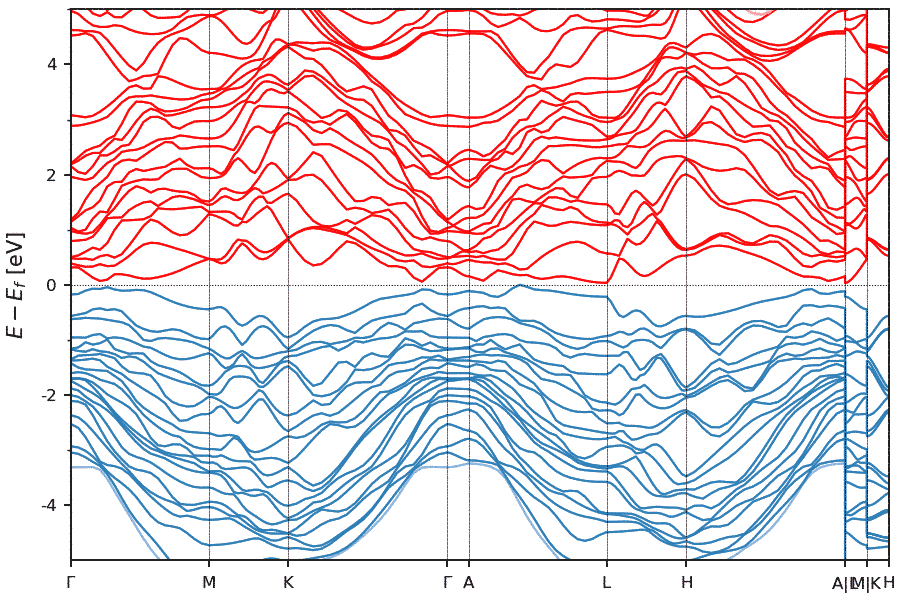}\\
\end{tabular}

\caption{\rTopoSEBR{4}}
\label{fig:rTopo_SEBR4}
\end{figure}

\begin{figure}[ht]
\centering
\begin{tabular}{c c}
\scriptsize{$\rm{W}_{2} \rm{C}$ - \icsdweb{619097} - SG 164 ($P\bar{3}m1$) - SEBR} & \scriptsize{$\rm{Bi}_{2} \rm{Mg}_{3}$ - \icsdweb{659569} - SG 164 ($P\bar{3}m1$) - SEBR}\\
\tiny{ $\;Z_{2,1}=0\;Z_{2,2}=0\;Z_{2,3}=0\;Z_4=2$ } & \tiny{ $\;Z_{2,1}=0\;Z_{2,2}=0\;Z_{2,3}=0\;Z_4=3$ }\\
\includegraphics[width=0.38\textwidth,angle=0]{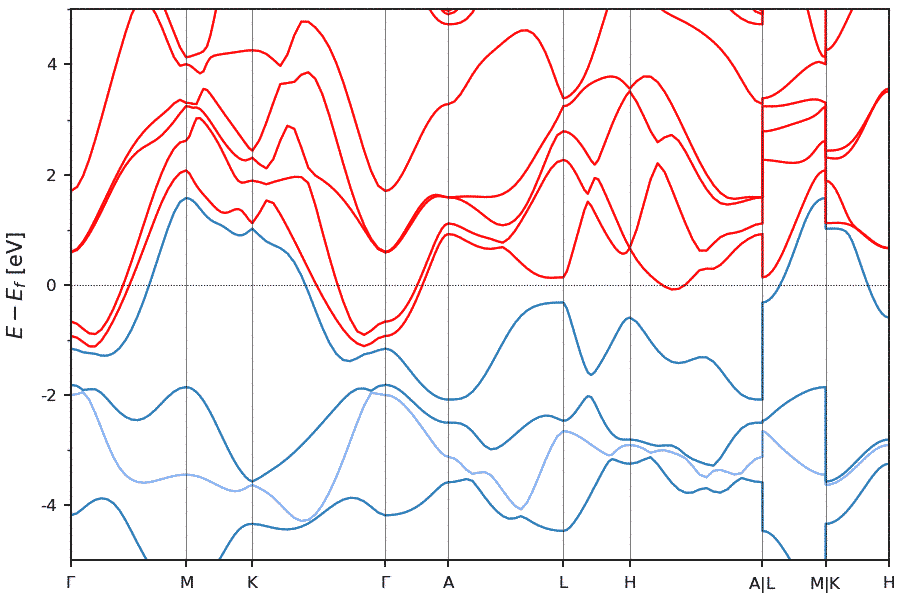} & \includegraphics[width=0.38\textwidth,angle=0]{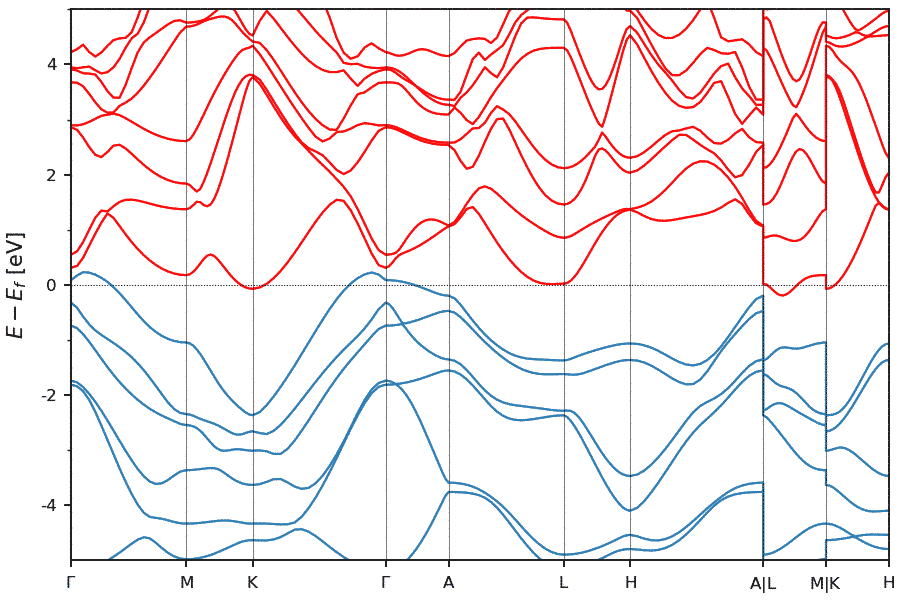}\\
\end{tabular}
\begin{tabular}{c c}
\scriptsize{$\rm{Zr} \rm{Cl}$ - \icsdweb{869} - SG 166 ($R\bar{3}m$) - SEBR} & \scriptsize{$\rm{Ta}_{2} \rm{S}_{2} \rm{C}$ - \icsdweb{23791} - SG 166 ($R\bar{3}m$) - SEBR}\\
\tiny{ $\;Z_{2,1}=1\;Z_{2,2}=1\;Z_{2,3}=1\;Z_4=0$ } & \tiny{ $\;Z_{2,1}=1\;Z_{2,2}=1\;Z_{2,3}=1\;Z_4=1$ }\\
\includegraphics[width=0.38\textwidth,angle=0]{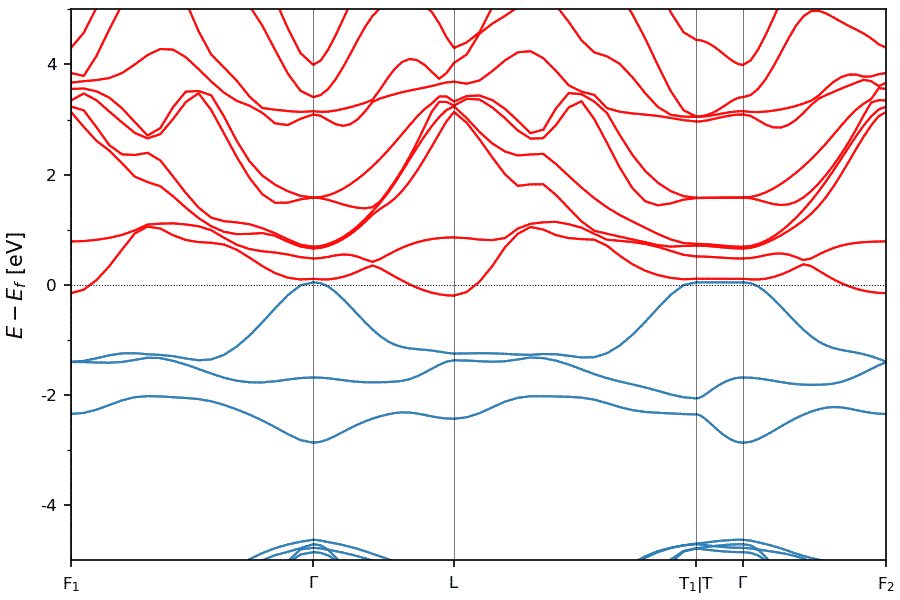} & \includegraphics[width=0.38\textwidth,angle=0]{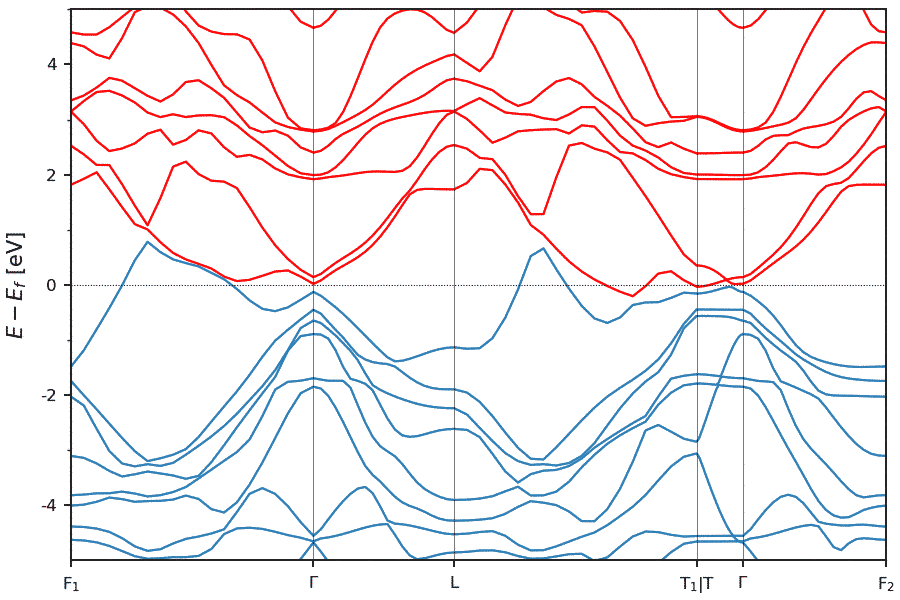}\\
\end{tabular}
\begin{tabular}{c c}
\scriptsize{$\rm{W} \rm{N}_{2}$ - \icsdweb{30364} - SG 166 ($R\bar{3}m$) - SEBR} & \scriptsize{$\rm{Te}$ - \icsdweb{52499} - SG 166 ($R\bar{3}m$) - SEBR}\\
\tiny{ $\;Z_{2,1}=0\;Z_{2,2}=0\;Z_{2,3}=0\;Z_4=3$ } & \tiny{ $\;Z_{2,1}=0\;Z_{2,2}=0\;Z_{2,3}=0\;Z_4=3$ }\\
\includegraphics[width=0.38\textwidth,angle=0]{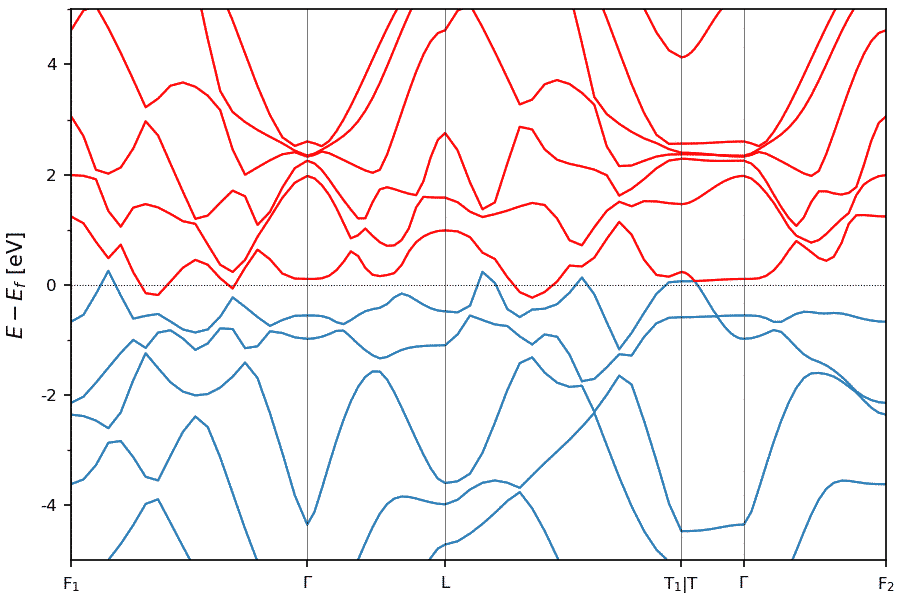} & \includegraphics[width=0.38\textwidth,angle=0]{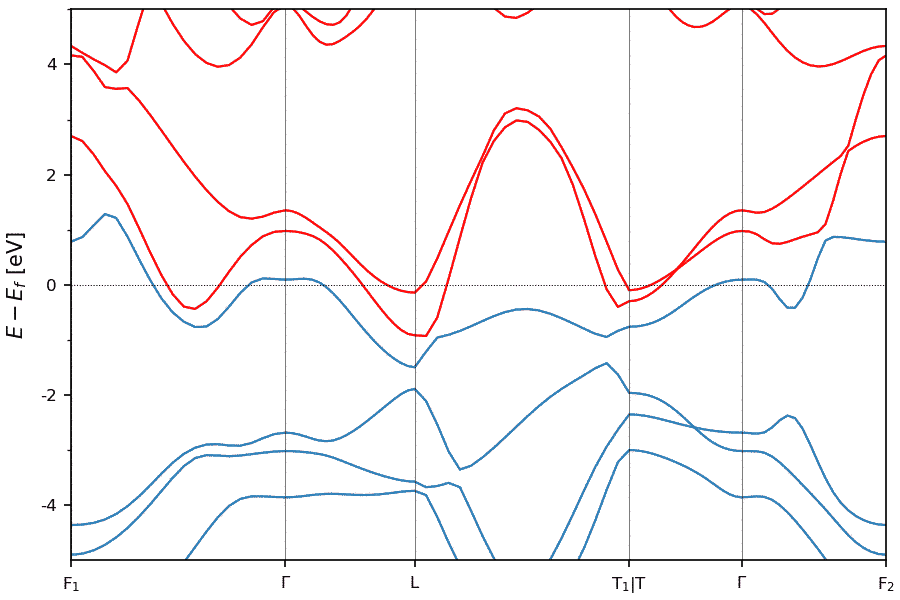}\\
\end{tabular}
\begin{tabular}{c c}
\scriptsize{$\rm{Bi}$ - \icsdweb{53797} - SG 166 ($R\bar{3}m$) - SEBR} & \scriptsize{$\rm{Sr} (\rm{Sn}_{2} \rm{As}_{2})$ - \icsdweb{82371} - SG 166 ($R\bar{3}m$) - SEBR}\\
\tiny{ $\;Z_{2,1}=0\;Z_{2,2}=0\;Z_{2,3}=0\;Z_4=2$ } & \tiny{ $\;Z_{2,1}=0\;Z_{2,2}=0\;Z_{2,3}=0\;Z_4=3$ }\\
\includegraphics[width=0.38\textwidth,angle=0]{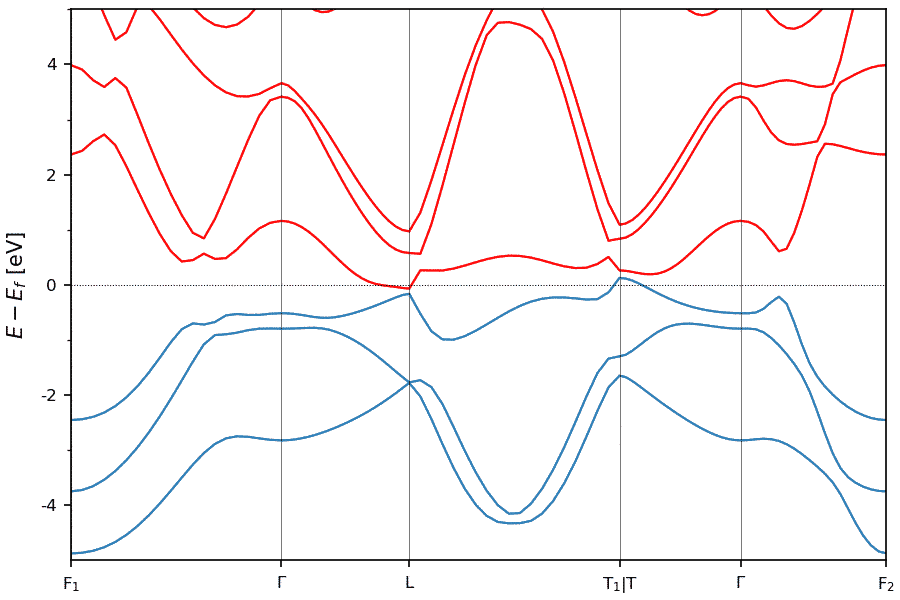} & \includegraphics[width=0.38\textwidth,angle=0]{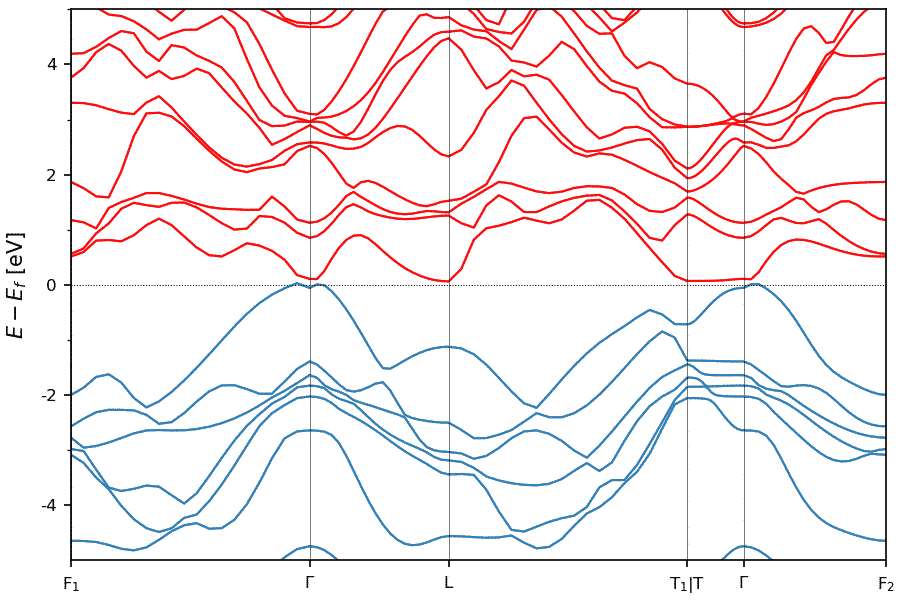}\\
\end{tabular}

\caption{\rTopoSEBR{5}}
\label{fig:rTopo_SEBR5}
\end{figure}

\begin{figure}[ht]
\centering
\begin{tabular}{c c}
\scriptsize{$\rm{Y}_{2} \rm{C}$ - \icsdweb{96188} - SG 166 ($R\bar{3}m$) - SEBR} & \scriptsize{$\rm{Ca} \rm{Si}_{2}$ - \icsdweb{248518} - SG 166 ($R\bar{3}m$) - SEBR}\\
\tiny{ $\;Z_{2,1}=1\;Z_{2,2}=1\;Z_{2,3}=1\;Z_4=1$ } & \tiny{ $\;Z_{2,1}=0\;Z_{2,2}=0\;Z_{2,3}=0\;Z_4=1$ }\\
\includegraphics[width=0.38\textwidth,angle=0]{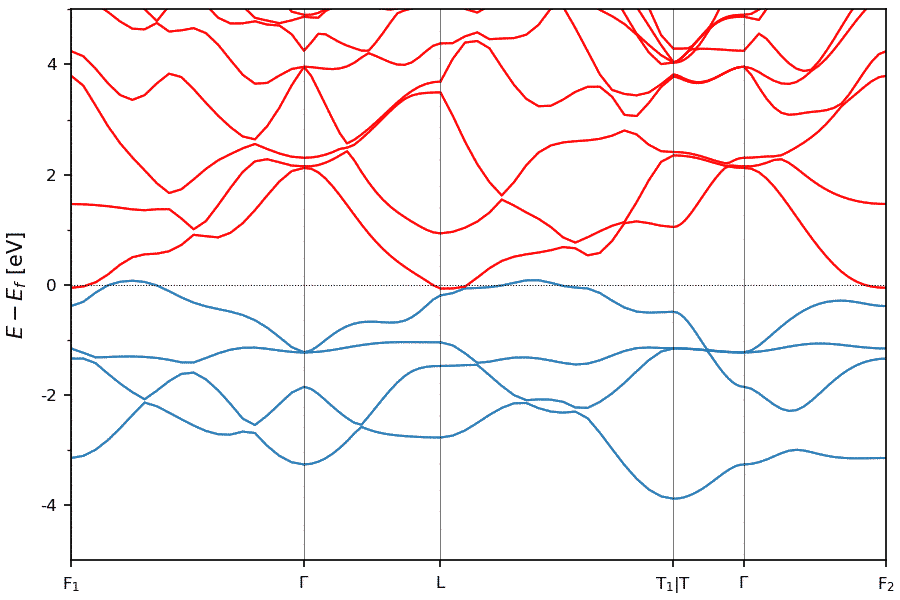} & \includegraphics[width=0.38\textwidth,angle=0]{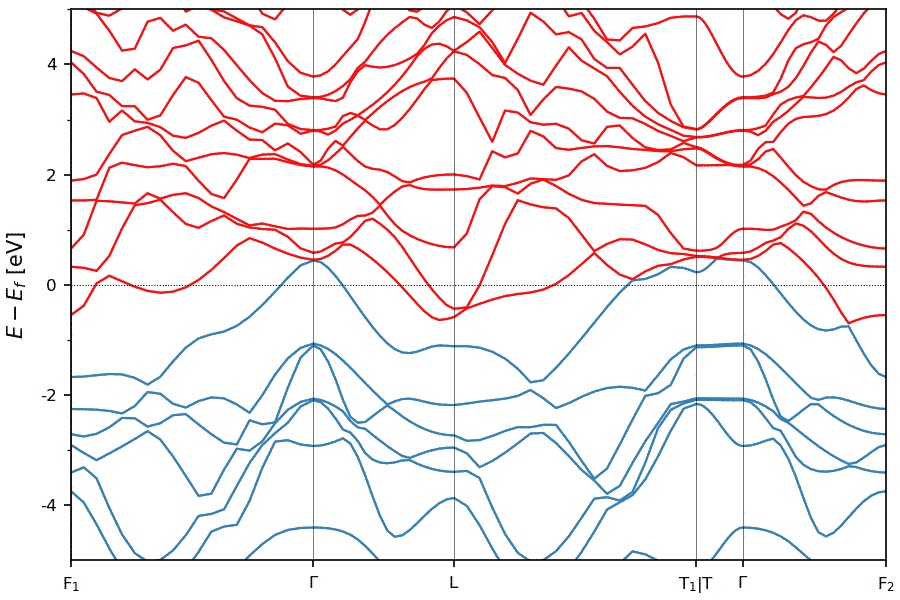}\\
\end{tabular}
\begin{tabular}{c c}
\scriptsize{$\rm{Ag}_{2} \rm{Ni} \rm{O}_{2}$ - \icsdweb{412279} - SG 166 ($R\bar{3}m$) - SEBR} & \scriptsize{$\rm{Bi}_{4} \rm{Se}_{3}$ - \icsdweb{617074} - SG 166 ($R\bar{3}m$) - SEBR}\\
\tiny{ $\;Z_{2,1}=1\;Z_{2,2}=1\;Z_{2,3}=1\;Z_4=1$ } & \tiny{ $\;Z_{2,1}=1\;Z_{2,2}=1\;Z_{2,3}=1\;Z_4=0$ }\\
\includegraphics[width=0.38\textwidth,angle=0]{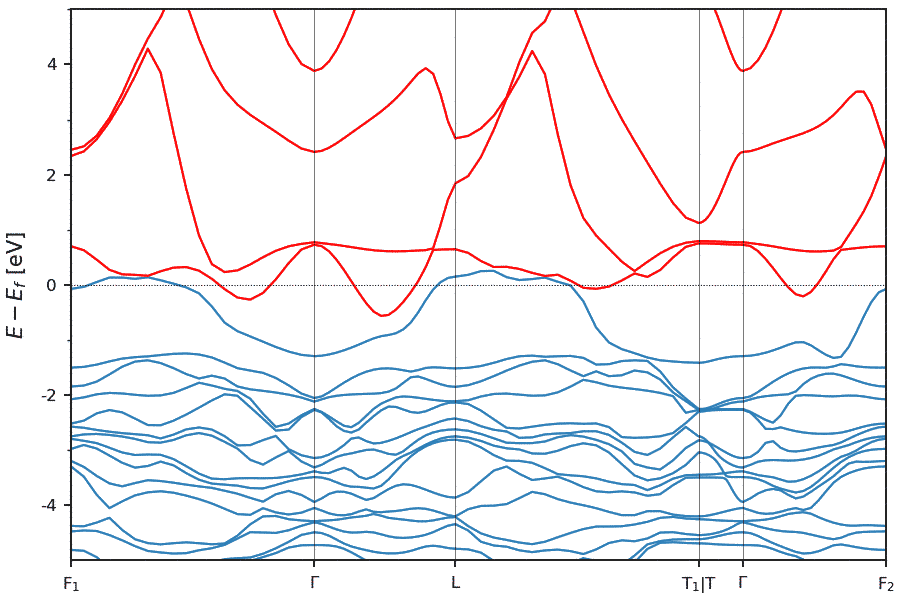} & \includegraphics[width=0.38\textwidth,angle=0]{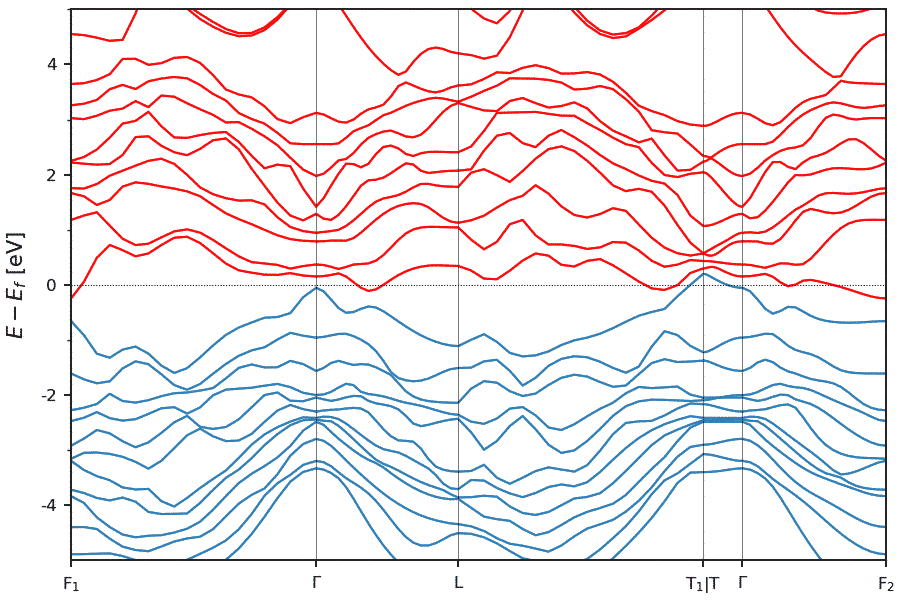}\\
\end{tabular}
\begin{tabular}{c c}
\scriptsize{$\rm{Pd}_{3} \rm{Tl}_{2} \rm{S}_{2}$ - \icsdweb{648762} - SG 166 ($R\bar{3}m$) - SEBR} & \scriptsize{$\rm{In}_{2} \rm{Te}_{3}$ - \icsdweb{657607} - SG 166 ($R\bar{3}m$) - SEBR}\\
\tiny{ $\;Z_{2,1}=1\;Z_{2,2}=1\;Z_{2,3}=1\;Z_4=0$ } & \tiny{ $\;Z_{2,1}=1\;Z_{2,2}=1\;Z_{2,3}=1\;Z_4=1$ }\\
\includegraphics[width=0.38\textwidth,angle=0]{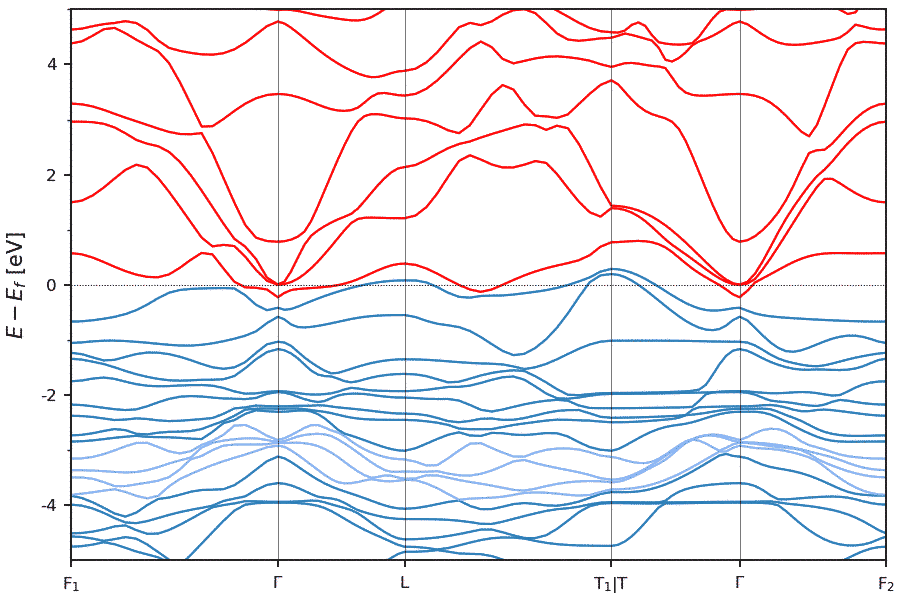} & \includegraphics[width=0.38\textwidth,angle=0]{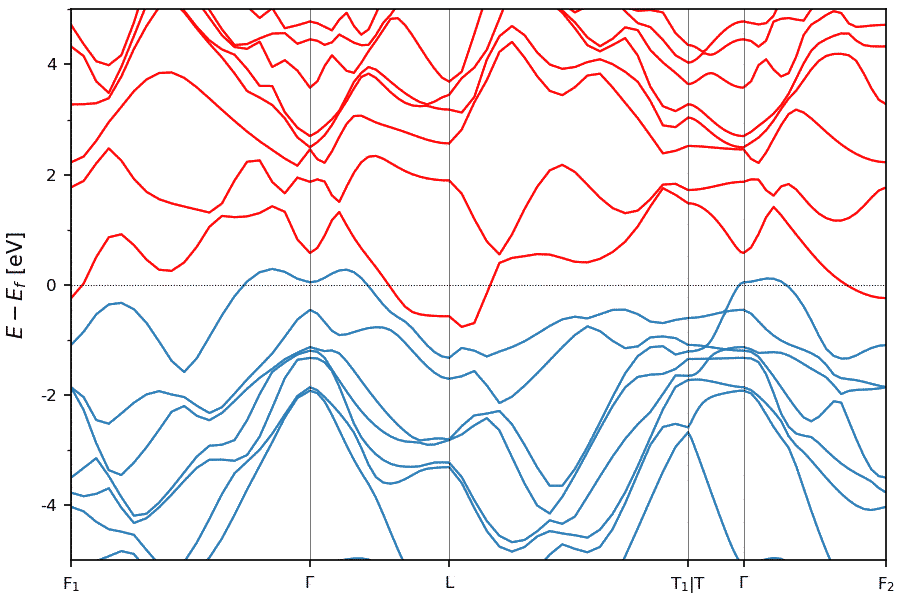}\\
\end{tabular}
\begin{tabular}{c c}
\scriptsize{$\rm{Ga}_{2} \rm{Te}_{3}$ - \icsdweb{657608} - SG 166 ($R\bar{3}m$) - SEBR} & \scriptsize{$\rm{Tl} \rm{Pd}_{3} \rm{H}$ - \icsdweb{247273} - SG 221 ($Pm\bar{3}m$) - SEBR}\\
\tiny{ $\;Z_{2,1}=1\;Z_{2,2}=1\;Z_{2,3}=1\;Z_4=1$ } & \tiny{ $\;Z_{2,1}=1\;Z_{2,2}=1\;Z_{2,3}=1\;Z_4=0\;Z_{4m,\pi}=3\;Z_2=0\;Z_8=0$ }\\
\includegraphics[width=0.38\textwidth,angle=0]{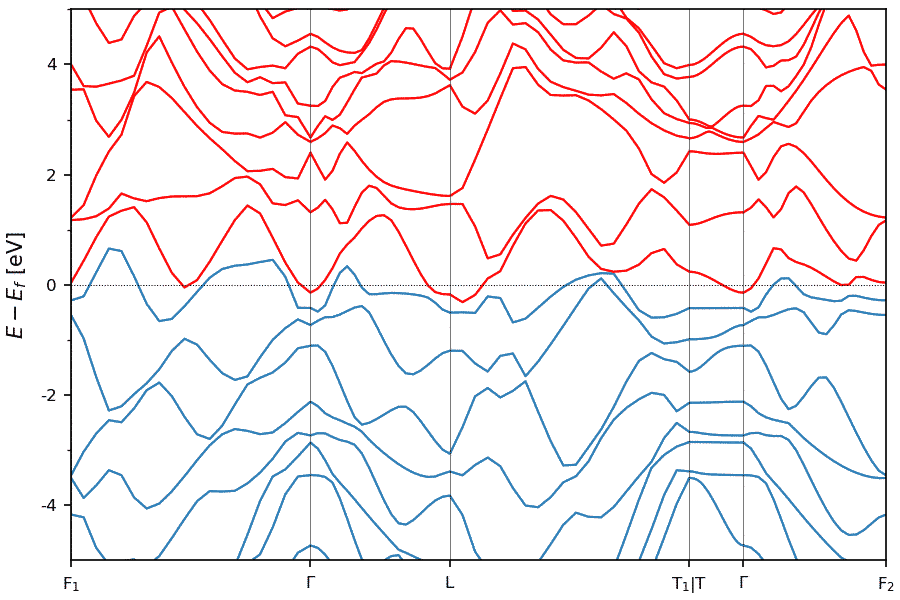} & \includegraphics[width=0.38\textwidth,angle=0]{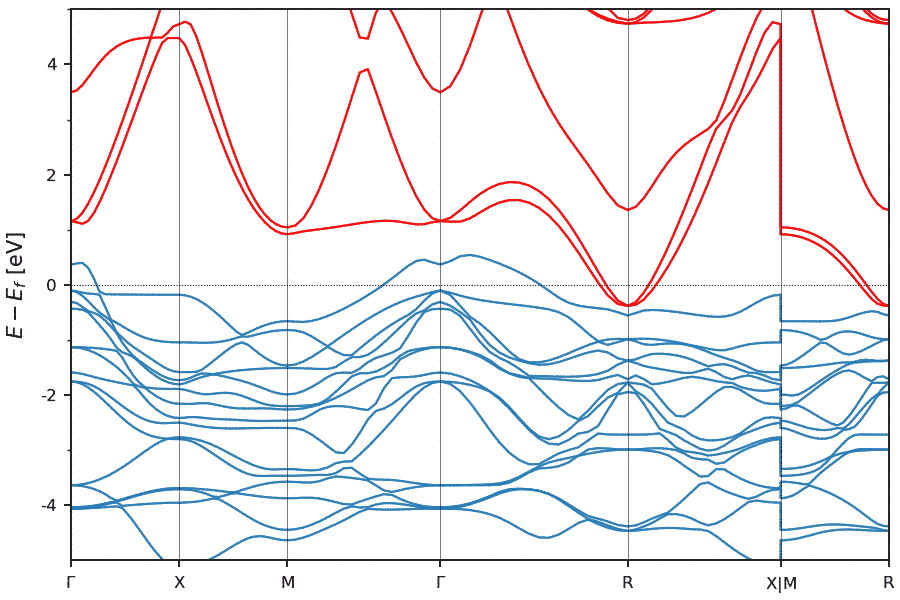}\\
\end{tabular}

\caption{\rTopoSEBR{6}}
\label{fig:rTopo_SEBR6}
\end{figure}

\clearpage

\subsection{Enforced Topological Semimetals}
\label{App:semimetals}

In this section, we will list the enforced semimetals with the smallest bulk Fermi pockets.  First, in \supappref{App:ESFD}, we will list the enforced semimetals whose nodal degeneracies are pinned to high-symmetry BZ points (ESFD), and then, in \supappref{App:ES}, we will list the enforced semimetals whose nodal degeneracies lie along high-symmetry BZ lines (ES).

\subsubsection{ESFD-Classified Semimetals}
\label{App:ESFD}

In this section, we list the enforced semimetals whose nodal degeneracies are pinned to high-symmetry BZ points (ESFD in the nomenclature established in Ref.~\onlinecite{AndreiMaterials}).  Most generally, ESFD phases can include both conventional (filling-)~\cite{WPVZ,WiederLayers,KramersNodalLine} enforced Dirac~\cite{SteveDirac,JuliaDirac,Steve2D,SteveMagnet} and Weyl~\cite{KramersWeyl} semimetals, as well as unconventional (multifold) fermion~\cite{DDP,NewFermions,KramersWeyl,RhSiArc,CoSiArc,PdSb2} semimetals.  Depending on whether or not the little groups of the ${\bf k}$ points of the enforced nodal degeneracies (as well as the full space group) contain rotoinversion symmetries, the ESFD semimetals identified in this section may further be classified as topologically chiral or achiral~\cite{KramersWeyl,RhSiArc}.  ESFD phases in structurally chiral crystals have in particular been highlighted for hosting monopole-like spin textures~\cite{KramersWeyl,ChiralTellurium}, exhibiting photoresponse signatures of topological chirality~\cite{FerNatCom,BarryMultifold,ReesMoorePhotoExp,LiangPennOpticalCoSi,LiangPennOpticalCoSi2}, and as possible venues for realizing topological superconductivity~\cite{AuBeChiralSC,KTLawMultifoldSC,TayRongKramersWeylSC}.

First, in Fig.~\ref{fig:strongSOC_3fold_ESFD}, we list the ESFD materials with the simplest bulk Fermi surfaces and threefold-degenerate spin-1 fermions~\cite{NewFermions} close to $E_{F}$.  Because the spin-1 fermions in the materials in Fig.~\ref{fig:strongSOC_3fold_ESFD} all lie at high-symmetry ${\bf k}$ points with rotoinversion (specifically $S_{4}$) symmetry, then the threefold degeneracies carry net-zero chiral charges, and do not exhibit associated topological surface Fermi arcs~\cite{KramersWeyl,RhSiArc}.  Next, in Figs.~\ref{fig:strongSOC_NOTCHIRAL_4fold_NewFermion_ESFD1},~\ref{fig:strongSOC_NOTCHIRAL_4fold_NewFermion_ESFD2},~\ref{fig:strongSOC_NOTCHIRAL_4fold_NewFermion_ESFD3},~\ref{fig:strongSOC_NOTCHIRAL_4fold_NewFermion_ESFD4},~\ref{fig:strongSOC_NOTCHIRAL_4fold_NewFermion_ESFD5},~\ref{fig:strongSOC_NOTCHIRAL_4fold_NewFermion_ESFD6},~\ref{fig:strongSOC_NOTCHIRAL_4fold_NewFermion_ESFD7},~\ref{fig:strongSOC_NOTCHIRAL_4fold_NewFermion_ESFD8},~\ref{fig:strongSOC_NOTCHIRAL_4fold_NewFermion_ESFD9}, and~\ref{fig:strongSOC_NOTCHIRAL_4fold_NewFermion_ESFD10}, we list the materials with well-isolated achiral fourfold-degenerate spin-3/2 fermions~\cite{NewFermions} close to $E_{F}$.  Then, in Figs.~\ref{fig:strongSOC_chiral_NewFermion_ESFD1},~\ref{fig:strongSOC_chiral_NewFermion_ESFD2},~\ref{fig:strongSOC_chiral_NewFermion_ESFD3},~\ref{fig:strongSOC_chiral_NewFermion_ESFD4},~\ref{fig:strongSOC_chiral_NewFermion_ESFD5}, and~\ref{fig:strongSOC_chiral_NewFermion_ESFD6}, we list the cubic ESFD materials with chiral crystal structures and enforced chiral fermions~\cite{KramersWeyl,NewFermions,RhSiArc,CoSiArc} close to $E_{F}$, which in most cases include chiral fourfold-degenerate spin-3/2 fermions and chiral sixfold-degenerate double-spin-1 fermions.  The materials in Figs.~\ref{fig:strongSOC_chiral_NewFermion_ESFD1},~\ref{fig:strongSOC_chiral_NewFermion_ESFD2},~\ref{fig:strongSOC_chiral_NewFermion_ESFD3},~\ref{fig:strongSOC_chiral_NewFermion_ESFD4},~\ref{fig:strongSOC_chiral_NewFermion_ESFD5}, and~\ref{fig:strongSOC_chiral_NewFermion_ESFD6} notably include members of the B20 chiral crystal family, such as CoSi [\icsdweb{189221}, SG 198 ($P2_{1}3$)], which have been shown in recent theoretical and experimental investigations to exhibit large topological surface Fermi arcs~\cite{RhSiArc,CoSiArc,CoSiObserveJapan,CoSiObserveHasan,CoSiObserveChina,AlPtObserve,PdGaObserve}.  In the following figures -- Figs.~\ref{fig:strongSOC_8fold_ESFD1},~\ref{fig:strongSOC_8fold_ESFD2}, and~\ref{fig:strongSOC_8fold_ESFD3} -- we then show the ESFD materials with eightfold-degenerate double Dirac fermions~\cite{DDP,NewFermions} close to $E_{F}$ that do not exhibit clear magnetic instabilities in DFT.  Specifically, in a number of ESFD compounds -- most notably candidate double Dirac semimetals such as CuBi$_2$O$_4$ [\icsdweb{15865}, SG 130 ($P4/ncc$)]~\cite{DDPMott1,DDPMott2,NewFermions} -- experimental investigations have revealed that the nodal degeneracies become gapped by electron-electron interactions, leading to correlated Mott -- or possibly more exotic -- insulating phases, even above the transition temperature for magnetic ordering.  The candidate double Dirac semimetals shown in Figs.~\ref{fig:strongSOC_8fold_ESFD1},~\ref{fig:strongSOC_8fold_ESFD2}, and~\ref{fig:strongSOC_8fold_ESFD3} exhibit less sharply peaked densities of states at $E_{F}$ than CuBi$_2$O$_4$, and may therefore be less susceptible to interaction-driven (semi)metal-insulator transitions.  Finally, in Figs.~\ref{fig:strongSOC_quadratic_other_ESFD1} and~\ref{fig:strongSOC_quadratic_other_ESFD2}, we list the ESFD materials with other forms of nodal fermions at $E_{F}$, such as fourfold degeneracies with quadratic dispersion.


\begin{figure}[ht]
\centering


\caption{The ESFD-classified topological semimetals with the simplest bulk Fermi surfaces and threefold-degenerate spin-1 fermions close to $E_{F}$.}
\label{fig:strongSOC_3fold_ESFD}
\end{figure}


\begin{figure}[ht]
\centering
%

\caption{\strongSOCNOTCHIRALfoldNewFermionESFD{1}}
\label{fig:strongSOC_NOTCHIRAL_4fold_NewFermion_ESFD1}
\end{figure}

\begin{figure}[ht]
\centering
%

\caption{\strongSOCNOTCHIRALfoldNewFermionESFD{2}}
\label{fig:strongSOC_NOTCHIRAL_4fold_NewFermion_ESFD2}
\end{figure}

\begin{figure}[ht]
\centering
%

\caption{\strongSOCNOTCHIRALfoldNewFermionESFD{3}}
\label{fig:strongSOC_NOTCHIRAL_4fold_NewFermion_ESFD3}
\end{figure}

\begin{figure}[ht]
\centering
%

\caption{\strongSOCNOTCHIRALfoldNewFermionESFD{10}}
\label{fig:strongSOC_NOTCHIRAL_4fold_NewFermion_ESFD10}
\end{figure}


\begin{figure}[ht]
\centering
%

\caption{\strongSOCchiralNewFermionESFD{1}}
\label{fig:strongSOC_chiral_NewFermion_ESFD1}
\end{figure}

\begin{figure}[ht]
\centering
%

\caption{\strongSOCchiralNewFermionESFD{2}}
\label{fig:strongSOC_chiral_NewFermion_ESFD2}
\end{figure}

\begin{figure}[ht]
\centering
%

\caption{\strongSOCchiralNewFermionESFD{3}}
\label{fig:strongSOC_chiral_NewFermion_ESFD3}
\end{figure}

\begin{figure}[ht]
\centering
%

\caption{\strongSOCchiralNewFermionESFD{6}}
\label{fig:strongSOC_chiral_NewFermion_ESFD6}
\end{figure}


\begin{figure}[ht]
\centering
%

\caption{\strongSOCeightfoldESFD{1}}
\label{fig:strongSOC_8fold_ESFD1}
\end{figure}

\begin{figure}[ht]
\centering
%

\caption{\strongSOCeightfoldESFD{2}}
\label{fig:strongSOC_8fold_ESFD2}
\end{figure}

\begin{figure}[ht]
\centering
%

\caption{\strongSOCeightfoldESFD{3}}
\label{fig:strongSOC_8fold_ESFD3}
\end{figure}


\begin{figure}[ht]
\centering
%

\caption{\strongSOCquadraticotherESFD{1}}
\label{fig:strongSOC_quadratic_other_ESFD1}
\end{figure}

\begin{figure}[ht]
\centering
%

\caption{\strongSOCquadraticotherESFD{2}}
\label{fig:strongSOC_quadratic_other_ESFD2}
\end{figure}

\clearpage

\subsubsection{ES-Classified Semimetals}
\label{App:ES}

In this section, we list in Figs.~\ref{fig:strongSOC_ES1},~\ref{fig:strongSOC_ES2},~\ref{fig:strongSOC_ES3},~\ref{fig:strongSOC_ES4},~\ref{fig:strongSOC_ES5}, and~\ref{fig:strongSOC_ES6} the enforced semimetals with the simplest bulk Fermi surfaces whose nodal degeneracies lie along high-symmetry BZ lines (ES in the nomenclature established in Ref.~\onlinecite{AndreiMaterials}).  Typically, the bulk Fermi pockets of ES materials are characterized by fourfold-degenerate, linearly dispersing Dirac fermions~\cite{ZJDirac,ZJDirac2,Khoury_JACS,SchoopAFM}.  However, it is also possible for ES materials to host chiral fermions with larger Chern numbers~\cite{ChenMultiWeyl,ZhijunMultiWeyl,ZahidMultiWeyl,StepanMultiWeyl,ScrewChiralWeyl}, threefold-degenerate ``nexus'' fermions~\cite{TripleChen,AlexeyTriple,ZahidTriple,WiederTripleSummary,MoPTriple,WCarc}, or more exotic mixtures of chiral and achiral fermions~\cite{DiracWeylYoungkuk}.  Almost all of the materials shown in Figs.~\ref{fig:strongSOC_ES1},~\ref{fig:strongSOC_ES2},~\ref{fig:strongSOC_ES3},~\ref{fig:strongSOC_ES4},~\ref{fig:strongSOC_ES5}, and~\ref{fig:strongSOC_ES6} are conventional Dirac semimetals, with two exceptions.  First, In$_2$ZnS$_4$ [\icsdweb{15636} and~\icsdweb{65725}, SG 160 ($R3m$)] in Fig.~\ref{fig:strongSOC_ES2} is a nexus fermion semimetal with narrowly separated threefold degeneracies.  Second, Ta$_2$ISe$_8$ [\icsdweb{35190}, SG 97 ($I422$)] in Fig.~\ref{fig:strongSOC_ES1} is a structurally chiral, quasi-1D Weyl semimetal that, as demonstrated in recent theoretical and experimental investigations, becomes gapped by a topological (axionic) charge-density wave when cooled just below room temperature~\cite{CDWWeyl,AxionCDWExperiment,TaSeIPRL,TaSeIDFTDagotto,TaSeIpressureSC,TaSeIZhijunDirac}.  Most notably, Figs.~\ref{fig:strongSOC_ES1} and~\ref{fig:strongSOC_ES5} respectively include the archetypal, experimentally confirmed Dirac semimetals Cd$_3$As$_2$ [\icsdweb{107918}, SG 137 ($P4_{2}/nmc$)]~\cite{ZJDirac,CavaDirac1,CavaDirac2,YulinCadmiumExp} and Na$_3$Bi [\icsdweb{26881}, SG 194 $P6_{3}/mmc$]~\cite{ZJDirac2,NaDirac,SYDiracSurface}.  As recently shown in Ref.~\onlinecite{HingeSM}, Dirac points with $4mm$ or higher symmetry also exhibit topological hinge states; because a significant number of the Dirac semimetals in Figs.~\ref{fig:strongSOC_ES1},~\ref{fig:strongSOC_ES2},~\ref{fig:strongSOC_ES3},~\ref{fig:strongSOC_ES4},~\ref{fig:strongSOC_ES5}, and~\ref{fig:strongSOC_ES6} have SGs whose point groups contain $4mm$, then the listed Dirac semimetals may also be classified as higher-order topological semimetals.  Finally, the results of Refs.~\onlinecite{HingeSM,TMDHOTI,TaylorToy,AshvinFragile2,VladHOFA,BitanHOFA,BitanHOFA2,TaylorWeylHOFA,OtherWeylHOFA,ZhijunAndreiElectride} also imply the presence of hinge states in additional ES materials without fourfold symmetry, including the materials shown in Figs.~\ref{fig:strongSOC_ES1},~\ref{fig:strongSOC_ES2},~\ref{fig:strongSOC_ES3},~\ref{fig:strongSOC_ES4},~\ref{fig:strongSOC_ES5}, and~\ref{fig:strongSOC_ES6} with threefold or sixfold rotational symmetries.


\begin{figure}[ht]
\centering
\begin{tabular}{c c}
\scriptsize{$\rm{Pd} \rm{Se}$ - \icsdweb{77895} - SG 84 ($P4_2/m$) - ES} & \scriptsize{$\rm{Ta}_{2} \rm{I} \rm{Se}_{8}$ - \icsdweb{35190} - SG 97 ($I422$) - ES}\\
\includegraphics[width=0.38\textwidth,angle=0]{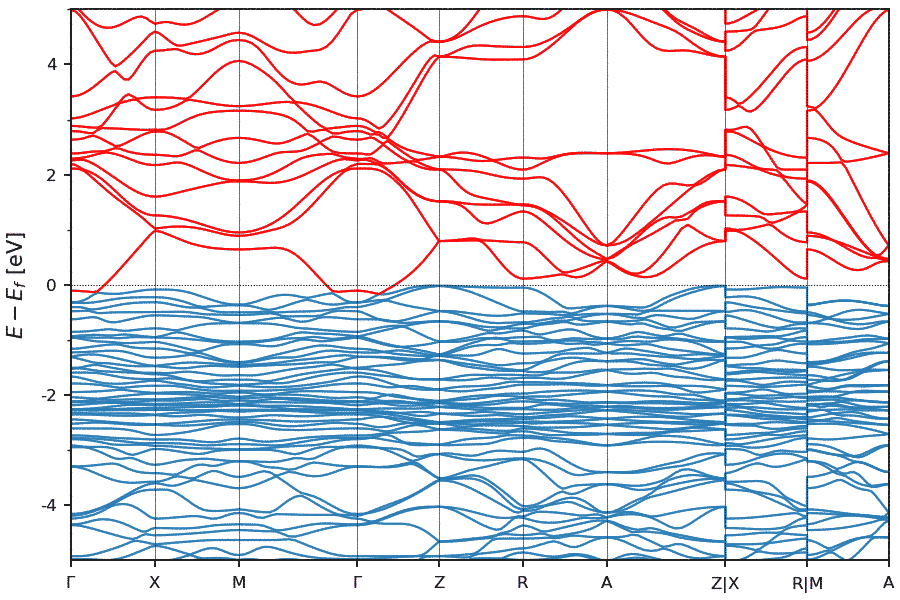} & \includegraphics[width=0.38\textwidth,angle=0]{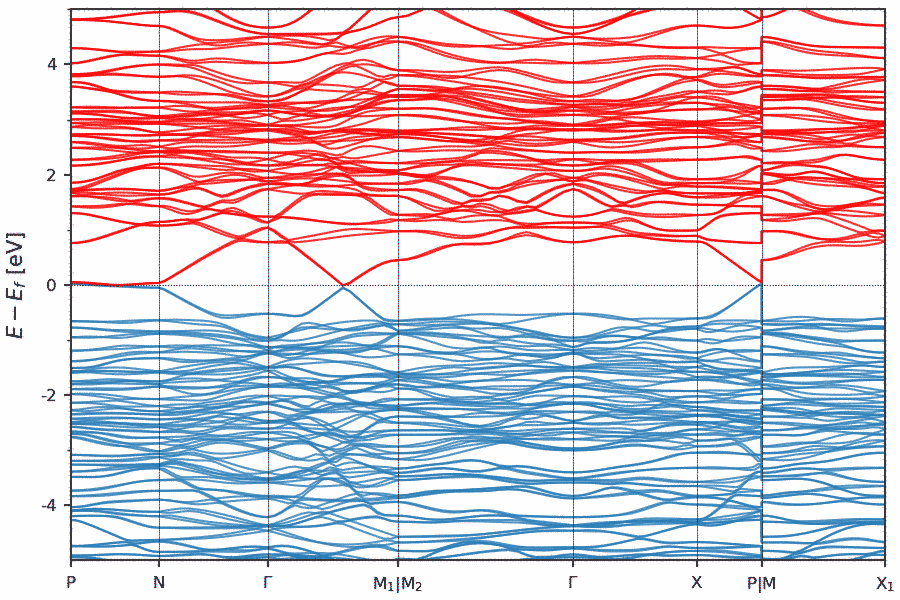}\\
\end{tabular}
\begin{tabular}{c c}
\scriptsize{$\rm{Rb} \rm{Ag}_{5} \rm{Se}_{3}$ - \icsdweb{50738} - SG 125 ($P4/nbm$) - ES} & \scriptsize{$\rm{Na} \rm{Cu} \rm{Se}$ - \icsdweb{12155} - SG 129 ($P4/nmm$) - ES}\\
\includegraphics[width=0.38\textwidth,angle=0]{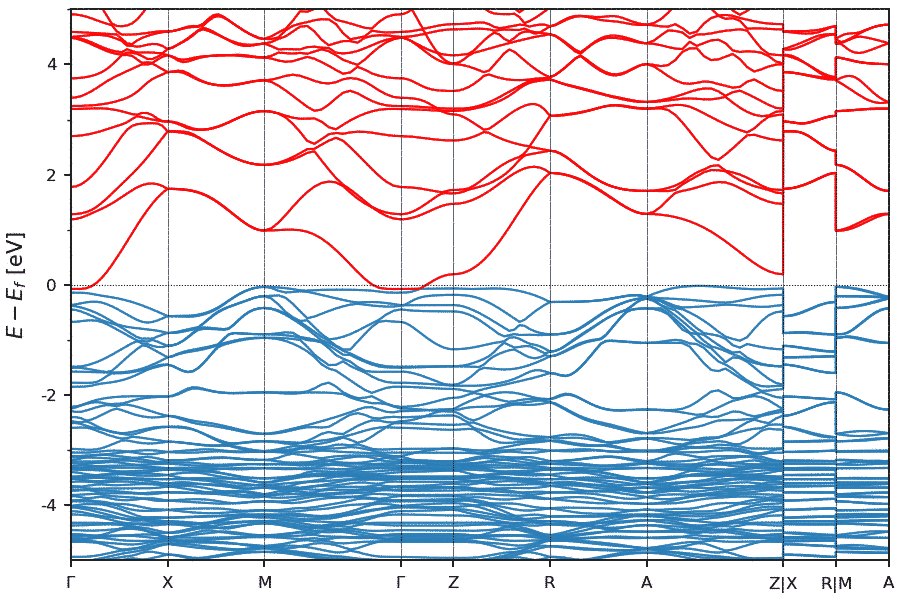} & \includegraphics[width=0.38\textwidth,angle=0]{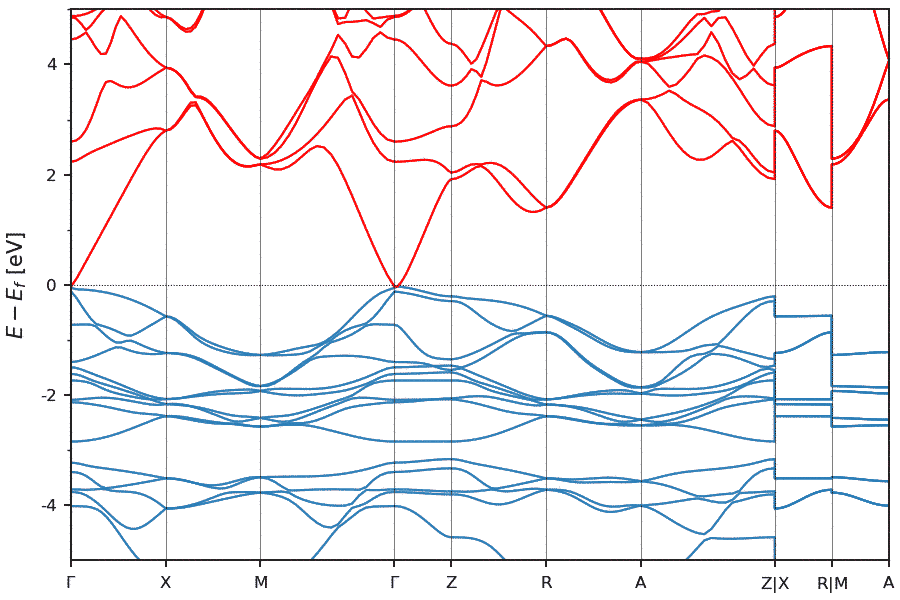}\\
\end{tabular}
\begin{tabular}{c c}
\scriptsize{$\rm{K} \rm{Mg} \rm{Bi}$ - \icsdweb{616748} - SG 129 ($P4/nmm$) - ES} & \scriptsize{$\rm{Bi}_{2} \rm{O}_{3}$ - \icsdweb{168807} - SG 132 ($P4_2/mcm$) - ES}\\
\includegraphics[width=0.38\textwidth,angle=0]{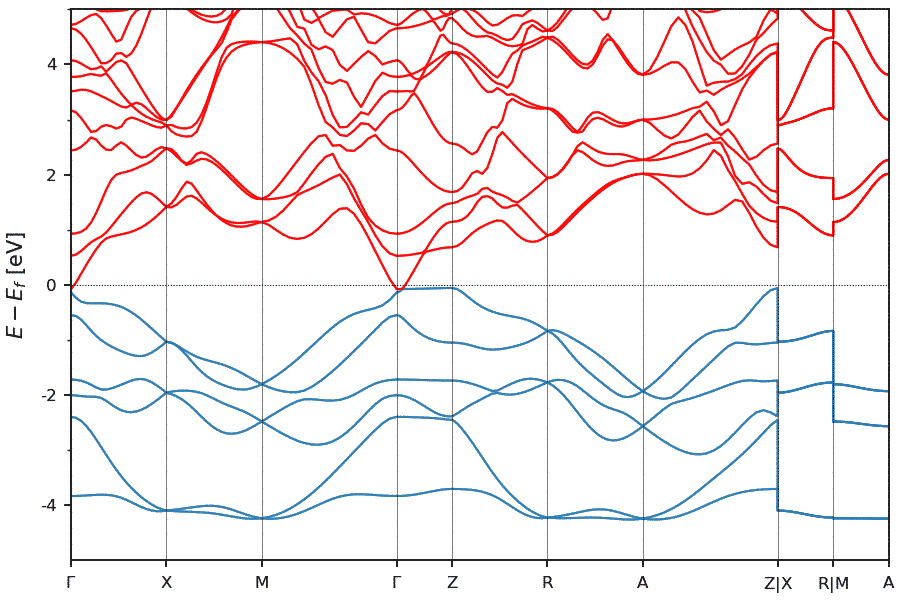} & \includegraphics[width=0.38\textwidth,angle=0]{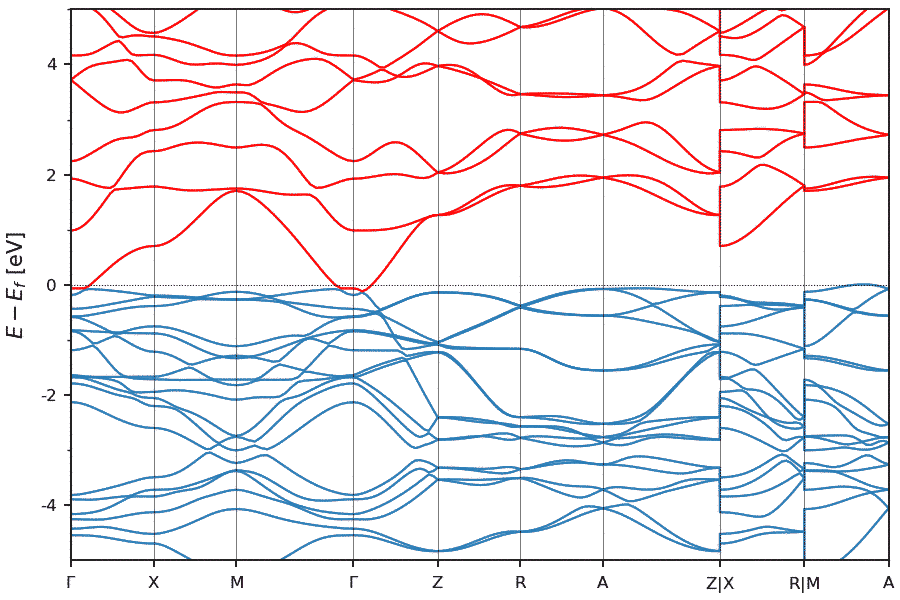}\\
\end{tabular}
\begin{tabular}{c c}
\scriptsize{$\rm{Mg} (\rm{Bi}_{2} \rm{O}_{6})$ - \icsdweb{50005} - SG 136 ($P4_2/mnm$) - ES} & \scriptsize{$\rm{Cd}_{3} \rm{As}_{2}$ - \icsdweb{107918} - SG 137 ($P4_2/nmc$) - ES}\\
\includegraphics[width=0.38\textwidth,angle=0]{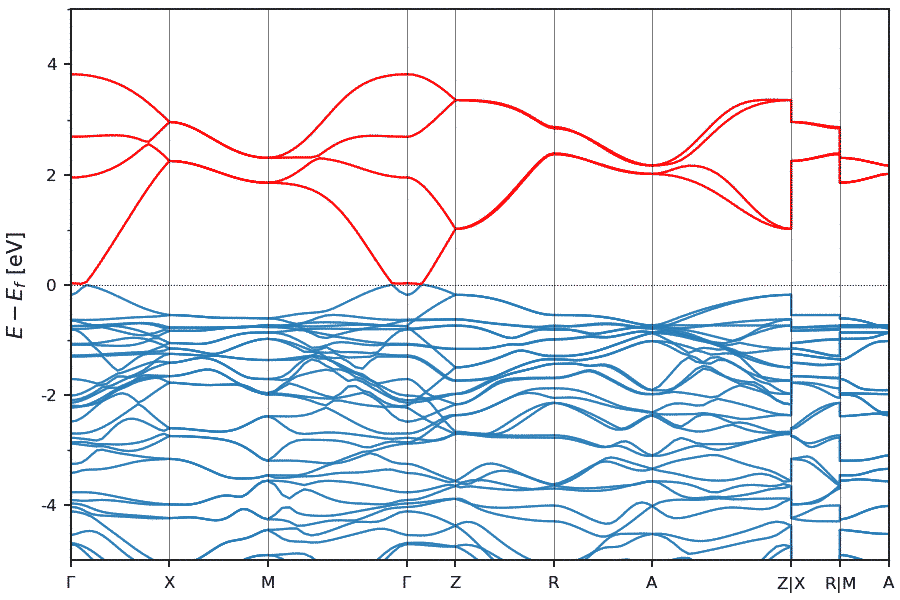} & \includegraphics[width=0.38\textwidth,angle=0]{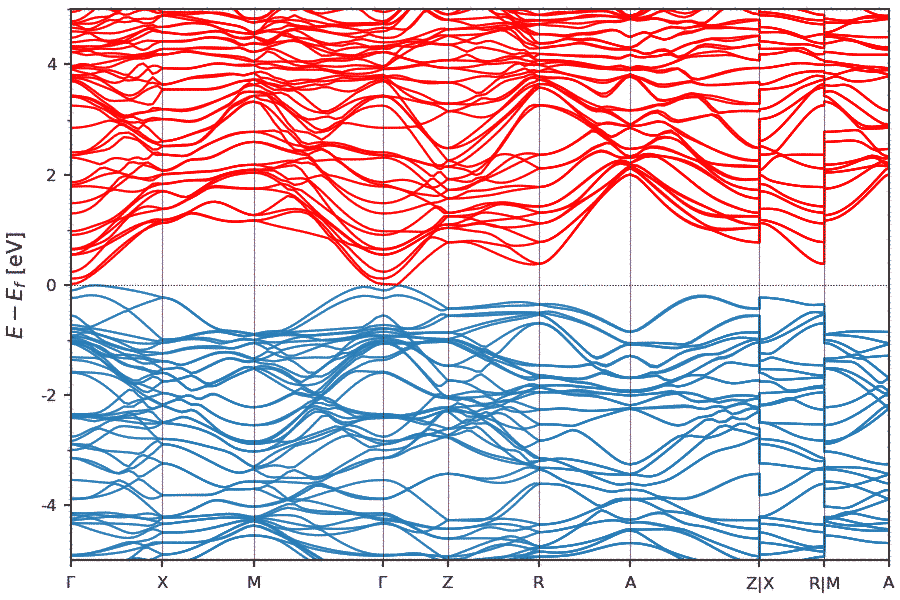}\\
\end{tabular}

\caption{\strongSOCES{1}}
\label{fig:strongSOC_ES1}
\end{figure}

\begin{figure}[ht]
\centering
\begin{tabular}{c c}
\scriptsize{$\rm{Cd}_{3} \rm{P}_{2}$ - \icsdweb{181134} - SG 137 ($P4_2/nmc$) - ES} & \scriptsize{$\rm{Al}_{3} \rm{V}$ - \icsdweb{58201} - SG 139 ($I4/mmm$) - ES}\\
\includegraphics[width=0.38\textwidth,angle=0]{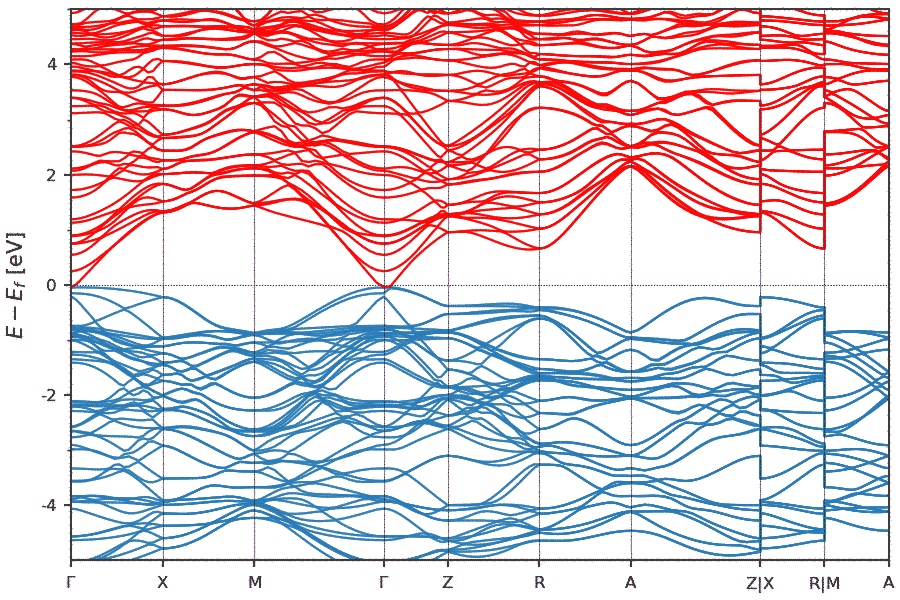} & \includegraphics[width=0.38\textwidth,angle=0]{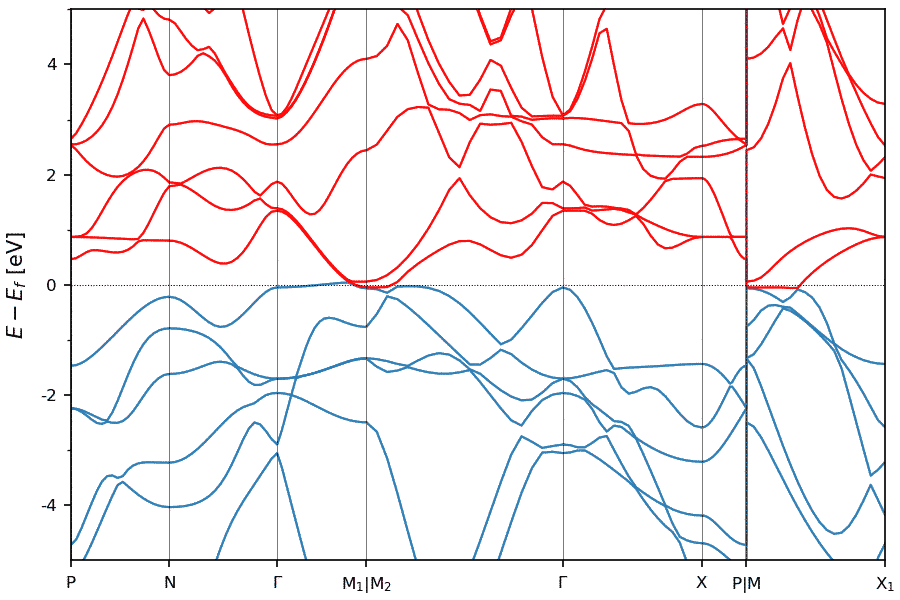}\\
\end{tabular}
\begin{tabular}{c c}
\scriptsize{$\rm{Au}_{2} \rm{Ti}$ - \icsdweb{58607} - SG 139 ($I4/mmm$) - ES} & \scriptsize{$\rm{Ga}_{3} \rm{Nb}$ - \icsdweb{103833} - SG 139 ($I4/mmm$) - ES}\\
\includegraphics[width=0.38\textwidth,angle=0]{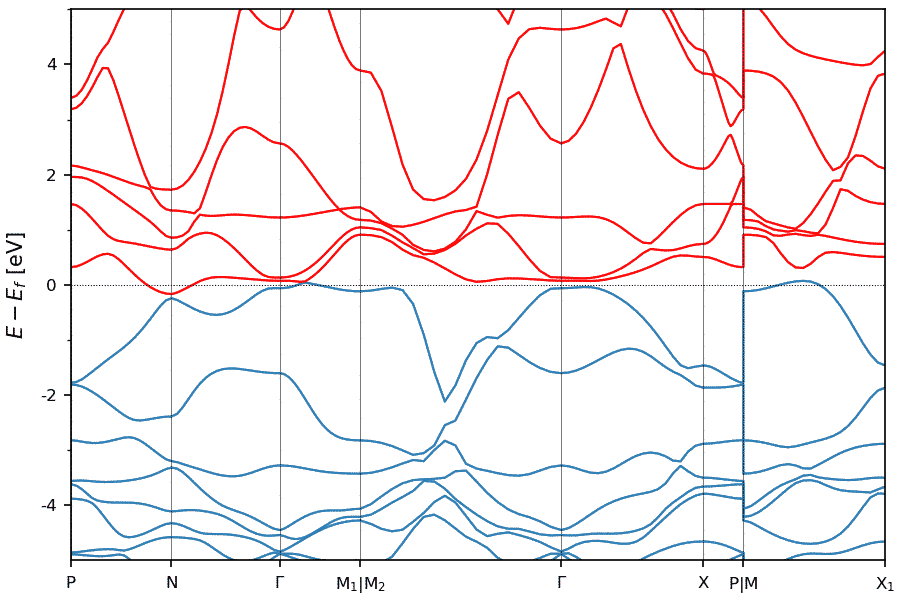} & \includegraphics[width=0.38\textwidth,angle=0]{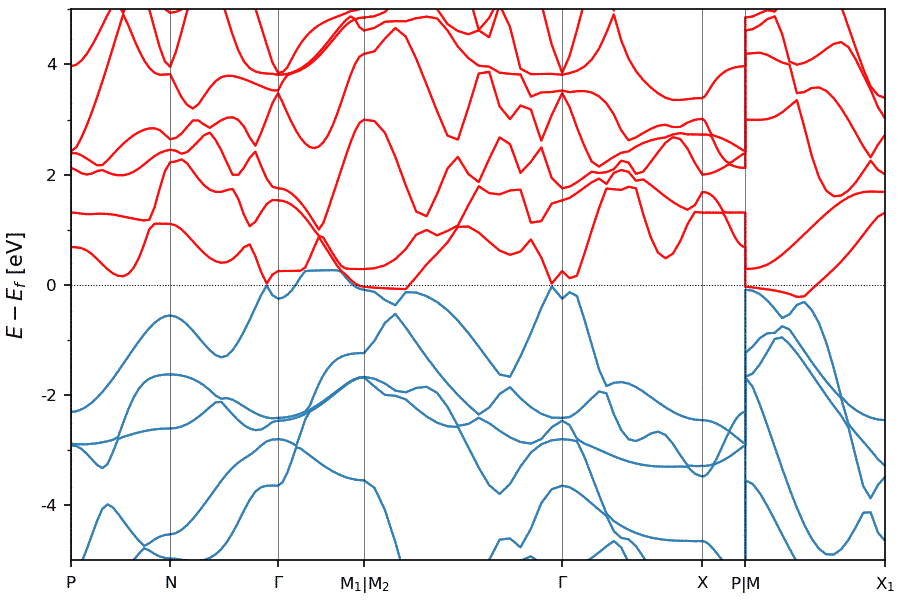}\\
\end{tabular}
\begin{tabular}{c c}
\scriptsize{$\rm{Al}_{3} \rm{Nb}$ - \icsdweb{608663} - SG 139 ($I4/mmm$) - ES} & \scriptsize{$\rm{Zn} \rm{In}_{2} \rm{S}_{4}$ - \icsdweb{15636} - SG 160 ($R3m$) - ES}\\
\includegraphics[width=0.38\textwidth,angle=0]{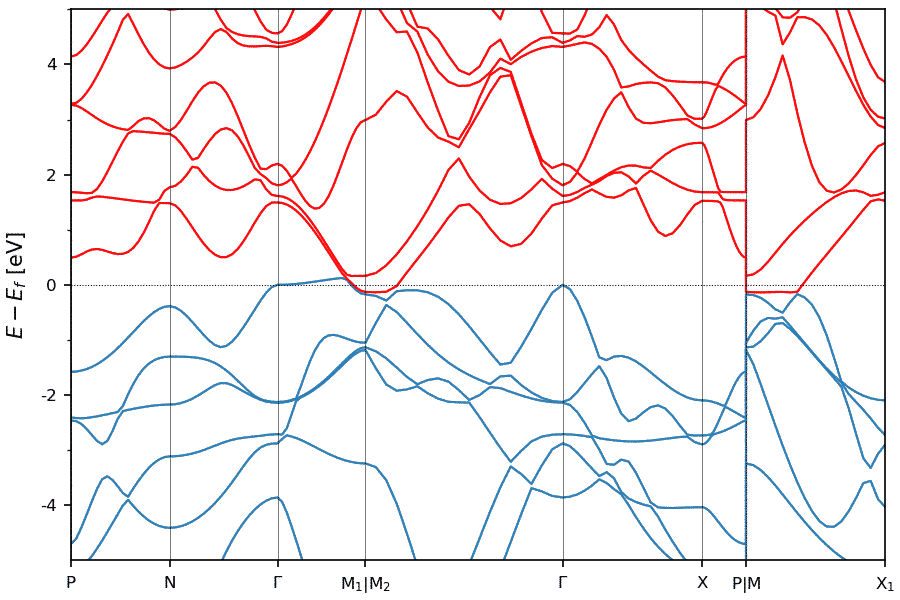} & \includegraphics[width=0.38\textwidth,angle=0]{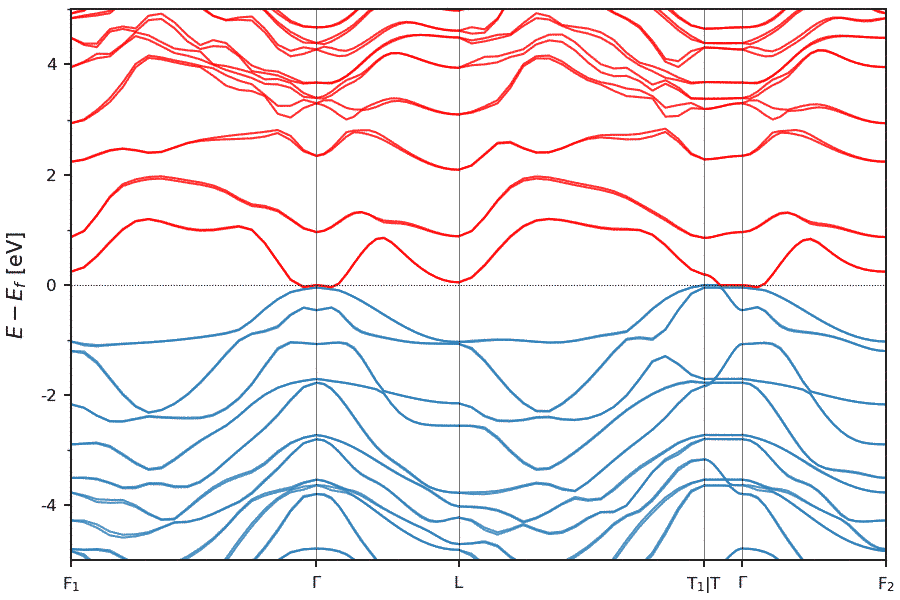}\\
\end{tabular}
\begin{tabular}{c c}
\scriptsize{$\rm{In}_{2} (\rm{Zn} \rm{S}_{4})$ - \icsdweb{65725} - SG 160 ($R3m$) - ES} & \scriptsize{$\rm{Ba} \rm{Mg}_{2} \rm{Bi}_{2}$ - \icsdweb{100049} - SG 164 ($P\bar{3}m1$) - ES}\\
\includegraphics[width=0.38\textwidth,angle=0]{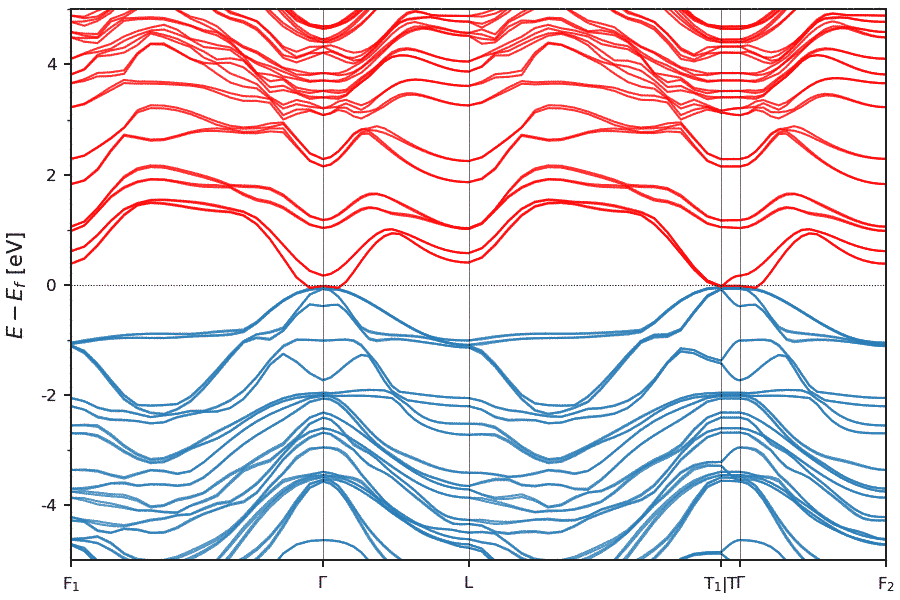} & \includegraphics[width=0.38\textwidth,angle=0]{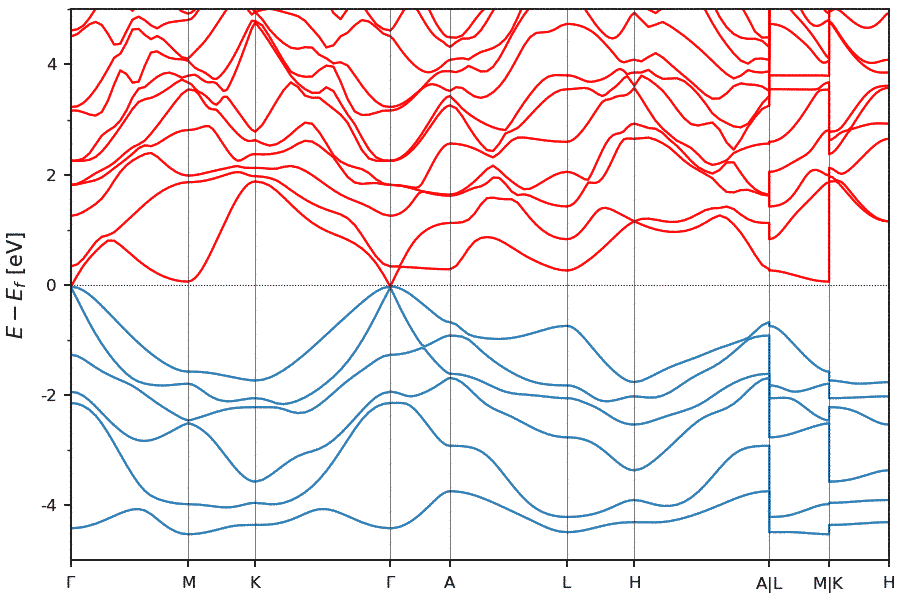}\\
\end{tabular}

\caption{\strongSOCES{2}}
\label{fig:strongSOC_ES2}
\end{figure}

\begin{figure}[ht]
\centering
\begin{tabular}{c c}
\scriptsize{$\rm{Li}_{3} \rm{As}$ - \icsdweb{610785} - SG 165 ($P\bar{3}c1$) - ES} & \scriptsize{$\rm{Bi}_{4} \rm{Te}_{3}$ - \icsdweb{30526} - SG 166 ($R\bar{3}m$) - ES}\\
\includegraphics[width=0.38\textwidth,angle=0]{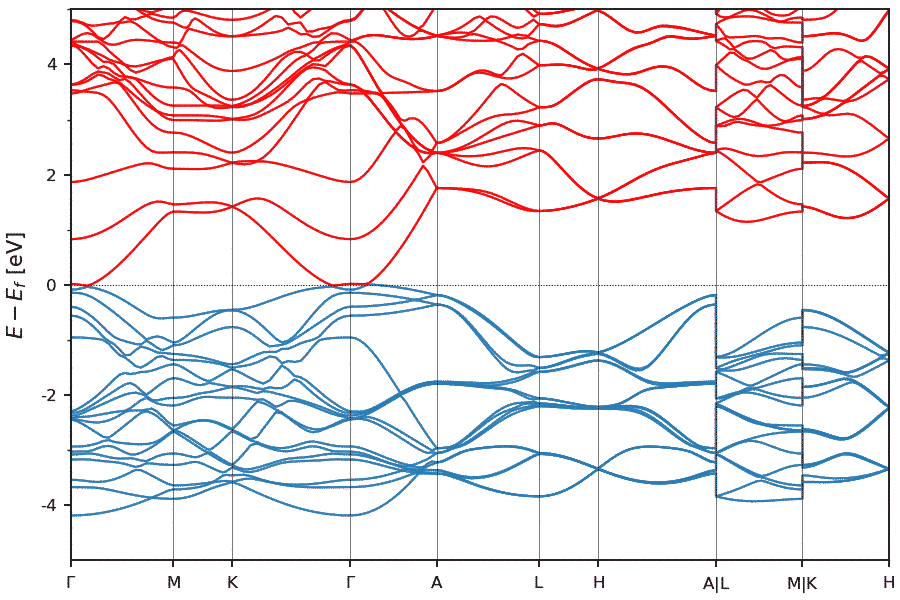} & \includegraphics[width=0.38\textwidth,angle=0]{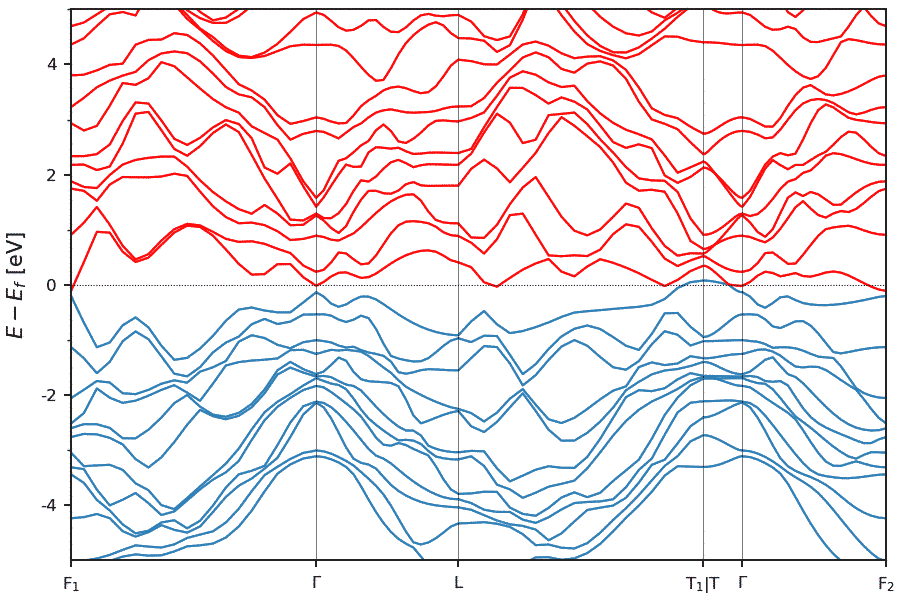}\\
\end{tabular}
\begin{tabular}{c c}
\scriptsize{$\rm{Cu} \rm{I}$ - \icsdweb{78266} - SG 166 ($R\bar{3}m$) - ES} & \scriptsize{$\rm{K} \rm{Cd}_{4} \rm{P}_{3}$ - \icsdweb{262033} - SG 166 ($R\bar{3}m$) - ES}\\
\includegraphics[width=0.38\textwidth,angle=0]{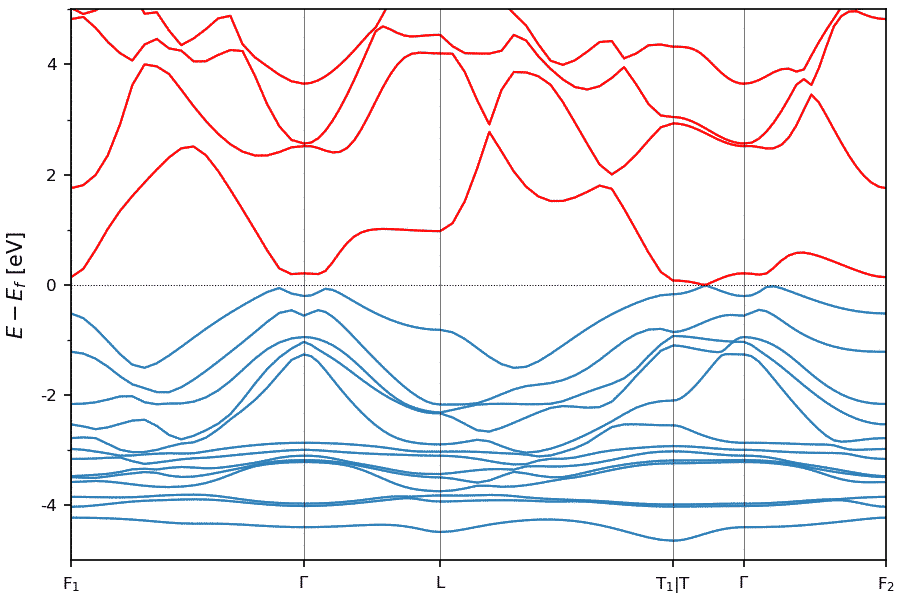} & \includegraphics[width=0.38\textwidth,angle=0]{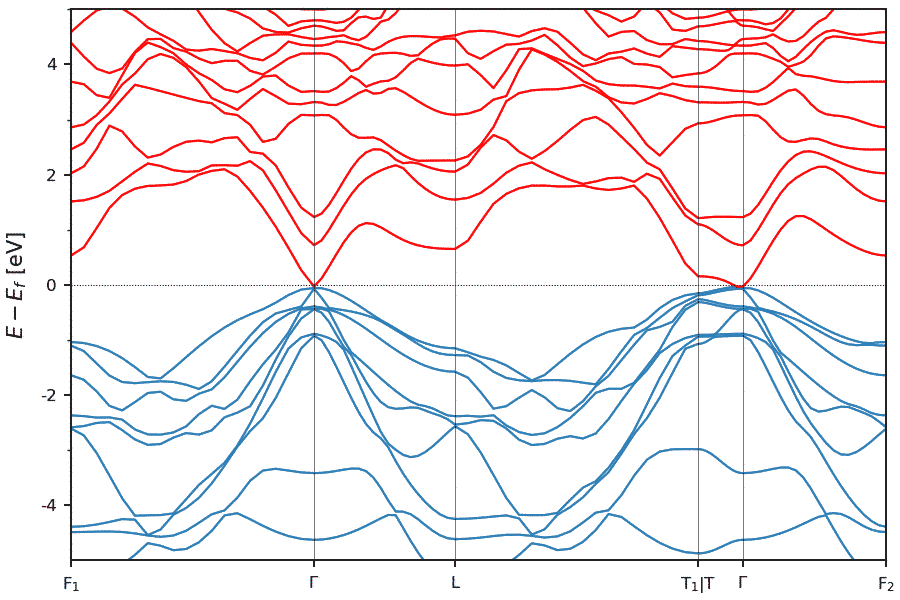}\\
\end{tabular}
\begin{tabular}{c c}
\scriptsize{$\rm{Na} \rm{Zn}_{4} \rm{As}_{3}$ - \icsdweb{262036} - SG 166 ($R\bar{3}m$) - ES} & \scriptsize{$\rm{In} \rm{Sn} \rm{Co}_{3} \rm{S}_{2}$ - \icsdweb{425137} - SG 166 ($R\bar{3}m$) - ES}\\
\includegraphics[width=0.38\textwidth,angle=0]{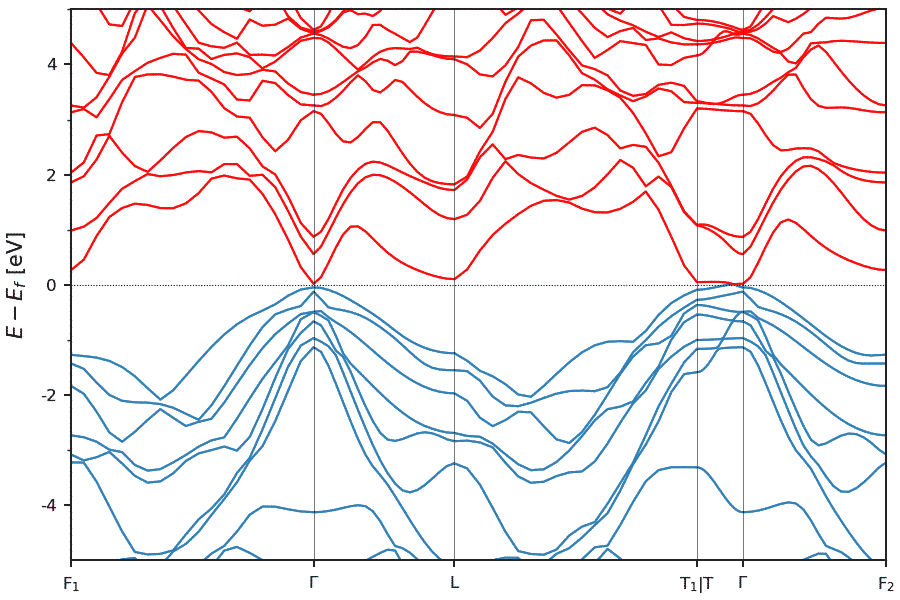} & \includegraphics[width=0.38\textwidth,angle=0]{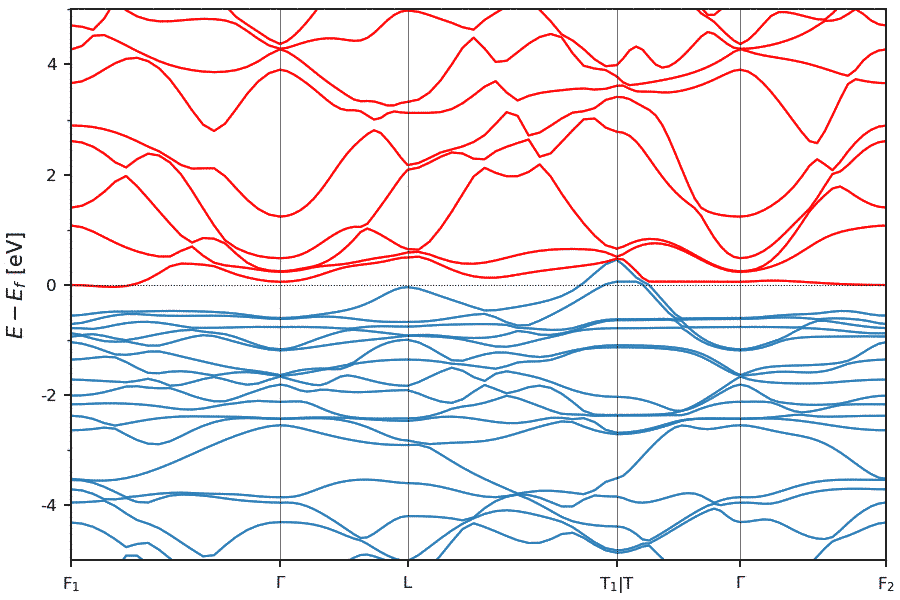}\\
\end{tabular}
\begin{tabular}{c c}
\scriptsize{$\rm{Ba} \rm{Mo}_{3} \rm{Te}_{3}$ - \icsdweb{603674} - SG 176 ($P6_3/m$) - ES} & \scriptsize{$\rm{Ba} \rm{Mo}_{3} \rm{Se}_{3}$ - \icsdweb{615982} - SG 176 ($P6_3/m$) - ES}\\
\includegraphics[width=0.38\textwidth,angle=0]{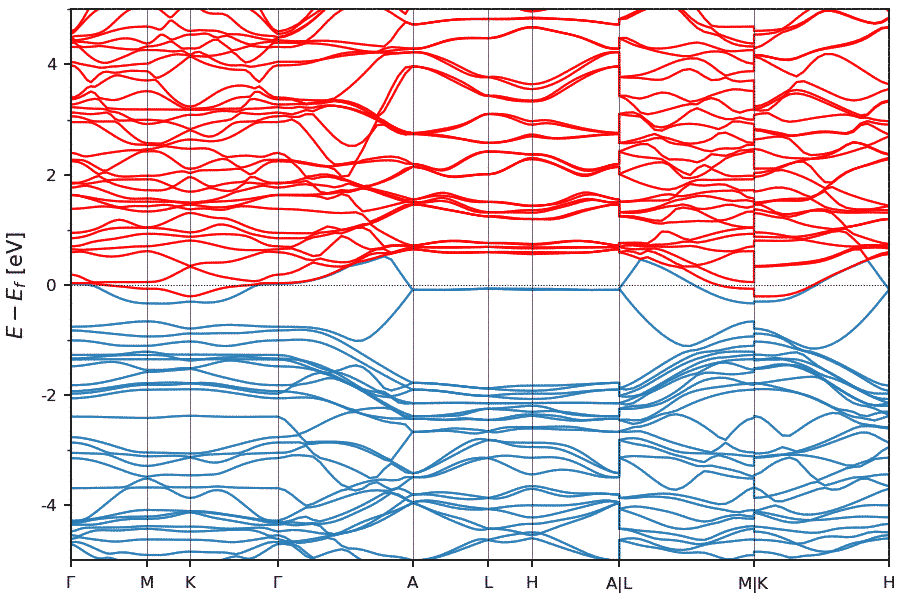} & \includegraphics[width=0.38\textwidth,angle=0]{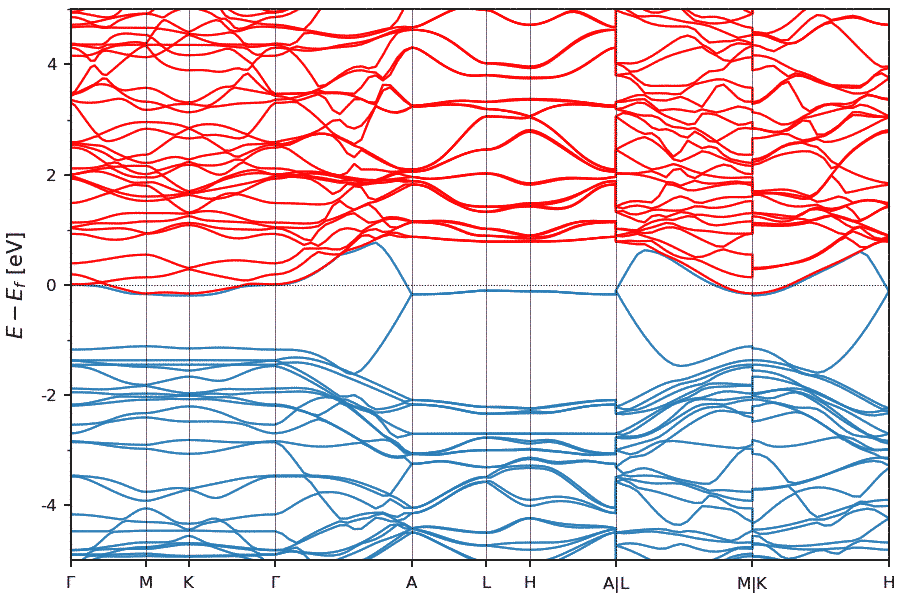}\\
\end{tabular}

\caption{\strongSOCES{3}}
\label{fig:strongSOC_ES3}
\end{figure}

\begin{figure}[ht]
\centering
\begin{tabular}{c c}
\scriptsize{$\rm{Na}_{3} \rm{As}$ - \icsdweb{182164} - SG 185 ($P6_3cm$) - ES} & \scriptsize{$\rm{Bi} \rm{K}_{3}$ - \icsdweb{409223} - SG 185 ($P6_3cm$) - ES}\\
\includegraphics[width=0.38\textwidth,angle=0]{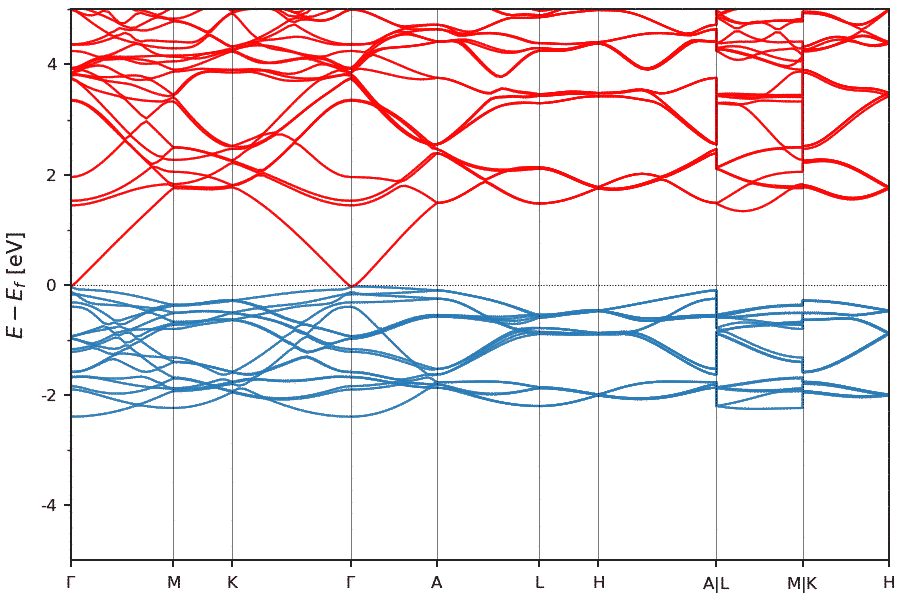} & \includegraphics[width=0.38\textwidth,angle=0]{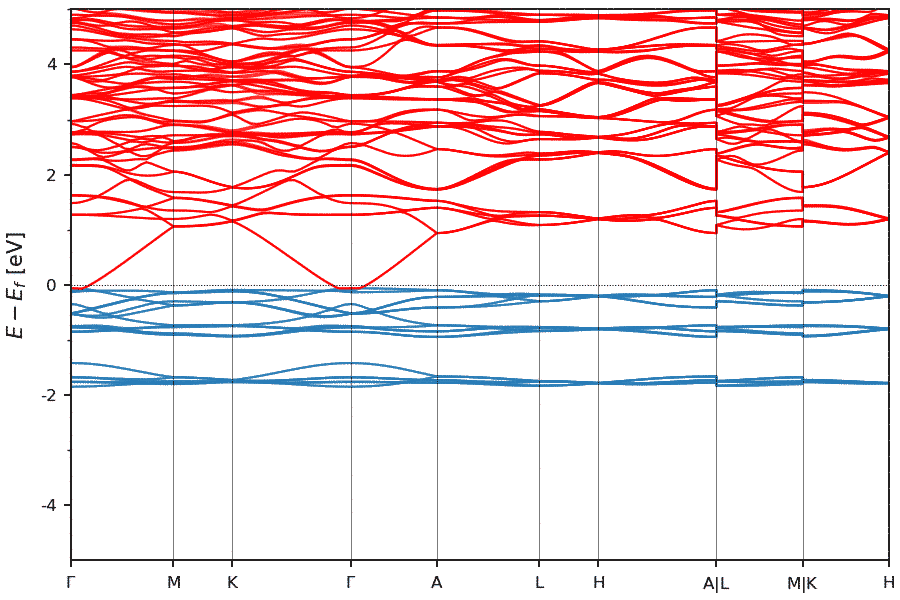}\\
\end{tabular}
\begin{tabular}{c c}
\scriptsize{$\rm{Y} \rm{Cu} \rm{Sn}$ - \icsdweb{416544} - SG 186 ($P6_3mc$) - ES} & \scriptsize{$\rm{Li}_{2} \rm{Na} \rm{N}$ - \icsdweb{92308} - SG 191 ($P6/mmm$) - ES}\\
\includegraphics[width=0.38\textwidth,angle=0]{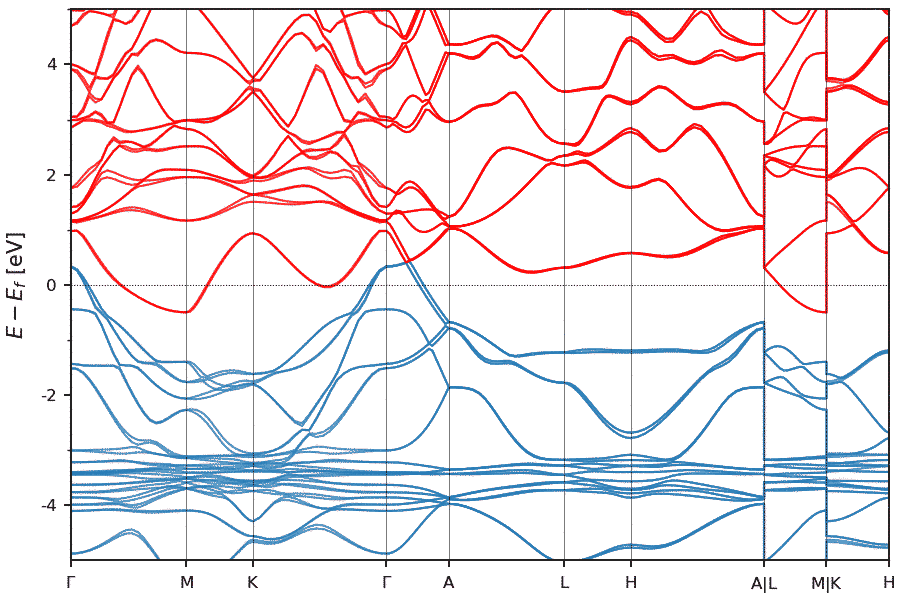} & \includegraphics[width=0.38\textwidth,angle=0]{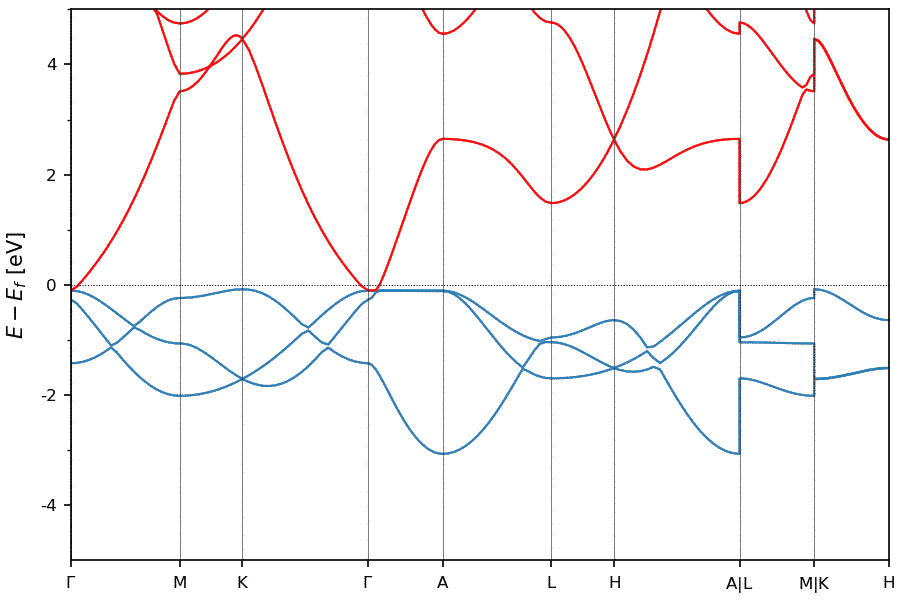}\\
\end{tabular}
\begin{tabular}{c c}
\scriptsize{$\rm{Li} \rm{Na}_{2} \rm{N}$ - \icsdweb{92310} - SG 191 ($P6/mmm$) - ES} & \scriptsize{$\rm{Na}_{3} \rm{N}$ - \icsdweb{165989} - SG 191 ($P6/mmm$) - ES}\\
\includegraphics[width=0.38\textwidth,angle=0]{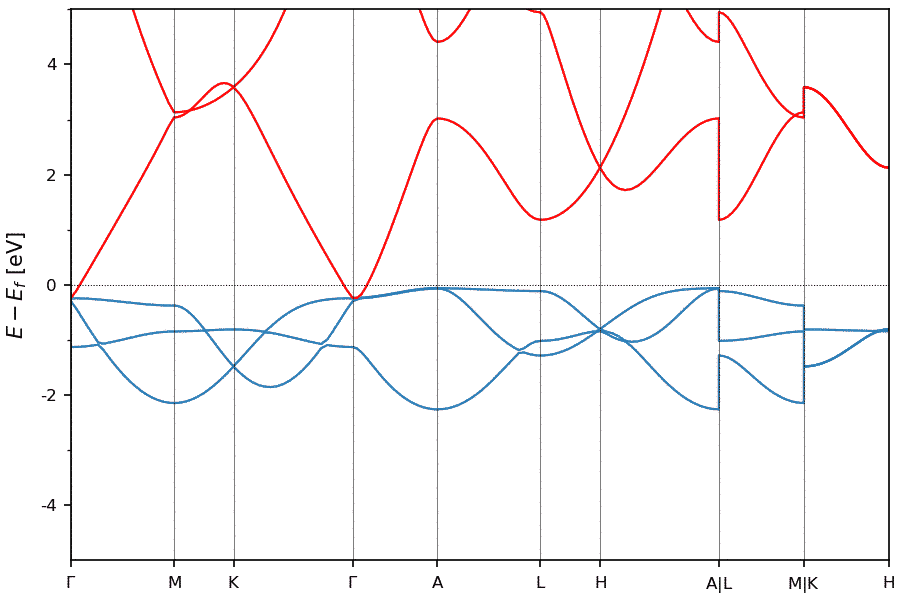} & \includegraphics[width=0.38\textwidth,angle=0]{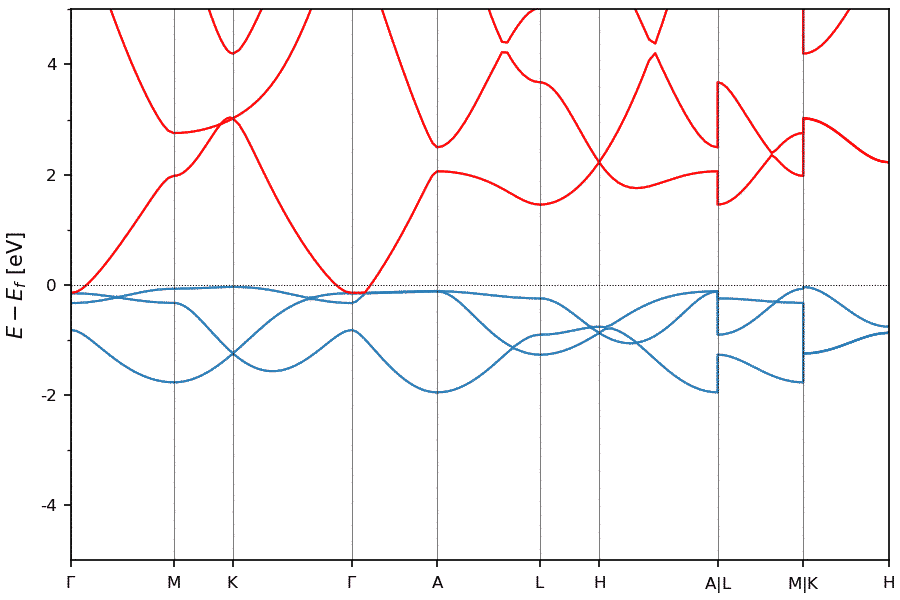}\\
\end{tabular}
\begin{tabular}{c c}
\scriptsize{$\rm{Zr}_{5} \rm{Pb}_{3} \rm{Se}$ - \icsdweb{604205} - SG 193 ($P6_3/mcm$) - ES} & \scriptsize{$\rm{Sn}_{3} \rm{Zr}_{5} \rm{S}$ - \icsdweb{656296} - SG 193 ($P6_3/mcm$) - ES}\\
\includegraphics[width=0.38\textwidth,angle=0]{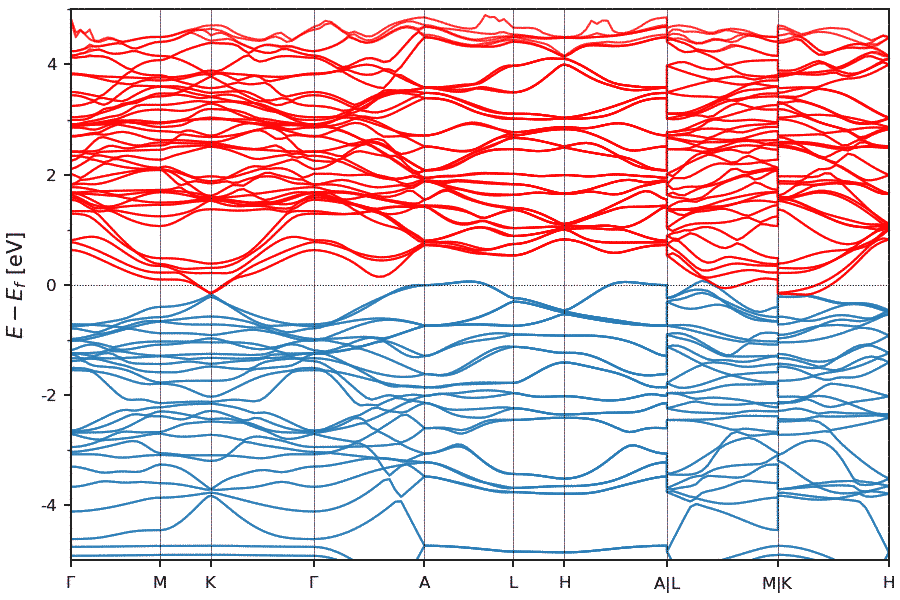} & \includegraphics[width=0.38\textwidth,angle=0]{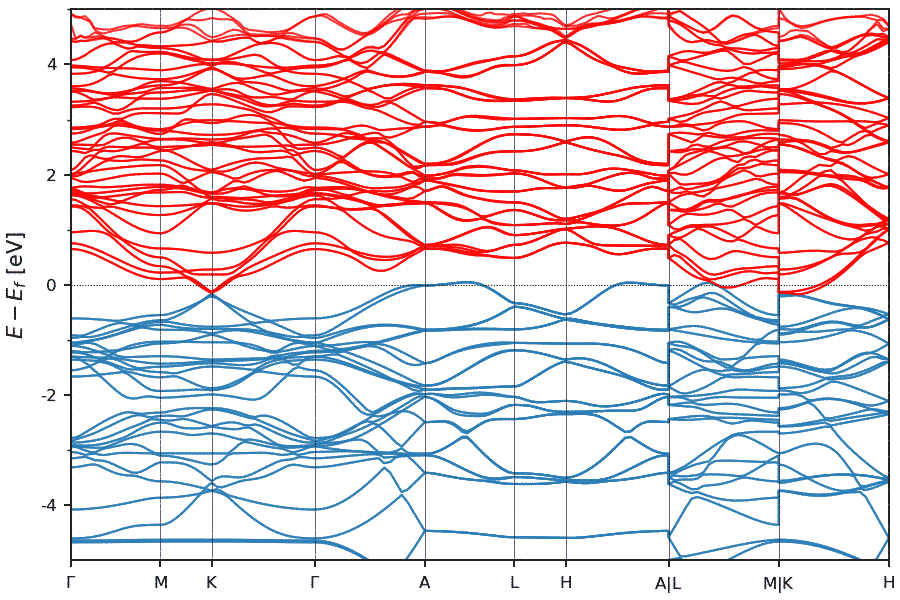}\\
\end{tabular}

\caption{\strongSOCES{4}}
\label{fig:strongSOC_ES4}
\end{figure}

\begin{figure}[ht]
\centering
\begin{tabular}{c c}
\scriptsize{$\rm{Sn}_{3} \rm{Zr}_{5} \rm{Se}$ - \icsdweb{656302} - SG 193 ($P6_3/mcm$) - ES} & \scriptsize{$\rm{Ba} \rm{Ag} \rm{As}$ - \icsdweb{8278} - SG 194 ($P6_3/mmc$) - ES}\\
\includegraphics[width=0.38\textwidth,angle=0]{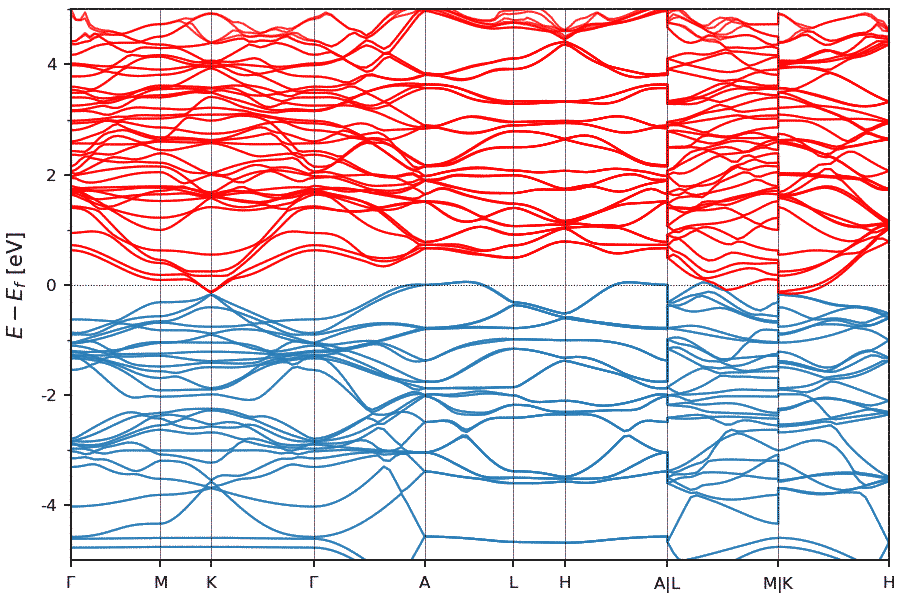} & \includegraphics[width=0.38\textwidth,angle=0]{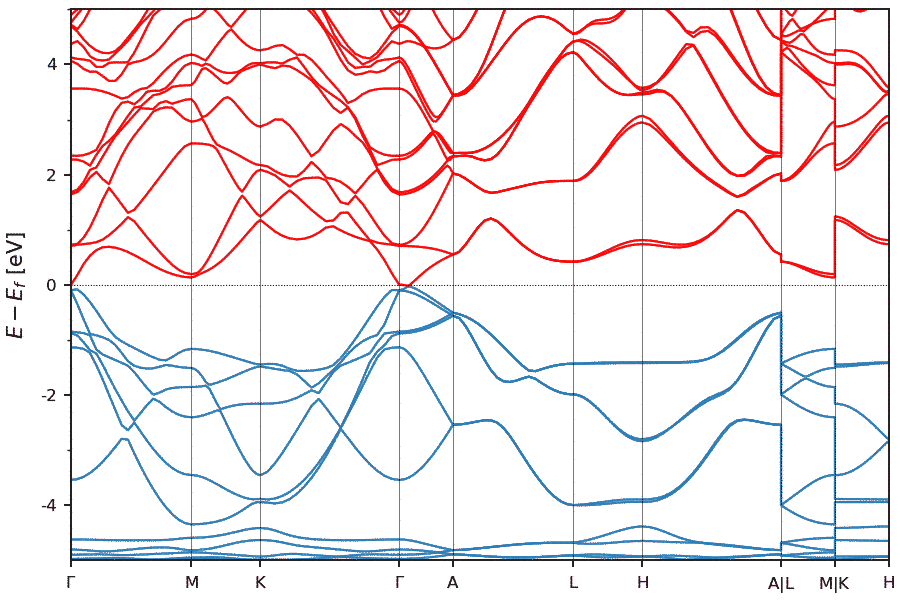}\\
\end{tabular}
\begin{tabular}{c c}
\scriptsize{$\rm{Na}_{3} \rm{Bi}$ - \icsdweb{26881} - SG 194 ($P6_3/mmc$) - ES} & \scriptsize{$\rm{K}_{3} \rm{Bi}$ - \icsdweb{26885} - SG 194 ($P6_3/mmc$) - ES}\\
\includegraphics[width=0.38\textwidth,angle=0]{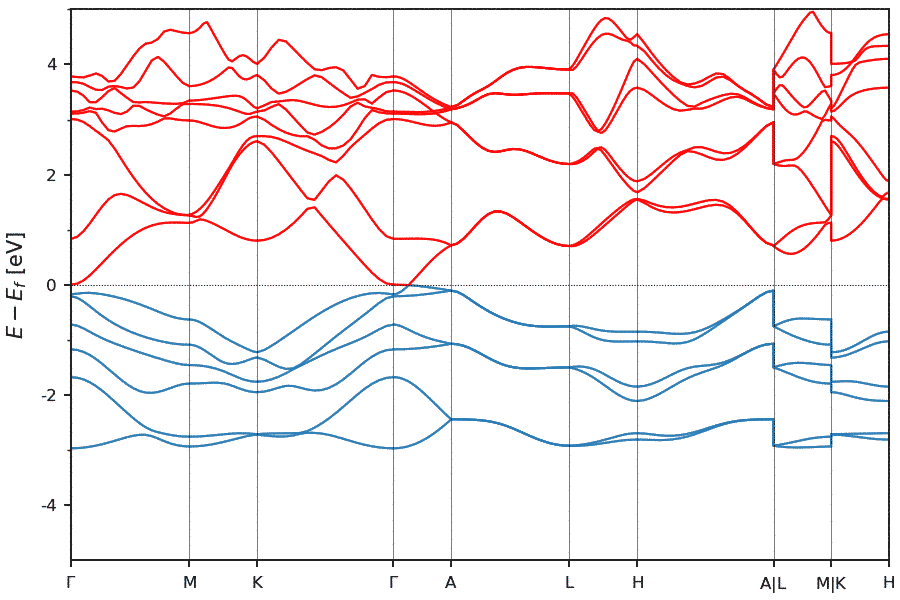} & \includegraphics[width=0.38\textwidth,angle=0]{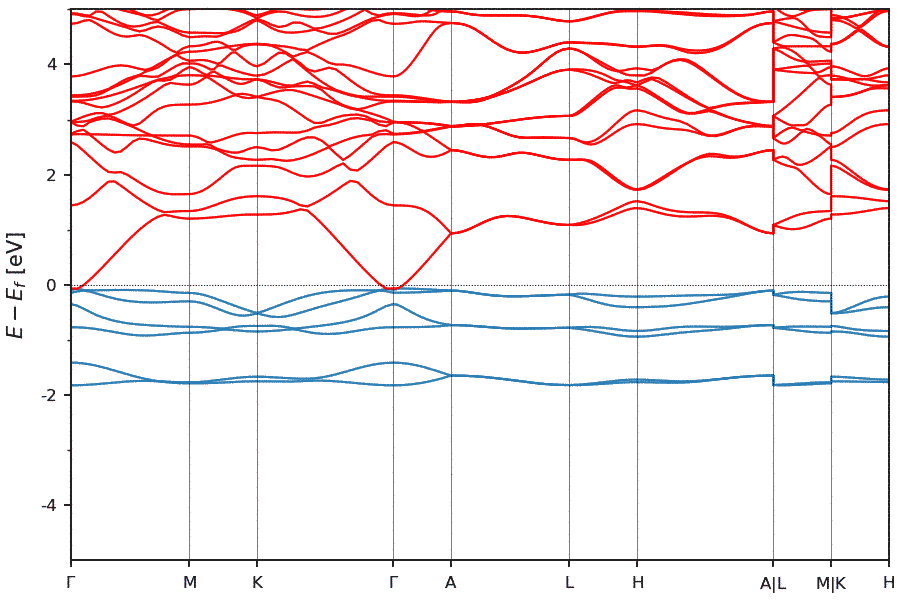}\\
\end{tabular}
\begin{tabular}{c c}
\scriptsize{$\rm{Ag} \rm{Ba} \rm{Bi}$ - \icsdweb{56978} - SG 194 ($P6_3/mmc$) - ES} & \scriptsize{$\rm{Bi} \rm{Rb}_{3}$ - \icsdweb{616995} - SG 194 ($P6_3/mmc$) - ES}\\
\includegraphics[width=0.38\textwidth,angle=0]{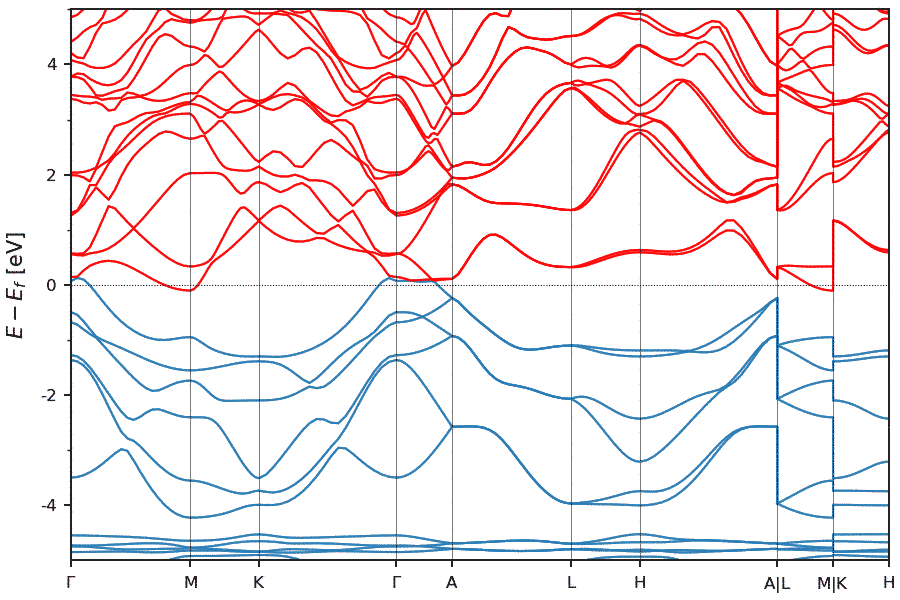} & \includegraphics[width=0.38\textwidth,angle=0]{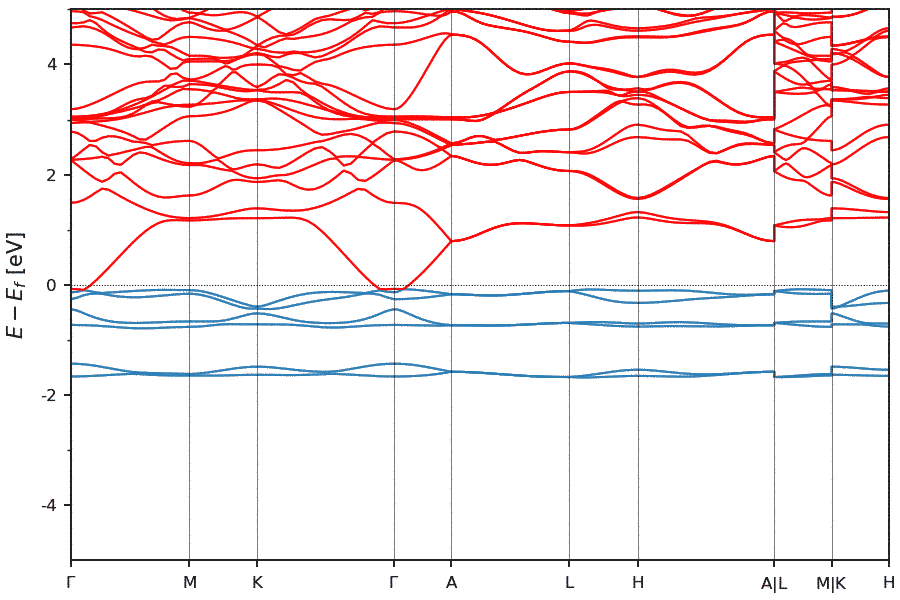}\\
\end{tabular}
\begin{tabular}{c c}
\scriptsize{$\rm{Al} \rm{Y}_{3} \rm{C}$ - \icsdweb{43869} - SG 221 ($Pm\bar{3}m$) - ES} & \scriptsize{$\rm{Y}_{3} \rm{Ga} \rm{C}$ - \icsdweb{56396} - SG 221 ($Pm\bar{3}m$) - ES}\\
\includegraphics[width=0.38\textwidth,angle=0]{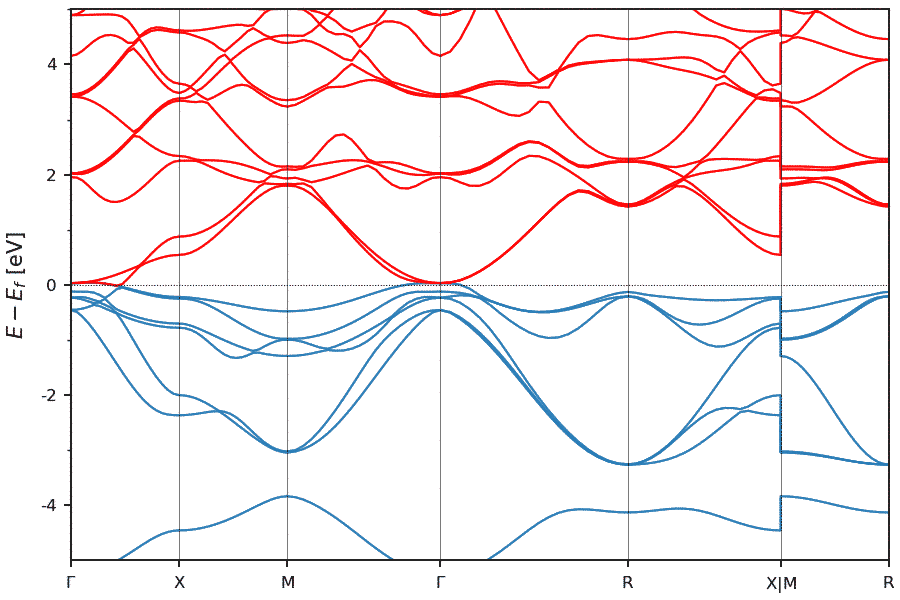} & \includegraphics[width=0.38\textwidth,angle=0]{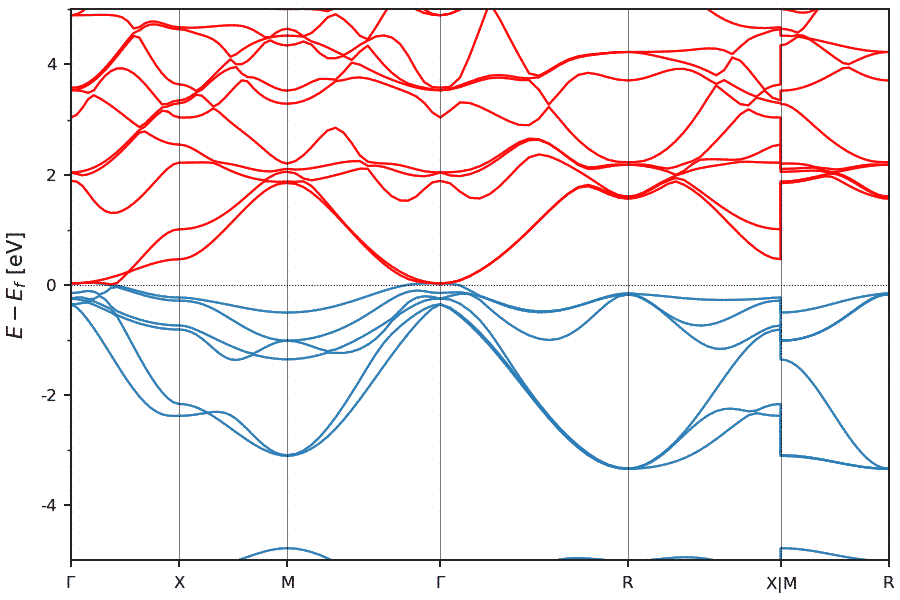}\\
\end{tabular}

\caption{\strongSOCES{5}}
\label{fig:strongSOC_ES5}
\end{figure}

\begin{figure}[ht]
\centering
\begin{tabular}{c c}
\scriptsize{$\rm{Al} \rm{Nb} \rm{Ni}_{2}$ - \icsdweb{58016} - SG 225 ($Fm\bar{3}m$) - ES} & \scriptsize{$\rm{Al} \rm{Ni}_{2} \rm{V}$ - \icsdweb{58071} - SG 225 ($Fm\bar{3}m$) - ES}\\
\includegraphics[width=0.38\textwidth,angle=0]{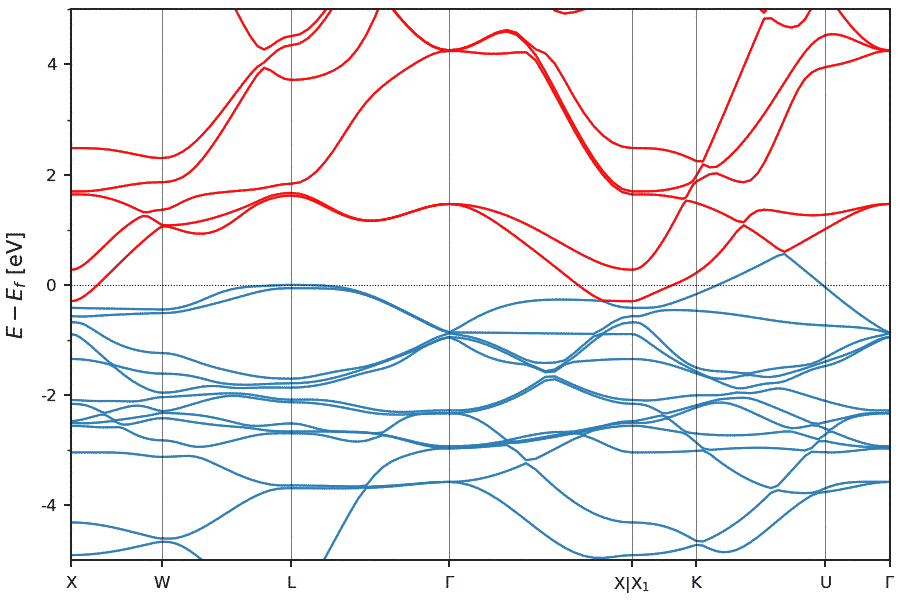} & \includegraphics[width=0.38\textwidth,angle=0]{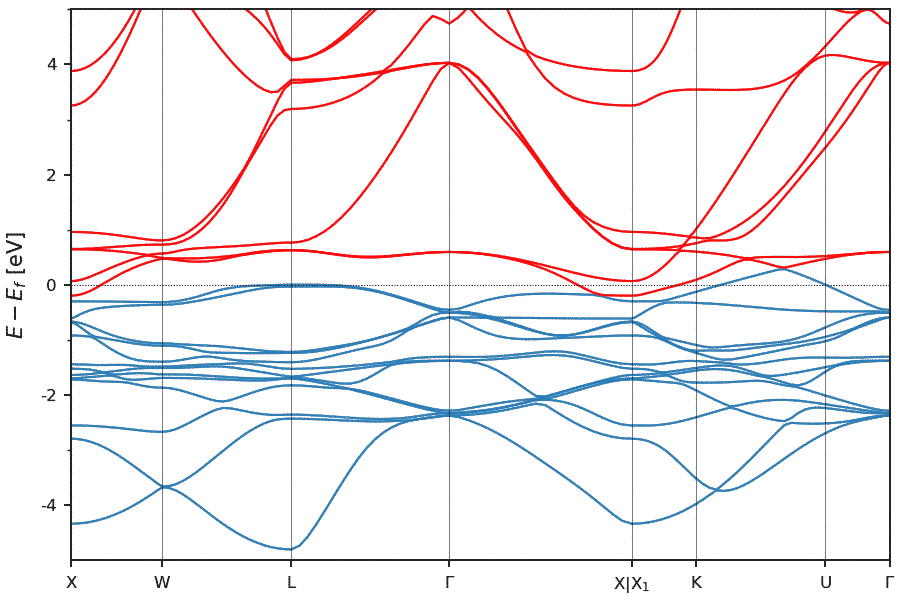}\\
\end{tabular}
\begin{tabular}{c c}
\scriptsize{$\rm{Cd} \rm{Cu}_{2} \rm{Zr}$ - \icsdweb{58961} - SG 225 ($Fm\bar{3}m$) - ES} & \scriptsize{$\rm{Cu}_{2} \rm{Zn} \rm{Zr}$ - \icsdweb{103161} - SG 225 ($Fm\bar{3}m$) - ES}\\
\includegraphics[width=0.38\textwidth,angle=0]{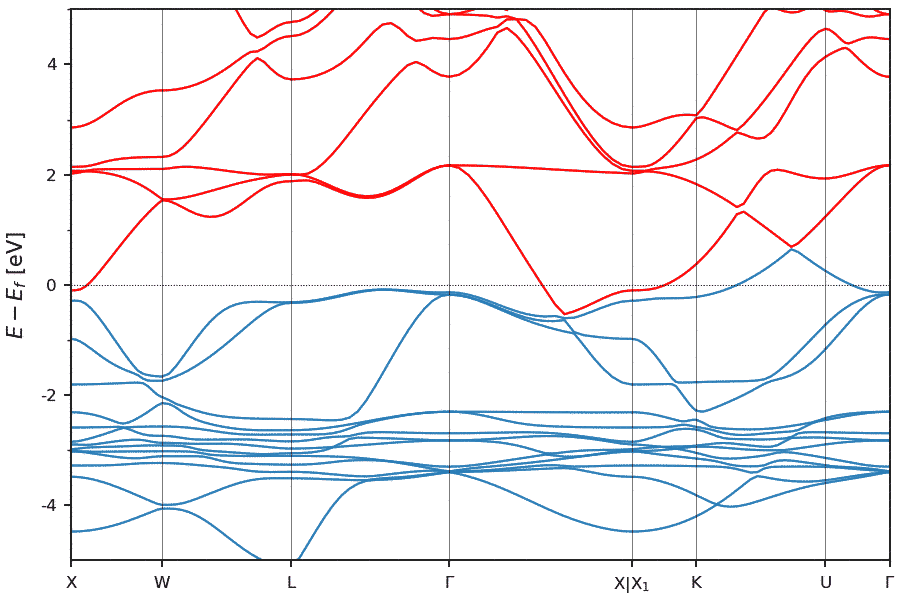} & \includegraphics[width=0.38\textwidth,angle=0]{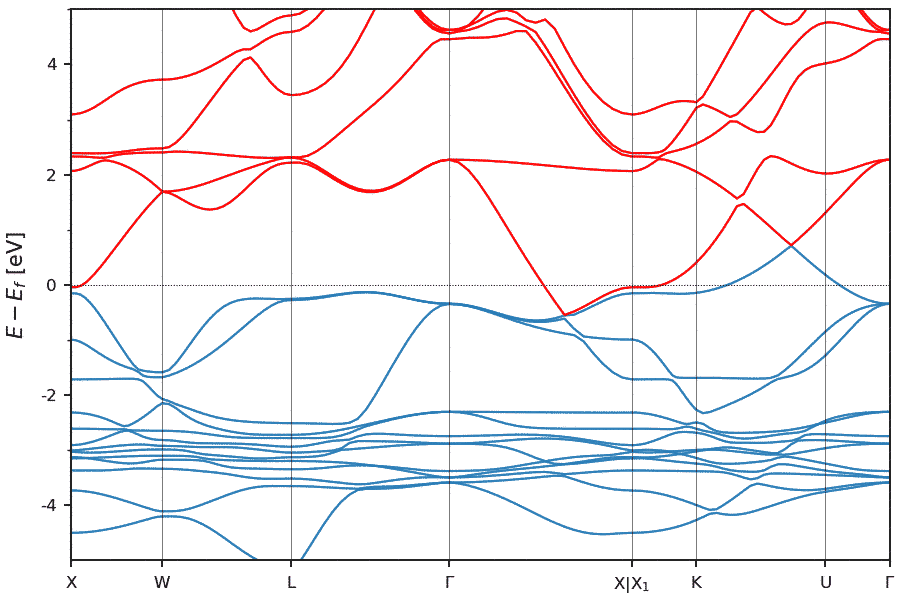}\\
\end{tabular}
\begin{tabular}{c c}
\scriptsize{$\rm{Ga} \rm{Ni}_{2} \rm{V}$ - \icsdweb{103892} - SG 225 ($Fm\bar{3}m$) - ES} & \scriptsize{$\rm{Ni}_{2} \rm{Sn} \rm{Ti}$ - \icsdweb{105369} - SG 225 ($Fm\bar{3}m$) - ES}\\
\includegraphics[width=0.38\textwidth,angle=0]{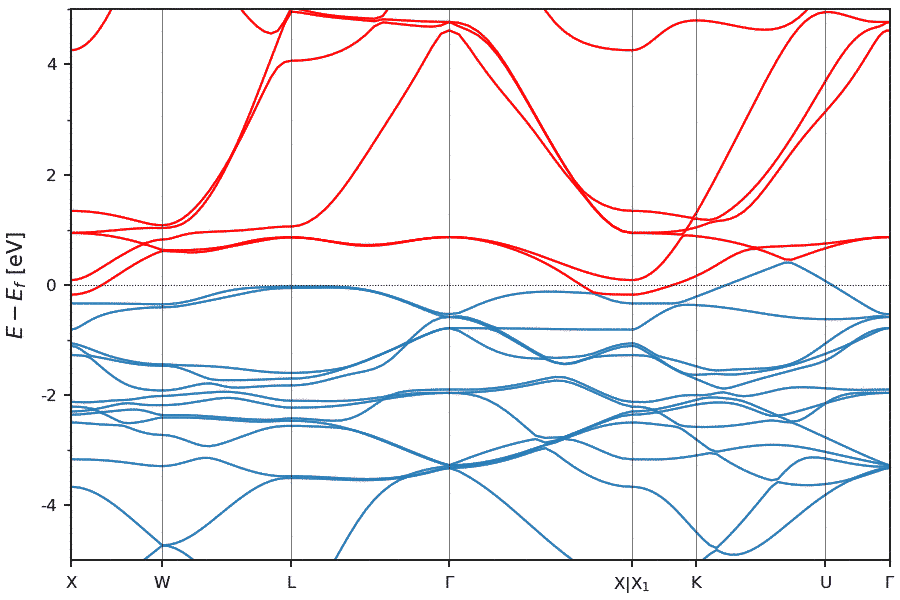} & \includegraphics[width=0.38\textwidth,angle=0]{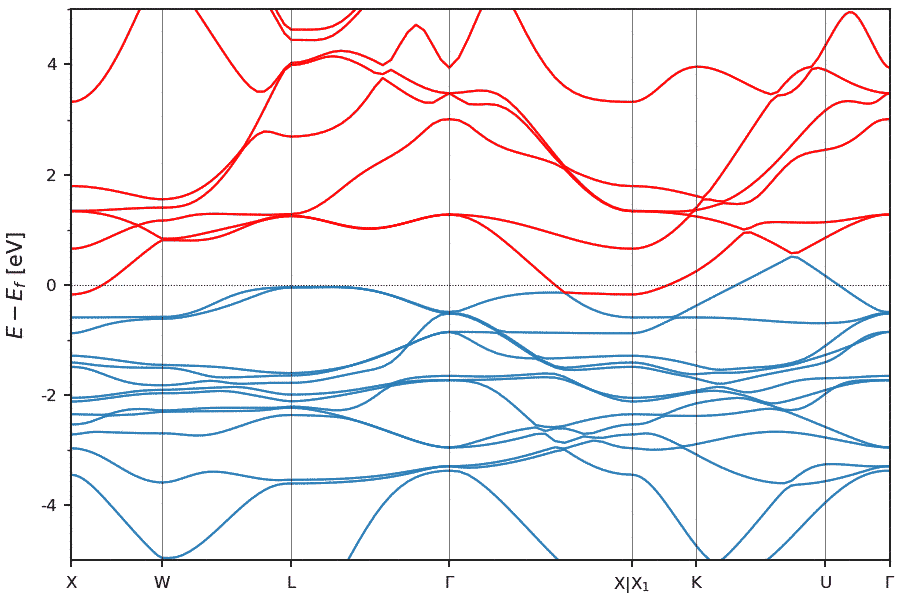}\\
\end{tabular}
\begin{tabular}{c c}
\scriptsize{$\rm{Ni}_{2} \rm{Sn} \rm{Zr}$ - \icsdweb{105383} - SG 225 ($Fm\bar{3}m$) - ES}\\
\includegraphics[width=0.38\textwidth,angle=0]{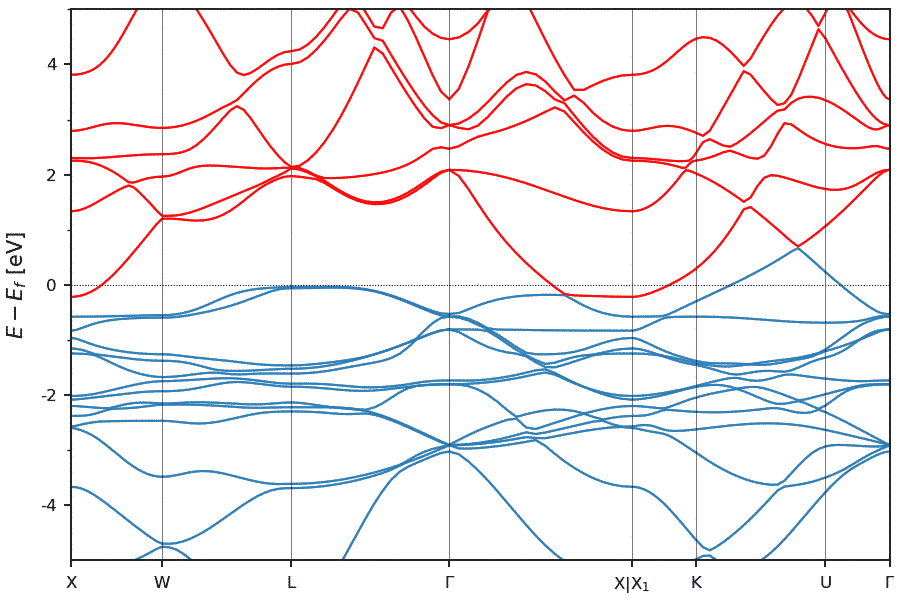}\\
\end{tabular}

\caption{\strongSOCES{6}}
\label{fig:strongSOC_ES6}
\end{figure}

\clearpage

\subsection{Semimetal-Insulator Transitions Driven by Spin-Orbit Coupling}
\label{App:transitions}

In this section, we show the materials with the largest gaps at $E_{F}$ or the fewest and smallest bulk Fermi pockets that are topological insulating in the presence of SOC and enforced or symmetry-indicated semimetals when the effects of SOC are neglected.  Specifically, when the strength of SOC is artificially tuned to zero without (un)inverting bands (\emph{i.e.} without changing the bulk band ordering), the bulk gaps of spinful TI, TCI, and HOTI phases necessarily close and the bulk transitions into a spinless (spin-degenerate) topological semimetal phase~\cite{YoungkukLineNode,ZhidaSemimetals,TMDHOTI}.  The prototypical example of an SOC-driven transition between a spinless topological semimetal phase without SOC and a TI phase with SOC occurs in graphene, which is an ESFD semimetal in the absence of SOC~\cite{QuantumChemistry,YoungkukLineNode,GrapheneReview}, and a 2D TI when the (weak) effects of SOC are taken into consideration~\cite{CharlieTI,KaneMeleZ2} (see \supappref{App:PhaseTransitionsNoSOCSOC}).  In Figs.~\ref{fig:ESFD_noSOC_to_TI1},~\ref{fig:ESFD_noSOC_to_TI2},~\ref{fig:ES_noSOC_to_TI1},~\ref{fig:ES_noSOC_to_TI2},~\ref{fig:NLCSM_noSOC_to_TI}, and~\ref{fig:SEBRSM_noSOC_to_TI}, we respectively show the spinful topological (crystalline) insulators that originate from weak-SOC ESFD, ES, NLC-symmetry-indicated (NLC-SM), and SEBR-symmetry-indicated (SEBR-SM) topological semimetals.  For concision, in Figs.~\ref{fig:ESFD_noSOC_to_TI1},~\ref{fig:ESFD_noSOC_to_TI2},~\ref{fig:ES_noSOC_to_TI1},~\ref{fig:ES_noSOC_to_TI2},~\ref{fig:NLCSM_noSOC_to_TI}, and~\ref{fig:SEBRSM_noSOC_to_TI}, we only show material band structures calculated incorporating the effects of SOC; the corresponding band structures and topological information calculated without SOC can be accessed for each material by clicking on the ICSD number listed above each plot.


\begin{figure}[ht]
\centering
\begin{tabular}{c c}
\scriptsize{$\rm{Pd} \rm{Te}_{2}$ - \icsdweb{42555} - SG 164 ($P\bar{3}m1$) - SEBR} & \scriptsize{$\rm{Ta}_{2} \rm{C}$ - \icsdweb{618840} - SG 164 ($P\bar{3}m1$) - SEBR}\\
\tiny{ $\;Z_{2,1}=0\;Z_{2,2}=0\;Z_{2,3}=0\;Z_4=3$ } & \tiny{ $\;Z_{2,1}=0\;Z_{2,2}=0\;Z_{2,3}=0\;Z_4=3$ }\\
\includegraphics[width=0.38\textwidth,angle=0]{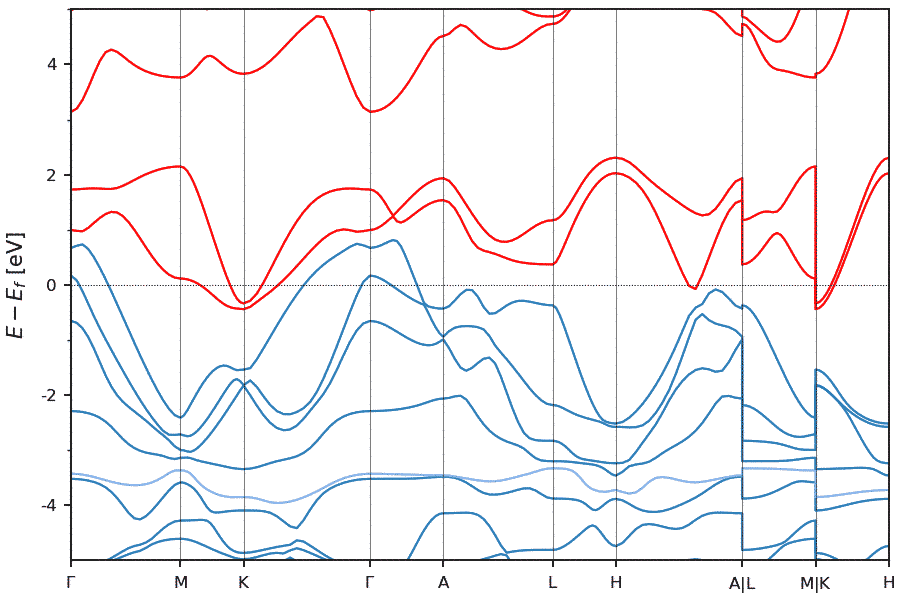} & \includegraphics[width=0.38\textwidth,angle=0]{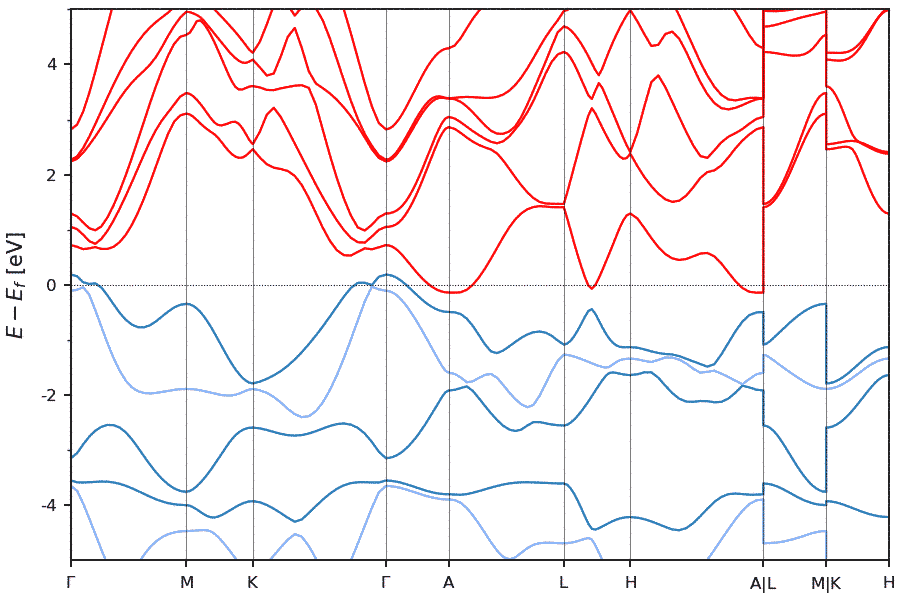}\\
\end{tabular}
\begin{tabular}{c c}
\scriptsize{$\rm{Hf} \rm{Ni}_{3}$ - \icsdweb{2414} - SG 166 ($R\bar{3}m$) - SEBR} & \scriptsize{$\rm{Cu} \rm{I}$ - \icsdweb{80259} - SG 166 ($R\bar{3}m$) - SEBR}\\
\tiny{ $\;Z_{2,1}=1\;Z_{2,2}=1\;Z_{2,3}=1\;Z_4=2$ } & \tiny{ $\;Z_{2,1}=0\;Z_{2,2}=0\;Z_{2,3}=0\;Z_4=3$ }\\
\includegraphics[width=0.38\textwidth,angle=0]{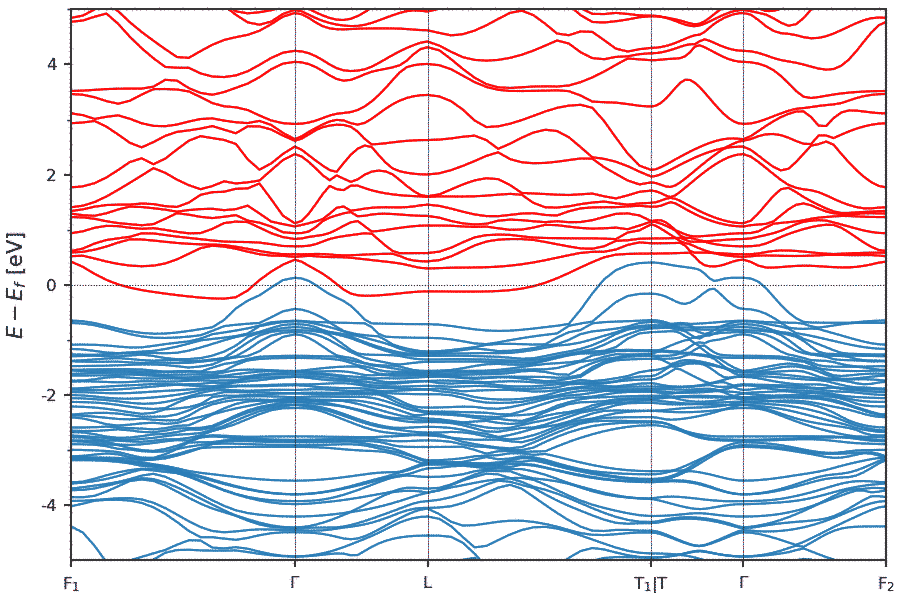} & \includegraphics[width=0.38\textwidth,angle=0]{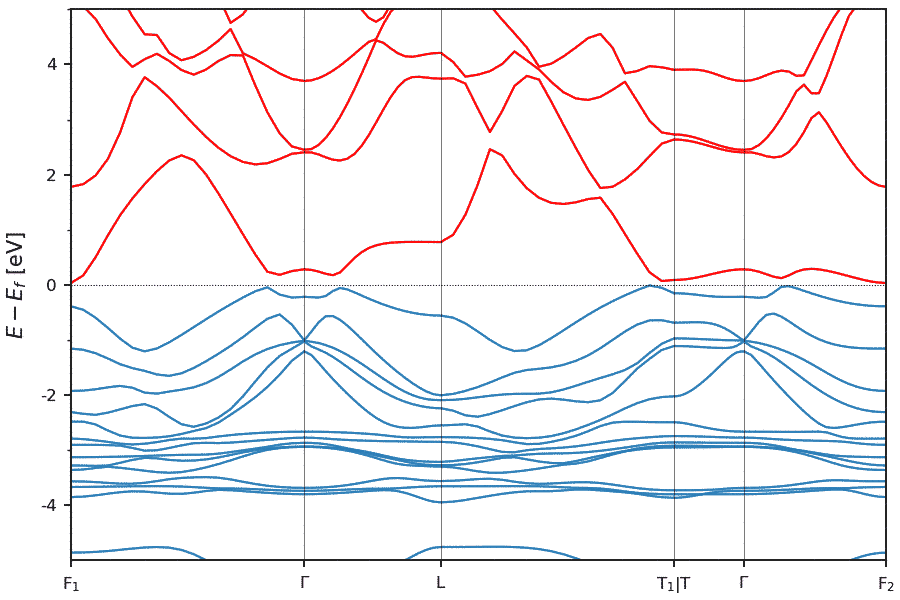}\\
\end{tabular}
\begin{tabular}{c c}
\scriptsize{$\rm{Sr} \rm{Ag}_{4} \rm{Sb}_{2}$ - \icsdweb{424311} - SG 166 ($R\bar{3}m$) - SEBR} & \scriptsize{$\rm{Sb}$ - \icsdweb{651490} - SG 166 ($R\bar{3}m$) - SEBR}\\
\tiny{ $\;Z_{2,1}=1\;Z_{2,2}=1\;Z_{2,3}=1\;Z_4=2$ } & \tiny{ $\;Z_{2,1}=1\;Z_{2,2}=1\;Z_{2,3}=1\;Z_4=3$ }\\
\includegraphics[width=0.38\textwidth,angle=0]{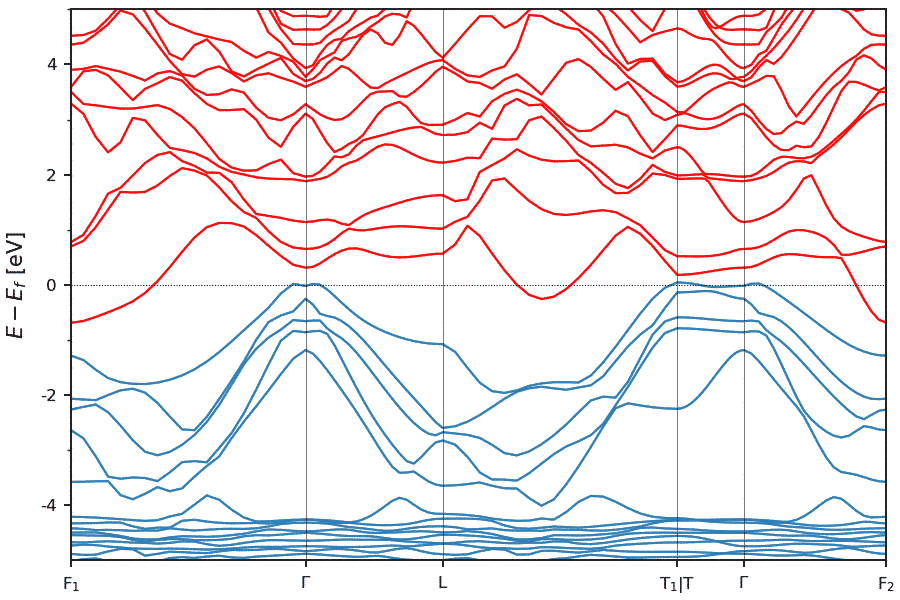} & \includegraphics[width=0.38\textwidth,angle=0]{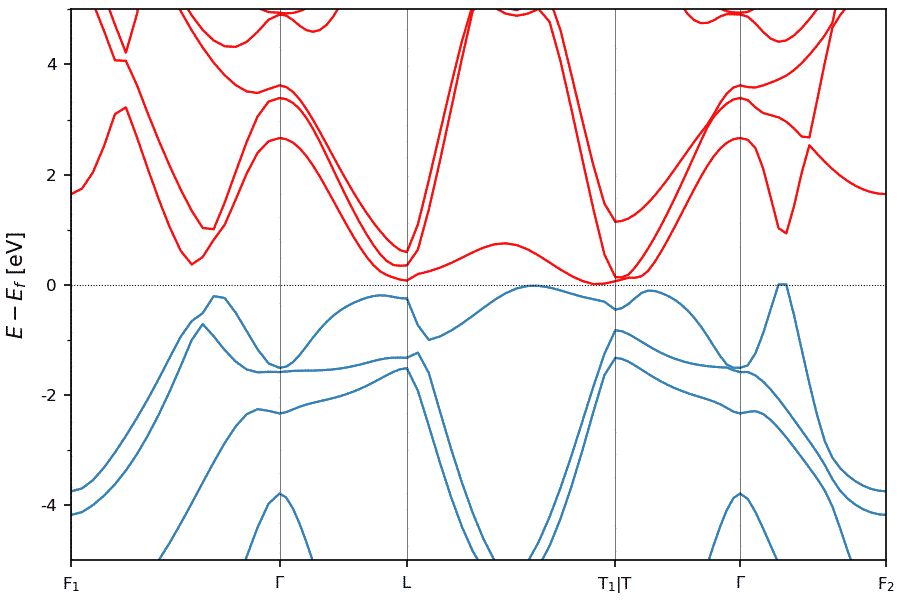}\\
\end{tabular}
\begin{tabular}{c c}
\scriptsize{$\rm{Hg} \rm{K} \rm{Sb}$ - \icsdweb{56201} - SG 194 ($P6_3/mmc$) - SEBR} & \scriptsize{$\rm{Ir}_{3} \rm{Sc} \rm{C}$ - \icsdweb{77044} - SG 221 ($Pm\bar{3}m$) - SEBR}\\
\tiny{ $\;Z_{2,1}=0\;Z_{2,2}=0\;Z_{2,3}=0\;Z_4=0\;Z_{6m,0}=2\;Z_{12}'=8$ } & \tiny{ $\;Z_{2,1}=0\;Z_{2,2}=0\;Z_{2,3}=0\;Z_4=3\;Z_{4m,\pi}=2\;Z_2=1\;Z_8=7$ }\\
\includegraphics[width=0.38\textwidth,angle=0]{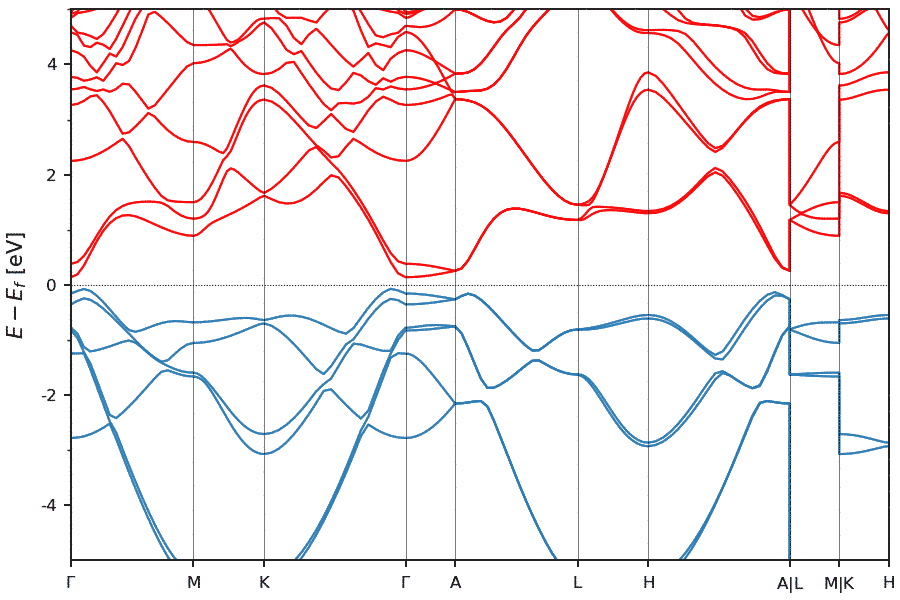} & \includegraphics[width=0.38\textwidth,angle=0]{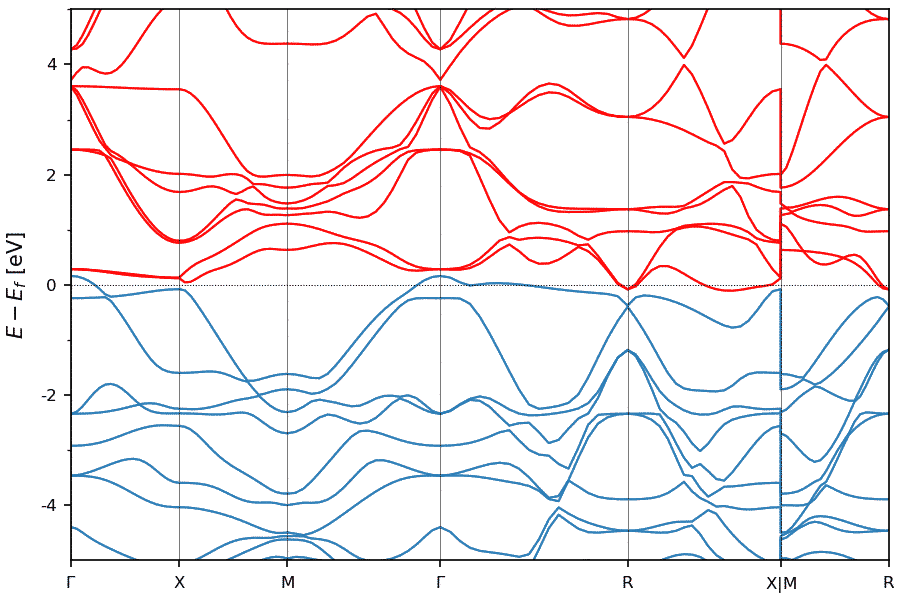}\\
\end{tabular}

\caption{\ESFDnoSOCtoTI{1}}
\label{fig:ESFD_noSOC_to_TI1}
\end{figure}

\begin{figure}[ht]
\centering
\begin{tabular}{c c}
\scriptsize{$\rm{Tl} \rm{Pd}_{3} \rm{H}$ - \icsdweb{247273} - SG 221 ($Pm\bar{3}m$) - SEBR} & \scriptsize{$\rm{Pb} \rm{Pt}_{3}$ - \icsdweb{648399} - SG 221 ($Pm\bar{3}m$) - SEBR}\\
\tiny{ $\;Z_{2,1}=1\;Z_{2,2}=1\;Z_{2,3}=1\;Z_4=0\;Z_{4m,\pi}=3\;Z_2=0\;Z_8=0$ } & \tiny{ $\;Z_{2,1}=1\;Z_{2,2}=1\;Z_{2,3}=1\;Z_4=2\;Z_{4m,\pi}=3\;Z_2=0\;Z_8=6$ }\\
\includegraphics[width=0.38\textwidth,angle=0]{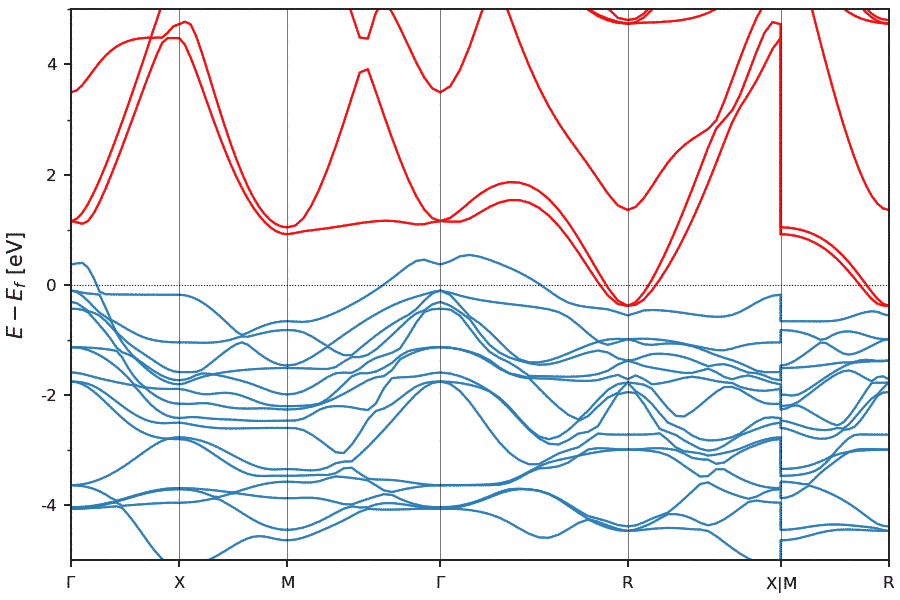} & \includegraphics[width=0.38\textwidth,angle=0]{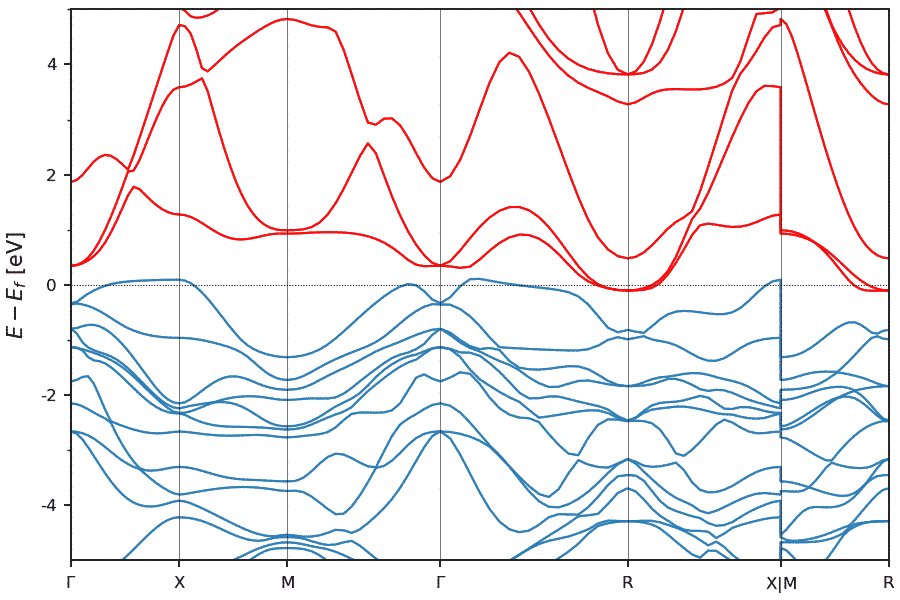}\\
\end{tabular}

\caption{\ESFDnoSOCtoTI{2}}
\label{fig:ESFD_noSOC_to_TI2}
\end{figure}


\begin{figure}[ht]
\centering
\begin{tabular}{c c}
\scriptsize{$\rm{Co} \rm{Sb}_{2}$ - \icsdweb{63553} - SG 14 ($P2_1/c$) - NLC} & \scriptsize{$\rm{Sb}_{2} \rm{Te}_{2}$ - \icsdweb{20459} - SG 164 ($P\bar{3}m1$) - SEBR}\\
\tiny{ $\;Z_{2,1}=1\;Z_{2,2}=0\;Z_{2,3}=0\;Z_4=3$ } & \tiny{ $\;Z_{2,1}=0\;Z_{2,2}=0\;Z_{2,3}=1\;Z_4=3$ }\\
\includegraphics[width=0.38\textwidth,angle=0]{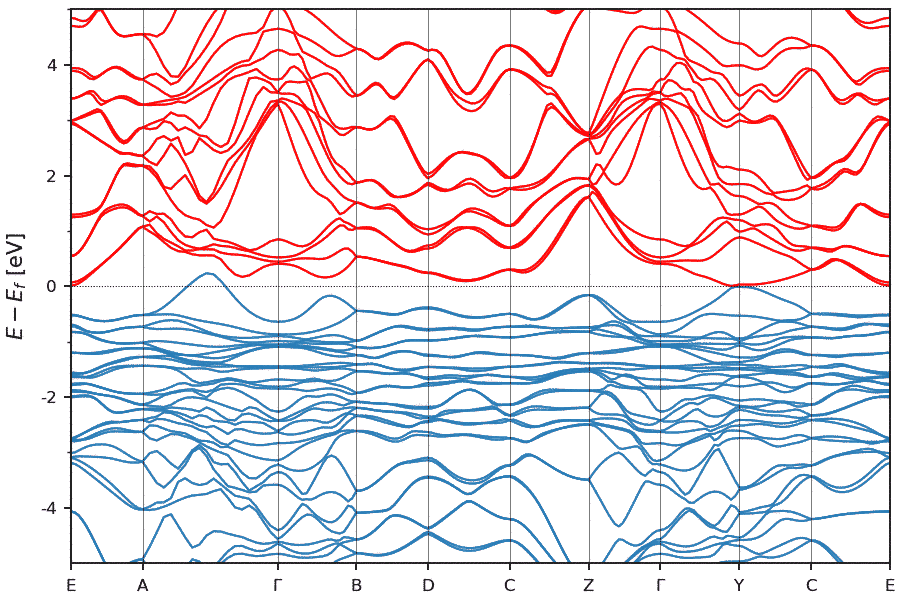} & \includegraphics[width=0.38\textwidth,angle=0]{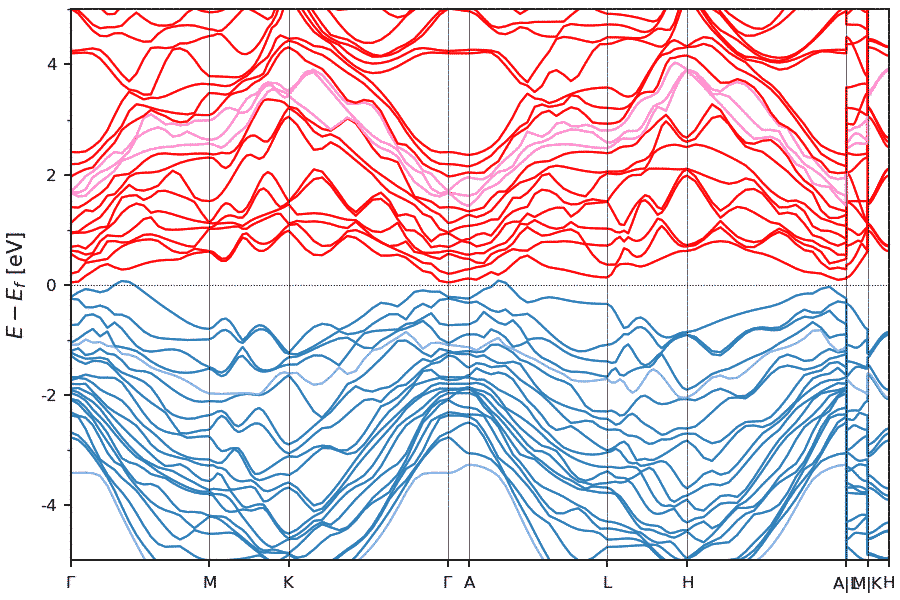}\\
\end{tabular}
\begin{tabular}{c c}
\scriptsize{$\rm{Sb}$ - \icsdweb{9859} - SG 166 ($R\bar{3}m$) - SEBR} & \scriptsize{$\rm{Ca} \rm{Ga}_{2} \rm{As}_{2}$ - \icsdweb{422526} - SG 166 ($R\bar{3}m$) - SEBR}\\
\tiny{ $\;Z_{2,1}=1\;Z_{2,2}=1\;Z_{2,3}=1\;Z_4=3$ } & \tiny{ $\;Z_{2,1}=1\;Z_{2,2}=1\;Z_{2,3}=1\;Z_4=3$ }\\
\includegraphics[width=0.38\textwidth,angle=0]{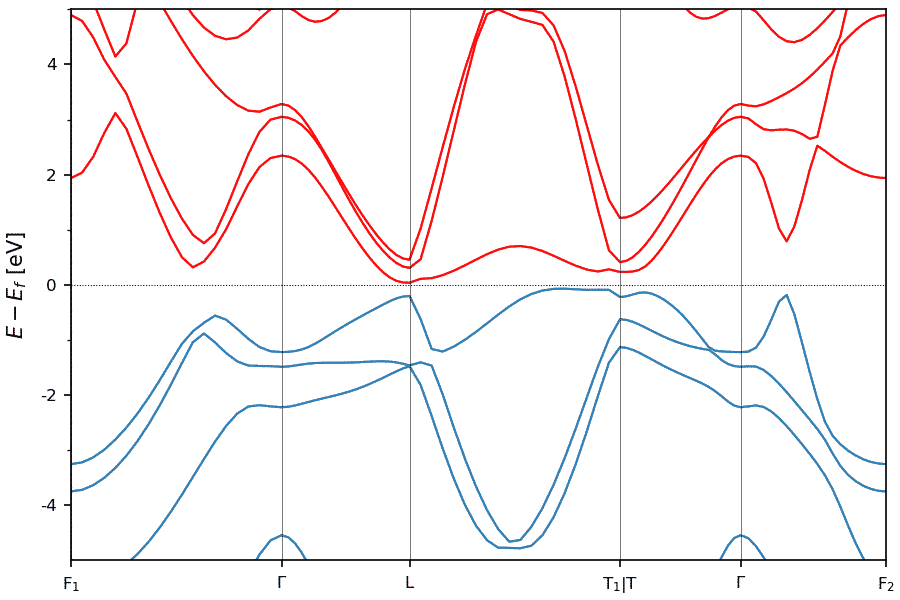} & \includegraphics[width=0.38\textwidth,angle=0]{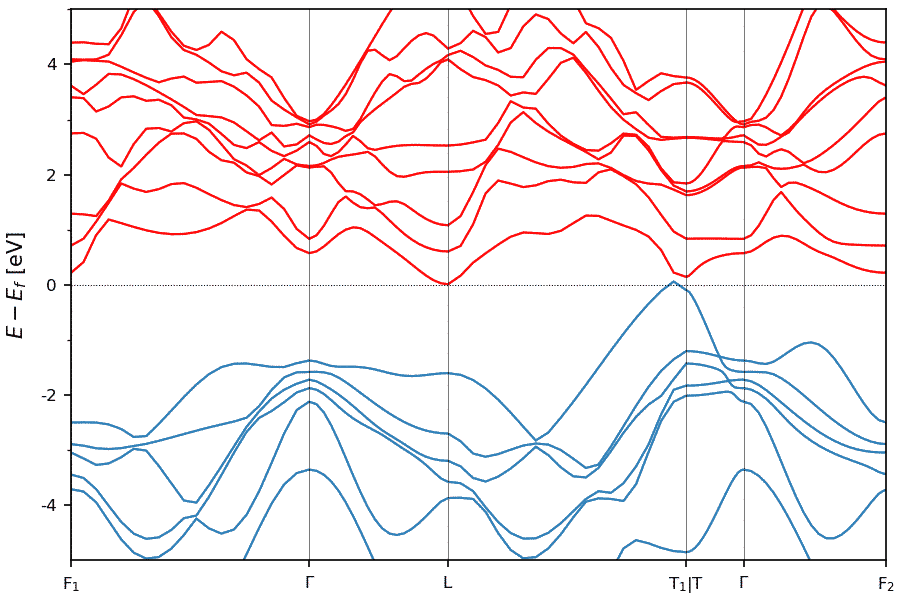}\\
\end{tabular}
\begin{tabular}{c c}
\scriptsize{$\rm{Sb} \rm{Tl} \rm{Te}_{2}$ - \icsdweb{651646} - SG 166 ($R\bar{3}m$) - SEBR} & \scriptsize{$\rm{Ca} \rm{Ag} \rm{P}$ - \icsdweb{10016} - SG 189 ($P\bar{6}2m$) - SEBR}\\
\tiny{ $\;Z_{2,1}=0\;Z_{2,2}=0\;Z_{2,3}=0\;Z_4=1$ } & \tiny{ $\;Z_{3m,0}=1\;Z_{3m,\pi}=0$ }\\
\includegraphics[width=0.38\textwidth,angle=0]{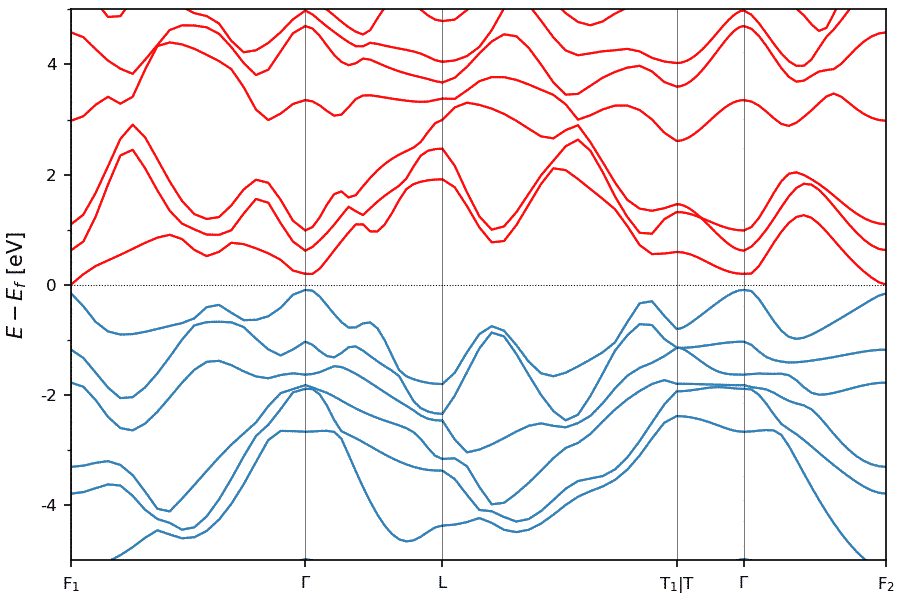} & \includegraphics[width=0.38\textwidth,angle=0]{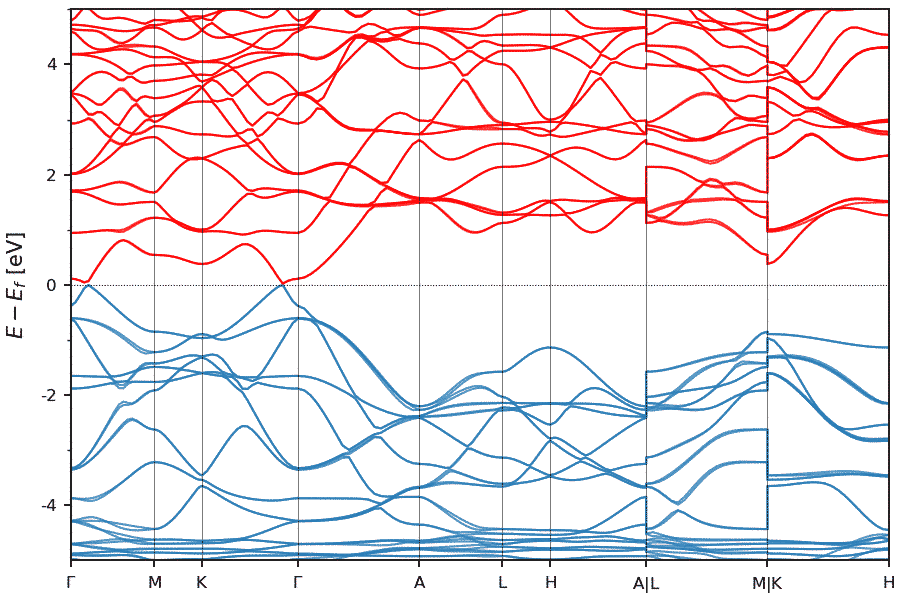}\\
\end{tabular}
\begin{tabular}{c c}
\scriptsize{$\rm{Sn} \rm{Se}$ - \icsdweb{52424} - SG 225 ($Fm\bar{3}m$) - SEBR} & \scriptsize{$\rm{Sn} \rm{S}$ - \icsdweb{651015} - SG 225 ($Fm\bar{3}m$) - SEBR}\\
\tiny{ $\;Z_{2,1}=0\;Z_{2,2}=0\;Z_{2,3}=0\;Z_4=0\;Z_2=0\;Z_8=4$ } & \tiny{ $\;Z_{2,1}=0\;Z_{2,2}=0\;Z_{2,3}=0\;Z_4=0\;Z_2=0\;Z_8=4$ }\\
\includegraphics[width=0.38\textwidth,angle=0]{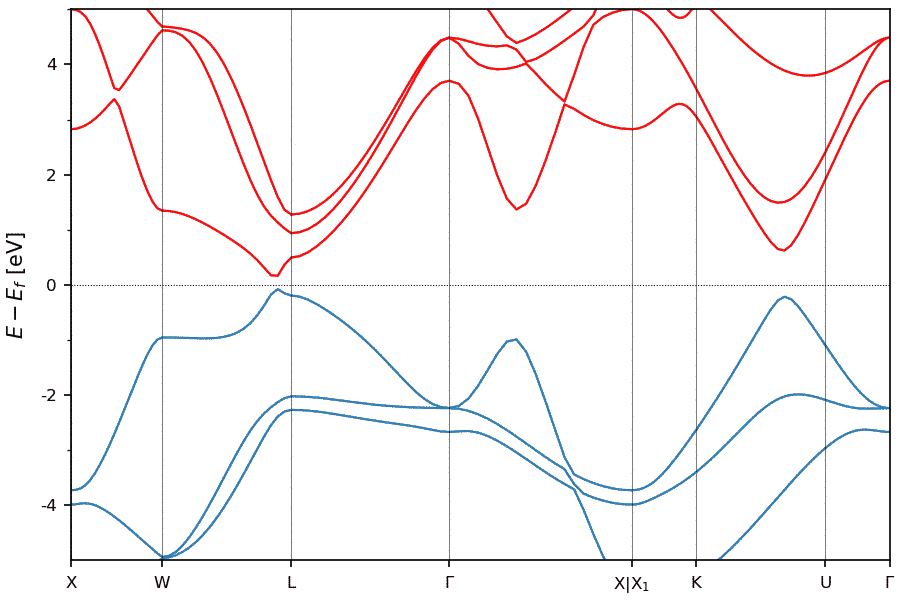} & \includegraphics[width=0.38\textwidth,angle=0]{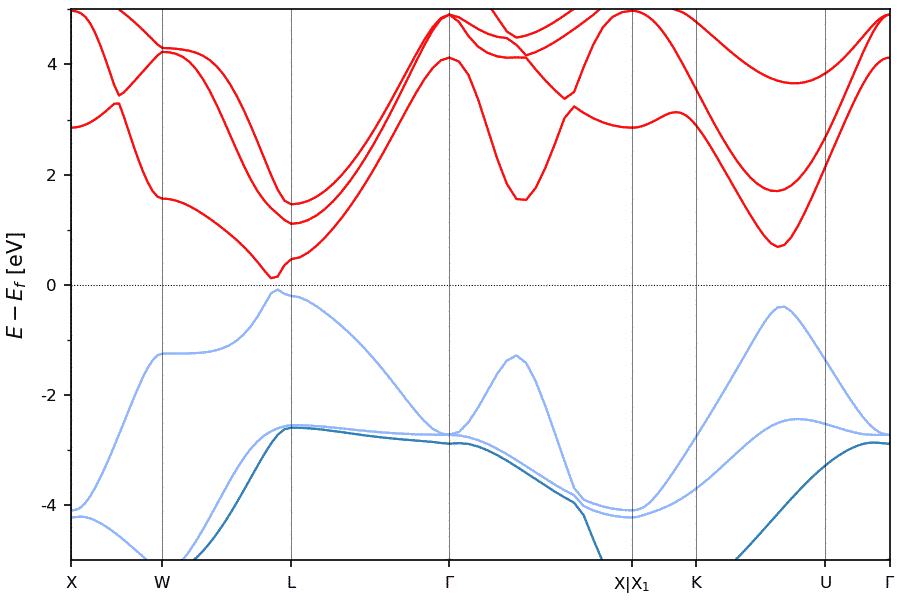}\\
\end{tabular}

\caption{\ESnoSOCtoTI{1}}
\label{fig:ES_noSOC_to_TI1}
\end{figure}

\begin{figure}[ht]
\centering
\begin{tabular}{c c}
\scriptsize{$\rm{Sn} \rm{Te}$ - \icsdweb{652759} - SG 225 ($Fm\bar{3}m$) - SEBR}\\
\tiny{ $\;Z_{2,1}=0\;Z_{2,2}=0\;Z_{2,3}=0\;Z_4=0\;Z_2=0\;Z_8=4$ }\\
\includegraphics[width=0.38\textwidth,angle=0]{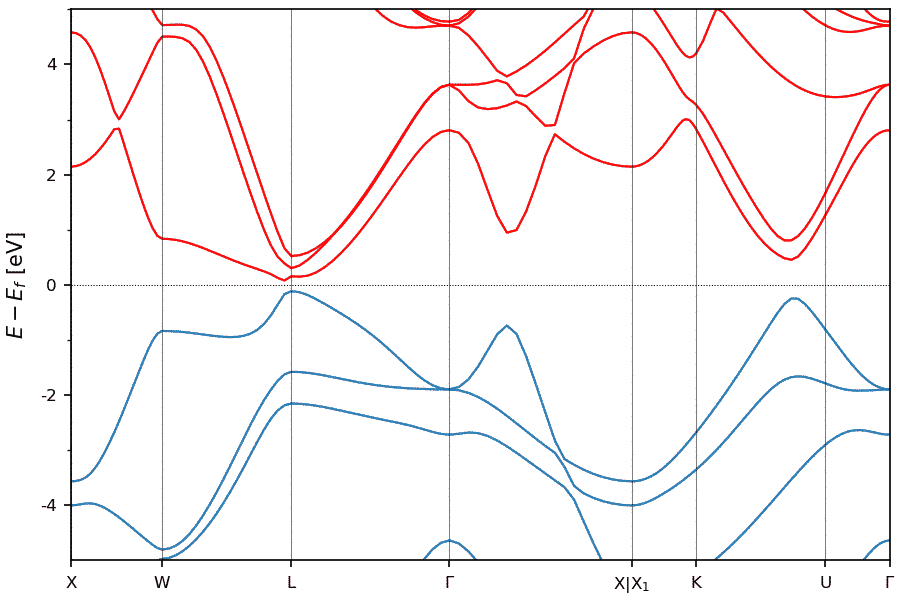}\\
\end{tabular}

\caption{\ESnoSOCtoTI{2}}
\label{fig:ES_noSOC_to_TI2}
\end{figure}


\begin{figure}[ht]
\centering
\begin{tabular}{c c}
\scriptsize{$\rm{Ca} \rm{As}_{3}$ - \icsdweb{193} - SG 2 ($P\bar{1}$) - NLC} & \scriptsize{$\rm{Ca} \rm{P}_{3}$ - \icsdweb{74479} - SG 2 ($P\bar{1}$) - NLC}\\
\tiny{ $\;Z_{2,1}=1\;Z_{2,2}=0\;Z_{2,3}=0\;Z_4=1$ } & \tiny{ $\;Z_{2,1}=0\;Z_{2,2}=1\;Z_{2,3}=0\;Z_4=1$ }\\
\includegraphics[width=0.38\textwidth,angle=0]{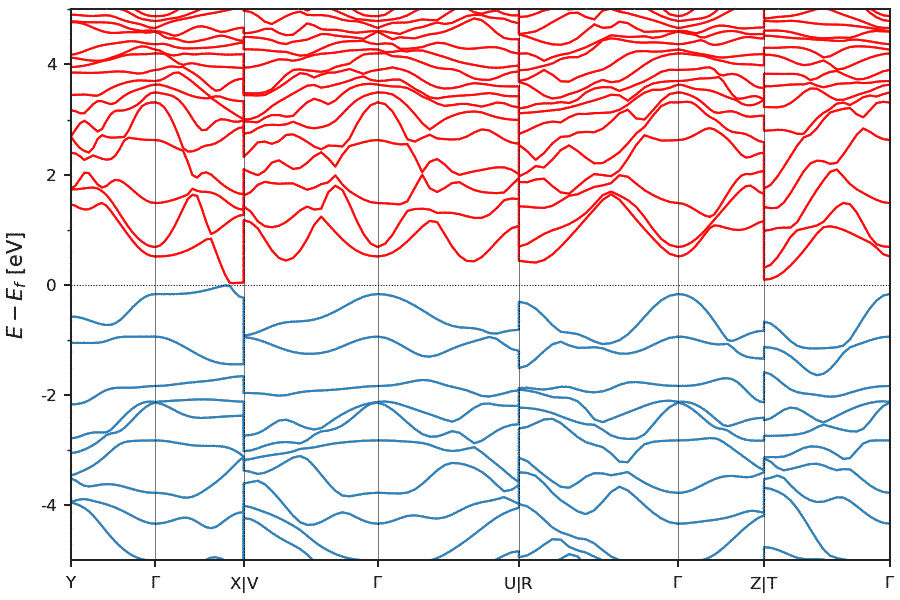} & \includegraphics[width=0.38\textwidth,angle=0]{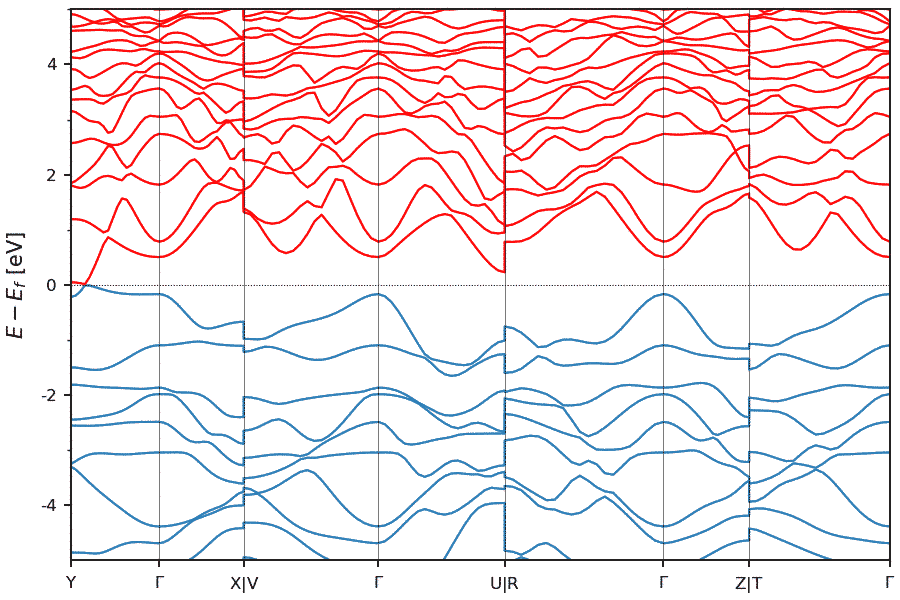}\\
\end{tabular}
\begin{tabular}{c c}
\scriptsize{$\rm{Ta}_{6} \rm{S}$ - \icsdweb{202564} - SG 2 ($P\bar{1}$) - NLC} & \scriptsize{$\rm{Rh}_{3} \rm{Ga}_{5}$ - \icsdweb{240179} - SG 2 ($P\bar{1}$) - NLC}\\
\tiny{ $\;Z_{2,1}=0\;Z_{2,2}=0\;Z_{2,3}=1\;Z_4=2$ } & \tiny{ $\;Z_{2,1}=1\;Z_{2,2}=0\;Z_{2,3}=0\;Z_4=0$ }\\
\includegraphics[width=0.38\textwidth,angle=0]{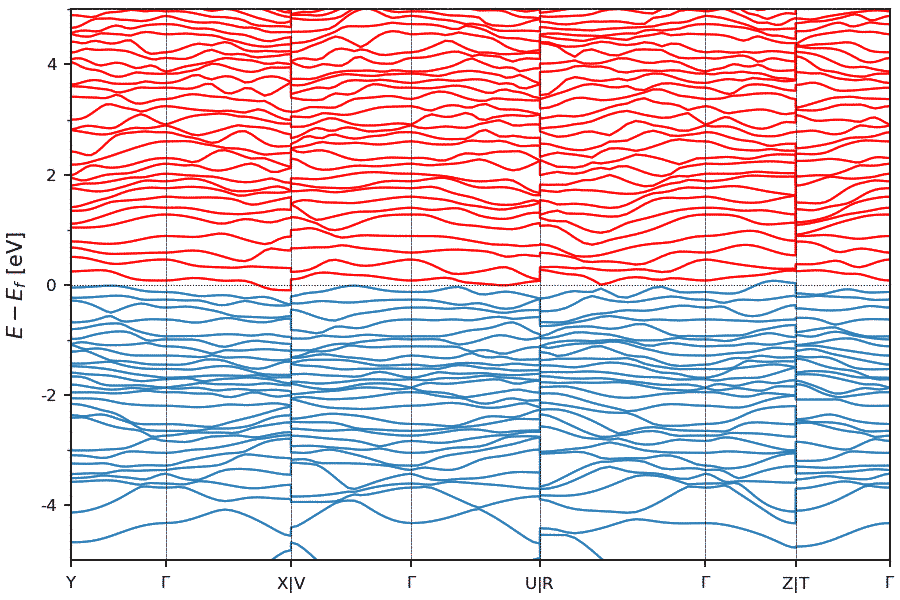} & \includegraphics[width=0.38\textwidth,angle=0]{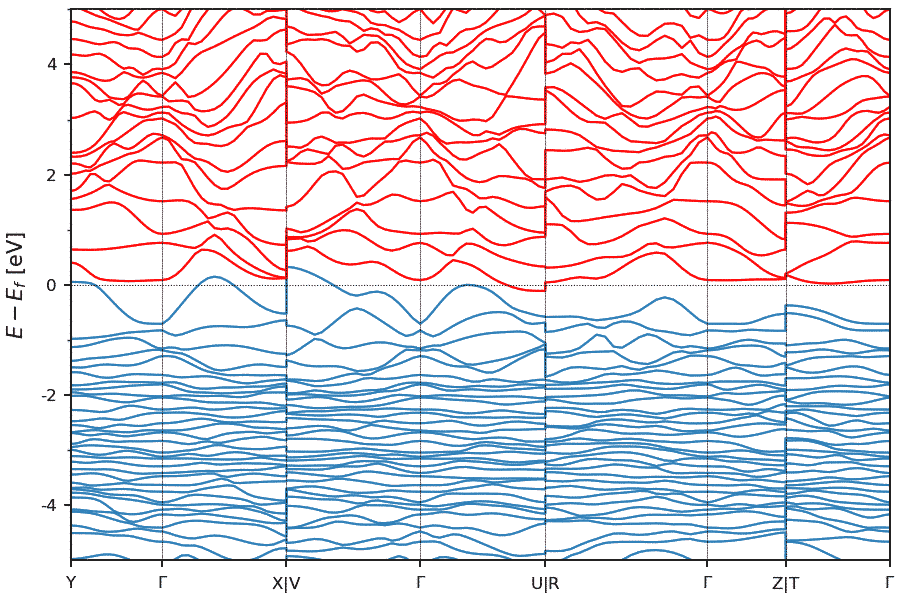}\\
\end{tabular}
\begin{tabular}{c c}
\scriptsize{$\rm{Al}_{2} \rm{Fe}_{3} \rm{Si}_{3}$ - \icsdweb{422342} - SG 2 ($P\bar{1}$) - NLC} & \scriptsize{$\rm{Y} \rm{As} \rm{Se}$ - \icsdweb{611398} - SG 14 ($P2_1/c$) - NLC}\\
\tiny{ $\;Z_{2,1}=0\;Z_{2,2}=1\;Z_{2,3}=0\;Z_4=3$ } & \tiny{ $\;Z_{2,1}=0\;Z_{2,2}=0\;Z_{2,3}=0\;Z_4=2$ }\\
\includegraphics[width=0.38\textwidth,angle=0]{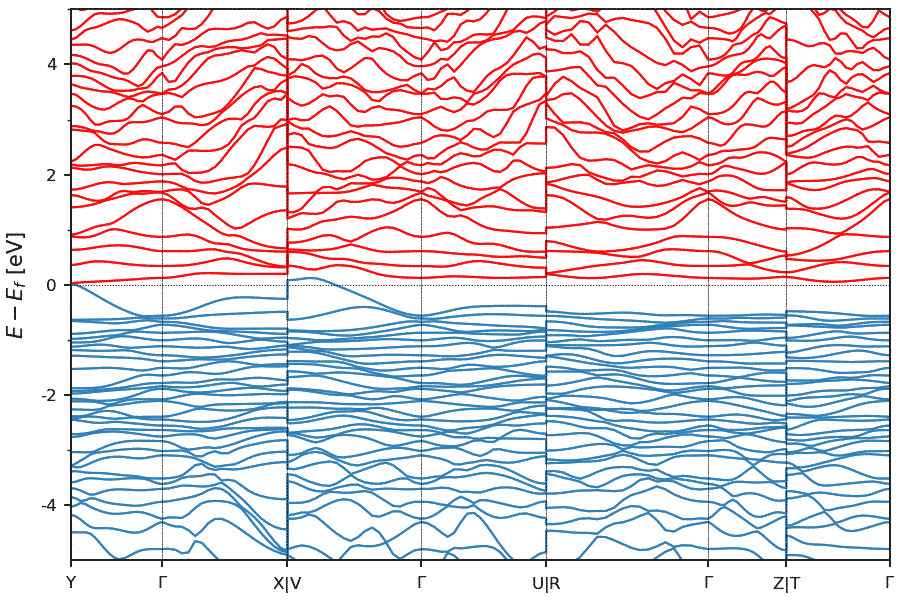} & \includegraphics[width=0.38\textwidth,angle=0]{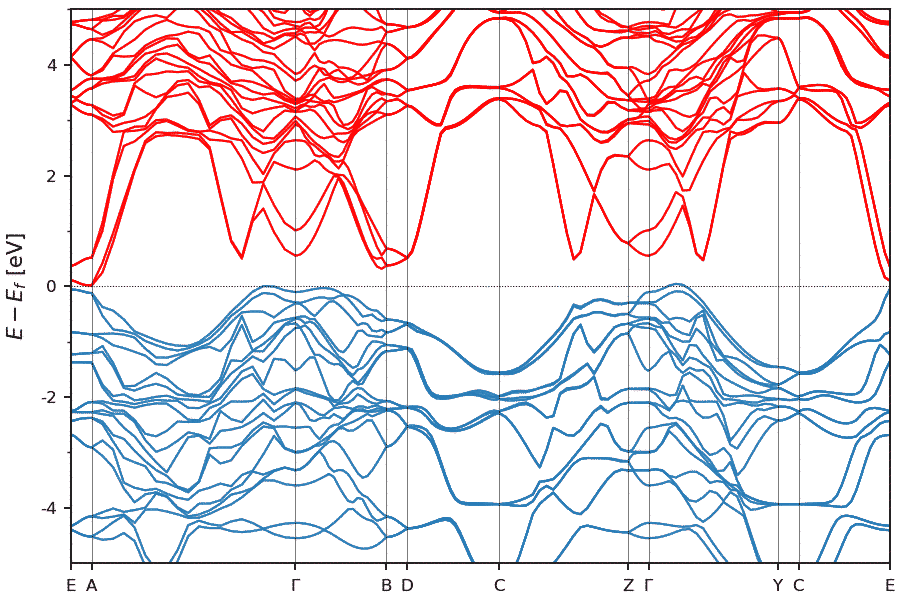}\\
\end{tabular}

\caption{The spinful topological (crystalline) insulators with the largest bulk gaps or the fewest Fermi pockets that are classified as NLC-SM topological semimetals when the effects of SOC are neglected.}
\label{fig:NLCSM_noSOC_to_TI}
\end{figure}


\begin{figure}[ht]
\centering
\begin{tabular}{c c}
\scriptsize{$\rm{Ni} \rm{Ti}_{3} \rm{S}_{6}$ - \icsdweb{26312} - SG 148 ($R\bar{3}$) - SEBR} & \scriptsize{$\rm{Tl}_{6} \rm{Te} \rm{O}_{12}$ - \icsdweb{37134} - SG 148 ($R\bar{3}$) - SEBR}\\
\tiny{ $\;Z_{2,1}=1\;Z_{2,2}=1\;Z_{2,3}=1\;Z_4=3$ } & \tiny{ $\;Z_{2,1}=0\;Z_{2,2}=0\;Z_{2,3}=0\;Z_4=3$ }\\
\includegraphics[width=0.38\textwidth,angle=0]{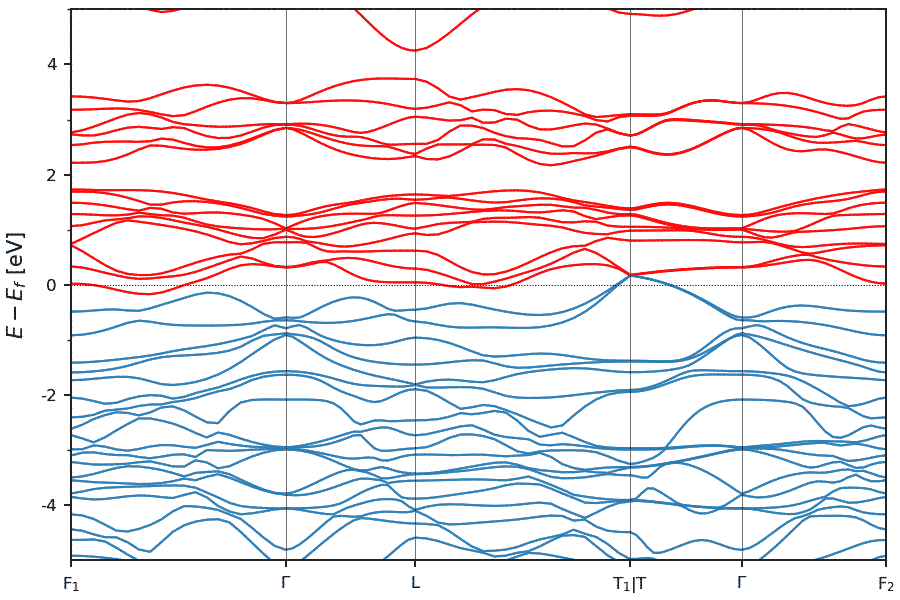} & \includegraphics[width=0.38\textwidth,angle=0]{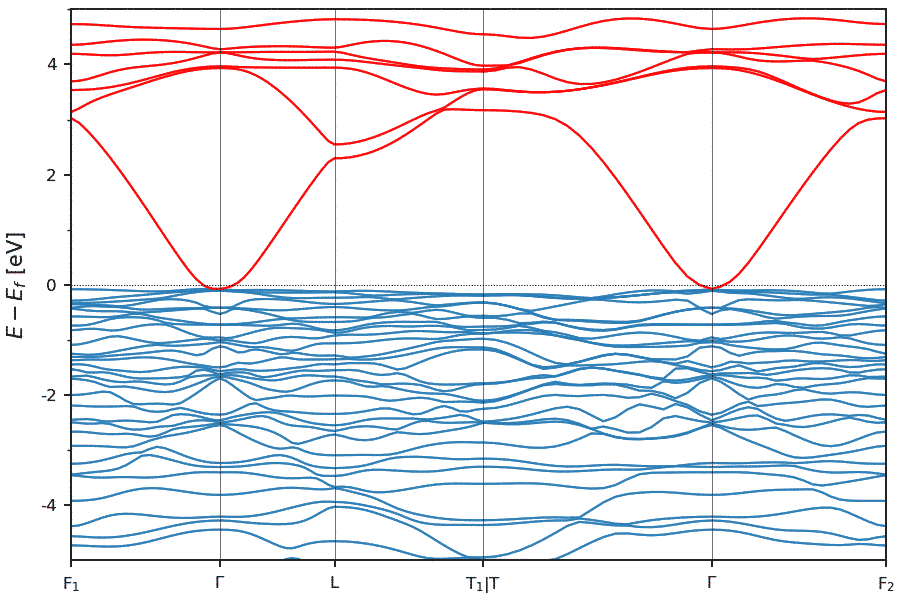}\\
\end{tabular}
\begin{tabular}{c c}
\scriptsize{$\rm{Mo}_{6} \rm{Te}_{8}$ - \icsdweb{59375} - SG 148 ($R\bar{3}$) - SEBR} & \scriptsize{$\rm{Mo}_{3} \rm{S}_{4}$ - \icsdweb{644244} - SG 148 ($R\bar{3}$) - SEBR}\\
\tiny{ $\;Z_{2,1}=1\;Z_{2,2}=1\;Z_{2,3}=1\;Z_4=3$ } & \tiny{ $\;Z_{2,1}=1\;Z_{2,2}=1\;Z_{2,3}=1\;Z_4=2$ }\\
\includegraphics[width=0.38\textwidth,angle=0]{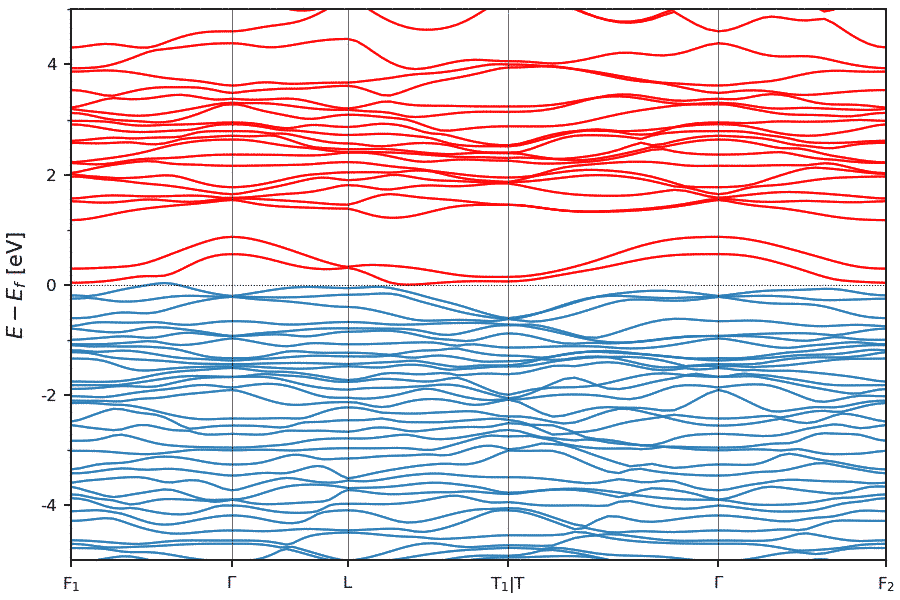} & \includegraphics[width=0.38\textwidth,angle=0]{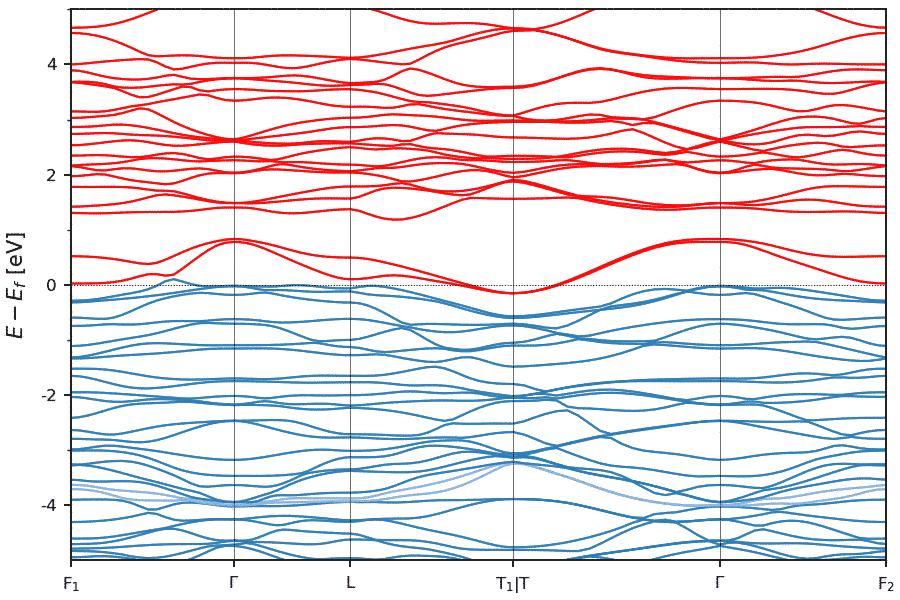}\\
\end{tabular}
\begin{tabular}{c c}
\scriptsize{$\rm{Mo}_{3} \rm{Se}_{4}$ - \icsdweb{644336} - SG 148 ($R\bar{3}$) - SEBR}\\
\tiny{ $\;Z_{2,1}=1\;Z_{2,2}=1\;Z_{2,3}=1\;Z_4=1$ }\\
\includegraphics[width=0.38\textwidth,angle=0]{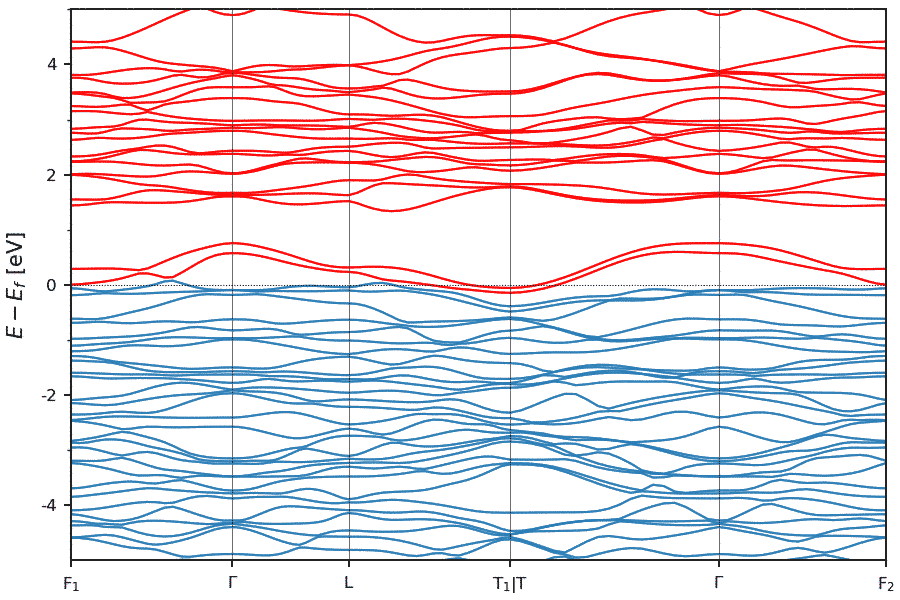}\\
\end{tabular}

\caption{The spinful topological (crystalline) insulators with the largest bulk gaps or the fewest Fermi pockets that are classified as SEBR-SM topological semimetals when the effects of SOC are neglected.}
\label{fig:SEBRSM_noSOC_to_TI}
\end{figure}

\clearpage

\subsection{Materials with Fragile Topological Bands at or Close to $E_{F}$}
\label{App:fragileBands}

In this section, we show the materials with the largest gaps at $E_{F}$ or the fewest and smallest bulk Fermi pockets that host groups of well-isolated bands at or near $E_{F}$ with symmetry-indicated fragile topology~\cite{AshvinFragile,JenFragile1,AshvinFragile2,BarryFragile,AdrianFragile,KoreanFragile,ZhidaFragile,FragileFlowMeta,ZhidaBLG,ZhidaFragile2,KoreanFragileInversion,DelicateAris,MBP_fragile}.  Specifically, as derived in Refs.~\onlinecite{JenFragile1,ZhidaFragile,KoreanFragileInversion}, in some cases, isolated groups of bands within the spectrum can be diagnosed as fragile topological, even if the entire valence (or conduction) manifold does not exhibit fragile topology.  In Figs.~\ref{fig:fragile_SEBR1},~\ref{fig:fragile_SEBR2},~\ref{fig:fragile_ESFD1},~\ref{fig:fragile_ESFD2},~\ref{fig:fragile_ES}, and~\ref{fig:fragile_LCEBR}, we respectively show representative examples of SEBR-, ESFD-, ES-, and LCEBR-classified materials with well-isolated fragile bands close to $E_{F}$ [we do not find any examples of NLC-classified topological (crystalline) insulators with few bulk Fermi pockets and well-isolated fragile bands close to $E_{F}$].  Though just over half of the materials listed in Figs.~\ref{fig:fragile_SEBR1},~\ref{fig:fragile_SEBR2},~\ref{fig:fragile_ESFD1},~\ref{fig:fragile_ESFD2},~\ref{fig:fragile_ES}, and~\ref{fig:fragile_LCEBR} were previously identified in Ref.~\onlinecite{ZhidaFragile} as hosting fragile bands, the remaining materials in this section have not been previously highlighted for hosting fragile topology.  The materials with well-isolated fragile bands not previously listed in Ref.~\onlinecite{ZhidaFragile} are Sb$_2$Te$_2$ [\icsdweb{20459}, SG 164 ($P\bar{3}m1$)], Ta$_2$S$_2$C [\icsdweb{23790}, SG 164 ($P\bar{3}m1$)], Pb$_2$Bi$_2$Se$_5$ [\icsdweb{30372}, SG 164 ($P\bar{3}m1$)], BiTe [\icsdweb{30525}, SG 164 ($P\bar{3}m1$)], As$_2$Ge$_5$Te$_8$ [\icsdweb{63174}, SG 164 ($P\bar{3}m1$)], Ta$_2$C [\icsdweb{409555}, SG 164 ($P\bar{3}m1$)], BiSe [\icsdweb{617073}, SG 164 ($P\bar{3}m1$)], and TiS$_2$ [\icsdweb{72042}, SG 227 ($Fd\bar{3}m$)] in Figs.~\ref{fig:fragile_SEBR1} and~\ref{fig:fragile_SEBR2}; TaSe$_2$ [\icsdweb{24313}, SG 164 ($P\bar{3}m1$)], BiTe [\icsdweb{44984}, SG 225 ($Fm\bar{3}m$)], and Bi$_4$Rh [\icsdweb{58854}, SG 230 ($Ia\bar{3}d$)] in Figs.~\ref{fig:fragile_ESFD1} and~\ref{fig:fragile_ESFD2}; SrZn$_2$Sb$_2$ [\icsdweb{12152}, SG 164 ($P\bar{3}m1$)], Sc$_2$C [\icsdweb{280743}, SG 164 ($P\bar{3}m1$)], and TaN [\icsdweb{105123}, SG 194 ($P6_{3}/mmc$)] in Fig.~\ref{fig:fragile_ES}; and TiS$_2$ [\icsdweb{91579}, SG 164 ($P\bar{3}m1$)] and Al$_5$C$_3$N [\icsdweb{36303}, SG 186 ($P6_{3}mc$)] in Fig.~\ref{fig:fragile_LCEBR}.

Finally, because recent works have demonstrated that insulators with fragile topological bands can exhibit anomalous corner modes~\cite{TMDHOTI,WiederAxion,HingeSM,AshvinFragile2,KoreanFragile,KoreanFragileInversion} and nontrivial twisted-boundary~\cite{ZhidaFragile2,FragileFlowMeta} and topological defect responses~\cite{WladCorners}, then the materials highlighted in this section represent new avenues for experimentally investigating novel topological response effects in solid-state systems.


\begin{figure}[ht]
\centering
\begin{tabular}{c c}
\scriptsize{$\rm{Sb}_{2} \rm{Te}_{2}$ - \icsdweb{20459} - SG 164 ($P\bar{3}m1$) - SEBR} & \scriptsize{$\rm{Ta}_{2} \rm{S}_{2} \rm{C}$ - \icsdweb{23790} - SG 164 ($P\bar{3}m1$) - SEBR}\\
\tiny{ $\;Z_{2,1}=0\;Z_{2,2}=0\;Z_{2,3}=1\;Z_4=3$ } & \tiny{ $\;Z_{2,1}=0\;Z_{2,2}=0\;Z_{2,3}=1\;Z_4=1$ }\\
\includegraphics[width=0.38\textwidth,angle=0]{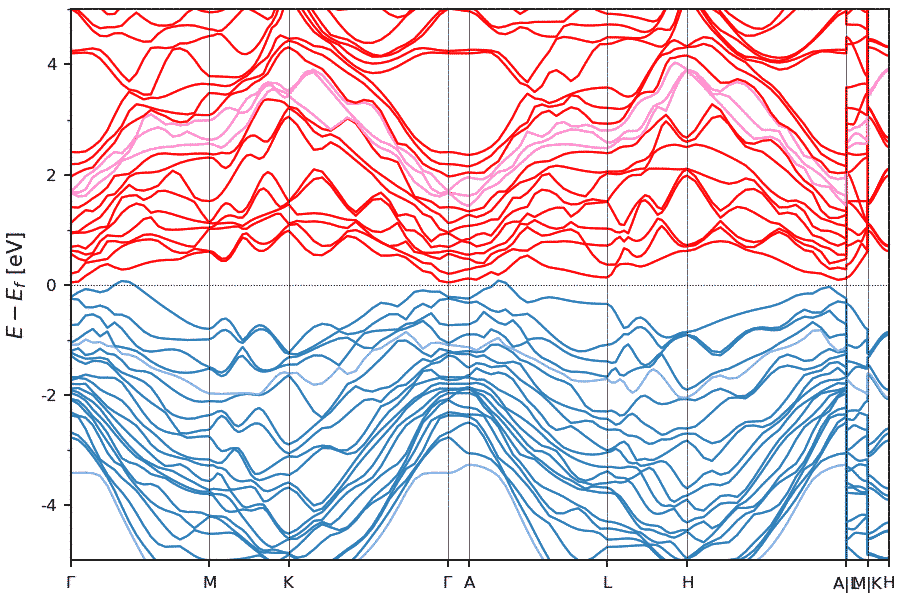} & \includegraphics[width=0.38\textwidth,angle=0]{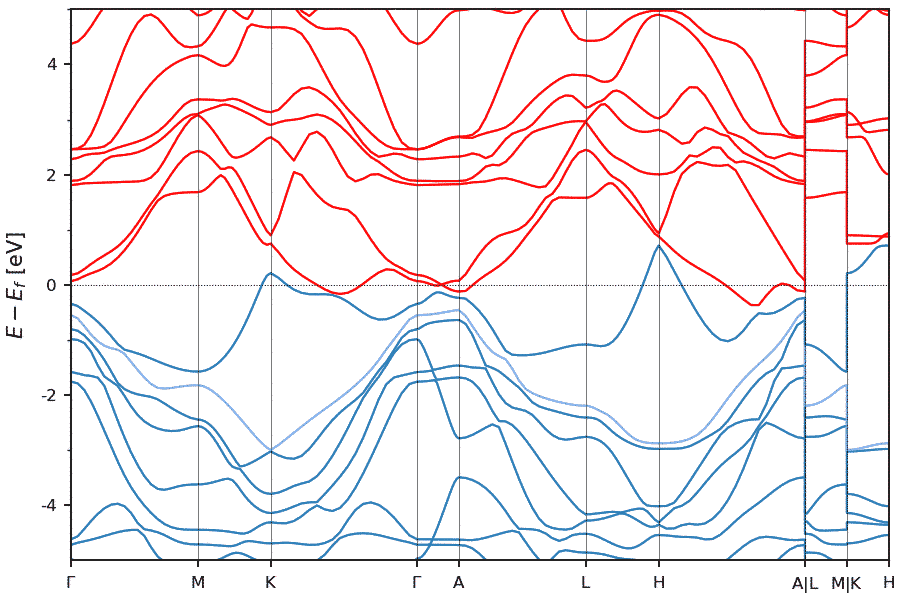}\\
\end{tabular}
\begin{tabular}{c c}
\scriptsize{$\rm{Pb}_{2} \rm{Bi}_{2} \rm{Se}_{5}$ - \icsdweb{30372} - SG 164 ($P\bar{3}m1$) - SEBR} & \scriptsize{$\rm{Bi} \rm{Te}$ - \icsdweb{30525} - SG 164 ($P\bar{3}m1$) - SEBR}\\
\tiny{ $\;Z_{2,1}=0\;Z_{2,2}=0\;Z_{2,3}=0\;Z_4=3$ } & \tiny{ $\;Z_{2,1}=0\;Z_{2,2}=0\;Z_{2,3}=1\;Z_4=0$ }\\
\includegraphics[width=0.38\textwidth,angle=0]{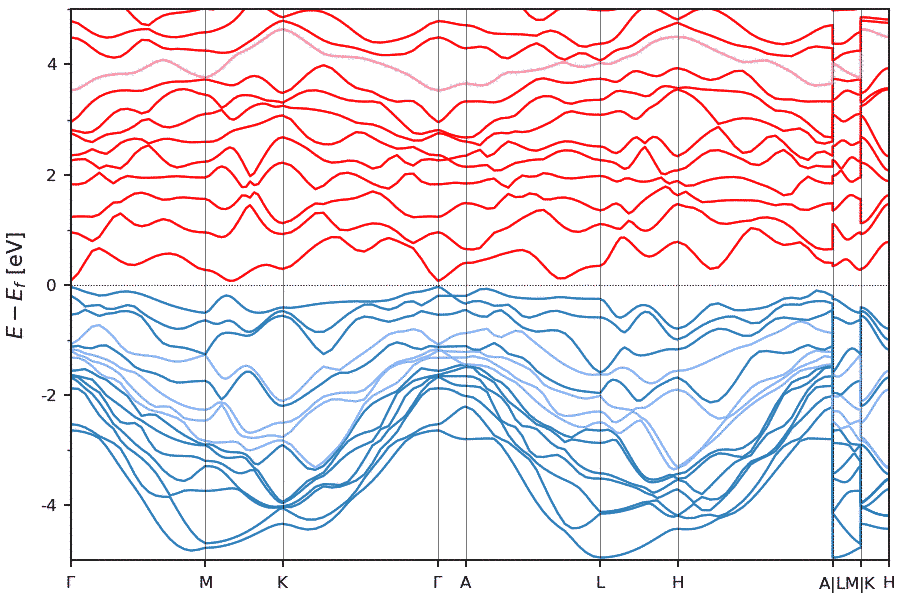} & \includegraphics[width=0.38\textwidth,angle=0]{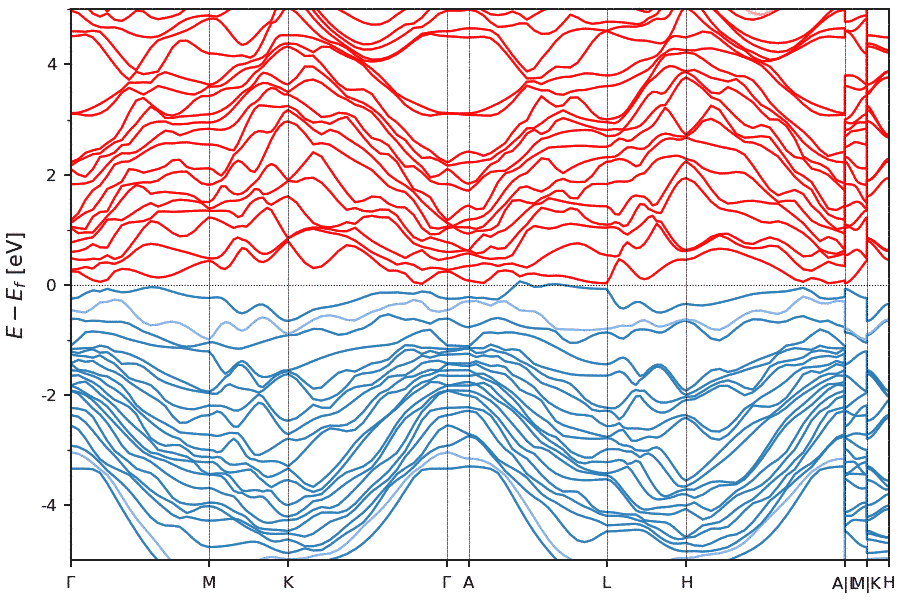}\\
\end{tabular}
\begin{tabular}{c c}
\scriptsize{$\rm{As}_{2} \rm{Ge}_{5} \rm{Te}_{8}$ - \icsdweb{63174} - SG 164 ($P\bar{3}m1$) - SEBR} & \scriptsize{$\rm{Ta}_{2} \rm{C}$ - \icsdweb{409555} - SG 164 ($P\bar{3}m1$) - SEBR}\\
\tiny{ $\;Z_{2,1}=0\;Z_{2,2}=0\;Z_{2,3}=1\;Z_4=1$ } & \tiny{ $\;Z_{2,1}=0\;Z_{2,2}=0\;Z_{2,3}=0\;Z_4=3$ }\\
\includegraphics[width=0.38\textwidth,angle=0]{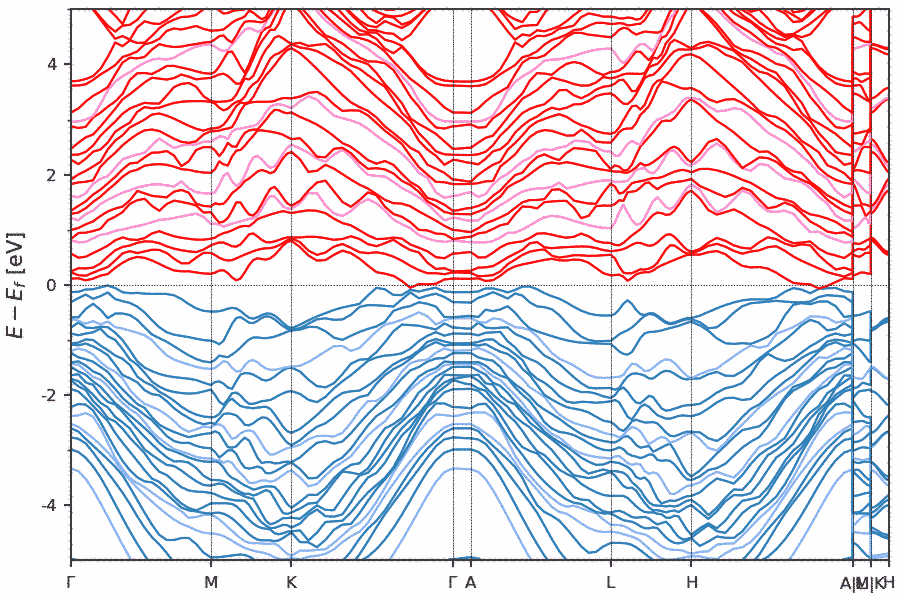} & \includegraphics[width=0.38\textwidth,angle=0]{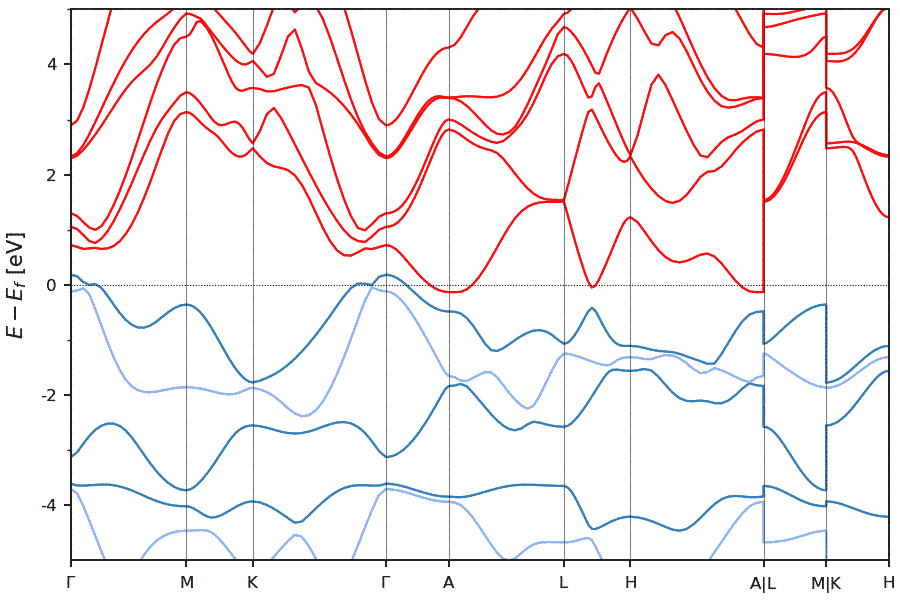}\\
\end{tabular}
\begin{tabular}{c c}
\scriptsize{$\rm{Bi} \rm{Se}$ - \icsdweb{617073} - SG 164 ($P\bar{3}m1$) - SEBR} & \scriptsize{$\rm{Pb}_{4} \rm{Se}_{4}$ - \icsdweb{238502} - SG 225 ($Fm\bar{3}m$) - SEBR}\\
\tiny{ $\;Z_{2,1}=0\;Z_{2,2}=0\;Z_{2,3}=1\;Z_4=0$ } & \tiny{ $\;Z_{2,1}=0\;Z_{2,2}=0\;Z_{2,3}=0\;Z_4=0\;Z_2=0\;Z_8=4$ }\\
\includegraphics[width=0.38\textwidth,angle=0]{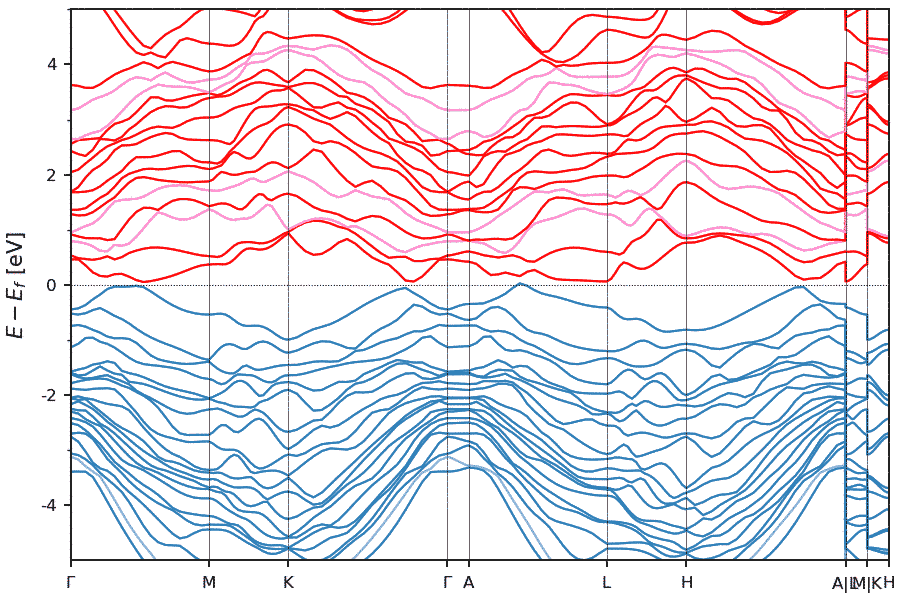} & \includegraphics[width=0.38\textwidth,angle=0]{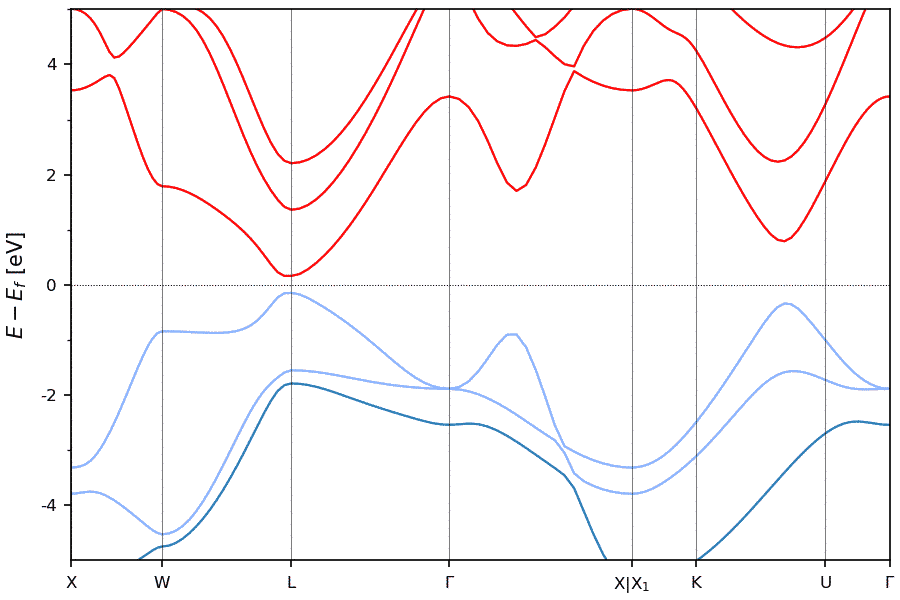}\\
\end{tabular}

\caption{\fragileSEBR{1}}
\label{fig:fragile_SEBR1}
\end{figure}

\begin{figure}[ht]
\centering
\begin{tabular}{c c}
\scriptsize{$\rm{Pb} \rm{S}$ - \icsdweb{250762} - SG 225 ($Fm\bar{3}m$) - SEBR} & \scriptsize{$\rm{Pb} \rm{Te}$ - \icsdweb{648615} - SG 225 ($Fm\bar{3}m$) - SEBR}\\
\tiny{ $\;Z_{2,1}=0\;Z_{2,2}=0\;Z_{2,3}=0\;Z_4=0\;Z_2=0\;Z_8=4$ } & \tiny{ $\;Z_{2,1}=0\;Z_{2,2}=0\;Z_{2,3}=0\;Z_4=0\;Z_2=0\;Z_8=4$ }\\
\includegraphics[width=0.38\textwidth,angle=0]{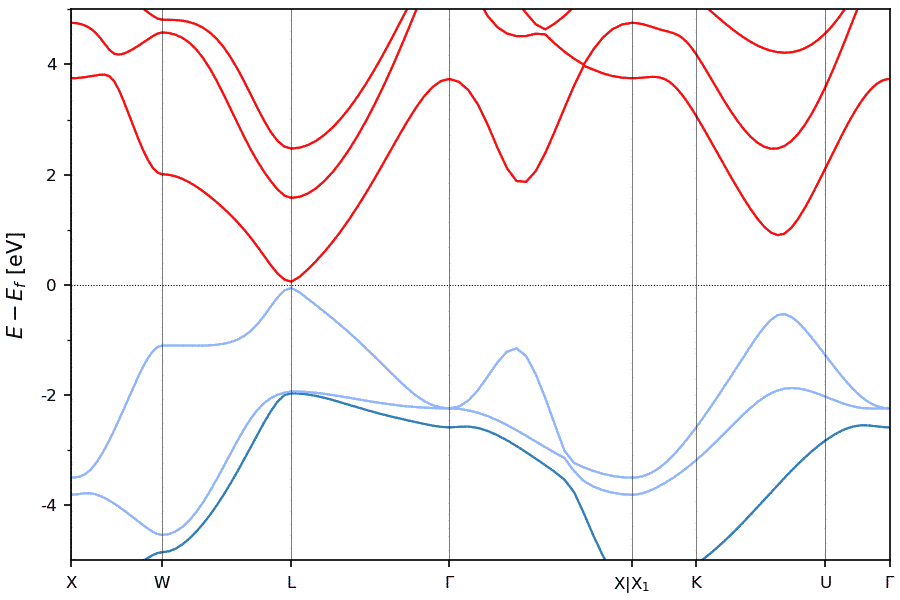} & \includegraphics[width=0.38\textwidth,angle=0]{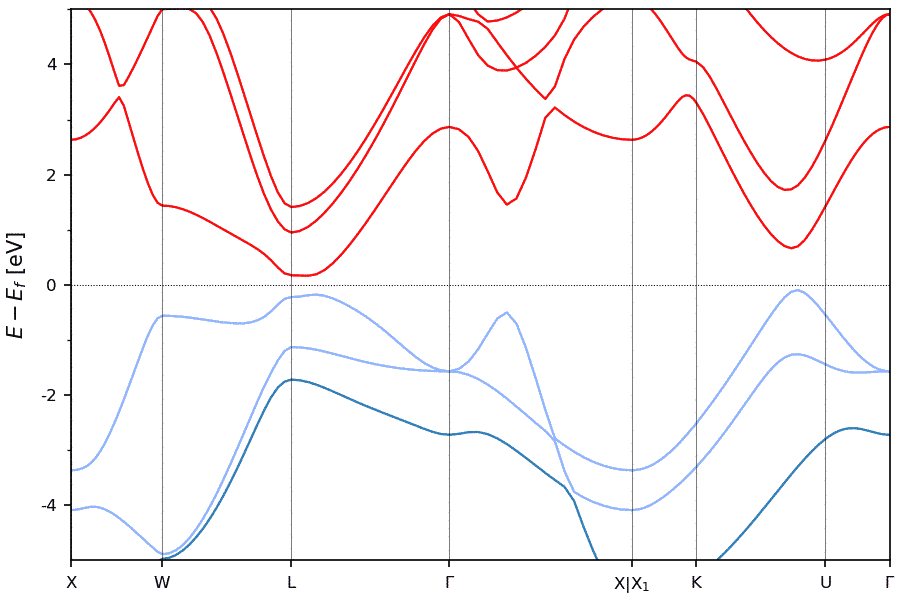}\\
\end{tabular}
\begin{tabular}{c c}
\scriptsize{$\rm{Sn} \rm{S}$ - \icsdweb{651015} - SG 225 ($Fm\bar{3}m$) - SEBR} & \scriptsize{$\rm{Ti} \rm{S}_{2}$ - \icsdweb{72042} - SG 227 ($Fd\bar{3}m$) - SEBR}\\
\tiny{ $\;Z_{2,1}=0\;Z_{2,2}=0\;Z_{2,3}=0\;Z_4=0\;Z_2=0\;Z_8=4$ } & \tiny{ $\;Z_{2,1}=0\;Z_{2,2}=0\;Z_{2,3}=0\;Z_4=1\;Z_2=1$ }\\
\includegraphics[width=0.38\textwidth,angle=0]{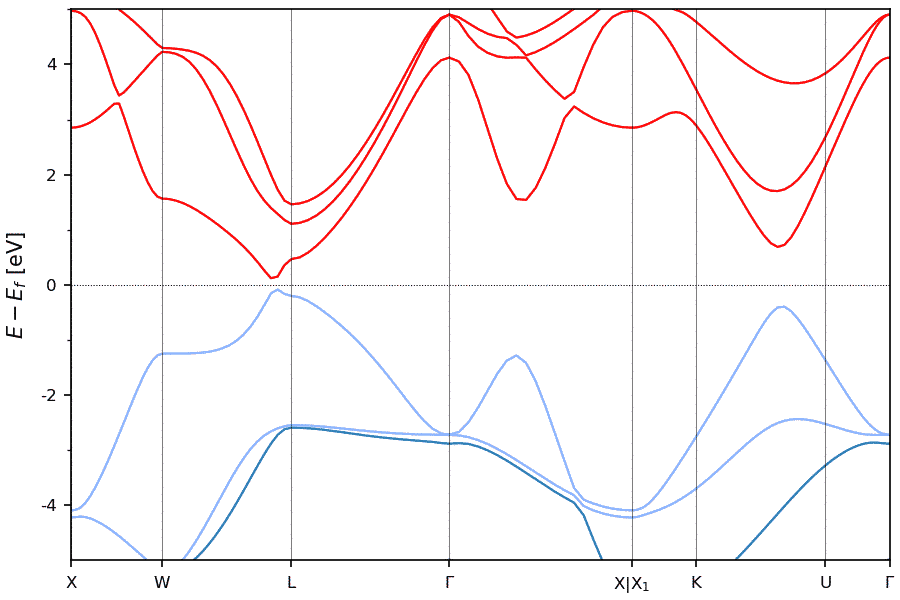} & \includegraphics[width=0.38\textwidth,angle=0]{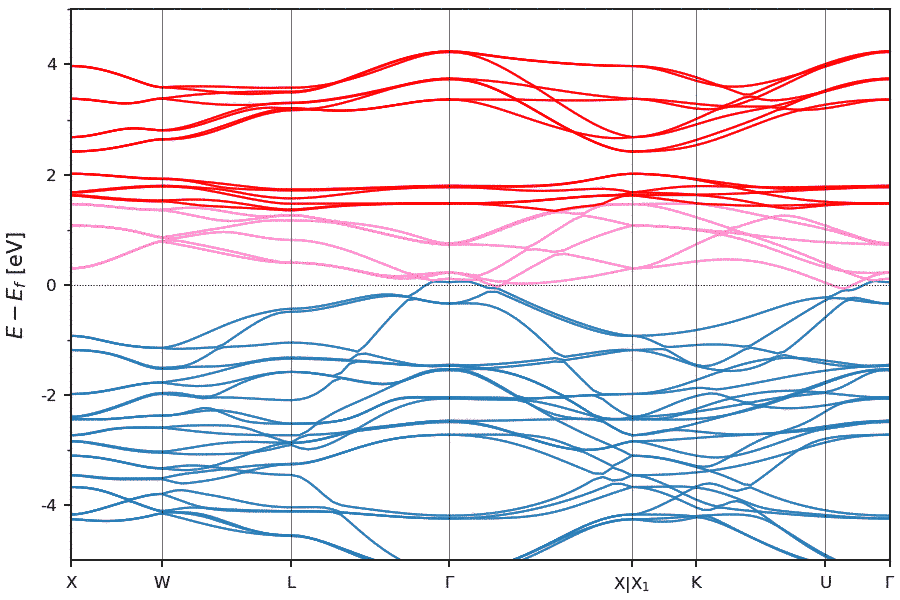}\\
\end{tabular}

\caption{\fragileSEBR{2}}
\label{fig:fragile_SEBR2}
\end{figure}



\begin{figure}[ht]
\centering
\begin{tabular}{c c}
\scriptsize{$\rm{Ta} \rm{Se}_{2}$ - \icsdweb{24313} - SG 164 ($P\bar{3}m1$) - ESFD} & \scriptsize{$\rm{Nb} \rm{Se}_{2}$ - \icsdweb{76576} - SG 164 ($P\bar{3}m1$) - ESFD}\\
\includegraphics[width=0.38\textwidth,angle=0]{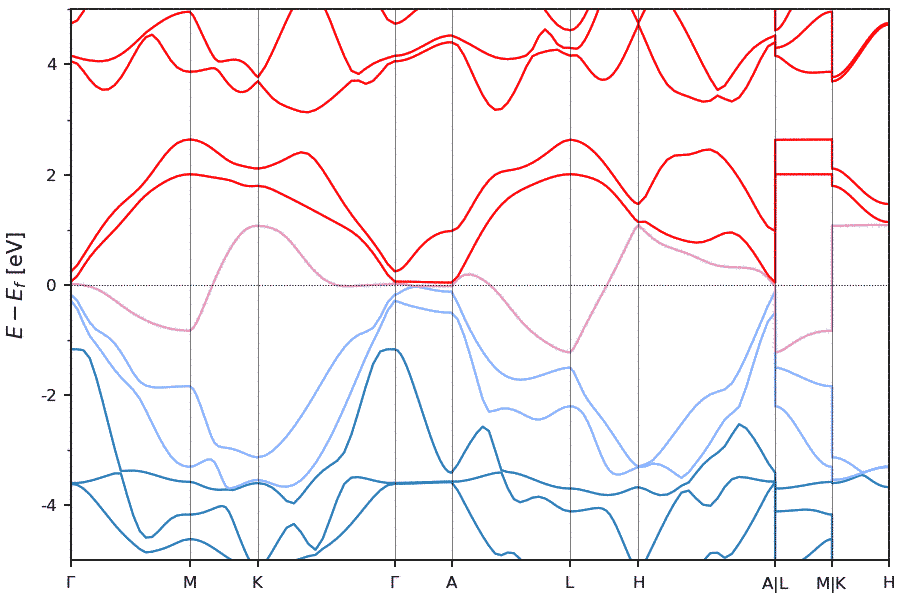} & \includegraphics[width=0.38\textwidth,angle=0]{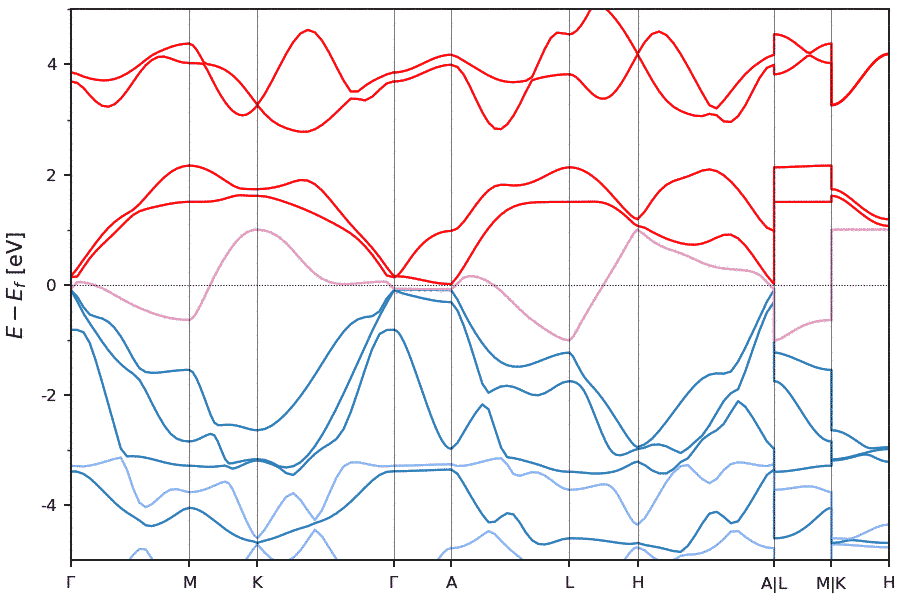}\\
\end{tabular}
\begin{tabular}{c c}
\scriptsize{$\rm{Li} \rm{Cd} \rm{As}$ - \icsdweb{609966} - SG 216 ($F\bar{4}3m$) - ESFD} & \scriptsize{$\rm{Sn} \rm{As}$ - \icsdweb{44063} - SG 225 ($Fm\bar{3}m$) - ESFD}\\
\includegraphics[width=0.38\textwidth,angle=0]{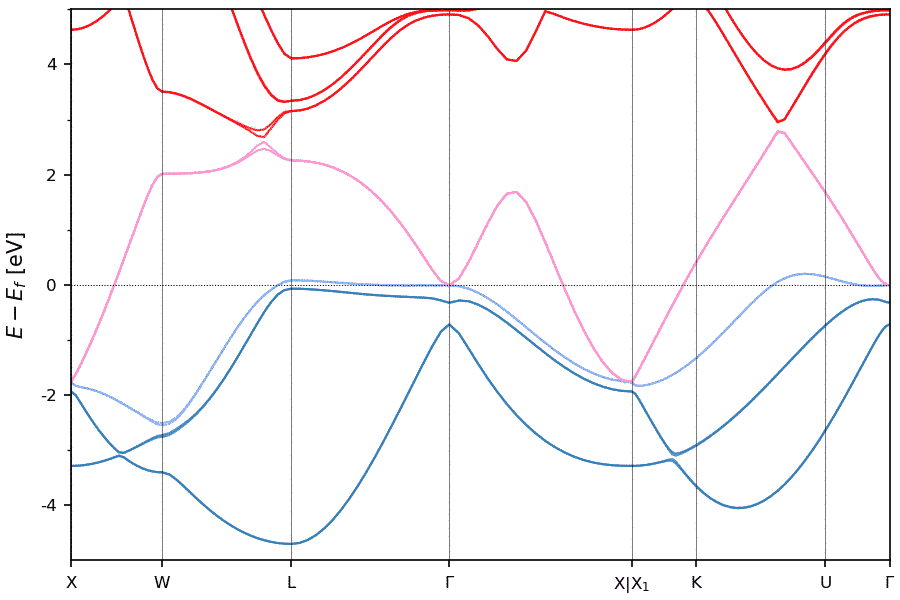} & \includegraphics[width=0.38\textwidth,angle=0]{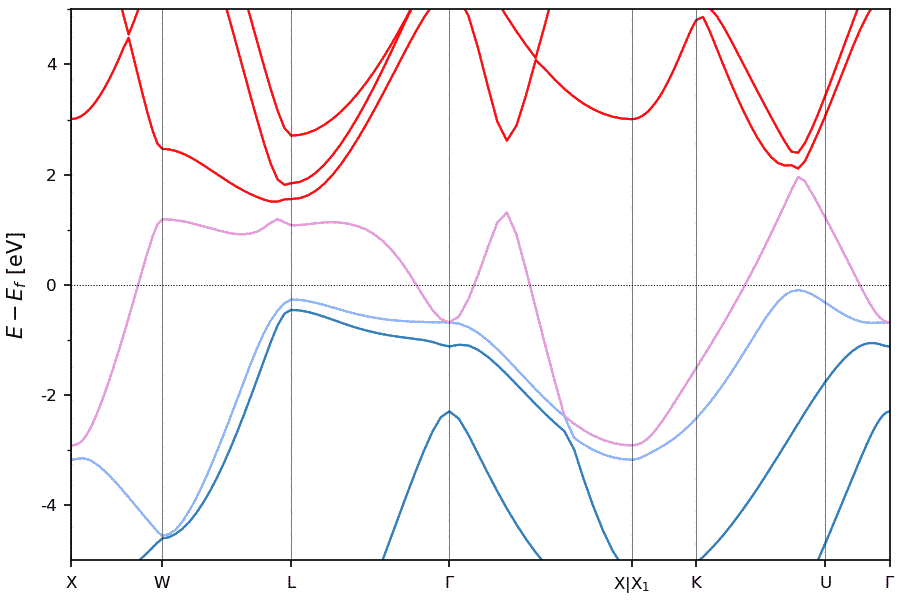}\\
\end{tabular}
\begin{tabular}{c c}
\scriptsize{$\rm{Bi} \rm{Te}$ - \icsdweb{44984} - SG 225 ($Fm\bar{3}m$) - ESFD} & \scriptsize{$\rm{Cd} \rm{Li}_{2} \rm{Ge}$ - \icsdweb{52803} - SG 225 ($Fm\bar{3}m$) - ESFD}\\
\includegraphics[width=0.38\textwidth,angle=0]{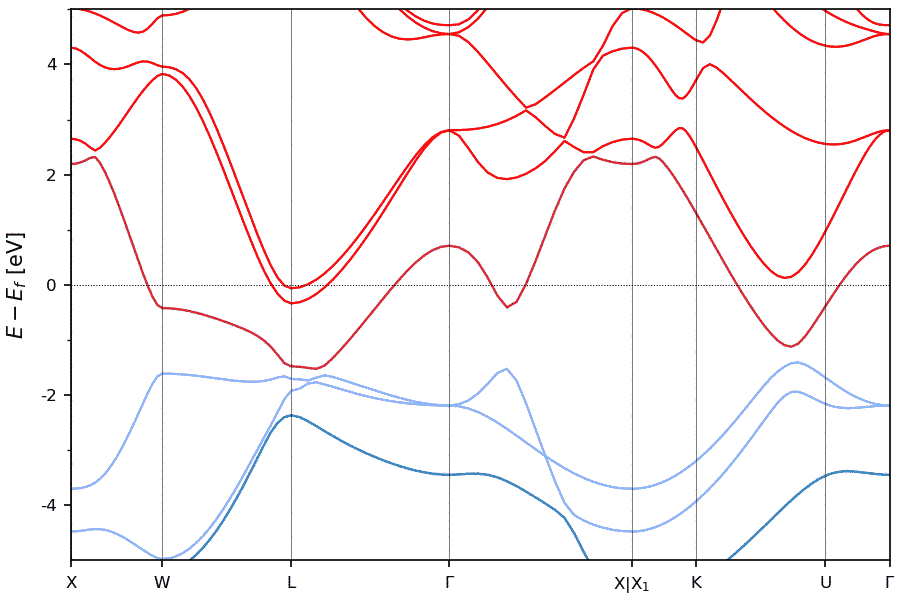} & \includegraphics[width=0.38\textwidth,angle=0]{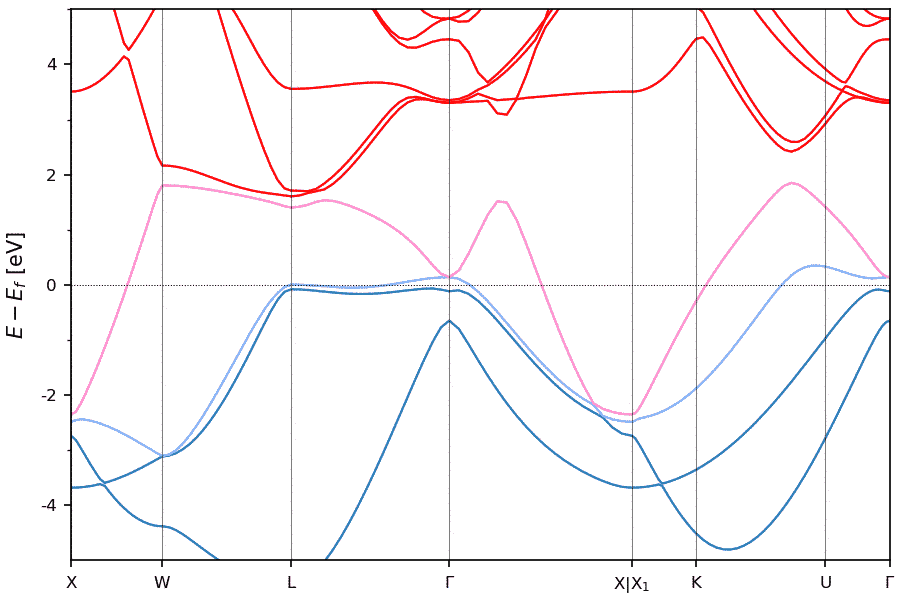}\\
\end{tabular}
\begin{tabular}{c c}
\scriptsize{$\rm{Sn} \rm{P}$ - \icsdweb{77786} - SG 225 ($Fm\bar{3}m$) - ESFD} & \scriptsize{$\rm{Mg}_{2} \rm{Pb}$ - \icsdweb{104846} - SG 225 ($Fm\bar{3}m$) - ESFD}\\
\includegraphics[width=0.38\textwidth,angle=0]{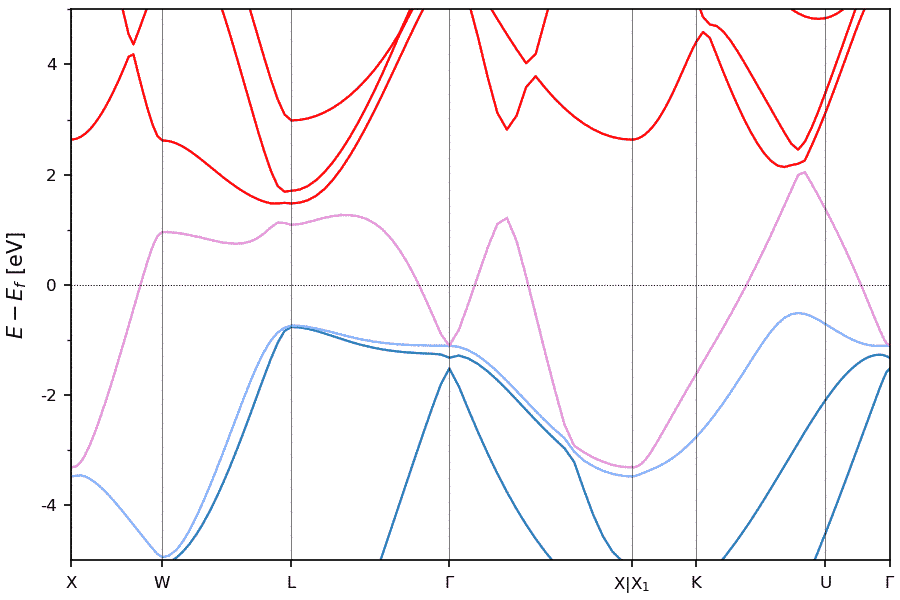} & \includegraphics[width=0.38\textwidth,angle=0]{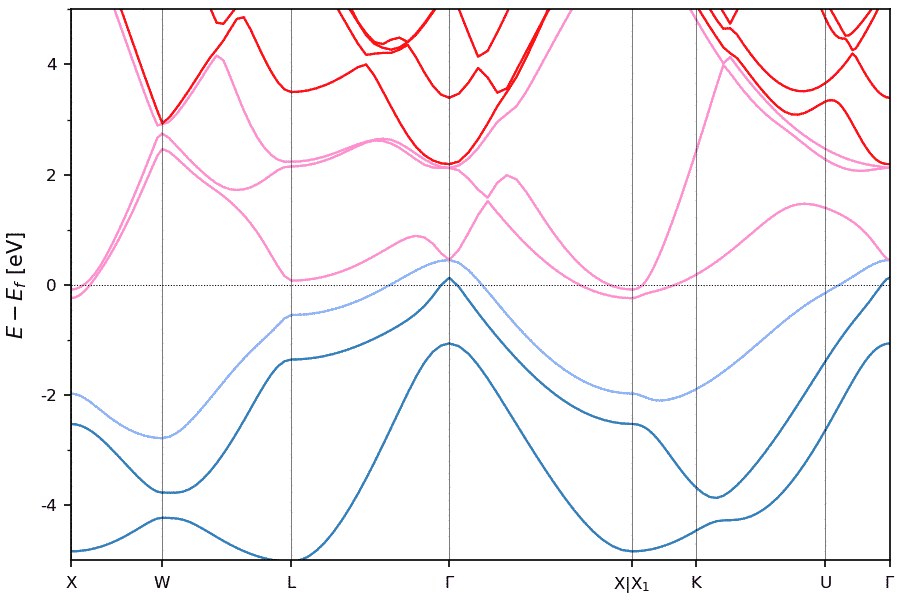}\\
\end{tabular}

\caption{\fragileESFD{1}}
\label{fig:fragile_ESFD1}
\end{figure}

\begin{figure}[ht]
\centering
\begin{tabular}{c c}
\scriptsize{$\rm{Ge}$ - \icsdweb{43422} - SG 227 ($Fd\bar{3}m$) - ESFD} & \scriptsize{$\rm{Bi}_{4} \rm{Rh}$ - \icsdweb{58854} - SG 230 ($Ia\bar{3}d$) - ESFD}\\
\includegraphics[width=0.38\textwidth,angle=0]{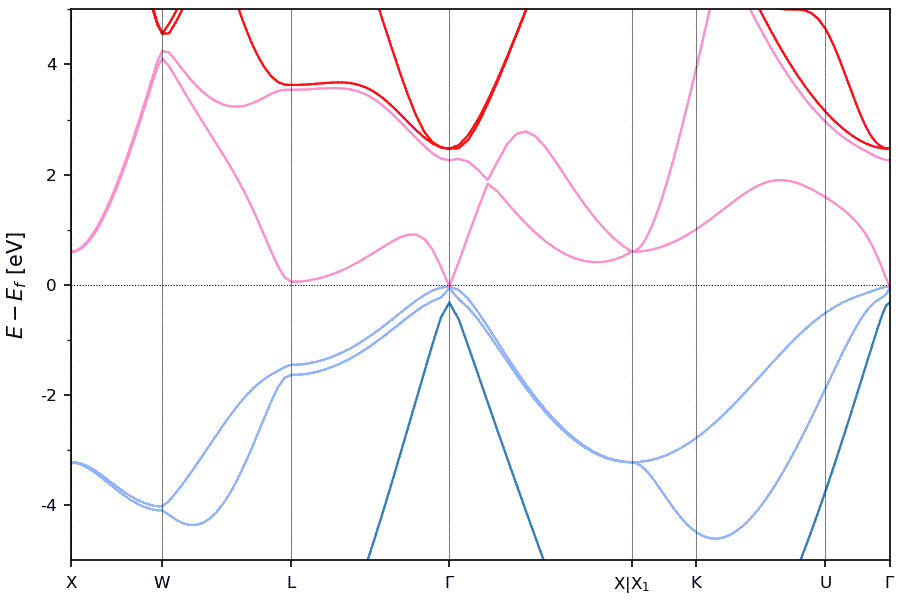} & \includegraphics[width=0.38\textwidth,angle=0]{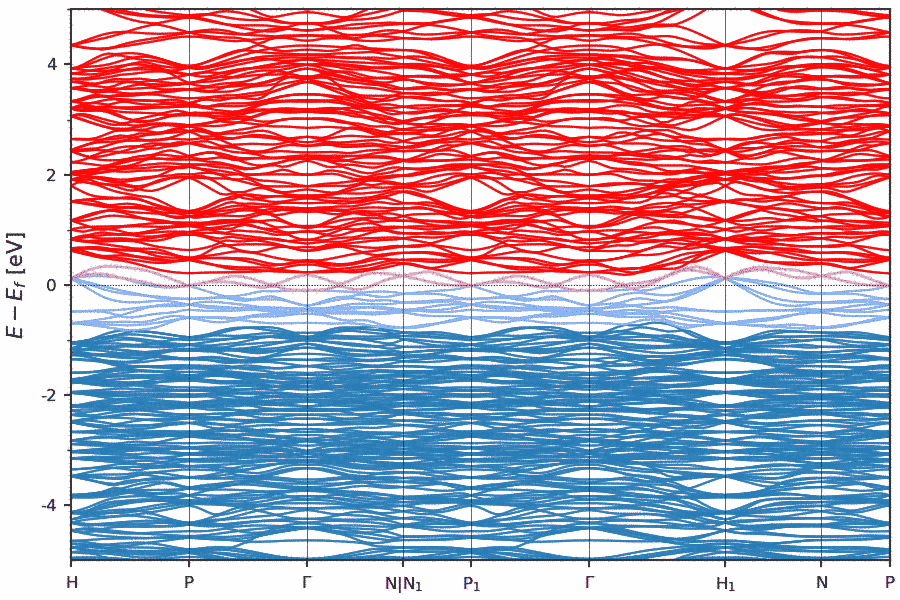}\\
\end{tabular}

\caption{\fragileESFD{2}}
\label{fig:fragile_ESFD2}
\end{figure}


\begin{figure}[ht]
\centering
\begin{tabular}{c c}
\scriptsize{$\rm{Mo} \rm{Ge}_{2}$ - \icsdweb{76139} - SG 139 ($I4/mmm$) - ES} & \scriptsize{$\rm{Sr} \rm{Zn}_{2} \rm{Sb}_{2}$ - \icsdweb{12152} - SG 164 ($P\bar{3}m1$) - ES}\\
\includegraphics[width=0.38\textwidth,angle=0]{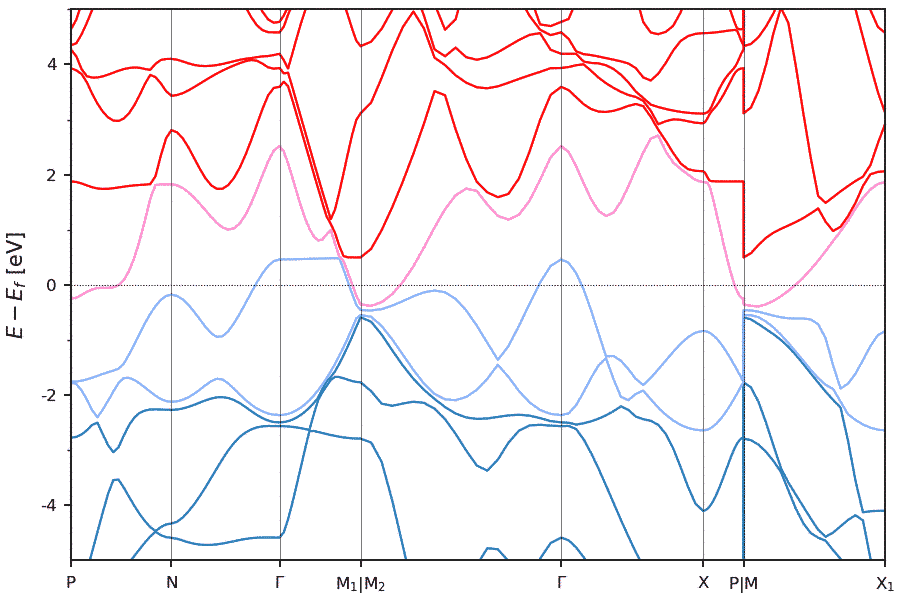} & \includegraphics[width=0.38\textwidth,angle=0]{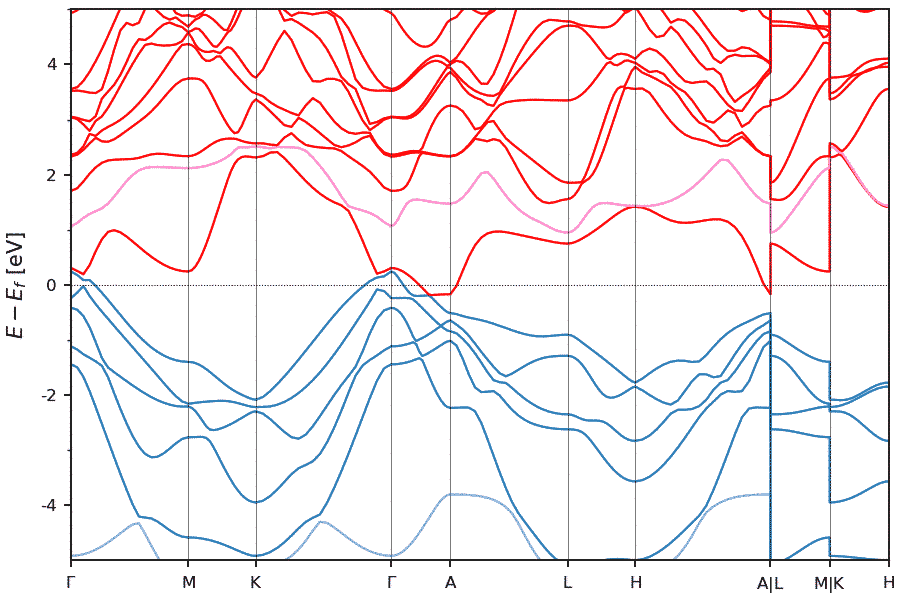}\\
\end{tabular}
\begin{tabular}{c c}
\scriptsize{$\rm{Sc}_{2} \rm{C}$ - \icsdweb{280743} - SG 164 ($P\bar{3}m1$) - ES} & \scriptsize{$\rm{Ta} \rm{N}$ - \icsdweb{105123} - SG 194 ($P6_3/mmc$) - ES}\\
\includegraphics[width=0.38\textwidth,angle=0]{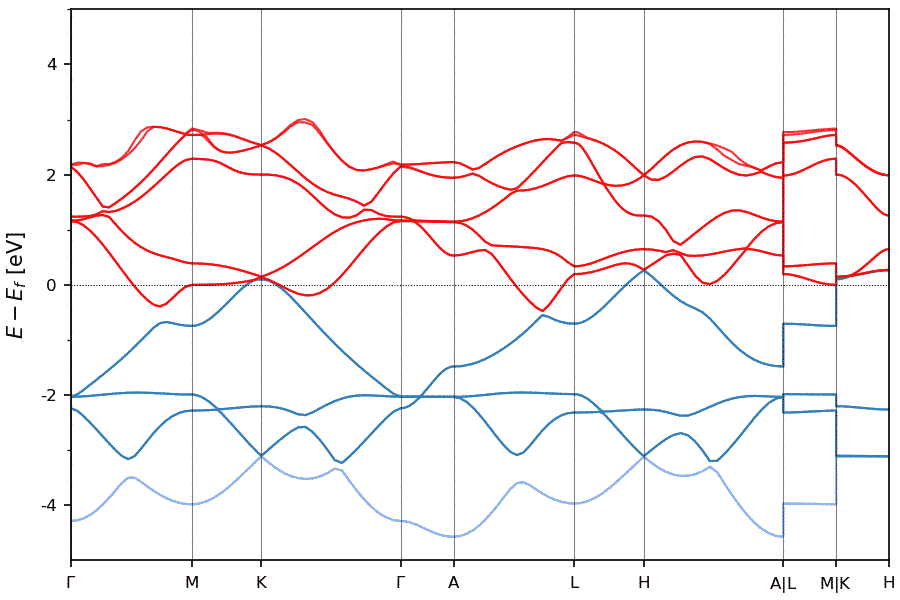} & \includegraphics[width=0.38\textwidth,angle=0]{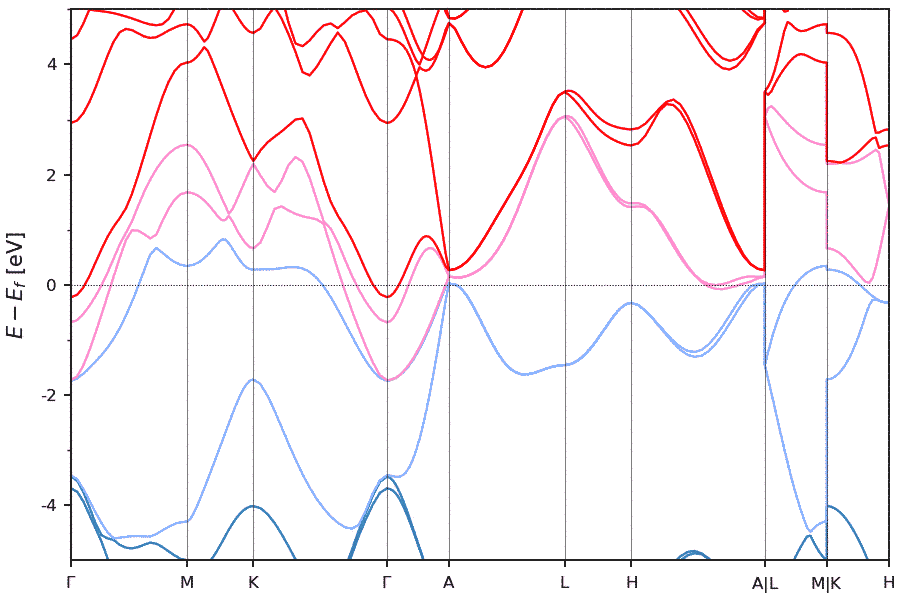}\\
\end{tabular}

\caption{The ES-classified topological semimetals with the simplest bulk Fermi surfaces and well-isolated fragile bands at or close to $E_{F}$.}
\label{fig:fragile_ES}
\end{figure}


\begin{figure}[ht]
\centering
\begin{tabular}{c c}
\scriptsize{$\rm{Cs} \rm{Sn} \rm{I}_{3}$ - \icsdweb{69995} - SG 127 ($P4/mbm$) - LCEBR} & \scriptsize{$\rm{Bi} \rm{Te} \rm{I}$ - \icsdweb{79364} - SG 156 ($P3m1$) - LCEBR}\\
\includegraphics[width=0.38\textwidth,angle=0]{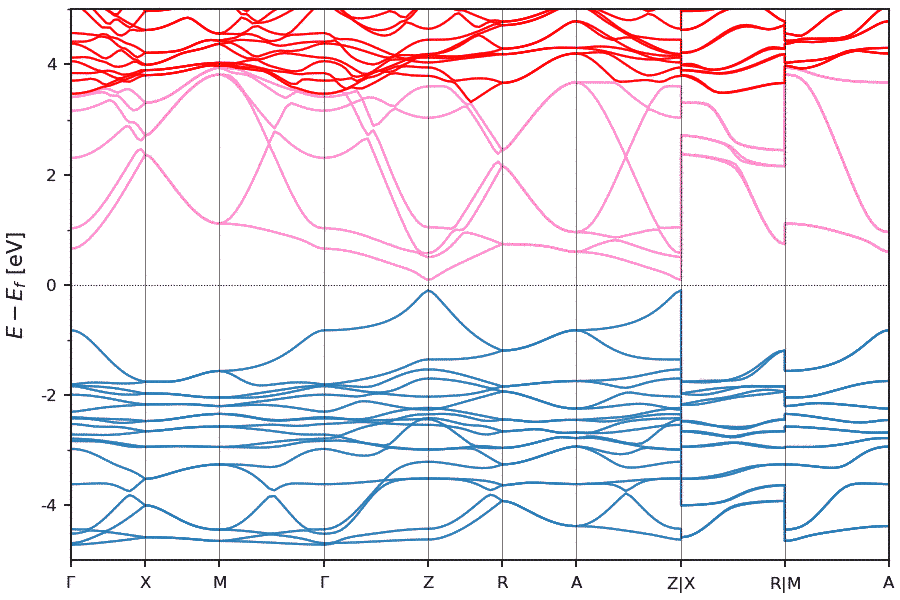} & \includegraphics[width=0.38\textwidth,angle=0]{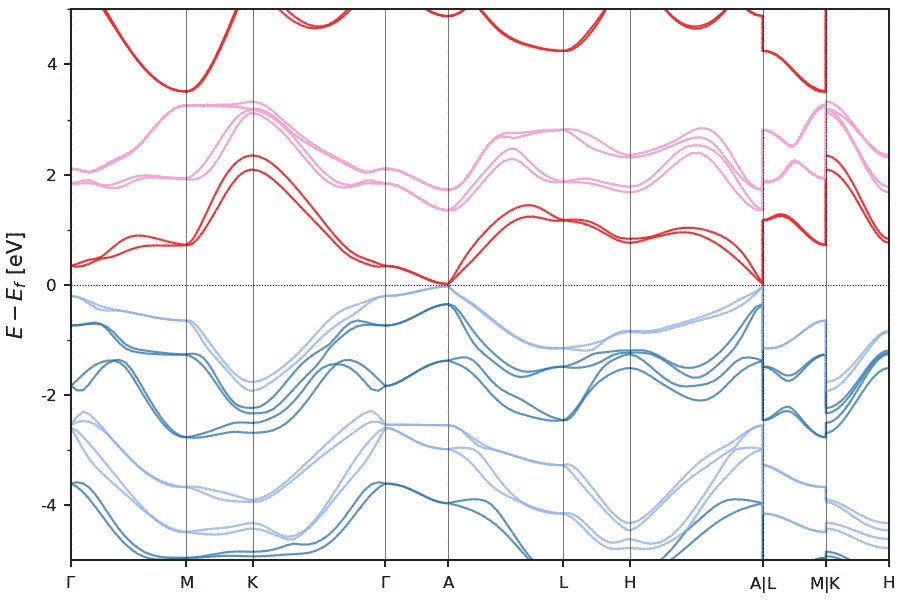}\\
\end{tabular}
\begin{tabular}{c c}
\scriptsize{$\rm{Ca} \rm{Al}_{2} \rm{Si}_{2}$ - \icsdweb{20278} - SG 164 ($P\bar{3}m1$) - LCEBR} & \scriptsize{$\rm{Ti} \rm{S}_{2}$ - \icsdweb{91579} - SG 164 ($P\bar{3}m1$) - LCEBR}\\
\includegraphics[width=0.38\textwidth,angle=0]{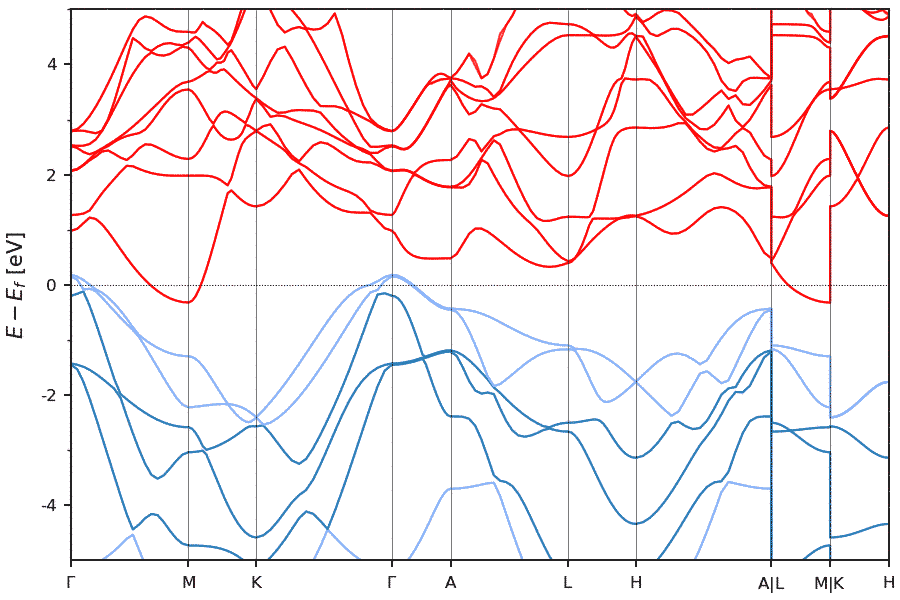} & \includegraphics[width=0.38\textwidth,angle=0]{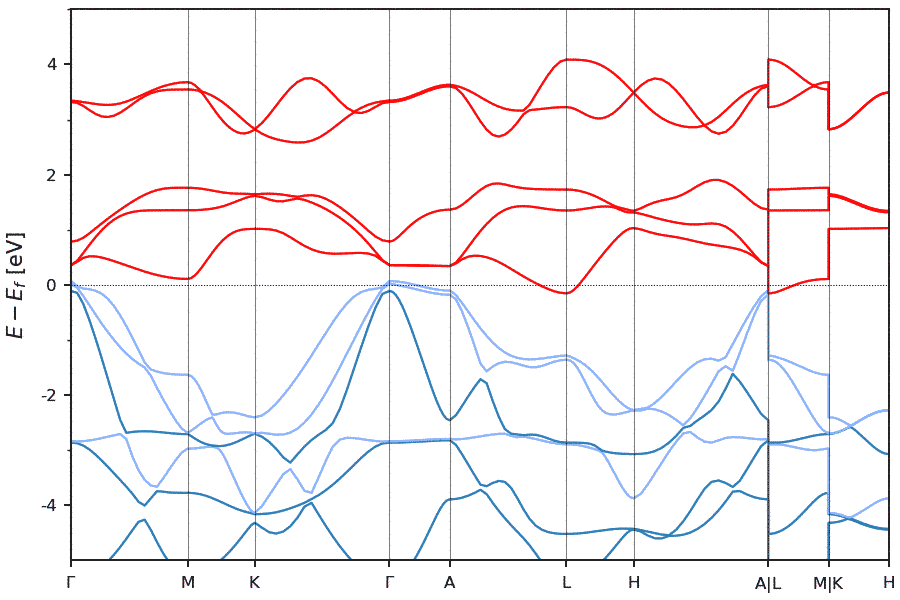}\\
\end{tabular}
\begin{tabular}{c c}
\scriptsize{$\rm{Sr} \rm{Al}_{2} \rm{Si}_{2}$ - \icsdweb{419886} - SG 164 ($P\bar{3}m1$) - LCEBR} & \scriptsize{$\rm{Hf} \rm{Se}_{2}$ - \icsdweb{638899} - SG 164 ($P\bar{3}m1$) - LCEBR}\\
\includegraphics[width=0.38\textwidth,angle=0]{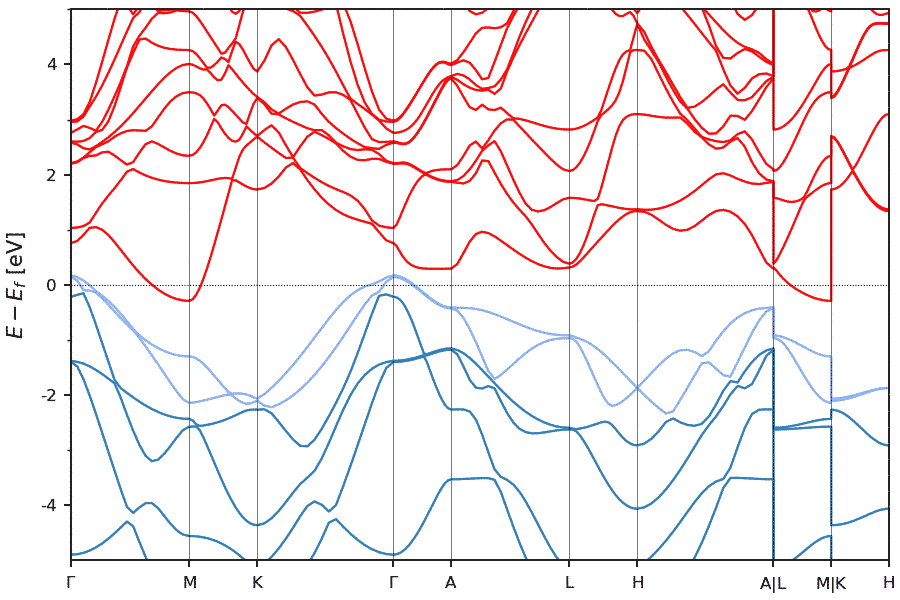} & \includegraphics[width=0.38\textwidth,angle=0]{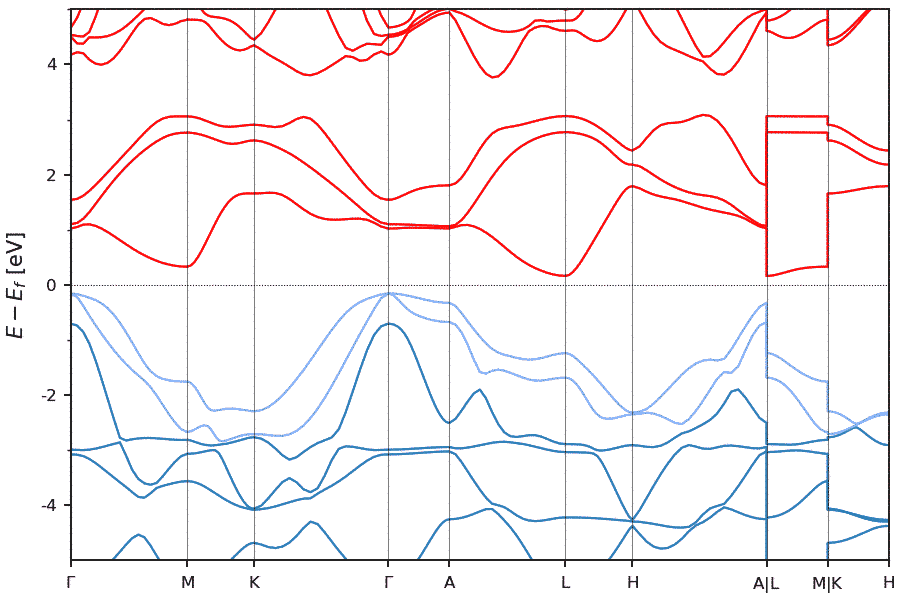}\\
\end{tabular}
\begin{tabular}{c c}
\scriptsize{$\rm{Al}_{5} \rm{C}_{3} \rm{N}$ - \icsdweb{36303} - SG 186 ($P6_3mc$) - LCEBR} & \scriptsize{$\rm{Bi}_{2} (\rm{Ru}_{2} \rm{O}_{7})$ - \icsdweb{73787} - SG 227 ($Fd\bar{3}m$) - LCEBR}\\
\includegraphics[width=0.38\textwidth,angle=0]{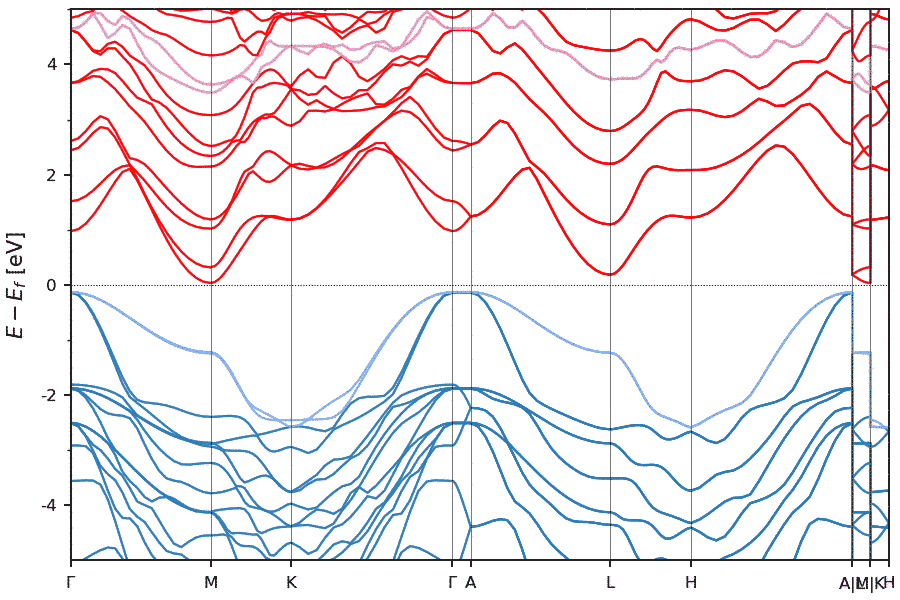} & \includegraphics[width=0.38\textwidth,angle=0]{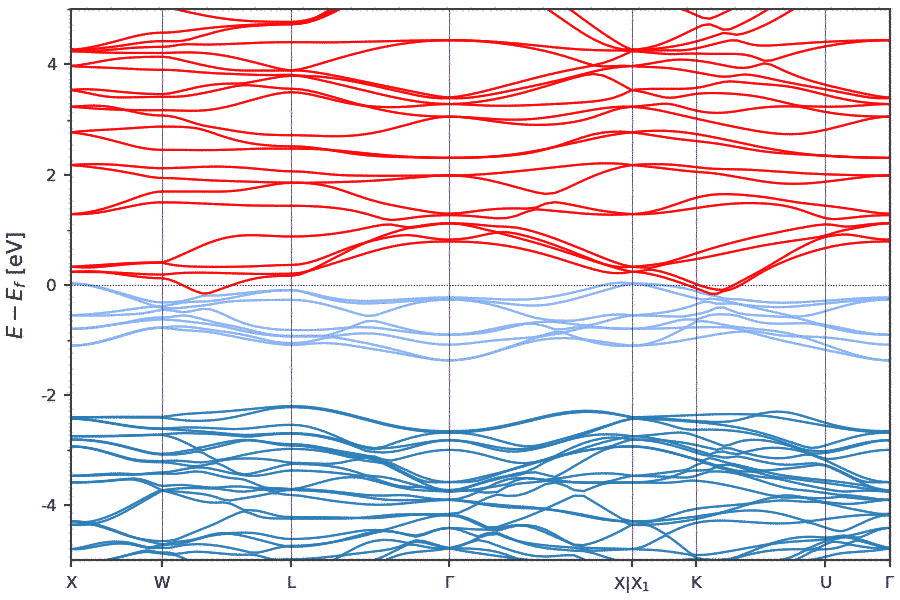}\\
\end{tabular}

\caption{Representative examples of LCEBR-classified insulators with trivial symmetry-indicated strong topology and well-isolated fragile bands at or close to $E_{F}$.}
\label{fig:fragile_LCEBR}
\end{figure}



\end{document}